\documentclass[10pt]{thesis_mw2}

\usepackage[usenames,dvipsnames,svgnames,table]{xcolor} 
\usepackage[obeyspaces,hyphens,spaces]{url}
\usepackage{thesisfrontmatter_HS}
\usepackage{pdfpages}

\definecolor{Blue}{rgb}{0.25, 0.41, 0.88}
\definecolor{Red}{rgb}{0.92,0.,0.}
\definecolor{darkorange}{rgb}{1.0,0.549,0.}
\definecolor{cobalt}{RGB}{44, 98, 120}
\definecolor{Mathematica1}{rgb}{0.368417, 0.506779, 0.709798}
\definecolor{Mathematica2}{rgb}{0.880722, 0.611041, 0.142051}
\definecolor{Mathematica3}{rgb}{0.560181, 0.691569, 0.194885}
\definecolor{Mathematica4}{rgb}{0.922526, 0.385626, 0.209179}
\definecolor{Mathematica5}{rgb}{0.528488, 0.470624, 0.701351}
\definecolor{Mathematica6}{rgb}{0.772079, 0.431554, 0.102387}
\definecolor{Mathematica7}{rgb}{0.363898, 0.618501, 0.782349}
\definecolor{Mathematica8}{rgb}{1, 0.75, 0}
\definecolor{Mathematica9}{rgb}{0.647624, 0.37816, 0.614037}
\definecolor{plotBlue}{RGB}{94, 130, 181}
\definecolor{plotRed}{RGB}{233, 85, 54}
\definecolor{plotGreen}{RGB}{142, 176, 50}
\definecolor{plotPurple}{RGB}{135, 120, 178}
\definecolor{cornellRed}{HTML}{B31B1B}
\definecolor{cornellBlue}{HTML}{0068AC}
\definecolor{cornellGreen}{HTML}{6EB43F}
\definecolor{dullpurple}{rgb}{0.431,0.188,0.534}
\definecolor{darkgreen}{rgb}{0.075,0.302,0.047}
\definecolor{darkergreen}{rgb}{0,0.196,0.125}
\definecolor{darkergreen2}{rgb}{0,0.294,0.188}
\definecolor{dullred}{rgb}{0.706,0.208,0.192}
\definecolor{darkred}{rgb}{0.545,0,0}
\definecolor{antiquefuchsia}{rgb}{0.57, 0.36, 0.51}
\definecolor{MaroonC}{rgb}{0,0.502,0.502}
\definecolor{dullblue}{rgb}{0,0.298,0.49}
\definecolor{blue3}{RGB}{31, 119, 180}
\definecolor{red3}{RGB}{	214, 39, 40}
\definecolor{orange3}{RGB}{255, 127, 14}
\definecolor{green3}{RGB}{44, 160, 44}

\usepackage{ifthen}
\usepackage{fancyhdr}
\usepackage[colorlinks=true
,urlcolor=dullblue
,anchorcolor=blue3
,citecolor=dullblue
,filecolor=blue3
,linkcolor=darkred
,menucolor=blue3
,pagecolor=blue3
,linktocpage=true
,pdfproducer=medialab
]{hyperref}
\usepackage{booktabs}
\usepackage[english, american, dutch]{babel}
\usepackage{amsmath, amssymb, amsbsy, amstext, amsthm, simplewick}
\usepackage{cleveref}
\usepackage{graphicx}
\usepackage{amsfonts}
\usepackage{enumitem}
\usepackage{upgreek}
\usepackage{framed}
\usepackage{tensor}
\usepackage{pifont}
\usepackage{latexsym, mathrsfs}
\usepackage{array}
\usepackage{xspace}
\usepackage{longtable}
\usepackage{multirow}
\usepackage{cases}
\usepackage{empheq}

\usepackage{bm}
\usepackage{amsmath}
\usepackage{bbm}
\usepackage{appendix}

\usepackage{braket}
\usepackage{yfonts}

\usepackage[load-configurations=astronomy, range-units=brackets, range-phrase=-, per-mode=reciprocal, mode=math]{siunitx}
\usepackage{threeparttable}
\usepackage{subfig}

\usepackage{ragged2e}

\usepackage{hhline}
\usepackage{array}
\newcolumntype{P}[1]{>{\centering\arraybackslash}p{#1}}
\usepackage{booktabs}
\usepackage{tikz}
\usetikzlibrary{calc,shadings,patterns,tikzmark,fadings}

\newcolumntype{C}[1]{>{\centering\let\newline\\\arraybackslash\hspace{0pt}}m{#1}}

\def\d{{\rm d}}
\def\l{{\ell}}

\def\r{{\bf r}}
\def\R{{\bf R}}

\def\p{{\bf p}}

\newcommand{\es}{\hspace{0.5pt}}

\makeatletter
\newlength{\apb@width}
\newcommand{\autoparbox}[2][c]{\settowidth{\apb@width}{#2}\parbox[#1]{\apb@width}{#2}}

\makeatother

\numberwithin{equation}{section}

\def\beq{\begin{equation}}
\def\eeq{\end{equation}}

\def\bea{\begin{eqnarray}}
\def\eea{\end{eqnarray}}

\def\d{{\rm d}}

\def\beq{\begin{equation}}
\def\eeq{\end{equation}}
\def\bea{\begin{eqnarray}}
\def\eea{\end{eqnarray}}

\def\d{{\rm d}}

\def\d{{\rm d}}
\def\l{{\ell}}

\def\tmp{bbh}
\def\gen{g}

\def\tmp{bbh}
\def\gen{g}

\newcommand{\ud}{\mathrm{d}}
\newcommand{\lab}[1]{{\mathrm{#1}}}
\newcommand{\mb}[1]{{\mathbf{#1}}}
\newcommand{\minus}{{\scalebox{0.75}[1.0]{$-$}}}
\newcommand{\sminus}{{\scalebox{0.5}[0.85]{$-$}}}
\newcommand{\indlab}[1]{{\scriptscriptstyle{(#1)}}}
\newcommand{\indlac}[1]{{#1}}

\def\R{{\bf{R}}}
\def\r{{\bf{r}}}

\DeclareRobustCommand{\SkipTocEntry}[4]{}

\usepackage{colortbl}
\definecolor{blue2}{cmyk}{1, 0.1, 0.1, 0}

\definecolor{pyBlue}{RGB}{31, 119, 180}
\definecolor{pyRed}{RGB}{214, 39, 40}
\definecolor{pyGreen}{RGB}{44, 160, 44}
\definecolor{pyBlue2}{RGB}{0, 111, 237}
\definecolor{pyRed2}{RGB}{224, 52, 36}

\newcommand{\red}[1]{\textcolor{pyRed}{#1}}
\newcommand{\blue}[1]{\textcolor{pyBlue}{#1}}

\newcommand{\green}[1]{\textcolor{pyGreen}{#1}}

\makeatletter
\def\Ddots{\mathinner{\mkern1mu\raise\p@
\vbox{\kern7\p@\hbox{.}}\mkern2mu
\raise4\p@\hbox{.}\mkern2mu\raise7\p@\hbox{.}\mkern1mu}}
\makeatother

\newcommand{\res}{\Omega} 
\newcommand{\Or}{\Omega_r}

\DeclareRobustCommand\mySingularity{\tikz[baseline=-3pt]{ \draw[black] (0ex,0ex) circle (0.8ex); \fill[black] (0, 0) circle (0.4ex);}}

\usepackage{chngcntr}
\usepackage{etoolbox}

\AtBeginEnvironment{subappendices}{%
}

\thesisfont{cmr} 
\thesissectionstyle{\rmfamily\bfseries}
\thesissectionsizes{\Large}{\large}{}

          \usepackage[explicit]{titlesec}
          \usepackage{lmodern}
          \newlength\chapnumb
          \newlength\chapnumbless
          \titleformat{\chapter}[block]
          {\normalfont\sffamily\scshape}{}{0pt}
          {\parbox[b]{\chapnumb}{
             \fontsize{50}{0}\selectfont\thechapter}
            \parbox[b]{\dimexpr\textwidth-\chapnumb\relax}{
              \raggedleft
				  \hfill{\huge#1 \\[5pt]}
              \rule{\dimexpr\textwidth-\chapnumb\relax}{0.4pt}
              }}      
         \titleformat{name=\chapter,numberless}[block]
          {\normalfont\sffamily\scshape}{}{0pt}
          {\parbox[b]{\chapnumbless}{
             \mbox{}}
            \parbox[b]{\dimexpr\textwidth-\chapnumbless\relax}{
              \raggedleft
              \hfill{\Huge#1}\\
              \rule{\dimexpr\textwidth-\chapnumbless\relax}{0.4pt}}}
              
\usepackage{fix-cm}
\newcommand\HUGE{\@setfontsize\Huge{38}{47}}

\title{{\Huge \sffamily Probing Particle Physics with Gravitational Waves \\[5pt]}}
\thesiscolofon{{    \noindent This work has been accomplished at the Institute for Theoretical Physics (ITFA) of the   University of Amsterdam (UvA). This research is funded by the Netherlands Organisation for Scientific Research (NWO).

\vfill
\par\vskip 12cm

\noindent\copyright{} Horng Sheng Chia, 2020\\[1em]
   \begin{minipage}[b]{\textwidth}
All rights reserved. Without limiting the rights under copyright reserved above, no part of this book may be reproduced, stored in or introduced into a retrieval system, or transmitted, in any form or by any means (electronic, mechanical, photocopying, recording or otherwise) without the written permission of both the copyright owner and the author of the book.
      \end{minipage}    
\vfill
      }}
\author{Horng Sheng Chia}
\placeofbirth{Sabah}
\theday{vrijdag}
\date{16 oktober 2020}
\thetime{11.00}

\rectormagnificus{prof. dr. ir. K.I.J. Maex}
\promotor{\committeeentry{prof. dr. D. Baumann}{Universiteit van Amsterdam}}
\copromotor{\committeeentry{dr. J.P. van der Schaar}{Universiteit van Amsterdam}}
\othercomitteemembers{ 
\committeeentry{prof. dr. G. Bertone}{Universiteit van Amsterdam},
\committeeentry{prof. dr. V. Cardoso}{Universidade T{\'e}cnica de Lisboa},
\committeeentry{prof. dr. C. Van Den Broeck}{Universiteit Utrecht},
\committeeentry{dr. S. Nissanke}{Universiteit van Amsterdam},
\committeeentry{dr. T. Hinderer}{Universiteit van Amsterdam},
}
\department{\small Faculteit der Natuurwetenschappen, Wiskunde en
Informatica}

\usepackage{pifont}
\usepackage[nottoc]{tocbibind}

\newcommand{\committeeentry}[2]{\begin{tabular*}{0.78\textwidth}{p{0.42\textwidth}l}#1 & #2\end{tabular*}}

\begin{document}
\selectlanguage{english}
{
\includepdf[fitpaper]{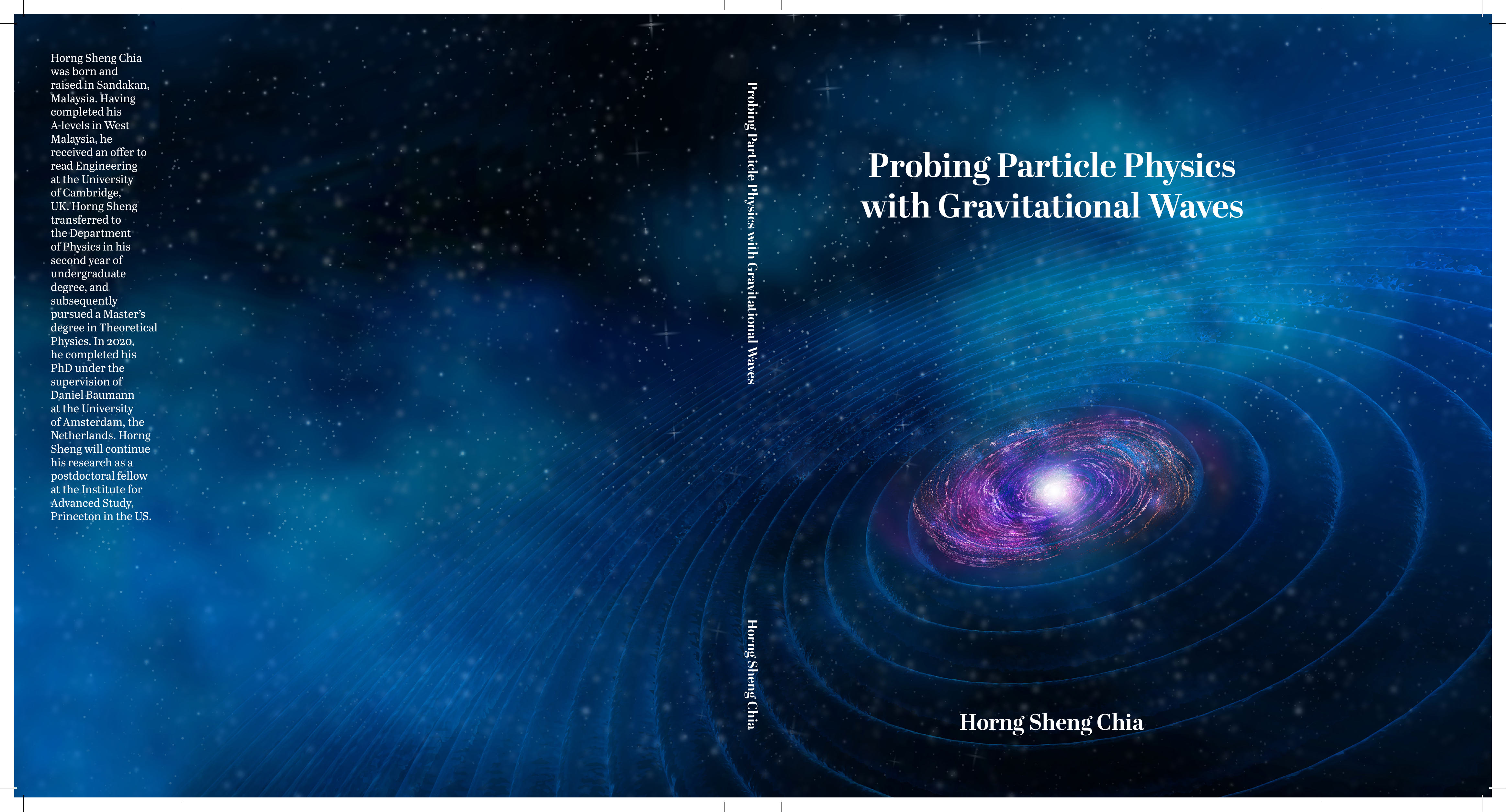}
\let\cleardoublepage\clearpage
\frontmatter
	\maketitle
	}
\basedon{Publications}{
{\scshape This thesis is based on the following publications:} \vskip 20pt

\begin{itemize}[]

{\fontfamily{cmr}\selectfont

\item [\cite{Baumann:2019eav}]

D. Baumann, H. S. Chia, J. Stout, and L. ter Haar, ``The Spectra of Gravitational Atoms", 
\href{https://iopscience.iop.org/article/10.1088/1475-7516/2019/12/006}{\textit{JCAP} \textbf{12} (2019) 006}, \href{https://arxiv.org/abs/1908.10370}{\ttfamily arXiv:1908.10370 [gr-qc]}. \vskip 5pt

Presented in Chapter {\bf 4}.\vskip 10pt

\item[\cite{Baumann:2018vus}]

D. Baumann, H. S. Chia, and R. A. Porto, ``Probing Ultralight Bosons with Binary Black Holes", \href{https://journals.aps.org/prd/abstract/10.1103/PhysRevD.99.044001}{\textit{Phys.~Rev.~D} \textbf{99}, 044001 (2019)}, \href{https://arxiv.org/abs/1804.03208}{\ttfamily arXiv:1804.03208 [gr-qc]}. Selected for Editors' Suggestion and \href{https://physics.aps.org/synopsis-for/10.1103/PhysRevD.99.044001}{Featured in Physics}. \vskip 5pt

Presented in Chapters {\bf 5} and {\bf 6}.\vskip 10pt

\item[\cite{Baumann:2019ztm}]

D. Baumann, H. S. Chia, R. A. Porto, and J. Stout, ``Gravitational Collider Physics", \href{https://journals.aps.org/prd/abstract/10.1103/PhysRevD.101.083019}{\textit{Phys.~Rev.~D} \textbf{101}, 083019 (2020)}, \href{https://arxiv.org/abs/1912.04932}{\ttfamily arXiv:1912.04932 [gr-qc]}. \vskip 5pt

Presented in Chapters {\bf 5} and {\bf 6}.\vskip 10pt

\item[\cite{Chia:2020psj}]

H. S. Chia and T. D. P. Edwards, ``Searching for General Binary Inspirals with Gravitational Waves", \href{https://iopscience.iop.org/article/10.1088/1475-7516/2020/11/033} {\textit{JCAP} \textbf{11} (2020) 033},  \href{https://arxiv.org/abs/2004.06729}{\ttfamily arXiv:2004.06729 [astro-ph.HE]}.  \vskip 5pt

Presented in Chapter {\bf 7}.

}

\end{itemize}

\newpage

}

\setcounter{tocdepth}{2}
\tableofcontents

\mainmatter

\chapter{Introduction}

The direct detection of gravitational waves has offered us a new window onto our Universe~\cite{Abbott:2016blz, TheLIGOScientific:2017qsa}. These ripples of spacetime, generated by some of the most powerful astrophysical events, carry vital information about physics and phenomena that are typically inaccessible by other observational means. This breakthrough has come at a time when many challenges in foundational aspects of physics, cosmology, and astrophysics remain unresolved. For instance, despite decades of active research, we still do not know what dark matter is~\cite{Aghanim:2018eyx}. In this thesis, I explore novel ways of probing new physics beyond the Standard Model of particle physics using current and future gravitational-wave observations.

\vskip 2pt 

Over the past several years, the LIGO and Virgo detectors have observed the gravitational waves emitted by multiple binary black hole and binary neutron star systems~\cite{LIGOScientific:2018mvr, Venumadhav:2019lyq, Nitz:2019hdf}. These detections have advanced our understanding of astrophysics and, at the same time, offer a wealth of new scientific opportunities. For example, by observing the gravitational waves of binary black holes, previously-unknown populations of astrophysical black holes were unveiled, with masses that are heavier than those expected from X-ray binaries~\cite{Remillard:2006fc}. The observed  binary signals have also motivated rapid advances in the field of ``precision gravity"~\cite{Porto:2016zng, Blanchet:2006zz, Porto:2016pyg, Levi:2018nxp, Schafer:2018kuf, Bern:2019crd}, which aims to construct highly-accurate template waveforms in order to reliably extract binary signals from noisy data. Furthermore, the simultaneous observations of gravitational waves and electromagnetic radiation from merging neutron stars inaugurated the field of multi-messenger astronomy~\cite{TheLIGOScientific:2017qsa, Cowperthwaite:2017dyu, Troja:2017nqp}. These neutron star signals have provided strong evidences for their role in manufacturing Nature's heavy elements, such as gold and platinum~\cite{Kasen:2017aa, Pian:2017aa}; in the future, they will offer independent measurements of the expansion rate of our Universe~\cite{Schutz:1986aa}. The examples given here represent only a few of the many achievements that have been made by the current network of gravitational-wave detectors. Moreover, next generation detectors will observe over a greater range of frequencies and with much better sensitivities, therefore having the potential to reveal much more about our Universe. 

\vskip 8pt

While it is clear that gravitational-wave observations will continue to transform astrophysics in many ways, it is less obvious how they can advance our understanding of particle physics. In the past, searches for new physics have been dominated by the high-energy frontier, whereby heavy particles are excited through scattering processes in particle colliders. While this way of probing new physics has been very successful, culminating in the development of the Standard Model~\cite{Aad:2012tfa, Chatrchyan:2012xdj, Tanabashi:2018oca}, it relies on the new particles having appreciable couplings with the particles involved in the collision processes. Traditional collider experiments are therefore blind to ``dark sectors" that interact very weakly with ordinary matter, even if the putative new degrees of freedom are light. However, by virtue of the equivalence principle, all forms of matter and energy must interact through gravity. Gravitational waves are therefore excellent probes of physics in the dark sector, as long as the associated time-dependent quadrupole moments are sufficiently large for the signals to be realistically detectable. This is the case if new types of compact objects exist in the dark sector and are present in inspiraling binary systems. Indeed, as I will elaborate shortly, various types of dark compact objects naturally arise in proposed extensions to the Standard Model. As long as these objects can be formed and are stable over astrophysical timescales, their impacts on the dynamics of binary systems and the associated gravitational waves could provide vital clues about new physics that is otherwise hard to detect. This way of probing new physics is complementary to other probes of particle physics in the \textit{weak-coupling frontier}~\cite{Essig:2013lka, Berlin:2018bsc}, which broadly seeks to develop creative methods to detect particles that couple very weakly to ordinary matter.

\vskip 2pt

One of the most well-motivated classes of particles at the weak-coupling frontier are ultralight bosons. These bosons are attractive examples of new physics because they are excellent candidates for dark matter and may resolve some outstanding problems of the Standard Model~\cite{Peccei:1977hh, Weinberg:1977ma, Wilczek:1977pj, Svrcek:2006yi, Arvanitaki:2009fg, Acharya:2010zx, Cicoli:2012sz, delAguila:1988jz, Goodsell:2009xc, Camara:2011jg}. These particles are characterized by their extremely small masses, corresponding to Compton wavelengths that can be of the order of astrophysical length scales. Remarkably, if the Compton wavelengths of these fields are comparable to or larger than the sizes of rotating black holes, they can be spontaneously amplified by extracting the energies and angular momenta of these black holes. This amplification mechanism, commonly known as \textit{black hole superradiance}~\cite{Zeldovich:1971a, Zeldovich:1972spj, Starobinsky:1973aij, Starobinsky:1974spj, Bekenstein:1973mi}, generates clouds of boson condensates that are gravitationally bounded to the black holes at their centers~\cite{Arvanitaki:2009fg, Arvanitaki:2010sy}. Since these bound states resemble the proton-electron structure of the hydrogen atom, they are often called ``gravitational atoms." This analogy is in fact not purely qualitative; rather, the equations that govern both types of atoms are actually mathematically identical.\footnote{For example, in the hydrogen atom, the electron wavefunction is bounded to the proton at its center through the $1/r$ Coulombic electrostatic potential. Similarly, in the gravitational atom, the boson cloud is attracted to the central black hole through the $1/r$ Newtonian gravitational potential.} As a result, the eigenstates of the gravitational atom have a Bohr-like energy spectrum as the hydrogen atom. Black hole superradiance is an excellent way of probing ultralight particles at the weak-coupling frontier because it only relies on the gravitational coupling between the black hole and the boson, with no assumptions made about other types of interactions of the fields. Furthermore, all properties of the gravitational atoms, such as their energy spectra and superradiant growth timescales, can be computed accurately from first principles, see e.g.~\cite{Detweiler:1980uk, Dolan:2007mj, Pani:2012vp, Pani:2012bp, Baryakhtar:2017ngi,Endlich:2016jgc}. Indeed, a central goal of this thesis is to fully explore the qualitative and quantitative aspects of these boson clouds~\cite{Baumann:2019eav}. Importantly, we will find that these computations reveal subtle features of the clouds that can have significant impacts on the clouds' phenomenologies.

\vskip 2pt

 \begin{figure}[t]
      	\centering
        \includegraphics[scale=0.82]{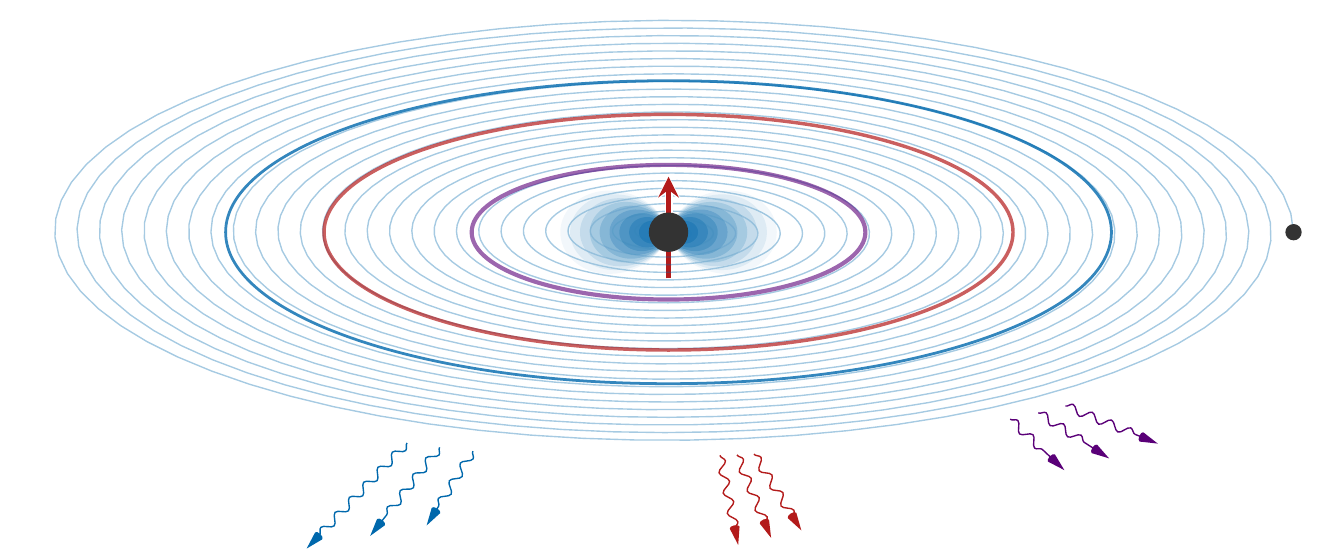}
        \caption{Illustration of the gravitational atom in a binary system. During the inspiral, the binary companion induces level mixings between different eigenstates of the cloud. This mixing becomes non-perturbative when the orbital frequency, which slowly increases due to gravitational-wave emission, matches the energy difference between the coupled states. During these resonant transitions (denoted by the colored ellipses), the cloud can significantly backreact on the binary’s orbit and affect the binary's gravitational-wave emissions. Furthermore, various types of finite-size effects associated to the cloud can also leave distinctive imprints on the phases of the binary waveforms.
         \label{fig:intro}}
    \end{figure}

The gravitational-wave signatures of these boson clouds are most significant when they are parts of binary systems~\cite{Baumann:2018vus, Baumann:2019ztm}. In particular, as I will describe in this thesis, the gravitational perturbation sourced by a binary companion can force the cloud to evolve in a highly dynamical manner. Due to the structural similarity between the hydrogen and the gravitational atoms, the dynamics of the cloud in a binary system is analogous to that of an atom shone by a laser beam. Many of the classic results in atomic physics, such as the presences of level mixings and selection rules, therefore also apply to the boson cloud. Moreover, these level mixings are dramatically enhanced when the orbital frequency of the binary matches the energy difference between different eigenstates in the spectrum. These resonance transitions can significantly backreact on the orbit, thereby affecting the gravitational waves emitted by the binary system (see Fig.~\ref{fig:intro} for an illustration). Since the clouds have large spatial extents, they can also affect in the gravitational waves of the binary through various so-called ``finite-size effects." A notable example is the spin-induced quadrupole moment of the cloud~\cite{Geroch:1970cd, Hansen:1974zz, Thorne:1980ru}, which characterizes its deformation from spherical symmetry due to its rotational motion. In addition, when the tidal force exerted by the binary companion deforms the shape of the cloud, the orbit would deviate from its normal evolution due to the associated change in the orbital binding energy and affect the flux emitted to asymptotic infinity~\cite{Flanagan:2007ix, Vines:2011ud}. Since the imprints of these finite-size effects on the binary waveforms are known in detail, a precise measurement of these effects in the waveforms provide clean probes of the nature of the gravitational atoms.

\vskip 2pt

The gravitational waves emitted by a binary system presents us with a unique window of opportunity to detect the existence of putative ultralight bosons in our Universe. Furthermore, these signatures offer us rare probes of the microscopic properties of the boson fields that form the clouds~\cite{Baumann:2018vus, Baumann:2019ztm}. In particular, the mass of the boson field is directly related to the resonance frequency of the gravitational waves emitted by the binary during the resonant transitions described above. On the other hand, the intrinsic spin of the boson field can be infered through interesting time-dependences in the finite-size imprints on the waveforms. This way of probing new particles is in fact directly analogous to the discipline of particle collider physics. This is because in an ordinary collider, a particle’s mass is determined by the energy at which the particle appears as a resonant excitation, while its spin is measured via the angular dependence of the final state. The analogy is made further apparent by the fact that the dynamics of the gravitational atoms near these resonance frequencies can be quantified by an S-matrix, making these systems effectively ``gravitational colliders"~\cite{Baumann:2018vus, Baumann:2019ztm}. Given its similarity with particle collider physics, we refer to this nascent discipline of probing ultralight bosons with binary systems as \textit{gravitational collider physics}.

\vskip 2pt

In addition to the gravitational atoms, other types of compact objects can also arise in many scenarios of new physics at the weak-coupling frontier. For example, if the new particles have Compton wavelengths that are much shorter than the sizes of astrophysical black holes, they behave essentially as point-like particles and can accrete around black holes to form high-density spikes~\cite{Gondolo:1999ef, Ferrer:2017xwm}. More exotic types of compact objects, such as boson stars~\cite{Kaup1968, Ruffini1969, Breit:1983nr, Colpi1986, Liebling:2012fv}, could also exist in our Universe. In all of these cases, when the dark compact objects are parts of binary systems, they can affect the binaries' gravitational-wave emissions through various dynamical effects, such as the finite-size effects described above for the gravitational atoms. To realistically search for these unusual types of binary systems, one would typically need to match the noisy observational data to template waveforms, which include the effects of the putative new physics~\cite{Thorne1980Lectures, Dhurandhar:1992mw, Cutler:1992tc}. This matched-filtering technique is, however, extremely sensitive to the phase coherence between the signal and template waveforms. A mismodelling of the signal template waveform can easily degrade its detectability~\cite{Cutler:1992tc}. In this thesis, I present an analysis whereby we assess the extent to which ordinary binary black hole template waveforms, which are used in the LIGO and Virgo search pipelines, would allow us to search for these binary systems. This represents an important first step towards understanding the limitations of existing search strategies to detect new physics with the gravitational waves of binary systems.

\vskip 2pt

The above discussion clearly demonstrates the potential of probing new physics using gravitational waves. While this thesis focuses primarily on the signatures from binary systems, other types of gravitational-wave signals, such as continuous monochromatic waves and stochastic gravitational-wave background, could also offer important clues about physics beyond the Standard Model. Since gravitational-wave science is a precision science, it is imperative that we continue the strong interplay between the developments of new theoretical ideas and novel search strategies. Current detectors are rapidly improving their observational capabilities, and we are seeing a flurry of proposals for developing future gravitational-wave detectors. Now is an opportune time to explore ways of maximizing the discovery potentials of these gravitational-wave measurements. While a detection of new physics would certainly transform our understanding of particle physics, even non-observations can be informative, as they would place meaningful bounds on new physics. All in all, I believe that this thesis represents an important step towards our goal of using gravitational-wave observations to probe particle physics at the weak-coupling frontier.

\newpage

\subsubsection*{Outline of this thesis}

\vskip -5pt
The rest of this thesis is organized as follows. In Chapter~\ref{sec:GWAstro}, I describe several aspects of gravitational-wave astrophysics. These include a discussion on the theoretical foundations of General Relativity and a summary of current and future gravitational-wave observatories. One of the main goals in this chapter is to elaborate on the dynamics of binary systems and the associated gravitational-wave emissions. In Chapter~\ref{sec:gravprobes}, I provide a broad overview of the weak-coupling frontier in particle physics. After a brief description of the Standard Model and motivations for new physics, I discuss several classes of weakly-coupled particles. I also introduce the concept of superradiance and the gravitational atom in this chapter, which will serve as important background material for subsequent chapters. While writing these two chapters, I had in mind a broad audience who may wish to learn the basics of gravitational-wave astophysics and particle physics. The particle physicists may therefore find Chapter~\ref{sec:GWAstro} helpful for delving further into the gravitational-wave literature, while the general relativists and astrophysicists may find the review on physics beyond the Standard model in Chapter~\ref{sec:gravprobes} useful.

The remaining chapters present the main results of this thesis. In Chapter~\ref{sec:spectraAtom}, I describe the analytic and numeric computations of the spectra of the gravitational atoms. In particular, I focus on the spectra of the scalar and vector gravitational atoms, demonstrating how our computations reveal important qualitative differences between the two types of atoms. In Chapter~\ref{sec:Collider}, I describe the dynamics of the gravitational atoms in binary systems. I pay special attention to the cloud's evolution when the orbital frequency of the binary matches the resonance frequencies associated to the cloud. I further elaborate on how these resonant dynamics are equivalent to the Landau-Zener transitions in quantum mechanics, with the physics effectively encoded in an ``S-matrix." In Chapter~\ref{sec:signatures}, I discuss the rich phenomenologies associated to the gravitational atoms in binary systems. These include the backreaction of the resonances on the orbit and various types of time-dependent finite-size imprints on the binary's waveform.  I also describe how the scalar and vector atoms can be distinguished through precise reconstructions of these gravitational waveforms. In Chapter~\ref{sec:search}, I present an analysis that quantify the extent to which binary black hole template waveforms can be used to detect new dark compact objects in binary systems. I first construct a phenomenological waveform that represents a wide range of new binary signals, and then compute the overlap between the binary black hole template waveforms and the phenomenological waveforms. Finally, I conclude and provide an outlook on future directions in Chapter~\ref{sec:Conclusions}. 

A number of appendices contains technical details that had been omitted in the main text. In Appendix~\ref{app:SpectralDetails}, I provide further details on the analytic and numeric computations of the spectra of the gravitational atoms. In Appendix~\ref{app:GP}, I elaborate on the gravitational perturbation sourced by a binary companion on the boson clouds.  In Appendix~\ref{app:LZ}, I present a formalism that encodes the dynamics of the Landau-Zener transition for an arbitary binary orbit configuration.  Finally, Appendix~\ref{app:AMTransfer} includes the derivation of the backreaction of the Landau-Zener transition on the orbit. 

\subsubsection*{Notation and conventions} 

In this thesis, I adopt the `mostly plus' signature $(-,+, +, +)$ for the metric. The covariant derivative with respect to the metric is represented by the symbol $\nabla$. Greek letters are used for spacetime indices ($\mu,\nu, \ldots$), while Latin letters will either stand for spatial indices ($i, j, \ldots$), label indices ($i, k, \ldots$), or vierbein indices ($a,b,\ldots$). To avoid confusion, I will sometimes wrap label indices in parentheses. Unless stated otherwise, I work in natural units with $G=\hbar = c = 1$.

I will adopt the Boyer-Lindquist coordinates $\{t, r, \theta, \phi \}$ for the Kerr black hole. Denoting  mass and specific angular momentum by $M$ and $a$, respectively, the line element is
\beq
\d s^2 = - \frac{\Delta}{\Sigma}\left(\d t - a \sin^2 \theta\, \d \phi \right)^2 + \frac{\Sigma}{\Delta} \d r^2 + \Sigma\hskip 1pt \d \theta^2 + \frac{\sin^2 \theta}{\Sigma}  \left(a \hskip 1pt \d t - (r^2 + a^2) \, \d \phi \right)^2\, , 
\eeq
where $\Delta \equiv r^2 -  2 Mr +a^2 $ and $\Sigma \equiv r^2 + a^2 \cos^2 \theta$. The roots of $\Delta$ determine the inner and outer horizons, located at $r_\pm = M \pm \sqrt{M^2 -a^2}$, and the angular velocity of the black hole at the outer horizon is $\Omega_H \equiv a/ 2 M r_+$. Dimensionless quantities, defined with respect to the black hole mass, are labeled by tildes; e.g.~$\tilde{a} \equiv a/M$ and $\tilde{r}_{\pm} \equiv r_\pm/M$.

The gravitational radius of a black hole is given by $r_g \equiv G M / c^2$.  On the other hand, the gravitational fine-structure constant is $\alpha \equiv r_g/\lambda_c$, where $\lambda_c \equiv \hbar / (\mu c)$ is the (reduced) Compton wavelength of a boson field with mass $\mu$. Quantities associated to the boson clouds will be denoted by the subscript~$c$. For example, the mass and angular momentum of the clouds are $M_c$ and $S_c$, respectively.  I use the subscript $*$ for quantities related to the binary companion, e.g. $M_*$ is its mass and $\{R_*, \iota_*, \varphi_*\}$ represent its spatial coordinates in the Fermi frame of the cloud.

The eigenstates of the scalar and vector atoms are denoted by $|\es n \es \ell \es m\rangle$ and $| \es n \es \ell \es j \es m\rangle$, with the integers $\{n, \ell, j, m \}$ labeling the principal, orbital angular momentum, total angular momentum, and azimuthal angular momentum numbers, respectively. Following the convention in atomic physics, we have~$n \geq \ell + 1$. I refer to vector modes that acquire a factor of $(-1)^j$ under a parity transformation as `electric modes,' and those that acquire a factor of $(-1)^{j+1}$ as `magnetic modes.' These modes are to be distinguished from odd and even modes, which, by my convention, receive a factor of $-1$ and $+1$ under parity, respectively.

\chapter{Gravitational-Wave Astrophysics} \label{sec:GWAstro}

Gravitational-wave astrophysics is a remarkably rich subject that involves theoretical studies of General Relativity and pioneering developments in many experimental techniques. Indeed, the direct detections of gravitational waves by the LIGO and Virgo observatories would not have been possible without the significant theoretical and experimental advancements in the field over the past several decades. At the same time, the field remains an active area of research with further developments required in order to maximally exploit the scientific opportunities offered by the current and future network of gravitational-wave detectors. 

In this chapter, I will describe both the theory and the observations underlying gravitational-wave astrophysics. In Section~\ref{sec:theoretical}, I review the foundational aspects of General Relativity and two of its most famous predictions: black holes and gravitational waves. In Section~\ref{sec:CBC}, I then describe the relativistic dynamics of compact binary systems and the associated gravitational-wave emissions. The literature on this subject is vast and can appear highly technical; my goal is to describe only the key qualitative features of these computations and include a summary of the state of the art. In Section~\ref{sec:detectors}, I summarize the status of current gravitational-wave detectors and their scientific achievements over the past few years. I also provide a broad overview of many existing proposals for future gravitational-wave observatories.

\section{Theoretical Foundations} \label{sec:theoretical}

General Relativity has revolutionized our understanding of gravity~\cite{Einstein1915}. It supersedes Newtonian gravity because many of the somewhat ad hoc features in the latter are explained by the fundamental principles of General Relativity. Furthermore, the predictions of General Relativity are in excellent agreement with virtually all observations made in cosmology and astrophysics, many of which cannot be accounted for in Newtonian physics~\cite{Will:2014kxa}. I will first review the basic elements of General Relativity, and then discuss black holes and gravitational waves as two of its most important solutions.

\subsection{General Relativity} \label{sec:GR}

General Relativity was developed out of the need for incorporating relativity into gravity. In Special Relativity, space and time are treated on equal footing. In addition, physical quantities are invariant under Lorentz transformations between different inertial frames. Nevertheless, these properties are absent in Newtonian gravity. To understand this concretely, let us consider the Poisson equation
\beq
\partial_i^2 U = 4 \pi \rho \, , \label{eqn:Poisson}
\eeq
where $\partial_i \equiv \partial/\partial x^i$ is a spatial derivative and $U$ is the gravitational potential sourced by an energy density $\rho$. Space and time are clearly treated differently in (\ref{eqn:Poisson}) because time derivatives are absent. A naive generalization of the Laplacian $\partial_i^2$ to the d'Alembertian $
\box \equiv -\partial_0^2 + \partial_i^2$ wouldn't be sufficient either because $\rho$ is not a Lorentz scalar. How, then, can one construct a relativistic theory of gravity?

A seemingly unrelated puzzle in Newtonian gravity offered Einstein the vital clue to tackle this problem. In Newtonian mechanics, the force experienced by a test particle in a gravitational potential is
\beq
m_I \frac{\d^2 x^i}{\d t^2} = - m_G \partial^i U \, , \label{eqn:NewtonEqmImA} 
\eeq 
where $m_I$ is the particle's inertial mass and $m_G$ is its gravitational mass. These two masses, \textit{a priori}, need not be the same: the former describes the particle's inertia to changing its motion, while the latter is a charge that quantifies its coupling to a gravitational field. 
Nevertheless, $m_I$ is found to be equal to $m_G$ in experiments and astrophysical observations.  Is there an explanation for this ``fine-tuning" between these two types of masses? 

Einstein intuited that the equality between the inertial and gravitational masses is not a coincidence, but rather reflects a fundamental property of gravity. The exact equality $m_I = m_G$ has profound consequences, as it implies that all test particles, regardless of their masses, experience the same gravitational acceleration. This is often formulated as the \textit{weak equivalence principle}, which states that the motion of a test particle in a gravitational field is independent of its mass and composition. All test particles with the same initial conditions must therefore travel on the same preferred class of trajectories called geodesics or free-falling trajectories. This universal behaviour of gravity makes it qualitatively distinct from other conventional forces, such as electromagnetism, where particles with different amounts of charge accelerate at different rates and hence propagate on different trajectories. In addition, the weak equivalence principle implies that gravity as experienced by a free-falling test particle necessarily vanishes. To guide our intuition, I use Einstein's famous thought experiment of an observer enclosed in a free-falling elevator. In this setup, the observer cannot distinguish whether she is free-falling on Earth or is floating in outer space because the walls of the elevator do not exert any reaction force. Since gravity is clearly absent in the latter case, one is led to conclude that this is similarly the case in the former.

Because the motion of test particles is independent of their masses, Einstein proposed what we experience as gravity is instead a fundamental property of the background spacetime on which test particles propagate. But what property of spacetime exactly? A crucial feature of the weak equivalence principle is that it is only applicable to test particles, but not spatially-extended objects.\footnote{More precisely, by a spatially-extended object I mean an object whose size is comparable or larger than the typical length scale on which the gravitational field in its vicinity varies. By contrast, a test particle has a size which is much smaller than this variation length scale and can therefore treated as point-like.} While a free-falling test particle cannot experience gravitation locally, an extended object does through tidal forces. The fact that gravity manifests itself differently in a ``local" region and a ``global" region of spacetime is reminiscent of curvature in geometry, whereby a globally curved surface appears flat in any local region. These insights led Einstein to conclude that gravity is a manifestation of curved spacetime.

Our discussion so far has been guided by the weak equivalence principle and by considering the motion of particles in a gravitational field. However, multiple generalizations of the weak equivalence principle can be formulated. For instance, one version stipulates that the universal coupling to gravity applies to all forms of matter and energy-momentum, not just to test particles. 
Another version, the so-called strong equivalence principle, states that the laws of Special Relativity are restored in all local free-falling frames. The strong equivalence principle therefore makes precise what I meant above by a ``locally flat region in curved spacetime" --- it refers to a local inertial frame whereby the background spacetime is Minkowskian. In what follows, these different versions of the equivalence principle will be referred universally as the ``equivalence principle."

The equivalence principle motivates generalizing the Minkowski spacetime in Special Relativity to a curved spacetime in General Relativity. One of the central objects in curved geometries is the proper length between two distinct points on the spacetime
\beq
\d s^2 = g_{\mu \nu} (x) \, \d x^\mu \d x^\nu \, , \label{eqn:lineelement}
\eeq
where $x^\mu$ is a choice of coordinate and $g_{\mu \nu}$ is the spacetime metric. In a local inertial frame, $g_{\mu \nu}$ reduces to the Minkowski metric, $ \eta_{\mu \nu} = \text{diag} (-1, +1, +1, +1)$. In that case, Special Relativity requires the proper length to remain invariant under Poincar\'e transformations. However, this principle no longer holds in General Relativity because the isometries of Minkowski spacetime are generally absent. Instead, the line element \eqref{eqn:lineelement} is viewed as a geometric object in General Relativity, whereby its proper length is invariant under \textit{coordinate transformations}. As a result, under the coordinate transformation $x^\mu \to x^{\prime \mu}$, the metric $g_{\mu \nu}$ must transform as 
\beq
g_{\mu \nu}(x^\prime) = \frac{\partial x^\rho}{\partial x^{\prime \hskip 1pt \mu}} \frac{\partial x^\sigma}{\partial x^{\prime \hskip 1pt \nu}} g_{\rho \sigma}(x) \, . \label{eqn:gencov}
\eeq
Crucially, unlike a Poincar\'e transformation, (\ref{eqn:gencov}) is a not a symmetry of General Relativity. Rather it represents the infinite amount of freedom we have in choosing the coordinates to describe the metric, with no changes whatsover in the underlying geometry of spacetime. Given its moral similarity with gauge transformations in field theory, I occasionally refer (\ref{eqn:gencov}) as the gauge transformation of $g_{\mu \nu}$.

The idea of invariance under a coordinate transformation is not restricted to geometric objects such as a line element. In fact, all laws of physics and physical observables are necessarily independent of the way in which we choose our coordinates to represent them. This is the foundational principle of General Relativity and is known as \textit{general covariance} or \textit{diffeomorphism invariance}. Mathematically, the laws of physics and physical observables must be tensorial, which means the forms of their expressions must remain unchanged under coordinate transformations. To illustrate the power of general covariance in constraining the possible forms of physical laws, it is instructive to consider the equation of motion of a test particle free-falling in a curved spacetime. This is given by the \textit{geodesic equation}:
\beq
\frac{\d v^\mu}{\d \tau} + \Gamma^{\mu}_{\alpha \beta} v^\mu v^\nu =0 \, , \label{eqn:geodesic}
\eeq
where $\tau$ is the particle's proper time, $v^\mu \equiv \d x^\mu / \d \tau$ is its four-velocity, and $\Gamma^{\mu}_{\alpha \beta} = g^{\mu \rho} (\partial_\alpha g_{\rho \beta} + \partial_\beta g_{\rho \alpha} - \partial_\rho g_{\alpha \beta} )/2$ is the Levi-Civita symbol. The first term $\d v^\mu / \d \tau$ acquires an additional term under a coordinate transformation. The Levi-Civita symbol is therefore introduced in (\ref{eqn:geodesic}) to cancel this term and ensure the geodesic equation is tensorial.

The equation (\ref{eqn:geodesic}) states that geodesics on curved spacetimes are equivalent to free-falling trajectories in gravity, which I argued above are special trajectories due to the weak equivalence principle. In Minkowski spacetime, where $\Gamma_{\alpha \beta}^\mu=0$, (\ref{eqn:geodesic}) restores the classic result that test particles move along straight lines in the absence of any forces. The Levi-Civita symbol, which depends on $g_{\mu \nu}$, therefore determines how a free-falling trajectory bends according to the geometry of its  background spacetime. In \S\ref{sec:dynamics}, we will see how the geodesic equation reduces to Newton's equation (\ref{eqn:NewtonEqmImA}) in the non-relativistic limit. Crucially, $m_I$ and $m_A$ make no appearance whatsoever in the geodesic equation. As such,  $m_I = m_A$ is naturally enforced in (\ref{eqn:NewtonEqmImA}). Newtonian gravity and the weak equivalence principle are hence naturally incorporated into General Relativity.

It is also important to examine the equation that governs the motion of a spatially-extended object in a curved spacetime. For simplicity, I model this object as consisting of two neighbouring test particles separated by the vector $\zeta^\mu $.  As both particles travel along their respective geodesics, the separation $\zeta^\mu $ changes according to the curvature of spacetime. More precisely, the evolution of $\zeta^\mu $ is given by the \textit{geodesic deviation equation}:
\beq
\frac{\mathrm{D}^2 \zeta^\mu}{\d \tau^2 } + {R^\mu}_{\nu \rho \sigma} v^\nu \zeta^\rho v^\sigma  = 0 \, , \label{eqn:geodesicdeviation}
\eeq
where $\mathrm{D} u^\mu / \d \tau = \d  u^\mu / \d \tau + {\Gamma^\mu}_{\alpha \beta} v^\alpha u^\beta$ is the covariant derivative of an arbitrary vector $u^\mu$ along an integral curve of the four-velocity $v^\mu$ (i.e.~an observer's worldline) and ${R^\mu}_{ \nu \rho \sigma} = \partial_\rho \Gamma^\mu_{\nu \sigma} - \partial_\sigma \Gamma^{\mu}_{\nu \rho} + \Gamma^\tau_{\nu \sigma} \Gamma^\mu_{\tau \rho} - \Gamma^\tau_{\nu \rho} \Gamma^\mu_{\tau \sigma}$ is the Riemann tensor.
In Minkowski spacetime, we have ${R^\mu}_{\nu \rho \sigma} = 0$ and $\zeta^\mu$ is either fixed or changes at a constant rate. This is simply a manifestation of the fact that geodesics in the Minkowski spacetime are straight lines. However, in a curved spacetime, ${R^\mu}_{\nu \rho \sigma} \neq 0$, the separation $\zeta^\mu$ can \textit{accelerate}. In other words, geodesics which are initially parallel to one another can either converge or diverge as they evolve. This relative acceleration is a manifestation of \textit{tidal forces} in gravity. The Riemann tensor is therefore essential in describing the curvature of spacetime and its gravitational field.

Through the geodesic equation (\ref{eqn:geodesic}) and the geodesic deviation equation (\ref{eqn:geodesicdeviation}), I have made precise the qualitative descriptions about the motions of particles and extended objects in curved spacetime. However, I have not discussed what sources curvature of spacetime to begin. This is described by the \textit{Einstein field equation}
\beq
R_{\mu \nu} - \frac{1}{2} R g_{\mu \nu} = 8 \pi T_{\mu \nu} \, , \label{eqn:Einstein}
\eeq
where $R_{\mu \nu} \equiv {R^{\alpha}}_{\mu \alpha \nu}$ is the Ricci tensor, $R \equiv {R^\alpha}_\alpha$ is the Ricci scalar, and $T_{\mu \nu} $ is the energy-momentum tensor. The left-hand side of (\ref{eqn:Einstein}) is also known as the Einstein tensor, $G_{\mu \nu}$. Since $G_{\mu \nu}$ represents the geometry of spacetime, while $T_{\mu \nu}$ describes the matter sector, the Einstein field equation dictates the way in which spacetime and energy-momentum interact with each other. In fact, this interaction is unique because the contracted Bianchi identity, $\nabla_{\mu} G^{\mu \nu}=0$, simultaneously enforces conservation of the energy-momentum tensor, $\nabla_\mu T^{\mu \nu} = 0$. While matter sources the curvature of spacetime, the spacetime in turn constrains the way in which matter can evolve. To quote Wheeler's famous dictum:  ``Matter tells spacetime how to curve, spacetime tells matter how to move."

Finally, I explicitly discuss how various principles of General Relativity are embedded in the Einstein field equation. Relativity is clearly obeyed by (\ref{eqn:Einstein}) because it consists of tensors defined on spacetime. In addition, since $T_{\mu \nu}$ encodes all forms of energy-momentum, (\ref{eqn:Einstein}) demonstrates their universal coupling in gravity, thereby manifesting the equivalence principle. Furthermore, the contracted Bianchi identity enforces four constraints on $g_{\mu \nu}$, which reflects the inherit redundancies in $g_{\mu \nu}$ due to our arbitrary choice of coordinates. As such, only six of the ten components of $g_{\mu \nu}$ are independent when $T_{\mu \nu} \neq 0$. As we shall see in \S\ref{sec:GW}, this is no longer the case in vacuum, $T_{\mu \nu} = 0$, where the metric only contains two independent degrees of freedom.

\subsection{Black Holes} \label{sec:BH}
 
Black holes are a remarkable prediction of General Relativity. Once considered a pure mathematical curiosity, they are now known to be ubiquitous in our Universe, even playing significant roles in astrophysical environments. In this subsection, I will review theoretical aspects of black holes. I will pay special attention to aspects which are relevant for this thesis.

Black holes are vacuum solutions of the Einstein field equation (\ref{eqn:Einstein}),
\beq
R_{\mu \nu} = 0 \, . \label{eqn:vacuumR}
\eeq
The static and spherically symmetric black hole solution was discovered by Schwarzschild in 1916, shortly after the Einstein's field equation was published~\cite{Schwarzschild1916, Droste1917}. The line element is
\beq
\d s^2 = - \left( 1 - \frac{2M}{r} \right) \d t^2 + \left( 1 - \frac{2M}{r} \right)^{-1} \d r^2 + r^2 \left( \d \theta^2 + \sin^2 \theta \d \phi^2 \right) \, ,  \label{equ:Schw}
\eeq
where $M$ is the mass of the black hole. The Schwarzschild solution (\ref{equ:Schw}) has two special locations: a singularity at $r=0$ that represents a divergence of the spacetime curvature, and the \textit{event horizon} at $r=2M$. The event horizon is defined as the null surface on which \textit{any} trajectory in the black hole interior, $r < 2M$, are unable to escape to the exterior, $r > 2M$. The gravitational pull of black holes is therefore so strong that its escape velocity exceeds the speed of light. The apparent singularity at $r=2M$ in (\ref{equ:Schw}) is merely a coordinate singularity. It can be removed by analytically extending the metric through coordinate systems which are adapted to null radial geodesics, such as the Eddington-Finkelstein coordinates. 

Remarkably, Birkhoff's theorem states that (\ref{equ:Schw}) is the unique solution of all spherically-symmetric vacuum spacetimes~\cite{Birkhoff, Jebsen}. The Schwarzschild metric is therefore also applicable to the exterior regions of many astrophysical objects, as spherical symmetry is often an excellent approximation. After all, it is for this reason that (\ref{equ:Schw}) could be used to accurately postdict the perihelion precession of Mercury around the Sun~\cite{Einstein1916mercury} and predict the bending of light by the moon in a solar eclipse~\cite{Eddington1920}. Furthermore, since staticity is not assumed in Birkhoff's theorem,  (\ref{equ:Schw}) remains applicable to a radially pulsating star. The metric in the interior of a star, on the other hand, depends on its energy-momentum tensor and can be solved using the Tolman-Volkoff-Oppenheimer equations~\cite{Tolman1939, OV1939}.

While the Schwarzschild solution was found shortly after the formulation of General Relativity, the solution for a rotating black hole was only derived by Roy Kerr half a century later~\cite{Kerr1963} (see~\cite{Kerr:2007dk} for a personal account of his discovery). In what follows, I will express the Kerr solution in Boyer-Lindquist coordinates~\cite{BL1967}, as it conveniently reduces to the Schwarzschild metric (\ref{equ:Schw}) when the black hole spin vanishes. With a slight abuse of notation, I also denote these coordinates by $\{ t, r, \theta, \phi\}$. The line element of a Kerr black hole with mass $M$ and angular momentum $J$ then reads
\beq
\d s^2 = - \frac{\Delta}{\Sigma}\left(\d t - a \sin^2 \theta\, \d \phi \right)^2 + \frac{\Sigma}{\Delta} \d r^2 + \Sigma\hskip 1pt \d \theta^2 + \frac{\sin^2 \theta}{\Sigma}  \left(a \hskip 1pt \d t - (r^2 + a^2) \, \d \phi \right)^2\, ,  \label{equ:Kerr}
\eeq
where $\Delta \equiv r^2 -  2 Mr +a^2$ and $\Sigma \equiv r^2 + a^2 \cos^2 \theta$. The spin parameter $a \equiv J/M$ is bounded by $0 \leq a \leq M$. The roots of $\Delta$ determine the inner and outer horizons, located at $r_\pm = M \pm \sqrt{M^2 -a^2}$. The inner horizon is inaccessible by an external observer, and can therefore be neglected for all practical purposes. The off-diagonal component $g_{0 \phi}$ in (\ref{equ:Kerr}) captures the rotational frame-dragging effect, also called the Lense-Thirring effect~\cite{Lens-Thirring1918}. From the point of view of an observer at infinity, a free-falling test particle with no angular momentum would co-rotate with the black hole. When this test particle approaches $r_+$, it co-rotates at the angular velocity of the black hole, $\Omega_{\rm H} = a/ 2 M r_+$.

In general, an observer can counteract the frame-dragging effect by acquiring angular momentum in the direction opposite to the black hole spin. However, near the black hole, there exists an \textit{ergoregion} in which all observers and even light \textit{must} co-rotate with the black hole. More formally, this is a region where a timelike Killing vector field at infinity turns spacelike. The ergoregion therefore lies within $r_+ <  r < r_{\rm erg}$, where $r_{\rm erg} = M + \sqrt{M^2 - a^2 \cos^2 \theta}$ is a root of $g_{00}$ in (\ref{equ:Kerr}). That a timelike vector must turn spacelike in the ergoregion is interesting because a test particle in this region can have negative energy, at least from the point of view of an observer at infinity. The Penrose process~\cite{Penrose:1969pc} is a thought experiment whereby a particle in the ergoregion disintegrates into a positive and a negative energy component, with the latter falling into the black hole. By local conservation of energy and momentum, the former must emerge from the ergoregion with larger energy and angular momemtum. As such, energy and angular momentum are extracted from the rotating black hole. Remarkably, there exists a limit to the amount of energy which can be extracted through this process. This is because the mass of the black hole can be decomposed into~\cite{Christodoulou:1970}
\beq
M^2 = M^2_{\rm irr} + \frac{J^2}{4 M_{\rm irr}^2}\, , \label{eqn:Mirr}
\eeq
where $M_{\rm irr}$ is the black hole's \textit{irreducible mass}. Heuristically, the irreducible mass can be interpreted as the ``rest mass" of the black hole, while the second term in (\ref{eqn:Mirr}) is its ``rotational energy," with only the latter extractable through the Penrose process. Though rather contrived, the Penrose process was the first concrete realization of an energy-extraction mechanism from a rotating black hole. As we shall see in Section~\ref{sec:atomreview}, a wave analog called \textit{black hole superradiance} can trigger an instability on a rotating black hole,  efficiently extracting energy and angular momentum from a rotating black hole.\footnote{I emphasize, however, that the analogy between the Penrose process and superradiance is qualitative at best. In particular, the ergoregion plays no role in superradiance. One can view this distinction as a difference in the scales involved: a particle in the Penrose process is point-like, whereas to trigger superradiance, the Compton wavelength of the wave must be larger than the size of the black hole.} Crucially, superradiance can occur spontaneously and is therefore realizable in astrophysical black holes.

The Schwarzschild and Kerr black holes are also distinct in their multipolar structures. As a consequence of spherical symmetry, a non-rotating black hole only sources a monopole. A Kerr black hole, on the other hand, generates an infinite tower of multipole moments. Generically, the multipole moments of a stationary object in General Relativity is organized through its \textit{mass moments}, $M_\ell$, and \textit{current moments}, $S_\ell$, where $\ell $ denotes all non-negative integers~\cite{Geroch:1970cd, Hansen:1974zz}. For an object that is reflection symmetric about its equatorial plane, like the Kerr black hole, the odd-$\ell$ mass moments and even-$\ell$ current moments vanish identically. Black holes are special in that all of its multipole moments can be succinctly described through the simple relation~\cite{Hansen:1974zz, Thorne:1980ru} 
\beq
M_\ell + i S_\ell = M \left( i a \right)^\ell \, . \label{eqn:BHmoments}
\eeq
The interpretation of this hierarchy of moments is clear: $M_0=M$ is the black hole's mass, $S_1 = M a$ is its spin, $M_2 = - Ma^2$ is its mass quadrupole, etc. Importantly, the multipolar structure of the black hole is uniquely fixed by $M$ and $a$. This is qualitatively distinct from a general astrophysical object, where the quadrupole and higher-order multipole moments typically depend on additional parameters that describe the internal structure of the object. 

The existence of the solutions (\ref{equ:Schw}) and (\ref{equ:Kerr}) does not preclude other black hole solutions. However, the uniqueness theorems~\cite{Israel1967, Carter:1971zc, Robinson:1975bv} state that the Schwarzschild and Kerr metrics are indeed the only solutions of black holes.\footnote{For electrovacuum spacetimes, black holes can also carry electric and magnetic charges (see~\cite{Chrusciel:2012jk} for a review on more general uniqueness theorems when other types of matter fields are present). Nevertheless, I will ignore charged black holes throughout this thesis, as they do not seem to be astrophysically relevant. I also do not consider black hole solutions in other spacetime dimensions.} As an astrophysical object collapses and forms a black hole, all of its parameters but $M$ and $a$ would disappear, thereby making (\ref{eqn:BHmoments}) the unique multipolar structure of all black holes in our Universe. This remarkable simplicity of black holes is often phrased more colloquially as black holes have ``no hair." 

Astrophysical black holes are often immersed in complicated environments and perturbed by surrounding matter. It is therefore important to examine the stability of black holes under these perturbations. To a good approximation, I describe the mode stability of black holes by considering their response when perturbed by an incident null wave.\footnote{There exists many notions of stability in General Relativity. In addition to mode stability, there are also linear and non-linear stabilities. See e.g.~\cite{Teukolsky:2014vca} for their differences and a summary of the stabilities of Kerr black holes.} In early pioneering works, it was shown that all modes decay for a Schwarzschild black hole~\cite{ReggeWheeler, Vishveshwara, Zerilli1970a}. These stable modes, also known as quasi-normal modes, correspond to the ringing spectra of the black hole.  On the other hand, the situation is richer for Kerr black holes. Through the Teukolsky equation~\cite{Teukolsky:1973ha, Press:1973zz, Teukolsky:1974yv}, it was found that Kerr black holes are only stable when its angular velocity satisfies $\Omega_{\rm H} < \omega/m$, where $\omega$ and $m$ are the frequency and azimuthal number of the mode --- conserved quantities due to the Kerr isometries. On the other hand, for a highly rotating black hole with 
\beq
\Omega_{\rm H} > \frac{\omega}{m} \, , \label{eqn:superradm1}
\eeq
the Kerr black hole is \textit{unstable} to a mode perturbation sourced by a massive field~\cite{Damour:1976kh, Detweiler:1980uk, Brito:2015oca}. This is precisely the superradiance phenomenon which I alluded to earlier. This instability, of course, does not prevent the formation of astrophysical black holes. Rather, as I shall elaborate in Section~\ref{sec:atomreview}, it provides a robust mechanism of coherently enhancing fields around black holes, making black holes interesting laboratories of physics beyond the Standard Model.

\subsection{Gravitational Waves} \label{sec:GW}

The Einstein field equation (\ref{eqn:Einstein}) involves a set of coupled partial differential equations which are hard to solve for generic dynamical spacetimes. Fortunately, analytic time-dependent solutions are attainable when a weak metric perturbation is considered on a fixed background spacetime. In this subsection, I first review the linear perturbation theory of General Relativity. I then describe gravitational waves as plane wave solutions of the linearized Einstein equations. I end by discussing how these waves are generated through accelerating quadrupole moments.

To simplify the following discussion, the background spacetime is fixed to be Minkowski space. This is an excellent approximation in many realistic scenarios, including the background spacetimes of gravitational-wave detectors and inspiraling binary systems. The metric can therefore be written as
\beq
g_{\mu \nu} = \eta_{\mu \nu} + h_{\mu \nu} \, , \label{eqn:linearpert}
\eeq
where $h_{\mu \nu}$ is a perturbation about Minkowski spacetime. In linear perturbation theory, $h_{\mu \nu}$ is assumed to be small.
In order for this 
perturbative scheme to be valid, we must choose a local inertial frame wherein the components of $h_{\mu \nu}$ are much smaller than unity.

The metric perturbation transforms as a spin-2 representation of the Poincar\'e group. It also transforms non-trivially under an infinitesimal coordinate transformation $x^\mu \to  x^{\prime \mu} = x^\mu + \xi^\mu$, where $\xi^\mu$ is an arbitrary function of $x^\mu$. In particular, the gauge transformation (\ref{eqn:gencov}) implies that
\beq
h_{\mu \nu} \to h^\prime_{\mu \nu} = h_{\mu \nu} -\left( \partial_\mu \xi_\nu + \partial_\nu \xi_\mu \right) \, , \label{eqn:lineardiff}
\eeq
at leading order in $\xi$, with $h^\prime_{\mu \nu} \equiv h_{\mu \nu}(x^\prime)$. Physical observables must therefore remain invariant under the gauge transformation (\ref{eqn:lineardiff}) in linearized General Relativity. 
For future convenience, I define the 
trace-reversed field, $\bar{h}_{\mu \nu} \equiv h_{\mu \nu} - \eta_{\mu \nu} h/2$. The gauge transformation (\ref{eqn:lineardiff}) then becomes $\bar{h}_{\mu \nu} \to \bar{h}^\prime_{\mu \nu}  = \bar{h}_{\mu \nu}  - \xi_{\mu \nu}$, where $\xi_{\mu \nu} \equiv \partial_\mu \xi_\nu  + \partial_\nu \xi_\mu  - \eta_{\mu \nu} \partial_\alpha \xi^\alpha$. 
We saw in (\ref{eqn:Einstein}) that the contracted Bianchi identity imposes four constraints on $g_{\mu \nu}$. A similar \textit{gauge condition}  must therefore be imposed on $\bar{h}_{\mu \nu}$ to remove its gauge redundancies. 
A convenient choice is the so-called \textit{harmonic gauge}, $\partial^\mu \bar{h}_{\mu \nu} = 0$, which imposes four gauge conditions on $\bar{h}_{\mu \nu}$, as desired, and dramatically simplifies the linear expansion of (\ref{eqn:Einstein}) to
\beq
\square \, \bar{h}_{\mu \nu} = - 16\pi T_{\mu \nu} \, , \label{eqn:linearEinstein}
\eeq
where $\square \equiv \eta^{\alpha \beta} \partial_\alpha \partial_\beta$ is the flat-space d'Alembertian. The linearized Einstein equation (\ref{eqn:linearEinstein}) therefore describes the sourcing of $\bar{h}_{\mu \nu}$ through a weak source $T_{\mu \nu}$. The harmonic gauge is also convenient because it automatically enforces $\partial_\mu T^{\mu \nu} = 0$, thereby preserving some structural similarity between the linear and full non-linear Einstein equations.

Before describing solutions of (\ref{eqn:linearEinstein}), it is critical to investigate the solutions of $\xi_\nu$ that enforce the harmonic gauge. Acting with a partial derivative on the the gauge transformation of $\bar{h}_{\mu \nu}$, we obtain $\partial^\mu \bar{h}_{\mu \nu} \to \partial^\mu \bar{h}^\prime_{\mu \nu} = \partial^\mu \bar{h}_{\mu \nu} - \square \xi_\nu$. The harmonic gauge is therefore imposed through the inhomogeneous solution $\xi_\nu = - \square^{-1} \left( \partial^\mu \bar{h}^\prime_{\mu \nu} \right)$, where $ \square^{-1} $ formally denotes the Green's function operator. In addition, we see that the harmonic gauge is preserved by a \textit{residual condition}
\beq
\square \xi_\nu = 0 \quad \implies  \quad \square \xi_{\mu \nu} = 0 \, , \label{eqn:residualgauge}
\eeq
where $\xi_{\mu \nu}$ is defined in the paragraph below (\ref{eqn:lineardiff}).
We see that these residual conditions are simple homogeneous wave equations. As such, the most general solution of $\xi_\nu$ consists of a superposition of the inhomogeneous solution described above and a plane wave solution
$\xi_\nu = C_\nu \, e^{i k_\alpha x^\alpha}$, where $C_{\nu}$ is an arbitrary constant vector and $k^\nu$ is null.\footnote{The most general solution to the wave equation is $\xi_\nu = C^+_\nu \, e^{+i k_\alpha x^\alpha} + C^-_\nu \, e^{-i k_\alpha x^\alpha}$. Here, I impose the condition that the wave propagates at a definite direction, such that either one of the terms vanishes. Without loss of generality, I choose $C_\mu^- = 0$. } Crucially, $\xi_\nu$ is an arbitrary function and hence no physical content can be attributed to $C_\nu$. As we shall see shortly, this gauge freedom in $C_\nu$ can be exploited to remove additional redundancies in vacuum solutions of $\bar{h}_{\mu \nu}$.

I now describe propagating plane wave solutions to (\ref{eqn:linearEinstein}) in the exterior region of a source. Setting $T_{\mu \nu}=0$, the linearized Einstein equation simplifies to the homogeneous wave equation $\square \bar{h}_{\mu \nu}=0$. Its solution reads
\beq
\bar{h}_{\mu \nu} = \text{Re}\big( H_{\mu \nu} e^{i k_\alpha x^\alpha}\big) \, , \label{eqn:GWplane}
\eeq
where $H_{\mu \nu}$ is a symmetric constant tensor. Because $k^\nu$ is null, gravitational waves propagate at the speed of light in vacuum. The harmonic gauge gives $k^\mu H_{\mu \nu}=0$, thereby forcing the components of $H_{\mu \nu}$ to lie in the plane transverse to the direction of propagation.

At this stage, one may erroneously conclude that $H_{\mu \nu}$ contains six independent degrees of freedom. Nevertheless, because $\square \left( \bar{h}_{\mu \nu} - \xi_{\mu \nu} \right) = 0$ is also a solution to the equation of motion in vacuum,  cf. (\ref{eqn:residualgauge}), one can freely choose the components of $C_\nu$ above to impose four additional constraints on $\bar{h}_{\mu \nu}$. An obvious choice is one that sets four of the components of $\bar{h}_{\mu \nu}$ to zero. This can be achieved through the \textit{transverse and traceless (TT) gauge}
 \beq
\bar{h}_{0 \mu} = 0 \, , \qquad  h=0 \, .  \label{eqn:TTgauge}
 \eeq
The transverse gauge, along with the harmonic gauge, only imposes three independent constraints. The traceless condition implies that $\bar{h}_{\mu \nu} = h_{\mu \nu} $. To impose the TT gauge onto (\ref{eqn:GWplane}), we can define the projection operator 
\beq
\Lambda_{ij, kl} \equiv P_{ik} P_{jl} - \frac{1}{2} P_{ij} P_{kl} \, , \qquad  P_{ij} (\hat{\textbf{n}}) \equiv \delta_{ij} - n_i n_j \, , \label{eqn:TTprojection}
\eeq
where $\hat{\textbf{n}}$ is a unit vector in the propagation direction of the gravitational wave. The solution in the transverse and TT gauge then is $h^{\rm TT}_{ij} = \Lambda_{ij, kl} \bar{h}_{kl}$. Without loss of generality, I take the wave to propagate in the $z$-direction, such that
\beq
h^{\rm TT}_{ij} = \begin{pmatrix} \phantom{+} h_+ & h_\times & 0 \phantom{+} \\
\phantom{+} h_\times & - h_+ & 0 \phantom{+} \\
\phantom{+} 0 & 0 & 0 \phantom{+}
\end{pmatrix}_{ij}  \cos \left[ \omega (t - z) \right] \, , \label{eqn:GWphysical}
\eeq
where $\omega$ is the gravitational-wave frequency. The components $h_+$ and $h_\times$ in (\ref{eqn:GWphysical}) therefore capture the two independent physical degrees of freedom of a gravitational wave. That a massless spin-2 field contains only two degrees of freedom is of course a classic result of Wigner's study of massless representations of the Poincar\'e group~\cite{Wigner1939}. Nevertheless, it is important to appreciate that Einstein did not notice these gauge artifacts at the time the solution was derived~\cite{Einstein1916, Weyl1922}. It was Eddington who later explicitly showed the unphysical nature of these gauge artefacts, as he found them to propagate at velocities that depend on the choice of coordinates; or as Eddington himself put it, travel at ``the speed of thought"~\cite{Eddington1922}. 

Despite the discovery of the solution (\ref{eqn:GWphysical}) in the 1920s, confusions over whether gravitational waves are physical persisted in the subsequent decades. For instance, Einstein believed (\ref{eqn:GWphysical}) to be an artifact of the linear approximation, and that no gravitational-wave solution would be found in the full theory. Upon finding a singularity in a non-linear cylindrical wave solution in 1936, he and Rosen initially concluded that gravitational waves were therefore unphysical, only to later retract from this conclusion once they understood that it was a coordinate singularity~\cite{EinsteinRosen1937}. Indeed, a recurring source of confusion over the decades had been the issue of coordinate or gauge dependences. For example, the harmonic and TT gauges, though mathematically convenient, lack clear physical interpretations. In addition, it was found that the energy-momentum tensor of a gravitational field is not invariant under the gauge transformation (\ref{eqn:lineardiff}), thereby casting further doubt on the physical nature of gravitational waves. These issues were discussed extensively in the historic Chapel Hill Conference, which had led to the resolutions of many of these problems~\cite{Pirani:1956wr, Bondi:1957dt, Bondi:1958aj}. Here, I will focus only on the issue of whether energy-momentum tensor of gravitational waves is physical.

The fact that the energy-momentum tensor of gravitational waves, $t_{\mu \nu}$, is not gauge invariant is a direct manifestation of the equivalence principle: one can choose a local inertial frame at any point in spacetime, where the gravitational field vanishes locally. While $t_{\mu \nu}$ is not well-defined locally, it certainly carries energy and momentum in an extended spacetime volume. In other words, one can show that the \textit{averaged} energy-momentum tensor over a spacetime region, $\langle t^{\mu \nu} \rangle$, is indeed gauge-invariant. Mathematically, this is because the averaging procedure allows us to perform integration by parts, which cancels the $\xi_\nu$-dependent terms that arise from the gauge transformation (\ref{eqn:lineardiff}). Physically, the averaging provides a meaningful separation between the short-wavelength metric perturbation and the long-wavelength background metric in (\ref{eqn:linearpert}). Since $\langle t^{\mu \nu} \rangle$ is gauge invariant, it is convenient to impose the harmonic and TT gauges, in which case one finds
\beq
\langle t^{\mu \nu} \rangle = \frac{1}{32\pi } \langle \partial^\mu h^{{\rm TT} ij} \partial^\nu h^{\rm TT}_{ij} \rangle \, .
\eeq
The averaged energy and momentum density are $\langle t^{00} \rangle = \langle t^{0z} \rangle =  \langle \dot{h}_+^2 + \dot{h}_\times^2 \rangle /16\pi$. Gravitational waves are therefore physical and can do work. In the limit where the averaged spacetime volume is taken to be infinite, a rigorous definition of total energy and momentum of the spacetime can be obtained. These general-relativistic quantities are called the ADM energy and  momentum~\cite{ADM1962}.

The plane-wave solution (\ref{eqn:GWphysical}) provides the foundation for more complicated gravitational-wave solutions. I now describe radiative solutions of (\ref{eqn:linearEinstein}) which are sourced by $T_{\mu \nu} \neq 0$. This inhomogeneous equation can be solved with a retarded Green's function, which imposes a ``no incoming radiation" boundary condition. In addition, since we are interested in the wave solution in the exterior of the source, I apply the TT gauge through the projection operator (\ref{eqn:TTprojection}) to remove the gauge redundancies in $h_{\mu \nu}$. The integral-solution therefore reads
\beq
h_{ij}^{\rm TT} (t, \textbf{x}) =  4  \, \Lambda_{ij, kl} (\hat{\textbf{n}}) \int \d^3 \textbf{y} \, \frac{T_{kl}\left(t - |\textbf{x} - \textbf{y}|,  \textbf{y} \right)}{|\textbf{x} - \textbf{y}|}  \, . \label{eqn:strain1}
\eeq
For simplicity, I will henceforth restrict myself to cases where an observed gravitational wave is far from the source,  i.e. $ r \equiv |\textbf{x}| \gg |\textbf{y}|$, as is typically the case in realistic astrophysical scenarios. This allows for a further simplication $|\textbf{x} - \textbf{y}| \approx r ( 1 -\textbf{y} \cdot \hat{\textbf{n}}  /r)$, where terms that decay as $\mathcal{O}(|\textbf{y}|^2/r^2)$ or faster can be ignored.

Generally, the source term in (\ref{eqn:strain1}) has a complicated spatial structure and moves in a dynamical manner. As such, it is more convenient to systematically study its properties through a multipole expansion. By performing a Taylor expansion about small values of $|\textbf{y}|$ in (\ref{eqn:strain1}), we find
\beq
\begin{aligned}
\hskip -3pt \int \d^3 \textbf{y} \, T_{k l} \left(t_r +  \textbf{y} \cdot \hat{\textbf{n}}, \textbf{y} \right) =\sum_{p=0}^\infty  \frac{n^{i_1} \cdots n^{i_p} }{p!} \frac{\d^p}{\d t^p} \left[ \int \d^3 \textbf{y} \,  y^{i_1} \cdots y^{i_p} T_{kl} (t, \textbf{y})  \right]_{\rm ret} , \hskip -10pt \label{eqn:stresstensor}
 \end{aligned}
\eeq
where $t_{r} \equiv t-r$ is the retarded time and the subscript means the quantities are evaluated at $t=t_r$. 
The term within the bracket in (\ref{eqn:stresstensor}) are the tensorial moments of the stress tensor $T_{ij}$. The physical interpretation of these terms, however, are clearer when they are rewritten in terms of the mass moments and current moments:\footnote{These definitions are the non-relativistic, though gauge-dependent, versions of the general-relativistic moments described in \S\ref{sec:BH}.}
\beq
\begin{aligned}
M^{i_1 \cdots i_p}(t_r) & \equiv \int \d^3 \textbf{y} \,  y^{ i_1} \cdots y^{i_p } \, T_{00} (t_r, \textbf{y})\, , \\ 
S_k^{i_1 \cdots i_p} (t_r) & \equiv \int \d^3 \textbf{y} \,  y^{ i_1} \cdots y^{i_p } \, T_{0k} (t_r, \textbf{y}) \, . \label{eqn:MJmoments}
\end{aligned}
\eeq
For example, the dominant source term in (\ref{eqn:stresstensor}) can be re-expressed as
\beq
\int \d^3 \textbf{y} \, T_{ij} (t_r, \textbf{y}) = \frac{1}{2} \left[ \ddot{M}_{ij} (t) \right]_{\rm ret} \, , \label{eqn:ddotM}
\eeq
where an overdot represents a derivative with respect to $t$.  The relation (\ref{eqn:ddotM}) is obtained through $\partial_\mu T^{\mu \nu}=0$, which is a consequence of the harmonic gauge, and integration by parts where the boundary terms vanish due to the finite spatial support of $T_{ij}$. Using the same techniques, one find that the higher-order terms in (\ref{eqn:stresstensor}) can be written as a sum of time-derivatives of the higher-order mass and current moments  (\ref{eqn:MJmoments}). The mass and current moments (\ref{eqn:MJmoments}) therefore completely characterize the multipolar structure of the source.

Because the projection operator (\ref{eqn:TTprojection}) satisfies $\Lambda_{ij, kl} \, \delta_{kl}=0$, it is conventional to remove the trace component of the moments in (\ref{eqn:strain1}). Defining the trace-free quadrupole moment $Q_{ij} \equiv M_{ij} - \delta_{ij} M_{kk}/3$, the leading-order gravitational-wave solution is therefore
\beq
h_{ij}^{\rm TT} (t, \textbf{x}) = \frac{2}{r} \Lambda_{ij, kl} (\hat{\textbf{n}})  \, \left[ \ddot{Q}_{kl} (t) \right]_{\rm ret}  \, . \label{eqn:ddotM2} 
\eeq
This clearly shows that an \textit{accelerating mass quadrupole} is the dominant source of gravitational waves. Importantly, monopole and dipole radiations are absent in General Relativity. While I have only shown this to be the case in linear theory, the conclusion remains unchanged in the full non-linear theory. In particular, the absence of monopole radiation is a consequence of Birkhoff's theorem~\cite{Birkhoff, Jebsen}, where all spherically spacetimes must admit the Schwarzschild solution which is time independent, cf. \S\ref{sec:BH}. On the other hand, dipole radiation is absent because, unlike electromagnetism, there are no opposite charges in gravity.

Using (\ref{eqn:ddotM2}), we can evaluate the power emitted by the accelerating source, $ P \equiv \int \d^3 \textbf{x}  \, \langle \hskip 1pt \dot{t}^{00} \hskip 1pt  \rangle $, where the averaging is performed over timescales that are much longer than the temporal variation of $Q_{ij}$. In addition, averaging the power emission over all gravitational-wave propagation directions, we find the averaged power loss
\beq
P = \frac{1}{5} \Big\langle \dddot{Q}_{ij} \dddot{Q}_{ij} \Big\rangle_{\rm ret}  \, . \label{eqn:quadrupoleformula}
\eeq
This is the famous \textit{Einstein quadrupole formula}~\cite{Einstein1918}. Similar expressions for the linear and angular momenta losses can also be obtained. Interestingly, while (\ref{eqn:quadrupoleformula}) was derived about a hundred years ago, its validity remained controversial until the 1980s, even after various conceptual issues of gravitational wave had been resolved at that time~\cite{Ehlers1976}. Observationally, the discovery of the Hulse-Taylor binary pulsar in 1974 had allowed for the first test of this formula through a measurement of the binary's orbital decay~\cite{Hulse:1974eb}. By the early 1980s, the orbital decay of the Hulse-Taylor binary was found to be in excellent agreement with (\ref{eqn:quadrupoleformula}), therefore providing strong observational confirmation of its validity~\cite{Taylor1982}. See~\cite{Kennefick:1997kb} for the history of this controversy.

\section{Compact Binary Coalescences} \label{sec:CBC}

Compact binary systems are one of the loudest and most important sources of gravitational waves in our Universe. Their inspiral motions can be highly relativistic, thereby sourcing large accelerating quadrupoles. Furthermore, binary systems are unique in that accurate computations of their gravitational waveforms, especially in the early inspiraling regime, are attainable. This makes an observed waveform a rich source of information about the binary's dynamics and the physics of its components.

\begin{figure}[t]
\centering
\includegraphics[scale=0.24, trim = 10 0 0 0]{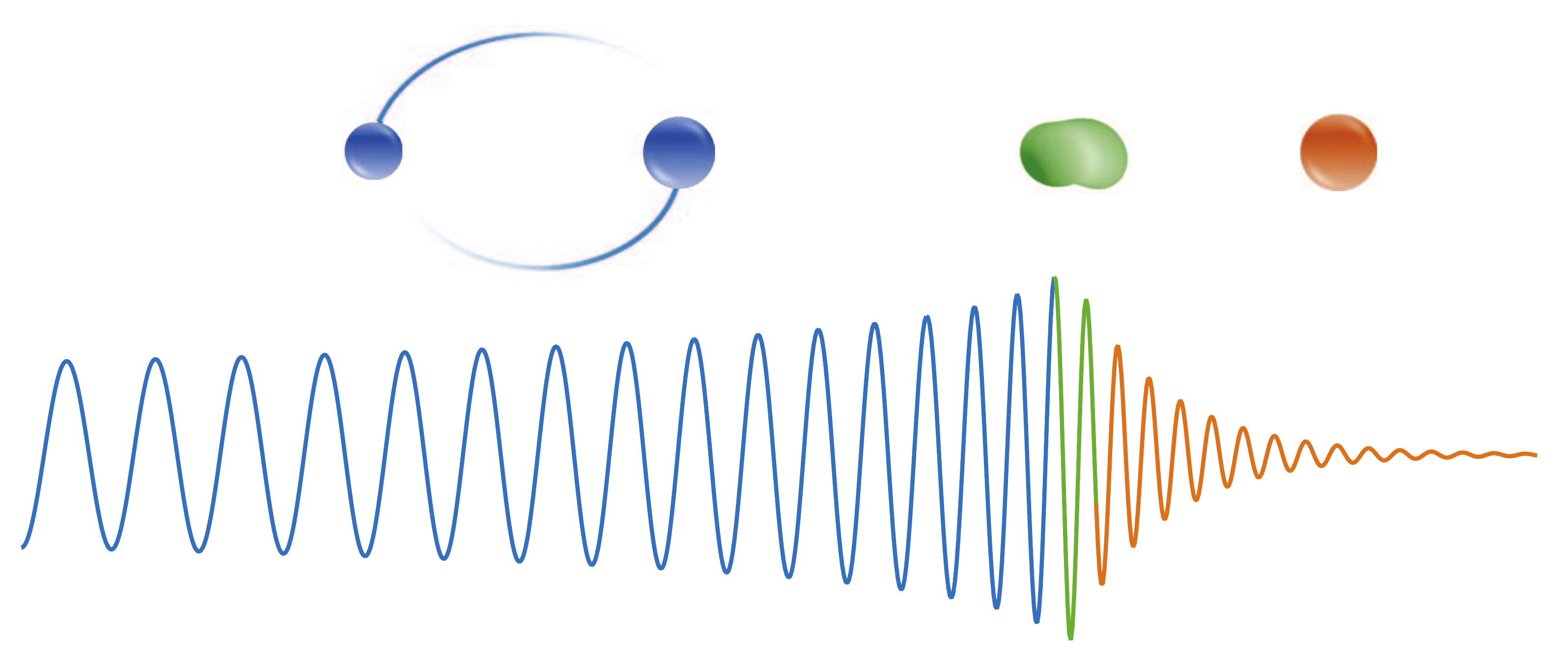}
\caption{Illustration of the inspiral (blue), merger (green), and ringdown (red) stages of a binary coalescence and its corresponding gravitational waveform. Figure adapted from~\cite{Antelis:2016icm}. }
\label{fig:IMR}
\end{figure}

I now describe the dynamics of a binary system in General Relativity. This includes both the conservative and dissipative effects on the orbital evolution. My main focus is on the dynamics in the early inspiraling stage, though I also briefly comment on the subsequent merger and ringdown regimes (see Fig.~\ref{fig:IMR} for the different regimes of a binary coalescence). I also discuss the gravitational waves emitted by binary systems. My primary goal there is to understand how various physical effects impact the waveforms of the gravitational waves emitted by the binary systems.

\subsection{Dynamics of Binary Systems} \label{sec:dynamics}

The dynamics of a two-body system in Newtonian mechanics is well understood since the publication of the \textit{Principia}. Nevertheless, relativistic corrections to a binary's Newtonian evolution are notoriously hard to compute, making them an active research problem to this date. After a brief description of how General Relativity 
reduces to Newtonian gravity, I review key facts of Keplerian orbits. I then discuss different perturbative schemes that have been developed to describe the relativistic dynamics of the two-body motion. A technical description of these perturbative methods is beyond the scope of my discussion. Rather, I aim to provide a qualitative sketch of the different physical effects at play, including those incurred by the binary components on the orbit. Crucially, my discussion includes binary systems whose components consist of general compact objects and not just black holes.

In what follows, I consider perturbative expansions of physical quantities, including the relative acceleration of the binary, $a^i$, its total energy, $E$, and its orbital angular momentum, $L^i$. It is hence convenient to schematically denote these quantites by $X \in \{ a^i, E, L^i \}$, such that
\beq
X  = X_{\rm N} + X_{\rm 1PN} + X_{\rm SO} + X_{\rm S_1 S_2} + X_{\rm Q} + X_{\rm T} + \cdots \, , \label{eqn:Xexpand}
\eeq
where the subscripts denote the Newtonian, first post-Newtonian, spin-orbit, spin-spin, quadrupole, and tidal deformability, respectively. I focus on these terms because they represent qualitatively new effects that impact the orbital dynamics.

\subsubsection*{Newtonian dynamics}

General Relativity reduces to Newtonian gravity when \textit{i)} the gravitational field is weak, and \textit{ii)} the sources move non-relativistically. More precisely, the former means that the background metric is well approximated by Minkowski spacetime; the latter implies that the spatial component of the four-velocity $v^\mu = \d x^\mu / \d \tau = (t, v^i) $ is suppressed,  $v^i \ll 1$.  With these approximations, the geodesic equation (\ref{eqn:geodesic}) reduces to the Newtonian equation of motion 
\beq
\frac{\d^2 x^i}{\d t^2} = - \partial^i U \, , \label{eqn:Newton}
\eeq
where the proper time $\tau$ is well-approximated by the coordinate time $t$ and $U = -h_{00}/2$ is the gravitational potential. In its classic ``$F=m a$" representation, (\ref{eqn:Newton}) would instead be written as (\ref{eqn:NewtonEqmImA}), which includes the inertial and gravitational masses. However, (\ref{eqn:Newton}) clearly demonstrates these masses are exactly the same, as required by the equivalence principle. Similarly, the Einstein field equation (\ref{eqn:Einstein}) reduces to the Poisson equation (\ref{eqn:Poisson}) in the weak-field and non-relativistic limit, in which case the time-time component of (\ref{eqn:Einstein}) dominates and the energy density is $\rho \equiv -T_{00}$.

Newton's equation (\ref{eqn:Newton}) applies to binary systems whose  components are well-approximated by point particles. We denote the masses of these components by $m_1$ and $m_2$. To solve the two-body problem, it is convenient to work in the center-of-mass frame, where the coordinates are given by the binary's barycenter position and the relative binary separation, $R^i = R \hat{R}^i$.\footnote{\label{footnote:Rgauge} The binary separation, $R$, though well-defined in Newtonian gravity, is ambiguous in General Relativity. In the rest of our discussion, physical quantities will still be written as functions of $R$, though we emphasize that these results are only valid in the center-of-mass frame. This gauge dependence can be eliminated by replacing $R$ with gauge-invariant quantities, such as the orbital frequency of the binary through the relativistic generalization of Kepler's law.} In this picture, the two-body system reduces effectively to a one-body description, whereby a fictitious particle with \textit{reduced mass}, $m_1 m_2 / (m_1 + m_2)$, moves in the gravitational potential sourced by the total mass $M_{\rm tot} = m_1 + m_2$. The potential experienced by this fictitious particle is therefore $U = - M_{\rm tot}/R$, while the relative acceleration, $a^i = 
\ddot R^i$, is given by
\beq
a^i_{\rm N} = - \frac{M_{\rm tot}}{R^2} \hat{R}^i \, , \label{eqn:CoMForce}
\eeq
where the subscript denotes Newtonian. We have simplified the problem, which initially has six components of spatial coordinates,  to a system with only three independent degrees of freedom. This is possible because of conservation of linear momentum, which implies that the motion of the barycenter is uniform and therefore trivial. Other conserved quantities include the energy and orbital angular momentum of the binary system, which are
\beq
\begin{aligned}
E_{\rm N} &= \nu M_{\rm tot} \left( \frac{1}{2} v^2 - \frac{M_{\rm tot}}{R} \right) , \\
L^i_{\rm N} &= \nu M_{\rm tot} \, \epsilon^{ijk} R_j v_k \, , 
\end{aligned}\label{eqn:EnLn}
\eeq
where $\nu \equiv m_1 m_2 / M^2_{\rm tot}$ is the symmetric mass ratio, $v \equiv |v^i|$ is the relative velocity of the binary, and $\epsilon_{ijk}$ is the Levi-Civita symbol. Crucially, conservation of $L^i_{\rm N}$ dictates that both its magnitude and direction do not evolve. The orbital plane, which is perpendicular to $L^i_{\rm N}$, is therefore fixed at all times. Depending on the ratio of $E_{\rm N}$ and $|L^i_{\rm N}|$, the orbit can have non-trivial eccentricity, $e$, with $e=0$ for a circular orbit and $)<e<1$ for an elliptical orbit.

Finally, we note that the orbital velocity of the binary and its gravitational potential are intrinsically related to each other. In particular, the \textit{virial theorem} states that the time-averaged (or equivalently the orbit-averaged) behaviour of the binary satisfies
\beq
v^2 = \frac{M_{\rm tot}}{R} \, . \label{eqn:virial}
\eeq
As we shall now see, relativistic corrections to the Newtonian dynamics can be computed through a perturbative expansion in $v$ or in $M_{\rm tot}/R$. The relation (\ref{eqn:virial}) will therefore be important for power counting the higher-order terms that appear in the expansions.

\subsubsection*{Relativistic dynamics}

\begin{figure}[t]
\centering
\includegraphics[scale=0.95, trim = 10 0 0 0]{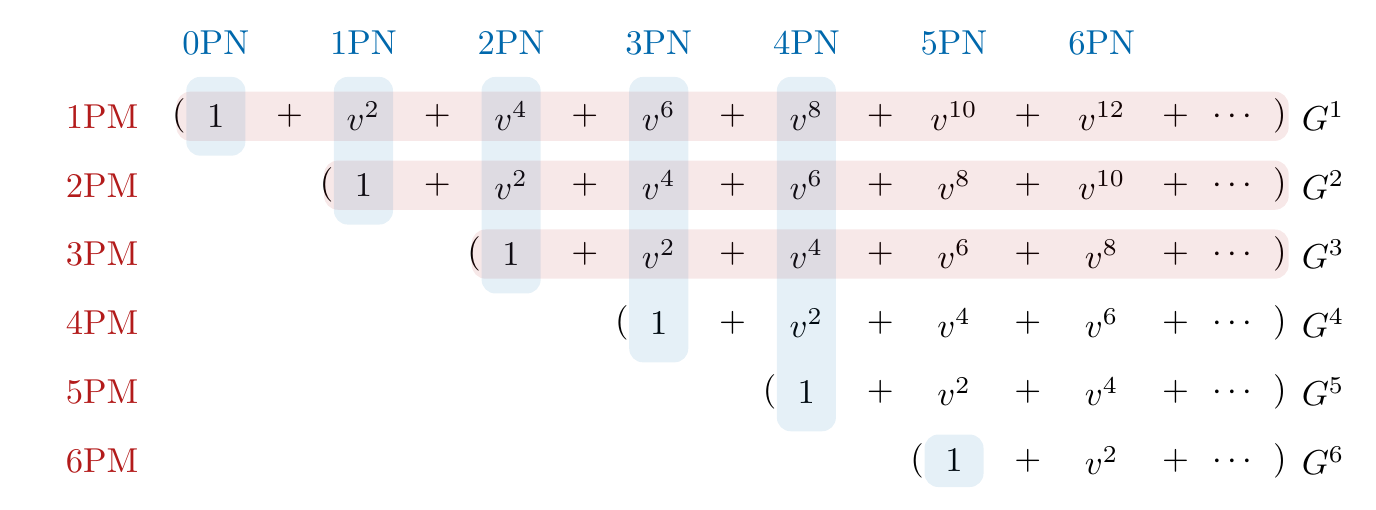}
\caption{Summary of known results for the conservative dynamics of point-particle binary systems in the PN and PM expansions. In particular, results up to 4PN and 3PM are fully known. Figure adapted from~\cite{Bern:2019crd}. }
\label{fig:PNPM}
\end{figure}

During the early inspiral stage of a binary, the gravitational field is weak and its motion remains slow. A perturbative expansion that includes relativistic dynamics can therefore be performed in this regime. When the binary components are treated as point particles, two choices of expansion parameters are possible: either in $v$, known as the \textit{post-Newtonian} (PN) expansion, or in Newton's constant $G$ (equivalently $M_{\rm tot}/R$ in units of $G=1$), known as the \textit{post-Minkowskian} (PM)  expansion.\footnote{Post-Newtonian terms at the $n$-th PN order are $v^{2n}$ suppressed compared to the leading Newtonian result. On the other hand, $n$-th order PM terms represent corrections to Minkowski spacetime at powers of $G^n$.} 
In fact, the virial theorem (\ref{eqn:virial}) implies that both expansions are a double series --- the perturbative series in either parameter would recover partial results of the other, cf. Fig.~\ref{fig:PNPM}.

In the PN expansion, a systematic expansion in $v \ll 1$ is performed on the metric perturbation $h_{\mu \nu}$, cf. (\ref{eqn:linearpert}). More precisely, this involves taking the non-relativistic limit in $T_{\mu \nu}$ and solving the linearized Einstein equation (\ref{eqn:linearEinstein}) iteratively in powers of $v$. In the point-particle approximation, the sizes of the binary components are taken to be comparable to their gravitational radii, $r_c \sim m$, in which case the virial theorem (\ref{eqn:virial}) implies that $r_c \sim R \hskip 1pt v^{2}$. On the other hand, the gravitational waves emitted by a binary has a typical wavelength $\lambda_{\rm gw} \sim R  \hskip 1pt  v^{-1}$. These scaling mean that the PN expansion can be viewed as an expansion that relies on the hierarchy $r_c \ll R \ll \lambda_{\rm gw}$.  A crucial feature of the PN expansion is that time derivatives are suppressed compared to spatial derivatives, $\partial_0 h_{\mu \nu}= v^i \partial_i h_{\mu \nu} \ll \partial_i h_{\mu \nu}$. As a result, retardation effects are treated as small corrections to instantaneous effects. This has important ramifications for the region of validity of the PN series: for small retarded times, we have
\beq
h_{\mu \nu}(t-r) = h_{\mu \nu}(t) - r \hskip 1pt \partial_0 h_{\mu \nu} + \cdots \, . \label{eqn:retardexp}
\eeq
Since $\partial_0 h_{\mu \nu} \sim h_{\mu \nu} / \lambda_{\rm gw}$, the series (\ref{eqn:retardexp}) diverges as the wave propagates over distances $r \gtrsim \lambda_{\rm gw}$. This motivates a separation of regions in the binary problem: the \textit{near zone}, $r \ll \lambda_{\rm gw}$, is a region where the PN expansion holds, while the \textit{wave zone}, $r \gg \lambda_{\rm gw}$, is where the approximation becomes invalid. By disregarding putative retardation effects, the PN expansion is only well suited to describe the \textit{conservative dynamics} of the binary. As we shall see shortly, this shortcoming is compensated by the PM expansion, which is valid in both regions.

The conservative dynamics of a point-particle binary system is symmetric under time reversal. As such, only terms that are suppressed by even powers of $v$ contribute to the relativistic corrections to the Newtonian dynamics. The first post-Newtonian correction was performed by Lorentz and Droste~\cite{Lorentz1937}, though the credit often goes to Einstein, Infeld and Hoffman~\cite{EIH1938} who generalized the result to a $N$-body system. It was found that the 1PN correction to the relative acceleration is
\beq
a^i_{\rm 1PN} = - \frac{M_{\rm tot}}{R^2} \left\{ \left[ (1+3\nu)v^2 - (4+2\nu)\frac{M_{\rm tot}}{R} - \frac{3}{2} \nu \dot{R}^2 \right] \hat{R}^i - (4-2\nu) \dot{R} \hskip 1pt v^i \right\} \, . \label{eqn:1PNforce}
\eeq
From (\ref{eqn:virial}), we see that terms in $a^i_{\rm 1PN}$ are $v^2$-suppressed relative to the Newtonian acceleration~(\ref{eqn:CoMForce}). A qualitatively new feature of (\ref{eqn:1PNforce}) is the presence of a force component along $v^i$, which perturbs the orbit from its Keplerian nature. In fact, this is precisely the origin of orbital precessions in General Relativity, such as that of Mercury around the Sun. It is also instructive to show the 1PN contributions to the energy and orbital angular momentum:
\beq
\begin{aligned}
E_{\rm 1PN} & = \nu M_{\rm tot} \left\{ \frac{3}{8} (1-3\nu) v^4 +  \frac{M_{\rm tot}}{2 R} \left[ (3+\nu) v^2  + \nu \dot{R}^2 + \frac{M_{\rm tot}}{R} \right]  \right\}  , \\
L^i_{\rm 1PN} & = L^i_{\rm N} \left[ \frac{1}{2} (1-3\nu) v^2 + (3+\nu) \frac{M_{\rm tot}}{R} \right]  . \label{eqn:E1pnL1pn}
\end{aligned}
\eeq
Comparing these expressions with their Newtonian counterparts (\ref{eqn:EnLn}), we see that the mass ratio $\nu$ first explicitly appears in these conserved quantities, up to an overall factor of $\nu M_{\rm tot}$. As we shall discuss in \S\ref{sec:waveforms}, a measurement of the 1PN contribution to the phases of gravitational waveforms would therefore allow us to infer the masses of both binary components. In addition, we find that $L^i_{\rm 1PN} $ is parallel to $L^i_{\rm N}$ --- a 1PN correction therefore does not change the orientation of the orbital plane.

Through heroic efforts over the past few decades, higher-order PN corrections to the conservative dynamics of point-particle binaries were computed via a myriad of techniques~\cite{ADM1962, Damour:1987b, Blanchet:2006zz, Buonanno:1998gg, nrgr}. Interestingly, it was found that up to 3PN order, the qualitative conclusions that we drew from the 1PN correction above remain unchanged~\cite{Ohta1973, Jaranowski:1997ky, Blanchet:2000nv, Damour:2001bu, Blanchet:2004ek, Foffa:2011ub, Schafer:2018kuf, nrgr}. This is despite the presence of various technical challenges posed at high PN orders, such as the emergence of ultraviolet (UV) divergences due to the point-particle approximation.\footnote{Logarithms that scale as $\propto \ln R$ generically appear in these high PN terms, as a result of regularizing the UV divergences. Nevertheless, as mentioned in footnote~\ref{footnote:Rgauge}, $R$ is a gauge-dependent quantity with ambiguous physical interpretation. Using (the PN-corrected) Kepler's law, $R$ can be replaced by the orbital frequency of the binary, which \textit{is} gauge invariant. In this case, one can show that these logarithms disappear, see e.g.~\cite{Blanchet:2006zz}.} To date, the point-particle dynamics has been completed up to 4PN order~\cite{Damour:2014jta, Damour:2016abl, Marchand:2017pir, Bernard:2017ktp, Foffa:2019rdf, Foffa:2019yfl}. In this case, a qualitatively new phenomenon emerges: due to the non-linearities of General Relativity, gravitational waves that were emitted by the binary in the past can backscatter onto the orbit in the future. This phenomenon, known as the \textit{tail effect}, first appears at 4PN order and contributes to the dynamics in a manner that is non-local in time~\cite{Blanchet:1987wq, Galley:2015kus, Marchand:2017pir, Foffa:2019yfl}. As a result, logarithmic terms of the form $v^8 \ln (v)$ appear in the conserved quantities $E_{\rm 4PN}$ and $L^i_{\rm 4PN}$. See~\cite{Bernard:2017ktp} for a summary of the explicit expressions of $a^i, E$ and $L^i$ up to 4PN order.

While the PN expansion is well-suited to describe the conservative dynamics in the near region, 
it is no longer valid in the wave region. This is especially important because wave retardation effects, which contribute to the \textit{dissipative dynamics} of the orbit, are missed in the PN series. This shortcoming is compensated by the PM expansion, which is valid in both the near zone and the wave zone. A classic treatment of the PM expansion involves the \textit{Landau-Lifshitz form} of the Einstein field equation~\cite{Landau1980Classical}. In particular, by introducing the inverse gothic (pseudo) metric, $\textsf{h}^{\mu \nu} \equiv \eta^{\mu \nu} -  \sqrt{-g} \hskip 1pt g^{\mu \nu}$, (\ref{eqn:Einstein}) can be written as
\beq
\square \, \textsf{h}^{\mu \nu} = - 16 \pi \tau^{\mu \nu} \, , \qquad \tau^{\mu \nu} = ( -g )\left( T^{\mu \nu} +  \Lambda^{\mu \nu} \right) \, ,  \label{eqn:relaxedEinstein}
\eeq
where $\square$ is the d'Alembertian in Minkowski spacetime, $\tau_{\mu \nu}$ is the effective (pseudo) energy-momentum tensor, and $\Lambda_{\mu \nu}$ is a tensor which consists of $\textsf{h}_{\mu \nu}$ and its derivative starting at the quadratic order $\sim (\partial \textsf{h} )^2$ (see \cite{Landau1980Classical} for its explicit expression). The Landau-Lifshitz form (\ref{eqn:relaxedEinstein}) must be supplemented by the constraint $\partial_\mu \tau^{\mu \nu} = 0$ to enforce the Bianchi identity. Because $\Lambda_{\mu \nu}$ only depends on $\textsf{h}_{\mu \nu}$, but not on $T_{\mu \nu}$, it is sometimes known heuristically as ``the energy-momentum tensor" of gravitational waves, though we described in \S\ref{sec:GW} that it is ill-defined without proper averaging. Expanding (\ref{eqn:relaxedEinstein}) to 1PM order about a Minkowski background, the contribution from $\Lambda_{\mu \nu}$ vanishes and $\textsf{h}_{\mu \nu}$ reduces to the trace-reversed metric $\bar{h}_{\mu \nu}$ introduced in \S\ref{sec:GW} --- we have recovered the linearized Einstein equation (\ref{eqn:linearEinstein}).  Higher-order PM expansions to the conservative dynamics, to which $\Lambda_{\mu \nu}$ does contribute, have been pursued over the past few decades~\cite{Bertotti:1956pxu, Kerr:1959zlt, Westpfahl:1979gu, Bel:1981be, Damour:2016gwp}. The state-of-the-art for the conservative dynamics is recently achieved at 3PM order~\cite{Bern:2019nnu, Bern:2019crd}. As demonstrated in Fig.~\ref{fig:PNPM}, results in the PM expansion are complementary to those obtained in the PN series.

The PM expansion is complementary to the PN expansion in the near zone. However, it is a necessity to describe the wave zone.
By \textit{matching} the conservative dynamics in the near zone with the gravitational waves propagating in the wave zone, we can infer the impact of a binary's evolution on the observed gravitational waves (see \S\ref{sec:GW} for more details at 1PM order). Conversely, this matching also allows us to study the backreaction of radiative effects on the orbit. The leading-order radiative-reaction acceleration is~\cite{Iyer:1995rn, Iyer:1993xi}\footnote{This expression is unique up to an additional gauge transformation in the radiative potential in the metric, see e.g.~\cite{poisson_will_2014} for a pedagogical discussion. Physical quantities, such as the orbit-averaged binding energy, are obviously not affected by such a gauge transformation.}
\beq
a^i_{\rm RR} = \frac{8 \nu}{5} \frac{M_{\rm tot}^2}{R^3} \left[ \left( 3 v^2 + \frac{17}{3} \frac{M_{\rm tot}}{R} \right) \dot{R} \hat{R}^i - \left(  v^2 + \frac{3 M_{\rm tot}}{R} \right) v^i \right]  . \label{eqn:RRforce}
\eeq
The physics of this radiative-reaction force is equivalent to the quadrupolar emission of gravitational waves~(\ref{eqn:quadrupoleformula}). Comparing (\ref{eqn:RRforce}) with the Newtonian acceleration (\ref{eqn:CoMForce}), we see that the dominant radiative-reaction force appears at 2.5PN order. Energy and angular momentum of the binary are therefore no longer conserved beyond this PN order, as they can dissipate through gravitational waves. In fact, simple dimensional analysis would have allowed us to predict the order at which dissipation first occurs: applying the quadrupole formula (\ref{eqn:quadrupoleformula}) to a binary system, we find that the power emitted by the binary is
\beq
P = \frac{32}{5} \nu^2 \left( M_{\rm tot} \Omega \right)^{10/3} \, , \label{eqn:quadbinary}
\eeq
where $\Omega$ is the orbital frequency of a circular binary and $v = ( M_{\rm tot} \Omega)^{1/3}$. Since acceleration is related to power through $P = \nu M_{\rm tot} a^i_{\rm RR} v^i $, we find $a_{\rm RR} \sim M_{\rm tot} R^{-2} v^5$, which is 2.5PN suppressed compared to the Newtonian acceleration. Since gravitational-wave dissipation is not symmetric under time reversal, radiative-reaction forces necessarily appear as odd powers of $v$. As it stands, relativistic corrections to (\ref{eqn:RRforce}) have been computed up to 3.5PN order~\cite{Pati:2002ux, Nissanke:2004er, Konigsdorffer:2003ue}. The dissipative contribution of the tail effect, like its conservative counterpart, appears at 4PN order.

\subsubsection*{Spins and finite-size effects}

So far, we have approximated the binary components as point-like particles. However, an astrophysical object generally has non-vanishing spin angular momentum. Depending on the internal structure of the object, it can also possess non-trivial quadrupole and higher-order multipole moments. Furthermore, it can acquire induced multipole moments in the presence of a gravitational perturbation. I will now describe how these different physical effects impact the dynamics of the orbit.

The dynamical evolution of a binary system is greatly enriched when its components have spins. We denote the spin vectors by $S^i_1\equiv m_1^2 \chi_1^i$ and $S^i_2 \equiv m_2^2 \chi_2^i$, where $\chi^i_{1,2}$ are the dimensionless spin parameters, which are bounded by $0 \leq | \chi^i_{1,2} | \leq 1$. Unless both spins are parallel to the orbital angular momentum of the binary, the mutual gravitational force of the binary exerts a torque on the spins. This torque is phenomenologically interesting because it leads to \textit{precessions} of the spins and the orbital plane. To understand this concretely, we consider the evolution of the total spin vector, $S^i = S^i_1 + S^i_2$, which satisfies~\cite{Barker75, HartleThorne1985, Kidder:1992fr, Kidder:1995zr}
\beq
\frac{\d S^i}{\d t} =  \frac{\epsilon^{ijk}}{2 R^3}  L^j_{\rm N} \Big[ 4 S^k + 3 \sigma^k \Big] + \frac{3 \hskip 1pt \epsilon^{ijk}}{R^3} \Big[ ( \hat{R}^l S_2^l ) ( \hat{R}^j S_1^k ) + ( \hat{R}^l S_1^l ) ( \hat{R}^j S_2^k ) \Big] \, , \label{eqn:Sprecession}
\eeq
where $\sigma^i \equiv ( m_2 / m_1) S_1^i +  (m_1/ m_2)  S_2^i $ is a mass-weighted spin parameter. The first term in (\ref{eqn:Sprecession}) is the \textit{spin-orbit interaction} and the second term is the \textit{spin-spin interaction}. We see that the total spin generally evolves in a very complicated manner. Nevertheless, the total angular momentum, $J^i = L^i + S^i$, is conserved---at least up to 2.5PN order, cf.~(\ref{eqn:RRforce}). This implies that the orbital angular momentum satisfies
$\d L^i / \d t = - \d S^i / \d t$, where both $L^i$ and $S^i$ precess around the fixed~$J^i$. To gain better intuition for the precessing motion, we simplify (\ref{eqn:Sprecession}) by taking $S^i_2=0$, such that the spin-spin interaction vanishes. In this case, the precession equations are
\beq
\frac{\d L^i}{ \d t} = \Omega_{\rm p} \, \epsilon^{ijk} \hat{J}^j L^k \, , \qquad \frac{\d S_1^i}{ \d t} = \Omega_{\rm p} \, \epsilon^{ijk} \hat{J}^j S_1^k \, , 
\eeq 
where $J^i \equiv J \hat{J}^i $ and $\Omega_{\rm p} = J (1+3M_{\rm tot} / m_1) / 2 R^3$ is the precession frequency. When the total angular momentum is dominated by the orbital angular momentum, we find that $\Omega_{\rm p} \sim \Omega \hskip 1pt v^2 < \Omega$ and the precession timescale is therefore longer than the orbital period. Crucially, the precession of $L^i$ leads to orbital precession of the binary.  Since gravitational waves are not emitted isotropically from a binary, a precessing orbit generates interesting modulations in the amplitude and phase of the observed waveform. Spin-spin precession can also occur when both $S^i_1$ and $S^i_2$ are non-vanishing. In that case, we infer from dimensional analysis, $|S^i| \sim \chi_i v |L^i| < |L^i|$, that the spin-spin precession timescale is even longer than that induced by the spin-orbit interaction.

It is instructive to consider how these spin effects contribute to the binding energy and orbital angular momentum of the binary. For the spin-orbit interaction, we have\footnote{The explicit expressions for $E_{\rm SO}$ and $L_{\rm SO}^i$ actually depend on the representative world line chosen for the binary system. This is similar to the ambiguity of the dipole moment of an object, which vanishes if the object's representative worldline is chosen to be that of its center of mass. Here, we present results evaluated along the world line that tracks the binary's centre-of-mass, as defined in the local comoving frame; see e.g.~\cite{poisson_will_2014} for a detailed discussion.}~\cite{Barker75, HartleThorne1985, Kidder:1992fr, Kidder:1995zr}
\beq
\begin{aligned}
E_{\rm SO}  &=  \frac{1}{R^3} L^i_{\rm N} \sigma^i \, , \\
 L^i_{\rm SO}  &= \nu \hskip 1pt \epsilon^{ijk} \epsilon^{klm} \left[ \frac{M_{\rm tot}}{R} \hat{R}^j \hat{R}^l \Big( 2 m_1^2 \hskip 1pt \chi ^m_1 + 2 m_2^2 \hskip 1pt  \chi^m_2 + \sigma^m \Big) - \frac{1}{2} v^j v^l \sigma^m   \right] \, . \label{eqn:Eso}
\end{aligned}
\eeq
A direct comparison between (\ref{eqn:Eso}) and (\ref{eqn:EnLn}) indicates that the spin-orbit interaction occurs at 1.5PN order. This can be understood intuitively from the fact that $L^i_{\rm N}$ couples to the gravitomagnetic potential of the Riemann tensor, which scales as $M_{\rm tot} v / R$ relative to the Newtonian potential. In addition, we find that $L^i_{\rm SO}$ is not necessarily parallel to $L^i_{\rm N}$. As the dominant component $L^i_{\rm N}$ precesses around $J^i$, $L^i_{\rm SO}$ can introduce an additional \textit{nutation} motion. On the other hand, the spin-spin contribution to the energy and orbital angular momentum are~\cite{Barker75, HartleThorne1985, Kidder:1992fr, Kidder:1995zr}
\beq
\begin{aligned}
E_{\rm S_1 S_2} &= \frac{\nu^2 M_{\rm tot}^4}{R^3} \left[  3 ( \hat{R}^i \chi^i_1 ) ( \hat{R}^i \chi^i_2 )  - \left( \chi^i_1 \chi^i_2 \right)  \right]   , \\
L^i_{\rm S_1 S_2} &= 0 \, . 
\end{aligned}\label{eqn:Ess}
\eeq
That this is a 2PN effect can be inferred straightforwardly from dimensional analysis, $|S^i| \sim \chi_i v |L^i| < |L^i|$. As it stands, higher-order corrections to the the spin-orbit interaction have been computed up to 3.5PN~\cite{Faye:2006gx, nrgrss, Bohe:2015ana} and 2PM orders~\cite{Bini:2017xzy, Bini:2018ywr}. On the other hand, the spin-spin interactions are known up to 4PN order~\cite{Levi:2016ofk} and 2PM order \cite{Vines:2017hyw, Bern:2020buy} (see~\cite{Porto:2016pyg, Levi:2018nxp} for reviews and further references therein).

In addition to the mass and spin parameters, a general astrophysical object is also characterized by its multipolar structure~\cite{Geroch:1970cd, Hansen:1974zz, Thorne:1980ru}. For simplicity, we only consider objects that are spherically symmetric when they are not spinning, such that they only acquire non-trivial multipole moments when their spins do not vanish.\footnote{This need not be the case, as a general astrophysical object can inherit higher-order permanent multipole moments, which are present even when the object is not spinning.} In this case, the dominant moment is given by the axisymmetric \textit{spin-induced quadrupole}, $Q$, which is often parameterized through~\cite{Poisson:1997ha}
\beq
    Q_a \equiv - \kappa_a \, m_a^3 \chi_a^2  \, , \label{eqn:Quad}
\eeq
 where $a = 1, 2$, the dimensionless quadrupole parameter, $\kappa$, quantifies the amount of shape deformation due to the spinning motion. The larger the (positive) value of $\kappa$, the more oblate the object is. Crucially, the value of $\kappa$ depends sensitively on the internal structure of the object. For instance, we saw in (\ref{eqn:BHmoments}) that Kerr black holes have $\kappa = 1$~\cite{Hansen:1974zz, Thorne:1980ru}, while $2\lesssim \kappa \lesssim 10$ for neutron stars, with the precise value depending on the nuclear equation of state~\cite{Laarakkers:1997hb, Pappas:2012ns}. As we shall see in later sections, more speculative objects such as superradiant boson clouds and boson stars can have $\kappa$ as large as $10^3$.

In a binary system, the quadrupole of a binary component couples to the tidal field sourced by its binary companion. This interaction provides an additional contribution to the binding energy of the orbit~\cite{Barker75,Poisson:1997ha}:
\beq
E_Q = - \frac{1}{2} \nu M_{\rm tot} \left( \frac{M_{\rm tot}}{R}\right)^3 \sum_{a=1}^2 \left( \frac{m_a}{M_{\rm tot}} \right)^2 \kappa_a  \left[ 3 \left( \hat{R}^i \chi_a^i \right)^2 - (\chi^i_a)^2 \right] \, . \label{eqn:Eq}
\eeq
Comparing this with the binding energy from the monopole (\ref{eqn:EnLn}), it is clear that this quadrupolar interaction is suppressed by $(M_{\rm tot}/R)^2$ and therefore contributes at 2PN order. That a quadrupole moment can also induce precession effects is well known in Newtonian mechanics, see e.g.~\cite{goldstein:mechanics}. Nevertheless, similar to the spin-spin interaction above, a quadrupole-induced precession occurs over timescales that are longer than the spin-orbit precession, and are therefore often neglected. While we only considered the quadrupole, higher-order multipoles can also contribute to the binding energy, though their effects are suppressed.

In addition to the spin-induced multipole moments, an astrophysical object can also develop additional multipole moments when it is perturbed by a gravitational field. In the case of a binary system, this occurs when the tidal field sourced by a binary companion, $\mathcal{E}_{ij}$, deforms the shape of the object. In reponse, the object acquires a \textit{tidally-induced quadrupole moment}, $\delta Q_{ij}$. When the tidal perturbation is adiabatic, the linear response of the object's moment to the tidal field is parameterized through the relation~\cite{Flanagan:2007ix}
\beq
\delta Q_{ij, a} \equiv - \lambda_a  \hskip 1pt  m_a^5  \mathcal{E}_{ij} \, , \label{eqn:love}
\eeq
where $\lambda$ is the object's (dimensionless) tidal deformability parameter, also known as the {\it Love number}. Remarkably, the Love numbers of a black hole vanish identically~\cite{Chia:2020yla, Binnington:2009bb, Damour:2009vw, Gurlebeck:2015xpa, Landry:2015zfa, Pani:2015hfa}. This conclusion holds for arbitrary values of black hole spin, for both the electric-type and magnetic-type perturbations, and to all orders in the multipole expansion of the tidal field~\cite{Chia:2020yla}. For neutron stars, $\lambda$ depends on the precise nuclear equation of state --- a measurement of this parameter through a phase binary neutron star waveform is indeed an active area of research, see \S\ref{sec:waveforms} for more discussion.

As binding energy is transfered to the binary components in order to deform their shapes, the orbit shrinks at an accelerating rate. This tidal effect therefore provides an additional channel to deepen the gravitational potential between the components. It contributes to $E$ in (\ref{eqn:Xexpand}) as follows~\cite{Flanagan:2007ix, Vines:2011ud}:
\beq
E_{\rm T} = - \frac{3}{2} \nu^2 M_{\rm tot} \left( \frac{M_{\rm tot}}{R} \right)^6 \, \sum_{a=1}^2  \left( \frac{m_a}{M_{\rm tot}} \right)^3 \lambda_a  \, . 
\eeq
This shows that the tidal deformability affects the orbital evolution at 5PN order, which is  a much higher-order effect than those described above. However, this does not necessarily mean that this effect is negligible, as $\lambda$ can be very large. As we shall see in \S\ref{sec:waveforms}, a high-PN order only implies that it is suppressed in the early inspiral stage of the binary but can be significant near the binary merger. Higher-order corrections of the tidal effects on the orbital dynamics have also been computed~\cite{Vines:2011ud, Banihashemi:2018xfb}.

\subsubsection*{Extreme mass ratio inspirals}

The PN and PM expansions assume that the self-gravity of a binary system is weak, so that the background spacetime is well-approximated by the Minkowski metric. However, this is no longer true when one of the binary components is much heavier than the other. In this case, strong gravitational effects of the heavier object are important even at leading order. The heavier object is usually taken to be a Kerr black hole because, in contrast to other astrophysical objects, it has no maximum mass. The mass ratio of these types of binary systems can easily be as small as $\nu \sim 10^{-5}$, and they are therefore commonly called \textit{extreme mass ratio inspiral} (EMRI) systems. 
The smallness of the mass ratio can be leveraged to  systematically expand the evolution equations in $\nu \ll 1$. Crucially, this perturbative expansion is complementary to the PN and PM expansions, as the results obtained at each order in $\nu$ necessarily restores partial results in $v$ and $G$ (see Fig.~\ref{fig:PNPM} for a similar relationship between the PN and PM series).

In the test particle limit, $\nu \to 0$, the motion of the smaller object traces a geodesic around the Kerr black hole~\cite{Bardeen1972}. However, the trajectory is no longer a geodesic when the self gravity of the object is non-negligible. This is because the object's motion now becomes a source of gravitational-wave emission, which backreacts on its orbit through a radiation-reaction force. These finite-$\nu$ corrections can be computed through the \textit{gravitational self-force} formalism (see~\cite{Barack:2018yvs} for a non-specialist review on this subject). To highlight some of the qualitative distinctions between the self-force theory and the PN and PM formalisms, we generalize the linearized Einstein equation (\ref{eqn:linearEinstein}) from Minkowski spacetime to an arbitrary background spacetime $g_{\mu \nu}^{(0)}$:
\beq
\square \, \bar{h}_{\mu \nu} + 2 R^{(0)}_{\mu \alpha \nu \beta} \bar{h}^{\alpha \beta} = - 16\pi T_{\mu \nu} \, , \label{eqn:linearEinsteinGeneral}
\eeq
where $R^{(0)}_{\mu \nu \rho \sigma}$ is the Riemann tensor of the background metric and $\bar{h}_{\mu \nu}$ is the trace-reversed metric perturbation, but now defined relative to $g_{\mu \nu}^{(0)}$. Modeling the smaller object as a point particle that sources $T_{\mu \nu}$ and denoting its coordinates by $z^\mu$, we can derive its equation of motion at first order in perturbation theory. This is known as the MiSaTaQuWa equation~\cite{Mino:1996nk, Quinn:1996am}
\beq
\frac{\text{D}^2 z^\mu}{\d \tau^2 } = - \frac{1}{2} g_{(0)}^{\mu \nu} \left( 2 h^{\rm tail}_{\nu \rho \sigma} - h^{\rm tail}_{\rho \sigma \nu} \right) u^\rho u^\sigma \, , \label{eqn:MiSaTaQuWa}
\eeq
where $h^{\rm tail}_{\mu \nu \rho} \equiv m \int  \d \tau \, \nabla_\rho G^{+}_{\mu \nu \alpha \beta } u^\alpha u^\beta$ is the \textit{gravitational tail field} generated by object's radiation reaction, $u^\mu \equiv \d z^\mu/\d \tau$ is the body's four-velocity, and $G^{+}_{\mu \nu \alpha \beta }$ is the retarded Green's function. The tail perturbation $h^{\rm tail}_{\mu \nu \rho}$, which depends on the object's entire worldline history, is exactly the same as that described in the PN expansion above. In that case, this effect only contributes at 4PN order. However, we see in (\ref{eqn:MiSaTaQuWa}) that the tail effect already appears at leading order and is, in fact, the only source of perturbation. This means the curvature of the background spacetime $g^{(0)}_{\mu \nu}$ is so strong that it can efficiently backscatter the gravitational waves emitted by the object's past to its future. The self force of an EMRI is therefore a radiation-reaction force that is dominated by the non-local tail effects.

The derivation of (\ref{eqn:MiSaTaQuWa}) marked only the beginning of the self-force program. For instance, it was shown that second order in perturbation theory must be achieved in order to construct template waveforms that are accurate enough to detect EMRIs through matched-filtering searches~\cite{Hinderer:2008dm}. At this perturbative order, the point-particle approximation leads to unphysical UV divergences and therefore introduces new technical challenges. This is certainly reminiscent of the similar problem of the point-particle approximation in the post-Newtonian expansion at 3PN order. In addition, the equations involved are coupled partial differential equations that are challenging to solve numerically. These problems motivated further developments in the self-force theory~\cite{Barack:2018yvs}. 
At present, the equation of motion has been pursued up to the second order~\cite{Gralla:2012db, Pound:2012nt, Pound:2017psq, Pound:2019lzj}. An overaching goal of the current self-force program is to develop numerically efficient methods to evolve the equation over long timescales, so that the gravitational waveforms emitted by EMRI systems can be readily obtained.

In addition to the formalism described above, the dynamics of EMRIs can also be computed using the \textit{Teukolsky equation}, which computes the linear perturbations on a fixed background black hole geometry (see e.g.~\cite{Sasaki:2003xr} for a review). Schematically, this approach involves solving the Teukolsky equation perturbatively in $\nu$, with the smaller object acting as a source term in the inhomogeneous equation~\cite{Teukolsky:1973ha, Mano:1996vt}. This approach is convenient because the Teukolsky equations are a set of ordinary differential equations, instead of partial differential equations, making them more computationally tractable. In addition, since the Teukolsky 
equation provides solutions for the gravitational waves radiated towards infinity, it is most useful for computing quantities such as the energy flux of the binary system. Indeed, this method has provided powerful consistency checks and even new results for the PN series in the radiative sector. For instance, at linear order in $\nu$ , recent endeavours had been able to compute the energy flux up to a remarkable 22PN order for a Schwarzschild black hole~\cite{Fujita:2012cm} and 11PN order for a Kerr black hole~\cite{Fujita:2014eta}. To derive the self-force correction to the orbit's conservative dynamics in this approach, one would have to employ metric reconstruct methods in the binary's near zone~\cite{Shah:2012gu, Keidl:2010pm}. The black hole perturbation theory formalism is therefore, in many ways, an important and complementary method of computing the self-force dynamics of EMRI systems.

Finally, we remark that the gravitational waves emitted by EMRIs provide a unique probe of the spacetime geometry of Kerr black holes. Specifically, by measuring the mass, spin, and higher-order moments of the black hole, we can directly test the ``no hair theorem" for black holes. A violation of this consistency condition, such as the measurement of a much larger quadrupole moment, would imply the presence of a more exotic compact object~\cite{Ryan:1995wh}. 

\subsubsection*{Binary mergers and ringdown}

So far, we have focused on the inspiral regime of the binary's evolution, where analytic predictions are attainable. Those perturbative schemes, however, break down in the last few orbits when strong non-linear gravitational effects dominate. In this case, \textit{numerical relativity} is necessary to resolve the detailed dynamics, which include both the merger and ringdown stages, cf.~Fig.~\ref{fig:IMR} for the different parts of a binary coalescence.

The goal in numerical relativity is to solve the ten coupled nonlinear partial differential equations of the Einstein field equation (\ref{eqn:Einstein}). Among the many technical challenges involved, the constraints imposed by the Bianchi identity, $\nabla_\mu G^{\mu \nu} = 0$, also known as the Hamiltonian and momentum constraints in the $3+1$ ADM formalism, must be enforced throughout the binary's evolution. In addition, choosing the appropriate gauge is essential in order to guarantee well-posedness of the binary setup, lest numerical errors due to finite resolution grow quickly in time. After decades of~attempts and technical developments, stable numeric evolutions of binary black hole mergers were finally achieved in 2005~\cite{Pretorius:2005gq, Baker:2005vv, Campanelli:2005dd}. Ever since this breakthrough,  many similar simulations were performed over a wider range of parameters and through different methods, see e.g.~\cite{Kidder:2000yq, Loffler:2011ay, Moesta:2013dna, Brugmann:2008zz, Clough:2015sqa}. Because these simulations are computationally expensive, they resolve at most the last $100$ orbits---see e.g.~\cite{Jani:2016wkt, Healy:2019jyf, Boyle:2019kee} for a catalogue of binary black hole simulations. Importantly, these results include both the merger and ringdown portions of the binary's evolution. In \S\ref{sec:waveforms}, we describe how these numerical results are combined with the analytic, inspiral part of the gravitational waveform, thereby constructing waveforms which incorporate the full evolution of the binary.

Last but not least, while the quasi-normal ringdown spectra emitted after a binary merger are natural outputs of numerical relativity simulations, they can also be computed analytically using black hole perturbation theory~\cite{Teukolsky:1973ha}. Schematically, the ringdown strain can be expressed parameterically as a linear superposition of damped sinusoids~\cite{Teukolsky:1973ha, Press:1971wr, Chandrasekhar:1975zza}
\beq
h(t) = \sum_{p=0}^\infty A_p \, e^{- i (\omega_p t + \phi_p ) - t/\tau_p} \, , 
\eeq
where $p$ denotes the overtone number, the quantities $\{A_p, \omega_p, \phi_p, \tau_p\}$ are the associated mode amplitude, frequency, phase, and decay timescale, respectively, and $t$ represents the time after a reference start time at merger. The lowest overtone, $p=0$, is  the longest-lived mode and has the largest amplitude. Crucially, the ringdown frequencies and decay timescales only depend on the mass and spin of the final black hole, and therefore offer a clean probe of the ``no hair theorem" of Kerr black holes described in \S\ref{sec:BH}~\cite{Dreyer:2003bv}. On the other hand, the amplitudes and phases depend on the geometry of the binary before merger, which dictates the degree to which these modes are excited~\cite{Lim:2019xrb}.

\subsection{Gravitational Waveforms} \label{sec:waveforms}

In \S\ref{sec:dynamics}, I described how various analytic approximations have been adopted to compute the dynamics of binary systems to high perturbative orders. These are not pure academic exercises --- rather, they are foundational in our efforts to search for the gravitational waves emitted by binary systems. This is because gravitational-wave signals are typically weak and 
buried in noisy data. The optimal way of recovering these signals is to \textit{matched filter} the data with very precise template waveforms~\cite{Thorne1980Lectures, Dhurandhar:1992mw, Cutler:1992tc}, which can only be constructed through a detailed understanding of the dynamics at play. As we shall see, a minimum accuracy of 3.5PN order in the phase of template waveforms is necessary to successfully perform matched filtering. My primary goal here is to sketch the methods used in constructing template waveforms. I will also demonstrate how 
gravitational waveforms are rich sources of information about the physics of the binary systems.

I first focus on the inspiral stage of the binary's evolution. For simplicity, I restrict myself to binaries in quasi-circular orbits. This is often a reasonable simplification because gravitational-wave emissions can efficiently damp away orbital eccentricities~\cite{PeterMatthews1963, Peter1964}. I also assume that the orbit evolves adiabatically, i.e. its tangential velocity dominates over the speed at which the orbital radius shrinks. This is an excellent approximation for the binary's inspiral stage and stops being valid only near its merger. In this case, Kepler's law implies that the velocity of the orbit is $v = ( M_{\rm tot} \Omega)^{1/3} $ and the quadrupole formula for a binary system (\ref{eqn:quadbinary}) is
\beq
P = \frac{32}{5} \nu^2 v^{10}  \, . \label{eqn:PbinaryQ}
\eeq
In order to construct waveforms up to a given PN order, $P$ must also be computed up to the same order (see discussion later). Higher-order radiative multipoles of the binary, such as its mass octupole and current quadrupole, must then be taken into account, cf.~\S\ref{sec:GW} for a related discussion. In addition, radiative tail effects, which begin to contribute to $P$ at 1.5PN, must be included.\footnote{The PN counting can sometimes be confusing and warrants further clarification. For instance, we mentioned in \S\ref{sec:dynamics} that the tail effect first appears at 4PN order in the binary's acceleration. In the dissipative sector, however, the PN counting has to be compared with the leading-order radiation-reaction acceleration, which appears at 2.5PN order. Since the latter contributes to the dominant energy flux (\ref{eqn:PbinaryQ}), the dissipative tail effect therefore appears as a 1.5PN correction in $P$.} At present, the point-particle energy flux is nearly completed up to 4PN order~\cite{Blanchet:1995fg, Blanchet:2001aw, Marchand:2020fpt}, while corrections from spin effects have been achieved up to various degrees of precision, see~\cite{Porto:2016pyg, Levi:2018nxp} for reviews.

Since $E$ is derived in the near zone, while $P$ is obtained in the wave zone, cf.~\S\ref{sec:dynamics}, we match the physics of these regions through the energy balance argument
\beq
P = - \left[ \hskip 1pt \frac{\d E}{\d t} \hskip 1pt  \right]_{\rm ret}  \label{eqn:Ebalance}\, .
\eeq
Solving (\ref{eqn:Ebalance}) allows us to determine the evolution of an orbit and its corresponding gravitational waveform. More precisely, the binary separation varies as $[ \hskip 1pt \dot{R} \hskip 1pt ]_{\rm ret} = - P / (\d E / \d R)$, while the phase of the gravitational waveform evolves as $[ \hskip 1pt \dot \phi_{\rm gw} \hskip 1pt ]_{\rm ret} = 2 \pi f  = 2 v^3/M_{\rm tot}$.\footnote{Here, we used the fact that for the dominant quadrupolar emission (\ref{eqn:PbinaryQ}) generates waves with a frequency that is twice the orbital frequency of the binary, $f = 2f_{\rm orb}$, with $\Omega = 2 \pi f_{\rm orb}$ and $v = (M \Omega)^{1/3}$. However, higher-order radiative multipoles in $P$ can generate waves with harmonics of $f_{\rm orb}$,  i.e.~$f = k f_{\rm orb}$ with $k \geq 1$, though their amplitudes are  suppressed~\cite{Blanchet:1996pi}. By ignoring these higher harmonics, we are implicitly working with a \textit{restricted waveform}. \label{footnote:circlev}} Using Kepler's law, these evolution equations can be written as differential equations in $[ \hskip 1pt \dot{v} \hskip 1pt ]_{\rm ret}$ and $[ \hskip 1pt \dot \phi_{\rm gw} \hskip 1pt ]_{\rm ret}$, whose solutions are 
\beq
\begin{aligned}
t_r(v) & = t_c + \int^{v_c}_{v} \d v^\prime \, \frac{\d E(v^\prime) / \d v^\prime}{P(v^\prime)}  \, , \\
 \phi_{\rm gw}(v) & = \phi_c + \frac{2}{M_{\rm tot} } \int^{v_c}_{v} \d v^\prime v^{\prime 3} \, \frac{\d E(v^\prime) / \d v^\prime}{P(v^\prime)} \, ,  \label{eqn:timedomainevolution}
\end{aligned}
\eeq
where $\phi_c$ and $t_c$ represent the phase and the retarded time at a reference velocity $v_c$. The time-domain gravitational waveforms as observed by detectors at time $t$ are $h(t) = h_+(t) F_+ + h_\times (t) F_\times$, where $F_{+, \times}$ are the detector's antenna-pattern functions. The polarizations of the gravitational waves schematically read
\beq
\begin{aligned}
h_+(t) = \mathcal{A}_+(t_r) \cos \left[ \phi_{\rm gw}(t_r) \right] \, , \qquad h_\times = \mathcal{A}_\times (t_r) \sin \left[ \phi_{\rm gw}(t_r) \right] \, , \label{eqn:hTime}
\end{aligned}
\eeq
where recall that $t_r \equiv t-r$, and $\mathcal{A}_{+, \times}$ are the strain amplitudes --- cf. (\ref{eqn:ddotM2}) for the relationship between $\mathcal{A}$ at leading order and the acceleration of the quadrupole moment. Higher PN corrections to $\mathcal{A}$, which also generate higher harmonics of $f$ (see footnote~\ref{footnote:circlev}), can be found in e.g.~\cite{Blanchet:1996pi}.

In principle, the solutions in (\ref{eqn:timedomainevolution}) can be obtained straightforwardly by substituting $E$ and $P$, each expanded to the required PN order, into the integrands. However, since the PN expansion is only an asymptotic series, there is an ambiguity in the way in which the ratio $(\d E/\d v ) / P$ can be treated. For instance, do we keep this ratio unexpanded, such that the integrand is given by a ratio of polynomials in $v$? Or, do we perform a Taylor expansion to the required PN order, with the denominator eliminated? Unfortunately, there is no clear resolution to this problem. This has led to multiple treatments of this ratio, thereby resulting in different PN waveform approximants, such as the Taylor T1, Taylor T2, Taylor T3, Taylor T4, and Taylor Et models; see e.g.~\cite{Damour:2000zb, Buonanno:2009zt} for a summary of their differences. Resummation methods have also been proposed to achieve better convergence~\cite{Damour:1997ub, Damour:2007xr}. In any case, it was found that these waveform approximants are in good agreement for low-mass binary system, though appreciable deviations can occur for large-mass binaries~\cite{Buonanno:2009zt}.

Since gravitational-wave data analysis is most conveniently performed in the frequency domain, it is desirable to Fourier transform (\ref{eqn:hTime}). This is not as straightforward as it may seem, because the signal only lasts for a finite period of time $-\infty < t < t_c + r$, while a Fourier transformation requires the integral domain to be $-\infty < t < \infty$. A common approach is to use the \textit{stationary phase approximation}~\cite{Thorne1980Lectures, Cutler:1994ys}, whereby the integral is evaluated at the value of $t$ that dominates the Fourier integral. More precisely, this is the time that minimizes the phase of the integrand, which oscillates quickly otherwise and therefore provides suppressed contributions. In this approximation, the strain amplitude in the frequency domain is schematically
\beq
\tilde{h} (f) = \mathcal{A} (f) f^{-7/6} e^{i \psi (f)} \, , \label{eqn:hfreq}
\eeq
where we extracted an overall factor of $f^{-7/6}$ for convenience. At leading Newtonian order, the amplitude $\mathcal{A}$ is then independent of $f$ and reads
\beq
\mathcal{A}_{+, \times} = \frac{1}{\pi^{2/3}}\left( \frac{5}{24}\right)^{1/2} \frac{\mathcal{M}^{5/6}}{r}  g_{+, \times} (\iota) \, , 
\eeq
where $\mathcal{M} \equiv \nu^{3/5} M_{\rm tot}$ is the \textit{chirp mass} of the binary, $r$ is the (luminosity) distance between the source and the detector, and the functions $g_+ = (1+\cos^2 \iota)/2$ and $g_\times = \cos \iota$ describe the angular distributions of the gravitational-wave polarizations, with $\iota$ being the inclination angle between the orbit's (Newtonian) orbital angular momentum vector and the detector's line of sight. Since higher-order corrections to $\mathcal{A}$ do not substantially affect matched filtering, they will be neglected for simplicity. On the other hand, the phase $\psi$ in the stationary-phase approximation is
\beq
\psi_+(f) = 2\pi f t_c  - \phi_{\rm gw} (f) - \frac{\pi}{4} \, ,
\eeq 
where $\phi_{\rm gw}$ is given by (\ref{eqn:timedomainevolution}) and $\psi_\times = \psi_+ + \pi/2$. Unlike in $\mathcal{A}$, high PN terms must be retained  in $\psi$ because match-filtering is extremely sensitive to the phase coherence between waveforms. A commonly-adopted frequency-domain waveform is the TaylorF2 waveform, where higher-order PN terms in the denominator of $\phi_{\rm gw}$ are Taylor expanded, so that the $v$-integral can performed analytically, see e.g.~\cite{Arun:2008kb}. The phase then is
\beq
\begin{aligned}
\psi^{\rm (F2)}_+ (f)  = & \,\, 2 \pi f t_c - \phi_c  + \frac{3}{128 \hskip 1pt \nu \hskip 1pt v^5} \Big[ \psi_{\rm PP} + \psi_{\rm S} + \psi_{\rm T} \Big] - \frac{\pi}{4} \, , \label{eqn:GWphase}
\end{aligned}
\eeq 
where $\psi_{\rm PP}, \psi_{\rm S}$ and $\psi_{\rm T}$ are the point-particle, spin-dependent, and the tidal contributions to the phase, respectively. Up to 3.5PN order, $\psi_{\rm PP}$ is~\cite{Arun:2004hn} 
\beq
\begin{aligned}
\hskip -5pt \psi_{\rm PP} = & 1 + \left( \frac{3715}{756} + \frac{55}{9} \nu \right) v^2 - 16 \pi  v^3  +  \bigg( \frac{15293365}{508032} + \frac{27145}{504} \nu  + \frac{3085}{72} \nu^2 \bigg) v^4 \\[2pt] & + \pi \left( \frac{38645}{756} - \frac{65}{9} \nu \right) \Big[ 1 + 3 \log ( v )  \Big] v^5  + \bigg[ \bigg( \frac{11583231236531}{4694215680} - \frac{640}{3} \pi^2 \\[2pt]
&  - \frac{6848 }{21}  \gamma_E \bigg)  + \left( \frac{2255 \pi^2}{12} - \frac{15737765635}{3048192}  \right) \nu + \frac{76055}{1728} \nu^2 - \frac{127825}{1296} \nu^3 \\[2pt]
&   - \frac{6848 }{21}  \log (4 v)  \bigg] v^6  + \pi \left( \frac{77096675}{254016} + \frac{378515}{1512} \nu - \frac{74045}{756} \nu^2 \right) v^7 \, , \hskip -10pt  \label{eqn:psiNS}
\end{aligned}
\eeq
where $\gamma_E$ is the Euler-Mascheroni constant. We see that the leading-order $v$-dependence in $\psi$ is $v^{-5}$, which formally diverges as $v \to 0$. Since matched filtering is only successful if the error in the waveform's phase is $\lesssim \mathcal{O}(1)$ radians, simple power counting suggests that terms up to 2.5PN order in $\psi$ must be retained. Nevertheless, since the coefficients of powers of $v$ can exceed order unity, as clearly shown in (\ref{eqn:psiNS}), a 2.5PN-accurate phase remains insufficient. Early estimates have instead shown that a minimum of 3.5PN accuracy is required for successful matched-filtering searches~\cite{Cutler:1992tc}.

To extract the parameters of the binary, notice that the leading $v$-dependence in (\ref{eqn:GWphase}) is proportional to the chirp mass of the binary, $(\nu v^5)^{-1} \propto \left( \mathcal{M} f \right)^{-5/3}$. A simultaneous measurement of $\psi$ and $\mathcal{A}$ would therefore allow us to infer the luminosity distance $r$, up to uncertainties in the orbit's inclination angle $\iota$. Moreover, a 1PN measurement of $\psi$ would allow us to determine $\nu$, thereby breaking the degeneracy between the component masses in $\mathcal{M}$. Interesting, we find that odd powers and logarithms of $v$ are present in (\ref{eqn:psiNS}), despite their absence in the point-particle conservative dynamics up to 3.5PN order, cf.~\S\ref{sec:dynamics}. These contributions arise from higher-order corrections to $P$, such as the dissipative tail effect.

For binary systems with spinning components, precession effects lead to modulations in the amplitude and the phase of the waveforms~\cite{Apostolatos:1994mx, Apostolatos:1995pj, Buonanno:2002fy}. For simplicity, we assume that the objects' spins are parallel to the (Newtonian) orbital angular momentum of the binary, i.e.~it is either aligned or anti-aligned, such that precession effects may be ignored. In this case, the spin-dependent phase can be decomposed as $\psi_{\rm S} = \psi_{\rm SO} + \psi_{\rm S_1 S_2} + \psi_{\rm S^2} + \psi_{\rm Q}$,  which are the spin-orbit, spin-spin, self-spin and spin-induced quadrupolar contributions. At their respective leading orders, these terms are~\cite{Poisson:1995ef, Poisson:1997ha, Mikoczi:2005dn, Arun:2008kb}
\beq
\begin{aligned}
\psi_{\rm SO} & = \sum_{a=1}^2 \left[ \frac{113}{12} \left(\frac{m_a}{M_{\rm tot}} \right)^2 + \frac{25}{4}\nu \right] \chi^i_a \hat{L}^i_{\rm N} \, v^3 \, , \\
 \psi_{\rm S_1 S_2} & = \nu \left[ - \frac{3605}{24} ( \chi_1^i \hat{L}_{\rm N}^i ) ( \chi_2^i \hat{L}_{\rm N}^i ) + \frac{1235}{24} \chi_1^i \chi_2^i \right] v^4 \, , \\
 \psi_{\rm S^2} & =  -  \frac{5}{48} \sum_{a=1}^2  \left( \frac{m_a}{M_{\rm tot}} \right)^2 \left[ 7 \chi^2_a - \left( \chi_a^i \hat{L}_{\rm N}^i \right)^2 \right]  v^4  \, ,  \\
\psi_{\rm Q}  & = - 50 \sum_{a=1}^2  \, \left( \frac{m_a}{M_{\rm tot}} \right)^2 \kappa_a \hskip 1pt  \chi_a^2 \, v^4 \, .  \label{eqn:spinphase}
\end{aligned}
\eeq
where $\hat{L}_{\rm N}^i$ is the unit vector of the orbital angular momentum, $L^i_{\rm N} = L_{\rm N} \hat{L}^i_{\rm N} $. We see that the spin-orbit interaction first appears at 1.5PN order, while the spin-spin and spin-induced quadrupole terms first appear at 2PN order, which are expected from their effects on the conservative dynamics, cf. \S\ref{sec:dynamics}. Crucially, there exists a new self-spin term $ \psi_{\rm S^2}$ at 2PN order, which arises from the current-quadrupolar contribution to $P$ in the dissipative sector~\cite{Mikoczi:2005dn}.\footnote{The spin-spin interaction that contributes to the phase is often denoted by spin(1)-spin(2), in order to distinguish it from these self-spinning terms, which are instead labeled by spin(1)-spin(1) and spin(2)-spin(2).} Higher-order corrections to these different effects can be found in~\cite{Mishra:2016whh, Krishnendu:2017shb}. Through precise measurements of the phases of waveforms, we can therefore determine the spins and quadrupole moments of the binary components.

The microphysical properties of non-spinning binary components can still be probed through their tidal deformabilities. In this case, the leading-order phase imprint is~\cite{Flanagan:2007ix}
\beq
\psi_{\rm T} = - 24 \sum_{a=1}^2 \left( \frac{m_a}{M_{\rm tot}} \right)^4 \left( 1 + 11\nu \frac{M_{\rm tot}}{m_a} \right) \lambda_a \, v^{10} \, . \label{eqn:TLNphase}
\eeq
While this 5PN effect may naively seem negligible, it is only suppressed in the early inspiral regime of the binary. Specifically, we saw in (\ref{eqn:GWphase}) that the leading $v$-dependence scales as $v^{-5}$. Terms that appear at $\gtrsim 2.5$PN can therefore become appreciable when $v$ approaches unity, before the adiabatic approximation breaks down. The dephasing (\ref{eqn:TLNphase}) can therefore be appreciable near the merger regime of the binary.

The analytic waveform described above only applies to the binary's inspiral stage, where $v$ is small and the adiabatic approximation holds. However, as the binary approaches merger, the adiabatic approximation breaks down and numerical relativity becomes necessary. Since it is important to construct template waveforms that are valid throughout the inspiral, merger, and ringdown stages, certain methods must be introduced in order to match the analytic and numeric waveforms. One of the most commonly adopted methods involves constructing \textit{phenomenological waveforms}~\cite{Husa:2015iqa, Khan:2015jqa}, whereby the Fourier-domain analytic waveform is matched with numerical relativity waveforms in an overlapping region. The matching in this overlapping region is performed by introducing phenomenological terms and finding the best-fitting parameters. An alternative approach is the \textit{effective-one-body} (EOB) method, which maps the conservative PN dynamics of a binary system to an effective system that involves a test particle orbiting a deformed Kerr metric~\cite{Buonanno:1998gg, Buonanno:2000ef}. In this case, numerical relativity results are used to calibrate certain free coefficients that exist in the associated effective Hamiltonian, see e.g.~\cite{Taracchini:2013rva}. Both the phenomenological and EOB waveforms have been employed extensively in existing gravitational-wave searches. A major goal in waveform modeling is to refine these models in order to increase their accuracy and include additional physics such as spins and finite-size effects, e.g.~\cite{Pratten:2020ceb, Thompson:2020nei, Matas:2020wab, Ossokine:2020kjp}.

\section{Gravitational-Wave Observatories} \label{sec:detectors}

\begin{figure}[b!]
\centering
\includegraphics[scale=0.75, trim = 10 0 0 0]{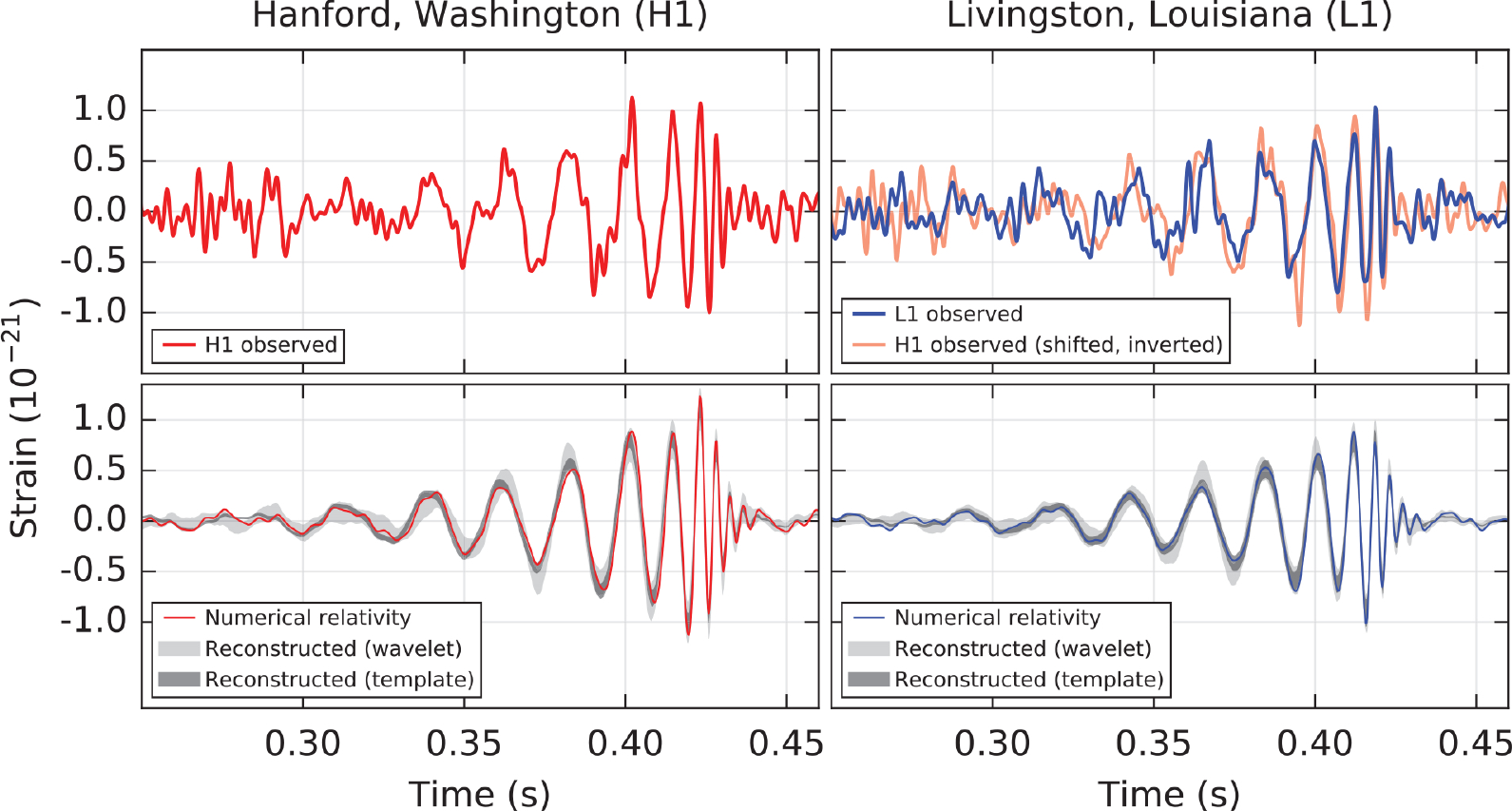}
\caption{The gravitational waveforms of GW150914, the first binary black hole merger event observed by LIGO. The top row demonstrates the strain data in the LIGO detectors, while the second row illustrates the best-fitting template waveform for the signal. Figure adapted from~\cite{Abbott:2016blz}. }
\label{fig:GW150914}
\end{figure}

In the previous sections, I reviewed theoretical aspects of black holes, gravitational waves, and the relativistic dynamics of binary systems. Remarkably, these 
predictions of General Relativity were all confirmed 
by the LIGO observatories in a binary black hole merger event in September 2015~\cite{Abbott:2016blz} (see Fig. \ref{fig:GW150914} for the signal waveforms). This discovery was followed by multiple detections of binary coalescences, including the historic discovery of a binary neutron star merger~\cite{TheLIGOScientific:2017qsa}. Here I review the status of current detectors and highlight some of their key scientific achievements to this date. In addition, I provide a summary of prospective future gravitational-wave detectors.

\subsection{Current Detectors} \label{sec:currentdetectors}

The currently operating gravitational-wave detectors include the two Advanced LIGO detectors in Hanford and Livingston, United States~\cite{TheLIGOScientific:2014jea}, as well as the Advanced Virgo detector in Pisa, Italy~\cite{TheVirgo:2014hva}.  These observatories are L-shaped Michelson interferometers whose laser arms are $4 \hskip 1pt $km and $3 \hskip 1pt $km long, respectively. With advanced engineering designs, such as Fabry-Perot cavities in the detector arms, LIGO and Virgo are able to measure differential changes in arm lengths to a precision of $\mathcal{O}(10^{-23}) \, \text{Hz}^{-1/2}$ in their most sensitive frequency range, approximate at $30 \hskip 1pt \text{Hz} - 2 \hskip 1pt  \text{kHz}$. A representative noise curve of Advanced LIGO at design sensitivity is shown in Fig.~\ref{fig:detectorcurves}. After the first detection of a binary black hole merger by LIGO in the first observing run (O1), the Virgo detector joined the subsequent runs O2 and O3. The addition of Virgo in this global network of observatories has allowed for significant improvements in the sky localization and signal detectability of gravitational-wave sources.

\vskip 4pt

The scientific achievements of LIGO and Virgo thus far are hard to overstate. To begin, the excellent agreement between the signal and template waveforms of binary black holes is yet another confirmation
of Einstein's theory of General Relativity as the
foundational description of gravity, even in the strongly dynamical regime. These observations have also provided a unique probe of various properties of black holes. For instance, by observing the quasi-normal modes in the ringdown stage of merged black holes, we were able to infer the nature of the black hole horizons~\cite{Dreyer:2003bv, Meidam:2014jpa, Isi:2019aib, Cardoso:2019rvt}. The observed binary black hole signals have also revealed previously unknown populations of astrophysical black holes, as the masses of these black holes lie beyond the expected mass range of those found in X-ray binaries~\cite{Remillard:2006fc}. A catalogue of binary black hole signals that were observed in O1 and O2, including both highly significant and marginal events, was published by the LIGO/Virgo collaboration~\cite{LIGOScientific:2018mvr}. See also~\cite{Venumadhav:2019lyq, Nitz:2019hdf} for additional binary events reported by independent research groups. 

\vskip 4pt

The observation of the first binary neutron star coalescence~\cite{TheLIGOScientific:2017qsa} has also brought about new scientific opportunities. Through a precise measurement of the tidal deformability parameter in the phase of the waveform, cf.~\S\ref{sec:waveforms}, we were able probe the nature of nuclear matter in the high-density regime~\cite{Abbott:2018exr}. Furthermore, the electromagnetic emissions from the neutron star merger, which accompany the gravitational-wave radiations in the earlier inspiral regime, offered an unprecedented opportunity for multi-messenger astronomy. For example, by comparing the relative propagation speeds between the gravitational waves and light in these events, significant constraints were put on various modified theories of gravity~\cite{Creminelli:2017sry, Ezquiaga:2017ekz, Baker:2017hug}. In addition, the electromagnetic spectrum of merged neutron stars provided strong evidence of their roles in manufacturing Nature's heavy elements, such as gold and platinum~\cite{Kasen:2017aa, Pian:2017aa}. Moreover, the electromagnetic counterpart provided a `standard siren' that allowed for a distance calibration to the source, thereby offering an independent measurement of the local Hubble parameter~\cite{Schutz:1986aa}.

\vskip 4pt

At the time of writing, the LIGO/Virgo O3 run was just completed. In addition to finding a new binary neutron star~\cite{Abbott:2020uma}, the collaboration discovered a binary black hole whose component masses are appreciably asymmetric, with the more massive black hole about four times heavier than its lighter counterpart~\cite{LIGOScientific:2020stg}. This is a qualitatively new signal because the asymmetric mass ratio allowed for a significant measurement of a higher-order radiative multipole moment in the gravitational-wave emission. As evident, every observing run thus far has brought about new scientific opportunities and this momentum has no sign of stopping. As it stands, O4 and O5 are planned to start operating in 2022 and 2025, respectively, each scheduled to observe for at least one calendar year~\cite{Aasi:2013wya}. The detectors should by then have been upgraded to Advanced LIGO Plus and Advanced Virgo Plus, whose sensitivities 
are few times better than those of current detectors. Plans for a more ambitious upgrade of the LIGO detectors, called the LIGO Voyager, are also well underway~\cite{Adhikari:2019zpy}. GEO 600, which is an operating 600m laser interferometer in Hanover, Germany is unlike to contribute significantly to the network of gravitational-wave detectors. Nevertheless, it will be tuned to focus on high frequency narrow-band sensitivity at a few kilohertz, possibly contributing to the understanding of neutron star merger physics where the signal may be loud~\cite{Dooley:2015fpa}.

\newpage

In addition to ground-based laser interferometers, there exists a global network of pulsar timing arrays (PTAs) which aims to detect a stochastic background of gravitational waves emitted by supermassive black hole inspirals. This collaboration, called the International Pulsar Timing Array~\cite{IPTA:2013lea},  consists of groups in Europe, North America and Australia~\cite{Kramer:2013kea, McLaughlin:2013ira, Hobbs:2013aka}. PTAs are sensitive to gravitational waves in the nanohertz frequency regime, approximately within $10^{-10} \hskip 1pt \text{Hz} - 10^{-8} \hskip 1pt  \text{Hz}$, see e.g. Fig. 1 of~\cite{Hazboun:2019vhv}.
The separation between PTAs and the observed pulsars play the same role as the detector arms in laser interferometers, but can now be as long as $10^{19}\, \text{m}$. Gravitational waves that propagate in the intermediate region would modulate the arrival times of the signals emitted by these pulsars, which are known to be very stable, periodic sources of radio waves. These arrays therefore seek to measure relative phase shifts between different pulsar arrival signals due to the modulations from gravitational waves~\cite{DetweilerPTA}. While no such gravitational-wave signal has been measured so far, a breakthrough is on the horizon~\cite{Lommen_2015}. 
The Square Kilometer Array (SKA), which is a trans-continental PTA with better sensitivities, is also expected to begin its observations
a few years later~\cite{Smits:2008cf}.

\subsection{Future Detectors} \label{sec:futuredetectors}

\begin{figure}[t]
\centering
\includegraphics[scale=0.45, trim = 0 10 0 0]{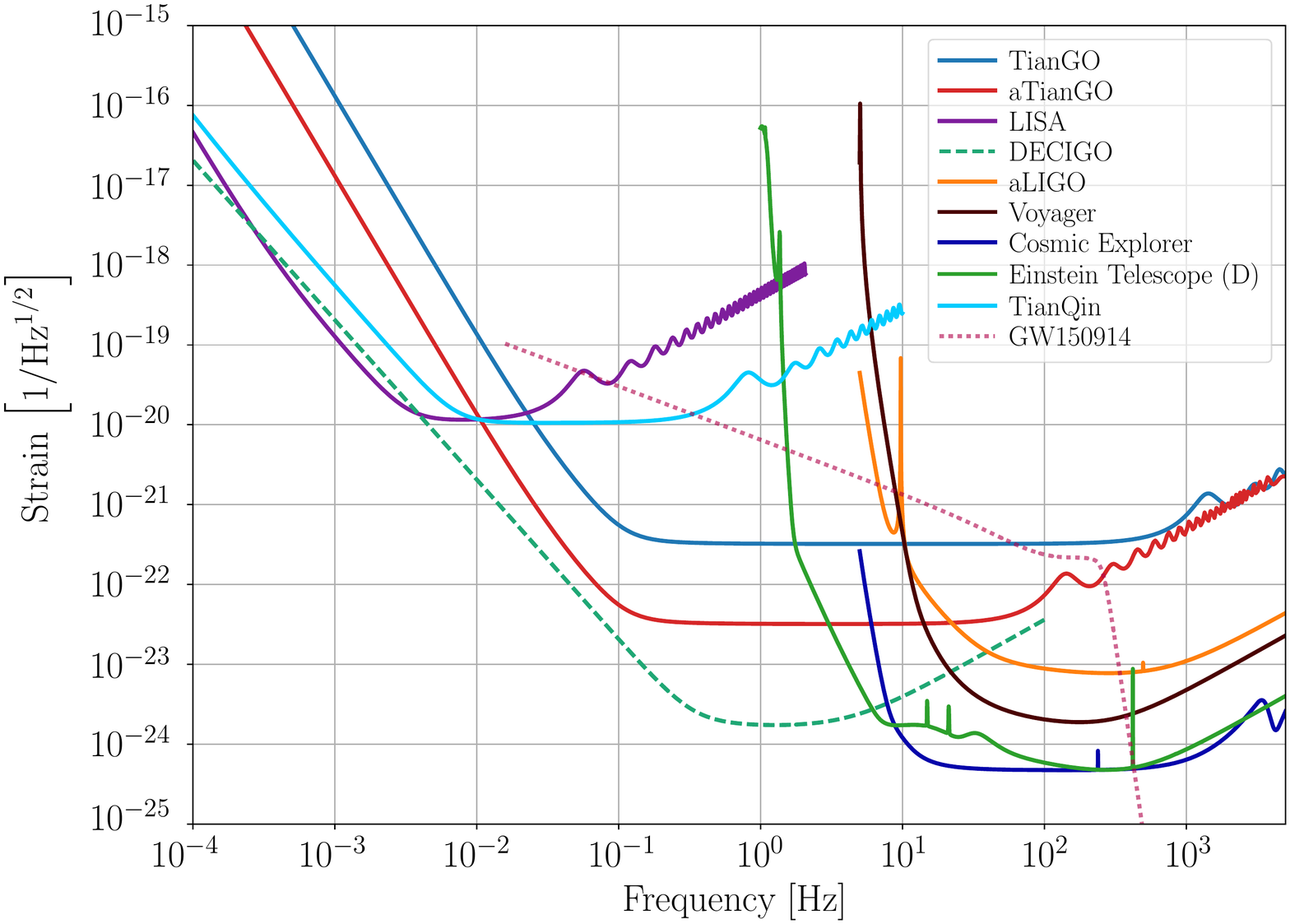}
\caption{The sensitivity curves of current and future gravitational-wave detectors. Figure adapted from~\cite{Kuns:2019upi}. For the PTA sensitivity curves, which have relatively low strain sensitivities and lie in the nanohertz regime, see e.g.~\cite{Hazboun:2019vhv}. }
\label{fig:detectorcurves}
\end{figure}

The successes of LIGO and Virgo and the promise of PTAs mark only the beginning of gravitational-wave astronomy. New ground-based detectors have already been constructed and will join LIGO and Virgo in future observing runs. In addition, there is now a flurry of proposals for developing next-generation gravitational-wave detectors. In what follows, I will summarize these prospective future gravitational-wave detectors, separating them into a few broad categories for convenience. A few representative noise curves in each category are illustrated in Fig.~\ref{fig:detectorcurves}.

\begin{itemize}
\item Current-generation ground-based detectors: in addition to the upgrades of LIGO and Virgo described above, two new Michelson interferometers are already constructed and will soon be operating: the Kamioka Gravitational Wave Detector (KAGRA) in Japan~\cite{Somiya:2011np} and LIGO-India~\cite{Unnikrishnan:2013qwa}. These detectors, whose arm lengths are $3$ km and $4$ km, respectively, will observe in the same frequency range as LIGO and Virgo, therefore contributing to a wider network of ground-based observatories. Crucially, these additional detectors will significantly improve the network's overall angular resolution and sensitivity to gravitational-wave sources. Like LIGO and Virgo, these detectors are expected to operate throughout the 2020s.

\item Next-generation ground-based detectors: in the 2030s, the Einstein Telescope is planned to be constructed in continental Europe. This is conceived as an underground laser interferometer with the shape of an equilateral triangle, each arm being $10$ km long~\cite{Sathyaprakash:2012jk}. In the United States, the Cosmic Explorer is envisioned to be a L-shaped Michelson interferometer with $40$ km arm lengths~\cite{Reitze:2019iox}. These detectors aim to have sensitivities that are an order-of-magnitude better than current advanced detectors.

\item Milli-hertz space-borne detectors: after decades of funding delay, the Laser Interferometer Space Antenna (LISA) has recently been selected by the European Space Agency as a mission to be launched in the early 2030s~\cite{2017arXiv170200786A}. With laser path lengths that are longer than $10^6$ km, this engineering marvel will probe gravitational waves in the milli-hertz regime, $10^{-4} \hskip 1pt {\rm Hz} - 10^{-2} \hskip 1pt {\rm Hz} $. Prospects for its realistic launch have also been bolstered by the success of the LISA Pathfinder, a precursor to the actual mission~\cite{2018PhRvD..98j2005A}. Similar space-based missions have also been proposed in China, including TianQin~\cite{Luo:2015ght} and Taiji~\cite{Guo:2018npi}. More futuristic milli-hertz detectors, such as the Big Bang Observer~\cite{Crowder:2005nr}, have also been envisioned. \\

\item Deci-hertz detectors: another unexplored gravitational-wave window is the deci-hertz frequency range, $10^{-2} \hskip 1pt {\rm Hz} - 1 \hskip 1pt {\rm Hz}$, which lies between the LISA and ground-based detector sensitivity bands. A primary detection method uses laser interferometry with ensembles of cold atoms playing the role of free-falling test masses~\cite{Dimopoulos:2008sv, Harms:2013raa, Canuel:2016pus, Badurina:2019hst}. Space-based laser interferometers, such as the Deci-hertz Interferometer Gravitational Wave Observatory (DECIGO) in Japan~\cite{Kawamura:2011zz} and TianGO in China~\cite{Kuns:2019upi}, have also been proposed.

\end{itemize}

This summary clearly demonstrates the exciting prospect of expanding the network of
gravitational-wave detectors in the near future and even several decades later. Together, these observatories would measure gravitational waves over ten orders of magnitude in frequency. The future of gravitational-wave astronomy is bright.

\chapter{Particle Physics at the Weak-Coupling Frontier} \label{sec:gravprobes}

In the previous chapter, I focused on General Relativity and its consequences for the gravitational sector of our Universe. This discussion excluded matter and the other known forces in Nature, which are instead described by the Standard Model of particle physics. Despite the excellent agreement between the Standard Model and the outcomes of many particle physics experiments, it is indisputable that physics beyond the Standard Model exists in our Universe. This new physics often involves the presence of new particles that couple very weakly to ordinary matter, making them hard to detect in particle colliders. Fortunately, by virtue of the equivalence principle, all particles must interact through gravity. The central goal of this thesis is to explore how gravitational waves could offer us universal probes of new physics. 

\vskip 4pt

In this chapter, I provide an overview of the weak-coupling frontier in particle physics. In Section~\ref{sec:BSM}, I review the Standard Model and summarize its various shortcomings. In Section~\ref{sec:weak-coupling}, I discuss several classes of new particles which are proposed to address these problems and, at the same time, are found to necessarily couple weakly to ordinary matter. In Section~\ref{sec:atomreview}, I discuss how rotating black holes are novel probes of putative ultralight bosons. This way of detecting new particles is attractive because it only relies on gravity --- the weakest of all forces. The discussion in this section will serve an important basis for Chapters~\ref{sec:spectraAtom}, \ref{sec:Collider}, and \ref{sec:signatures}.

\section{Physics Beyond the Standard Model} \label{sec:BSM}

The Standard Model is the quantum field theory that governs the fundamental interactions (except gravity)  between all known particles in our Universe (see Fig.~\ref{fig:SM}). Despite its excellent agreement with the outcomes of particle physics experiments, there is now overwhelming evidence that it is far from a complete description of Nature.

\subsection{The Standard Model} \label{sec:SM}

The Standard Model is built upon the isometries of Minkowski spacetime and a set of non-Abelian gauge symmetries. The spacetime symmetry dictates that the particles of the theory transform as unitary irreducible representations of the Poincar\'e group~\cite{Wigner1939}. On the other hand, the gauge symmetry governs the way in which these particles interact with one another. Specifically, the gauge group of the Standard Model is $SU(3)_C \otimes SU(2)_L \otimes U(1)_Y$, where $SU(3)_C$ is associated to quantum chromodynamics (QCD), the theory of the strong nuclear force~\cite{Gross:1973id, Gross:1973ju, Politzer:1973fx}, and $SU(2)_L \otimes U(1)_Y$ describes the electroweak sector~\cite{Glashow:1961tr, Weinberg:1967tq, Salam:1968rm}. More precisely, the subscript $C$ denotes the color charge of QCD, while $L$ and $Y$ stands for left-handedness and hypercharge respectively. The left-handedness of $SU(2)_L$ is especially interesting because it implies that the right-handed particles do not couple to this subgroup. The Standard Model is therefore chiral, i.e.~it is not invariant under a parity transformation, P. As we shall see, the theory is also not invariant under time reversal, T, and charge conjugation, C. Nevertheless, it must respect the combined CPT symmetry of any local quantum field theory in Minkowski spacetime.

\begin{figure}[t]
\centering
\includegraphics[scale=0.27, trim = 0 15 0 15]{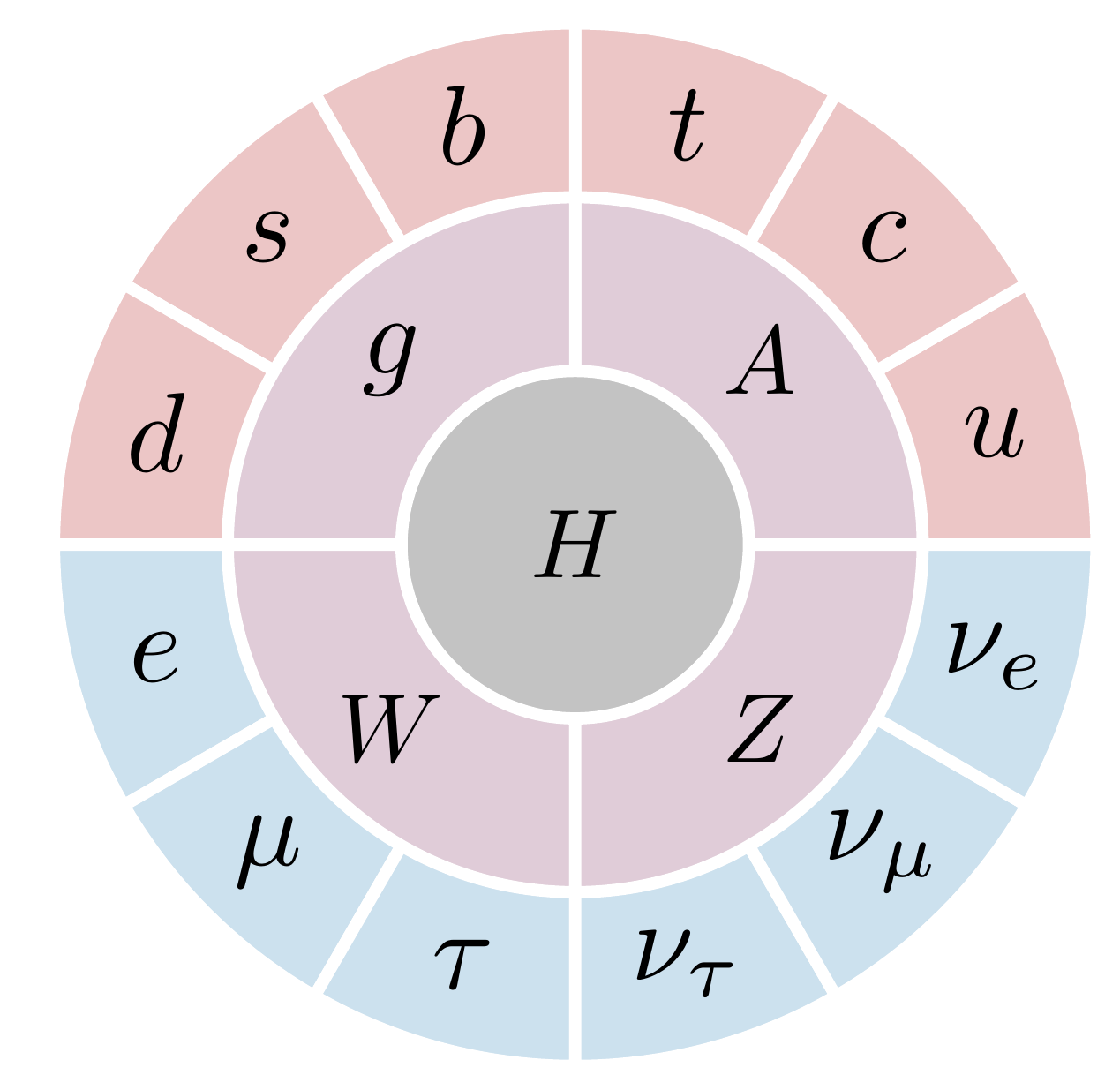}
\caption{The Standard Model particles in a nutshell. The red, blue, purple, and gray categories represents the quarks, leptons, gauge bosons, and the Higgs boson, respectively.}
\label{fig:SM}
\end{figure}

The particle content of the Standard Model can be organized into four broad categories: \textit{i)} the gauge bosons, which mediate the forces; \textit{ii)} the quarks, which interact via both the strong and electroweak forces; \textit{iii)} the leptons, which only couple to the electroweak sector; and \textit{iv)} the Higgs boson, which mediates the spontaneous symmetry breaking of the electroweak sector. The gauge bosons are massless vector (spin-1) fields while the quarks and leptons are spin-1/2 fermions. The Higgs boson is the only scalar (spin-0) field in theory. See Fig.~\ref{fig:SM} for a summary of the Standard Model particles.

The gauge group of the Standard Model itself does not inform us how these particles interact with one another. Instead, we must also know the representations under which they transform in this group. For instance, the gauge bosons transform as adjoint representations of each of the $SU(3)_C$, $SU(2)_L$, and $U(1)_Y$ subgroups. The self-interactions of the gauge bosons are therefore given by the structure constant of the subgroup's Lie algebra. On the other hand, the matter fields transform as fundamental (and anti-fundamental) representations of the respective subgroups. For instance, in QCD, the quarks are organized into two types of triplet representations: the up-type quarks, $(u, c, t)$, and the down-type quarks, $(d, s, b)$. On the other hand, in the electroweak sector, the left-handed components of the quarks are organized into three generations of doublet representations: $ (u, d)_L$, $(c, s)_L$, and $ (t, b)_L$. By contrast, the right-handed quarks transform as trivial representations under $SU(2)_L$, though they have non-vanishing hypercharges. Similarly, the left-handed leptons are organized into three flavor generations: the electron and its neutrino $ (\nu_e, e^- )_L$, its muonic counterpart $(\nu_\mu, \mu^- )_L$, and its tau-flavour counterpart $ (\nu_\tau, \tau^- )_L$. Most notably, there are only right-handed electron, muon, and tau leptons in the Standard Model, but no right-handed neutrinos. Because all quarks and leptons are Dirac spinors (except for the neutrinos, which could be either Dirac or Majorana fermions), they have antiparticle counterparts with opposite charges. Finally, the Higgs field also transforms as a doublet representation under $SU(2)_L \otimes U(1)_Y$.

The symmetries and particle content 
uniquely fix the Lagrangian of the Standard Model.\footnote{Here, we focus only on the renormalizable part of the Standard Model Lagrangian. In other words, only operators with mass dimensions four or below are considered in (\ref{eqn:SMlagrangian}).  Higher-dimensional operators can be included in an effective field theory treatment, and are suppressed by the scale of new physics~\cite{Weinberg:1978kz, Weinberg:1979sa, Buchmuller:1985jz, Lehman:2014jma, Henning:2015alf}. \label{footnote:SMEFT}} Schematically, it reads
\beq
\mathcal{L}_{\rm SM} =  - \frac{1}{4} X_{\mu \nu} X^{\mu \nu} + i \bar{\Psi} \gamma^\mu D_\mu \Psi + \left( \bar{\Psi}_i Y_{ij} \Psi_j H + h.c. \right) + \left| D_\mu H \right|^2  - V(H) \, , \label{eqn:SMlagrangian}
\eeq
where $X_{\mu \nu}$ represents the gauge field strengths, $D_{\mu}$ is the covariant derivative with respect to the gauge fields, $\Psi$ denotes the quarks and leptons in their appropriate representations, $Y_{ij}$ is the Yukawa matrix, $h.c.$ denotes the Hermitian conjugate, $H$ is the complex Higgs field, and $V(H) = m_H^2 |H |^2 + \lambda_H |H |^4 $ is the Higgs potential. Crucially, the charges of the matter fields are appropriately assigned such that the Lagrangian (\ref{eqn:SMlagrangian}) is invariant under the gauge transformations. The interactions between the matter fields and the gauge fields are implicitly encoded in $D_\mu$. Since the weak interaction only couples to left-handed particles and right-handed antiparticles, it violates C and P maximally. Furthermore, it also violates CP because the Yukawa matrix in the quark sector contains complex elements.

At high energies, the gauge symmetry of the Standard Model is fully restored. In this limit, all particles, except the Higgs boson, are massless.  However, below the electroweak scale at $246$~GeV, the Higgs boson acquires a non-trivial vacuum expectation value, thereby triggering spontaneous symmetry breaking~\cite{Higgs:1964pj, Guralnik:1964eu, Higgs:1964ia, Englert:1964et} in the electroweak sector $SU(2)_L \otimes U(1)_Y \to U(1)_{EM}$, where $U(1)_{EM}$ is the gauge group of quantum electrodynamics (QED). Two important implications emerge from this phenomenon. Firstly, all fermions, except the neutrinos, become massive, with their masses proportional to the sizes of the corresponding elements in $Y_{ij}$. Secondly, the gauge bosons split into the massive $W^{\pm}$ and $Z$ bosons, which mediate the weak nuclear force, and the photon $A$. Crucially, the large masses of the $W^{\pm}$ and $Z$ bosons at $ \approx 100$ GeV mean that the weak interaction is a short-range force. This short-range behavior is effectively modeled as point-like in Fermi theory, which is the effective field theory of the weak nuclear force at low energies. On the other hand, $A$ is massless and mediates the usual long-range electromagnetic force in QED.

The strong nuclear force displays characteristic features that are absent in the electroweak sector. One of the key properties of the theory is asymptotic freedom, which means the coupling strength of the strong force, $g_s$, increases at high energies and decreases at low energies. This is qualitatively distinct from QED, where the electromagnetic coupling strength runs in the opposite manner. As a result, there exists a typical energy scale, also known as the QCD scale $\approx 100$ MeV,  below which $g_s$ exceeds order unity and QCD becomes strongly interacting. At low energies, quarks and gluons therefore hadronize to form baryons and mesons, which are composite particles with no color charges. In fact, quarks and gluons are believed to be so strongly bound that they cannot be observed as free particles. This confinement behaviour, though not proven mathematically, has been demonstrated by numerical simulations in lattice QCD, see e.g.~\cite{Kondo:2014sta, Ratti:2018ksb} for recent reviews. Another important feature of QCD is its topologically non-trivial vacuum manifold. In particular, vacuum solutions in the theory are separated by energy barriers constructed from gluon gauge field configurations. Tunnelling events through these energy barriers are known as instantons~\cite{tHooft:1976snw, tHooft:1976rip}, and can contribute to important non-perturbative dynamics in QCD.

The success of the Standard Model is ultimately grounded in its excellent agreement with the outcomes of many experiments. Over the past few decades, particle physics had been predominantly probed through the {\it high-energy frontier}, in which heavy particles are produced through relativistic collisions of light elementary particles. 
This way of probing new physics has allowed us to observe virtually all of the particles and forces predicted by the Standard Model. Impressively, the Large Hadron Collider (LHC) has confirmed the validity of the theory up to approximately $10$ TeV~\cite{Aad:2012tfa, Chatrchyan:2012xdj, Tanabashi:2018oca}, which is nearly two orders of magnitude larger than the electroweak scale. Furthermore, the Standard Model has been shown to be correct by many high-precision measurements of the theory. This is demonstrated most impressively through the anomalous magnetic moment of the electron, $a_e$, whose theoretical and experimental results have been compared to an exquisite precision~\cite{Aoyama:2014sxa}:
\beq
\begin{aligned}
a_e & = 0.001159652181643(25)  \, , \qquad \text{(theory)} \\
a_e & = 0.00115965218073(028) \, , \qquad \text{(experiment)} \, .
\end{aligned}
\eeq
Other Standard Model parameters, though measured to a lesser degree of precision, have also demonstrated excellent agreements between theory and experiment~\cite{Tanabashi:2018oca}.

While the Standard Model successfully  describes the strong, weak, and electromagnetic forces, gravity is notably absent in the theory. As we have seen in \S\ref{sec:GR}, the latter is instead described by the Einstein field equation (\ref{eqn:Einstein}), whose dynamical degree of freedom is  the metric $g_{\mu \nu}$. These four known fundamental forces of Nature are represented succinctly by the action
\beq
S = \int \d^4 x \sqrt{-g}  \,\left[  \frac{1}{2} M_{\rm pl}^2  R  + \mathcal{L}_{\rm SM} \right] \, , \label{eqn:EH+SM}
\eeq
where $M_{\rm pl} = ( \hbar c /8\pi G )^{1/2} = 2.4 \times 10^{18} \, \text{GeV}$ is the (reduced) Planck scale. The first term is called the Einstein-Hilbert action, which yields the Einstein field equation when it is varied with respect to $g_{\mu \nu}$. Unlike $\mathcal{L}_{\rm SM}$, the Einstein-Hilbert term is not renormalizable. A ultraviolet completion is therefore necessary in order to describe gravity at high energies. In fact, both terms in the action (\ref{eqn:EH+SM}) are only empirically verified at low energies $\lesssim 10$ TeV and should therefore be viewed as an effective field theory of a high-energy theory, plausibly one that unifies all of the four forces. New physics that appear at high energies could be incorporated in (\ref{eqn:EH+SM}) through higher-dimensional operators that respect diffeomorphism invariance and the symmetries of the Standard Model. From dimensional analysis, these operators must be suppressed by the energy scales at which new physics appear. As a result, their phenomenological implications at low energies are also typically suppressed.

\subsection{Motivations for New Physics} \label{sec:Motivation}

Despite the successes of the Standard Model, multiple observational evidences point towards the need for new physics. In addition, theoretical considerations suggest the presence of several flaws in the theory, which could be resolved through the introduction of new physical principles. In this subsection, I will present a summary of key motivations for physics beyond the Standard Model.

A stunning realization of modern physics is that most of the matter in our Universe is dark. Through various astrophysical and cosmological observations, we have discovered that the Standard Model particles only constitute $15\%$ of the total matter in our Universe, while the remaining $85\%$ is a mysterious form of \textit{dark matter} (see e.g.~\cite{Bertone:2004pz} for a review). Dark matter cannot be the Standard Model particles because it is non-baryonic, i.e. it doesn't seem to interact with the strong, weak, and electromagnetic forces at all. Nevertheless, thanks to the equivalence principle, we have been able to detect it through its gravitational interactions. For instance, dark matter halos are known to be responsible for the flattening of galaxy rotation curves~\cite{Rubin:1970zza, Roberts1975, Rubin:1980zd, Persic:1995ru}, such as that of the Milky Way~\cite{Iocco:2015xga}. Dark matter is also essential in explaining the intricate features of the acoustic oscillations of the cosmic microwave background radiation~\cite{Aghanim:2018eyx}. Although we understand what dark matter does in our cosmos, we still do not know what it is. Given the intricate structure of the Standard Model, it is not unreasonable to speculate that dark matter also consists of a 
rich spectrum of particles, possibly even mediated by many hidden forces. In Section~\ref{sec:weak-coupling}, I will review several important classes of dark matter. Indeed, a central goal of this thesis is explore how different types of dark matter can be probed through the gravitational waves emitted by astrophysical systems, such as binary black holes.

Matter, which includes both the Standard Model particles and dark matter, only makes up $30\%$ of the total energy budget of our Universe. Remarkably, the majority of the Universe's energy density is contained in a form of negative-pressure fluid that is commonly known as \textit{dark energy}. Dark energy is a critical component of our cosmic menu because it drives the accelerated expansion of our Universe~\cite{Riess:1998cb, Perlmutter:1998np}. In the $\Lambda$CDM model~\cite{Aghanim:2018eyx}, dark energy is well-described by a cosmological constant, i.e.~a uniform energy density that permeates all of spacetime. This effect can be incorporated into the action (\ref{eqn:EH+SM}) through the inclusion of a dimension-zero operator,\footnote{That is, this term does not depend on any fields or derivatives.}  $\Lambda_{ cc}^4 \sim (1 \, \text{meV})^4$. Because this constant term is used to renormalize the zero-point energy in a quantum field theory, the cosmological constant can be interpreted as the energy density of vacuum. Classically, the relatively small value of $\Lambda_{cc}$ compared to other known energy scales in our Universe poses no conceptual problem. Nevertheless, as we shall shortly describe, in quantum field theory this presents us with a severe theoretical challenge.

Although the Standard Model is an extremely successful theory, one of its predictions is actually wrong. Specifically, it predicts that all neutrinos are massless, which contradicts many experimental outcomes that showed the presence of \textit{neutrinos masses} (at least for two of the neutrino species). For instance, neutrino-oscillation experiments~\cite{Fukuda:1998mi, Ahmad:2001an, Ahmad:2002jz} have confirmed that the flavour eigenstates of neutrinos do not coincide with their mass eigenstates, which is only possible if they have non-vanishing masses. One of the ways in which neutrinos can acquire masses is through the introduction of right-handed sterile neutrinos, which only couple to the Standard Model neutrinos~\cite{Abazajian:2012ys}. In this case, the smallness of the total neutrino masses, i.e. at $\sim 50 $ meV, can be explained through their couplings to very heavy sterile neutrinos, in what is usually called the see-saw mechanism~\cite{Minkowski:1977sc, Yanagida:1979as, Mohapatra:1979ia, Schechter:1980gr}. Alternatively, without enlarging the particle content of the theory, the neutrino masses can be explained by including the Weinberg operator~\cite{Weinberg:1979sa} into the Standard Model Lagrangian (\ref{eqn:SMlagrangian}). Schematically, this operator reads
\beq
\mathcal{L}_{\rm SM} \supset \frac{1}{\Lambda_{\nu}} \bar{\Psi} H^\dagger H \Psi + h.c. \, , \label{eqn:WeinbergOperator}
\eeq
where $\Psi$ denotes the left-handed lepton fields and $\Lambda_{\nu}$ is a new high-energy scale. Interestingly, by choosing $\Lambda_\nu \sim 10^{14}$ GeV, which is close to the energy scales of Grand Unified Theories~\cite{Georgi:1974sy, Pati:1974yy, Fritzsch:1974nn}, electroweak symmetry breaking would lead to the correct total neutrinos masses. Furthermore, the Weinberg operator is in fact the only dimension-five operator in the effective field theory of the Standard Model~\cite{Weinberg:1978kz, Weinberg:1979sa, Buchmuller:1985jz, Lehman:2014jma, Henning:2015alf}. It is remarkable that the only flawed prediction of the Standard Model is naturally corrected by incorporating the only leading non-renormalizable term into (\ref{eqn:SMlagrangian}).

\vskip 2pt

Another shortcoming of the Standard Model is its inability to explain the dominance of matter over antimatter in our Universe. From the Sakharov conditions~\cite{Sakharov_1991}, we know that \textit{baryogenesis} must be driven by non-equilibrium dynamics and violate C, CP, and baryon number (see e.g.~\cite{Cline:2006ts} for a pedagogical review on the subject). Intriguingly, all of these conditions are actually satisfied by the Standard Model: the Universe is out of equilibrium as it cools, C- and CP-violating terms are present in the electroweak sector, and baryon number is violated by an anomaly~\cite{Adler:1969gk, tHooft:1976rip}. However, the amount of C, CP, and baryon number violations in the Standard Model are insufficient to explain the observed matter-antimatter asymmetry observed in our Universe. Furthermore, baryogenesis is necessarily a strong first-order phase transition, while the electroweak phase transition is of second-order nature~\cite{Kajantie:1996qd}. The need to explain the matter-antimatter asymmetry in our Universe is therefore another strong motivation for introducing physics beyond the Standard Model.

\vskip 2pt

One of the most profound observations in cosmology is the apparent acausal correlations between different regions of our Universe. For example, measurements of the cosmic microwave background (CMB) have shown that the average temperatures of widely-separated parts on our celestial sphere are identical~\cite{Penzias:1965wn}. This is despite the fact that, in the hot Big Bang paradigm, different regions of the early Universe did not have enough time to thermalize with one another before recombination, which is the epoch when the CMB was emitted. Furthermore, the statistics of the CMB density fluctuations were found to display subtle correlations over length scales that were larger than the horizon at recombination~\cite{Smoot:1992td, Spergel:2003cb}. The most popular resolution to this \textit{horizon problem} involves the introduction of an inflationary epoch before the hot Big Bang, when the Universe expanded at an accelerated rate~\cite{Guth:1980zm, Linde:1981mu, Albrecht:1982wi} (see~\cite{Baumann:2009ds} for a pedagogical introduction). In this picture, different patches of the CMB were actually in causal contact with one another in the pre-inflationary epoch. Crucially, the statistics of quantum mechanical fluctuations that were generated during inflation agree excellently with those of the primordial density fluctuations inferred from the CMB~\cite{Guth:1982ec, Bardeen:1983qw}. That inflation offers a causal mechanism to generate the correct initial conditions of our Universe makes it a leading paradigm in modern cosmology. Importantly, none of the Standard Model particles including the Higgs, at least in its minimal form in (\ref{eqn:SMlagrangian}), can be responsible for driving the inflationary epoch.

So far, we have described observational evidences for physics beyond the Standard Model. However, a number of theoretical considerations also hint at the presence of new physics. One of the key conceptual problems of the Standard Model concerns the unstable hierarchy between the electroweak scale and other higher energy scales. Specifically, the \textit{hierarchy problem} is the theoretical observation that loop corrections to the Higgs boson mass, $m_H$, are quadratically divergent with respect to the masses of putative heavy particles that appear above the electroweak scale (see e.g.~\cite{Schwartz:2013pla} for details of this computation). The parameter $m_H$ is therefore extremely sensitive to physics at the ultraviolet (UV). In fact, it would be natural for $m_H$ to be lifted by UV physics to a value that is of the order of the new high scale. This lifting can, of course, be avoided if there exists counterterms in the UV theory that renormalize these divergences. Nevertheless, the precision at which these counterterms must be fine-tuned in order to maintain the small Higgs mass at $m_H = 125$ GeV can be unusually high. For instance, assuming $M_{\rm pl}$ is the only energy scale above the electroweak scale, as is the case in the action (\ref{eqn:EH+SM}),  the counterterms must be fine tuned at the level of $\sim (m_H/M_{\rm pl})^2 \sim 10^{-34}$. This fine-tuning problem suggests the presence of new physics near the electroweak scale, which motivated the construction of many beyond the Standard Model proposals such as low-scale supersymmetry~\cite{Dimopoulos:1981zb, Sakai:1981gr, Dimopoulos:1981yj, Ibanez:1981yh} and large extra dimensions~\cite{ArkaniHamed:1998rs, Randall:1999ee, Randall:1999vf}. Importantly, the hierarchy problem does not apply to the Standard Model fermions and gauge bosons, as their relatively small masses are protected by approximate chiral and gauge symmetries respectively, which become exact when their masses vanish~\cite{tHooft:1979rat}. On the other hand, no such approximate symmetry manifestly exists for the Higgs boson.

An even more severe fine-tuning problem exists for the cosmological constant. To begin, quantum corrections to $\Lambda_{cc}$ would generate UV divergences that scale quartically  with respect to a higher energy scale. Comparing this with the quadratic scaling for the Higgs mass, $\Lambda_{cc}$ is much more sensitive to corrections by UV physics.\footnote{Heuristically, these UV scalings can be understood from the fact that the degree of divergence of any operator in a Lagrangian is equal to the spacetime dimension minus the operator's mass dimension~\cite{Weinberg:1995mt}. The UV divergence of the $\Lambda^4$ term, which is a dimension-zero operator, therefore scales with the forth power of a high energy scale. On the other hand, $m_H^2 H^\dagger H$, which is a dimension-two operator, scales quadratically. \label{footnote:UVscale}} Furthermore, the fine-tuning problem is exacerbated by the fact that we believe we know what the UV physics above $\Lambda_{cc}$ is --- it's the Standard Model! Unfortunately, loop corrections to $\Lambda_{cc}$ from the vacuum fluctuations of the Standard Model particles do not cancel each other. As a result, the relevant counterterms must already be fine tuned at least at the level of $\sim (\Lambda_{cc} / m_H)^4 \sim 10^{-56}$. In the limit where the cosmological constant is sensitive to physics at the Planck scale, the fine tuning would be at a staggering level of $\sim (\Lambda_{cc} / M_{\rm pl})^4 \sim 10^{-120}$. The apparent instability of $\Lambda_{cc}$ to a miniscule change to its value is known as the \textit{cosmological constant problem}~\cite{Weinberg:1988cp, Padmanabhan:2002ji}. The lack of generally-accepted theoretical mechanisms that resolve this issue has led to anthropic reasonings, which posit that the observed value of $\Lambda_{cc}$ in our Universe is the result of a selection effect in a much larger Multiverse~\cite{Weinberg:1987dv, Bousso:2000xa}.

The smallness of another parameter in the Standard Model also offers a tentative hint for new physics. In QCD, the following P- and CP-violating term should, in principle, be present in the Lagrangian (\ref{eqn:QCDphase}),
\beq
\mathcal{L}_{\rm SM} \supset \theta \hskip 1pt \frac{  g_s^2 }{32\pi^2} G_{\mu \nu} \tilde{G}^{\mu \nu} \, ,  \label{eqn:QCDphase}
\eeq
where $G_{\mu \nu}$ is the gluon field strength, $\tilde{G}_{\mu \nu} \equiv \epsilon_{\mu \nu \rho \sigma} G^{\rho \sigma}/2 $ is its dual field strength, with $\epsilon_{\mu \nu \rho \sigma}$ the fully antisymmetric tensor, and $\theta$ is called the strong CP phase. The term (\ref{eqn:QCDphase}) is usually ignored in the Lagrangian because it is a total divergence and therefore does not affect the classical equation of motion. However, it plays an important role in the quantum mechanical structure of the QCD vacuum and hence cannot be ignored. \textit{A priori}, $\theta$ can take any value between $0$ and $2\pi$. Since all other dimensionless parameters in the Standard Model have values that are $\mathcal{O}(1)$, one naturally expects this to also be the case for $\theta$ as well. However, experimental measurements of the neutron electric dipole moment constrain this parameter to be $|\theta| < 10^{-10}$~\cite{Baker:2006ts, Afach:2015sja}, which is many orders of magnitude smaller than the generic expectation. This lack of observed CP violation in the strong interaction is also known as the \textit{strong CP problem}. Despite the apparent similarity with the hierarchy and cosmological constant problems, we emphasize that the smallness of $\theta$ is of a distinct nature, as it does not represent any sensitivity to loop corrections introduced by physics in the ultraviolet. In \S\ref{sec:UltralightBosons}, I will describe one attractive resolution to this problem, which involves promoting $\theta$ to a dynamical field, such that the smallness of $\theta$ is naturally achieved by the field relaxing towards the minimum of its potential.

Finally, the Standard Model does not include gravity. A complete description of quantum gravity is perhaps the most ambitious goal in theoretical physics. In \S\ref{sec:SM}, we described how the action (\ref{eqn:EH+SM}) should only be viewed as an effective field theory that is valid at low energies $\lesssim 10$ TeV. This is especially true for the Einstein-Hilbert action, which is non-renormalizable and must therefore be ultraviolet-completed at high energies. Perhaps the most promising candidate of \textit{quantum gravity} is string theory, which posits that the particles that we observe are excitations of elementary strings. String theory is a remarkably rich subject and has dominated much of theoretical physics over the past few decades (see~\cite{Green:1987sp, Green:2012pqa, Polchinski:1998rq, Polchinski:1998rr, Becker:2007zj} for classic textbooks). Indeed, the phenomenology that arises in string theory could potentially resolve many of the deficiencies of the Standard Model described above~\cite{stringpheno2012, Baumann:2014nda}. For the purpose of this thesis, it suffices to remark only on a few key features of the theory. Firstly, all versions of string theory require the presence of extra compact spatial dimensions, whose topological properties are generally incredibly rich~\cite{Candelas:1985en, Douglas:2006es}. Secondly, the quantized string spectrum necessarily includes many new particles, such as antisymmetric tensor fields. The Kaluza-Klein reduction of these higher-dimensional fields over a compact manifold would generate new particles in our observable four-dimensional Universe, possibly forming a component of dark matter. Finally, string theory is also a theory that includes solitonic objects, such as D-branes, world-sheet instantons, and gauge-theory instantons~\cite{tHooft:1976rip, Dine:1986zy, Becker:1995kb, Kallosh:1995hi, Polchinski:1995mt}. These objects would introduce non-perturbative effects, which affect the low-energy physics by giving these new particles small but non-vanishing masses.

\section{The Weak-Coupling Frontier} \label{sec:weak-coupling}

\begin{figure}[t]
  \centering
  \includegraphics[scale=0.85, trim = 0 0 0 0]{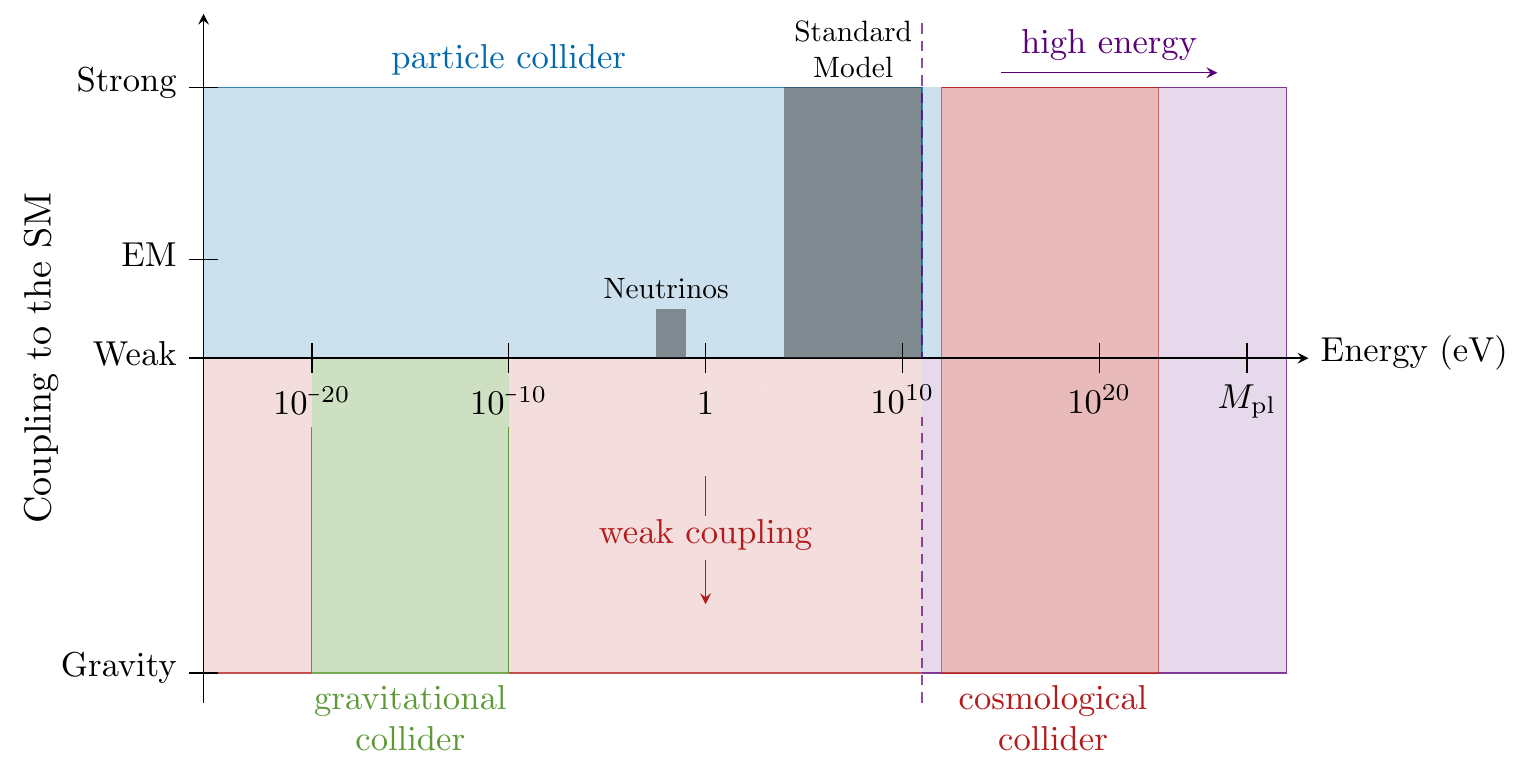}
  \caption{Particle colliders are blind to new particles that are either too heavy or too weakly-coupled to be produced in sufficient numbers. Very massive particles (up to $10^{14}$\,GeV) can be created in the ``cosmological collider" that was active during inflation~\cite{Chen:2009zp, Baumann:2011nk, Noumi:2012vr, Flauger:2013hra,Arkani-Hamed:2015bza, Arkani-Hamed:2018kmz, Baumann:2019oyu}.  In this thesis, I explore the weak-coupling frontier through the gravitational waves emitted by binary systems. In particular, in Chapters~\ref{sec:spectraAtom}, \ref{sec:Collider}, and \ref{sec:signatures}, I focus on the signatures of ultralight bosons within the mass range $[10^{-20}, 10^{-10}]$\,eV, which can form boson clouds around rotating astrophysical black holes. Remarkably, the dynamics of these clouds in binary systems are described by an ``S-matrix" and are sensitive to both the masses and intrinsic spins of the bosons. In direct analogy with ordinary colliders, I will therefore refer to these binary systems as ``gravitational colliders". In Chapter~\ref{sec:search}, I discuss how other types of weakly-coupled particles could form dark compact objects in our Universe. These objects can similarly be probed through the gravitational waves emitted when they are parts of binary systems.} 
  \label{fig:col1}
\end{figure}

In \S\ref{sec:Motivation}, I described multiple evidences for physics beyond the Standard Model in our Universe. Proposed solutions to these problems often involve the introduction of new degrees of freedom that have escaped our detections thus far. Over the past few decades, searches for new particles have been dominated by the high-energy frontier, where heavy particles appear as resonances in scattering processes of colliders, such as the LHC. While this way of probing new physics have helped to uncover the fundamental building blocks of Nature~\cite{Aad:2012tfa, Chatrchyan:2012xdj, Tanabashi:2018oca}, it relies on having appreciable interactions with the Standard Model particles involved in the collision processes. In other words, traditional collider experiments are blind to ``dark sectors" which couple very weakly to ordinary matter, even if the associated new particles are very light~\cite{Essig:2013lka, Berlin:2018bsc}. Roughly speaking, if these new particles interact with the Standard Model in a way that is weaker than the weak nuclear force, they can easily escape collider searches because of the small interaction cross sections (see Fig.~\ref{fig:col1}). In that case, we must find creative ways of detecting new light particles at the weak-coupling frontier. In what follows, I will motivate several generic classes of new particles, which I universally refer to as ``dark matter", although these particles could also simultaneously solve the other problems described in \S\ref{sec:Motivation}. I will also briefly mention how they are being searched for by many different types of experiments, and point the reader to references for existing constraints.

\subsection{Particle Dark Matter} \label{sec:ParticleDM}

In this subsection, I focus on dark matter that behaves like ``particles", in that they do not display wave-like properties over astrophysical length scales. As a rough fiducial guide, I consider dark matter that is heavier than the total mass of neutrinos $\sim 1$ meV, which are the lightest particles in the Standard Model, to fall within this category. These particles would have Compton wavelengths that are shorter than $ 0.1$~mm, or equivalently have de Broglie wavelengths that are shorter $0.1$~m if they follow the typical velocity distribution of dark matter in astrophysical environments. The discussion below will be guided by the types of forces---weak, electromagnetic and gravitational---through which dark matter could couple to the Standard Model.

A popular class of dark matter are the \textit{weakly-interacting massive particles} (WIMPs) --- hypothetical particles that interact with ordinary matter through the weak nuclear force (see~\cite{Arcadi:2017kky} for a recent review). One of the most attractive features of the WIMP paradigm is that the predicted thermal relic abundance agrees with the observed dark matter density in our Universe~\cite{Srednicki:1988ce, Gondolo:1990dk}. Furthermore, since the interactions of WIMPs are naturally tied to the electroweak scale, they can be easily embedded into many theories that are proposed to address other problems of the electroweak sector, such as the hierarchy problem. While the details of WIMPs can vary with different theories, a generic feature of all models is the presence of an interaction that mediates two-two scattering processes with the Standard Model particles. Schematically, this is captured by the effective operator
\beq
\mathcal{L}_\chi \supset c_{\chi} \chi \chi \hskip 1pt \psi \psi \, , \label{eqn:WIMP}
\eeq
where $\chi$ is the dark matter particle, $\psi$ denotes a Standard Model particle, and $c_{\chi}$ is the interaction coefficient, which is typically suppressed by the electroweak scale. Indeed, (\ref{eqn:WIMP}) provides the basis for most of the current experimental searches for WIMPs. For instance, direct-detection methods are designed to detect the elastic scattering process $\chi + \psi \to \chi + \psi$ through the recoils of stable nuclei~\cite{Akerib:2016vxi, Cui:2017nnn, Aprile:2018dbl}. Interestingly, these experiments will soon approach the ``neutrino floor", below which they would instead observe the background atmospheric and solar neutrinos and hence become insensitive to WIMPs with weaker couplings~\cite{Schumann:2019eaa}. On the other hand, indirect-detection searches aim to detect the final products of inelastic scattering (annihilation) processes $\chi + \chi \to \psi + \psi$, where $\psi$ can be gamma rays, cosmic rays, or neutrinos~\cite{FermiLAT2009, Adriani:2008zr, Abbasi:2012ws}. Through these observations, we are beginning to constrain the annihilation cross section of WIMP dark matter~\cite{Gaskins:2016cha}.

Dark matter could also potentially carry small amounts of electric charge. These \textit{charged dark matter} or \textit{milli-charged particles}~\cite{Goldberg:1986nk, Cheung:2007ut, Feldman:2007wj}, would couple feebly with the Standard Model particles through the electromagnetic force. Denoting the ratio of the dark matter electric charge to the elementary charge by $Q$, the relevant terms in the dark matter Lagrangian are
\beq 
\mathcal{L}_{Q} \supset  -\frac{1}{4} F_{\mu \nu} F^{\mu \nu} + Q e J^\mu_{\rm D} A_{\mu}  \, , \label{eqn:MCP}
\eeq
where $A_\mu$ is the electromagnetic field, $F_{\mu \nu}$ is its field strength, $e$ is the elementary charge, and $J_{\rm D}^\mu$ is the charge current of dark matter. Because charged dark matter must be invariant under a $U(1)_{EM}$ gauge transformation, it must either be a complex scalar or a Dirac fermion, in which case the interaction in (\ref{eqn:MCP}) is directly analogous to scalar and spinor QED, respectively. Unlike the electric charges of the Standard Model particles,  $Q$ need not be quantized. In addition, unless the charged dark matter is chiral, the value of $Q$ does not interfere with anomaly cancellation of the Standard Model. Constraints for this type of dark matter have been obtained through astrophysical and collider searches, see e.g.~\cite{Dimopoulos:1989hk, Davidson:2000hf, Essig:2013lka, Berlin:2018bsc}. Interestingly, dark matter with $Q \sim 1$ could still contribute a sizable fraction to the total dark matter density, if the dark matter mass falls within the approximate range $[10^{5},10^{11}]$ GeV~\cite{DeRujula:1989fe, Chuzhoy:2008zy}.

In the previous example, dark matter was electrically charged and therefore coupled to electromagnetism. However, the dark sector could also interact with ordinary matter through the electromagnetic force if photons kinetically mix with a hypothetical \textit{dark photon}~\cite{Holdom:1986eq, Okun:1982xi, Galison:1983pa}. Denoting the dark photon by $A^\prime_\mu$, the most general renormalizable Lagrangian is
\beq
\mathcal{L}_{A^\prime} = -\frac{1}{4} F_{\mu \nu} F^{\mu \nu} + 
e J_{\rm SM}^\mu A_\mu -\frac{1}{4} F^\prime_{\mu \nu} F^{\prime \mu \nu} - \frac{\epsilon}{2} F^\prime_{\mu \nu} F^{\mu \nu}  - \frac{1}{2} \mu^2 A^\prime_\mu A^{\prime \mu}  \, , \label{eqn:darkphotonLagrangian}  
\eeq    
where $ J_{\rm SM} $ is the charge current sourced by the Standard Model particles, $F^\prime_{\mu \nu}$ is the dark photon field strength, and $\epsilon \ll 1$ is the kinetic mixing parameter. The physical interpretation of the kinetic mixing is clearer if we perform the gauge transformation $A_{\mu } \to A_{\mu} - \epsilon A^\prime_{\mu}$, in which case the mixing term is removed while the second term in (\ref{eqn:darkphotonLagrangian}) generates an additional source term $\epsilon e J^\mu_{\rm SM} A^\prime_\mu$. This interaction is therefore directly analogous to that in (\ref{eqn:MCP}), except that it is the Standard Model fermions (and not the photon) that couple to the dark sector through suppressed electric charges. A recent review on this subject, which includes discussions of massless dark photons and mixings with the $U(1)_Y$ hypercharge gauge group, can be found in~\cite{Fabbrichesi:2020wbt}. In \S\ref{sec:UltralightBosons}, I will describe current constraints on dark photons in the $\mu$-$\epsilon$ parameter space. Importantly, we shall find that existing constraints apply predominantly to heavy particle-like dark photons but are largely absent in the ultralight regime.

Finally, dark matter particles can also gravitate and form astrophysical bound states. For instance, they could be accreted around black holes over astrophysical timescales and develop high-density minispikes~\cite{Gondolo:1999ef, Ferrer:2017xwm}. In addition, they could also form dark compact objects such as boson stars and solitons~\cite{Kaup1968, Ruffini1969, Breit:1983nr, Colpi1986, Lee1987-2, Lee1987-3} through a wide variety of astrophysical processes~\cite{Seidel:1993zk, Grasso:1990zg, Schunck:2003kk}. If these objects are part of binary systems, they would greatly enrich the binaries' dynamical evolution, such as through the finite-size effects that we described in Section~\ref{sec:CBC}. This would in turn leave significant imprints in the gravitational waves emitted by the binary system, thereby allowing us to probe the internal structure of these objects through the observed waveforms. In Chapter~\ref{sec:search}, I will elaborate on probing particle dark matter through these dark compact objects and their associated gravitational-wave signatures.

\subsection{Ultralight Bosons} \label{sec:UltralightBosons}

Dark matter could also consist of new degrees of freedom that are light, so light that their Compton and de Broglie wavelengths are of the order of astrophysical length scales. This type of dark matter is qualitatively distinct from the particle dark matter described above, as they can display wave-like phenomenology over large scales, which are hard to observe if their masses are large. The Compton wavelength and the de Broglie wavelength of a massive field, denoted by $\lambda_c$ and $\lambda_{db}$ respectively, are
\beq
\lambda_c \simeq 2 \hskip 1pt \text{km} \left( \frac{10^{-10} \hskip 1pt \text{eV}}{\mu} \right)   \, , \quad \lambda_{db} \simeq 3000 \hskip 1pt \text{km} \left(  \frac{10^{-10} \hskip 1pt  \text{eV}}{\mu}  \right) \left( \frac{200 \hskip 1pt \text{km s}^{-1}}{u} \right) \, , \label{eqn:wavelengths}
\eeq
where $\mu$ and $u$ are the field's mass and average velocity. We therefore see that dark matter with masses smaller than order $10^{-10}$ eV would display wave-like properties over scales that are larger than the sizes of solar-mass black holes. Boson fields with such small masses are especially interesting because can occupy a narrow region in phase space in large numbers. The resulting boson state is therefore a classical coherent oscillation that acts like a pressureless fluid, which is precisely the property of cold dark matter, albeit only over length scales that are longer than the corresponding de Broglie wavelength.\footnote{Ultralight bosons with masses of the order of $10^{-22}$ eV or smaller, also known as fuzzy dark matter, have been studied extensively in the literature~\cite{Hu:2000ke, Hui:2016ltb}.
This mass range is interesting because it corresponds to bosons having de Broglie wavelengths that are longer than order 10 kpc, which are the length scales probed by most astrophysical and cosmological observations. It is therefore possible to distinguish them from cold dark matter through their distinct phenomenologies at these large scales. In order to probe the wave-like properties of ultralight bosons that are heavier than $10^{-22}$eV, one would need to observe astrophysical events at shorter scales. In Section~\ref{sec:atomreview}, we shall see how these heavier bosons can trigger novel effects around astrophysical black holes.} In this subsection, I discuss several well-motivated classes of ultralight boson fields. I organize our discussion according to the intrinsic spins of the boson fields.

Theoretical solutions to the strong CP problem are surprisingly scarce. Perhaps the most appealing proposal involves the introduction of a new global $U(1)$ symmetry, also called the Peccei-Quinn symmetry, to the Standard Model~\cite{Peccei:1977hh, Weinberg:1977ma, Wilczek:1977pj}. In this scenario, the Goldstone boson that arises from spontaneous breaking of the Peccei-Quinn symmetry removes $\theta$ in (\ref{eqn:QCDphase}) through the boson's dynamical evolution to the minimum of its potential. This Goldstone mode is commonly known as the \textit{axion}, $a$, and the energy scale at which Peccei-Quinn symmetry breaking occurs, $f$, is called the axion decay constant. Below the Peccei-Quinn and electroweak scales, the effective action for the axion at leading-order in $a/f$ is
\beq
\mathcal{L}_{a} = - \frac{1}{2} \partial_\mu a \, \partial^\mu a + \frac{a}{f} \frac{g_s^2}{32\pi^2} G_{\mu \nu} \tilde{G}^{\mu \nu} + \frac{\partial_\mu a}{f} j^\mu_a + \frac{1}{4} \, a \, g_{a \gamma \gamma} F_{\mu \nu} \tilde{F}^{\mu \nu} \, , \label{eqn:axionLagrangian}
\eeq
where $j^\mu_a$ is an axial current sourced by the quarks~\cite{Srednicki:1985xd}, $F_{\mu \nu}$ is the electromagnetic field strength, $\tilde{F}_{\mu \nu}$ is its dual, and $g_{a \gamma \gamma} \propto f^{-1}$ is a coupling constant. The Lagrangian (\ref{eqn:axionLagrangian}) enjoys the continuous shift symmetry $a \to a + \text{const}$, as is the case for all Goldstone modes. However, non-perturbative effects introduced by the instantons in QCD break this continuous shift symmetry to a discrete shift symmetry~\cite{tHooft:1976snw, tHooft:1976rip, Sikivie:1982qv}. The effective potential of the axion below the QCD scale, which is obtained by integrating out the quarks and gluons in the second and third terms of (\ref{eqn:axionLagrangian}), therefore has the following structure:
\beq
\mathcal{L}_{a} \supset  - \Lambda_{a}^4 \Big[ 1 - \cos \left( a/2\pi f \right) \Big] + \cdots \, , \label{eqn:Axionpotential}
\eeq
where $\Lambda_{a}$ is proportional to the QCD scale and we have ignored multi-instanton contributions. The potential (\ref{eqn:Axionpotential}) clearly has the discrete symmetry $a \to a + 2\pi n f$, for arbitrary integers~$n$. Crucially, the minimum of the potential is located at $a=0$. The axion can therefore resolve the strong CP problem by dynamically relaxing to its potential minimum.

\begin{figure}[t]
\centering
\includegraphics[scale=0.58, trim = 30 15 0 0]{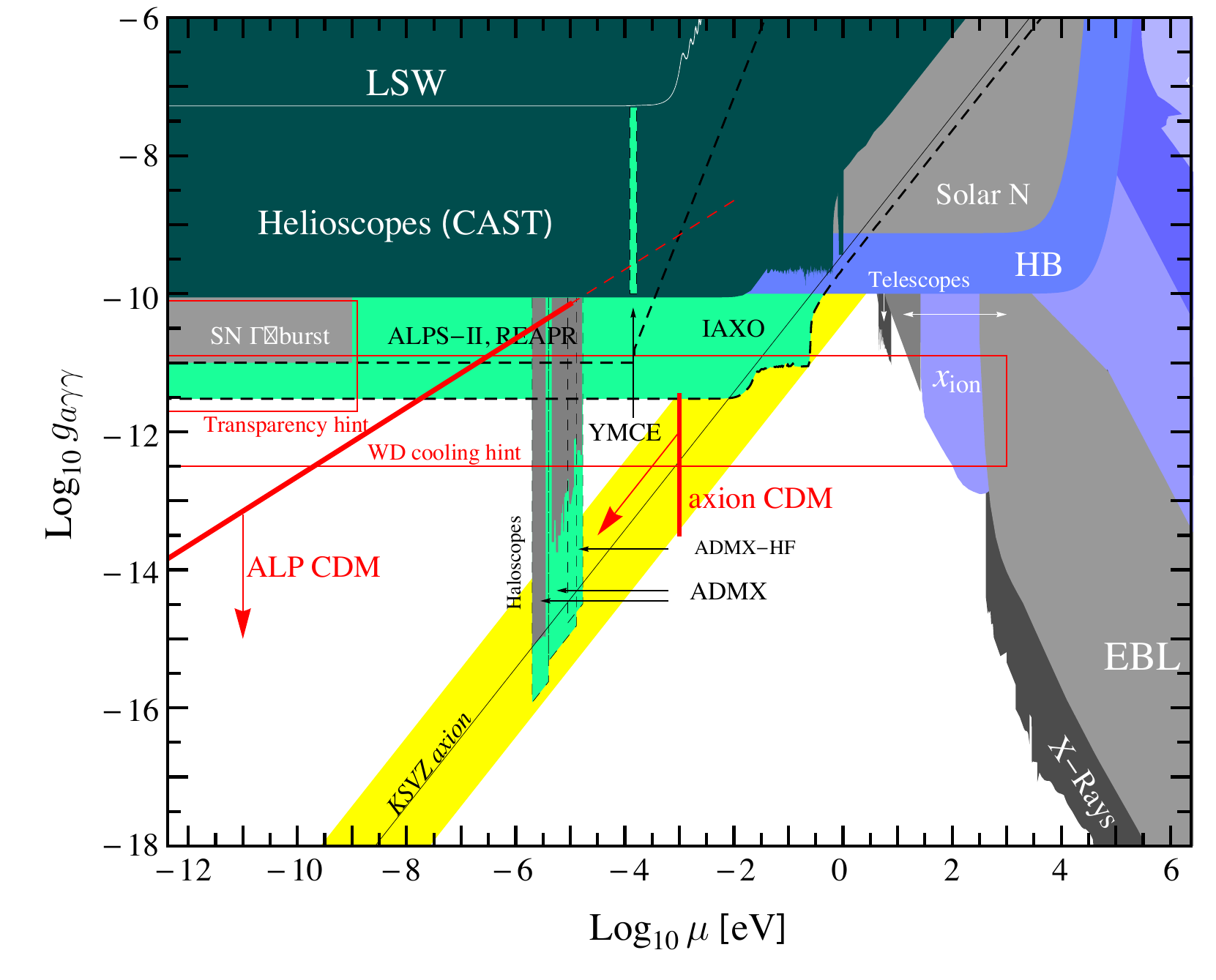}
\caption{Illustration of the constraints on the axion in the $g_{a \gamma \gamma}$-$\mu$ parameter space. Details of these various experimental constraints can be found in e.g.~\cite{Carosi:2013rla}. The yellow diagonal band denotes the predicted window for the QCD axion. Importantly, this figure shows it is challenging to constrain these new particles in the small-mass and weak-coupling regime. Figure adapted from~\cite{Ringwald:2012cu}.}
\label{fig:Axion}
\end{figure}

In the late Universe, the QCD axion oscillates around the minimum of its potential with a small amplitude. In that case, we can approximate the potential (\ref{eqn:Axionpotential}) by performing a Taylor expansion about $a = 0$ to quadratic order. The quadratic potential provides a notion of the QCD axion mass, which is~\cite{diCortona:2015ldu, DiVecchia:1980yfw}
\beq
\mu = 5.7 \times 10^{-6} \, \text{eV} \left(\frac{10^{12} \hskip 1pt \text{GeV}}{f} \right) \, . \label{eqn:QCDaxionmass}
\eeq
Crucially, this shows that $\mu$ and $f$ are inversely proportional to each other. Stellar-cooling constraints have already placed a lower bound on $f \gtrsim 10^{10}$ GeV~\cite{Friedland:2012hj, Giannotti:2017hny}, which is much higher than the electroweak scale. As a result, the QCD axion must be ultralight. In the limit where $f$ exceeds the energy scales of Grand Unified Theories $\sim 10^{16}$ GeV, the axion mass can even be as small as $\lesssim 10^{-10}$ eV. At present, most of our constraints on the QCD axion rely on the photon-axion coupling term in (\ref{eqn:axionLagrangian}). These constraints are summarized in Fig.~\ref{fig:Axion}, where the yellow band illustrates the predicted parameter ranges for the QCD axion. Note that, despite the precise relationship between $\mu$ and $f$ in (\ref{eqn:QCDaxionmass}), the finite thickness of the QCD axion window in Fig.~\ref{fig:Axion} reflects an order-one model dependence in the coefficient $g_{a \gamma \gamma}$~\cite{Kim:1979if, Shifman:1979if, Dine:1981rt, Zhitnitsky:1980tq}. See e.g.~\cite{Wu:2019exd, Barth:2013sma} for similar constraints that rely on the axion-electron and axion-neutron couplings.

Ultralight axions are also generic predictions of string theory and many beyond-the-Standard-Model scenarios with extra dimensions.\footnote{These axions are commonly called axion-like particles, or ALPs in short, in order to distinguish them from the QCD axion. Nevertheless, such a distinction is not important in this thesis. Unless stated otherwise, I will therefore broadly refer all of them simply as ``axions."} In particular, they arise from the Kaluza-Klein reduction of $p$-form gauge potentials (i.e.~massless antisymmetric tensor fields with $p$ number of indices) over $p$-cycles of compact spatial dimensions~\cite{Witten:1984dg, Svrcek:2006yi}. In string theory, these $p$-form fields arise naturally as part of the particle spectrum of the quantized string. These axions can be expressed more precisely as 
\beq
a_i(x^\mu) = \int_{\Sigma_{p}} \d^m y \, A_{p} \left( x^\mu, y^m \right) \, , \label{eqn:Stringaxion}
\eeq
where $x^\mu$ and $y^m$ are the coordinates of the four-dimensional spacetime and the compact space respectively, $A_p$ is the $p$-form gauge field, and $\Sigma_p$ denotes the compact manifold with non-trivial $p$-cycles. The label $i$ indexes the list $i=1, 2, \cdots N$, where $N$ is the number of non-trivial $p$-cycles on $\Sigma_p$. The relation (\ref{eqn:Stringaxion}) therefore illustrates that the number of axions is equal to the number of non-trivial $p$-cycles on $\Sigma_p$.\footnote{To unpack this rather complicated language, it is instructive to consider the example of a torus, which is a two-dimensional surface with two 1-cycles (the two circles that cannot be continuously deformed to one other) and one 2-cycle (the torus's interior volume). The Kaluza-Klein reduction of a 1-form, which is a higher-dimensional analog of the electromagnetic gauge field, would therefore produce two axions, while a 2-form would instead yield a single axion.} In string theory, the compact manifolds are six-dimensional surfaces with rich topologies and have large values of $N$~\cite{Candelas:1985en, Douglas:2006es}. As a result, a natural prediction of string theory is the existence of \textit{many} axions, possibly dozens or even hundreds of them, in our Universe~\cite{Arvanitaki:2009fg, Acharya:2010zx, Cicoli:2012sz}.

The axions which we just described enjoy a continuous shift symmetry to all orders in perturbation theory, as a direct consequence of the higher-dimensional gauge invariance of the $p$-forms. However, like the QCD axion, non-perturbative effects break this continuous shift symmetry to a discrete shift symmetry. In string theory, these non-perturbative effects can arise from a myriad of sources, including D-branes, world-sheet instantons, gauge-theory instantons, and so on~\cite{tHooft:1976rip, Dine:1986zy, Becker:1995kb, Kallosh:1995hi, Polchinski:1995mt}. The low-energy effective action of these axions therefore takes the following form:
\beq
\mathcal{L}_a = - \frac{1}{2} K_{ij} \, \partial_\mu a^i \partial^\mu a^j - \sum_{i=1}^N \Lambda_i^4 \Big[ 1 - \cos \left( {Q_i}^j a_j / 2 \pi f \right)  \Big] \, , 
\eeq
where $K_{ij}$ is the internal metric that describes the couplings between the axions, $\Lambda_i$ are dynamically generated scales, and ${Q_i}^j$ parameterizes the effective decay constant $\sim f/ |Q_{i}^j |$ of each of the axion. Crucially, $\Lambda_i \propto M_s \hskip 1pt e^{-S_i}$ is exponentially suppressed by the non-perturbative effects, where $S_i \gg 1$ is the sum of the action of these effects and $M_s$ is a high-energy scale, typially taken to be the string scale. As a result, the axion masses are generically expected to be extremely small (see e.g.~\cite{Demirtas:2018akl} for the distribution of axion mass for a given choice of compact manifold). Crucially, unlike (\ref{eqn:QCDaxionmass}), the masses need not be related to decay constants at all. The plausible parameter ranges for these axions are therefore significantly wider than for the QCD axion. For instance, the constraints outside the QCD axion window in Fig.~\ref{fig:Axion} should be viewed as applicable to these axions as well. Similar constraints obtained via the axion-electron and axion-neutron couplings~\cite{Wu:2019exd, Barth:2013sma} are also applicable.  Crucially, in all of these cases, there are virtually no constraints in the ultralight and weakly-interacting regime.

\begin{figure}[t]
\centering
\includegraphics[scale=0.58, trim = 3 10 0 0]{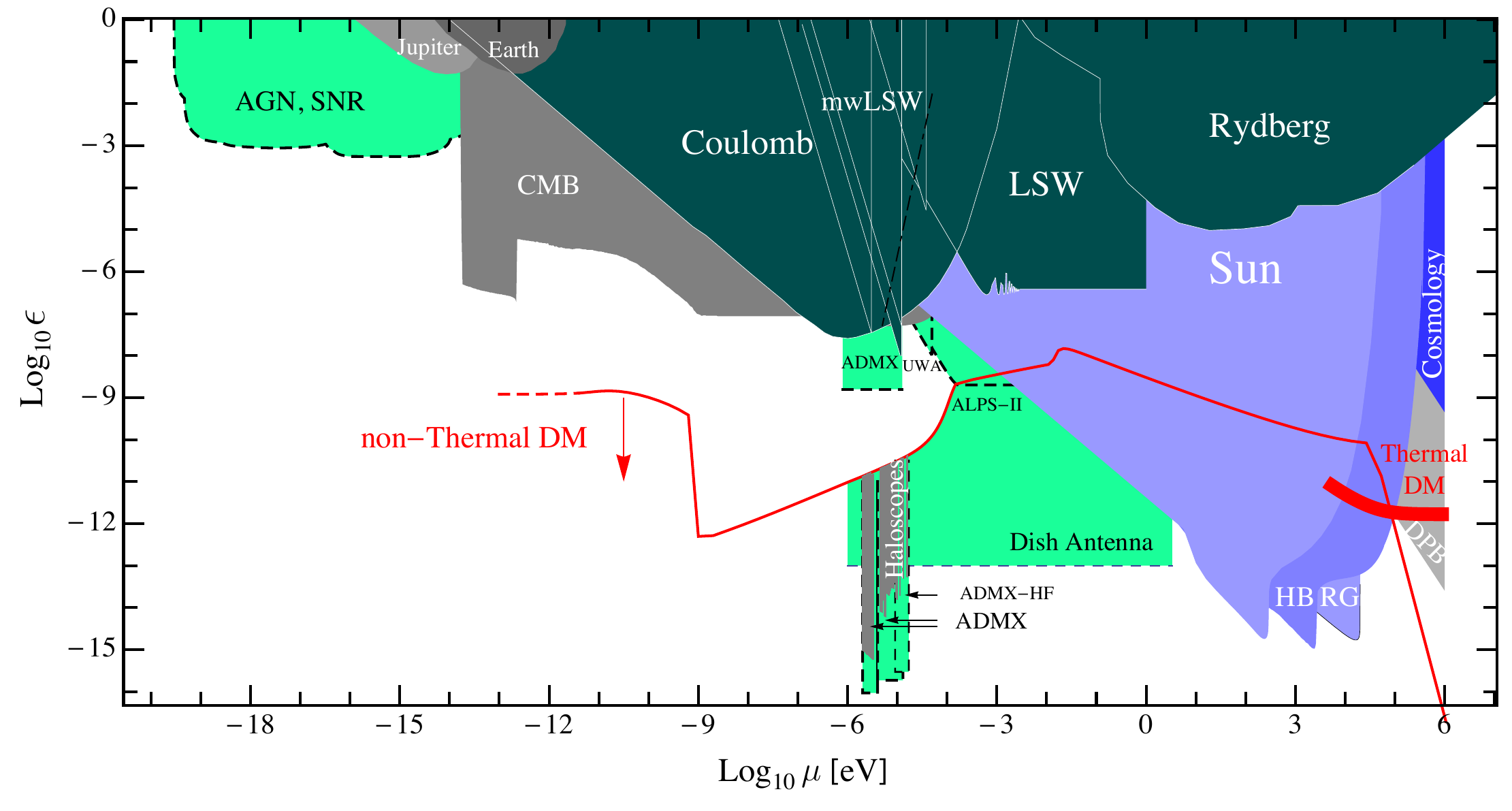}
\caption{Illustration of the constraints on dark photons. Details of these observational constraints can be found in e.g.~\cite{Essig:2013lka}. Similar to the axions, the the small-mass and weak-coupling parameter regime is virtually unconstrained by these observations. Figure adapted from~\cite{Essig:2013lka}. }
\label{fig:DarkPhoton}
\end{figure}

While we have focused on ultralight scalar fields so far, ultralight vector fields can also arise in many scenarios of physics beyond the Standard Model. Dark photons, which we have described in \S\ref{sec:ParticleDM}, constitute a well-motivated class of such new particles. These vector fields can acquire their masses through the Higgs mechanism, in which case $\mu \sim \epsilon \, m_H$ and $m_H$ is the Higgs scale associated to either the electroweak sector or the dark sector. The masses are naturally suppressed by the fact that both $m_H$ for the dark sector and $\epsilon$ can be extremely small~\cite{delAguila:1988jz, Goodsell:2009xc, Camara:2011jg}. In addition, if dark photon couples to a strongly-interacting dark sector, the associated dynamical symmetry breaking would naturally generate a vector mass that is exponentially suppressed compared to a high-energy scale~\cite{Reece:2018zvv}. In any case, a small vector mass is technically natural because gauge symmetry is restored in the massless limit, making ultralight vector fields ubiquitous in many particle physics models. Current constraints on the dark photon in the $\epsilon$-$\mu$ parameter space are shown in Fig.~\ref{fig:DarkPhoton} (see~\cite{Fabbrichesi:2020wbt} for a recent review on these experimental constraints). Crucially, this figure shows that dark photons with low masses, i.e. $\mu \lesssim 10^{-6}$ eV, are hard to detect by existing experiments.

Finally, we remark that many other types of ultralight bosons with spins have been proposed in the literature. For instance, dark radiation, which are ultralight vector fields that completely decouple from the Standard Model gauge group, also arise in string compactifications and many physics beyond the Standard Model scenarios~\cite{Goodsell:2009xc, Camara:2011jg}. In these cases, none of the constraints shown in Fig.~\ref{fig:DarkPhoton} would apply. Ultralight massive tensor fields could also exist in theories that modify gravity, in which case they must couple to Einstein gravity in a non-linear manner~\cite{Hinterbichler:2011tt, deRham:2014zqa}. The only realistic way of detecting these new ultralight bosons would therefore be through their gravitational couplings, such as in cosmology or through black holes. 

\section{Gravitational Atoms} \label{sec:atomreview}

In \S\ref{sec:UltralightBosons}, we saw that ultralight bosons are well-motivated particles that exist in many proposed extensions to the Standard Model. Interestingly, if these bosons have masses from $10^{-10}$ eV to $10^{-20}$ eV, they can be spontaneously amplified around rotating astrophysical black holes (see Fig.~\ref{fig:col1} for a comparison with other energy scales in our Universe). This is the case because the wave amplification mechanism, also known as \textit{black hols superradiance}, is most efficient when the Compton wavelength of the field is of the order of the gravitational radius of the black hole. 
Bosons in the stated mass range have Compton wavelengths that range from about $1$ km to $10^{10}$ km, which coincide with the sizes of known black holes in our Universe. Superradiance would generate in a coherent bosonic configuration that is gravitationally bounded to the rotating black hole. Remarkably, these bound states resemble the proton-electron structure of the hydrogen atom, and are therefore often called ``gravitational atoms". I will first describe the qualitative features of the superradiance phenomenon and then sketch some of the key properties of the gravitational atoms.

\subsection{Black Hole Superradiance} \label{sec:superrad}

Superradiance is the phenomenon where the amplitude of a bosonic wave is magnified due to the presence of dissipation.\footnote{By contrast, superradiant amplification cannot occur for fermionic fields~\cite{Unruh:1973bda, Unruh:1974bw, Manogue1988}. This is because the amplification process increases the occupation number of a single state, which is nevertheless forbidden by the Pauli exclusion principle.} In fact, it is a purely kinematic effect that pervades many areas of physics, though in subtly different guises (see~\cite{Bekenstein:1998nt, Brito:2015oca} for reviews). Superradiance was first understood in the context of Cherenkov radiation~\cite{Cerenkov:1934zz}, which is emitted by a medium due to the superluminal inertial motion of a charged particle that propagates within it. More precisely, this radiation is emitted when the linear velocity of the particle, $v$, exceeds the phase velocity of the radiation in the medium~\cite{Frank:1937fk}
\beq
v > \frac{c}{N} \, ,  \label{eqn:linearCherenkov}
\eeq
where $c$ is the speed of light in vacuum (restored for clarity) and $N > 1$ is the medium's index of refraction. In this case, the electromagnetic interaction between the charged particle and the atoms in its surrounding region provides the source of transfering energy between the particle and the medium. As the particle's superluminal motion triggers the formation of a radiation shock front, its kinetic energy is transfered to excite the nearby atoms in the medium. Cherenkov radiation is therefore the radiation that is spontaneously emitted by the relaxation of these excited atoms. Crucially, the amplitude of this emission can be large due to the bosonic nature of photons, which allows for a large occupation number of the excited state.

Superradiance can also be triggered by the rotational motion of an axisymmetric body. This was first conceived by Zeldovich through his thought experiment of the scattering an electromagnetic wave onto a rotating metal cylinder~\cite{Zeldovich:1971a, Zeldovich:1972spj}. In this example, the conducting nature of the cylinder's surface provides a mechanism of transfering of energy and angular momentum between the cylinder and the wave. In general, the incident wave would be partially reflected and partially absorbed by the cylinder. However, if the cylinder is highly rotating, the incident wave can instead extract energy and angular momentum from the cylinder, resulting in a reflected wave that has a larger amplitude. This only occurs when the angular velocity of the cylinder, $\Omega_{\rm c}$, exceeds the angular phase velocity of the incident wave 
\beq
\Omega_{\rm c} > \frac{\omega}{m} \, , \label{eqn:superradcylinder}
\eeq
where $\omega$ and $m$ are the frequency and azimuthal angular momentum number of the wave. Indeed, this kinematic condition is merely the rotational analog of (\ref{eqn:linearCherenkov}). While we have described rotational superradiance in a scattering process, electromagnetic waves can also be spontaneously emitted due to the quantum-mechanical fluctuations of these modes in the vacuum~\cite{Unruh:1974bw}. Although rotational superradiance is theoretically well-understood, this way of amplifying light has not been experimentally realized because it is difficult to spin cylinders up to the order of the frequency of electromagnetic waves. Instead, rotational superradiance has been detected through the amplification of phonons by draining vortex flows in water tanks~\cite{Torres:2016iee}, which have been proposed to be analog systems for black holes~\cite{Unruh:1980cg, Weinfurtner:2010nu}.

Remarkably, rotational superradiance can also be excited by highly-rotating black holes~\cite{Starobinsky:1973aij, Starobinsky:1974spj, Bekenstein:1973mi}. In fact, black holes are in many ways ideal objects to trigger this amplification process. To begin, the event horizon of a black hole is a perfect absorber and therefore provides the most efficient mechanism of transfering energy and angular momentum with its environment. Furthermore, this process only relies on the gravitational interaction between the black hole and the fields in its vicinity. By virtue of the equivalence principle, all boson fields in our Universe that satisfy the superradiance condition can therefore experience this amplification.\footnote{In practice, however, not all bosons would amplify appreciable over reasonable astrophysical timescales. See \S\ref{sec:Hatom} for more details about the rates of amplification.} Crucially, these fields can even be spontaneously amplified through their quantum-mechanical fluctuations in the vacuum~\cite{Unruh:1974bw}. The kinematic condition for black hole superradiance is similar to (\ref{eqn:superradcylinder}), except the angular velocity is replaced by that of the black hole:
\beq
\Omega_{\rm H} > \frac{\omega}{m} \, , \label{eqn:superradiant0}
\eeq
where $\omega$ and $m$ are now the frequency and azimuthal angular momentum of any boson field and not necessarily those of light. The condition (\ref{eqn:superradiant0}) demonstrates that only modes that co-rotate with the black hole ($m > 0$) can grow, while counter-rotating modes ($m < 0$) would be decay into the black hole. Crucially, because the Kerr metric (\ref{equ:Kerr}) is perfectly axisymmetric, these growing and decaying modes do not mix with one another around isolated black holes. Recall that (\ref{eqn:superradiant0}) was also mentioned in \S\ref{sec:BH} in the context of the Teukolsky equation, which describes the perturbation of a black hole by a massless wave. Here I emphasize that (\ref{eqn:superradiant0}) also applies to massive perturbations, with $\omega$ dependent on the boson mass (see \S\ref{sec:Hatom} for more details).

It is perhaps counter-intuitive that waves can extract energy and angular momentum from a black hole --- isn't the event horizon supposed to be a one-way membrane in which everything that enters it cannot escape? Rotating black holes are in this sense special because, from the point of view of an asymptotic observer at infinity, ingoing waves that satisfy (\ref{eqn:superradiant0}) have negative energies, which are equivalent to fluxes of outgoing positive energy. Superradiance is therefore qualitatively similar to the Penrose process described in \S\ref{sec:BH}, in which case the energy of a black hole is extracted because particles can have negative energies inside the black hole's ergoregion. Indeed, the very idea that black holes can actually lose energy has proven to be very influential in the history of physics. For instance, superradiance was in fact the key inspiration behind the formulation of Hawking radiation~\cite{Hawking:1974sw} (see~\cite{sunyaev2004zeldovich} for Thorne's personal account on visiting Zeldovich along with Hawking in the early 1970s). The latter is nevertheless qualitatively distinct from superradiance in two important ways: Hawking emission is a quantum-mechanical phenomenon and it can be radiated even for Schwarzschild black holes.

\subsection{Hydrogenic Bound States} \label{sec:Hatom}

We described that black hole superradiance can in principle occur for both massive and massless boson fields. However, this process is only efficient for massive boson fields because the boson's mass, $\mu$, sources an additional potential barrier at distances far away from the black hole (see e.g. Fig. 7 of~\cite{Arvanitaki:2010sy}). In the superradiant-scattering picture, this barrier would reflect waves that have already been amplified back to the black hole and hence trigger more superradiance. As a result, the wave's amplitude further increases while the black hole spins further down. This process would repeat continuously until the angular velocity of the black hole saturates the inequality (\ref{eqn:superradiant0}). The final state of this black-hole instability is a cloud of boson condensate that is gravitationally bounded to a slowly-rotating black hole. Depending on the efficiency of the superradiant extraction, the mass of the cloud can range up to about $30\%$ of the mass of the black hole~\cite{Christodoulou:1970}. This way of extracting energy and angular momentum efficiently from a black hole is a natural realization of the ``black-hole bomb" thought experiment~\cite{Press:1972zz}. Importantly, this instability does not occur for a massless field, which does not source a reflection barrier and therefore only experiences superradiance as a single event.

The key parameter underlying the superradiance phenomenon of a massive field is the ratio of the gravitational
radius of the rotating black hole, $r_g$, to the (reduced) Compton wavelength of the field, $\lambda_c$. This is often called the gravitational fine-structure constant, whose typical values are of order
\beq
\alpha \equiv \frac{G M \mu}{\hbar c} \simeq 0.04 \left(\frac{M}{60 M_\odot}\right) \left(\frac{\mu}{10^{-13}\, \lab{eV}}\right) \, . \label{eqn:alphaDef}
\eeq
This parameter plays a similar role as the fine-structure constant of QED and dictates virtually all properties of the boson clouds, including their sizes and energy spectra. As we shall see, superradiance occurs when $\alpha$ is smaller than order unity. From (\ref{eqn:alphaDef}), we find that ultralight bosons with masses from $10^{-20}$ eV to $10^{-10}$ eV are therefore efficiently amplified by astrophysical black holes in our Universe, which are believed to have masses approximately from $1 M_{\odot}$ to $10^{10} M_{\odot}$.

Although superradiance can occur for bosons for any spin, in what follows I only focus on boson clouds made of scalar and vector fields. I will also assume that the self interactions of the fields are negligible, such that the only interaction in these systems is the gravitational coupling between the boson's mass and the Kerr metric. There is an extensive literature on superradiance of minimally-coupled scalar~\cite{Detweiler:1980uk, Dolan:2007mj, Arvanitaki:2010sy, Yoshino:2012kn, Dolan:2012yt, Okawa:2014nda, Yoshino:2014, Baumann:2018vus, Brito:2014wla, Arvanitaki:2014wva, Arvanitaki:2016qwi, Brito:2017wnc, Brito:2017zvb} and vector fields~\cite{Witek:2012tr, Pani:2012vp, Pani:2012bp, Endlich:2016jgc, East:2017ovw, East:2017mrj, East:2018glu, Baryakhtar:2017ngi, Cardoso:2018tly, Dolan:2018dqv, Baumann:2019eav, Siemonsen:2019cr}. Here I only review some of the key properties of these clouds that are known in the literature before the works presented in subsequent chapters were pursued.

\subsubsection*{Scalar clouds}

Scalar boson clouds are the most well-studied examples of the gravitational atom. They are bound-state solutions of the Klein-Gordon equation of the scalar field $\Phi$ around the spacetime of a rotating black hole
\beq
\left(g^{\alpha \beta} \nabla_\alpha \nabla_\beta - \mu^2 \right) \Phi (t, \textbf{r}) = 0 \, , \label{equ:KG} 
\eeq
where $g_{\alpha \beta}$ is the Kerr metric (\ref{equ:Kerr}). As described above, these solutions are remarkably similar to the states of the hydrogen atom in quantum mechanics. To make this manifest, it is useful to consider the ansatz~\cite{Arvanitaki:2010sy}
\beq
\Phi(t, \bm{r}) = \begin{dcases}  \frac{1}{\sqrt{2\mu}} \left[ \psi (t, \bm{r})\, e^{-i \mu t} + \psi^* (t, \bm{r}) \, e^{+ i \mu t}  \right] & \text{$\Phi$ real}\,, \\[4pt]
\frac{1}{\sqrt{2 \mu}} \, \psi (t, \bm{r})\, e^{-i \mu t}  & \text{$\Phi$ complex}\,, \end{dcases}   \label{equ:NRansatz}
\eeq
where $\psi$ is a complex scalar field which varies on a timescale that is longer than $\mu^{-1}$, and is therefore often called the non-relativistic field. Substituting this ansatz into (\ref{equ:KG}) and expanding in powers of $\alpha$, the Klein Gordon equation reduces to a Schr\"{o}dinger-like equation
\begin{equation}
  i \frac{\partial}{\partial t} \psi (t, \textbf{r}) =  \left( -\frac{1}{2\mu} \nabla^2  - \frac{\alpha}{r} + \Delta V\right) \psi (t, \textbf{r}) \, , \label{eqn:NonRelScalar}
\end{equation}
where $\Delta V$ represents higher-order corrections in $\alpha$.  At~leading order, this is identical to the Schr\"{o}dinger equation for the hydrogen atom, whose eigenfunctions are labeled by the principal ``quantum" number $n$, the orbital angular momentum number~$\ell$, and the azimuthal angular momentum number~$m$.\footnote{The rigorous definitions of these quantum numbers, in the context of the (approximate) isometries of the Kerr metric, will be discussed in \S\ref{sec:TensorKerr} in detail.} These quantum numbers satisfy the inequalities $n \geq \ell+1$, $\ell \geq 0$, and $ \ell \geq |m|$. The bound-state solutions of the scalar cloud at leading order are thus given by
 \beq
\psi_{n \ell m}(t, \textbf{r} ) = R_{n \ell}(r) Y_{\ell m}(\theta, \phi ) \, e^{- i (\omega_{n \ell m}-\mu) t}  \, , \label{eqn:ScalarEigenstate}
\eeq
where $Y_{\ell m}$ are the scalar spherical harmonics and $R_{n \ell}$ are the hydrogenic radial functions 
\beq
R_{n \ell} (r) = A \, e^{- \sqrt{\mu^2 - \omega^2} \hskip1pt r } r^\ell L^{(2\ell+1)}_{n-\ell-1} \left( 2\sqrt{\mu^2 - \omega^2}\hskip1pt r \right)\, , \label{eqn:RadialHydrogen}
\eeq
with $A$ the amplitude of the field and $L$ the associated Laguerre polynomial. For notational simplicity, we will denote the normalized eigenstates (\ref{eqn:ScalarEigenstate}) by $|n \es \ell \es m\rangle$, with $\langle n \es \ell \es m | n' \es \ell' \es m'\rangle = \delta_{n n'} \delta_{\ell \ell'} \delta_{m m'}$. The overall amplitude $A$, determined by the total mass of the cloud, will be restored when necessary. Notice that for small values of $\alpha$, following the analogy with the hydrogen atom, the radial profile peaks at the ``Bohr radius'' 
\beq
r_\lab{c} \equiv (\mu \alpha)^{-1} \label{equ:BohrRadius}\, .
\eeq
The cloud is thus concentrated at a distance which is larger than both the gravitational radius of the black hole and the Compton wavelength of the~field.

\begin{figure}[t]
    \centering
    \begin{minipage}{0.45\textwidth}
        \centering
        \includegraphics[scale=0.9]{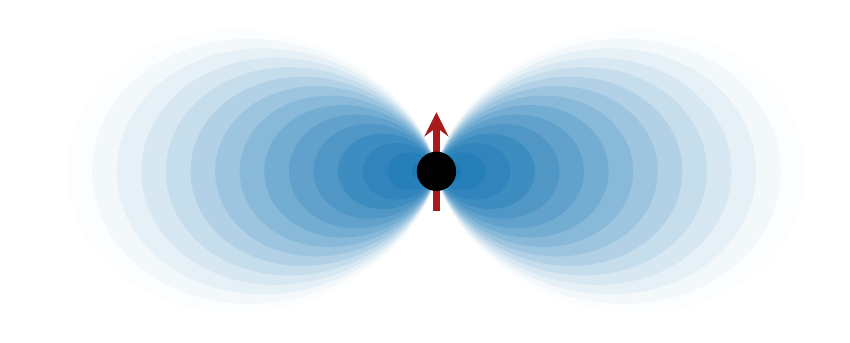}  
        \end{minipage}\hfill
    \begin{minipage}{0.45\textwidth}
        \centering
        \includegraphics[scale=0.9]{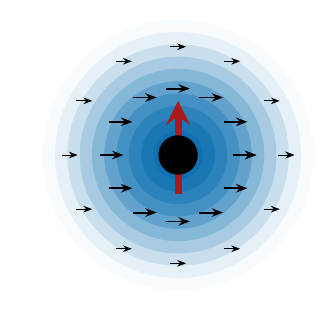} 
    \end{minipage}
    \vspace{-0.25cm}
    \caption{Illustration of the dominant growing modes of the scalar ($2p$) and vector ($1s$) gravitational atoms. The small arrows denote the intrinsic spin of the vector field, which allows for superradiant growth even for states without orbital angular momentum.}
    \label{fig:atom}
\end{figure}

Despite the similarity between the hydrogen and the gravitational atoms,  there is an important difference between them. While the former has wavefunctions that are regular at $r = 0$, the latter must satisfy purely ingoing boundary conditions at the black hole's event horizon. As a consequence, the eigenstates of a boson cloud have eigenfrequencies which are generally complex,
\beq
\omega = E + i \es \Gamma \, ,
\eeq
where $E$ and $\Gamma$ denote the energies and instability rates, respectively.  Before the work presented in Chapter~\ref{sec:spectraAtom} was pursued, the explicit expressions for $E$ and $\Gamma$ are known only at leading order. In particular, they are~\cite{Detweiler:1980uk, Arvanitaki:2010sy} 
\begin{align}
E_{n \ell m}  &= \mu \left( 1 - \frac{\alpha^2}{2n^2}  + \mathcal{O}\left(\alpha^3\right) \right)   , \label{eqn:scalarspectrum0} \\
\Gamma_{n \ell m} &= 2 \tilde{r}_+ C_{n \ell} \, g_{\ell m}(\tilde{a}, \alpha,\omega) \, (m \Omega_\lab{H}   - \omega_{n \ell m})\hskip 1pt \alpha^{4\ell+5}  \, , \label{eqn:ScalarRate0}
\end{align}
where the numerical coefficients $C_{n \ell}$ and $g_{\ell m}$ can be found in \S\ref{sec:Summary}, while $\tilde{r}_+ \equiv r_+/M$ and $\tilde{a} \equiv a/M$ are the dimensionless black hole outer horizon and spin. Following the analogy with the hydrogen atom, we will refer (\ref{eqn:scalarspectrum0}) as the Bohr $(\Delta n \neq 0)$ energy levels of the gravitational atom. From (\ref{eqn:ScalarRate0}), we clearly see that a scalar mode only grows when the superradiance inequality (\ref{eqn:superradiant0}) is satisfied. The dominant growing mode is the $|n \es  \ell \es m\rangle  = |2\es 1\es 1\rangle$ state, with $\Gamma_{211} \propto \alpha^9/M$ (see Fig.~\ref{fig:atom} for an illustration of this dominant $2p$ scalar configuration). Because $\Gamma_{211}$ depends so sensitively on $\alpha$, the growth timescale of the scalar cloud can easily vary from few minutes to over astrophysical timescales. In particular, the typical growth timescale is
\beq
\Gamma^{-1}_{211} \simeq \frac{10^6 \, \text{yrs}}{\tilde{a}} \left( \frac{M}{60 M_\odot} \right) \left( \frac{0.019}{\alpha} \right)^{9} \, . \label{eqn:211growthtime}
\eeq
By demanding the scalar cloud to grow appreciably within the age of the Universe, we can place an observationally-relevant lower bound of $\alpha \gtrsim 0.005$. In addition, the superradiance condition $m \hskip 1pt \Omega_{\rm H} > \omega$ also places an upper bound of $\alpha <0.5$, which is obtained by assuming a maximally-spinning black hole. A black hole can therefore in principle probe bosons with masses over two orders of magnitude, although the amplification process is more efficient for larger values of $\alpha$.

While the spectra (\ref{eqn:scalarspectrum0}) and (\ref{eqn:ScalarRate0}) are identical for both real or complex scalar fields, their stability properties may be different. Because the energy-momentum tensor of a real scalar field is always time dependent and non-axisymmetric, clouds made out of real fields are a continuous source of nearly-monochromatic gravitational waves~\cite{Arvanitaki:2010sy}. The power of this emission is given by~\cite{Arvanitaki:2010sy, Yoshino:2014}
\beq
\begin{aligned}
P_{n \ell} = D_{n \ell} \left(\frac{M_c}{M} \right)^2 \alpha^{4\ell+10} \, , \label{eqn: Schwarzschild strain 2}
\end{aligned}
\eeq
where $M_c \simeq \mu A^2$ is the mass of the cloud and the coefficient $D_{n \ell}$ can be found in~\cite{Yoshino:2014}. Complex fields, on the other hand, can have certain configurations that suppress this gravitational-wave emission. This occurs when the real and imaginary components of the field are in the same eigenstate, but their relative phase is such that they produce time-independent and axisymmetric configurations. Whether this is realized in a specific case depends on the initial conditions of the superradiant growth of the cloud. In either case, since the timescale for this gravitational-wave emission is much longer than the timescale of the superradiance instability, the former doesn't inhibit the formation of the cloud~\cite{Yoshino:2014}. For instance, the typical lifetime of the $\ket{211}$ mode for a real scalar field is~\cite{Arvanitaki:2010sy, Yoshino:2014}
\beq
T_{211} \simeq 10^8 \, \text{yrs} \left( \frac{M}{60 M_\odot} \right) \left( \frac{0.07}{\alpha} \right)^{15} \, , \label{eqn:211lifetime}
\eeq
which is much longer than its typical growth timescale (\ref{eqn:211growthtime}). However, depending on the values of $M$ and $\alpha$, the clouds may still deplete on cosmological and astrophysical timescales.

\newpage

Our discussion above summarizes the leading-order behaviour of the scalar cloud. However, an important gap in the literature was that modes with different values of $\ell$ and $m$ remain degenerate in (\ref{eqn:scalarspectrum0}). To fully understand the qualitative features of the scalar atom, it is important that we pursue higher-order corrections to $E$, up to the order in $\alpha$ at which the degeneracies between all modes in the spectrum are broken. This is not just a pure academic exercise; rather, we shall find in this thesis that these small energy splittings would play crucial roles in determining the evolution of the clouds when they are part of binary systems. The rigorous analytic and numeric computations of these corrections will be shown in Chapter~\ref{sec:spectraAtom}. Continuing the analogy with the hydrogen atom, we will refer to the energy splittings that differ by $\Delta \ell \neq 0$ and $\Delta m \neq 0$ as fine and hyperfine splittings, respectively. These results are summarized in \S\ref{sec:Summary}, and the full scalar spectrum is illustrated in Fig. \ref{fig:ScalarSpectra}.

\subsubsection*{Vector clouds} 

The vector clouds are bound-stated solutions of the Proca equation of a massive vector field $A_\mu$ around the Kerr metric,
\beq
\left(g^{\alpha \beta} \nabla_\alpha \nabla_\beta - \mu^2 \right) A^\mu = 0 \, , 
\label{equ:Proca}
\eeq
which must be supplemented by the Lorenz condition  $ \nabla_\mu A^\mu = 0$. Because the Lorenz condition completely determines $A^0$, we may focus entirely on the spatial components $A^i$. Just like in the scalar case, it is convenient to introduce the analog of the ansatz (\ref{equ:NRansatz}) for $A^i$, where the non-relativistic field is now denoted by $\psi^i$. Subtituting this ansatz into (\ref{equ:Proca}) and expanding in powers of $\alpha$, the Proca equation becomes~\cite{Baryakhtar:2017ngi} 
\beq
i \frac{\partial }{\partial t} \psi^i (t, \textbf{r})  =  \left[ -\frac{\delta^{il}}{2\mu} \nabla^2  - \delta^{il} \, \frac{\alpha }{r}   + \Delta V^{il}   \right] \psi^l (t, \textbf{r})  \, , \label{eqn:NonRelVector}
\eeq
where the terms in (\ref{eqn:NonRelVector}) share the same interpretation as their scalar analogs in (\ref{eqn:NonRelScalar}). Since spherical symmetry is restored when $\alpha \ll 1$, the vector eigenstates can still be characterized by definite angular momenta numbers in this limit. It is therefore straightforward to express the leading-order eigenfunctions as follows
\beq
\begin{aligned}
\psi_{n\ell j m}^i (t, \textbf{r}) & = R_{n \ell}(r) Y^i_{\ell, j m} (\theta, \phi) \, e^{- i (\omega_{n \ell j m} -\mu)t} \,  , 
\label{eqn:FarLOAsol}
\end{aligned}
\eeq
where $R_{n \ell}$ is the hydrogenic radial function (\ref{eqn:RadialHydrogen}) and the angular function $Y_{\ell, j m}^i$ is the pure-orbit vector spherical harmonic~\cite{Baryakhtar:2017ngi, Thorne:1980ru}. Compared to the scalar solution (\ref{eqn:ScalarEigenstate}), we see the vector cloud has a more elaborate angular structure. In particular, the eigenstates of massive vector fields are now labeled by four ``quantum numbers'' $\{ n, \ell, j, m \}$, where $n$ and $\ell$ have the same interpretation as in the scalar case, and $j = \ell\pm1$, $\ell$ is the total angular momentum with $m$ its azimuthal component. That $j$ can take three different values for a given $\ell$ is a direct manifestation of the fact that a massive vector field has three independent degrees of freedom. Since (\ref{eqn:FarLOAsol}) acquires a factor of $(-1)^{\ell+1}$ under a parity transformation, the $j = \ell \pm 1$ states are called electric-type modes, while the $j = \ell$ states are magnetic-type.  For notational simplicity, we will denote the normalized states (\ref{eqn:FarLOAsol}) by $|n \es \ell \es j \es m\rangle$.

Like the scalar atom, the boundary condition at the black hole horizon implies the vector clouds have complex eigenfrequencies. Prior to the computations performed in Chapter~\ref{sec:spectraAtom}, the energies and instability rates of the vector quasi-bound states are found to be~\cite{Baryakhtar:2017ngi}
\begin{align}
E_{n \ell j m} &= \mu \left(  1 -\frac{\alpha^2}{2n^2} + \mathcal{O} \left( \alpha^3 \right) \right) , \label{eqn:vectorspectrum0}\\
\Gamma_{n\ell jm} & \propto  \left( m \Omega_H - \omega_{n \ell j m} \right) \alpha^{2 \ell + 2 j + 5} \, . \label{eqn:VectorRates0}
\end{align}
At leading order, the vector energy is the same as its scalar counterpart (\ref{eqn:scalarspectrum0}). This is to be expected, since both the scalar and vector fields have the same Coulomb-like potential in their equations of motion, cf. (\ref{eqn:NonRelScalar}) and (\ref{eqn:NonRelVector}). On the other hand, the vector instability rate is only known up to the superradiance condition and the $\alpha-$scaling, which depends on both $\ell$ and $j$. Through this $\alpha$-scaling, we find that the dominant growing mode for the vector field is $|1 \es 0\es 1\es 1\rangle$, which has $\Gamma_{1011} \propto \alpha^7/M$  (see Fig.~\ref{fig:atom} for an illustration of this dominant $1s$ vector configuration). This is enhanced, by two powers of $\alpha$, with respect to the dominant growing mode for the scalar field. In particular, the typical growth timescale of the dominant vector mode is
\beq
\Gamma^{-1}_{1011} \simeq \frac{10^6 \, \text{yrs}}{\tilde{a}} \left( \frac{M}{60 M_\odot} \right) \left( \frac{0.0033}{\alpha} \right)^7 \, . 
\eeq
This shows that vector fields with $\alpha \gtrsim 0.0005$ can grow significantly within the age of the Universe. Combined with the upper bound $\alpha < 0.5$ derived from the superradiance condition, we find a black hole can probe ultralight vector fields with masses spanning over three orders of magnitude, which is significantly wider than the scalar case. As in the scalar caase, the vector clouds could also dissipate through the monochromatic gravitational-wave emission if their configurations have time-dependent and non-axisymmetric components. Unfortunately, an analytic expression like (\ref{eqn: Schwarzschild strain 2}) for the vector emission power is not known. Nevertheless, there had been attempts to estimate the lifetime of the dominant $\ket{1011}$ mode through numerical relativity results and analytic determination of its $\alpha$-scaling~\cite{Baryakhtar:2017ngi, East:2017mrj, East:2018glu}. In this case, it was found that the typical lifetime of the vector $1s$ mode is
\beq
T_{1011} \simeq 10^8 \, \text{yrs} \left( \frac{M}{60 M_\odot} \right) \left( \frac{0.01}{\alpha} \right)^{11} \, .
\eeq
Compared to (\ref{eqn:211lifetime}), the dominant vector cloud dissipates much faster than its scalar counterpart for the same values of $M$ and $\alpha$.

The coefficient of (\ref{eqn:VectorRates0}) was not known in the literature because the Proca equation is hard to solve near the black hole horizon, which is necessarily for a rigorous computation for $\Gamma_{n \ell j m}$. Previous attempts had at least been made to calculate the coefficient for the dominant $1s$ mode with different approximation schemes~\cite{Pani:2012bp, Endlich:2016jgc,Baryakhtar:2017ngi}, though these results did not agree with one other (see Table 1 of~\cite{Cardoso:2018tly} for a summary). In Chapter~\ref{sec:spectraAtom}, I resolve this issue by rigorously deriving the analytic expression for $\Gamma_{n \ell j m}$ that is valid for all vector modes. In addition, like the scalar case, I also derive the analytic results for the higher-order corrections to the vector energies (\ref{eqn:vectorspectrum0}), up to the order in $\alpha$ at which the degeneracies between all modes in the vector spectrum are lifted (see Fig. \ref{fig:VectorSpectra} for an illustration of the vector spectrum). Numeric computations of both the vector energies and instability rates are also demonstrated, in which case we find excellent agreement between the analytic and numeric results. These results, along with those obtained for the scalar atom, are summarized in \S\ref{sec:Summary}. Importantly, they represent the state of the art and complete our qualitative and quantitative understandings of both the scalar and vector clouds.

Before proceeding to subsequent chapters, it is important to emphasize the key qualitative distinctions betweeen the scalar and the vector atoms. Firstly, because the scalar field has no intrinsic spin, it can only excite superradiance by having $\ell > 0$. The centrifugal barrier generated by the orbital angular momentum therefore causes the scalar modes to peak far away form the black hole, as for the dominant $2p$ state. On the other hand, a vector field has intrinsic spin and can trigger superradiance without relying on orbital angular momentum. Vector states with $\ell=0$, such as the dominant $1s$ mode, therefore have non-negligible support near the black hole and are sensitive to strong gravity effects. A second important difference between the two types of atoms is the plurality of the number of states in their spectra. As we shall find in Chapter~\ref{sec:spectraAtom}, the vector field contains many more nearly-degenerate energy levels in its spectrum. This simple fact would turn out to play an important role in distinguishing the phenomenologies of the scalar and vector clouds when they are part of binary systems. In particular, level transitions which are induced by the binary companion typically would only involve two levels for the scalar atom, while multiple levels can simultaneously participate in the vector case. As I will elaborate in Chapters~\ref{sec:Collider} and \ref{sec:signatures}, a two-level and a multi-level transition would leave qualitatively different imprints in the gravitational waves emitted by the binary system.

\chapter{The Spectra of Gravitational Atoms} \label{sec:spectraAtom}

We saw in \S\ref{sec:Hatom} that the leading-order behaviours of the scalar and vector gravitational atoms resemble the Bohr-like structure of the hydrogen atom. However, higher-order corrections to these solutions are also of interest, as they play important roles in the dynamics of the clouds when they are part of binary systems (Chapter~\ref{sec:Collider}). The evolution of the cloud in turn perturbs the orbital motion of the binary system, thereby affecting the gravitational waves emitted by the binary (Chapter~\ref{sec:signatures}). In this chapter, I present the work done in~\cite{Baumann:2019eav}, where we investigated the spectra of the scalar and vector gravitational atoms in detail. Our principal goal was to accurately compute the energy splittings between the eigenstates of the clouds --- the analog of the fine and hyperfine structure of the hydrogen atom --- and the instability rates for both types of fields.

\section{Overview and Outline}

\begin{figure}[t]
\centering
\includegraphics[scale=0.92, trim = 0 0 0 0]{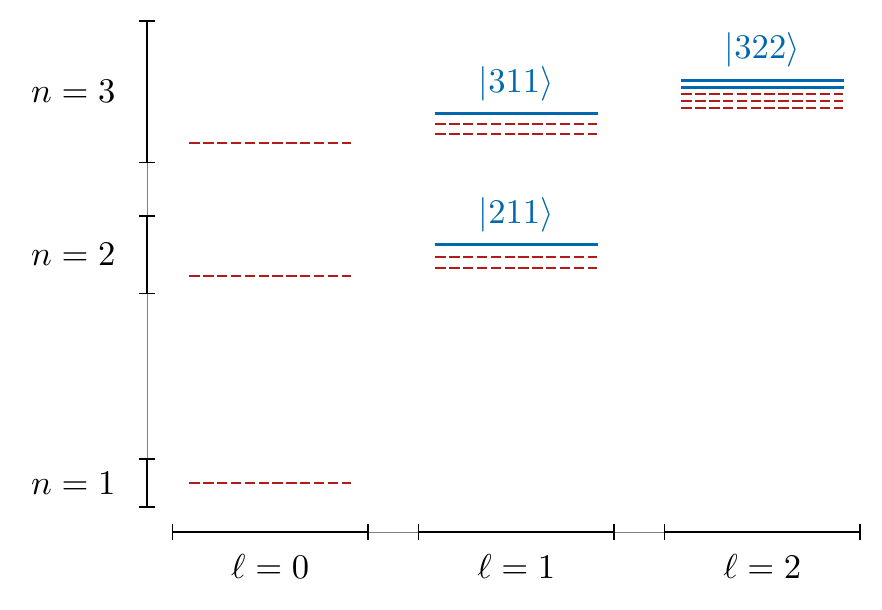}
\caption{Schematic illustration of the energy spectrum for a massive scalar field. Each state is labeled by the  quantum numbers $|\es n \es \ell \es m\rangle$ as described in \S\ref{sec:Hatom}. The solid blue lines are growing modes, while the dashed red lines are decaying modes. The dominant scalar mode is $\ket{211}$.} 
\label{fig:ScalarSpectra}
\end{figure}

When the gravitational fine structure constant (\ref{eqn:alphaDef}) is small, $\alpha \ll 1$, the spectra can be treated perturbatively as an expansion in powers of $\alpha$. Using such an expansion and the separability of the Klein-Gordon equation (\ref{equ:KG}), we solve for the energy spectrum~\cite{Baumann:2018vus, Baumann:2019eav} and instability rates~\cite{Detweiler:1980uk, Baumann:2019eav} of a massive scalar field, up to the order in $\alpha$ at which the degeneracies between all modes in the spectrum are broken (see Fig.~\ref{fig:ScalarSpectra} for an illustration of the scalar field spectrum). A similar computation for the massive vector field had been previously attempted~\cite{Endlich:2016jgc, Baryakhtar:2017ngi, Pani:2012vp, Pani:2012bp}, though not with full rigor. Until recently, the main technical obstruction in the vector analysis had been the fact that the Proca equation (\ref{equ:Proca}) could not be decomposed into separate
 angular and radial equations. However, in~\cite{Krtous:2018bvk, Frolov:2018ezx}, a separable ansatz was discovered for the electric modes  of a massive vector field on the Kerr background.\footnote{In~\cite{Dolan:2018dqv}, it was shown that a special class of magnetic modes are contained within this ansatz (see Appendix~\ref{app:details}). However, it is not yet clear whether it contains every magnetic mode, nor is it clear how to recover them.}  We use this ansatz to compute the spectrum and instability rates of a Proca field perturbatively. As described in \S\ref{sec:Hatom}, a special feature of superradiantly-generated vector clouds is that the dominant growing mode is a $1s$ state, which has non-negligible support near the black hole, see Fig.~\ref{fig:atom}. These modes are then especially sensitive to the near-horizon geometry of the black hole, and this sensitivity necessitate going beyond ordinary perturbation theory to  determine the spectrum.

While the magnetic modes of the Proca field are separable on the Schwarzschild background \cite{Rosa:2011my}, a separable ansatz in the Kerr spacetime remains elusive,
 and thus rigorous analytic results are difficult to achieve for this part of the spectrum. Fortunately, all degeneracies between the states in the spectrum are broken at linear order in the spin of the black hole, and the Schwarzschild ansatz still separates the Proca equation at this order~\cite{Pani:2012bp,Pani:2012vp}. Assuming that the fine and hyperfine structure of the magnetic modes are similar to their electric counterparts, we use this small-spin expansion to derive the magnetic spectrum perturbatively in $\alpha$. Furthermore,  we `derive' their leading-order instability rates using educated guesswork, and are thus able to attain perturbative results for the most phenomenologically relevant aspects of the spectrum, for all modes of the Proca field.

To test the analytic results, and to move beyond the limit of small $\alpha$, we compute the spectra of scalar and vector clouds numerically (see also e.g.~\cite{Dolan:2007mj,Pani:2012bp, Cardoso:2018tly, Dolan:2018dqv}). Formulating this as a nonlinear eigenvalue problem allow us to attain highly-accurate results for the spectrum with little computational cost. Our method also has the advantage that it does not rely upon a separable ansatz, which allows us to rigorously determine the magnetic spectrum and check our analytic guesswork for its instability rates. Our method is reliable even for very small values of $\alpha$, where precise numeric results are typically difficult to achieve without separability~\cite{Pani:2012bp, Cardoso:2018tly}, and we apply it to both the dominant growing modes and other phenomenologically relevant states. We pay special attention to the numerical obstacles that can potentially destroy the accuracy of a solution, and describe how to avoid them. This new formulation thus provides a robust and flexible method for finding the quasi-bound state spectra for arbitrary ultralight tensor fields about any stationary black hole, without relying on separability of the equations of motion.

The outline of this chapter is as follows: in Section~\ref{sec:preliminaries}, I describe the Klein-Gordon and Proca equations around rotating black holes. I also present a summary of our analytic and numeric results in \S\ref{sec:Summary}. In Section~\ref{sec:analytic}, I discuss how the method of matched asymptotic expansion is used to compute the spectrum of energy eigenvalues and the associated instability rates, for both the scalar field and the electric modes of the vector field. In addition, I describe the computation for the energy spectra of the magnetic modes at linear order in the black hole spin $\tilde a$ and motivate their conjectured instability rates. In Section~\ref{sec:numeric}, I present our numerical methods. Our techniques do not assume separability of the equations of motion and, in principle, work for ultralight fields of arbitrary spin. Numerous technical details are relegated to Appendix~\ref{app:SpectralDetails}.

To avoid notational clutter, I will set $\mu \equiv 1$ in this chapter, so that times and distances are measured relative to the typical oscillation timescale and Compton wavelength of the fields. The appropriate factors of $\mu$ will be restored when presenting explicit results, as in \S\ref{sec:Summary}.

\section{Scalars and Vectors around Kerr}
\label{sec:preliminaries}

This section outlines the fundamentals of computing the quasi-bound state solutions of the Klein-Gordon equation (\ref{equ:KG}) and Proca equation (\ref{equ:Proca}) around the Kerr background. In \S\ref{sec:TensorKerr}, we provide a rigorous discussion about the tensor representations of Kerr spacetime, thereby making precise the definitions of the ``quantum numbers" mentioned in \S\ref{sec:Hatom}. In \S\ref{sec:separable}, we review the separability of the equations of motion in the Kerr geometry. A summary of both the analytic and numeric results is presented in \S\ref{sec:Summary}, with the details of their derivations relegated to the subsequent sections.

\subsection{Tensor Representations} 
\label{sec:TensorKerr}

We begin with a discussion of general tensor fields $T_{\mu \nu \dots \rho}$ on the Kerr background. We are interested in quasi-bound states solutions, which are `purely ingoing' at the outer horizon and vanish at infinity. As in atomic physics, these solutions will be characterized by a set of discrete `quantum numbers' describing the energy, the orbital angular momentum and the intrinsic spin of the states. A precise definition of these quantum numbers is complicated by the fact that the spin of black hole breaks spherical symmetry. We will label our states by the quantum numbers they attain in the flat-space limit, where the spin of the black holes  can be ignored.

In order to define the flat-space limit, it is convenient to rescale the temporal and radial coordinates, $t \mapsto t \mu^{-1}$ and $r \mapsto r \mu^{-1}$, so that the metric (\ref{equ:Kerr}) becomes
\beq
\begin{aligned}
\mu^2 \d s^2 = & - \frac{\Delta}{\Sigma}\left(\d t - \alpha \hskip 1pt \tilde{a} \sin^2 \theta\, \d \phi \right)^2 + \frac{\Sigma}{\Delta} \d r^2 \\
&  + \Sigma \hskip 1pt \d \theta^2 + \frac{\sin^2 \theta}{\Sigma}  \left(\alpha \hskip 1pt \tilde{a} \hskip 1pt \d t - (r^2 + \alpha^2 \tilde{a}^2) \, \d \phi \right)^2\, , \label{equ:Kerr2}
\end{aligned}
\eeq
where $\tilde{a} \equiv a / M$ and 
\beq
\begin{aligned}
\Delta &\equiv r^2 -  2 \alpha r +\alpha^2 \tilde{a}^2\, ,  \\ 
\Sigma &\equiv r^2 + \alpha^2 \tilde{a}^2 \cos^2 \theta\, .
\end{aligned}
\eeq
All physical quantities are now measured in units of the Compton wavelength of the field, $\mu^{-1}$, which we henceforth set to $\mu^{-1} \equiv 1$. 
Notice that the spin parameter $\tilde{a}$ in (\ref{equ:Kerr2}) always appears in the combination~$\alpha \hskip 1pt \tilde{a}$, so spherical symmetry is approximately restored when $\alpha \ll 1$. In fact, in the limit $\alpha \to 0$, the line element (\ref{equ:Kerr2}) reduces to that of Minkowski spacetime, while at linear order in $\alpha$ it becomes the Schwarzschild solution. Spin-dependent terms, such as those corresponding to the Lense-Thirring effect, only appear at quadratic order in~$\alpha$. 
As far as the dynamics of the field is concerned, we therefore expect the effects of spin to be subleading compared to the gravitational potential sourced by $M$.

The eigenstates of the field are labeled by a discrete set of quantum numbers that reflect the (approximate) isometries of the background metric. To identify these quantum numbers, we first note that the Kerr geometry has two Killing vectors  
		\begin{equation}
			k_t \equiv - i \frac{\partial}{\partial t} \, , \qquad k_{z} \equiv - i \frac{\partial}{\partial \phi} \, , \label{eq:kerrIsometries}
		\end{equation}
representing the fact that the metric (\ref{equ:Kerr2})  is both stationary and axisymmetric. 
The equations of motion for an arbitrary tensor field $T_{\mu \nu \cdots \rho}$ on the Kerr background will inherit the isometries~(\ref{eq:kerrIsometries}). If these equations are linear, we may decompose their solutions in terms of states with definite frequency and azimuthal angular momentum,\footnote{Our analysis will focus entirely on complex scalar and vector fields, as the complex representations of these isometries are much simpler than for real fields. Because the equations of motion (\ref{equ:KG}) and (\ref{equ:Proca}) are linear, a real solution can be generated by simply taking the real part of a complex solution, and so the spectra are identical.}
\begin{equation}
    \begin{aligned}
        \pounds_{t} \hskip 2pt T_{\mu \nu \cdots \rho} &= -\omega  \hskip 2ptT_{\mu \nu \cdots \rho}\, , \\  \pounds_{z}  \hskip 2ptT_{\mu \nu \cdots \rho} &= +m  \hskip 2pt T_{\mu \nu \cdots \rho}\,, 
        \end{aligned} \label{eqn:DefiniteEigen}
    \end{equation}
where $\pounds_t \equiv \pounds_{k_t}$ and $\pounds_z \equiv \pounds_{k_z}$ are the Lie derivatives with respect to the isometries (\ref{eq:kerrIsometries}). In the Schwarzschild limit $\tilde{a} \to 0$, the geometry gains two more Killing vector fields $k_x$ and $k_y$ which, together with $k_z$, form an $\lab{SO}(3)$ algebra. 
We can then expand the field into its temporal and  spatial components, which we further expand in representations of this algebra, i.e. eigentensors of the \textit{total angular momentum} operator 
      \begin{equation}
      \pounds^2  \hskip 2pt T_{i k \cdots l} \equiv \left(\pounds_x^2 + \pounds_y^2 + \pounds_z^2 \right) T_{i k \cdots l} = j(j+1)  \hskip 2pt T_{i k \cdots l}\, , \label{eq:quadCasimir}
    \end{equation}
    where $j$ denotes the total angular momentum. Furthermore, in the flat-space limit, $\alpha \to 0$, solutions also have definite orbital angular momentum $\ell$ (cf.~Appendix \ref{app:harmonics}), so that they can be characterized by the quantum numbers $\omega$, $\ell$, $j$ and $m$. We will label such states by $|\es n\es \ell \es j \es m \rangle$, where we have introduced an integer quantum number $n$ that indexes the discrete quasi-bound state frequencies $\omega_n$. Since total and orbital angular momenta are the same for the scalar field, we will label its states by $|\es n \es \ell \es m \rangle$. States still have definite total angular momentum in the Schwarzschild limit, but indefinite orbital angular momentum. At finite $\tilde{a}$, these states no longer have definite total angular momentum. Nevertheless, we will still label our states by $|\es n \es \ell \hspace{0.5 pt} j \es m \rangle$ with the understanding that these quantum numbers regain their physical meaning as $\alpha \to 0$.

  Finally, the Kerr metric is invariant under the parity transformation 
  \beq
  {\cal P}:\,\,\, (\theta, \phi) \mapsto (\pi - \theta, \phi + \pi)\, . \label{equ:parity}
  \eeq
  This helps us to further organize the spectrum into states with definite parity. Solutions to the Klein-Gordon and Proca equations with definite parity are invariant under (\ref{equ:parity}) up to a sign, and, with our conventions, we have $\mathcal{P} |\es n \es \ell \es m \rangle = (-1)^{\ell} | \es n \es \ell \es m \rangle$ and $\mathcal{P} |\es n \es \ell \es j \es m \rangle = (-1)^{\ell +1} | \es n \es \ell \es j \es m \rangle$ for scalars and vectors, respectively. See Appendix~\ref{app:harmonics} for a more detailed discussion.

\subsection{Separable Equations of Motion} \label{sec:separable}

We will now cast the Klein Gordon equation (\ref{equ:KG}) and the Proca equation (\ref{equ:Proca}) around Kerr black holes into sets of coupled ordinary differential equations. These separable forms will be essential in our analytical treatment in Section~\ref{sec:analytic}, though they are not necessary in our numerical analyses in Section~\ref{sec:numeric}.

\subsubsection*{Massive scalar fields} 

We first specialize to the case of a massive scalar field around a Kerr black hole. Since the Klein-Gordon equation is relatively simple, this will serve as a useful illustration of our approach without the technical distractions that arise in the vector analysis.

\begin{figure}[t]
\centering
\includegraphics[scale=0.92, trim = 0 10 0 5]{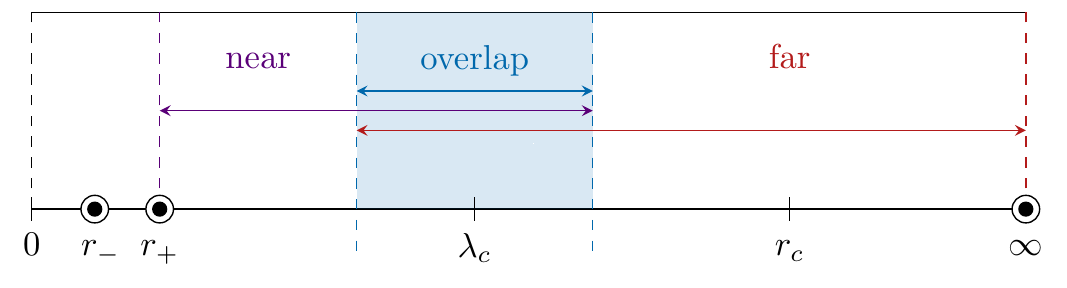}
\caption{Illustration of the near and far regions used in our perturbative treatment for
the massive scalar field, where $r_\pm$ are the inner and outer horizons of the black hole, $\lambda_c = \mu^{-1}$ is the Compton wavelength of the field, and $r_c = \left(\mu \alpha\right)^{-1}$ is the typical Bohr radius of the quasi-bound state. The two asymptotic solutions are matched in the overlap region.
}
\label{fig:ScalarZones}
\end{figure}

The Klein-Gordon equation (\ref{equ:KG}) is separable in Boyer-Lindquist coordinates through the ansatz~\cite{Brill:1972xj, Carter:1968}  
\begin{align}
\Phi \left(t, \textbf{r} \right) = e^{- i \omega t + i m \phi} R(r) S(\theta)  \, . \label{eqn:ScalarAnsatz}
\end{align}
The spheroidal harmonics $S(\theta)$ obey the differential equation\footnote{To avoid clutter, we suppress the dependence on the angular quantum numbers, i.e.~$S\equiv S_{\ell m}$ and $\Lambda \equiv \Lambda_{\ell m}$.} 
\begin{equation}
	\left(-\frac{1}{\sin \theta} \frac{\ud}{\ud \theta}\left(\sin \theta \frac{\ud}{\ud \theta}\right) - c^2 \cos^2 \theta + \frac{m^2}{\sin^2 \theta}\right)S = \Lambda S \, , \label{eqn: spheroidal harmonic equation}
\end{equation}
where $\Lambda$ is the angular eigenvalue and the spheroidicity parameter is 
\beq
c^2 \equiv - \alpha^2 \tilde{a}^2( 1-\omega^2 ) \,.  \label{equ:cc}
\eeq
The radial equation is a {confluent Heun equation},\footnote{The Heun equation is a generalization of the hypergeometric equation with four, instead of three, regular singular points. By merging  
two of these singularities, we arrive at the confluent Heun equation, which is similarly a generalization of the confluent hypergeometric equation (\ref{eqn:scalarRadialLO}) that determines the wavefunctions of the hydrogen atom. The angular equation (\ref{eqn: spheroidal harmonic equation}) is also a confluent Heun equation. For a detailed treatment, see \cite{Andre:1995heun}.} 
\beq
\begin{aligned}
0 \,=\,  \frac{1}{R\, \Delta} \frac{\ud}{\ud r} \!\left(\!\Delta \frac{\ud R}{\ud r}\right)  &\ - \frac{\Lambda}{\Delta} - \left(1 - \omega^2\right) + \ \frac{P_+^2}{(r-r_+)^2} + \frac{P_-^2}{(r - r_-)^2} \\
&\ -\frac{A_+}{(r_+ - r_-) (r -r_+)} +\frac{A_-}{(r_+-r_-)(r-r_-)} \, , \label{eqn: scalar radial equation}
\end{aligned}
\eeq
where we have defined the following coefficients 
\beq
\begin{aligned}
A_{\pm}  &\equiv P_+^2 + P_-^2 + \gamma^2 + \gamma_\pm^2   \quad {\rm and} \quad 
P_{\pm}  \equiv \frac{\alpha (\tilde{a} m - 2 \hskip 1pt  r_{\pm} \omega)}{r_+ - r_-} \, , 
\end{aligned}
  \label{equ:AP}
\eeq
with
\beq
\begin{aligned}
\gamma^2 & \equiv \frac{1}{4} (r_+ - r_-)^2 ( 1 - \omega^2) \, , \\
\gamma_\pm^2 & \equiv  \left[ \alpha^2( 1 - 7 \omega^2 )  \pm  \alpha (r_+ - r_-)(1 - 2 \omega^2)  \right]  . \label{eqn:defgamma}
\end{aligned}
\eeq
We are interested in quasi-bound states, i.e.~solutions to (\ref{eqn: scalar radial equation}) 
that are purely ingoing at the horizon and vanish at infinity:
\begin{equation}
		R(r) \propto \begin{cases}
			(r - r_+)^{i P_+} & \text{as } r \to r_+ \\
			\exp(-\sqrt{1-\omega^2}\hskip 1pt r) & \text{as } r \to \infty
		\end{cases}\, . \label{eqn:bc}
	\end{equation}
As is typical for eigenvalue problems, these two boundary conditions can only be simultaneously satisfied
for specific values of~$\omega$. Unfortunately, while solutions to both (\ref{eqn: spheroidal harmonic equation}) and (\ref{eqn: scalar radial equation}) are well known, closed-form expressions for their eigenvalues are not. However, in the limit $\alpha \ll 1$, we can construct a perturbative expansion of the spectrum. This perturbative approach  is complicated by the presence of terms in (\ref{eqn: scalar radial equation}) that diverge as $r \to r_+$. 
These terms represent \emph{singular perturbations} in $\alpha$, i.e.~if we naively expand them in powers of $\alpha$, an infinite number of terms become relevant as $r \to r_+$. Luckily, this is not a disaster. The role of these singular terms is simply to modify the characteristic scale on which $R(r)$ varies.  
As we will see explicitly in Section~\ref{sec:analytic}, the radial function satisfies 
	\begin{equation}
		\frac{1}{R}\frac{\ud R}{\ud r} \sim \begin{cases}
			\alpha^{-1} & \text{as } r \to r_+ \\
			\alpha & \text{as } r \to \infty
		\end{cases}\, .
	\end{equation}
	This motivates splitting the interval $[r_+, \infty)$ into a `near region' and a `far region' (see Fig.\,\ref{fig:ScalarZones}). 
	In the near region, the $(r-r_+)^{-2}$ pole in (\ref{eqn: scalar radial equation}) dominates and an approximate solution to the differential equation for $R(r)$ can be obtained by dropping the subleading terms.  In the far region, on the other hand, the non-derivative terms are dominated by the constant $(1 - \omega^2)$ term. 
By dropping the sub-leading contributions to (\ref{eqn: scalar radial equation}) in each region, we can then construct approximate analytic solutions in the near and far regions, order-by-order in $\alpha$. These solutions are then matched in the `overlap region,' allowing a determination of the
frequency eigenvalue~$\omega$, as an expansion in powers of $\alpha$. The details of this \emph{matched asymptotic expansion} will be presented in Section~\ref{sec:analytic}. For the angular problem, the perturbative solutions are valid over the entire angular domain and no matching is necessary.

To gain an intuitive understanding of the behavior of the solutions at leading order, it is instructive to first consider the equation of motion obeyed by the far-zone radial function, $R^{\rm far}_0$. At leading order, this reads
\beq
\left[ -\frac{1}{2r^2}\frac{\d}{\d r} \left( r^2 \frac{\d }{\d r}\right) - \frac{\alpha}{r} + \frac{\ell(\ell+1)}{2r^2} +\frac{1-\omega^2}{2} \right] R^{\text{far}}_0 = 0 \, . \label{eqn:scalarRadialLO}
\eeq
This is analogous to the time-independent Schr\"odinger equation for the hydrogen atom, where the radial-gradient term arises from the kinetic energy, the $1/r$ term corresponds to a Coulomb-like central potential with gravitational coupling constant $\alpha$, the $1/r^2$ term is its centrifugal barrier with orbital angular momentum number $\ell$, and the constant term captures the exponential behavior of the quasi-bound states at $r \to \infty$. We therefore expect the energy spectrum of all modes to be Bohr-like at leading order, with the radial function
\beq
R^{\text{far}}_0(r) \propto e^{-\sqrt{1-\omega^2} \, r} r^\ell L^{(2\ell+1)}_{n-\ell-1}\big( 2\sqrt{1-\omega^2} \, r\big) \, , \label{eqn: scalar radial far LO}
\eeq
as in (\ref{eqn:RadialHydrogen}). The exponential behavior in (\ref{eqn: scalar radial far LO}) implies that the solution has a typical Bohr radius given by $r_c \equiv (\mu \alpha)^{-1}$, as we found in (\ref{equ:BohrRadius}). Since (\ref{eqn:scalarRadialLO}) only captures the physics in the far region, it does not describe physical effects that depend on the boundary condition at the event horizon, such as the instability rates of the energy eigenstates. To study these effects, we match the near- and far-zone solutions. By pushing this procedure to higher orders in $\alpha$, we determine the complete spectrum of the scalar field. A schematic illustration of this spectrum appears in Fig.~\ref{fig:ScalarSpectra}.

\subsubsection*{Massive vector fields} 

The task of solving for the quasi-bound state solutions of vector fields is more involved because the matching involves three different regions, and the equation of motion is not obviously separable. The latter problem was addressed in~\cite{Frolov:2018ezx} and \cite{Pani:2012bp} where separable ans\"atze were found for the electric modes and magnetic modes (in the limit of small black hole spin), respectively. In Section~\ref{sec:analytic}, we use these results to derive the spectrum of vector quasi-bound states perturbatively.

Consider the following ansatz for a vector field on the Kerr background~\cite{Krtous:2018bvk, Frolov:2018ezx}
\begin{align}
A^\mu = B^{\mu \nu}\nabla_\nu Z \, , \quad {\rm with} \quad Z(t,\r) = e^{-i\omega t + i m \phi} R(r) S(\theta) \, , \label{eqn: Proca ansatz}
\end{align}
where $R$ and $S$ are the radial and angular functions.\footnote{We emphasize that $R$ and $S$ are not the actual radial and angular profiles
of the vector field, since the tensor~$B^{\mu \nu}$ depends on both $r$ and $\theta$. However, we still retain the terminology `radial' and `angular function,' and denote them by $R$ and $S$, to draw a direct parallel with the scalar. 
 The precise relationships between $R$, $S$ and the profile of the vector field $A_\mu$
 in the far zone will be given in \S\ref{sec:ProcaLO}. \label{footnote:ActualRS}}  The polarization tensor $B^{\mu \nu}$ is defined by
\begin{align}
B^{\mu \nu} \left( g_{\nu \sigma} + i \lambda^{-1} h_{\nu \sigma}\right) = \delta^\mu_\sigma \, , \label{eqn:Bdef}
\end{align}
where $\lambda$ is generally a complex parameter, which we will refer to as the angular eigenvalue, and $h_{\mu \nu}$ is the principal tensor of the Kerr spacetime~\cite{Frolov:2017kze}.\hskip 1pt\footnote{To simplify many of the following equations, we take the angular eigenvalue $\lambda$ to be the inverse of that in~\cite{Krtous:2018bvk,Frolov:2018ezx}.}
An explicit expression for $B^{\mu \nu}$ is given in Appendix~\ref{app:details}.  The three independent degrees of freedom of the vector field can be organized in terms of
$j = \ell \pm 1, \ell$. 
Since the $j = \ell \pm 1$ modes acquire a  factor of $(-1)^j$ under a parity transformation, they are called \emph{electric modes}, while the $j = \ell $ modes are the {\it magnetic modes}~\cite{Thorne:1980ru}. In Section~\ref{sec:analytic}, we will show that the ansatz (\ref{eqn: Proca ansatz}) captures all of the electric modes of the vector field.\footnote{In~\cite{Dolan:2018dqv}, it was found that the ansatz (\ref{eqn: Proca ansatz}) restores at least a subset of the magnetic modes in a special limit (see Appendix~\ref{app:details} for further details). It remains unclear, however, how to write all magnetic modes in separable form. In this thesis, I will instead utilize a different ansatz, which, in the limit of small black hole spin, provides a separable equation for all of the magnetic modes.}

Substituting (\ref{eqn: Proca ansatz}) into the Proca equation, we obtain 
differential equations for $S(\theta)$ and $R(r)$. The angular equation reads~\cite{Dolan:2018dqv} 
\beq
\begin{aligned}
\frac{1}{\sin \theta} \frac{\d}{\d\theta} \! \left( \sin\theta \frac{\d S}{\d \theta} \right)  + & \left( c^2 \cos^2 \theta - \frac{m^2}{\sin^2 \theta} +  \Lambda  \right) S \\
& = \frac{2 \alpha^2 \tilde{a}^2 \cos \theta}{\lambda^2 q_\theta} \left( \sin \theta \frac{\d }{\d \theta} + \lambda \,\sigma \cos \theta \right)S \, , \label{equ:ProcaS}
\end{aligned}
\eeq
where the spheroidicity parameter $c^2$ was
 defined in (\ref{equ:cc}), and 
  \beq
\begin{aligned}
q_\theta & \equiv 1 - \alpha^2 \tilde{a}^2  \lambda^{-2} \cos^2 \theta  \, , \\[4pt]
\sigma &\equiv \omega + \alpha \hskip 1pt \tilde{a} \hskip 1pt \lambda^{-2} (m - \alpha \tilde{a} \hskip 1pt \omega)  \, , \\[4pt]
 \Lambda &\equiv \lambda \left( \lambda - \sigma \right) + 2 \alpha \tilde{a} \hskip 1pt m \omega - \alpha^2 \tilde{a}^2 \omega^2 \, . \label{eqn:ProcaSigmaLambda}
\end{aligned}
\eeq
The radial equation becomes
\beq
\begin{aligned}
0 = & \frac{\d ^2 R}{\d r^2} + \left( \frac{1}{r-r_+} + \frac{1}{r-r_-} - \frac{1}{r-\hat{r}_+} - \frac{1}{r-\hat{r}_-} \right) \frac{\d R}{\d r} \\
& + \bigg( - \frac{\Lambda}{\Delta} - (1 - \omega^2) +  \frac{P_+^2}{(r-r_+)^2} + \frac{P_-^2}{(r-r_-)^2}  - \frac{A_+}{(r_+-r_-) (r-r_+) } \\
&  \ \ \ \ \ \ + \frac{A_-}{(r_+-r_-)(r-r_-)} - \frac{ \phantom{\mathcal{B}} \lambda \es \sigma r }{\Delta \left( r- \hat{r}_+ \right)} - \frac{\phantom{\mathcal{B}} \lambda \es \sigma r}{\Delta \left(r- \hat{r}_-\right)}\bigg)\, R \, , \label{eqn:ProcaR}
\end{aligned}
\eeq
where $\hat{r}_\pm  \equiv \pm i \lambda$ depend on the angular eigenvalue  
and the parameters $P_\pm$ and $A_\pm$ were defined in (\ref{equ:AP}). The presence of the additional poles at $r = \hat{r}_\pm$, makes this equation considerably more complicated than the corresponding  equation~(\ref{eqn: scalar radial equation}) in the scalar case.

\begin{figure}[]
\centering
\includegraphics[scale=0.87, trim = 3 0 0 0]{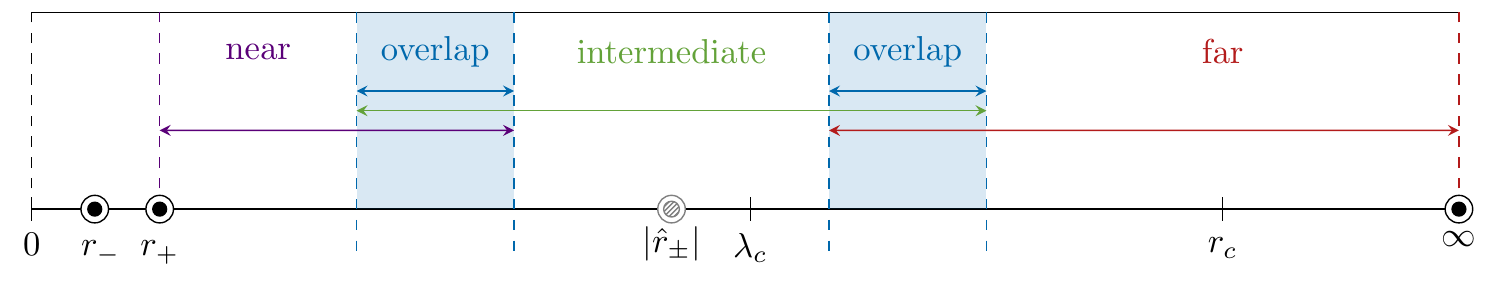}
\caption{
Illustration of the different regions used in our perturbative
 treatment for the massive vector field. The additional poles at $\hat{r}_\pm$ along the imaginary axis reduce the radii of convergence of the near- and far-zone solutions, so that there is no region where they overlap. Matching therefore requires  the intermediate region.} 
\label{fig:Vectorzones}
\end{figure}

In principle, the task of determining the spectrum of electric modes of the vector quasi-bound states is the same as for the scalar. However, due to the additional poles at $r=\hat{r}_\pm$,  
the widths of the near and far regions are now much smaller and these regions
 no longer overlap. To match the asymptotic expansions in the near and far regions,
 we must 
 introduce an {\it intermediate region}\footnote{Physically, the necessity of the intermediate region arises because the dynamics in the near and far regions depend on the angular momentum of the vector field in distinct ways.  
As we will see in Section~\ref{sec:analytic}, while the dynamics in the near region is sensitive to the total angular momenta of the field (including its intrinsic spin), the dynamics in the far region depends only on its orbital angular momentum. The intermediate region thus smoothly interpolates between these two behaviors. } that overlaps with both 
 (see Fig.\,\ref{fig:Vectorzones}) and construct a solution that serves as a bridge. The matching is then performed in a two-step procedure, first between the far and intermediate regions, and then between the intermediate and near regions.

In the Schwarzschild limit, solutions have definite total angular momentum and parity. Because the vector spherical harmonic $Y^{i}_{j,\, j \es m}$ has opposite parity to $Y^i_{j \pm 1,\, j \es m}$ and the scalar harmonic $Y_{j\es m}$, it completely decouples from all other angular modes. This magnetic mode,
\beq
A^i (t, \textbf{r} ) = r^{\minus 1} R(r)\,Y^i_{j, \, j\es m} (\theta, \phi)\,e^{- i \omega t }  \, , \label{eqn:ProcaMagneticAnsatz}
\eeq
is thus completely separable in the Schwarzschild limit~\cite{Rosa:2011my}, and \cite{Pani:2012vp,Pani:2012bp} showed that this persists to linear order in $\tilde{a}$. We can thus use this ansatz to determine the magnetic spectrum in the limit of small spin. 

\begin{figure}[t]
\centering
\makebox[\textwidth][c]{\includegraphics[scale=0.8, trim = 0 0 20 0]{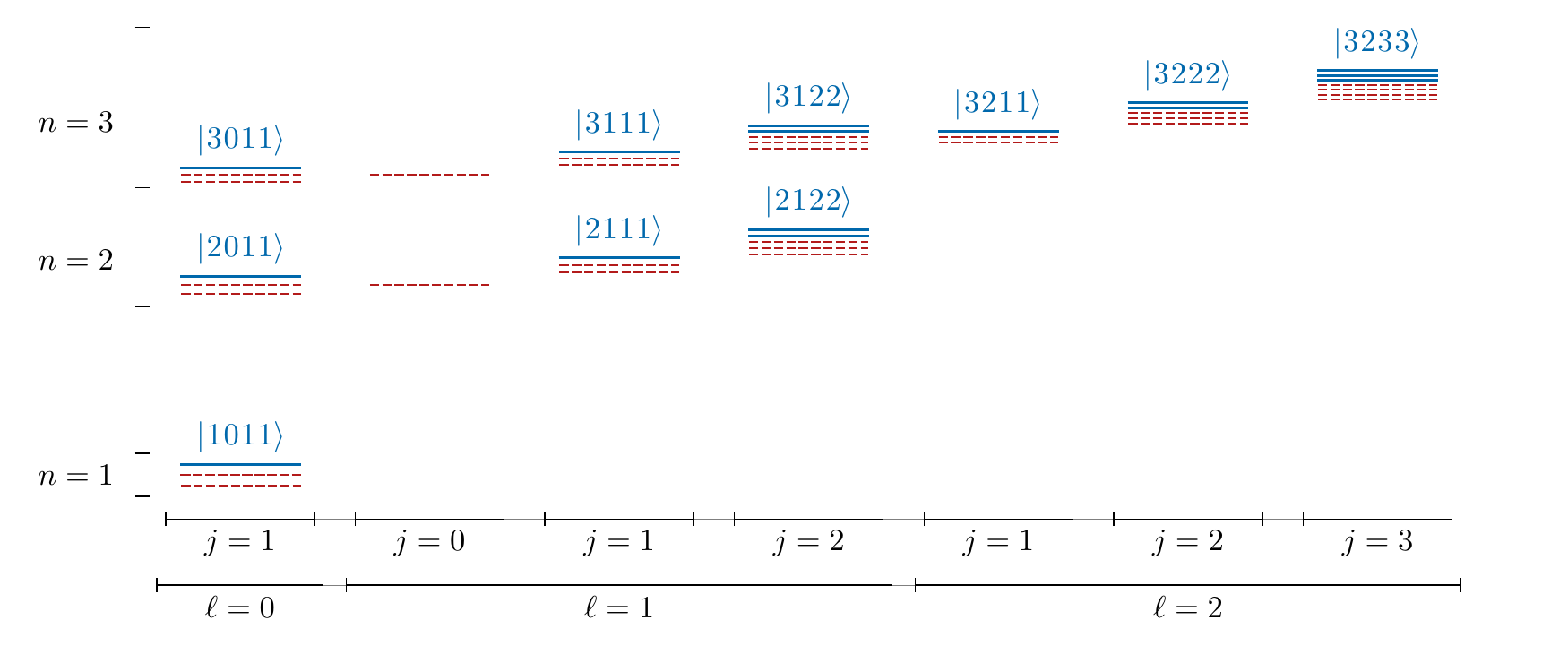}}
\caption{Schematic illustration of the energy spectrum for a massive vector field. Each state is labeled by the quantum numbers $|\es  n \es  \ell \es   j \es   m\rangle$. The dominant vector mode is $\ket{1011}$. } 
\label{fig:VectorSpectra}
\end{figure}

Substituting (\ref{eqn:ProcaMagneticAnsatz}) into the Proca equation and expanding 
to first order in $\tilde{a}$, the radial function $R(r)$ satisfies~\cite{Pani:2012vp, Pani:2012bp}
\beq
\begin{aligned}
 & 0  =  \frac{\d^2 R}{\d r^2 } + \left( \frac{1}{r-\check{r}_+} - \frac{1}{r - \check{r}_-} \right) \frac{\d R}{\d r} \ + \bigg( - \frac{\check{\Lambda} }{\Delta} - \left(1 - \omega^2\right)  \\
&\ + \ \frac{\check{P}_+^2}{(r-\check{r}_+)^2} + \frac{\check{P}_-^2}{(r - \check{r}_-)^2} -\frac{\check{A}_+}{(\check{r}_+ - \check{r}_-) (r -\check{r}_+)} +\frac{\check{A}_-}{(\check{r}_+- \check{r}_-)(r- \check{r}_-)} \bigg) R \, ,\label{eqn:MagneticRadial}
\end{aligned}
\eeq
where $\check{\Lambda} = j(j+1)$ and $\check{P}_\pm, \check{A}_\pm$, $\check{r}_\pm$ represent $P_\pm$, $A_\pm$, $r_\pm$ expanded to linear order in $\tilde{a}$, respectively. Specifically, the position of the inner and outer horizons, $\check{r}_- = 0$ and $\check{r}_+ = 2 \alpha$, have shifted in this approximation, and we expect that (\ref{eqn:MagneticRadial}) does not accurately describe the near-horizon behavior of these magnetic modes. When expanded to linear order in $\tilde{a}$, the scalar radial equation~(\ref{eqn: scalar radial equation}) differs from (\ref{eqn:MagneticRadial}) only in its $\ud R/\ud r$ coefficient. In Section~\ref{sec:analytic}, we will discuss the error this small-spin approximation introduces. To compute the magnetic spectra for arbitrary spin, we must still solve the Proca equation (\ref{equ:Proca}) numerically. For a schematic illustration of the vector field spectrum, see Fig.~\ref{fig:VectorSpectra}.

\subsection{Summary of Results} 
\label{sec:Summary}

In Sections \ref{sec:analytic} and \ref{sec:numeric}, we derive the spectra of scalar and vector quasi-bound states around Kerr black holes in detail. Here, we summarize our main results.

We write the frequency eigenvalues as 
\beq
\omega \hskip 1pt \equiv \hskip 1pt   E + i \hskip 0.5 pt \Gamma \hskip 2pt  \equiv \hskip 1pt  \mu \, \sqrt{1 - \frac{\alpha^2}{\nu^2}}  \, , \label{eqn:generalspectrum}
\eeq
where the real part represents the energy $E$  and the imaginary part determines the instability rates $\Gamma$. 
At leading order, we expect the energy spectrum to be Bohr-like, so that
the real part of $\nu$ is 
an integer $n$. As we will see below, it is also convenient to work with $\nu$ because the fine and hyperfine structure can be read off directly from its $\alpha$-expansion. 

\begin{figure}[t]
  \begin{center}
    \makebox[\textwidth][c]{\includegraphics[scale=0.8, trim = 0 0 0 0]{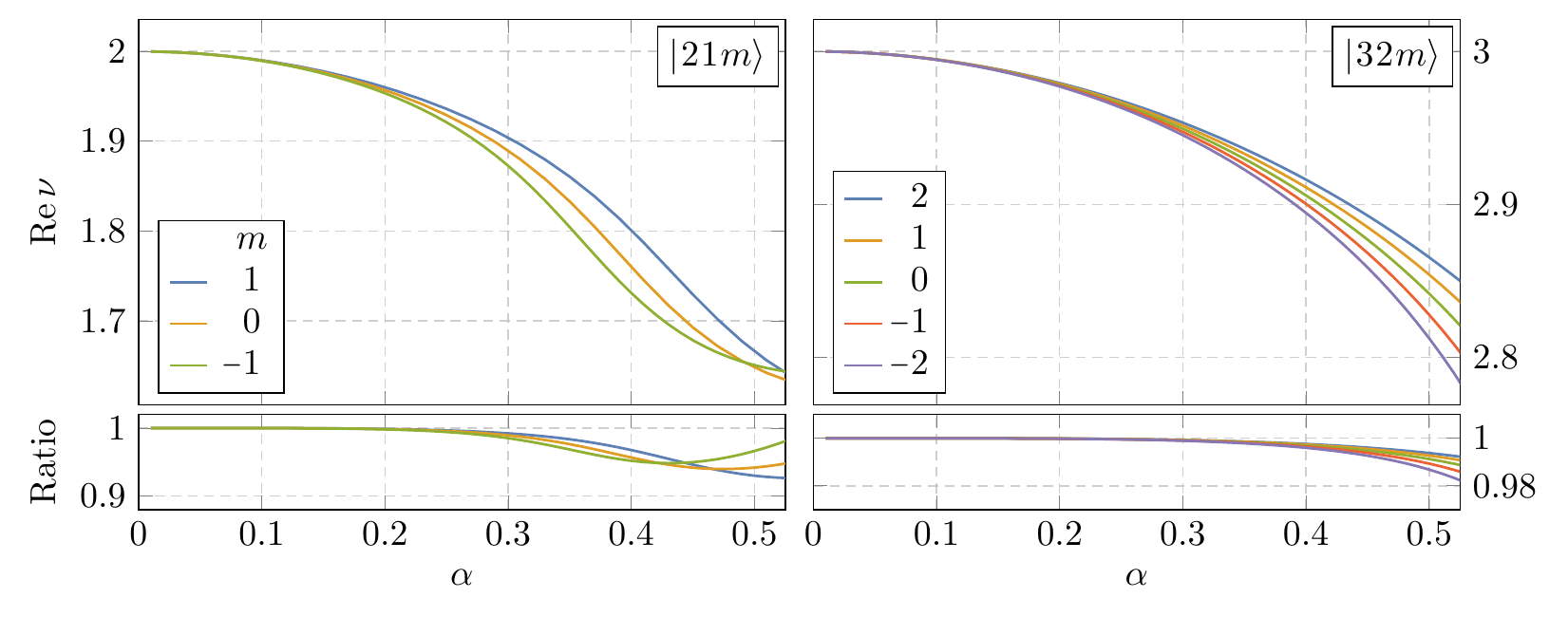}}
    \caption{Numeric results of the energy eigenvalues for scalar field eigenstates $\ket{n \ell m}$, for~$\tilde{a} = 0.5$. The lower panels show the ratio of the numeric results to the analytical predictions in~(\ref{eqn:scalarspectrum}). The fine and hyperfine structures are clearly seen in $\lab{Re}\, \nu$, and we find that (\ref{eqn:scalarspectrum}) is an excellent approximation for $\alpha \lesssim 0.2$.} \label{fig:ScalarSpectrumPlots}
  \end{center}
\end{figure}

For the scalar field, the energy eigenvalues are 
\beq
\begin{aligned}
 E_{n \ell m}   &= \mu \left( 1 - \frac{\alpha^2}{2n^2} - \frac{\alpha^4}{8n^4} + \frac{f_{n \ell} }{n^3}\hskip 2pt  \alpha^4 +  \frac{h_{\ell }  }{n^3} \hskip 2pt \tilde a m \hskip 1pt   \alpha^5 + \cdots \right)  , \\[2pt]
\lab{Re}\,(\nu_{n \ell m})  &= n + f_{n \ell} \hskip 1pt \alpha^2 + h_{\ell} \hskip 1pt \tilde a m \hskip 1pt \alpha^3 + \cdots \, , \label{eqn:scalarspectrum}
\end{aligned} 
\eeq
where the principal quantum number $n$ are integers that satisfy $n \geq \ell + 1$, and
\beq
\begin{aligned}
f_{n \ell} &\equiv  - \frac{6}{2\ell + 1} + \frac{2}{n}  \, , \\
h_{\ell} &\equiv \frac{16}{ 2 \ell \left( 2\ell+1 \right) \left( 2\ell+2 \right)}\, . \label{eqn:scalarspectrumfh}
\end{aligned}
\eeq
The first three terms of $E_{n \ell m}$ describe the constant mass term, the hydrogen-like Bohr energy levels and the relativistic corrections to the kinetic energy. The terms proportional to $f_{n\ell}$ and $h_{\ell}$ are the fine-structure $(\Delta \ell \neq 0)$ and hyperfine-structure $(\Delta m \neq 0)$ splittings, respectively.  
As the second line in (\ref{eqn:scalarspectrum}) suggests, these splittings can be more easily extracted from the real part of~$\nu_{n \ell m}$. Numeric results for $\lab{Re}\, \nu$, and their comparison with the perturbative approximations~(\ref{eqn:scalarspectrum}), are shown in Fig.~\ref{fig:ScalarSpectrumPlots} for representative scalar modes; cf.~Fig.~\ref{fig:ScalarSpectra} for a schematic illustration of the scalar spectrum.

\begin{figure}[t]
  \begin{center}
    \makebox[\textwidth][c]{\includegraphics[scale=0.8, trim = 0 0 0 0]{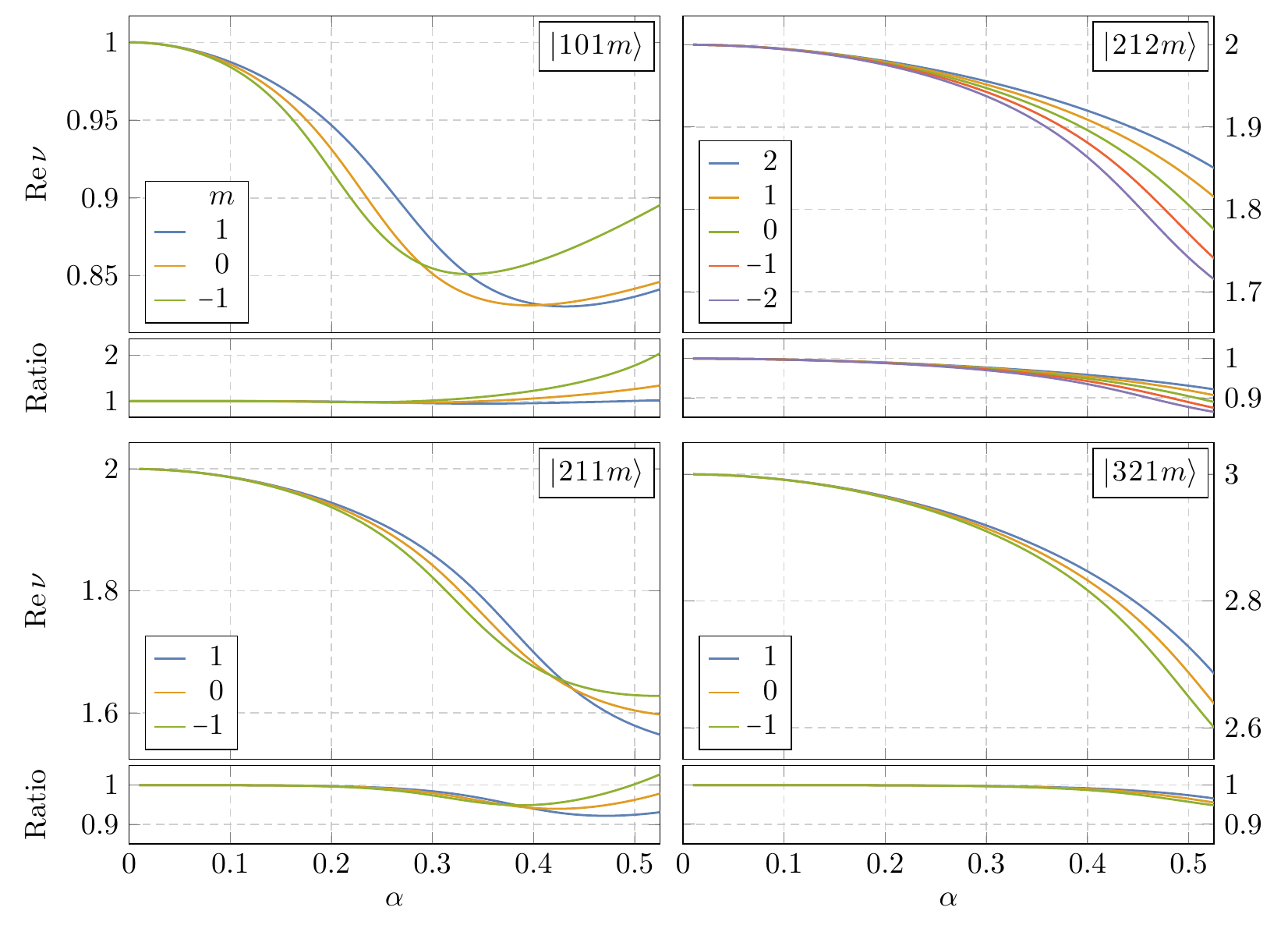}}
    \caption{Numeric results of the energy eigenvalues for vector field eigenstates $\ket{n \ell jm}$, for~$\tilde{a} = 0.5$. The lower panels show the ratio of the numeric results to our perturbative results in~(\ref{eqn:vectorspectrumGeneral}).} 
    \label{fig:VectorSpectrumPlots}
  \end{center}
\end{figure}

For the vector field, we study the electric and magnetic modes separately using the ans{\"a}tze (\ref{eqn: Proca ansatz}) and (\ref{eqn:ProcaMagneticAnsatz}). However, we find that the energy eigenvalues for \emph{all} vector modes can be written as
\beq
\begin{aligned}
E_{n \ell j m}  &= \mu \left(  1 -\frac{\alpha^2}{2n^2} - \frac{\alpha^4}{8n^4} + \frac{f_{n \ell j} }{n^3} \hskip 2pt \alpha^4 + \frac{h_{ \ell j } }{n^3}  \hskip 2pt \tilde a  m  \hskip 1pt \alpha^5 + \cdots\right)   , \\[2pt]
\lab{Re}\,(\nu_{n \ell j m})  &= n + f_{n \ell j} \hskip 1pt \alpha^2 + h_{\ell j } \hskip 1pt \tilde a \hskip 1pt m  \alpha^3  + \cdots\, , \label{eqn:vectorspectrumGeneral}
\end{aligned}
\eeq
with the coefficients
\beq
\begin{aligned}
f_{n \ell j} & =  - \frac{4 \left( 6 \hskip 1pt \ell \hskip 1pt j  + 3 \ell + 3 j + 2 \right) }{\left( \ell + j \right)  \left( \ell + j + 1 \right) \left( \ell + j + 2 \right)  }+\frac{2}{n}  \,, \\[-9pt]   \\[-9pt]
h_{\ell j} & = \frac{16}{ \left( \ell + j \right) \left( \ell + j + 1 \right) \left( \ell + j +2 \right) } \, , \label{eqn:vectorspectrumfh}
\end{aligned}
\eeq
for $j = \ell \pm 1, \ell$. Notably, we find that the spectrum of the vector field is qualitatively similar to the scalar case~(\ref{eqn:scalarspectrum}); cf.~Fig.~\ref{fig:VectorSpectra} for a schematic illustration of the vector spectrum. Numeric results for $\lab{Re}\, \nu$ for representative electric and magnetic modes, and their comparison with (\ref{eqn:vectorspectrumGeneral}), 
are shown in Fig.~\ref{fig:VectorSpectrumPlots}.

  \begin{figure}
    \begin{center}
      \hspace{0 pt}\includegraphics[scale=0.86, trim = 0 0 0 0]{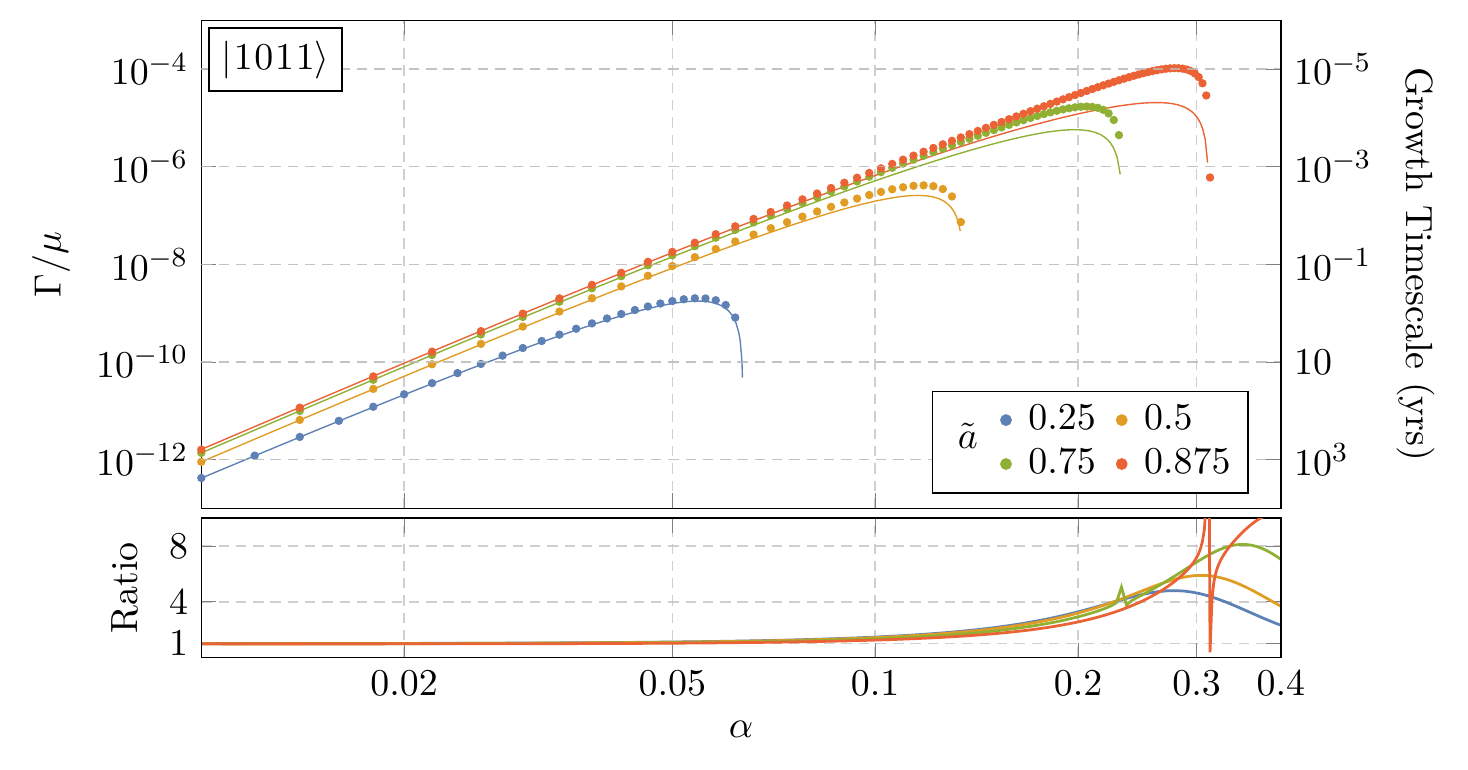}
      \caption{Superradiant growth rates for the dominant vector mode $\ket{1011}$, for different values of the black hole spin $\tilde a$. The growth timescale is for a Proca field of mass $\mu = \hbar/10^{-9}\,\lab{yr}\approx 2.1 \times 10^{-14}\,\lab{eV}$. The points are numeric data, while the solid lines are the approximation 
      (\ref{eqn:VectorRates}). The lower panel shows the ratio of our numerics and analytics, and includes the decaying regime, $\Gamma < 0$, at larger values of $\alpha$.} \label{fig:VectorDominant}
    \end{center}
  \end{figure}

\begin{figure}
  \begin{center}
    \makebox[\textwidth][c]{\hspace{-0cm}\includegraphics[scale=0.81, trim = 0 0 0 0]{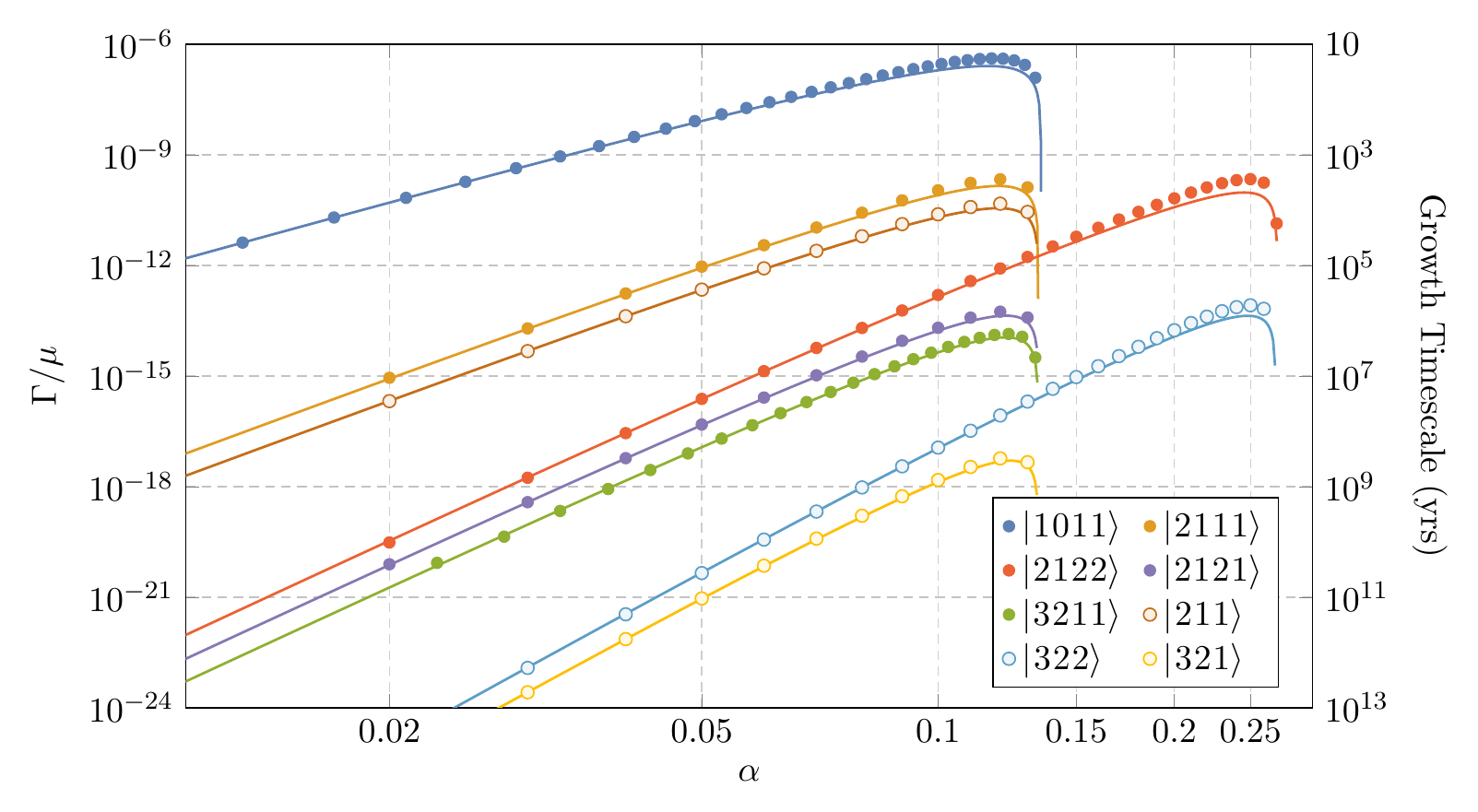}}
    \caption{Superradiant growth rates for selected scalar and vector modes, for $\tilde a =0.5$. The growth timescale is for $\mu = \hbar/10^{-9}\,\lab{yr} \approx 2.1 \times 10^{-14}\,\lab{eV}$. The points represent numeric data, while the solid lines are our perturbative predictions (\ref{eqn:ScalarRate}) and  (\ref{eqn:VectorRates}). \label{fig:growthRates}}
  \end{center}
\end{figure}

We also compute the instability rates for these scalar and vector quasi-bound states. 
For the scalar field, we find (see also~\cite{Detweiler:1980uk, Baryakhtar:2017ngi, Pani:2012bp})
\beq
\Gamma_{n \ell m} = 2 \tilde{r}_+ C_{n \ell} \, g_{\ell m}(\tilde{a}, \alpha,\omega) \, (m \Omega_H - \omega_{n \ell m})\hskip 1pt \alpha^{4\ell+5}  \, , \label{eqn:ScalarRate}
\eeq
where we have defined
\begin{align}
C_{n \ell} &\equiv  \frac{2 ^{4\ell+1} (n+\ell)!}{  n^{2\ell+4} (n-\ell-1)! } \left[ \frac{\ell !}{(2\ell)! (2\ell+1)!} \right]^2  \, , \label{eqn:ScalarRateCoeff} \\
g_{\ell m}(\tilde{a}, \alpha, \omega) &\equiv \prod^{\ell}_{k=1} \left( k^2  \left( 1-\tilde{a}^2 \right) + \left( \tilde{a} m - 2 r_+ \hskip 1pt  \omega \right)^2  \right)  . \label{eqn:RateCoeffProd}
\end{align}
The dominant growing mode is $|\es n \es \ell \es m \rangle = |\es 2 \es 1 \es 1\rangle$, and hence $\Gamma_{211} \propto \mu \, \alpha^8$~\cite{Dolan:2007mj,Detweiler:1980uk},\footnote{For the normalization of the dominant growing mode, we find $C_{21} = 1/48$, which differs from the result in~\cite{Detweiler:1980uk} by a factor of $2$. This discrepancy was also observed in the appendices of~\cite{Baryakhtar:2017ngi, Pani:2012bp}.} where we have used $\Omega_H \sim \mu \alpha^{-1}$.  Numeric results for the growth rates $\Gamma$ of the fastest growing scalar states are shown in Fig.~\ref{fig:growthRates}, along with their comparison to the analytic approximations (\ref{eqn:ScalarRate}).

The instability rates for the vector modes take a very similar form. In particular, we find that they are
\beq
\Gamma_{n\ell jm} = 2 \tilde{r}_+ C_{n \ell j } \, g_{j m}(\tilde{a}, \alpha, \omega)  \left( m \Omega_H - \omega_{n \ell j m} \right) \alpha^{2 \ell + 2 j + 5} \, , \quad  \text{for $j = \ell \pm 1, \ell$}\, ,  \label{eqn:VectorRates}
\eeq	
where the coefficient, 
\beq
\begin{aligned}
 C_{n \ell j}\equiv & \frac{2^{2\ell + 2 j +1} (n+\ell)!}{n^{2\ell+4}(n-\ell-1)!} \\
 & \times  \left[ \frac{(\ell)!}{(\ell + j )!(\ell + j+1)!}\right]^2 \left[ 1 + \frac{ 2  \left( 1+ \ell  - j   \right) \left(1   - \ell  +  j \right)  }{\ell + j}\right]^2 \, , \label{eqn:VectorRatesCoeff}
 \end{aligned}
\eeq
is valid for all $j = \ell \pm 1, \ell$,
and the function $g_{jm}$ is obtained by replacing $\ell$ by $j$ in (\ref{eqn:RateCoeffProd}). The $\alpha$-scaling in (\ref{eqn:VectorRates}) was also found 
in~\cite{Pani:2012vp, Pani:2012bp, Endlich:2016jgc, Baryakhtar:2017ngi}. Notice that (\ref{eqn:VectorRates}) is also valid for the scalar rate~(\ref{eqn:ScalarRate}), if we set $j =\ell$ and multiply $C_{n \ell j}$ by a factor of $j^2/(j+1)^2$.  The dominant vector growing mode is $|\es n \es \ell \es j \es m \rangle = |\es 1 \es 0 \es 1 \es 1\rangle$ and has $\Gamma_{1011} \propto \mu \hskip 1pt \alpha^6$~\cite{Endlich:2016jgc, Baryakhtar:2017ngi, Cardoso:2018tly, East:2017ovw, East:2017mrj, East:2018glu, Pani:2012vp, Pani:2012bp, Dolan:2018dqv} (see Table 1 of~\cite{Cardoso:2018tly} for a summary of $\Gamma_{1011}$ obtained through other methods).
This is faster than the dominant growing mode for the scalar field. In Fig.~\ref{fig:VectorDominant}, we show our numeric results for the growth rate $\Gamma_{1011}$ at different values of the black hole spin $\tilde{a}$, and compare it with (\ref{eqn:VectorRates}). In Fig.~\ref{fig:growthRates}, we compare the numeric results for the growth rates of other electric and magnetic vector modes to their perturbative approximations (\ref{eqn:VectorRates}). For comparison, we also include the dominant scalar modes.

Unfortunately, the separable ansatz (\ref{eqn:ProcaMagneticAnsatz}) for the magnetic mode does not describe the Proca field in the near-horizon geometry of the Kerr black hole, and thus cannot be used to rigorously derive a perturbative expression for the magnetic instability rates. In our framework, the result (\ref{eqn:VectorRates}) for $j = \ell$ was obtained through an educated guess and, in the absence of numerical evidence or an alternative approximation scheme, should be taken as purely conjectural. Having said that, it both matches the result of~\cite{Baryakhtar:2017ngi} and is in excellent agreement with our numeric results (c.f.~Figures~\ref{fig:growthRates}~and~\ref{fig:magInstability}), and so we include it as a guide for phenomenology. 

\section{Analytical Computation of the Spectra}
\label{sec:analytic} 

In this section, we compute in detail the energy spectra and instability rates of quasi-bound states of massive scalar and vector fields around rotating black holes, using the method of matched asymptotic expansions~\cite{bender1999advanced, holmes1998introduction}. The conceptual challenge of the computation can be illustrated using the scalar field, while avoiding the technical heft of the vector. We thus begin with the analysis of a scalar field~\cite{Starobinsky:1973aij,Detweiler:1980uk} in \S\ref{sec:MAE} and \S\ref{sec:ScalarAnalytic}, before moving to the vector case in \S\ref{sec:ProcaLO}.

In what follows, it will be convenient to write the frequency eigenvalue as 
\beq
\omega \equiv \sqrt{1 - \frac{\alpha^2}{\nu^2}}
 \, , \label{eqn:nudef}
\eeq
where the real part of $\nu$ is an integer for the leading-order Bohr spectrum. We will compute $\lab{Im}\, \nu$ to leading order in $\alpha$ and $\lab{Re}\, \nu$ to order $\alpha^3$, as all degeneracies are broken at this order.

\subsection{Matched Asymptotic Expansion}  \label{sec:MAE}

We begin by illustrating the essential features of a matched asymptotic expansion through the simpler scalar field radial equation (\ref{eqn: scalar radial equation}). Since perturbative solutions of (\ref{eqn: scalar radial equation}) cannot capture the boundary conditions at $r=r_+$ and $r \to \infty$ simultaneously, a matched asymptotic expansion is necessary to compute the eigenvalues $\nu$. Schematically, we approximate the radial function separately in both the near and far regions (cf.~Fig.~\ref{fig:ScalarZones}), and match these solutions in the region of overlap.

To define the \emph{near region}, it is convenient to introduce the rescaled radial coordinate
\beq
z \equiv \frac{r-r_+}{r_+ - r_-} =  \frac{r- \alpha \hskip 1pt \tilde{r}_+}{\alpha \left( \tilde{r}_+ - \tilde{r}_- \right)} \, , \label{eqn: z variable}
\eeq
where $\tilde{r}_\pm \equiv r_\pm / \alpha$, so that $\tilde{r}_\pm \sim \mathcal{O}(1)$. In this coordinate, the inner and outer horizons are mapped to $z=-1$ and $z=0$, respectively.  As discussed in~\S\ref{sec:separable}, the radial function varies much more rapidly  in the near region than in the far region, $\ud R/\ud r \sim \alpha^{-1}$, and the reason we change coordinates is to accommodate this rapid change in the $\alpha$-expansion. The radial equation  (\ref{eqn: scalar radial equation}) then becomes
\beq
\begin{aligned}
\frac{1}{R \, z(z+1)}\frac{\d}{\d z} \! & \left( \! z(z+1) \frac{\d R}{\d z}\right)  - \frac{\Lambda}{z(z+1)} \\
&  - 4 \gamma^2  + \frac{P_+^2}{z^2} + \frac{P^2_-}{(z+1)^2} - \frac{A_+}{z} + \frac{A_-}{z+1} = 0   \, , \label{eqn: Heun-z}
\end{aligned}
\eeq
where the coefficients were defined in (\ref{equ:AP}) and (\ref{eqn:defgamma}). 
We then expand (\ref{eqn: Heun-z}) in powers of $\alpha$, with $z$ kept fixed. By comparing the dominant and subdominant terms at each order in $\alpha$, we find that this region covers the range $0 \leq z \lesssim \alpha^{-2}$, cf.~Fig.~\ref{fig:ScalarZones}. 
Near the outer horizon~at $z=0$,  the $1/z^2$ pole dominates and the radial solution scales as 
\beq
R(z) \,\sim\, B_1 \hskip 1pt z^{ i P_{+}} + B_2 \hskip 1pt  z^{- i P_+} \, , \mathrlap{\qquad z \to 0\, ,} \label{eqn:scalar radial outer horizon}
\eeq 
where $B_1$ and  $B_2$ are integration constants. The purely ingoing boundary condition at the event horizon requires that $B_2 = 0$. Similarly, approaching the inner horizon at $z=-1$, we find that $\lim_{z \to \sminus 1} \, R(z) \, \sim \, (z+1)^{\pm i P_-}$. However, unlike the outer horizon, the inner horizon is not physically accessible, and hence no boundary condition has to be imposed on it.

To identify the \textit{far region}, it is useful to use the alternative coordinate
\beq
x \equiv 2 \sqrt{1 - \omega^2} \, (r-r_+) = 4 \gamma z \, ,  \label{eqn: x variable}
\eeq
where $\gamma$ was defined in (\ref{eqn:defgamma}). In this coordinate, the near horizon region, $z \sim 1$, is mapped to $x \sim \alpha^2$, and the Bohr radius $r_c \sim  \alpha^{-1} $ is at~$x \sim 1$. In the far region, we expand the radial equation in powers of $\alpha$, while keeping $x$ fixed. As illustrated in Fig.~\ref{fig:ScalarZones}, the far region spans~$\alpha^2 \lesssim x < \infty$. The radial solution then behaves as 
\beq
\hskip -60pt R(x) \,\sim\, B_3 \hskip 1pt  e^{- x/2} \, x^{-1 + \nu - 2 \alpha^2/\nu} + B_4 \hskip 1pt  e^{+ x/2} \, x^{-1 -\nu + 2 \alpha^2/\nu} \, ,  \mathrlap{\qquad x \to \infty \, ,}\label{eqn: scalar radial infinity}
\eeq
where $\nu$ was defined in (\ref{eqn:nudef}). Since we are interested in quasi-bound states with $\omega < \mu$, the radial function should vanish at large distances. We therefore set $B_4 = 0$.

To perform matched asymptotic expansions at high orders in $\alpha$, we expand the radial and angular functions\hskip 1pt\footnote{It is essential that $P_+$ is held fixed, since an expansion of $P_+$ in powers of $\alpha$ would change the boundary condition~(\ref{eqn:scalar radial outer horizon}) at the horizon. Indeed, the main reason this matched asymptotic expansion is needed at all is that, regardless of which order in $\alpha$ they appear, \emph{all} terms multiplying the $z^{-2}$ pole of (\ref{eqn: Heun-z}) become arbitrarily important as we approach the outer horizon, $z \to 0$. It is thus crucial that we do not disturb this pole in our $\alpha$-expansion, lest our approximation becomes arbitrarily bad, to
a point where we must impose a different boundary condition. \label{footnote:FixedPp}} 
\beq
X = \sum_k \alpha^k X_{\indlac{k}}\, , \label{eqn:SumX} 
\eeq
where $X = \{R^{\rm near}, R^{\rm far},S\}$. Schematically, the equations of motion become
\beq
\square \, X = \Big[ \square^{\indlab{0}} + \alpha \, \square^{\indlab{1}}  + \alpha^2 \, \square^{\indlab{2}}  +  \cdots \Big] \Big[ X_{\indlac{0}} + \alpha  X_{\indlac{1}}  + \alpha^2  X_{\indlac{2}} +  \cdots \Big] =0 \, , \label{eqn:XExpansion}
\eeq
where the $\alpha$-expansion of the differential operators $\square$ includes expansions of the angular and energy eigenvalues: 
\beq
\Lambda = \sum_k \alpha^k \Lambda_{\indlac{k}}\, , \qquad \nu = \sum_k \alpha^k \nu_{\indlac{k}}\, .
\eeq
In order to lift the degeneracies between all modes of the spectrum, it is 
sufficient to go to order $\alpha^3$ in $\Lambda$ and $\lab{Re}\,\nu$.  We will then solve these equations order by order in $\alpha$, imposing the above boundary conditions at each order. Finally, we match the solutions in the overlap region to determine $\nu$.

\subsection{Scalar Field Solutions} 
\label{sec:ScalarAnalytic} 

Using the formalism outlined in \S\ref{sec:MAE}, we now proceed to solve the leading-order and higher-order solutions for the massive scalar field. 

\subsubsection*{Leading-order solution} 

We expect the leading-order solution to yield the hydrogenic spectrum, although unlike the hydrogen atom, these states will be quasi-stationary. We must therefore also compute their instability rates.

At leading order, the angular equation (\ref{eqn: spheroidal harmonic equation}) is
\beq
 \left[\frac{1}{\sin \theta} \frac{\d}{\d \theta}\left( \sin \theta \frac{\d}{\d\theta} \right) - \frac{m^2}{\sin^2 \theta} + \Lambda_\indlac{0} \right]\!S_\indlac{0} = 0 \, .
\eeq
Setting $\Lambda_0 = \ell(\ell+1)$, and imposing regular boundary conditions at the antipodal points $\theta = 0$ and $\pi$, the solutions are given by the associated Legendre polynomials $S_\indlac{0} = P_{\ell m}(\cos \theta)$. This is expected,
since the spheroidal harmonics reduce to the ordinary spherical harmonics in the limit $\alpha \to 0$.

The leading-order near- and far-zone radial equations are
\begin{align}
\bigg[  \frac{\d^2}{\d z^2 } + \left( \frac{1}{z} + \frac{1}{z+1} \right)\frac{\d}{\d z} -  \frac{\ell(\ell+1)}{z(z+1)} \nonumber \hskip 80pt & \\
 + \frac{P_+^2}{z^2} + \frac{P^2_+}{(z+1)^2} - \frac{2P_+^2}{z} + \frac{2P_+^2}{z+1} \bigg] R^{\rm near}_\indlac{0}  &= 0 \, , \label{eqn:scalarLONear}  \\ 
\left[ \frac{\d^2 }{\d x^2 } + \frac{2}{x}\frac{\d }{\d x } + \frac{\nu_0}{x}   - \frac{\ell(\ell+1)}{x^2} - \frac{1}{4}  \,\right] \! R_\indlac{0}^{\mathrlap{\lab{far}}\hphantom{near}}  &= 0 \, .
\label{eqn:hydrogenic}
\end{align}
Imposing the correct boundary conditions,
we obtain
\begin{align}
R^{\rm near}_\indlac{0}(z)   &= \mathcal{C}^{\rm near}_0 \, \left( \frac{z}{z+1} \right)^{i P_+ } \!\!\!{}_2 F_1(-\ell, \ell+1, 1-2 i P_+, 1+z) \, , \label{eqn:scalarRadialLOnear} \\[4pt]
R^{\rm far}_\indlac{0}(x) &= \mathcal{C}^{\rm far}_{0} \, e^{-x/2} x^{\ell}\, U( \ell+1 -\nu_0, 2+2\ell, x) \, , \label{eqn:scalarRadialLOfar}
\end{align}
where ${}_2 F_1$ is the hypergeometric function and $U$ is the 
confluent hypergeometric function of the second kind. For integer values of $\nu_0 \geq \ell+1$, the functions $U$ become Laguerre polynomials.

The widths of the near and far regions can be determined by comparing the terms in (\ref{eqn: Heun-z}) that have been neglected when writing  (\ref{eqn:scalarLONear}) and (\ref{eqn:hydrogenic}), to those that have been kept.\footnote{The widths we quote for the near, far, and overlap regions are parametric statements, and so will be unaffected by higher-order corrections.}
 We find that the near and far regions are valid for $z \lesssim \alpha^{-2}$ and $x \gtrsim \alpha^2$, respectively. This means that the two regions have an overlapping region than spans the range $ \alpha \lesssim r \lesssim \alpha^{-1}$ (see Fig.~\ref{fig:ScalarZones}).  
Since (\ref{eqn:scalarRadialLOnear}) and (\ref{eqn:scalarRadialLOfar}) are approximations of the same function, they must agree over the entire overlap region. To match the solutions, it is convenient to introduce the following \textit{matching coordinate}\footnote{The $\alpha$-expansions of $R_{0}^{\rm near}(\xi)$ and $R_{0}^{\rm far}(\xi)$ are equivalent to taking the limits $z \to \infty$ and $x \to 0$, respectively. While the latter is more commonly adopted in the literature (see e.g.~\cite{Starobinsky:1973aij, Detweiler:1980uk}), matching in terms of $\xi$ has the advantage that it is organized solely in powers of $\alpha$. As we shall see later, this is a particularly useful way of organizing the matching at higher orders. 
}
\beq
\xi \equiv \frac{x}{\alpha^\beta} = \frac{2 (\tilde{r}_+ - \tilde{r}_-) }{\nu } \frac{z}{\alpha^{\beta - 2}} \, , \qquad \text{with} \qquad 0 < \beta < 2 \, . \label{eqn:IntCoord}
\eeq
Requiring the $\alpha$-expansions of the near and far-zone solutions to match, while keeping $\xi$ fixed, will fix the free coefficients of the solutions and thus determine $\nu$.

In terms of the coordinate~$\xi$, the  $\alpha$-expansion of the near-zone solution is\hskip 1pt\footnote{To aid the reader, we group terms that are matched in the overlapping region by color.} 
\beq
\begin{aligned}
R^{\text{near}}_0(\xi)  \sim  \tilde{\mathcal{C}}_{0}^{\rm near} \bigg[ & \red{(\alpha^{\beta-2} \xi)^\ell}  \left( 1  + \cdots  \right) \\ &+ \blue{(\alpha^{\beta-2} \xi )^{-\ell-1}} \, \mathcal{I}_\ell   \left( \frac{2 (\tilde{r}_+ - \tilde{r}_-)}{\nu} \right)^{2\ell+1}  \left( 1  + \cdots  \right) \bigg]\, , \,\, \mathrlap{\,\, \alpha \to 0 \, ,} \label{eqn: scalar matching near2}
\end{aligned}
\eeq
where the ellipses denote expansions in powers of $(\alpha^{\beta-2}\xi)^{-1}$ with real coefficients. In the following, it will only be important that the omitted terms are purely real. We have absorbed an overall coefficient into the rescaling $\mathcal{C}_{0}^{\rm near} \to \tilde{\mathcal{C}}_{ 0}^{\rm near} $ 
and defined
\beq
\begin{aligned}
\mathcal{I}_\ell & \equiv   - i P_+  \frac{(\ell!)^2}{(2\ell)! (2\ell+1)!} \prod_{k=1}^\ell \left( k^2 + 4P_+^2\right)   , \label{eqn:ImI}
\end{aligned}
\eeq
which is purely imaginary. 
Similarly, the $\alpha$-expansion of the far-zone solution~is 
\beq
\begin{aligned}
R^{\text{far}}_0(\xi)  \sim   \mathcal{C}_{0}^{\rm far} \bigg[ & \blue{( \alpha^\beta \xi)^{-\ell-1}}  \, \frac{\Gamma(2\ell+1)}{\Gamma(\ell+1-\nu_0)} \left( 1 +  \cdots \right)  \\ & +  \, \red{(\alpha^\beta \xi)^\ell} \, \frac{\mathcal{K}_\ell (\nu_0)}{\Gamma(2\ell+2)\Gamma(-\ell-\nu_0)}  \left( 1+  \cdots \right) \\
& \quad +  ( \alpha^\beta \xi)^\ell \log ( \alpha^\beta \xi ) \,  \frac{1}{\Gamma(2\ell+2)\Gamma(-\ell-\nu_0)} \left( 1 + \cdots \right) \bigg]\, , \mathrlap{\qquad \alpha \to 0 \, ,} \label{eqn: scalar matching far} 
\end{aligned}
\eeq
where the ellipses denote expansions in powers of $\alpha^{\beta}\xi$, and we have defined the $\nu_0$-dependent constant
\beq
\mathcal{K}_\ell (\nu_0) \equiv \gamma_E - \psi(2\ell+2) + \psi(\ell+1-\nu_0)  +  \sum_{k=1}^{2\ell+1} \frac{(-1)^{k+1} \Gamma(2\ell+2) \Gamma(-\ell + \nu_0)}{ 2^{k} \hskip 1pt k \hskip 1pt \Gamma(\nu_0-\ell+k) \Gamma(2\ell+2-k) } \, , \label{eqn:Kconstant}
\eeq
where $\psi(z) = \Gamma'(z)/\Gamma(z)$ is the digamma function and $\gamma_E$ the Euler-Mascheroni constant. Since logarithmic terms are absent in (\ref{eqn: scalar matching near2}), they do not play a role in the matching at leading order. However, these terms will appear at higher orders in the near-zone expansion and will ultimately match 
the logarithmic term in (\ref{eqn: scalar matching far}).

Matching the common terms in (\ref{eqn: scalar matching near2}) and (\ref{eqn: scalar matching far}) allows us to solve for the frequency eigenvalue~$\nu_0$. Although there are $2\ell+1$ such terms, 
only the $\red{\xi^\ell}$ and $\blue{\xi^{-\ell-1}}$ terms need to be matched at leading order. This is because they contain the dominant behaviors of $R^{\rm near}_0$ and $R^{\rm far}_0$ in the limit $\alpha \to 0$. The remaining terms are suppressed by powers of $\alpha$ and, like the logarithmic term above, can only be matched consistently when higher-order corrections to $R_0^{\rm near}$ and $R_0^{\rm far}$ are taken into account. Matching the coefficients of $\red{\xi^\ell}$ and $\blue{\xi^{-\ell-1}}$, and taking their ratio,
we obtain the following matching condition 
\beq
\frac{\Gamma(-\ell-\nu_0) \hskip 1pt \nu_0^{2\ell+1}}{\Gamma(\ell +1-\nu_0) \hskip 1pt \mathcal{K}_\ell(\nu_0)} = \alpha^{4\ell+2}\left(   \frac{\big[ 2 (\tilde{r}_+ - \tilde{r}_-)\big]^{2\ell\mathrlap{+1}}}{(2\ell!) (2\ell+1)!} \,{\cal I}_\ell + \cdots \right) , \label{eqn:ScalarMatching}
\eeq
where the ellipses denote terms that are purely real.
We emphasize that, although we have so far only expanded the equations of motion at leading-order,  higher powers of $\alpha$ arise in (\ref{eqn:ScalarMatching}) due to the hierarchy of the coordinates $x/z \sim \alpha^2$.  This is why the leading-order result for $\lab{Im}\,\nu$ can be $\mathcal{O}(\alpha^{4 \ell +2})$.

Solving (\ref{eqn:ScalarMatching}) for $\nu_0$, we get
\beq
\begin{aligned}
\nu_0 = & n + i P_+  \big[ 2\alpha^2 (\tilde{r}_+ - \tilde{r}_-)\big]^{2\ell+1} \\
& \times  \frac{(n+\ell)!}{n^{2\ell+1}(n-\ell-1)!}\left[ \frac{\ell!}{(2\ell )! (2\ell+1)!}\right]^2 \prod_{k=1}^\ell \left( k^2 + 4P_+^2\right)  ,  \label{eqn:NuLOScalar}
\end{aligned}
\eeq
where $n$ is an integer, with $ n \geq \ell + 1$. Substituting (\ref{eqn:NuLOScalar}) into (\ref{eqn:nudef}), and restoring a factor of $\mu$, the energy spectrum reads\footnote{From the point of view of the matching procedure, the real and imaginary parts of the 
eigenvalue arise from the matching of the coefficients of $\xi^\ell$ and $\xi^{-\ell-1}$, respectively. This can be seen directly from (\ref{eqn: scalar matching near2}), since the coefficient of $\xi^\ell$ is purely real while that of $\xi^{-\ell-1}$ has an imaginary part.} 
\beq
\omega_{n \ell m} = \mu \left( 1 - \frac{\alpha^2}{2n^2} \right)  + i \Gamma_{n \ell m} \, ,   \label{eqn: Bohr energy1}
\eeq
where the instability rate is given by (\ref{eqn:ScalarRate}). 
The real part of (\ref{eqn: Bohr energy1}) shows that the system has the expected hydrogen-like spectrum. However, due to the nontrivial boundary condition at the horizon, the imaginary part of (\ref{eqn:NuLOScalar}) is non-vanishing and the energy eigenstates are only quasi-stationary. For $m \Omega_H > \omega$, the instability rate is positive and superradiant growth occurs.

\subsubsection*{Higher-order corrections} 

Next, we compute higher-order corrections to $\nu$. We focus only on the corrections to the real part of the spectrum, since all degeneracies in the imaginary part have already been broken at leading order. It is convenient to rearrange the $\alpha$-expansion of the equations of motion (\ref{eqn:XExpansion}) into the following form 
\beq
\square^{\indlab{0}} \hspace{-0.5pt} X_{\indlac{i}} = -  \sum_{k=0}^{i-1} \square^{\indlab{i-k}} X_{\indlac{k}} \equiv J^X_{\indlac{i}} \, , \label{eqn:Jdef}
\eeq
so that the solutions of order $k < i$ are sources for the solution at order $i$.
Expanding the angular equation (\ref{eqn: spheroidal harmonic equation}) in powers of $\alpha$, we obtain 
 $J_{\indlac{i}}^\theta = 0$ and  $\Lambda_{\indlac{i}} = 0$, for $i = 1, 2, 3$, which means that the angular eigenvalue and eigenstate are uncorrected up to third order:
\beq
\Lambda = \ell(\ell+1) + \mathcal{O}\big(\alpha^4\big)  \, , \qquad  S(\theta) = P_{\ell m}(\cos \theta) + \mathcal{O}\big(\alpha^4\big)  \, . \label{eqn:ScalarHigherAngular}
\eeq
This is to be expected, since any deviation between $P_{\ell m}$ and the spheroidal harmonics, which are the exact solutions of (\ref{eqn: spheroidal harmonic equation}), is parametrized by the spheroidicity parameter $c^2 \sim \alpha^4$. 
To the order in $\alpha$ that we are working in, we can therefore use the leading-order angular solutions.

The asymptotic expansions of the radial functions need to be treated more carefully.
We will distinguish between modes with $\ell=0$ and $\ell \ne 0$.
For the former, the radial function peaks near the horizon and the solution is especially sensitive to the near-horizon geometry.

\paragraph{$\boldsymbol{\ell=0}$ modes} 

While the matching procedure is conceptually the same as before, the system of equations (\ref{eqn:Jdef}) is
more challenging to solve, because the source terms $J_i^X$ do not vanish.
 We solve these inhomogeneous equations via the method of \textit{variation of parameters}, where the general solutions contain integrals over $J_i^X$, with the integration limits appropriately chosen such that the boundary conditions in the respective regions are satisfied.
We relegate all technical details to Appendix~\ref{app:details}, and provide a more qualitative description here.

We perform the matching at higher orders by converting the radial coordinates $x$ and $z$ to the matching coordinate (\ref{eqn:IntCoord}), and expanding the resulting functions in the limit $\alpha \to 0$, just as we did at leading order. However, since $x$ and $z$ also appear in the integration limits of the integrals mentioned above, their asymptotic expansions must be systematically organized such that no spurious divergences appear as $\alpha \to 0$; see Appendix~\ref{app:details}. At finite order in $\alpha$, we expect only a finite number of the terms in $R^\lab{near}(\xi)$ and $R^\lab{far}(\xi)$ to match as $\alpha \to 0$.
For example, for the $\ell=0$ mode, only the terms $\xi$, $\xi^0$, $\log \xi$ and $\xi^{-1}$ are shared between the  near- and far-zone solutions at order $\alpha^2$. We solve for the energy eigenvalues by matching the coefficients of these terms. Although there are generally more terms to match than unknowns, the matching is only consistent if all of these common terms match.
For the $\ell=0$ mode, this procedure yields
\beq
E_{n 0 0} = \mu \left( 1 - \frac{\alpha^2}{2n^2} -\frac{\alpha^4}{8n^4} +  \frac{\left( 2-6n \right)\alpha^4 }{n^4}   \right)  . \label{eqn:l0ScalarSpectrum}
\eeq
Since $\ell=0$ implies $m=0$, there is no hyperfine splitting $\propto m \tilde{a} \hskip 2pt \alpha^5$.

\paragraph{$\boldsymbol{\ell \neq 0}$ modes} Although the method of matched asymptotic expansion described so far is robust and correctly captures all boundary conditions of the radial equation, the procedure quickly becomes cumbersome for modes with arbitrary quantum numbers $\{n, \ell, m \}$. 
Fortunately, since the typical Bohr radii of the $\ell \neq 0$ modes are peaked in the far region, most of their support is far from the horizon. The spectra of the $\ell \neq 0$ modes can therefore be obtained by naively extending the radial solution in the far region (\ref{eqn: scalar radial far LO}) towards $x \to 0$ and assuming that it is regular at the origin. While this approximation does not capture the instability rates, it is sufficient for determining the real parts of the energy eigenvalues~$\nu_i$, at least up to the order of interest.\footnote{We have also performed a matched asymptotic expansion for the $\ell=1$ mode, and we found agreement with the stated approximation up to order $\alpha^5$. Since the approximation gets better for larger values of $\ell$, we expect the results obtained through this simplification also hold for all other $\ell \neq 0$ modes.} As shown explicitly in  Appendix~\ref{app:details}, this leads to the expression (\ref{eqn:scalarspectrum}) for the scalar field energy eigenvalue. This result also agrees with (\ref{eqn:l0ScalarSpectrum}) obtained for $\ell=0$ through the matched asymptotic expansion.\footnote{In \eqref{eqn:scalarspectrum}, the hyperfine structure naively diverges for the $\ell=0$ mode. However, since this mode necessarily has $m=0$, it does not actually receive a contribution from the hyperfine splitting.} These higher-order contributions have been computed in \cite{Baumann:2018vus} in the non-relativistic limit. Our results here, which make no such assumption, agrees with the previous result.

\subsection{Vector Electric Modes} 
\label{sec:ProcaLO}

Having illustrated our approach for the case of a scalar field, we are now ready to attack the more technically challenging vector field. We will first treat the electric modes of the field here, before moving on to the magnetic modes in the next subsection.

Our electric analysis relies on 
the separable ansatz (\ref{eqn: Proca ansatz}) introduced in~\cite{Krtous:2018bvk, Frolov:2018ezx}. Just like for 
the scalar field, we can transform the radial equation (\ref{eqn:ProcaR}) into near and far regions through the variables $z$ and $x$, defined in~(\ref{eqn: z variable}) and~(\ref{eqn: x variable}), respectively. The asymptotic behavior of the radial function as $z \to 0 $ and $ x \to \infty $ is 
\begin{align}
R(z) &\,\sim\, B_1 \hskip 1pt z^{ i P_{+}} + B_2 \hskip 1pt  z^{- i P_+} \,,  \mathrlap{\hspace{3.62cm} z \to 0\, ,}  \hskip 30pt  \label{eqn:vector radial outer horizon} \\
R(x) &\,\sim\, B_3 \hskip 1pt  e^{- x/2} \, x^{+\nu - 2 \alpha^2/\nu} + B_4 \hskip 1pt  e^{+ x/2} \, x^{-\nu + 2 \alpha^2/\nu}\,,  \mathrlap{\hspace{0.25cm} x \to \infty \, .}  \hskip 30pt \label{eqn: vector radial infinity}
\end{align}
The near horizon behavior is similar to that of the scalar in (\ref{eqn:scalar radial outer horizon}), since the residues of the $1/z^2$ poles are identical.  The asymptotic behavior in the far region, on the other hand, differs from that of the scalar, c.f.~(\ref{eqn: scalar radial infinity}), because the coefficients of the $\d R/\d r$ term in the radial equations differ as $r \to \infty$.
The purely ingoing boundary condition at the horizon requires that $B_2 = 0$, and for a quasi-bound state solution we must set $B_4=0$.

A crucial difference between the analysis for scalar and vector fields is the presence of additional poles at $r=\hat{r}_{\pm}$ in the radial equation (\ref{eqn:ProcaR}). 
Since these poles are located at an $\mathcal{O}(1)$ distance from the origin of the complex-$r$ plane, the widths of the near and far regions are now much smaller. Indeed, we find that the near and far regions cover $z \lesssim \alpha^{-1}$ and $x \gtrsim \alpha$, and therefore do not overlap. This means that the matching between these regions must be performed indirectly through an intermediate region (see Fig.~\ref{fig:Vectorzones}). To describe this intermediate region, it is convenient to introduce the following coordinate
\beq
y \equiv r - r_+ \, . \label{eqn: y variable}
\eeq
We find that the intermediate region spans the range $\alpha \lesssim y \lesssim \alpha^{-1}$, which overlaps with both the near and far regions and therefore allows the two-step matching procedure.  
Since no boundary conditions need to be imposed in the intermediate region, the coefficients are determined by matching the near- and far-zone solutions.

To solve the coupled differential equations (\ref{equ:ProcaS}) and (\ref{eqn:ProcaR}) order-by-order in $\alpha$, we now expand all relevant quantities in powers of $\alpha$. The system of equations are schematically organized as in~(\ref{eqn:XExpansion}), where $X$ now includes the intermediate radial function $R^{\rm int}$. The parameters appearing in the differential operators, $\lambda$ and $\nu$, are also expanded in powers of $\alpha$.

\subsubsection*{Leading-order solutions}

The leading-order angular equation for the electric modes is
\beq
\left[ \frac{1}{\sin \theta} \frac{\d}{\d \theta} \left( \sin \theta \frac{\d}{\d \theta} \right) - \frac{m^2}{\sin^2 \theta} + \lambda_0 (\lambda_0 - 1) \right] S_0 = 0 \, . \label{eqn: ProcaAngular}
\eeq
Since the vector field has intrinsic spin, the quantum number $m$ now describes the \textit{total angular momentum} projected along the spin of the black hole (see Appendix~\ref{app:harmonics}). The solutions are the associated Legendre polynomials $P_{jm}$, provided that the separation constant obeys $\lambda_0(\lambda_0 - 1) = j(j+1)$. 
In the far region, we may identify $j$ as the total angular momentum of solution.
This yields the following two solutions
\beq
\lambda_0^+   = - j \quad {\rm and} \quad \lambda_0^- = j+1 \, .   \label{eqn:lambda0sol}
\eeq
To understand the physical interpretation of these solutions, it is instructive to consider the leading-order form of the ansatz $A^\mu \equiv B^{\mu \nu} \nabla_\nu Z$. Using the results of Appendix~\ref{app:details}, we find
\begin{align}
A^{\mu}_{\indlab{0}}  = \frac{\lambda_0}{\lambda_0^2  + r^2} & \begin{pmatrix}
\minus \lambda_0 & \hphantom{\minus}i r  & \hphantom{\,\,}0 \hphantom{\,\,} & \hphantom{\,\,}0 \hphantom{\,\,} \\
 \minus i r & \lambda_0 & 0 & 0 \\
 0 & 0 & 0 & 0 \\
 0 & 0 & 0 & 0 \\
\end{pmatrix}\begin{pmatrix}\hphantom{\,} \partial_t \hphantom{\,} \\ \partial_r \\ \partial_\theta \\ \partial_\phi \end{pmatrix} Z_0 \\
& +  \frac{1}{r^2} \begin{pmatrix}
 \hphantom{\,\,}0\hphantom{\,\,} & \hphantom{\,\,}0\hphantom{\,\,} & \hphantom{\,\,}0\hphantom{\,\,} & 0 \\
 0 & 0 & 0 & 0 \\
 0 & 0 & 1 & 0 \\
 0 & 0 & 0 & \lab{csc}^{2}\theta \\
\end{pmatrix}\begin{pmatrix}\hphantom{\,} \partial_t \hphantom{\,} \\ \partial_r \\ \partial_\theta \\ \partial_\phi \end{pmatrix} Z_0 \, ,  \label{eqn: ansatz a0 explicit}
\end{align}
which is diagonal in angular gradients. At large distances $ r \gg \lambda_0$, the spatial components of the vector field simplify to
\beq
A^i_{\indlab{0}} \propto r^{\minus 1}  R_0^{\rm far}(r) \left( r \hskip 1pt \partial^i Y_{jm} -\lambda_0 \,Y_{jm} \hat{r}^i \right) e^{- i \omega t}, \label{eqn:Aisol}
\eeq
where $Y_{jm}$ are the scalar spherical harmonics, and we have normalized the basis vectors with the appropriate scale factors $\{1, r, r \sin \theta \}$ in spherical coordinates. Using the relationships between the `pure-orbital' and `pure-spin' vector spherical harmonics in flat space~\cite{Thorne:1980ru},\footnote{See Appendix~\ref{app:harmonics} for a discussion of vector spherical harmonics in spherically symmetric spacetimes.} 
\beq 
\begin{aligned}
Y^i_{j-1, jm} & = \frac{1}{\sqrt{j(2j+1)}} \left[  r  \hskip 1pt \partial^i Y_{jm} + j Y_{jm} \hat{r}^i \right]  , \\
Y^i_{j+1, jm} & = \frac{1}{\sqrt{(j+1)(2j+1)}} \left[  r  \hskip 1pt \partial^i Y_{jm}  - (j+1) Y_{jm} \hat{r}^i \right] , \label{eqn:PureOrbit-PureSpin}
\end{aligned}
\eeq
we see that the solutions (\ref{eqn:Aisol}) with eigenvalues $\lambda_0^{\pm}$ correspond to the  $j=\ell\pm1$ electric modes of the vector field. Furthermore, (\ref{eqn:Aisol}) relates $R_0^{\rm far}$ to the  actual radial profile of the vector field at large distances, with the additional power of $r^{-1}$ accounting for the different asymptotic behaviors of (\ref{eqn: scalar radial infinity}) and (\ref{eqn: vector radial infinity}). This means that the usual hydrogen-like intuition still applies for all spatial components of the vector field~\cite{Endlich:2016jgc, Baryakhtar:2017ngi}.\hskip 1pt\footnote{This is perhaps most obvious if we rewrite the Proca equation as the system of coupled scalar equations~(\ref{eq:procaEqDecompMain}). In the far-field limit, the mixings vanish, and the Proca equation reduces to a set of uncoupled scalar equations~(\ref{eqn: scalar radial equation}).}

The near, intermediate and far-zone radial equations are 
\begin{align}
\bigg[  \frac{\d^2}{\d z^2 } + \left( \frac{1}{z} + \frac{1}{z+1} \right)\frac{\d}{\d z} -  \frac{\lambda_0 (\lambda_0 - 1)}{z(z+1)} \hskip 80pt \nonumber &  \\
+ \frac{P_+^2}{z^2} + \frac{P^2_+}{(z+1)^2} - \frac{2P_+^2}{z} + \frac{2P_+^2}{z+1} \bigg] R^{\rm near}_0 &  = 0  \, , \label{eqn:ProcaNear} \\
\left[ \frac{\d^2}{\d y^2} +\left(  \frac{2}{y} - \frac{2 y}{  \lambda_0^{2} +  y^2 } \right) \frac{\d}{\d y} - \frac{ \lambda_0 \left( \lambda_0 - 1 \right)}{ y^2} - \frac{2\lambda_0}{ \lambda_0^2 + y^2} \right] \, R_0^{\rm int}\ &= 0 \, , \label{eqn:ProcaInt} \\
\left[ \frac{\d^2 }{\d x^2 } + \frac{\nu_0}{x}   - \frac{\lambda_0(\lambda_0+1)}{x^2} - \frac{1}{4}   \right]\, R_0^{\rm far}\ & = 0\, . \label{eqn:VectorFarF}
\end{align}
We see that the combination of $\lambda_0$ that appears in (\ref{eqn:ProcaNear}) is the same as that in (\ref{eqn: ProcaAngular}). This means that the near-horizon behavior is sensitive to the total angular momentum of the vector field, instead of just its orbital angular momentum. This is to be expected since the near-horizon limit resembles a `massless' limit of the vector field, and Teukolsky's equation for a massless gauge field in this same limit manifestly depends on the spin of the field~\cite{Teukolsky:1973ha}. Furthermore, we find that the dependence on $\lambda_0$ in (\ref{eqn:ProcaNear}) differs from that in (\ref{eqn:VectorFarF}). Since the $\alpha \to 0$ limit in the far region is equivalent to taking the flat-space limit, the far region is therefore only sensitive to the orbital angular momentum of the field. 
Indeed, solving 
$\lambda_0 \left( \lambda_0 + 1 \right) = \ell(\ell+1) $, we obtain
\beq
\lambda_0^+ = -(\ell+1) \quad {\rm and} \quad \lambda_0^- = \ell \, . \label{eqn:lambda0solL}
\eeq
Comparing (\ref{eqn:lambda0sol}) and (\ref{eqn:lambda0solL}), we find that the eigenvalues $\lambda_0^\pm$ correspond to the $j = \ell \pm 1$ modes, which agrees with our analysis above.

Imposing the correct boundary conditions, we find 
\begin{align}
R^{\rm near}_0(z)   &\,=\, \mathcal{C}_{0}^{\rm near}  \left( \frac{z}{z+1} \right)^{i P_+} \!\!\!\!{}_2 F_1(-j, j+1, 1-2 i P_+, 1+z) \, , \label{equ:Near0} \\[4pt]
R^{\rm int}_0(y)   &\,=\, \mathcal{C}_{0}^{\rm int} y^{- \lambda_0}  + \mathcal{D}_{0}^{\rm int} \, y^{-1 + \lambda_0} \left[ \lambda_0^2 \hskip 1pt (2\lambda_0 +  1 ) + (2 \lambda_0-1) \, y^2 \right]  ,\label{equ:Int0} \\[10pt]
R^{\rm far}_0(x) &\,=\, \mathcal{C}_{0}^{\rm far} \, e^{-x/2} x^{\ell+1}\, U( \ell+1 -\nu_0, 2+2\ell, x)\, . \label{equ:Far0}
\end{align}
Since the intermediate region depends on both $j$ and $\ell$, the powers of the polynomials in (\ref{equ:Int0}) are determined by the angular eigenvalues (\ref{eqn:lambda0sol}). As we shall see, the fact that the term proportional to $\mathcal{D}_{0}^{\rm int}$ contains two distinct $y$-dependences is crucial for the matching with the asymptotic expansions of the near and far-zone solutions. In the following, we will replace $\ell$ by $j$ via the substitution $j = \ell \pm 1$. This is because the azimuthal number $m$ is interpreted as $m_j$, so the eigenstates of the vector field are characterized by $j$ instead of $\ell$.

Since the solutions (\ref{equ:Near0}), (\ref{equ:Int0}) and (\ref{equ:Far0}) are different approximations of the same function, they must agree in the regions of overlap. Although the near and far regions do not overlap with each other, they each overlap with the intermediate region. It is therefore convenient to introduce the following matching coordinates
\beq
\begin{aligned}
\xi_1 & \equiv \frac{x}{\alpha^{\beta}} = \frac{2}{\nu} \frac{y }{\alpha^{\beta-1}} \, , \hskip 65pt \qquad 0 < \beta < 1 \, , \\[4pt]
\xi_2 & \equiv \frac{2(\tilde{r}_+ - \tilde{r}_-)}{\nu} \frac{z}{\alpha^{\beta-2}} = \frac{2}{\nu} \frac{y }{\alpha^{\beta-1}}  \, , \qquad 1 < \beta < 2 \, . \label{eqn:IntCoordVector}
\end{aligned}
\eeq
We match these solutions by demanding that the intermediate-zone solution has the same $\xi_1$ dependence as the far-zone solution, and the same $\xi_2$ dependence as the near-zone solution. This forces the intermediate solution to agree with the solutions in the near and far regions in the regions of overlap. This matching procedure thus fixes the coefficients and determines the frequency eigenvalues.

In terms of $\xi_2$, the $\alpha$-expansion of the near-zone solution is 
\beq
\begin{aligned}
R^{\text{near}}_0  \,\sim\,  \tilde{\mathcal{C}}_{0}^{\rm near} \bigg[ & \red{(\alpha^{\beta - 2} \xi_2)^j}  \left[ 1  + \cdots  \right] \\
&+\, \blue{(\alpha^{\beta - 2} \xi_2 )^{-j-1}} \, \mathcal{I}_j  \left( \frac{2 (\tilde{r}_+ - \tilde{r}_-)}{\nu} \right)^{2j+1}  \left[ 1  + \cdots  \right] \bigg] \, , \,\, \mathrlap{\,\, \alpha \to 0 \, ,} 
\end{aligned}
\label{eqn: vector matching near2}
\eeq
where $\mathcal{I}_j$ was defined in (\ref{eqn:ImI}), with $\ell$ replaced by $j$. For the far and intermediate-zone solutions, we have to treat the $j = \ell\pm1$ modes separately.

Substituting $\ell = j-1$ into (\ref{equ:Far0}) and expanding in powers of $\alpha$, the far-zone solution becomes 
\beq
\begin{aligned}
R^{\text{far}}_0  \,\sim\,   \mathcal{C}_{0}^{\rm far} \bigg[ & \green{( \alpha^\beta \xi_1)^{-j+1}}  \, \frac{\Gamma(2j-1)}{\Gamma(j-\nu_0)} \left[ 1 +  \cdots \right]  \\
& +  \, \red{(\alpha^\beta \xi_1)^{j}} \, \frac{\mathcal{K}_{j-1} (\nu_0)}{\Gamma(2j)\Gamma(1-j-\nu_0)}  \left[ 1+  \cdots \right] \\
& +  ( \alpha^\beta \xi_1)^j \log ( \alpha^\beta \xi_1 ) \,  \frac{1}{\Gamma(2j)\Gamma(1-j-\nu_0)} \left[ 1 + \cdots \right] \bigg]\, , \quad \alpha \to 0 \, , \label{eqn: vector matching far 1} 
\end{aligned}
\eeq
where $\mathcal{K}_{j-1}$ was defined in (\ref{eqn:Kconstant}), with the indices appropriately replaced. Although logarithmic terms are absent in (\ref{eqn: vector matching near2}), they will appear at higher orders in the near-zone expansion, such that they can be matched with the logarithmic term in (\ref{eqn: vector matching far 1}). Substituting $\lambda_0 = \lambda_0^+$ into (\ref{equ:Int0}), the intermediate-zone solution reads
\beq
\begin{aligned}
R^{\rm int}_0   \,  =\, & \mathcal{\tilde{C}}_{0}^{\rm int} \red{(\alpha^{\beta - 1}  \xi )^{j}} \\
& + \mathcal{\tilde{D}}_{0}^{\rm int}  \left[ 4 j^2(2j-1) \blue{(\alpha^{\beta - 1}  \xi )^{-j -1}} + (2j+1) \, \nu^2\, \green{(\alpha^{\beta - 1}  \xi )^{-j+1}} \right]  , \label{eqn: vector matching int 1} 
\end{aligned}
\eeq
where we have rescaled the coefficients $\mathcal{C}_{0}^{\rm int} \to \mathcal{\tilde{C}}_{0}^{\rm int}$ and $\mathcal{D}_{0}^{\rm int} \to \mathcal{\tilde{D}}_{0}^{\rm int}$ for future convenience. For the matchings with the far and near zones, we will use $\xi = \xi_1$ and $\xi = \xi_2$, respectively.

To determine the correct matching, we consider the behavior of the two asymptotic expansions (\ref{eqn: vector matching near2}) and (\ref{eqn: vector matching far 1}) as $\alpha \to 0$. From the perspective of the near-zone solution (\ref{eqn: vector matching near2}), this limit `zooms' in on the overlap of the near and intermediate regions, as seen `from' the near region. The $\red{\xi_2^j}$ term then dominates (\ref{eqn: vector matching near2}), and it must match the corresponding $\red{\xi^j}$ term in (\ref{eqn: vector matching int 1}). Similarly, if we take $\xi = \xi_1$ in the intermediate region (\ref{eqn: vector matching int 1}), then the $\alpha \to 0$ limit zooms in on the overlap between the intermediate and far regions, as seen from the intermediate region. In this case, $\red{\xi_1^j}$ also dominates, and so it must match the same term in (\ref{eqn: vector matching far 1}). Thus, as we move from the near to intermediate to far regions, the dominant behavior is always $\red{\xi^j}$, and its coefficients in the three different regions must match.

We may also work in reverse. In the far region, the $\alpha \to 0$ limit zooms into its overlap with the intermediate region, but this time as seen from the far region. The dominant behavior now is $\green{\xi_1^{-j+1}}$, and this must match with the corresponding term in (\ref{eqn: vector matching int 1}). Similarly, taking $\xi = \xi_2$ in (\ref{eqn: vector matching int 1}), the $\alpha \to 0$ limit zooms into its overlap with the near region, though this time as seen from the intermediate region. We find that the $\blue{\xi_2^{-j-1}}$ dominates the intermediate solution and so is matched with its partner in (\ref{eqn: vector matching near2}). As we move from the far region, through the intermediate, and into the near region, the dominant behavior changes and thus the resulting matching condition is more nontrivial.

Taking the ratio of these two matching relations, we obtain
\beq
\frac{\Gamma(1-j-\nu_0) \, \nu_0^{2j-1} }{\Gamma(j-\nu_0) \mathcal{K}_{j-1} (\nu_0)} = \alpha^{4j} \left(    \frac{(2j+1) \, \left[ 2 (\tilde{r}_+ - \tilde{r}_-) \right]^{2j+1}}{\left[ 2 j (2j-1)!\right]^2 }\,\mathcal{I}_j  + \cdots  \right)  .\label{eqn:VectorplusRatio} 
\eeq
We again emphasize that, although we have expanded the equations of motion only to leading order, higher powers of $\alpha$ appear in (\ref{eqn:VectorplusRatio}) from $x/y \sim y/z \sim \alpha$. Solving for the leading-order real and imaginary parts of $\nu_0$, and restoring factors of $\mu$, we find 
\beq
\omega_{n \ell j m} = \mu \left( 1 - \frac{\alpha^2}{2n^2} \right) + i \Gamma_{n \ell j m}\, , \label{equ:omega+}
\eeq
where the instability rate is given by \eqref{eqn:VectorRates}. The dominant $j=\ell+1$ growing mode, $|\es 1 \es 0 \es 1 \es 1 \rangle$, has a growth rate $\Gamma_{1011} \propto \mu \, \alpha^6$ that is much larger than the dominant growing mode in the scalar case.  
The $j = \ell -1$ electric mode has the same energy spectrum (\ref{equ:omega+}), but a different instability rate $\Gamma_{n\ell jm}$;~cf.~(\ref{eqn:VectorRates}). Its dominant growing mode $|\es 3 \es 2 \es 1 \es 1 \rangle$ has a significantly suppressed growth rate, $\Gamma_{3211} \propto \mu \, \alpha^{10}$, compared to the the dominant $j= \ell +1$ mode.

\subsubsection*{Higher-order corrections}

Finally, we compute the higher-order corrections to the real part of the frequency eigenvalues. 
The following is a sketch of the computation, with details relegated to  Appendix~\ref{app:details}.

Due to the presence of additional $\theta$-dependent terms on the right-hand side of (\ref{equ:ProcaS}), the higher-order angular equations are now harder to solve than in the scalar case. In particular, these terms induce new cross couplings in the angular eigenstates
\beq
\begin{aligned}
S(\theta) &= P_{j m}(\cos \theta) +  \Delta S(\theta) +  \mathcal{O}(\alpha^4) \, , \label{eqn: Electric angular eigenstate order2} \\[4pt]
\Delta S(\theta) &= \Big( \alpha^2 \tilde{a}^2  b_{j-2}  + \alpha^3 \tilde{a}^3 c_{j-2}  \Big) \, P_{j-2, m}(\cos \theta) \\
& + \Big( \alpha^2 \tilde{a}^2 b_{j+2}  + \alpha^3 \tilde{a}^3  c_{j+2} \Big)  \,  P_{j+2, m}(\cos \theta) \, , 
\end{aligned}
\eeq
where the coefficients $b_{j \pm 2}$ and $c_{j \pm 2}$ are given in Appendix~\ref{app:details}. Strictly speaking, $j$ is no longer a good quantum number at order $\alpha^2$. However, as we discussed in \S\ref{sec:TensorKerr}, an approximate notion of total angular momentum still exists---especially in the $\alpha \to 0$ limit---and we continue to label our states with $j$, even though it has no precise physical meaning. 
The angular eigenvalues $\lambda$ for $j = \ell \pm 1$, expanded up to order $\alpha^3$, are also given in 
Appendix~\ref{app:details}. 
Substituting these results for $\lambda$ into the radial equations allows us solve for the energy eigenvalues at high orders.

It is also instructive to compute the higher-order corrections to the actual vector field configuration (\ref{eqn:Aisol}) in the far zone 
\beq
A^i_{\indlab{1}} \propto r^{-1}R_0^{\rm far} \, i \tilde{a} \alpha   \,  \left(   \cos \theta \,  \epsilon^{ikl} r^k  \partial^l \, Y_{jm} -  \lambda_0\sin \theta \, Y_{jm} \, \hat{\phi}^i \right) e^{- i \omega t} \, . \label{eqn:AisolCorrection}
\eeq
Despite the presence of a magnetic vector spherical harmonic in (\ref{eqn:AisolCorrection}),
the vector field configuration is still of the electric type, since the $\cos \theta$ factor acquires a factor of $(-1)$ under a parity transformation. The presence of the $\hat{\phi}^i$-term in (\ref{eqn:AisolCorrection}) is a manifestation of the fact that spherical symmetry is already broken at this order in $\alpha$.

\paragraph{$\boldsymbol{\ell = 0}$ modes} We first consider the dominant growing mode, with $j=1$ and $\ell=0$. Since the radial wavefunctions are peaked near the horizon, they are most sensitive to strong gravity effects. This sensitivity manifests itself through the failure of ordinary perturbation theory, which diverges for this mode. It is not clear how to regulate these divergences, and thus this technology of matched asymptotic expansions is necessary to derive this mode's fine and hyperfine structure.  The matching procedure at higher orders is the same as that illustrated for the scalar case above, except that the matching between the near and far-zone solutions must be performed through the intermediate zone. 
 We find that these higher-order intermediate-zone solutions take the form of simple linear combinations of powers and logarithms of $y$, which can be matched with the corresponding terms found in the asymptotic expansions of the near and far-zone solutions.
Performing the matched asymptotic expansion up to third order, we find 
\beq
\begin{aligned}
E_{n01m}  & =   \mu \left( 1 -\frac{\alpha^2}{2n^2} - \frac{\alpha^4}{8n^4} +  \frac{\left(  6 -10n \right) \alpha^4 }{3n^4}  \, + \frac{8  \tilde{a} m \alpha^5}{3n^3} \right)  .\label{eqn:VectorMain}
\end{aligned}
\eeq
Since the vector field has intrinsic spin, the $\ell=0$ modes now have hyperfine splittings $\propto \tilde{a} m \es \alpha^5$.

\paragraph{$\boldsymbol{\ell \neq 0}$ modes}  Since the $\ell \neq 0$ modes peak in the far zone, the calculation of the higher-order corrections of their spectra can be simplified by naively extrapolating the far-zone solution (\ref{equ:Far0}) towards the horizon $x \to 0$ and imposing regular boundary conditions there.
The far-zone radial function and the eigenvalues are then solved as an expansion in powers of $\alpha$. This leads to the result (\ref{eqn:vectorspectrumGeneral}) for the energy eigenvalues. Remarkably, we find that (\ref{eqn:vectorspectrumGeneral}), obtained through this simplified treatment, agrees with (\ref{eqn:VectorMain}), which was derived more rigorously through matching (see Appendix~\ref{app:details} for a more detailed discussion).

\subsection{Vector Magnetic Modes} \label{sec:MagneticVector} 

In principle, the task of solving for the magnetic mode spectrum involves a straightforward application of the machinery developed in previous sections to the relevant separated radial and angular equations. Unfortunately, these equations are not yet known, except at linear order in $\tilde{a}$~\cite{Pani:2012vp,Pani:2012bp}. In this section, we discuss the extent to which the 
approximate radial equation (\ref{eqn:MagneticRadial}) can be used to derive the energy spectrum and instability rates. Our discussion will be brief and mostly qualitative, since our solutions are similar to those found in previous sections.

\begin{figure}
  \begin{center}
     \makebox[\textwidth][c]{\includegraphics[scale=0.82]{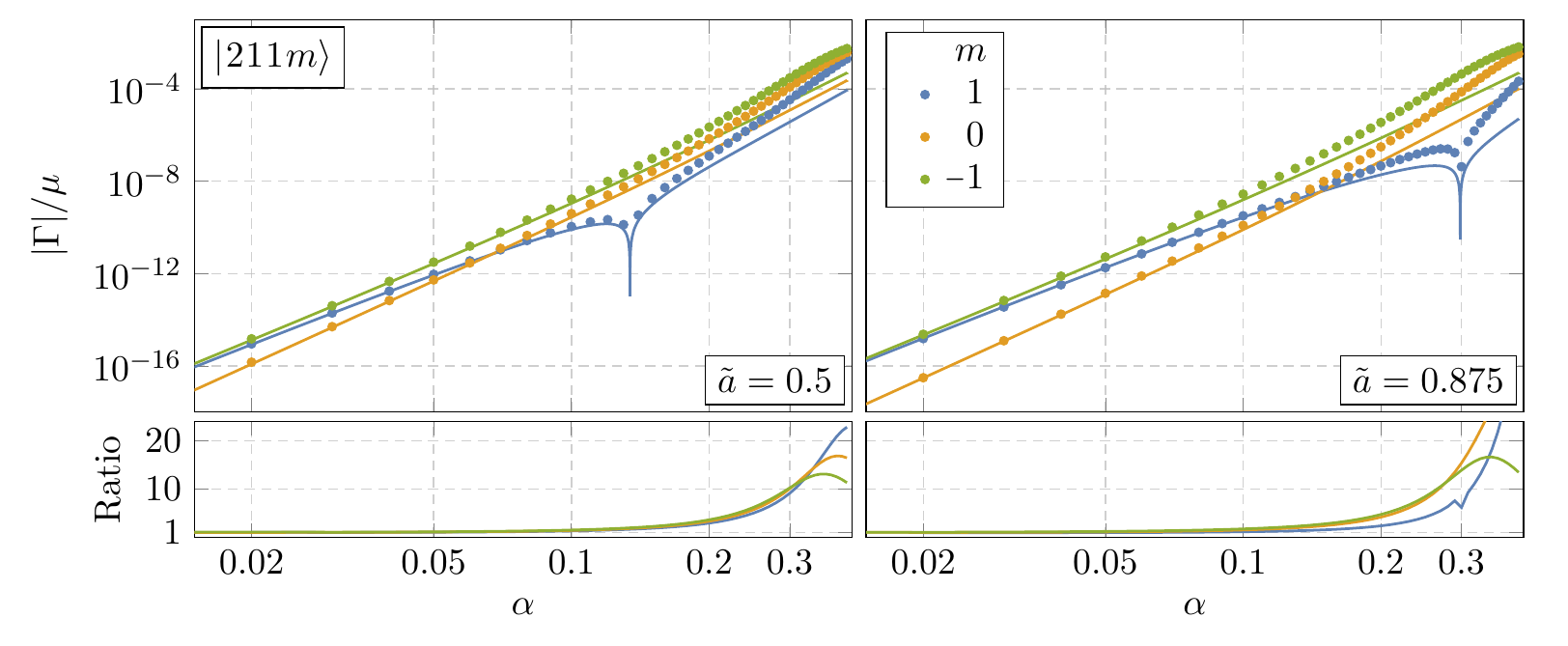}}
    \caption{Instability rates for the magnetic mode $|\es 2 \es 1 \es 1 \es m\rangle$, for $\tilde{a} = 0.5$ (left) and $\tilde{a} = 0.875$ (right). We compare our numeric results (denoted by points) with the conjectured form of the instability rate (solid lines), and plot the ratio of  numerics to analytics in the lower panels.     \label{fig:magInstability}}
  \end{center}
\end{figure}

\subsubsection*{Leading-order solution}

In the far region, the electric  (\ref{eqn:VectorFarF}) and magnetic (\ref{eqn:MagneticRadial}) radial equations are of the same form at leading order,
with $\lambda_0(\lambda_0 + 1)$ replaced with $j(j+1)$~\cite{Rosa:2011my, Pani:2012bp}. The magnetic quasi-bound states therefore take the same form 
as their electric counterparts in the far region (\ref{eqn:Aisol}), and 
\begin{equation}
A^i_\indlab{0} \propto r^{-1} R_0^\lab{far}(r) \, Y_{j,\es j \es m}^i(\theta, \phi) \,e^{-i \omega t}\,.
\end{equation} 
As expected, 
these magnetic modes are therefore also hydrogenic~\cite{Endlich:2016jgc,Baryakhtar:2017ngi} and the spectrum is Bohr-like to leading order in $\alpha$.

In the previous sections, we derived the leading-order instability rates by finding the near- and far-zone solutions, and matched 
them in the overlap region. One might hope to do the same with~(\ref{eqn:MagneticRadial}) to derive the magnetic instability rates to linear order in $\tilde{a}$. However, 
since the radial equation was derived in the far region, extending it into the near region is not as innocuous as it naively seems. This is because the near region---and especially the boundary condition at the outer horizon---is sensitive to the full nonlinear spin-dependence of the metric. As we discussed in Footnote~\ref{footnote:FixedPp}, getting these important terms wrong in our perturbative expansion can cause our approximation 
to deviate strongly from the actual solution, and it is not clear how to infer the correct behavior from a far-zone solution at linear order in $\tilde{a}$.\footnote{The authors of \cite{Pani:2012bp,Pani:2012vp} were able to derive an accurate expression for the magnetic instability rate to linear order in $\tilde{a}$, by matching the solution in the far region to a near-zone solution that is obtained by extrapolating (\ref{eqn:MagneticRadial}) toward the outer horizon $r_+ = 2 \alpha$. However, they needed to discard terms like $\check{P}_+^2$, which contains both constant and linear-in-$\tilde{a}$ pieces, to obtain a finite result. In effect, discarding these divergent terms 
imposes the correct near-horizon behavior.}

We will instead guess the form of the magnetic instability rate and check this guess against our numerics. 
It is natural to assume that the magnetic instability rates take the same functional form as the rates for the scalar field~(\ref{eqn:ScalarRate}) and the electric modes of the vector field~(\ref{eqn:VectorRates}). The overall normalization and the dominant $\alpha$-scaling can be fixed by demanding that this ansatz matches the Schwarzschild limit~\cite{Rosa:2011my}, and thus we arrive at the conjectural instability rate~(\ref{eqn:VectorRates}) with~(\ref{eqn:VectorRatesCoeff}). 
As we show in Fig.~\ref{fig:magInstability}, we find our guess to be in excellent agreement with our numeric results for small $\alpha$, even at high values of $\tilde{a}$.  However, we emphasize that this formula represents our best educated guess for the magnetic instability rate at arbitrary spin and is {\it not} rigorously derived. A conclusive analytic result can only be found by solving a separable equation valid for all $\tilde{a}$.

\subsubsection*{Higher-order corrections}

\begin{figure}
  \begin{center}
    \includegraphics[scale=0.82, trim=7 0 0 0]{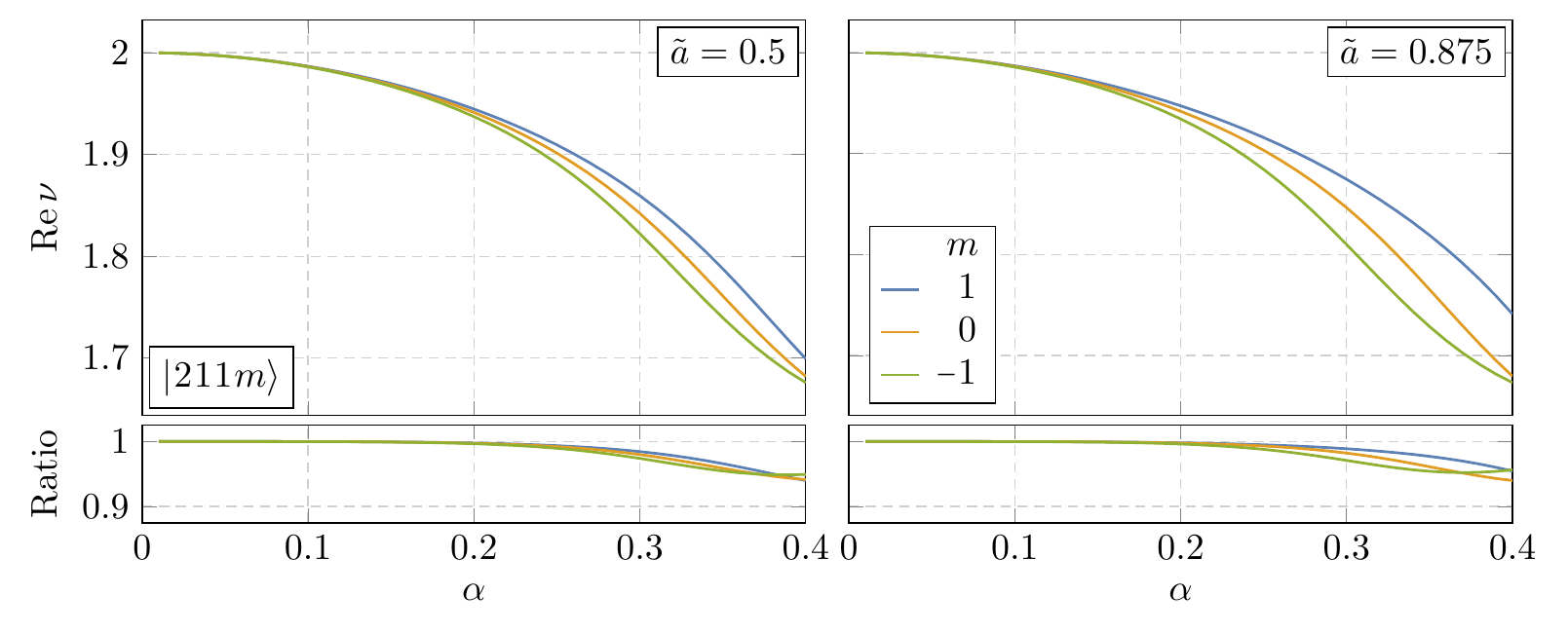}
    \caption{\label{fig:magneticReal} Numeric results for the spectra of the magnetic mode $|\es 2 \es 1 \es 1 \es m \rangle$, for $\tilde{a} = 0.5$ (left) and $\tilde{a} = 0.875$ (right). The ratios of the numeric results to their perturbative predictions (\ref{eqn:vectorspectrumGeneral}) are shown in the panels below.} 
  \end{center}
\end{figure}

Since the linear-spin approximation 
does not fully capture the leading-order behavior of the near region, we cannot use (\ref{eqn:MagneticRadial}) to perform a matched asymptotic expansion at high orders in $\alpha$. Fortunately, the magnetic modes have non-vanishing orbital angular momentum and are thus peaked far away from the horizon. 
 As we have illustrated for both the scalar and the electric modes of the vector, we may derive the energy spectrum of these modes to $\mathcal{O}(\alpha^3)$ by extending the far-zone radial solutions (\ref{equ:Far0}) toward the horizon, imposing regular boundary conditions, and solving for $\nu$ perturbatively in powers of $\alpha$. This then yields the fine and hyperfine structure of the magnetic modes to linear order in $\tilde{a}$. However, following the pattern suggested by the electric modes in the spectrum, we expect the fine structure to be independent of $\tilde{a}$, and the hyperfine structure to be proportional to $m \tilde{a}$,
so that our results for the magnetic energy spectrum should be valid for arbitrary spin. 
In Figure~\ref{fig:magneticReal}, we compare the approximation~(\ref{eqn:vectorspectrumGeneral}) to our numeric results and find that it is very accurate even at large $\tilde{a}$, suggesting that our extrapolation of the fine and hyperfine structure to arbitrary $\tilde{a}$ is correct.

\section{Numerical Computation of the Spectra}
\label{sec:numeric}

The analysis of the previous section required $\alpha$ to be small and we only derived rigorous results for the scalar field and the electric modes of the vector field.
In this section, we solve for the quasi-bound state spectra numerically, providing results for arbitrary values of $\alpha$ and $\tilde{a}$ including the magnetic modes of the vector field. We will thus be able to use these results to determine when the perturbative approximations summarized in \S\ref{sec:Summary} break down.

To make contact with the literature, we will begin by reviewing the continued fraction method for determining the scalar spectrum in \S\ref{sec:contFrac}. We will discuss the main limitations of the method, including why precise results for $\alpha \ll 1$ are difficult to achieve numerically. We then show how these difficulties can be surmounted by reformulating the problem as a nonlinear eigenvalue problem in \S\ref{sec:NLEP}. In \S\ref{sec:Separable}, we use this alternative formulation to solve for the scalar spectrum, and show how to modify the techniques to solve for the separated Proca equations (\ref{equ:ProcaS}) and (\ref{eqn:ProcaR}). As we will find, the Chebyschev polynomials provide a particular convenient set of basis to perform these numeric computations. This method yields an efficient, accurate, and robust method for computing the spectra at arbitrary $\alpha$ and $\tilde{a}$. To attack the magnetic spectrum, we cannot rely on a separable ansatz. Fortunately, the techniques of the previous section are flexible enough not to rely on one. In \S\ref{sec:nosep}, we describe how to formulate the non-separated equations of motion for both the scalar and vector fields as nonlinear eigenvalue problems. This allows us to accurately and robustly determine the entire quasi-bound state spectrum for a Proca field around a Kerr black hole. 

   \subsection{Continued Fraction Method} \label{sec:contFrac}
    
    In \S\ref{sec:separable}, we expanded the field $\Phi$ in spheroidal harmonics and obtained the radial differential equation (\ref{eqn: scalar radial equation}). For the numerical analysis, it will be convenient to write the radial function as
      \begin{equation}
        R(r) = \frac{(r - r_+)^{i P_+}}{(r - r_-)^{i P_-}} B(x)\, , \label{eq:contFracAnsatz}
      \end{equation}
      where $x = 2\sqrt{1-\omega^2}(r -r_+)$, as in \S\ref{sec:MAE} and (\ref{eqn: x variable}). For quasi-bound state solutions, $B(x)$ approaches a constant at the horizon, $x \to 0$, and decays exponentially $B(x) \sim e^{-x/2}$ at spatial infinity, $x \to \infty$. Any function with these properties can be represented by a linear combination of associated Laguerre polynomials $L_{k}^{(\rho)}(x)$ multiplied by $e^{-x/2}$,
      \begin{equation}
        B(x) = \sum_{k = 0}^{\infty} b_k \,e^{-x/2} L_{k}^{(\rho)}(x)\,. \label{eq:lagExp}
      \end{equation}
      In general, the expansion in (\ref{eq:lagExp}) transforms the scalar radial equation (\ref{eqn: scalar radial equation}) into a five-term recursion relation for the coefficients $b_k$. However, by choosing $\rho = 2 i P_+$, 
      one obtains the three-term recursion relation
      \begin{equation} 
        \alpha_{k} b_{k+1} + \beta_{k} b_{k} + \gamma_{k} b_{k - 1} = 0 \,,\label{eq:recurrence}
      \end{equation}
      where, in the notation of \S\ref{sec:separable}, we have defined
      \beq
    \begin{aligned}
      \alpha_{k} &\equiv (k+1) (k - c_1 + c_2 + 2 + 2 i P_+ - 4 \gamma ) \, , \\
      \beta_{k} &\equiv -2 k^2 + \left(c_1 - 2 (1 + c_2 + 2 i P_+)\right)k - c_2(1 + 2 i P_+) - c_3\, , \\
      \gamma_{k} &\equiv (k+2 i P_+)(k+ c_2 - 1)\, ,
    \end{aligned}
    \eeq
     with 
      \beq
    \begin{aligned}
    \hspace{0.7cm}  c_1 &\equiv 2(1 + i(P_+ - P_-) - 2 \gamma)\, ,\\
      c_2 &\equiv 1 + i(P_+ - P_-) + \frac{1}{4 \gamma}(\gamma_+^2 - \gamma_-^2) \, , \\
      c_3 &\equiv \left(P_+ - P_-\right)^2 - i(P_+ - P_-) + 2 \gamma(1+2 i P_+) + \gamma^2 + \gamma_+^2 + \Lambda\, .
    \end{aligned}
    \eeq
   For any initial data $b_{0}$ and $b_{1}$,\footnote{By convention, we define $b_{-1} \equiv 0$.} we can iteratively solve (\ref{eq:recurrence}) for  $b_{k}$ and thereby find a solution to the radial equation (\ref{eqn: scalar radial equation}).\footnote{An alternative recursion relation was derived in \cite{Dolan:2007mj} by performing a different mode expansion. Although the detailed forms of these coefficient functions change if we use this expansion, our discussion does not.} However, for generic values of $\omega$, solutions to (\ref{eq:recurrence}) diverge as $k \to \infty$, so that $B(x)$ will grow---rather than decay---exponentially at spatial infinity. This is the discrete analog of the fact that the boundary conditions (\ref{eqn:bc}) can only be satisfied simultaneously at special values of $\omega$ and that, for generic $\omega$, the solutions that satisfy the ingoing boundary condition at $r=r_+$ diverge as $r \to \infty$.

    It is only for special values of $\omega$---the quasi-bound state frequencies---that the recurrence relation (\ref{eq:recurrence}) admits a \emph{minimal solution}, which is finite both as $k \to 0$ and $k \to \infty$ \cite{Gautschi:1967cat}. Denoting the two linearly-independent solutions of (\ref{eq:recurrence}) as $f_k$ and $g_k$, the solution $f_k$ is  minimal  if
    \begin{equation}
      \lim_{k \to \infty} \frac{f_k}{g_k} = 0\, .
    \end{equation}
    A theorem by Pincherle \cite{Gautschi:1967cat} states that the recurrence relation (\ref{eq:recurrence}) admits a minimal solution if and only if $\omega$ solves the continued fraction equation 
    \begin{equation}
      \frac{\beta_0}{\alpha_0} = \cfrac{\gamma_1}{\beta_1 - \cfrac{\alpha_1 \gamma_{2}}{\beta_{2} - \cfrac{\alpha_{2} \gamma_{3}}{\beta_{3} - \cdots}}}\, . \label{eq:contFrac}
    \end{equation}
Determining the quasi-bound state spectrum is therefore reduced to the much simpler problem of finding the roots of a transcendental equation. This method has been used to find the quasi-normal modes of massless perturbations about Kerr black holes~\cite{Leaver:1985ax} and the quasi-normal modes and quasi-bound state spectra of massive fields~\cite{Dolan:2007mj, Rosa:2011my, Pani:2012bp, Dolan:2015eua}.

\begin{figure}[t]
      \begin{center}
        \includegraphics[scale=0.91, trim = 0 0 0 0]{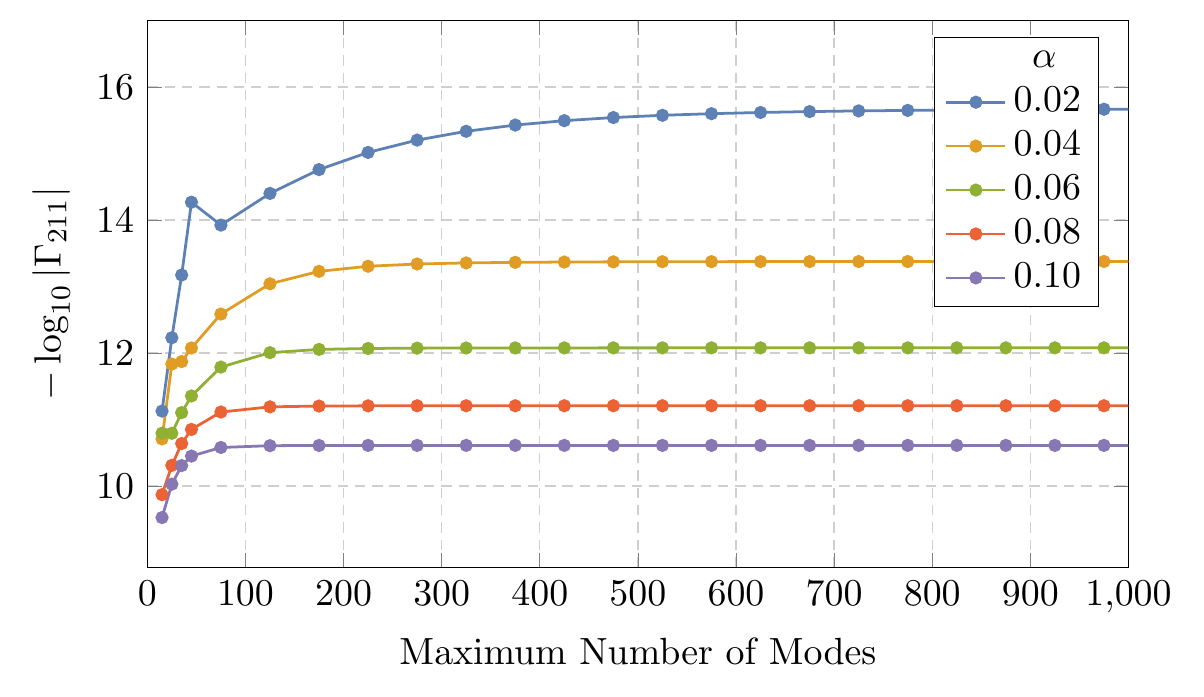}
        \caption{The imaginary part of $\omega$ slowly converges for small $\alpha$, requiring about $1000$ modes (terms in the continued fraction) to achieve accurate results for $\alpha \sim 0.02$. The results shown in the figure are for the $\ell = 1$, $m = 1$ mode, with $\tilde{a} = 0.5$, but similar conclusions apply to other states. Identical results apply to the recursion relation used in \cite{Dolan:2007mj}. \label{fig:contFracResults} }
      \end{center}
    \end{figure}

    As will become apparent, sensitivity to numerical errors is a common challenge in finding the 
     spectrum for a massive field about Kerr, and it is important to use methods that tamp down numerical errors as much as possible. For instance, the fact that the coefficients $b_k$ diverge exponentially quickly when $\omega$ is not exactly a quasi-bound state frequency means that, unless properly accounted for, numerical errors will grow at the same rate. Solving for $\omega$ typically involves starting with an initial seed and stepping towards the solution through a sequence of intermediate frequencies, and large errors introduced during these intermediate steps can impact the accuracy of the `solution.'
    As seen in Fig.\,\ref{fig:contFracResults}, one needs to include several hundred of the coefficients $b_k$ to accurately compute the decay rates $\Gamma_{n\ell m}$, so this exponential growth in error---which scales with the number of coefficients---can be an enormous problem. The major advantage of phrasing this as the continued fraction (\ref{eq:contFrac}) is that there exist numerically robust methods---for instance, the modified Lentz method \cite{Press:2007nr}---for efficiently and accurately evaluating the continued fraction to a specified precision, which can then be passed to standard root-finding methods (i.e.~Newton-Raphson) to find the eigenfrequencies.

  The main issue with the continued fraction method is its rigidity---one must first separate the partial differential equation and then find a basis of functions, like $e^{-x/2} L_{k}^{(2 i P_+)}$, that reduces the scalar radial equation (\ref{eqn: scalar radial equation}) to a three-term recurrence relation.\footnote{One can sometimes convert a five-term recurrence relation into a three-term relation using Gaussian elimination~\cite{Cardoso:2005vk}. However, it will be more convenient to bypass (\ref{eq:contFrac}) entirely. } This is problematic for two reasons. The first reason is simply that it is not always possible to find such a basis---for instance, it is not clear that there exists a separable ansatz for the magnetic modes of a vector field, or that there exists a basis of functions that reduces (\ref{eqn:ProcaR}) to a relation like (\ref{eq:recurrence}).  The second reason is more subtle. 
    We have seen, in Fig.\,\ref{fig:contFracResults}, that several hundred modes are needed to achieve accurate results for the growth rate of the main superradiant mode $|\es2\es1\es1\rangle$. But the radial profile of this mode does not oscillate wildly and is instead quite smooth. So, why are all of these modes needed? The problem is that, as we discussed in \S\ref{sec:separable}, the solution in the near region varies on extremely short scales, $\Delta x \sim \alpha^2$, while the basis functions $e^{-x/2} L_{k}^{(2 i P_+)}(x)$ naturally vary on much larger scales, $\Delta x \sim 1$. A large number of modes are therefore needed to approximate this (relatively) rapid behavior in the near region, and the number of required modes increases the smaller this region gets.  As we will discuss shortly, the numerical error often scales with the size of a problem, and thus it can be computationally difficult to access reliable results at small $\alpha$ with these methods.

\subsection{Nonlinear Eigenvalue Problem}  
\label{sec:NLEP}
    
The rigidity of the continued fraction method presents a serious obstacle to efficiently and accurately computing the quasi-bound state spectrum of scalar and vector fields. However, we can make progress by recognizing that the recurrence relation (\ref{eq:recurrence}) is analogous to the infinite-dimensional matrix equation
        \begin{equation}
          \mathcal{M}(\omega)\, \mb{b} = 
          \begin{pmatrix} 
            \beta_0 & \alpha_0 & 0 & 0 & 0 &\dots\\
            \gamma_1 & \beta_1 & \alpha_1 & 0 & 0 &\dots\\
            0 & \gamma_2 & \beta_2 & \alpha_2 & 0 &\dots\\
            0 & 0 & \gamma_3 & \beta_3 & \alpha_3 &\dots\\
            \vdots & \vdots & \vdots & \vdots & \vdots & \ddots
          \end{pmatrix}
          \begin{pmatrix}
            \,b_0 \,\\
            b_1 \\
            b_2 \\
            b_3 \\
            \vdots
          \end{pmatrix} = 0\,, \label{eq:nlev}
        \end{equation}
        whose elements are nonlinear functions of $\omega$. When we decompose $B(x)$ onto a basis of functions that individually satisfy the boundary conditions (\ref{eqn:bc}) and then truncate to a finite set of size $N+1$, this becomes a \emph{nonlinear eigenvalue problem}, i.e. we must find pairs $\omega$ and $\mb{b}$ such that (\ref{eq:nlev}) is satisfied~\cite{Dias:2015nua}. For a recent review of the numerical techniques developed to attack these problems, see \cite{Guttel:2017cup}.

       There are two sources of error in using (\ref{eq:nlev}) to solve for $\omega$. The first is truncation error---in order to actually solve this matrix equation on a computer, we must truncate the representation of $B(x)$ to a finite number of coefficients. We similarly needed to truncate the number of `levels' in the continued fraction (\ref{eq:contFrac}) to actually evaluate it. This truncation introduces an error, as $B(x)$ is now \emph{approximated} by a smaller collection of functions. As we increase the size of this collection, we are able to more faithfully represent $B(x)$ and, as seen in Fig.\,\ref{fig:contFracResults}, this truncation error will decrease. We can then estimate the accuracy of our solution by how sensitive it is to changes in~$N$.
      The second source of error is numerical error. Floating point arithmetic is inherently noisy, as there are round off errors incurred after every operation (e.g.~addition and multiplication), and this noise can be amplified by careless numerics. Unfortunately, this error grows with the number of operations performed, and thus with the number of coefficients we include in (\ref{eq:nlev}). This is potentially disastrous, since, if we try suppress truncation error by increasing $N$, we could be overrun by this numerical error. We will specifically choose a representation of (\ref{eq:nlev}) to help sooth these numeric problems. In the following, we will motivate and explain these choices, and discuss them in more detail in Appendix~\ref{app:cheb}.

        Regardless of representation, numerical errors can also creep in when we try to numerically solve (\ref{eq:nlev}), and it is important to use methods that avoid such instabilities. For instance, perhaps the simplest way to determine the quasi-bound state spectrum is to find the roots of the equation $\lab{det}\,  \mathcal{M}(\omega) = 0$. Unfortunately, this is not feasible for large $N$. In order for the determinant to vanish, there must be sensitive cancellations between a large number of operations, and accumulated roundoff errors can totally destroy the accuracy of our solution. This is a common problem for any nearly-singular matrix---a rule of thumb is that relative errors are amplified by the so-called condition number, the ratio of the matrix's largest and smallest eigenvalues, and this condition number diverges for exactly the frequencies we are solving for. For small $\alpha$ and large matrices, this numerical error easily dominates the instability rates and energy splittings.

        We will instead use nonlinear inverse iteration \cite{Guttel:2017cup,Dias:2015nua}, a form of Newton's method applied directly to $\mathcal{M}(\omega)\mb{b} = 0$ that iteratively solves for both $\omega$ and $\mb{b}$. This method circumvents the numerical instability caused by these nearly-singular matrices because this error is amplified much more along the singular direction---i.e. the one we are interested in---than any other. That is, these errors only change the length of the solution $\mb{b}$ and not its direction, and so their effect is nullified. In practice, this method converges both quickly and accurately as long as one has a good initial guess for the pair $(\omega, \mb{b})$.

      \subsection{Chebyshev Interpolation}
\label{sec:Separable}

We will now apply the algorithm sketched in \S\ref{sec:NLEP} to the scalar and electric vector fields in their separable forms.
In the next subsection, we will see how this method can be easily extended to the nonseparable ans\"atze, thus including the vector magnetic modes in our analysis. 

\subsubsection*{Scalar}

As shown in (\ref{eqn:ScalarAnsatz}), the scalar modes are separable into radial and angular functions. It is convenient to write the radial function as
        \begin{equation}
        R(r) = \left(\frac{r - r_+}{r - r_-}\right)^{i P_+} \!\!\!(r - r_-)^{-1+ \nu - 2 \alpha^2/\nu} e^{-\alpha(r-r_+)/\nu} B(\zeta)\,, \label{eq:pseudoSpecPeelOff}
        \end{equation}
      where the asymptotic behavior shown in  (\ref{eqn:scalar radial outer horizon}) and (\ref{eqn: scalar radial infinity}) has been extracted explicitly. The remaining function $B(\zeta)$ is defined on the finite interval $\zeta \in [-1, 1]$ via a map $\zeta(r)$. Different choices of $\zeta(r)$ will be discussed below. We will work with $\nu$, defined in (\ref{eqn:nudef}), instead of $\omega$. The function $B(\zeta)$ satisfies a linear differential equation of the form
      \begin{equation}
        \mathcal{D}_{\nu}[B(\zeta)] \equiv \left(\frac{\partial^2}{\partial \zeta^2} + \mathcal{C}_1(\nu, \zeta) \frac{\partial}{\partial \zeta} + \mathcal{C}_2(\nu, \zeta) \right) B(\zeta) = 0\,, \label{eq:zetaEq}
      \end{equation}
     where the precise form of the functions\footnote{These functions also depend on $\alpha$, $\tilde{a}$, and the orbital and azimuthal quantum numbers
     $\ell$ and $m$, respectively. We will suppress this dependence, as it does not play a crucial role in our story.} $\mathcal{C}_i(\nu, \zeta)$ depend on our choice of $\zeta(r)$. 
 The function $B(\zeta)$ will satisfy the correct boundary conditions if it approaches a constant 
 at both the outer horizon ($\zeta = -1$) and spatial infinity ($\zeta = 1$). Because~(\ref{eqn: scalar radial equation}) has no singularities between the outer horizon and spatial infinity, $B(\zeta)$ is necessarily a smooth function for all $\zeta \in [-1, 1]$, and we may represent it in a variety of ways. For computational simplicity, we will use Chebyshev polynomials of the first kind---defined by $T_{n}(\cos t) = \cos n t$ and described at length in Appendix~\ref{app:cheb}---but our conclusions will largely be independent of this choice.

    Above, we derived the recurrence relation (\ref{eq:recurrence}) by projecting both $B(x)$ and the radial equation~(\ref{eqn: scalar radial equation}) onto Laguerre functions.  We could now mimic this procedure 
    by first constructing a \emph{polynomial projection} of $B(\zeta)$ in terms of the Chebyshev polynomials,
    \begin{equation}
      B_N(\zeta) = \sum_{k = 0}^{N} b_k T_k(\zeta)\,, \label{eq:bWrong}
    \end{equation}
    which is a degree-$N$ polynomial that 
     is guaranteed to converge $\lim_{N \to \infty} B_{N}(\zeta) = B(\zeta)$, since $B$ is smooth on $\zeta \in [-1, 1]$ and the Chebyshev polynomials form a complete set. 
  Projecting the radial equation (\ref{eq:zetaEq}) onto the Chebyshev polynomials, we would obtain  the matrix equation $\sum_k \mathcal{M}_{n k}(\nu) b_k = 0$,
where\hskip 1pt\footnote{This formula must be multiplied by a factor of $1/2$ when $n = 0$.}
    \begin{equation}
      \mathcal{M}_{n k}(\nu) \equiv \frac{2}{\pi} \int_{-1}^{1}\!\ud \zeta\, \frac{T_{n}(\zeta)\, \mathcal{D}_\nu[T_{k}(\zeta)]}{\sqrt{1- \zeta^2}}\,. \label{eq:specMatrix}
    \end{equation}
    The matrix given by \eqref{eq:specMatrix} is similar to that in (\ref{eq:nlev}), though it has many more non-zero elements. We could again pass this to a solver to determine the quasi-bound state spectrum, but unfortunately this formulation of the problem proves (cf.~Appendix~\ref{app:cheb}) to be numerically unstable, it is difficult to reduce truncation errors without numerical errors biting back.

   To circumvent this problem, we will represent $B(\zeta)$ not by its Chebyshev coefficients, but by its values at the \emph{Chebyshev nodes}
    \begin{equation}
      \qquad\qquad\qquad \zeta_n = \cos\left(\frac{\pi(2 n + 1)}{2 N + 2}\right) , \quad \text{where} \quad n = 0, 1, \dots, N\,. \label{eq:chebPoints}
    \end{equation}
Introducing the associated \emph{cardinal polynomials}  $p_k(\zeta)$---degree-$N$ polynomials that are defined by $p_k(\zeta_n) = \delta_{nk}$ (cf.~Appendix~\ref{app:cheb})---we may rewrite (\ref{eq:bWrong}) as 
    \begin{equation}
      B_N(\zeta) = \sum_{k = 0}^{N} B(\zeta_k) \,p_k(\zeta)\,. \label{eq:bRight}
    \end{equation}
Note that this is nothing more than a reorganization of (\ref{eq:bWrong}) in a different basis of degree $N$ polynomials such that the values of $B(\zeta)$ appear explicitly. Furthermore, because the operator $\mathcal{D}_\nu[B(\zeta)]$ appearing in (\ref{eq:zetaEq}) also defines a smooth function on $\zeta \in [-1, 1]$, we can also represent it by its values at these specified points. This means that we can approximate (\ref{eq:zetaEq}) by the matrix equation 
    \beq
    \sum_{k = 0}^{N} \mathcal{M}_{n k}(\nu) B(\zeta_k) = 0\,,
    \eeq
     where
    \begin{equation}
       \mathcal{M}_{nk}(\nu) \equiv p_k^{\hskip 1pt \prime \prime}(\zeta_n) + \mathcal{C}_1(\nu, \zeta_n)\,p_k^{\hskip 1pt \prime}(\zeta_n) + \mathcal{C}_{2}(\nu, \zeta_n) \,\delta_{nk}\,.\label{eq:scalarMatrix} 
    \end{equation}
    In practice, this formulation proves to be both simpler and more numerically robust than (\ref{eq:specMatrix}), as long as $p_k^{\hskip 1pt \prime \prime}(\zeta_n)$ and $p_k^{\hskip 1pt \prime}(\zeta_n)$ are computed using (\ref{eq:cardDeriv}) and (\ref{eq:cardDDeriv}), respectively. See Appendix~\ref{app:cheb} for more details.

    The specific form of the mapping $\zeta(r)$ can dramatically affect the truncation error. Just like the convergence of a Laurent series about a point is set by the largest \emph{circular} domain of analyticity, the convergence of the interpolation (\ref{eq:bRight}) is set by the largest \emph{ellipsoidal} domain of analyticity about the interval $\zeta \in [-1, 1]$. Specifically, it can be shown (cf.~Appendix~\ref{app:cheb}) that the largest disagreement between the function and its Chebyshev interpolation anywhere on the interval scales as
    \begin{equation}
      \|B- B_N\| \sim \mathcal{O}\big(\rho^{-N}\big)\,. \label{equ:conv}
    \end{equation}
The parameter $\rho$ measures the size\footnote{Concretely, $\rho$ is the sum of the semi-major and semi-minor axes of the ellipse with foci at $\zeta = \pm 1$. The parametric form of this ellipse is given by (\ref{eq:bernsteinEllipse}) and it is depicted in Fig.~\ref{fig:bernstein}.} of the largest ellipse with foci at $\zeta = \pm 1$ inside which $B(\zeta)$ is analytic. The interpolation thus converges more slowly the closer a singularity of $B(\zeta)$ is to the interval $\zeta\in[-1,1]$, as measured by these ellipses.

     The radial equation (\ref{eqn: scalar radial equation}) has an additional singularity at $r = r_-$, which 
     maps to a singularity in $B(\zeta)$ at $\zeta_- = \zeta(r_-)$. This is why our choice of $\zeta(r)$ can affect the truncation error. If this mapping does not place $\zeta_-$ far from the interval $\zeta \in [-1, 1]$, this singularity can drastically increase the number of modes $N$ needed to accurately approximate the solution. This is also why we avoided introducing factors of $r$ in the ansatz (\ref{eq:pseudoSpecPeelOff}), as they could introduce inessential singularities into $B(\zeta)$ and potentially affect convergence.

     Our choice of mapping $\zeta(r)$ will also determine how the interpolation points (\ref{eq:chebPoints}) sample the radial domain $[r_+, \infty)$. As we have argued in \S\ref{sec:contFrac}, one of the reasons the Laguerre basis performed so poorly was that it naturally varies on scales much larger than the width of the near region and so it has trouble approximating the behavior there. This is easier to see if we note that this Laguerre basis samples the function $B(x)$ at the zeros of $L_{N+1}^{(2 i P_+)}(x)$ \cite{NIST:DLMF}, whose smallest root scales as $\mathcal{O}(N^{-1})$. We would thus need to include $\mathcal{O}(\alpha^{- 2})$ modes to accurately sample the boundary layer $x \sim \mathcal{O}(\alpha^2)$, in good agreement with the numerical experiments shown in Fig.\,\ref{fig:contFracResults}. One of the benefits of the Chebyshev basis is that the interpolation points (\ref{eq:chebPoints}) have a much higher density $\Delta \zeta \sim \mathcal{O}(N^{-2})$ near the boundary than in the interior $\Delta \zeta \sim \mathcal{O}(N^{-1})$, which makes it easier for the basis to resolve phenomena in the near region. Still, we must be careful that the mapping we choose does not obstruct this useful behavior.

    The simplest choice that maps the triplet $r =(r_-, r_+, \infty)$ to $\zeta = (-\infty, -1, 1)$ is 
    \begin{equation}
      \zeta_1(r) = \frac{r - 2 r_+ + r_-}{r - r_-} \, . \label{equ:z1}
    \end{equation} 
A drawback of this mapping is that the Bohr peak at $r_\lab{c} \sim \nu/\alpha$ is mapped to $\zeta_1(r_\lab{c}) \sim 1 - \mathcal{O}(\alpha^2)$,
    while the middle of the interval $\zeta = 0$ corresponds to $(2 r_+ - r_-) = \mathcal{O}(\alpha)$. This creates the opposite problem as with the expansion into Laguerre polynomials---we are now sampling the near-horizon region very well, but at the expense of the far region. To ensure the far region is also well-sampled, we again require a relatively large number of modes $N \sim \alpha^{-1}$, although the properties of the Chebyshev nodes help alleviate this problem somewhat compared to the expansion (\ref{eq:lagExp}). In practice, this map works fairly well for the $n = \ell+1$ quasi-bound states, as they have very little structure in the far region.

    An alternative mapping that avoids this problem is
    \begin{equation}
      \zeta_2(r) = \frac{r - \sqrt{4 r_+(r-r_-) + r_-^2}}{r- r_-}\,. 
      \label{equ:z2}
    \end{equation}
This maps the Bohr radius to $\zeta_2(r_\lab{c}) \sim 1 - \mathcal{O}(\alpha)$, which means that we only need $N \sim \mathcal{O}(\alpha^{-1/2})$ modes to resolve the far region.  While the singularity at $r_-$ is only mapped to the finite point ${\zeta_- = -(2 r_+ - r_-)/r_-}$, it is still displaced far enough from the interval that it does not dramatically affect convergence. For instance, $\rho > 2$ in (\ref{equ:conv}) as long as $\tilde{a} \lesssim 0.96$ and so $\zeta_2$ is an effective map except in the extremal limit. In principle, there are better maps that send $r_-$ infinitely far from the interval and more equally distribute the interpolation points (\ref{eq:chebPoints}) across the near, intermediate, and far regions. However, the map $\zeta_2$ works well enough in practice that we will not pursue others.

\begin{figure}[t]
      \begin{center}
        \includegraphics[scale=0.81, trim=10 0 0 0]{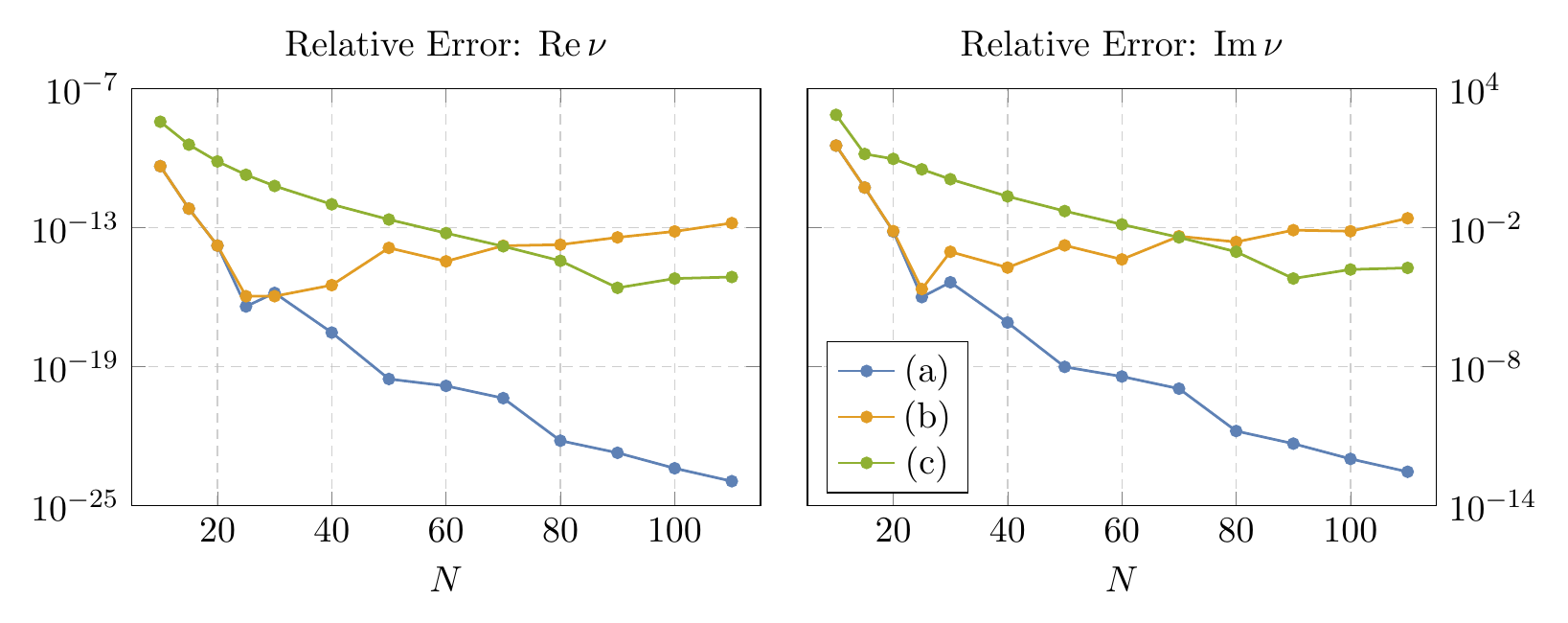}
        \caption{Comparison between the two mappings $\zeta_1(r)$ and $\zeta_2(r)$ defined in (\ref{equ:z1}) and (\ref{equ:z2}), as a function of the number of modes $N$. The displayed lines are:
    {(a)}\,{\color{Mathematica1}[blue]} $\zeta_2$ with 60 digits of precision, {(b)}\,{\color{Mathematica2}[orange]} $\zeta_2$ with machine precision, {(c)}\,{\color{Mathematica3}[green]} $\zeta_1$ with either 60 digits of precision or with machine precision. The data shown in the figures is for $\alpha = 0.01$, $\tilde{a} = 0.5$, and $|n \es \ell \es m \rangle = |2 \es 1 \es 1\rangle$. All relative errors are measured with respect to the high precision $\zeta_2$ result with $N = 120$.
\label{fig:errorsScalar}}
      \end{center}
    \end{figure}

A comparison between the two maps $\zeta_1$ and $\zeta_2$ is shown in Fig.\,\ref{fig:errorsScalar}. We see that we can achieve very accurate results for the parameter $\nu$ using only $N \sim 30$ modes.  For comparison, reaching  a similar level of accuracy using the continued fraction method requires $N \sim 10^4$ modes. Clearly, the map $\zeta_2$ converges much more quickly than $\zeta_1$, and becomes limited by the resolution $\epsilon \sim 10^{-16}$ of machine precision numbers (the `machine epsilon') very quickly.

This method was used to determine the spectra of the $|\es2\es 1 \es m\rangle$ and $|\es 3 \es 2 \es m \rangle$ modes in Figures~\ref{fig:ScalarSpectrumPlots}~and~\ref{fig:growthRates}. In particular, the extremely small decay rates $\Gamma/\mu \sim 10^{-24}$ of the $|\es 3 \es 2 \es m\rangle$ modes are accurately computed using only $N= 60$. In contrast, the continued fraction method requires $N \sim 2\times 10^{4}$ modes to achieve comparable precision! With the scalar now solved, we can turn our attention to the vector electric modes, whose equations of motion are separable.

    \subsubsection*{Vector electric modes}
    
     Clearly, the methods described above also apply to the separated vector equation~(\ref{eqn:ProcaR}), albeit with one small wrinkle. Our discussion there relied on our ability to accurately compute the eigenvalues of the scalar angular equation (\ref{eqn: spheroidal harmonic equation}), and we skipped past this complication because various software packages already natively compute these quantities.\footnote{For instance, the eigenvalues of the spheroidal harmonics and their derivatives can be computed to arbitrary precision using Mathematica's \texttt{SpheroidalEigenvalue}.} 
     This is not the case for the Proca angular equation (\ref{equ:ProcaS}). 

     \begin{figure}[t]
        \begin{center}
          \includegraphics[scale=0.85, trim=10 0 0 0]{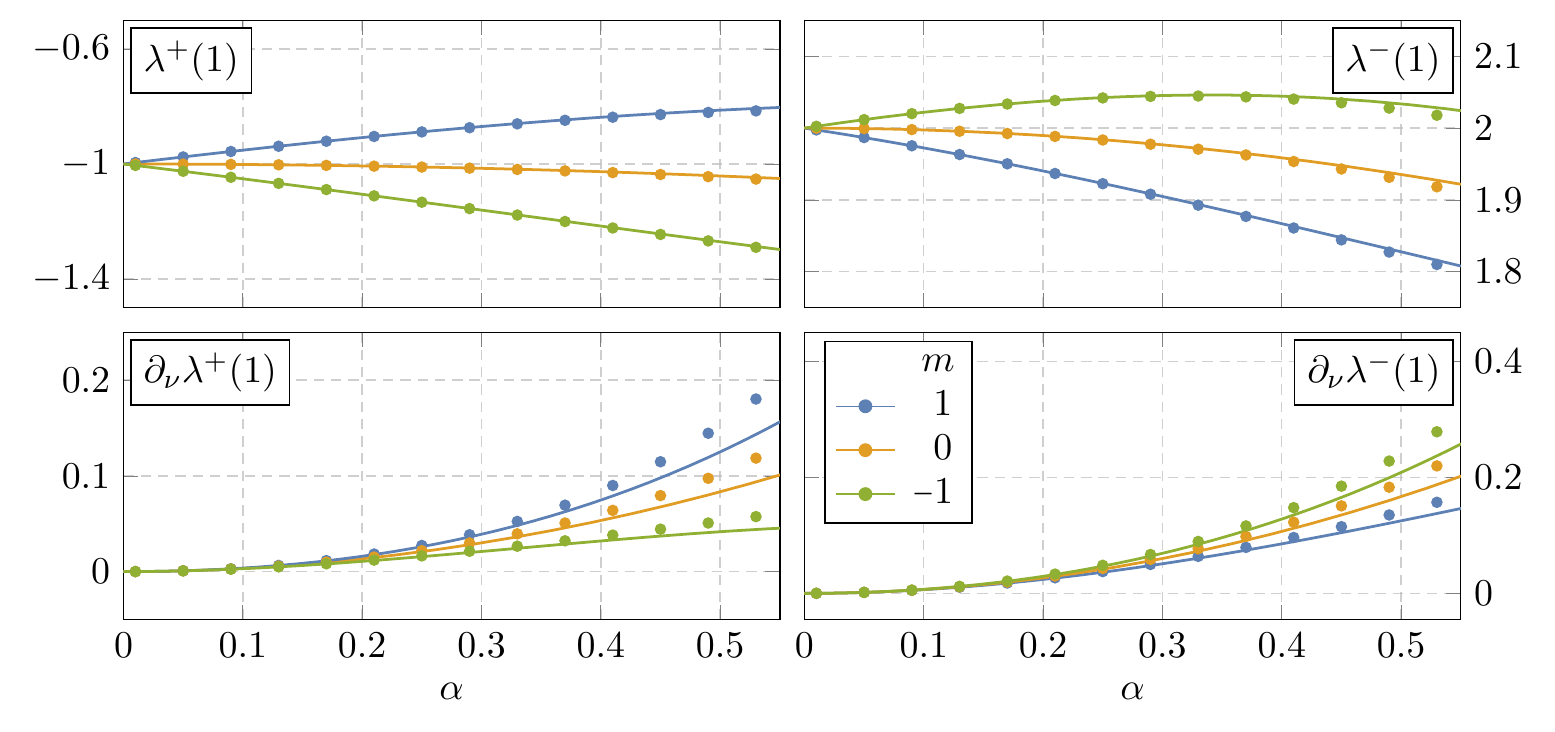}
          \caption{Comparison between the numeric (dots) and perturbative (solid) results for the angular eigenvalues $\lambda^{\pm}$ (top row) and their derivatives with respect to $\nu$ (bottom row) for $\tilde{a} = 0.5$, $j=1$, and $\nu =1$. The perturbative expressions (\ref{eqn:lambdaPlusMinus}) only compute the derivatives $\partial_\nu \lambda^{\pm}$ to next-to-leading order and thus deviate from the numeric results significantly at large $\alpha$. \label{fig:vectorAngular} }
        \end{center}
      \end{figure}

     Fortunately, (\ref{equ:ProcaS}) is also a nonlinear eigenvalue problem which we can solve using the techniques of the previous section. Defining $\zeta = \cos \theta$ and expanding the angular function in the Chebyshev cardinal polynomials,
     \begin{equation}
        S(\zeta) = \sum_{k = 0}^{N} S(\zeta_k)\, p_k(\zeta)\,,
     \end{equation}
     we convert the angular equation (\ref{equ:ProcaS}) into another finite-dimensional matrix equation
     \begin{equation}
        \sum_{k = 0}^{N} \mathcal{A}_{n \es k}(\lambda, \nu) S(\zeta_k) \equiv \mathcal{A}(\lambda, \nu) \,\mb{S} = 0\,. \label{eq:angMatSol}
     \end{equation}
   For a given $\nu$, we can then pass the matrix $\mathcal{A}_{n \es k}(\lambda, \nu)$ to a solver to find~$\lambda$ as a function of~$\nu$. Furthermore, since nonlinear inverse iteration will require the derivative $\partial_\nu \mathcal{M}_{n\es k}(\nu)$ of the radial matrix, we must also find the derivative of the angular eigenvalues $\lambda'(\nu)$. By differentiating~$\lab{det}\, \mathcal{A}_{n\es k}(\lambda(\nu), \nu) = 0$ with respect to $\nu$, we can rewrite this derivative in terms of derivatives of the matrix $\mathcal{A}$,
     \begin{equation}
        \lambda'(\nu) = -\frac{\lab{tr}\left(\mathcal{A}^{-1}\, \partial_\nu \mathcal{A}\right)}{\lab{tr}\left(\mathcal{A}^{-1}\partial_\lambda \mathcal{A}\right)} = -\frac{\mb{S}_\lab{L}^\top \!(\partial_\nu \mathcal{A})\mb{S}^{\phantom{\top}}_\lab{R}}{\mb{S}_\lab{L}^\top \!(\partial_\lambda \mathcal{A})\mb{S}^{\phantom{\top}}_\lab{R}}\, ,
      \end{equation}
     where the second equality follows since the trace is dominated by the zero left and right eigenvectors, $\mb{S}_\lab{L}$ and $\mb{S}_\lab{R}$, when evaluated at the eigenvalue $\lambda(\nu)$. We can thus calculate $\lambda'(\nu)$ for very little additional computational cost. In Fig.\,\ref{fig:vectorAngular}, we compare our numeric results with their perturbative approximations (\ref{eqn:lambdaPlusMinus}).

      With the angular eigenvalue in hand, the techniques discussed in \S\ref{sec:Separable} apply almost unchanged. We introduce a map $\zeta(r)$ from $r\in[r_+, \infty)$ to the finite interval $\zeta \in [-1, 1]$ and rewrite the radial function as
      \begin{equation}
        R(r) = \left(\frac{r - r_+}{r - r_-}\right)^{i P_+} \!\!\!\left(r - r_-\right)^{\nu - 2 \alpha^2/\nu} e^{-\alpha(r - r_+)/\nu} B(\zeta)\,,
      \end{equation} 
      so that $B(\zeta)$ approaches a constant at the endpoints of the interval $\zeta = \pm 1$. We need not worry about the poles at $r = \hat{r}_\pm$ in (\ref{eqn:ProcaR}), which will not affect convergence because $R(r) \sim C_\pm + D_\pm (r - \hat{r}_\pm)^2$ is analytic for $r \to \hat{r}_\pm$. Expanding in the Chebyshev cardinal polynomials $p_k(\zeta)$ and sampling the radial equation (\ref{eqn:ProcaR}) at the Chebyshev nodes, we find a finite-dimensional matrix $\mathcal{M}_{n \es k}(\nu)$ that can then be passed to a solver to determine the quasi-bound state spectrum.

\subsection{Solving Without Separability } 
\label{sec:nosep}
  
Having rephrased the problem as a nonlinear eigenvalue problem in the previous subsections affords us the flexibility to use a nonseparable ansatz. While this is mainly relevant for the Proca equation, for which a separable ansatz is not known in general, we will first illustrate the technique using the scalar field. It can then be applied to all of the vector modes with minimal complication.

\subsubsection*{Scalar}

    In the Schwarzschild limit, $\tilde{a} \to 0$, spherical symmetry is restored in the Kerr geometry. Moreover, since $\tilde{a}$ always appear dressed by factors of $\alpha$, spherical symmetry is only weakly broken when $\alpha \ll 1$. This mean that, for $\alpha \ll 1$, we can use the weakly broken symmetry to organize our ansatz for the scalar field $\Phi$ and decompose $\Phi$ into scalar spherical harmonics. The general idea is to first decompose the Klein-Gordon equation into operators that act naturally on the scalar spherical harmonics, and then to use the methods of the previous section to convert it into a finite-dimensional matrix equation.

     Our first step is to rewrite the Klein-Gordon equation using the isometries (\ref{eq:kerrIsometries}) and the total angular momentum operator $\pounds^2$, described in detail in Appendix \ref{app:harmonics}. One can show that the operator 
    \begin{equation}
    \begin{aligned}
      \bigg[  \Sigma \nabla^2 + \pounds^2  - & \left(\Sigma +  \frac{2 \alpha r(r^2 + \alpha^2 \tilde{a}^2)}{\Delta}\right)  \pounds_t^2  \\
      &  - \frac{\alpha^2 \tilde{a}^2}{\Delta} \pounds_z^2 - \frac{4 \alpha^2 \tilde{a} r}{\Delta} \pounds_t \pounds_z \bigg] \Phi   = \partial_r\left(\Delta \partial_r \Phi\right) \label{eq:purelyRadial}
      \end{aligned}
    \end{equation}
    is purely radial when acting on a scalar field $\Phi$. By assuming that $\Phi$ has definite frequency, $\pounds_t \Phi = - \omega \Phi$, and azimuthal angular momentum, $\pounds_z \Phi =  m \Phi$, the Klein-Gordon equation reduces to 
    \beq
    \begin{aligned}
        0 & \, =\ \,  \frac{1}{\Delta}\partial_r \left(\Delta \partial_r \Phi \right) - \frac{1}{\Delta}\left(\pounds^2 + \alpha^2 \tilde{a}^2 (1 - \omega^2) \cos^2 \theta\right) \Phi  + \bigg(-(1 - \omega^2) \\
        & + \frac{P_+^2}{(r - r_+)^2} + \frac{P_-^2}{(r - r_-)^2} + \frac{A_{-}}{(r_+ - r_-) (r - r_-)} - \frac{A_+}{(r_+ - r_-)(r - r_+)}\bigg)\Phi\, \label{eq:startingPoint} 
    \end{aligned}
    \eeq
    with minimal effort.
    We recognize the spheroidal harmonic equation (\ref{eqn: spheroidal harmonic equation}) 
    \begin{equation}
      \left(\pounds^2 + \alpha^2 \tilde{a}^2(1 - \omega^2) \cos^2 \theta\right) S = \Lambda S \,,
    \end{equation}
    and so (\ref{eq:startingPoint}) is of the same form as~(\ref{eqn: scalar radial equation}), yet without needing the separable ansatz~(\ref{eqn:ScalarAnsatz}). 
    Mimicking (\ref{eq:pseudoSpecPeelOff}), we then strip $\Phi$ of its asymptotic behavior and  decompose it into scalar spherical harmonics,
    \begin{equation}
      \Phi= e^{-i \omega t} \left(\frac{r - r_+}{r - r_-}\right)^{i P_+}\!\!\! (r - r_-)^{-1 + \nu - 2 \alpha^2/\nu} e^{-\alpha(r-r_+)/\nu} \sum_{\ell} B_\ell(\zeta) \,Y_{\ell m}(\theta, \phi)\,, \label{eq:scalarSHDecomp}
    \end{equation}
    where the sum ranges over even or odd values of $\ell$ for parity even or odd modes, respectively.

    With this ansatz, we can project the Klein-Gordon equation (\ref{eq:startingPoint}) onto the scalar spherical harmonics to find 
    \beq
    \begin{aligned}
      & \left(\frac{\partial^2}{\partial \zeta^2} + \mathcal{C}_1(\nu, \zeta) \frac{\partial}{\partial \zeta} + \mathcal{C}_2(\nu, \zeta) \right) \hskip 1pt B_{\ell}(\zeta) \\
      &   \hskip 40pt + \mathcal{C}_{\ell, \ell+2}(\nu, \zeta)\,B_{\ell+2}(\zeta) + \mathcal{C}_{\ell, \ell-2}(\nu, \zeta) B_{\ell-2}(\zeta) = 0\,,
      \label{eq:decompScalar}
           \end{aligned} 
    \eeq
    where the functions $\mathcal{C}_i$ again depend on our choice of $\zeta(r)$, but now have explicit $\ell$ dependence. Importantly, the $\cos^2 \theta$ term in (\ref{eq:startingPoint}) breaks spherical symmetry and thus couples different spherical harmonics to one another, through the function $\mathcal{C}_{\ell,\ell'}$. By expanding the radial functions $B_\ell(\zeta)$ in cardinal polynomials, as in (\ref{eq:bRight}), and sampling at the interpolation points (\ref{eq:chebPoints}), this system of equations can be rewritten as the matrix, 
    \begin{equation}
      \begin{aligned}
      \mathcal{M}_{n \ell; k \ell'}(\nu) = \ &\big(p_{k}^{\hskip 1pt \prime \prime}(\zeta_n) + \mathcal{C}_1(\nu, \zeta_n) \hskip 1pt p^{\hskip 1pt \prime}_k(\zeta_n) + \mathcal{C}_2(\nu, \zeta_n) \hskip 1pt  \delta_{nk}\big) \delta_{\ell \ell'} \\
      &+ \mathcal{C}_{\ell, \ell+2}(\nu, \zeta_n) \hskip 1pt  \delta_{nk} + \mathcal{C}_{\ell, \ell-2} (\nu, \zeta_n) \hskip 1pt  \delta_{nk}\,,
      \end{aligned} \label{eq:scalarNonSepMat}
    \end{equation}
    which can then be passed to a solver to find the bound state spectrum.

To fit (\ref{eq:scalarNonSepMat}) on a computer, it is necessary both to sample a finite number of radial points and to include only a finite number of angular modes. This angular truncation is an additional source of error that must be controlled. Fortunately, both the Klein-Gordon and Proca equations enjoy an approximate spherical symmetry that is restored in the Schwarzschild limit---this is why we expanded (\ref{eq:scalarSHDecomp}) in terms of scalar spherical harmonics, as opposed to another complete set of functions. In this basis, the $\mathcal{C}_{\ell,\ell\pm 2}$ mixings in (\ref{eq:scalarNonSepMat}) are proportional to $\alpha^2 \tilde{a}^2$. For the main superradiant state $|\es 2 \es 1 \es 1 \rangle$, the angular truncation error roughly scales as $(\alpha \tilde{a})^{2 L}$ if we include only $\ell = 1, 3, \dots, 2L + 1$ in the expansion (\ref{eq:scalarSHDecomp}). However, we must emphasize that this estimate of the truncation error is only accurate at small $\alpha$ or $\tilde{a}$. Of course, the scalar (or vector) spherical harmonics are a complete set and so may represent an arbitrary scalar (or vector) field configuration. As long as we include `enough' of these angular modes, we are guaranteed to faithfully represent any field configuration. In fact, one can show from their recurrence relation \cite{Abramowitz:1965} that the coefficients of the spherical harmonic decomposition of the spheroidal harmonics decay faster than exponentially, independent of $\alpha \tilde{a}$. This reflects the intuition that low-energy solutions to the Klein-Gordon equation are relatively smooth, and the same will be true for the Proca field. In practice, this method is as fast\footnote{Though the matrix (\ref{eq:scalarNonSepMat}) passed to the nonlinear eigenvalue solver is generally much larger than (\ref{eq:scalarMatrix}), this is generally balanced by no longer needing to compute the spheroidal harmonic eigenvalue and (more importantly) its derivative with respect to $\nu$.} as that using the separable ansatz in \S\ref{sec:Separable} and can be made just as accurate. Applying this algorithm reproduces the results presented in \S\ref{sec:Summary}.

  \subsubsection*{Vector}
  
  A similar strategy works for the Proca equation, although we will encounter
   additional technical challenges. As for the case of the scalar, we will first rewrite the equations of motion using operators that act simply on either the radial or angular directions. We will then decompose the temporal and spatial components of the vector field into scalar and vector spherical harmonics, and use this decomposition to convert the equations of motion into a matrix equation. In the following, we provide a mostly qualitative overview of our techniques, keeping the many technical details confined to Appendix~\ref{app:numericalDetails}.

  One complication, compared with the scalar case, is that the Proca equation (\ref{equ:Proca})~needs to be supplemented by the Lorenz condition, $\nabla^\mu A_\mu = 0$. We can do this either by first solving $\nabla^\mu A_\mu = 0$ for the temporal component 
  $A_t$ and then substituting it into the Proca equation to find an equation purely in terms of the spatial components $A_i$, or by including the Lorenz constraint as an additional `block' of the matrix equation. We will choose the latter, since it is more flexible and generally easier to implement.

  With this in mind, our first step is to rewrite both the Proca equation and Lorenz constraint using operators that act simply in either the radial or angular directions. The operator (\ref{eq:purelyRadial}) we used in the previous section is, unfortunately, only radial when acting on a scalar. It will thus be extremely convenient to expand the vector field along a carefully chosen tetrad\hskip 1pt\footnote{We will use tetrad indices $a, b, \dots$ to run from $0$ to $3$, and indices $i,j,\dots$ to run from $1$ to $3$. By convention, repeated indices are summed over.}  $A_\mu = A_a f\indices{^a_\mu}$, so that we may instead work with the four scalar fields $A_a$ instead of the four-vector $A_\mu$. As detailed in Appendix~\ref{app:numericalDetails}, we require that this tetrad is stationary, $\pounds_t f\indices{^a_\mu} = 0$.  We will take $f\indices{^0_\mu}\ud x^\mu \propto \ud t$ to be purely temporal with no angular dependence and the $f\indices{^i_\mu}$ to have definite total and azimuthal angular momentum---their explicit form is given in (\ref{eq:formFieldDefs}).

   We may then use this decomposition with  (\ref{eq:purelyRadial}) to rewrite the Proca equation as
  \beq
  \begin{aligned}
         & 0  =\  \bigg(\frac{1}{\Delta} \partial_r(\Delta \partial_r) - \frac{1}{\Delta} \left(\pounds^2 + \alpha^2 \tilde{a}^2 \left(1 - \omega^2\right) \cos^2 \theta \right)\!\bigg)A_b + \bigg(- (1 - \omega^2)   \\
        & + \frac{P_+^2}{(r - r_+)^2} + \frac{P_-^2}{(r -r_-)^2} - \frac{A_+}{(r - r_+)(r_+ - r_-)} + \frac{A_-}{(r -r_-)(r_+ - r_-)}\bigg) A_b \\[6pt]
        & + \mathcal{S}\indices{_b^a} A_a + \mathcal{Q}\indices{_b^a}\, \pounds_z A_a + \mathcal{R}\indices{_b^a} \,\partial_r A_a + \mathcal{P}\indices{_b^a} \,\mathcal{D}_+ A_a + \mathcal{Z}\indices{_b^a} \,\mathcal{D}_0 A_a + \mathcal{M}\indices{_b^a} \,\mathcal{D}_- A_a\,, \label{eq:procaEqDecompMain}
  \end{aligned} 
  \eeq
  where we have assumed that the vector field has definite frequency $\pounds_t A_\mu = -\omega A_\mu$ and azimuthal angular momentum $\pounds_z A_\mu = m A_\mu$. We recognize that (\ref{eq:procaEqDecompMain}) is simply four copies of the scalar equation (\ref{eq:startingPoint}), coupled together through the mixing matrices $\mathcal{S}$, $\mathcal{Q}$, $\mathcal{R}$, $\mathcal{P}$, $\mathcal{Z}$ and $\mathcal{M}$, whose precise form can be found in (\ref{eq:mixingMatrices}) and depend on our choice of tetrad. These mixing matrices encode the `vector-ness' of the scalars $A_a$, while $\mathcal{D}_{\pm}$ and $\mathcal{D}_0$---cf.~(\ref{eq:dpm}) and~(\ref{eq:d0})---are purely angular operators that act simply on the scalar spherical harmonics. The Lorenz constraint can be similarly decomposed,
  \begin{equation}
  0 = \mathcal{T}^0 A_0 + \mathcal{S}^i A_i + \mathcal{R}^i \,\partial_r A_i + \mathcal{P}^i \,\mathcal{D}_+ A_i + \mathcal{Z}^i\, \mathcal{D}_0 A_i + \mathcal{M}^i \,\mathcal{D}_- A_i\,, \label{eq:lorenzEqDecompMain}
  \end{equation}
  in terms of the mixing vectors defined in (\ref{eq:mixingVectors}).

  As we explain in Appendix~\ref{app:boundaryConditions}, we also choose the $f\indices{^a_\mu}$ so that the scalars $A_a$ have the same asymptotic behavior,  (\ref{eqn:scalar radial outer horizon}) and (\ref{eqn: scalar radial infinity}), as the scalar field $\Phi$. Then, by mimicking the scalar decomposition (\ref{eq:scalarSHDecomp}), we strip the $A_a$ of their asymptotic behavior and decompose  the temporal component $A_0$ into scalar spherical harmonics and the spatial components $A_i$ into one-form harmonics
  \begin{align}
    A_0 &= e^{-i \omega t} \left(\frac{r - r_+}{r - r_-}\right)^{i P_+}\!\! \nonumber \\ 
    & \hskip 40pt (r- r_-)^{-1 + \nu - 2 \alpha^2/\nu} e^{-\alpha(r - r_+)/\nu} \sum_{j} B_{0,j}(\zeta)\, Y_{j m}(\theta, \phi)\,, \label{eq:pseudoVecTemp} \\
    A_i &= e^{-i \omega t} \left(\frac{r - r_+}{r - r_-}\right)^{i P_+}\!\! \nonumber \\
    & \hskip 40pt (r -r_-)^{-1+\nu- 2\alpha^2/\nu} e^{-\alpha(r - r_+)/\nu} \sum_{\ell,j} B_{\ell j}(\zeta)\, Y_{i}^{\ell, j m}(\theta,\phi)\,. \label{eq:pseudoVecSpat}
  \end{align}
   In the scalar case, we found that only components with even (odd) parity could couple to one another. This is a reflection of the fact that the Kerr metric is invariant under parity transformations, and so only scalar harmonics with the same parity couple to one another. This also applies to the Proca field. As discussed in Appendix \ref{app:harmonics}, the parity of the scalar and vector spherical harmonics in (\ref{eq:pseudoVecTemp}) and (\ref{eq:pseudoVecSpat}) are $(-1)^j$ and $(-1)^{\ell+1}$, respectively. When we solve for the spectrum of a parity even or odd mode, we can restrict the expansions (\ref{eq:pseudoVecTemp}) and (\ref{eq:pseudoVecSpat}), halving the number of angular terms we must include to faithfully represent the Proca field.

   \begin{figure}
        \begin{center}
          \makebox[\textwidth][c]{\includegraphics[scale=0.8]{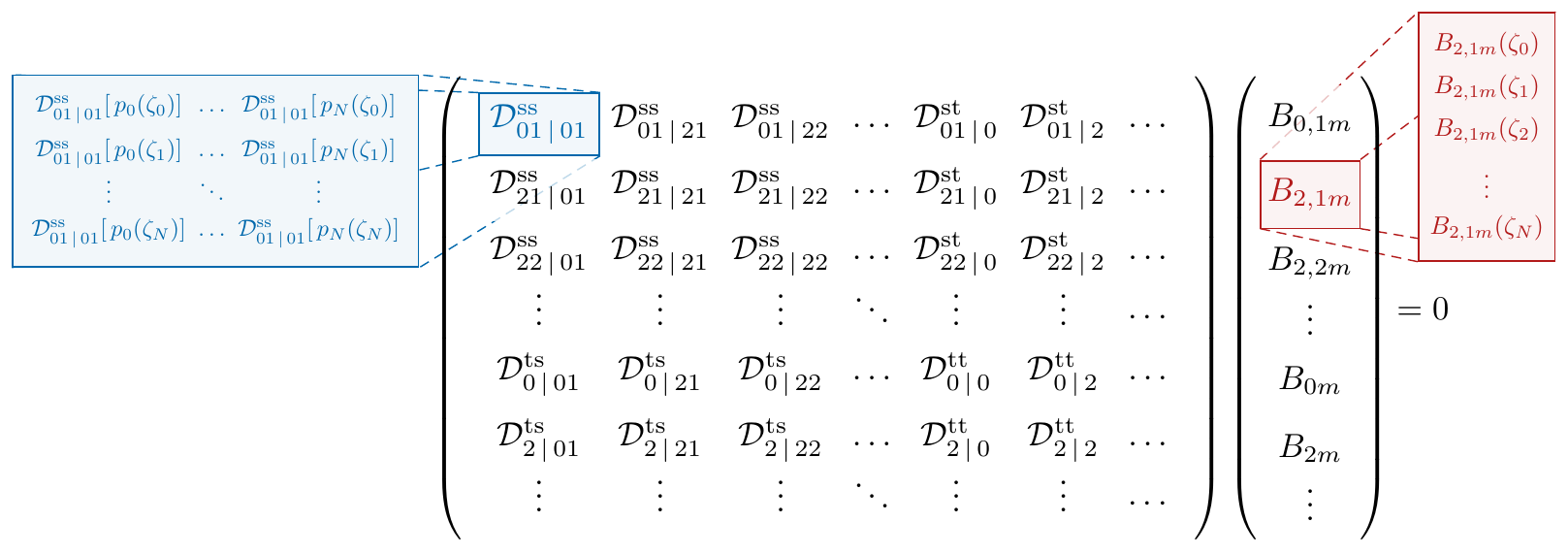}}
          \caption{Structure of the matrix equation for the parity even modes. Each angular block forms a separate radial sector, which is sampled at $N+1$ points $\{\zeta_k\}$.\label{fig:vectorMatrixStructure}}
        \end{center}
      \end{figure}

   Following the scalar case, we project both (\ref{eq:procaEqDecompMain}) and (\ref{eq:lorenzEqDecompMain}) onto the different scalar and vector harmonics to find the following system of equations,
   \begin{align}
        \sum_{\ell',j'} \mathcal{D}^{\lab{ss}}_{\ell j\,|\,\ell' j'}[B_{\ell', j' m}(\zeta)] + \sum_{j'} \mathcal{D}^{\lab{st}}_{\ell j\, |\, j'}[B_{j'm}(\zeta)] &= 0\,, \label{eq:procaProject} \\
          \sum_{\ell',j'} \mathcal{D}^{\lab{ts}}_{j\, |\, \ell' j'}[B_{\ell', j' m}(\zeta)] + \sum_{j'} \mathcal{D}^{\lab{tt}}_{j\, | \, j'}[B_{j' m}(\zeta)] &= 0\,, \label{eq:lorenzProjectMain}
    \end{align}
    which can be written as the matrix equation schematically depicted in Fig.~\ref{fig:vectorMatrixStructure}, where we have introduced a collection of operators defined in (\ref{eq:sphericalMatrices}). This system of equations is the vector analog of (\ref{eq:decompScalar}), though it appears more complicated because we have explicitly separated the spatial and temporal components. As before, this matrix may then be passed to a nonlinear eigenvalue solver to determine the quasi-bound state spectrum. This method reproduces the quasi-bound state frequencies of the electric modes computed using the separable ansatz in~\S\ref{sec:Separable}. More importantly, it accurately computes the energy spectrum of the magnetic modes $|\es 2 \es 1 \es 1 \es m \rangle$, shown in Figures~\ref{fig:VectorSpectrumPlots}~and~\ref{fig:magneticReal}, and provides a crucial check of the magnetic instability rate (\ref{eqn:VectorRates}), displayed in Figures~\ref{fig:growthRates}~and~\ref{fig:magInstability}.

\section{Summary}

In this work, we have computed the quasi-bound state spectra of massive scalar and vector fields around rotating black holes, both analytically and numerically. The results of this chapter are summarized in \S\ref{sec:Summary}. The main challenge has been the fact that the fields vary rapidly in the  near-horizon region of the black hole, which causes ordinary perturbative approximations to fail and many numerical methods to become unreliable.
To address this issue in our analytical treatment, we performed a perturbative expansion in $\alpha$ and constructed independent solutions in different asymptotic regions  of the black hole spacetime. The spectrum is then determined by demanding that these solutions match in their regions of shared validity. For the scalar field, the asymptotic expansions in the near and far regions were matched directly, while, for the vector field, these expansions could only be matched indirectly, via a solution in an additional intermediate region. This reflected the fact that the vector field in the near and far regions depend on different types of angular momentum, which a matched solution must smoothly interpolate between. 

\newpage

Our perturbative analysis relied on the separability of the equations of motion, and thus did not apply to all magnetic modes of the vector field. However, by working at linear order in the black hole spin $\tilde{a}$, we derived perturbative results for the energy spectrum of the magnetic modes, which we argued plausibly extend to arbitrary values of $\tilde{a}$. Furthermore, we provided an educated guess for the magnetic instability rate by demanding that it takes the same functional form as the electric instability rates, and fixing its undetermined overall coefficient and dominant $\alpha$-scaling
using its known Schwarzschild limit.

\vskip 2pt

To check our analytic results and to study all modes at large values of $\alpha$, we computed the spectra numerically. The rapid variation of the fields in the near-horizon region presents serious numerical obstacles which can potentially destroy the accuracy of a solution. We described how to avoid these pitfalls and presented a formulation of the problem that accurately computes the quasi-bound state frequencies without relying on a separable ansatz. In principle, this formulation can be extended to ultralight fields of arbitrary spin about any stationary spacetime, although it works best if spherical symmetry is approximately restored at large distances. These numeric results provided a valuable check of our analytic approximations, which we found to accurately predict the energy eigenvalues and instability rates for both electric and magnetic modes as long as  $\alpha \lesssim 0.2$, even at large $\tilde{a}$. 

\vskip 2pt

The spectra of the scalar and vector gravitational atoms are shown schematically in Figures~\ref{fig:ScalarSpectra} and \ref{fig:VectorSpectra}. Importantly, we find that the vector field's intrinsic spin allows for many more nearly degenerate states than the scalar. The results of this chapter will provide an essential input for the subsequent chapters in this thesis,  where we will distinguish the phenomenologies of the scalar and vector clouds through their dynamics when they are parts of binary systems.

\chapter{Gravitational Collider Physics} \label{sec:Collider}

In the previous chapter, we studied the spectra of the scalar and vector gravitational atoms to high precision. Those computations were not purely academic exercises; as I will demonstrate in this chapter and the next, the intricate fine and hyperfine splittings significantly impact the dynamics of the atoms when they are part of binary systems. The results presented in these two chapters are based on the collaborative works~\cite{Baumann:2018vus, Baumann:2019ztm}.

\section{Overview and Outline}

A binary companion enriches the atom's evolution by mixing different levels in the spectra through its gravitational perturbation. Like in atomic physics, not all levels couple to one another as the mixings must obey certain selection rules. One of the most dramatic phenomenon occurs when the orbital frequency of the binary matches the energy difference between different levels, leading to resonant transitions between the corresponding states. Since the orbit of the binary shrinks due to its gravitational-wave emission, we must also carefully take the time dependence of the orbital frequency into account. This time dependence, however small, has an important effect on the dynamics of the cloud near a resonance, leading to an analog of the Landau-Zener transition in quantum mechanics~\cite{Landau, Zener}.

\vskip 4pt 

A key result of this chapter is that the resonant transitions in the scalar atom are qualitatively distinct from those for the vector atom. In particular, for scalar clouds in quasi-circular orbits, the relevant mixings are predominantly between two states. In this case, the dynamics near the resonance are rather simple and can be studied in detail.  However, qualitatively new features appear when multiple, nearly degenerate states are involved in the transition, which is the case for vector clouds.  When the binary moves slowly through the resonance band, the Landau-Zener transition could result in a complete transfer of population from the initial state to a linear combination of allowed states. While the transition populates only a single state for scalars, a vector cloud can evolve into a superposition of states. In the latter case, the superposition lead to ``neutrino-like oscillations" between different vector shapes and therefore leave intersting time-dependent imprints in the finite-size effects of these objects. 
   
   \vskip 4pt 
   
 \begin{figure}
      	\centering
        \includegraphics[scale=0.9, trim=25 0 0 0]{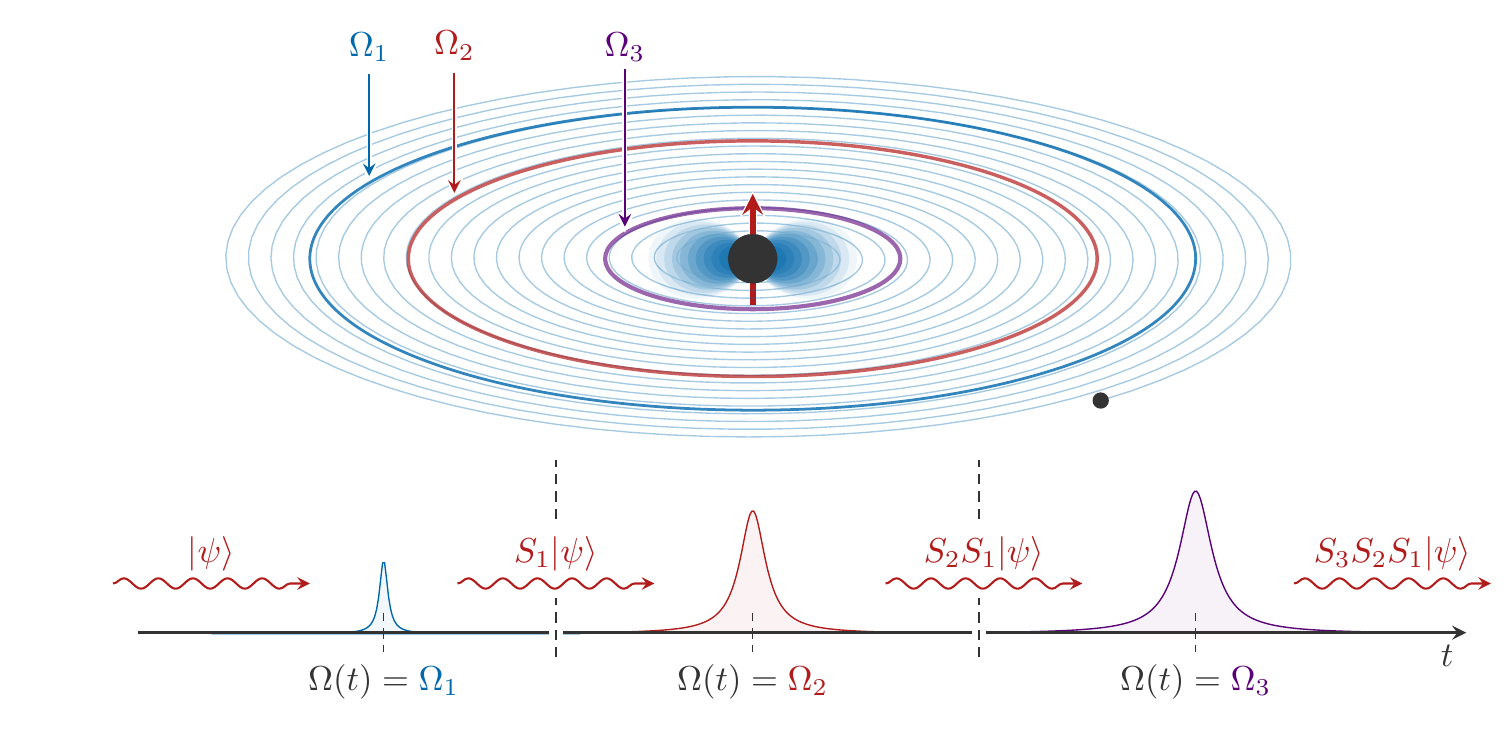}
        \caption{The orbit of a quasi-circular inspiral slowly scans through orbital frequencies $\Omega(t)$. The behavior of the cloud during the inspiral can be decomposed into a set of resonances at frequencies $\Omega_i$. Each resonance is characterized by an S-matrix, $S_i$, which describes the evolution of the system through that event. The S-matrix that describes the entire inspiral is simply the product of all individual S-matrices,~$S_\lab{tot} = \prod_i S_i$.  \label{fig:inspiral}}
    \end{figure}
   
Remarkably, the study of these clouds in binary systems is similar to the discpline of particle collider physics in many ways. Firstly, because the transitions are localized in time (or frequency), they can also be treated as ``scattering events" and described individually by an S-matrix. This S-matrix formalism will be particularly useful when we consider multiple sequential resonances, as it allows us to follow the state of the cloud through a sequence of resonances. In particular, given an initial state $|i\rangle$, we find that the final state after $N$ resonances can be written as 
\begin{equation}
|f \rangle  = \prod_{n=1}^N S_n\, |i \rangle \, , \nonumber
\end{equation} 
where $S_n$ is a unitary operator that evolves the state across the $n$-th resonance, see Fig.~\ref{fig:inspiral}. Given the differences in the interaction Hamiltonian and eigenstates, the state of the cloud after multiple resonances will be different for scalar and vector clouds. Secondly, as I have just described and will explore in more detail in Chapter~\ref{sec:signatures}, these scatterings are sensitive to the masses and spins of the ultralight bosons. The gravitational waves emitted by the binary during these resonances therefore carry vital information about the nature of the boson field in the system. A similar analogy, between collider physics and the imprints of massive particles in cosmological correlation functions, was drawn in \cite{Arkani-Hamed:2015bza} (see Fig.~\ref{fig:col1} for the relevant energy scales probed).\footnote{In the case of the ``cosmological collider", particles are produced as `resonances', when their Compton wavelength exceeds the cosmological horizon during inflation.  The properties of the new particles in the intermediate state are reflected in the momentum scaling and angular behavior in the soft limits of correlation functions~\cite{Chen:2009zp, Baumann:2011nk, Assassi:2012zq, Chen:2012ge, Noumi:2012vr, Assassi:2013gxa, Flauger:2013hra, Arkani-Hamed:2015bza, Lee:2016vti, Baumann:2017jvh, Kumar:2017ecc, An:2017hlx}.} Taking inspiration from cosmology, we will refer to this nascent discipline in gravitational-wave science as the ``gravitational collider physics", which complements other searches for new light particles, in the lab~\cite{Essig:2013lka}, astrophysics~\cite{Raffelt:1996wa} and cosmology~\cite{Abazajian:2016yjj, Baumann:2016wac, Brust:2013xpv}.

\vskip 4pt 

The plan of this chapter is as follows: in Section~\ref{sec:Binary}, I describe the gravitational perturbations generated by the companion when the clouds are part of binary systems. In addition, I show how this perturbation induces mixings between different states in the spectrum. In~Section~\ref{sec:gcollider}, I discuss the dynamics of this level mixing. We will find that the resonant transitions can be described as scattering events, and derive the transition ``probabilities" between coupled states. Furthermore, I explain how this S-matrix approach is particularly well suited to chain together a sequence of multiple resonances. The appendices contain additional details: In Appendix~\ref{app:GP}, I elaborate on the gravitational perturbations induced by the binary companion. In Appendix~\ref{app:LZ}, I review the analytic solution for the two-level Landau-Zener transition. I also discuss how Floquet theory can be used to deal with more general orbital configurations.

\newpage
\section{Gravitational Atoms in Binaries} 
\label{sec:Binary}

The presence of a binary companion introduces a perturbation to the dynamics of the cloud. We start, in \S\ref{sec: gravitational perturbation}, by discussing the nature of this gravitational perturbation. After a brief outline of the geometry of the problem, we will concentrate on the form of the metric perturbation in the cloud's free-falling frame. We will also briefly discuss the possibility of mass/energy transfer between the cloud and the companion. In \S\ref{sec:gravlevelmix}, we then describe how this perturbation can induce transitions between different levels of the spectrum. The~time dependence of these perturbations and the conditions for exciting resonant transitions are discussed in~\S\ref{sec:TimeDependentTidalMoments}.

\subsection{Metric Perturbation} \label{sec: gravitational perturbation}

 \begin{figure}[b!]
\centering
\includegraphics[scale=0.9, trim = 0 10 0 10]{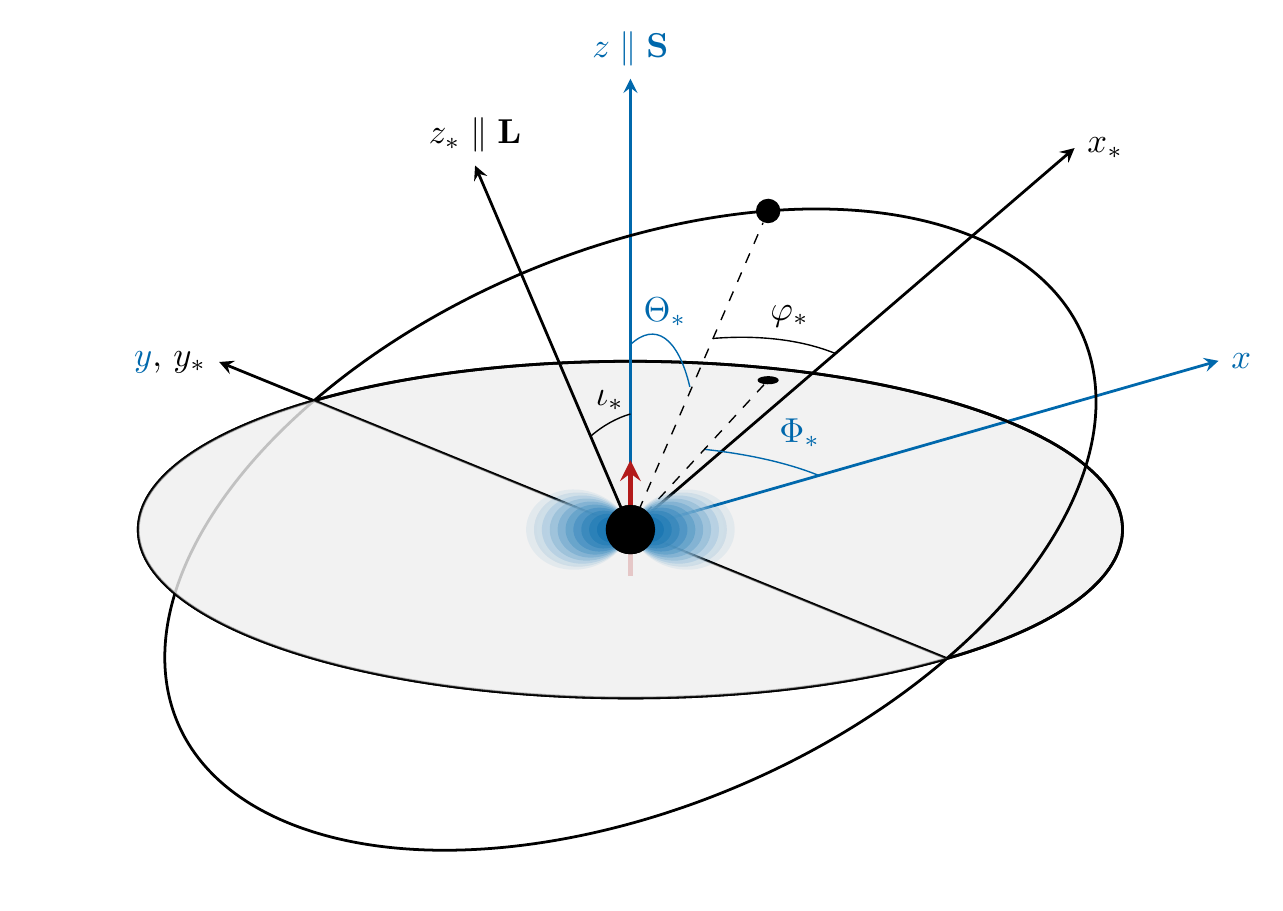}
\caption{Illustration of the coordinates used to describe the binary system. The Cartesian bases $\{x, y, z \}$ and $\{ x_*, y_*, z_* \}$ are adapted to the cloud's equatorial plane and the binary's orbital plane, respectively. The equatorial plane (gray) intersects the orbital plane at an inclination angle~$\iota_*$, and we have chosen
the pericenter to be located on the $x_*$-axis, such that $\varphi_*$ denotes the true anomaly of the binary. } 
\label{fig:BinaryPlane}
\end{figure}

To investigate the dynamics of a boson cloud in a binary system, we must work in the cloud's Fermi comoving frame \cite{Baumann:2018vus}. The relative motion of the companion is most conveniently described by the coordinates $\textbf{R}_* \equiv \{R_*, \iota_*, \varphi_*\}$, where $R_*$ is the 
separation between the members of the binary, $\iota_*$ is the \textit{inclination} angle between the orbital plane and the cloud's equatorial plane, and $\varphi_*$ is the \textit{true anomaly}, which represents the angle between the binary separation and the pericenter of the orbit (see  Fig.~\ref{fig:BinaryPlane}). 
 These angles are related to the usual angular spherical coordinates of the binary companion, $\{ \Theta_*, \Phi_*\}$, via 
\beq
 \begin{aligned}
 \cos \Theta_* & = \sin \iota_* \cos  \varphi_* \, , \\
  \tan \Phi_* & = \sec \iota_* \tan  \varphi_*  \, . \label{eqn:Angles}
 \end{aligned}
 \eeq
Since $0 \leq \Theta_* \leq \pi$, we can choose $-\pi/2 \leq \iota_* \leq \pi/2$, such that the spin of the cloud, $\textbf{S}_c $, projected on the $z_*$-axis of the orbital frame, satisfies $\textbf{z}_* \cdot \textbf{S}_c \geq 0$. While $\varphi_*$ increases in magnitude during the orbital evolution, the sign of $\varphi_*$ determines the orientation of the orbit. 
Orbits with $\varphi_* > 0$ are {\it co-rotating}, while those with $\varphi_* < 0$ are {\it counter-rotating}.

\subsubsection*{Free-falling clouds}

The perturbed metrics for compact objects in external fields have been studied extensively in the literature, see e.g.~\cite{Landry:2015zfa} and references therein.  The binary companion induces a time-dependent perturbation to the metric, such that $g_{\mu\nu} = g^{(0)}_{\mu\nu} +  h_{\mu\nu}$, where $g^{(0)}_{\mu\nu}$ is the unperturbed Kerr spacetime. In general, the metric perturbation $h_{\mu\nu}$ consists of two separate components: {\it i}\hskip 1pt) a direct contribution from the gravitational potential due to the mass of the companion, $M_*$, and {\it ii}~) the response of the cloud to tidal deformations. In this chapter, I will focus only on the former type of perturbation, and delay a discussion about the latter case to Chapter~\ref{sec:signatures}.

In Fig.~\ref{fig:BinaryPlane}, the Fermi comoving coordinates of the cloud are denoted by $\{ t, x^i \}$. Rewriting these coordinates to spherical coordinates $ \r \equiv \{ r,  \theta,  \phi\}$,\footnote{Since the center-of-mass of the cloud is the same as that of the isolated central black hole, these Fermi coordinates coincide with the Boyer-Lindquist coordinates $\{t, r, \theta, \phi \}$ at leading order in the post-Newtonian expansion.}  the trace-reversed metric component $\bar{h}^{00}$ is
\beq 
\begin{aligned}
\bar{h}^{00}( t,  r) =  4 M_* \sum_{\ell_*\neq 1} \sum_{|m_*| \leq \ell_*} & \mathcal{E}_{\ell_* m_*} (\iota_*, \varphi_*)  Y_{\ell_* m_*}  (\theta, \phi) \\
& \times \Bigg( \frac{r^{\ell_*}}{R^{\ell_* + 1}_*} \Theta (R_* - r) + \frac{R_{*}^{\ell_*}}{r^{\ell_* + 1}_*} \Theta (r - R_*)  \Bigg)   \,, \label{eqn:h00Expansion}
\end{aligned}
\eeq
where $\mathcal{E}_{\ell_* m_*}$ is the tidal tensor, $Y_{\ell_* m_*}$ is the scalar spherical harmonic, and $\Theta$ is the Heaviside step-function. The explicit form of the tidal moments $\mathcal{E}_{\ell_* m_*}$ depend on the geometry of the binary (see Appendix~\ref{app:TidalMoments} for a detailed discussion). Crucially, the dipole moment $\ell_*=1$ is absent in the Fermi frame of the cloud \cite{Baumann:2018vus}, such that the leading tidal interaction is provided by the gravitational {\it quadrupole} $\ell_*=2$. That a dipole moment is absent in gravity is in fact a well-known consequence of the equivalence principle. In principle, a different choice of observers/coordinates --- not free-falling with the cloud  --- could lead to the appearance of extra terms in the potential, for instance a {\it dipole}. However, as we show in Appendix~\ref{sec:fictitious},  this fictitious dipole eventually cancels.

In this thesis, I will concentrate on the cloud's dynamics when the binary companion is located {\it outside} of the cloud ($R_* > r_c$), where the metric perturbation is dominated by the first term in \eqref{eqn:h00Expansion}. However, when the binary separation approaches the Bohr radius of the cloud, $R_* \sim r_c$, the second term in \eqref{eqn:h00Expansion} can provide non-negligible support to the perturbation. 

\subsubsection*{Mass transfer} \label{sec: Roche-Lobe}

\begin{figure}[t!]
\centering
\includegraphics[scale=0.145, trim = 0 220 0 100 ]{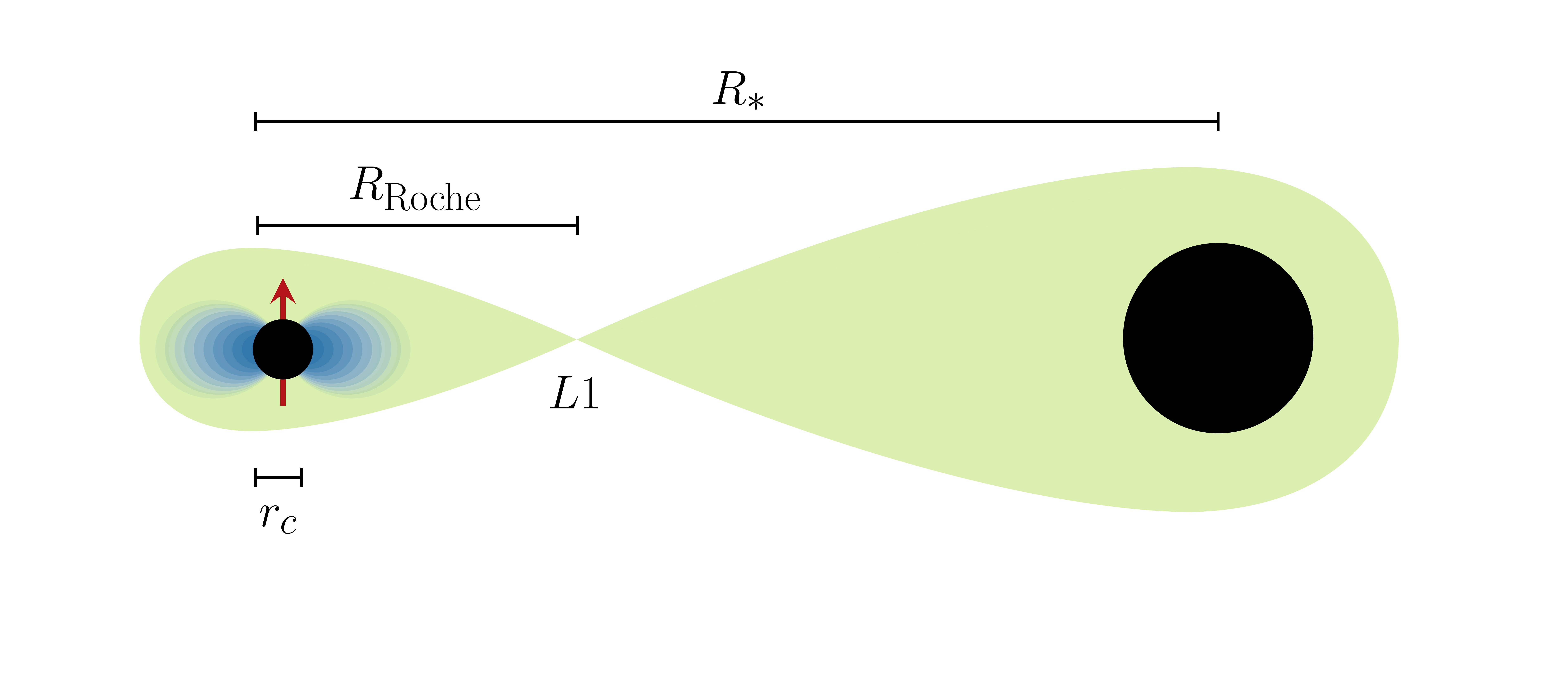}
\caption{Illustration of the Roche lobes for a binary with $q \gg 1$. As the separation decreases, the Roche lobe of the cloud  begins to shrink. At the critical value $R_{*,{\rm cr}}$, given by \eqref{Rochemass}, the size of the cloud $r_c$ exceeds $R_{\rm Roche}$,
and mass transfer starts to occur. } 
\label{fig:Roche}
\end{figure}

While we focus primarily on the cloud's dynamics when $R_* \gtrsim r_c$, the mutual gravitational attraction between the bodies can also induce transfer of mass/energy when $R_* \lesssim r_c$. Specifically, this happens when $r_c$ exceeds the Lagrange point, $L1$, located in between the two objects of the binary. The equipotential surface with the same gravitational potential as $L1$ is called the Roche lobe (see Fig.~\ref{fig:Roche}). 

Mass transfer from the cloud to the companion happens when $r_c \gtrsim R_{\rm Roche}$.
Using Eggleton's fitting formula~\cite{Eggleton:1983rx}, this can be converted into a critical orbital separation  
\beq
R_{*, \rm cr} \equiv   \left( \frac{0.49 \, q^{-2/3}}{0.6 \, q^{-2/3} + \ln \left( 1+q^{-1/3} \right)}  \right)^{-1} \, r_c \,.\label{Rochemass}
\eeq
This phenomenon becomes particularly important when $q \gg 1$, since the Roche lobe surrounding the cloud is then relatively small. In this limit, we find
\beq
R_{*,\rm cr} \simeq 2q^{1/3} r_c\,.\label{masst2}
\eeq 
Throughout this thesis, I will impose $R_{*} > R_{*,\rm cr}$, such that mass transfer can be ignored.

\subsection{Gravitational Level Mixings}
 \label{sec:gravlevelmix}
 
The gravitational perturbation described in the previous subsection are encoded by additional potential terms in the Schr\"{o}dinger equations,~(\ref{eqn:NonRelScalar}) or~(\ref{eqn:NonRelVector}), for scalar and vector fields, respectively. At leading order, these terms are~ \cite{Baumann:2018vus, Baumann:2019ztm}
\begin{align}
V_* &=  - \frac{1}{4} \mu \bar{h}^{00} \mathrlap{\qquad \quad (\text{scalar})\, ,} \label{eqn:ScalarPert} \\
V_*^{il}  &=  - \frac{1}{4} \mu   \bar{h}^{00} \delta^{il}   \mathrlap{\qquad (\text{vector})\, ,} \label{eqn:VectorPert} 
\end{align}
where $\bar{h}_{00}$ is given in (\ref{eqn:h00Expansion}). As we shall see, the presence of the $\ell_* \geq 2$ terms enables the mixing between different states of the spectrum.

In principle, we could also include couplings in (\ref{eqn:ScalarPert}) and (\ref{eqn:VectorPert}) that are higher order in~$\alpha$, such as those arising from interaction terms with spatial and temporal gradients acting on the non-relativistic fields, or from the post-Newtonian corrections to the metric perturbation.\footnote{Notice that gradients acting on the non-relativistic field are $\alpha$-suppressed, $\partial_i \psi \sim \mu \alpha \psi$ and $\partial_0 \psi  \sim \mu \alpha^2 \psi$. Similarly, the virial theorem demands that the companion's velocity is roughly $v \sim \sqrt{( M + M_*)/R_*}$, so that velocity-dependent PN corrections are also $\alpha$-suppressed whenever $R_* \gtrsim r_c$ \cite{Baumann:2018vus}.\label{footnote:alpha-scaling}}
These subleading potential terms are given explicitly in Appendix~\ref{app:HigherOrder}, and we defer a detailed study of their impact on the evolution of the cloud to future work.

\subsubsection*{Selection rules}

The perturbation (\ref{eqn:h00Expansion}) induces level mixings between different states of the gravitational atom. However, not all levels couple to one another as certain {\it selection rules} must be obeyed. As we shall see, the associated selection rules are different for scalar and vector transitions.

To deduce the allowed transitions for both scalar and vector clouds, we must compute the overlap \cite{Baumann:2018vus, Baumann:2019ztm}
\beq
\langle a' | V_*(t) | a \rangle = - M_* \, \mu \sum_{\ell_* \neq 1} \sum_{m_* \leq |\ell_*|} \mathcal{E}_{\ell_* m_*} \!\times I_{r} \times I_{\Omega}  \, , \label{eqn:MatrixElement}
\eeq
where $V_*$ schematically represents both (\ref{eqn:ScalarPert}) and (\ref{eqn:VectorPert}), and the states $|a\rangle$ and $|a^\prime\es \rangle$ can  denote either the scalar $|n \es \ell \es m\rangle$ or vector $|n  \es \ell \es  j \es m\rangle$ eigenstates. The radial integral in both cases is
\begin{align}
I_{r} & \equiv \int_0^{R_*} \!\d r \, r^2 \left( \frac{r^{\ell_*}}{R_*^{\ell_* + 1}} \right) R_{n^\prime \ell^\prime } R_{n \ell }, + \int_{R_*}^{\infty} \!\d r \, r^2 \left( \frac{R_*^{\ell_*}}{r^{\ell_* + 1}} \right) R_{n^\prime \ell^\prime } R_{n \ell} \, , 
\label{eqn:I_r}
\end{align}
where $R_{n \ell}$ are the radial functions defined in (\ref{eqn:RadialHydrogen}). However, the angular integrals $I_{\Omega}$ are different for the scalar and vector:
\begin{align}
I_{\Omega} &\equiv \int \!\d \Omega \,\, Y^*_{\ell^\prime m^\prime} (\theta, \phi)\, Y_{\ell_* m_*} (\theta, \phi) \, Y_{\ell m} (\theta, \phi) \mathrlap{\quad \quad \qquad \ \  (\text{scalar})\, ,} \hskip 40pt \label{eqn:I_Omega_scalar} \\
I_{\Omega} &\equiv \int \d \Omega \ Y_{\ell_* m_*} (\theta, \phi)\, \textbf{Y}^*_{\ell^\prime, j^\prime m^\prime }(\theta, \phi)  \cdot \textbf{Y}_{\ell, j m} (\theta, \phi)  \mathrlap{\qquad (\text{vector})\, .}  \hskip 40pt  \label{eqn:I_OmegaVec}
\end{align}
The first integral (\ref{eqn:I_Omega_scalar}) is only non-vanishing when the following selection rules are satisfied~\cite{Baumann:2018vus} 
\beq
\begin{aligned}
\text{({\rm S1})} & \hskip 8pt -m^\prime + m_* + m = 0 \, , \\
\text{({\rm S2})} & \hskip 8pt  \ell + \ell_*  + \ell^\prime = 2p \, ,  \  \text{for} \ p\in \mathbb{Z} \, , \\
\text{({\rm S3})} & \hskip 8pt  |\ell - \ell^\prime | \le \ell_* \le \ell + \ell^\prime  \, .
\label{eqn:GeneralTransitionRulesScalar}
\end{aligned}
\eeq
Recall that the fastest growing mode $\ket{211}$ has $\l=m=1$, while the fastest decaying mode $\ket{100}$ has $\l^\prime=m^\prime=0$. In this case, the above selection rules would require $\l_*= \pm  m_*=1$, namely a {\it dipole} coupling, which is absent in \ref{eqn:h00Expansion}. For the quadrupole, the two fastest decaying modes that can couple to this dominant mode are $\l^\prime =1, m^\prime=-1$ and $\l^\prime=1, m^\prime=0$. Since these rules are obtained purely from the angular dependence of the eigenfunctions, they apply equally to the fundamental mode ($ n=2$) and the overtones ($n \ge 3$).

Similarly, the integral (\ref{eqn:I_OmegaVec}) implies the following selection rules for the vector cloud:
\beq
\begin{aligned}
\text{(V1)} & \hskip 8pt -m^\prime + m_{*} + m = 0 \, , \\
\text{(V2)} & \hskip 8pt  \ell + \ell_*  + \ell^\prime = 2p \, , \  \text{for} \ p\in \mathbb{Z} \, ,  \\
\text{(V3)} & \hskip 8pt  |\ell - \ell^\prime | \le \ell_* \le \ell + \ell^\prime  \, , \\
\text{(V4)} & \hskip 8pt  |j - j^\prime | \le \ell_* \le j + j^\prime  \, .
 \label{eqn:GeneralTransitionRulesVector2}
\end{aligned}
\eeq
While (V1) -- (V3) are similar to (S1) -- (S3), there is a new rule~(V4) for the vector, which  reduces to (S3) when $j = \ell$.\footnote{There are special cases where $I_\Omega=0$ even when the selection rules (V1) -- (V4) are naively satisfied. These occur when the inner product between the vector spherical harmonics in the integrand (\ref{eqn:I_OmegaVec}) vanishes, i.e. when $\ell = \ell^\prime, m = m^\prime= 0$, and the inner product is taken between an electric $j = \ell \pm 1 $ and a magnetic $j^\prime = \ell^\prime $ vector spherical harmonic.}  
The rule (V3) implies that the dominant state $|1 \es 0 \es 1 \es 1\rangle$ of the vector cloud  can only transition to  modes with $\ell^\prime \geq 2$. We describe the phenomenological consequences of the different selection rules for scalar and vector clouds in Section~\ref{sec:unravel}.

\subsection{Dynamical Perturbation} 
\label{sec:TimeDependentTidalMoments}

We now discuss how the binary's orbital motion forces the gravitational perturbation on the cloud to evolve in time, which leads to novel dynamical effects.

\subsubsection*{Shrinking orbits}

As the coordinates of the binary $\mb{R}_*(t) = \{R_*(t), \iota_*(t), \varphi_*(t)\}$ evolve, so does the metric perturbation (\ref{eqn:h00Expansion}).
For general orbit, the overlap in (\ref{eqn:MatrixElement}) is (see Appendix~\ref{app:TidalMoments} for details)
\beq
\bra{a} V_*(t) \ket{b} \equiv \sum_{ m_\varphi\in \mathbb{Z}} \eta_{ab}^{(m_\varphi)} (R_*(t), \iota_*(t)) \, e^{-i m_\varphi \varphi_* (t)} \, , \label{eqn:etaDef}
\eeq
where $\eta_{ab}^{(m_\varphi)}$ characterizes the strength of the perturbation, and the oscillatory factors $e^{-i m_\varphi \varphi_* (t)}$ arise from the tidal moments $\mathcal{E}_{\ell_* m_*}$.\footnote{As we will illustrate in Appendix~\ref{app:TidalMoments}, the sum over $m_\varphi$ can be understood as a sum over polarizations. Depending on the inclination $\iota_*$ of the orbit, different polarizations may contribute.} Crucially, the presence of these oscillatory terms means that the perturbation acts like a periodic driving force, which greatly enriches the dynamics of the boson cloud.  In principle, the couplings $\eta_{ab}^{(m_\varphi)}$ receive contributions from \emph{all} multipoles in the expansion (\ref{eqn:h00Expansion}). However, we will concentrate on the $\ell_* = 2$ quadrupole coupling, which dominates the perturbation.

It is therefore necessary to understand the behavior of $\varphi_*(t)$, including the effects induced by the shrinking of the orbit due to gravitational-wave emission.
 In terms of the instantaneous orbital frequency, $\Omega(t) > 0$, we have \cite{Baumann:2018vus}
\beq
\varphi_*(t) = \pm \int_{0}^t \d t^\prime \, \Omega(t^\prime) \, ,  \label{eqn:GeneralTrueAnomaly}
\eeq
where $t \equiv 0$ is an initial reference time and the upper (lower) sign denotes co-rotating (counter-rotating) orbits. In general, $\varphi_*(t)$ and $\Omega(t)$ evolve in a complicated manner. However, for quasi-circular equatorial orbits, the orbital frequency is determined by~\cite{PeterMatthews1963}
\beq
\frac{\d \Omega}{\d t} = \gamma \left(\frac{\Omega}{\Omega_0} \right)^{11/3} \, , \label{eqn:Omega-Quasi-Circular}
\eeq
where $\Omega_0$ is a reference orbital frequency and $\gamma$ is the rate of change due to the gravitational-wave emission,
\beq
\gamma \equiv \frac{96}{5} \frac{q}{(1+q)^{1/3}} (M \Omega_0)^{5/3} \Omega^2_0\, , \label{eqn:circleRate}
\eeq
with $q \equiv M_* / M$ the mass ratio of the black holes.

When the orbital frequency does not change appreciably near $\Omega_0$, the solution of (\ref{eqn:Omega-Quasi-Circular}) can be approximated by
\beq
\Omega(t) = \Omega_0 + \gamma t \, , \label{eqn:OmegaLinear}
\eeq
where we have dropped nonlinearities that become important on the timescale $t \sim \Omega_0/\gamma$.
The effects of a time-periodic perturbation on the gravitational atom, with fixed frequency $\Omega_0$, were studied in~\cite{Baumann:2018vus}, where \textit{Rabi oscillations} were found. The oscillations are enhanced when $\Omega_0$ matches the difference between two energy levels of the clouds, such that resonances are excited. 
However, as we discuss in~Section~\ref{sec:gcollider}, the presence of a new timescale associated to the shrinking of the orbit introduces additional coherent effects that qualitatively change the behavior of the cloud during the resonant transition.

\subsubsection*{Resonances}

As we will elaborate in Section~\ref{sec:gcollider}, resonant transitions can be excited when the orbital frequency, or its overtones, matches the energy difference $\Delta E_{ab} \equiv E_a - E_b$ between two states $|a \rangle$ and $| b \rangle$ connected by the gravitational perturbation,
  \begin{equation}
    \dot{\varphi}_*(t) = \pm \Omega(t) = \frac{\Delta E_{ab}}{m_\varphi}\,. \label{eqn:resCondition}
  \end{equation}
  The weak gravitational perturbation is resonantly enhanced at these frequencies, and can force the cloud to evolve into an entirely different configuration. In most of this thesis, I will focus on quasi-circular equatorial orbits for which $m_\varphi = \Delta m_{ab}$.\footnote{In Appendix~\ref{app:TidalMoments}, we show that, for equatorial orbits, the summation over $m_\varphi$ in (\ref{eqn:etaDef}) reduces to a single term with $m_\varphi = m_*$. The selection rules (S1) and (V1) then imply that $m_\varphi = \Delta m_{ab} \equiv m_a - m_b$. \label{footnote:equatorial}} This restriction is convenient, because fewer transitions are allowed by the selection rules, and transitions between a pair of states may only occur at a single frequency. The resonance condition (\ref{eqn:resCondition}) then simply becomes $\dot{\varphi}_* = \Delta E_{ab} / \Delta m_{ab}$.

\begin{figure}
      	\centering
        \includegraphics[scale=1]{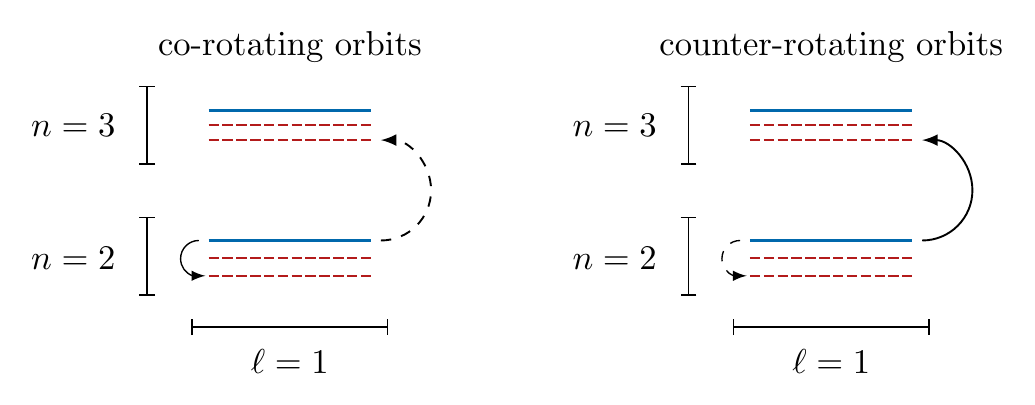}
        \caption{Illustration of the Bohr transition and the hyperfine transition of the dominant scalar $\ket{211}$ mode, in co-rotating and counter-rotating orbits. Solid arrows represent the allowed resonant transitions, while dashed arrows denote perturbative level mixings.  \label{fig:covscounter}}
   \end{figure}

  Since $\Delta E_{ab}/\Delta m_{ab}$ can be either positive or negative, the orientation of the orbit determines which resonances are excited. Specifically, co-rotating orbits can only excite resonances for which $\Delta E_{ab}$ and $\Delta m_{ab}$ have the same sign, while counter-rotating orbits require the signs to be opposite. Resonances do not occur if either $\Delta E_{ab} = 0$ or $\Delta m_{ab} = 0$, since energy and angular momentum must be transferred in the process. 
These constraints, in addition to the selection rules, dictate whether or not a transition is allowed. As a concrete example, we consider the transitions of the dominant scalar mode $ \ket{211}$, whose selection rules restrict it to couple only to the $\ket{21-1}$ and $\ket{31-1}$ decay modes in its immediate vicinity. In this case, co-rotating orbits can only excite resonances between the $\ket{211}$ and $\ket{21-1}$ modes. Since these states have hyperfine energy splittings, we refer to these type of resonances as \textit{hyperfine resonances}.  On the other hand, counter-rotating orbits would excite resonances between the $\ket{211}$ and $\ket{31-1}$ states. Similarly, because these states have Bohr energy splittings, we will refer to these types of resonances as \textit{Bohr resonances}. In either case, the $\ket{211}$ mode would still couple perturbatively with the other allowed decaying states, though these mixings are significantly suppressed compared to the resonant transitions. These descriptions are summarized in Fig.~\ref{fig:covscounter}. In what follows, we will denote a general resonance frequency by
  \begin{equation}
    \res_{ab} = \left|\frac{\Delta E_{ab}}{\Delta m_{ab}}\right| , \label{eqn:resCondition2}
  \end{equation}
  though I stress that not all transitions between states are accessible during a particular inspiral.

In order to understand the dynamics of the cloud near the resonances, we will thus set the reference frequency to be $\Omega_0 \equiv \res_{ab}$, such that (\ref{eqn:OmegaLinear}) becomes 
  \begin{equation}
    \Omega(t) = \res_{ab} + \gamma_{ab} \hskip 1pt t\,, \label{eqn:ResonanceLinear}
  \end{equation}
where $\gamma_{ab}$ is (\ref{eqn:circleRate}) evaluated at $\Omega_0=\res_{ab}$. The linear approximation (\ref{eqn:ResonanceLinear}) is justified if the fractional change in the orbital frequency is small throughout the transition. We find that this is indeed the case for virtually all transitions, for both the scalar and vector clouds. When there is no risk of confusion, we will often label a general resonance frequency as $\Omega_r$.

Before moving on, it will be useful to understand the order of magnitudes and scalings of the  relevant quantities involved in these transitions. For concreteness, let us consider Bohr transitions between states with different principal quantum numbers $n_a$ and  $n_b$.  Compared to the resonance frequency~$\res_{ab}$, the typical values of $\eta_{ab}$ and $\gamma_{ab}$ are
  \begin{align}
  \frac{\gamma_{ab}}{\res_{ab}^2} & \simeq 3.2 \times 10^{-6} \, \frac{q}{(1+q)^{1/3}} \left( \frac{\alpha}{0.07}\right)^5 \left( \frac{2}{|\Delta m_{ab}|} \right)^{5/3} \left| \frac{1}{n^2_a} - \frac{1}{n^{ 2}_b } \right|^{5/3} \, , \label{eqn:gamma} \\[4pt]
  \frac{\eta_{ab}}{ \res_{ab}} & \simeq 0.3 \left( \frac{R_{ab}}{0.3} \right) \frac{q}{1+q} \, ,\label{eqn:Eta}
  \end{align}
where we have assumed that the transition is mediated by the quadrupole, $\ell_* = 2$, and introduced $R_{ab}$, a dimensionless coefficient that characterizes the overlap between different states in the spectrum. For the relevant states, we typically have $R_{ab} \lesssim 0.3$.\footnote{We assumed that the binary separation is much larger than the Bohr radius of the cloud, such that the first term in (\ref{eqn:I_r}) dominates the $q$-scaling. The general dependence on $q$ is actually more complicated, and is properly taken into account in the numeric results of Section~\ref{sec:unravel}.}

Because the transitions do not happen instantaneously, an initial state can, in principle, resonate with several states at the same time. This will occur if both the selection rules allow it and the couplings $\eta_{ab}$ are strong enough to excite states that are slightly off-resonance. In practice, the resonance between states $a$ and $b$ has a \emph{bandwidth} set by the coupling~$\eta_{ab}$.\footnote{This implies that the timescale of the transition is $\Delta t \sim \eta_{ab}/\gamma_{ab}$. The linear approximation (\ref{eqn:ResonanceLinear}) is therefore justified during the transition as long as $\eta_{ab}/\res_{ab} \ll 1$, which in light of (\ref{eqn:Eta}) is always the case.} Whenever the resonance frequency of a different state $c$ falls within this bandwidth, it may also participate in the transition.

Remarkably, Bohr transitions for the scalar atom typically involve only two states, whereas boson clouds with vector fields can access multiple states. 
To illustrate this point, it is instructive to consider transitions mediated by the quadrupole 
in counter-rotating orbits. Since perturbations have $m_*=\pm 2$ (see Appendix~\ref{app:TidalMoments}), we are thus restricted to transitions with $\Delta E_{ab} > 0$, which must 
satisfy $\Delta m_{ab} = -2$. Due to the selection rule  (\ref{eqn:GeneralTransitionRulesScalar}), the scalar $|2\es 1\es 1 \rangle$ state  
only couples to the $| 3\es 1\es \, \minus 1 \rangle$ state. 
In contrast, the vector $| 2\es 1\es 2 \es 2 \rangle$ state
 simultaneously resonates with all three $|3\es 1 \es j \es 0 \rangle$ modes with the same azimuthal quantum number, 
since their energy differences, of order $\mu \alpha^4$, are small compared to the size of the bandwidth $\sim \mu \alpha^2$. While I have illustrated this difference with a specific example, we find that virtually all relevant scalar and vector transitions follow this behavior.

As we shall see momentarily, there are qualitative differences for transitions involving more than two states. I study next how these differences manifest themselves in the dynamics of the cloud during the resonance, and how we can encode it into an S-matrix formalism. We return to this point in Section \ref{sec:unravel}, where we discuss in more detail how the different evolution trees for scalar and vector clouds can help us measure the spin of the ultralight particles.

\section{The Gravitational Collider} 
\label{sec:gcollider}

The goal of this section is to characterize the dynamics of the cloud under a perturbation whose frequency gradually increases with time, as in the binary system. We will find that a type of ``collision event'' (or resonance) occurs when the orbital frequency matches the energy difference between two or more states of the cloud.  As we will show, the dynamics of the cloud through this event can be captured by an S-matrix, which describes how a state defined far before the resonance evolves long after it has passed it. The evolution throughout the entire inspiral can then be described as a series of scattering events, with the S-matrix formulation providing a convenient and simple description of a generally complicated process.

 A general bound state can be written as $|\psi(t)\rangle = \sum_a c_a (t) | a \rangle$, where $a$ ranges over all states in the spectrum.  In this eigenstate basis, the Schr\"odinger equations (\ref{eqn:NonRelScalar}) and (\ref{eqn:NonRelVector}) reduce to
 \begin{equation}
      i \es \frac{\ud c_a}{\ud t} = \sum_{a} \mathcal{H}_{ab}(t) \, c_b\, ,\label{eqn:Schr}
    \end{equation}
 where the Hamiltonian $\mathcal{H}_{ab} = E_a \delta_{ab} + V_{ab}(t)$ splits into a constant, diagonal matrix of eigenstate energies, (\ref{eqn:scalarspectrum}) and (\ref{eqn:vectorspectrumGeneral}), and a time-varying off-diagonal piece encoding the gravitational mixings induced by the companion, (\ref{eqn:MatrixElement}). Conservation of the total occupation density demands that $\sum_{a} |c_a(t)|^2 = 1$. We describe the qualitative behavior of (\ref{eqn:Schr}) in this section.

We begin, in \S\ref{sec:twoState}, with the simple case of a two-state system. 
  I show that the dynamics is characterized by the Landau-Zener (LZ) transition, as it occurs in quantum mechanics.
 In~\S\ref{sec:MultiState}, we generalize the formalism to a multi-state system.  
 We work out in detail the case of a three-state system, which captures all qualitative features of the generic case. Finally, in \S\ref{sec:Smatrix}, we discuss how the same evolution can also be described by an S-matrix, which provides a convenient way to describe a sequence of LZ transitions.

\subsection{Two-State Transitions} 
\label{sec:twoState}

To illustrate the main features of the cloud's dynamics, it is convenient to first study a simple two-level system. This is more than just a pedagogical device, since the transitions in  scalar atoms typically involve only two states (see \S\ref{sec:TimeDependentTidalMoments}).

To avoid unnecessary complications, we will truncate the level mixing (\ref{eqn:etaDef}) to a single quadrupolar interaction term. This is equivalent to assuming that the companion travels on a large, equatorial, quasi-circular orbit, so that the dominant gravitational perturbation connecting the different energy eigenstates oscillates with a definite phase.  Specifically, if the states $|1\rangle$ and $|2\rangle$ have azimuthal angular momenta $m_1$ and $m_2$, with $\Delta m_{21} \equiv m_2 - m_1$, the gravitational mixing (\ref{eqn:etaDef}) takes the form 
    \begin{equation}
      V_{12}(t) \equiv \langle 1 | V_*(t) | 2 \rangle = \eta_{12}(t)\hskip 2pt e^{i \Delta m_{21} \hskip 1pt \varphi_*(t)}\, ,
    \end{equation}
    where the strength of the interaction $\eta_{12}(t)$ varies slowly in time. As discussed in \S\ref{sec:TimeDependentTidalMoments}, the true anomaly $\varphi_*(t)$ for quasi-circular orbits near the resonances can be approximated as\hskip 1pt\footnote{To avoid clutter, we have dropped the subscripts on $\res_{ab}$ and $\gamma_{ab}$.}
    \begin{equation}
      \varphi_*(t) = \pm \int_{0}^{t}\!\ud t'\, \Omega(t') = \pm \left(\res \hskip 1pt t + \frac{\gamma}{2} \hskip 1pt t^2 \right) , \label{eq:linearizedOmeg}
    \end{equation} 
    where the sign depends on the orientation of the orbit. 
 It will be instructive to first consider the simplest case, where (\ref{eqn:Schr}) describes two states separated by an energy gap $\Delta E \equiv E_2 - E_1 > 0$, with $\Delta m_{21} \equiv \Delta m$ and constant $ \eta_{12} \equiv \eta$. 
    The Hamiltonian in the ``Schr\"{o}dinger frame'' then reduces~to 
    \begin{equation}
      \mathcal{H} = \begin{pmatrix} -\Delta E/2 & \eta \hskip 1pt e^{i  \Delta m \es \varphi_*(t)} \\ \eta \hskip 1pt e^{-i \Delta m \es \varphi_*(t)} & \Delta E/2 \end{pmatrix} . \label{eqn:Two-State-Schr-H}
    \end{equation}
    While the off-diagonal terms oscillate rapidly, they do so at a slowly increasing frequency, and it will be useful to work in a \emph{dressed frame} that isolates this slow behavior.  Said differently, while the companion's motion in a fixed frame is complicated, 
     its motion simplifies drastically if we rotate along with it. We thus define a time-dependent unitary transformation
    \begin{equation}
      \mathcal{U}(t) = \begin{pmatrix} e^{i \Delta m \es \varphi_*/2} & 0 \\ 0 & e^{-i \Delta m \es \varphi_*/2} \end{pmatrix} , \label{eq:timeDependentUnitary}
    \end{equation}
    so that the Schr\"{o}dinger frame coefficients can be written as $c_{a}(t) = \mathcal{U}_{ab}(t) d_{b}(t)$. 
Assuming a co-rotating orbit, such that $\dot{\varphi}_*(t) = \Omega(t)$, the dressed frame coefficients $d_a(t)$ then evolve according to the following Hamiltonian,
    \begin{equation}
      \mathcal{H}_\lab{D}(t) = \mathcal{U}^\dagger \mathcal{H} \hskip 1pt \mathcal{U} - i\hskip 1pt  \mathcal{U}^\dagger \frac{\ud \mathcal{U}}{\ud t} = \begin{pmatrix} 
          (\Delta m \es \Omega(t)-\Delta E)/2 & \eta \\
          \eta & -(\Delta m \es \Omega(t) - \Delta E)/2
      \end{pmatrix} . \label{eq:dressedFrameHam}
    \end{equation}
Note the useful fact that the structure of the transformation (\ref{eq:timeDependentUnitary}) implies that the magnitudes of the coefficients  in the Schr\"{o}dinger and dressed frames are equal, $|c_a(t)|^2 = |d_a(t)|^2$. For this reason, we will abuse notation and use $|a \rangle$ to denote the energy eigenstates in  both the Schr\"{o}dinger and dressed frames. Since our main interest will be in the overall population of the states $|a\rangle$, we only need to work with dressed frame quantities.

\begin{figure}
      	\centering
        \includegraphics[scale=0.91]{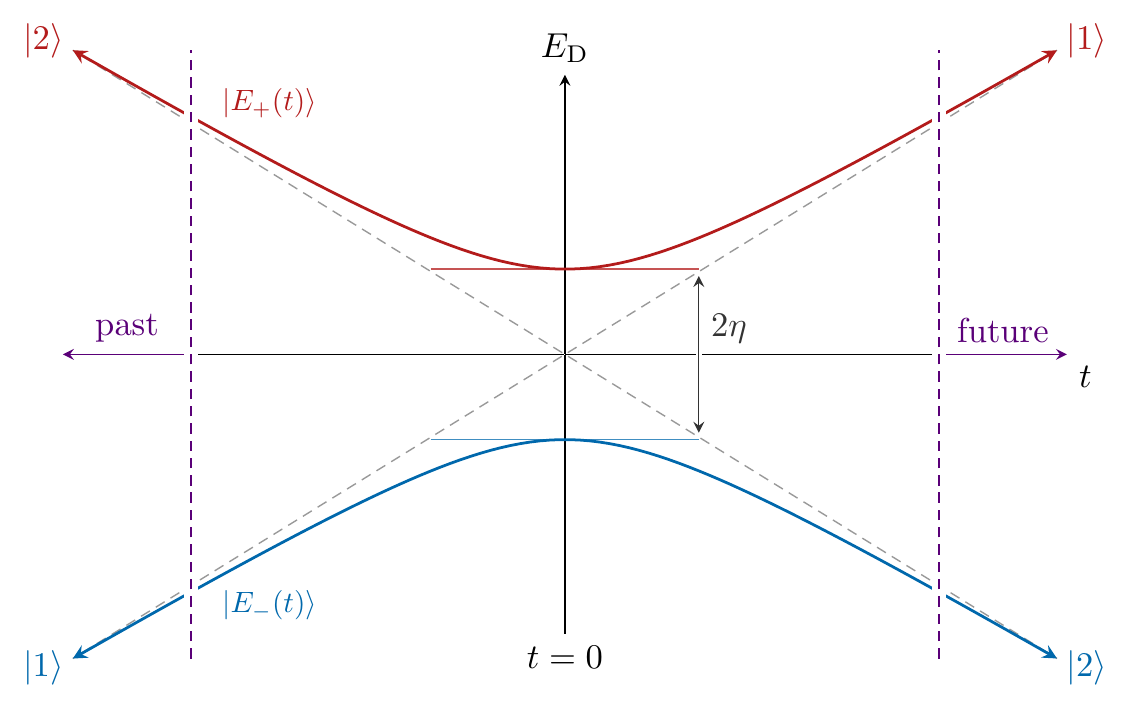}
        \caption{Instantaneous energy eigenvalues of the dressed frame Hamiltonian (\ref{eq:dressedFrameHam}) as a function of time. If $\eta = 0$, these energies cross when the frequency of the perturbation matches the energy difference between the two states, $\Omega(t) = \Delta E$. For non-vanishing $\eta$, these levels avoid each other. \label{fig:2by2eigen}}
   \end{figure}

  It is clear from (\ref{eq:dressedFrameHam}) that there is a ``collision'' or ``resonance'' time  $t_r \equiv 0$ when the frequency of the companion matches the energy difference between these two states,
  \begin{equation}
    \Omega(t_r) = \frac{\Delta E}{\Delta m} \,,
  \end{equation}
  and the diagonal entries of the dressed frame Hamiltonian vanish. Note that, since $\Omega(t)$ is always positive, a resonance in a co-rotating orbit can only happen if $\Delta E$ has the same sign as $\Delta m$ (see \S\ref{sec:TimeDependentTidalMoments} for more discussion). 
 To reduce clutter, we will set $\Delta m = 1$ in what follows.\footnote{Restoring factors of $\Delta m$ will simply mean replacing $\gamma \to \left| \Delta m \right| \gamma$.} With these simplifications, the dressed frame Hamiltonian is
  \begin{equation}
    \mathcal{H}_\lab{D}(t) =  \frac{\gamma t}{2} \begin{pmatrix} 
     1 & 0 \\
      0 & -1
      \end{pmatrix}  + \eta \begin{pmatrix} 
      0 & 1 \\
      1& 0
      \end{pmatrix} , \label{eq:dressedFrameHamSimp}
  \end{equation}
whose instantaneous energy eigenvalues $E_{\pm}(t) = \pm \sqrt{(\gamma t/2)^2 + \eta^2}$ I depict in Fig.~\ref{fig:2by2eigen}. The asymptotic states shown in the figure are $\ket{1} \equiv (1, 0)$ and $\ket{2} \equiv (0, 1)$.

  We can use this figure to understand qualitatively how the instantaneous energy eigenstates  evolve in time. In the far past, $t \to -\infty$, the first term in (\ref{eq:dressedFrameHamSimp}) dominates and the instantaneous eigenstates are simply 
  \begin{equation}
    |E_{+}(-\infty) \rangle = \binom{0}{1}   \, , \qquad |E_-(-\infty)\rangle = -\binom{1}{0} \,.
    \label{equ:before}
  \end{equation}
  As we then move toward the collision, $t \to 0$, these states become nearly degenerate, the second term dominates, and the eigenvalues $E_\pm(t)$ are forced to repel. In the far future, $t \to + \infty$, the first term in (\ref{eq:dressedFrameHamSimp}) again dominates, and so the eigenstates take the same form. However, this repulsion event forces the states to have permuted their identities, so that in the far future
  \begin{equation}
    |E_{+}(+\infty) \rangle = -|E_{-}(-\infty)\rangle  \, ,\qquad  |E_-(+\infty) \rangle = |E_{+}(-\infty)\rangle \,.
  \label{equ:after}
  \end{equation}
  This can also be seen explicitly from the exact form of the time-dependent eigenstates,
  \begin{equation}
    \begin{aligned}
      |E_{\pm}(t) \rangle &= \mathcal{N}^{-1}_\pm \left(\gamma t/2 \pm \sqrt{(\gamma t/2)^2 + \eta^2}\,, \eta \right) ,
       \label{eq:2by2eigensystem}
        \end{aligned}
    \end{equation}
 where $ \mathcal{N}_\pm(t)$ is the appropriate normalization.

As long as the evolution is adiabatic---meaning that the dressed frame Hamiltonian $\mathcal{H}_\lab{D}(t)$ evolves slow enough---the system tracks its instantaneous eigenstates.  If the system begins its life in the dressed frame's $\ket{1}$
 state, this implies that there is a \emph{complete} transfer of population into the $\ket{2}$ 
state after the resonance, cf.~Fig.~\ref{fig:2by2eigen}. This is a key characteristic of the LZ transition for a two-level system.  
  
  \begin{figure}
    \centering
      \includegraphics[scale=0.82, trim=8 0 0 0]{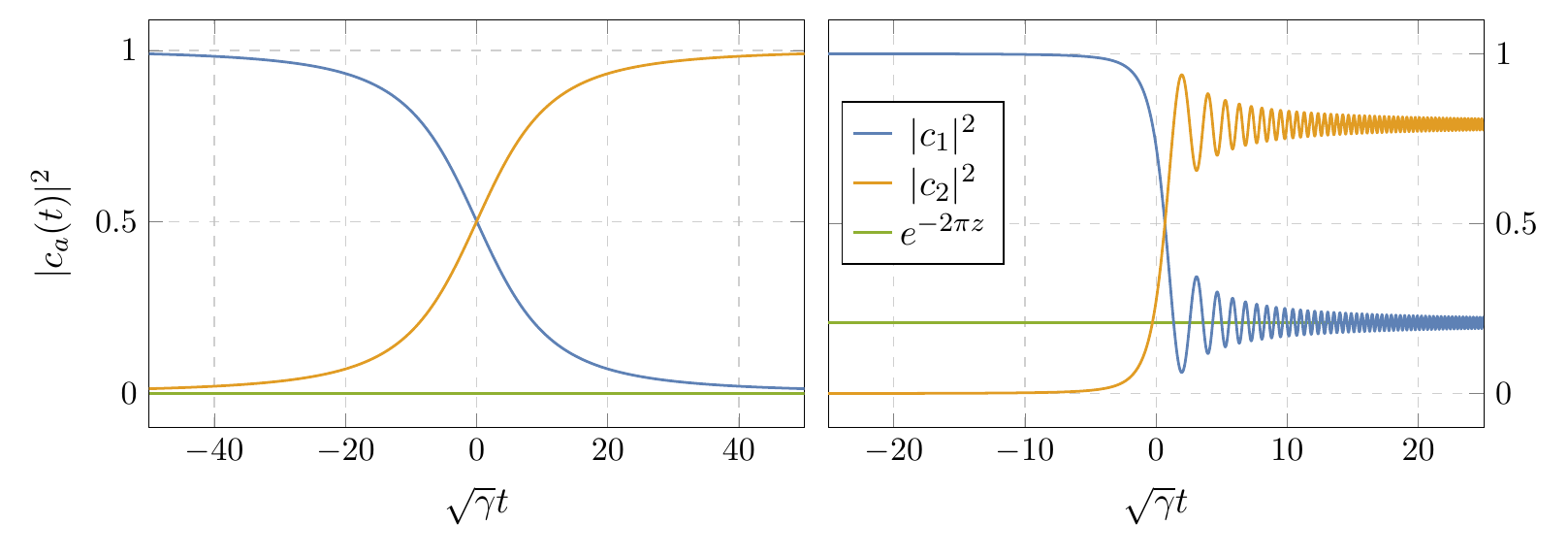}
      \caption{Adiabatic (\emph{left}) and non-adiabatic (\emph{right}) behavior of the Schr\"{o}dinger frame coefficients for transitions with Landau-Zener parameters $z = 25$ and $z = 1/2$, respectively. \label{fig:ad2by2}}
  \end{figure}

The model described by (\ref{eq:dressedFrameHamSimp}) has the advantage that it can be solved exactly, even away from the adiabatic regime.  As detailed in Appendix~\ref{app:twoState}, assuming that the state initially fully occupies the lowest energy eigenstate $|\psi(-\infty)\rangle \propto |E_{-}(-\infty)\rangle$, the population contained in the other eigenstate long after the transition is  
\begin{equation}
    |\langle E_+(\infty) | \psi(\infty)\rangle|^2  = \exp\left(-2 \pi z\right) , \label{eq:lzProb}
\end{equation}
where we have defined the (dimensionless) Landau-Zener parameter \cite{Landau,Zener,landau2013quantum} 
\begin{equation}
  z \equiv \frac{\eta^2}{\gamma}\, , \label{eqn:LZParam}
\end{equation}
which measures how (non-)adiabatic a given transition is.

In an adiabatic transition, with $z \gg 1$, the dressed frame Hamiltonian evolves slowly enough that we can ignore the other instantaneous eigenstate entirely. Intuitively, the system has enough time (as measured by $\eta$) to respond to a change in the dressed frame Hamiltonian, so that any fluctuations in the state's energy can decay.
As shown in the left panel of Fig.~\ref{fig:ad2by2}, an adiabatic transition causes the system to completely transfer its population from one state to another, a process known as \emph{adiabatic following}. Since the system tracks an instantaneous eigenstate, the magnitudes of the Schr\"{o}dinger frame coefficients change smoothly, and from the explicit expression (\ref{eq:2by2eigensystem}) we see that this transition happens on a timescale set by 
  \begin{equation}
    \Delta t \sim \frac{2 \eta}{\gamma}\,. \label{eq:lzTimeScale}
  \end{equation}
Intuitively, there is a finite ``resonance band'' of width $\Delta \Omega \sim 2 \eta$ in which this transition is active, and (\ref{eq:lzTimeScale}) is the time spent moving through it.

A non-adiabatic transition, with $z \lesssim 1$, is qualitatively different (see the right panel of Fig.~\ref{fig:ad2by2}). Before passing through the resonance (i.e.~before the ``scattering event''), the system remains in an instantaneous eigenstate and the Schr\"{o}dinger frame coefficients change smoothly. During the transition, however, the system will partly evolve into the other instantaneous eigenstate, not dissimilar from particle production via high-energy scattering. After this event, the system exists in a linear combination of eigenstates and they oscillate among one another at a frequency set by the energy difference $E_{+} - E_{-} \sim \gamma t$.
These oscillations eventually decay away on a timescale again set by (\ref{eq:lzTimeScale}), so that the Schr\"{o}dinger (dressed) frame coefficients have a well-defined limit as $t\to \infty$. Finally, if the transition is extremely non-adiabatic, $z \ll 1$, the system has no time to respond to the changing Hamiltonian, and so its state is unaffected by the~transition.

  \subsection{Multi-State Transitions}
  \label{sec:MultiState}
  
  A  qualitatively new feature appears when multiple, nearly degenerate states are involved in a transition. To illustrate this, we study a three-state extension of (\ref{eq:dressedFrameHamSimp}), described by the following dressed frame Hamiltonian
  \begin{equation}
    \mathcal{H}_\lab{D}(t) = \frac{\gamma t}{2} \begin{pmatrix} 
    1 & 0 & 0 \\
    0 & \minus 1 & 0 \\
    0 & 0 & \minus 1 \end{pmatrix} + 
    \begin{pmatrix}
      0 & \eta_{12} & \eta_{13} \\
      \eta_{12} & 0 & \eta_{23} \\
      \eta_{13} & \eta_{23} & 0 
    \end{pmatrix}. \label{eq:3state}
  \end{equation}
As I described in \S\ref{sec:TimeDependentTidalMoments}, this type of multi-state transition will be relevant for vector clouds, as they tend to have multiple states that are nearly degenerate in energy and connected via the gravitational perturbation. Indeed, (\ref{eq:3state}) describes any transition which involves a single state interacting with two other states that have the same azimuthal angular momentum and unperturbed energy. However, our focus on the three-state system is primarily pedagogical, as vector cloud transitions tend to involve four or more states. Fortunately, we will find that the main features of these more complicated multi-state transitions can be explained using the simpler three-state model.

  \begin{figure}[t]
          \centering
            \includegraphics[scale=0.91]{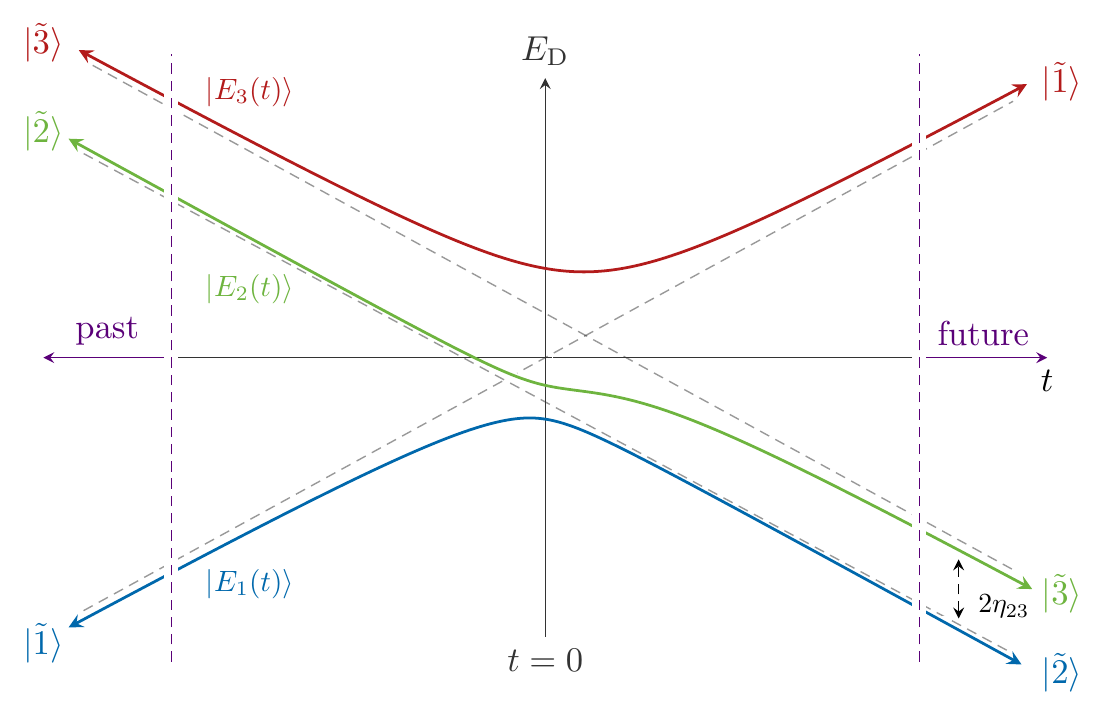}
            \caption{ Evolution of the instantaneous eigenstates of a coupled three-state system. \label{fig:threeStateEigVals}}
  \end{figure}

\begin{figure}
      \includegraphics[scale=0.82, trim=7 0 0 0]{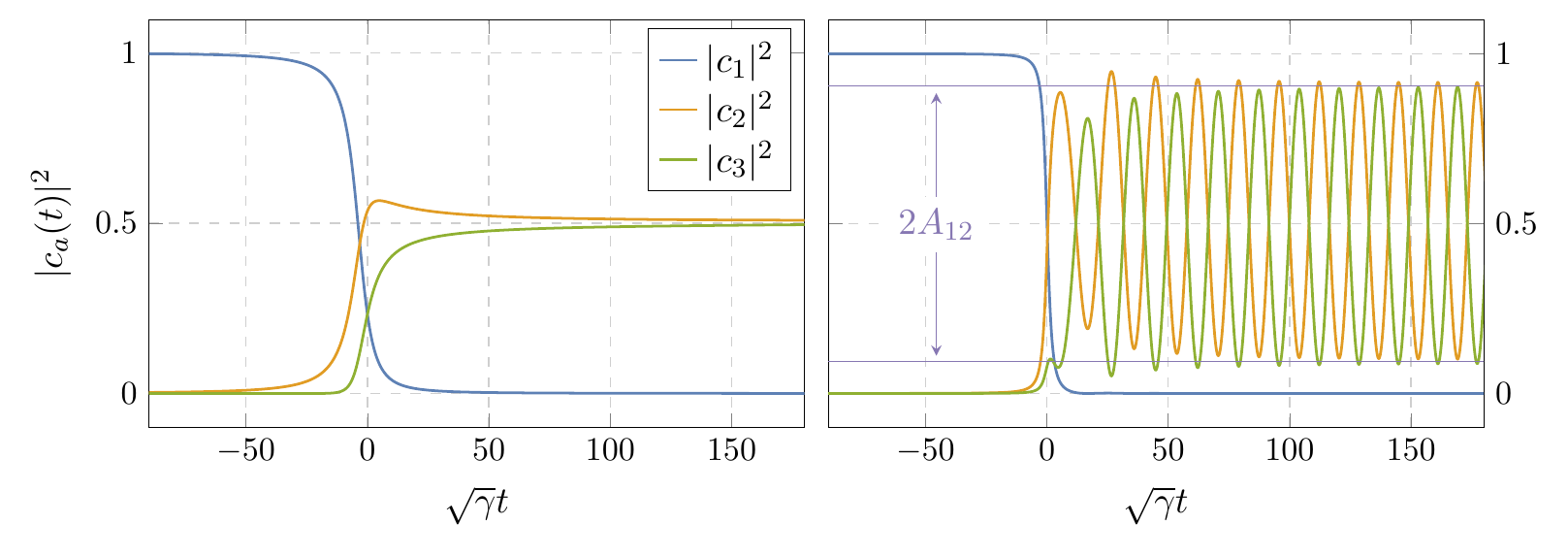}
      \caption{
      Adiabatic (\emph{left})
       and non-adiabatic (\emph{right}) evolution of the three-state system. Specifically, the adiabatic and non-adiabatic evolution are simulated with $(\eta_{12}, \eta_{13}, \eta_{23}) = (2.5, 1.5, 0.25)$ and $(\eta_{12}, \eta_{13}, \eta_{23}) = (4.75, 0.75, 3.5)$, respectively, in units of $\sqrt{\gamma}$.  The final state in the right panel oscillates with a frequency of about $2 \eta_{23}$.  } \label{fig:3state}
  \end{figure}

  It is again helpful to plot the instantaneous energy eigenvalues as a function of time, shown in~Fig.~\ref{fig:threeStateEigVals}. At late times, we see that there are two instantaneous energy eigenstates, $|E_1(t) \rangle$ and $|E_2(t)\rangle$, whose energy difference approaches a \emph{constant} as $t \to +\infty$. This is significant because, if the motion is non-adiabatic enough to excite the system into a combination of these two instantaneous eigenstates, there can be large, coherent oscillations with a \emph{fixed} frequency set by the interaction strength~$\eta_{23}$. Such oscillations are indeed seen in the right panel of Fig.~\ref{fig:3state}. This is impossible in a two-state transition (\ref{eq:dressedFrameHamSimp}), as the difference in energies necessarily diverges as $t \to +\infty$, and any oscillatory behavior induced by non-adiabaticity rapidly increases in frequency and decays away, as seen in the right panel of  Fig.~\ref{fig:ad2by2}.

  Let us understand this transition more quantitatively. At very early times, the first term in (\ref{eq:3state}) again dominates. However, there is now a degenerate subspace that is only lifted by $\eta_{23}$ and we cannot ignore it. It will thus be convenient to explicitly diagonalize this subspace.  I therefore introduce a new basis of states $|\tilde{a}\rangle$, which is related to the dressed frame basis $|a\rangle$ by  
  \begin{equation}
    |\tilde{1} \rangle = |1\rangle \,, \quad |\tilde{2}\rangle = \frac{1}{\sqrt{2}} \left(- |2 \rangle + |3 \rangle \right) , \quad\text{and}\quad |\tilde{3} \rangle = \frac{1}{\sqrt{2}} \left(|2 \rangle + |3 \rangle \right) .
  \end{equation}
  In this basis, the dressed frame Hamiltonian becomes 
    \begin{equation}
    \tilde{\mathcal{H}}_\lab{D} = \begin{pmatrix} 
      \gamma t & \tilde{\eta}_{12} & \tilde{\eta}_{13} \\
      \tilde{\eta}_{12} & - \gamma t - \eta_{23} & 0 \\
      \tilde{\eta}_{13} & 0 & - \gamma t + \eta_{23} 
    \end{pmatrix} , \label{eq:diagDressFrameHam}
  \end{equation}
  where we have defined the effective couplings
  \begin{equation}
    \tilde{\eta}_{12} = \frac{1}{\sqrt{2}} (\eta_{12} - \eta_{13}) \qquad \text{and} \qquad \tilde{\eta}_{13} = \frac{1}{\sqrt{2}} (\eta_{12} + \eta_{13})\,. \label{eqn:EffectiveEta}
  \end{equation}
  We see that the role of the coupling $\eta_{23}$ is to break the degeneracy between $|\tilde{2}\rangle$ and $|\tilde{3}\rangle$, and force $|\tilde{1}\rangle$ to become nearly degenerate with $|\tilde{2}\rangle$ at a different time than it becomes degenerate with~$|\tilde{3}\rangle$. As we will argue in~\S\ref{sec:Smatrix}, we can thus treat this three-state transition as a combination of two-state transitions.

  In the asymptotic past, the instantaneous energy eigenstates are then
  \begin{equation}
\hskip -5pt    |E_{1}(\minus \infty) \rangle =  \left(\begin{array}{@{\mkern3mu} r @{\mkern3mu}} 1 \\ 0 \\ 0 \end{array}\right), \hskip 4pt |E_{2}(\minus \infty) \rangle = \frac{1}{\sqrt{2}}\left(\begin{array}{@{\mkern-2mu} r @{\mkern3mu}} 0 \\ -1 \\ 1 \end{array}\right),  \hskip 4pt |E_3(\minus \infty) \rangle = \frac{1}{\sqrt{2}}\left(\begin{array}{@{\mkern3mu} r @{\mkern3mu}} 0 \\ 1 \\ 1 \end{array}\right),
  \end{equation}
  and from Fig.~\ref{fig:threeStateEigVals} we see that they permute in the asymptotic future,
  \begin{equation}
  \hskip -5pt  |E_{1}(+\infty) \rangle \propto |E_{2}(\minus \infty)\rangle , \hskip 4pt |E_{2}(+\infty) \rangle \propto |E_{3}(-\infty)\rangle, \hskip 4pt  |E_{3}(+\infty)\rangle \propto |E_{1}(\minus \infty)\rangle\,.
  \end{equation}
  Such a cyclic permutation is present in any transition which involves one state going into many degenerate states, as is visually apparent from Fig.~\ref{fig:threeStateEigVals}.

Where do the large, coherent oscillations in Fig.~\ref{fig:3state} come from?
Let us focus on the transition mediated by $\tilde{\eta}_{12}$, with associated LZ parameter $\tilde{z}_{12} \equiv \tilde{\eta}_{12}^2/\gamma$. Note that, even if all of the original LZ parameters are very large, $z_{ab}  = \eta_{ab}^2/\gamma \gg 1$, the transition may still be non-adiabatic if there is a cancellation between $\eta_{12}$ and $\eta_{13}$ in (\ref{eqn:EffectiveEta}), such that $\tilde{z}_{12} \lesssim 1$. I assume that the transition is fast enough ($\tilde{z}_{12} \sim 1$) to excite $|E_{2}(t)\rangle$, but slow enough ($\tilde{z}_{13} \gg 1$) that we can ignore  $|E_3(t)\rangle$. If the initial state is $|\psi(\minus \infty)\rangle \propto |E_{1}(\minus \infty)\rangle$, then at late times it becomes 
  \beq
  \begin{aligned}
    |\psi(t)\rangle \xrightarrow{\ t\to \infty \ }   & e^{-i \gamma t^2/2 - i \eta_{23} t}\sqrt{1 - e^{-\pi \tilde{z}_{12}}}\, |E_1(\infty)\rangle \\
    &  + e^{-i \gamma t^2/2 + i \eta_{23} t + i \delta} e^{-\pi \tilde{z}_{12}/2}\, |E_{2}(\infty)\rangle \,,
    \end{aligned}
  \eeq
  where the phase difference $\delta$ depends on $\gamma$ and $\eta_{ab}$. We thus see that, in the asymptotic future, there are oscillations in the Schr\"{o}dinger frame populations with frequency $2 \eta_{23}$.
  For example, we have 
  \begin{equation}
    |c_{2}(t)|^2 = \frac{1}{2} -  A_{12} \cos(2 \eta_{23} t + \delta)\, , \label{eqn:NeutrinoOsc}
  \end{equation}
  where $A_{12} = e^{-\pi \tilde{z}_{12}/2} \sqrt{1 - e^{-\pi \tilde{z}_{12}}}$. This amplitude achieves its maximum $A_{12} = \tfrac{1}{2}$ when $\tilde{z}_{12} = \tfrac{1}{\pi} \log 2$, and we see that these oscillations disappear both when the transition is either very adiabatic, so that the  state $|E_2(t)\rangle$ is never excited, or very non-adiabatic, so that the state $|E_1(t)\rangle$ does not survive.

  These slow oscillations will persist even if the state $|E_{3}(t)\rangle$ is excited by the transition, i.e. when $\tilde{z}_{12} \sim \tilde{z}_{13} \sim 1$. In this case, there will then be transient oscillations, like those seen in the two-state system and in the right panel of Fig.~\ref{fig:ad2by2}, which will eventually decay and leave only these fixed frequency oscillations. 
  However, if the transition is extremely non-adiabatic, such that both $\tilde{z}_{12}$ and $\tilde{z}_{13} \ll 1$, the system will always almost entirely occupy the state $|E_{3}(+\infty)\rangle = |E_{1}(\minus \infty)\rangle$ and the system will evolve very little in time.

  The adiabatic and (mildly) non-adiabatic evolution of this three-state system are shown in the left and right panels of Fig.~\ref{fig:3state}, respectively. The slow  oscillations created by a somewhat non-adiabatic transition are seen on the right. On the left, the completely adiabatic transition creates a final state that is a linear superposition of Schr\"{o}dinger frame states, i.e.~energy eigenstates. It is a special feature of this three-state system that the relative final state populations are independent of the coupling $\eta_{23}$. Indeed, if four or more states are involved in the transition, the final state populations do depend on the couplings $\eta_{ab}$, since the asymptotic energy eigenstates clearly depend on how the dressed frame Hamiltonian's degenerate subspace is lifted.

 We find that multi-state transitions generically yield final states that are superpositions of the dressed basis states (and thus the energy eigenstates). Almost all of the transitions that we will analyze in Section~\ref{sec:unravel} will be adiabatic, so this superposition can be easily approximated by an eigenvector of the coupling matrix $\eta_{ab}$, restricted to the degenerate subspace. As we mentioned at the outset of this section, these multi-state transitions qualitatively distinguish scalar and vector clouds, and we will exploit this to infer the spin of the boson from gravitational-wave observations in \S\ref{sec:modulatingfinite} and \S\ref{sec:final}.

\subsection{An S-Matrix Approach} 
\label{sec:Smatrix}

In \S\ref{sec:twoState}, we found that a two-state system initially occupying the lowest instantaneous energy eigenstate $|\psi(\minus \infty)\rangle \propto |E_{-}(\minus \infty)\rangle$ could transition into the other instantaneous eigenstate with an occupation ``probability''~(\ref{eq:lzProb}) with a well-defined asymptotic limit. Given the system's causal structure---the nature of the transition provides a well-defined notion of both asymptotic past and asymptotic future---it is tempting to interpret (\ref{eq:lzProb}) as defining the element of a scattering matrix, 
   \begin{equation}
       |\langle E_+(+\infty)| E_-(\minus \infty)\rangle|^2 = |S_{+-}|^2 = e^{-2\pi z}\,.
   \end{equation}
   However, a rigorous definition of the scattering matrix must be more subtle, as the three-state system does not necessarily have a well-defined, static final state. For instance, what is the scattering matrix describing the coherent oscillations in the right panel of Fig.~\ref{fig:3state}?

   The goal of this section is three-fold. We first generalize our discussion of the LZ transition beyond the simplified toy models like (\ref{eq:dressedFrameHamSimp}) and toward a description of the boson cloud throughout the entire binary inspiral. We will then introduce an ``interaction picture,'' leading to an interaction Hamiltonian that is well-localized in time. This is crucial for constructing a well-defined S-matrix.  We will leverage this localization to argue that different resonances during the inspiral decouple from one another, allowing us to ignore all but the finite number of states involved in a particular resonance. The behavior of the boson cloud during the full inspiral can thus be described as a series of localized ``scattering events,'' each of which can be analyzed individually. 
   
    \subsubsection*{Adiabatic decoupling}

      Let us first consider the simplest inspiral configurations: large, equatorial, quasi-circular orbits for which the Hamiltonian (\ref{eqn:Schr}) takes the form
      \begin{equation}
        \mathcal{H}_{ab} = E_{a} \delta_{ab} + \eta_{ab}(t) e^{-i \Delta m_{ab} \varphi_*(t)}\, ,\label{eq:multiStateDefPhase}
      \end{equation}
where $\Delta m_{ab} = m_a - m_b$ is the difference in azimuthal angular momentum between the states $|a \rangle$ and $|b\rangle$ and $\eta_{ab}(t)$ is the matrix of cross-couplings. 
The latter evolves slowly in time. Unlike in the previous sections, we will not truncate (\ref{eq:multiStateDefPhase}) to a finite set of states $\{|a\rangle\}$, but instead argue that such a truncation is possible. The indices in this equation should thus be understood to run over all states in the atomic spectrum.

      As before, we may define a time-dependent unitary transformation 
      \begin{equation}
        \mathcal{U}_{ab}(t) = e^{-i m_a \varphi_*(t)} \delta_{ab} \, ,\label{eq:multiDressedUnitary}
      \end{equation}
      that moves our system into a frame that rotates along with the binary companion. This isolates the ``slow'' motion responsible for LZ transitions and discards the perturbation's distracting fast motion. In this dressed frame, the Hamiltonian is given by
      \begin{equation}
        \left(\mathcal{H}_\lab{D}\right)_{ab} = \delta_{ab}\left(E_a  - m_a \dot{\varphi}_*\right)+ \eta_{ab}\,. \label{eq:multiStateDressedHam}
      \end{equation}
      This is analogous to (\ref{eq:dressedFrameHam}), though here we have not shifted by an overall reference energy.

      Because the gravitational perturbations $\eta_{ab}$ are always small compared to the energy eigenvalues~$E_a$, the instantaneous eigenstates $|E_{i}(t) \rangle$ are almost always\footnote{This argument fails for states that form a nearly-degenerate subspace that is predominantly lifted by the gravitational perturbation $\eta_{ab}$, as in the three-state toy model in \S\ref{sec:MultiState}. As we saw there, the perturbation cannot be ignored and the instantaneous eigenstates far away from the transition do not asymptote to a single dressed basis state. However, the following discussion also applies to these degenerate subspaces---as long as the system is not excited into a subspace, it decouples from the dynamics.} well-approximated by the gravitational atom's energy eigenstates---that is, $|E_{i}(t) \rangle \approx |a\rangle$ for some dressed state $a$.  We will 
      denote the state of the cloud by $|\psi (t)\rangle$. If the system evolves adiabatically, its population in each instantaneous eigenstate, $|\langle E_i(t) | \psi(t)\rangle|^2$, remains unchanged. Since these instantaneous eigenstates are well-approximated by the dressed states $|a \rangle$, the cloud's population in each dressed state, $|\langle a | \psi(t)\rangle|^2$, similarly remains unchanged. This means that, unless the system already has an appreciable population in a particular dressed state, we can simply ignore it.\footnote{Strictly speaking, this is only true at leading order, and this state does have a small effect on the states we do not ignore. However, this small effect can be incorporated using standard perturbative techniques.}

    This will always be the case unless there is a point in time when a subspace becomes nearly degenerate with the subspace the system occupies, in which case the approximation $|E_{i}(t)\rangle \approx |a \rangle$ fails. Because of the assumed hierarchy between the energies $E_a$ and the couplings $\eta_{ab}$, this can only happen when the instantaneous frequency $\Omega(t)$ is such that the diagonal elements for two or more states are approximately equal, 
    \begin{equation}
      \dot{\varphi}_* \equiv \pm \Omega(t) \approx \frac{E_a - E_b}{m_a - m_b}\,. \label{eq:resonanceCondition}
    \end{equation}
    To determine which states we need to keep track of in a given frequency range, we therefore simply need to find for which states this \emph{resonance condition} is met and include them with the states that the system already occupies.

    Critically, the width of the resonance defined by (\ref{eq:resonanceCondition}) is roughly set by $|\Delta \Omega| \sim \eta_{ab}$. Once
     these states go through a resonance, they very quickly decouple and the magnitudes of their coefficients become non-dynamical. Hence, if the resonances are widely separated, we may consider them as a sequence of events and analyze them individually. Associated to each event is thus an S-matrix, which describes how the system evolves through the resonance.

    This logic can be extended to more general orbital configurations. 
    As we discuss in Appendix~\ref{app:adFloTheo}, each pair of states is generically connected by a perturbation that oscillates at multiple frequencies.  The dressed frame transformation (\ref{eq:multiDressedUnitary}) must then be generalized beyond a simple unitary transformation in order to find a slowly varying dressed frame Hamiltonian like (\ref{eq:multiStateDressedHam}). This can be done, but since it requires a technology that would distract from our main story, we do not describe it here. Instead, we relegate this generalization to Appendix~\ref{app:adFloTheo}. Fortunately, the above discussion also applies there,
     \emph{mutatis mutandis}. To avoid getting bogged down by unnecessary technical details, we thus continue with the simplified setup described by~(\ref{eq:multiStateDefPhase}).

  \subsubsection*{Interaction picture} 

    We must define a scattering matrix with respect to a set of states which do not evolve in the asymptotic past or future. Since the dressed frame's instantaneous eigenstates are stationary (up to a phase) under adiabatic motion, it is natural to define an \emph{interaction frame} by expanding the general state $|\psi(t)\rangle$ as 
    \begin{equation}
      |\psi(t) \rangle = \sum_{i} e_i(t)\, e^{-i \int^t \!\ud t' \, E_i(t')} |E_{i}(t)\rangle\,, \label{eq:instantBasisExpansion}
    \end{equation}
    where the sum runs over all instantaneous eigenstates. From our previous discussion, we know that this expansion is approximately the same as the expansion in terms of dressed states
    \begin{equation}
      |\psi(t) \rangle = \sum_{a} c_a(t)\, e^{-i \int^t \!\ud t' \, (E_a - m_a \dot{\varphi}_*(t'))} |a \rangle\,, \label{eq:dressedBasisExpansion}
    \end{equation}
    at least far from any resonance. We can think of (\ref{eq:dressedBasisExpansion}) as the analog of the interaction picture in textbook quantum field theory in which one works with free particle states, while (\ref{eq:instantBasisExpansion}) is the analog of the  renormalized state basis. While either can be used, our focus will be on (\ref{eq:instantBasisExpansion}) since the dynamics are clearer in that basis.

  \begin{figure}
  		\centering
        \includegraphics[scale=0.9, trim=0 0 0 0]{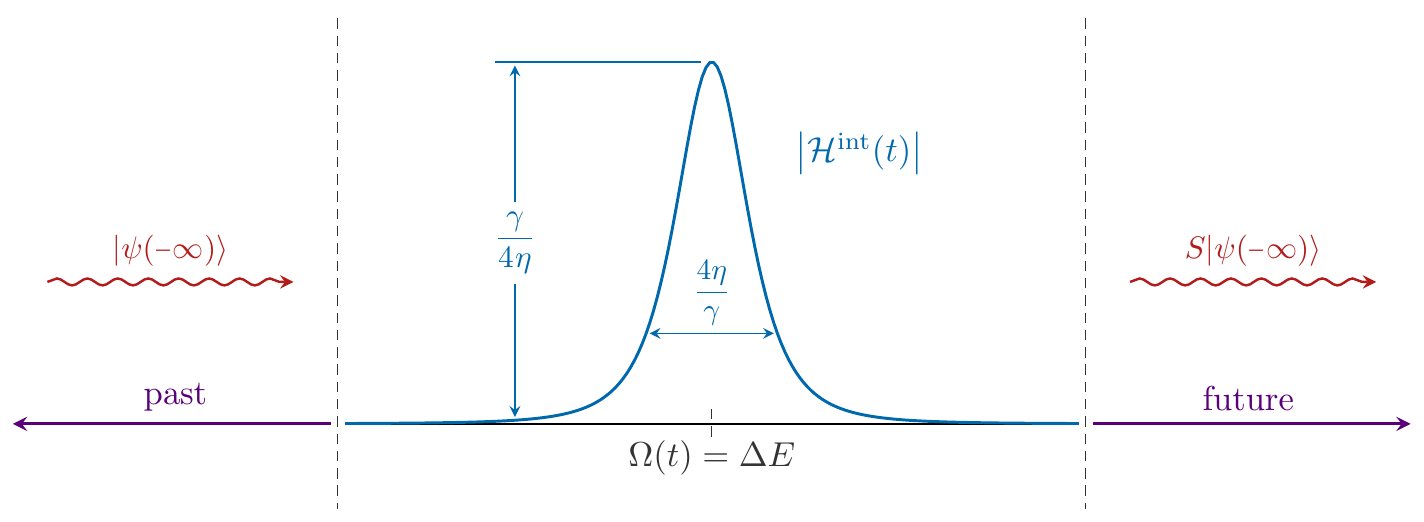}
        \caption{Interaction Hamiltonian for the two-state system as a function of time. We see that the interaction is localized over a width $4 \eta/\gamma$ near the resonance at $\Omega(t) = \Delta E$. \label{fig:interactionPotential}}
    \end{figure}

    The Schr\"{o}dinger equation simplifies in the instantaneous eigenstate basis (\ref{eq:instantBasisExpansion}) to 
    \begin{equation}
      i \frac{\ud e_i}{\ud t} = \mathcal{H}_{ij}^{\lab{int}}(t) \,e_j(t)\,,
    \end{equation}
    where we have defined the \emph{interaction Hamiltonian}\footnote{In general, we should also  include the adiabatic phase factor $e^{i \delta}$ in (\ref{eq:interactionPotential}), with $\delta = i \int^t\!\ud t' \, \langle E_{i}(t') | \partial_{t'} E_{i}(t')\rangle$. However, we can normalize the instantaneous eigenstates $|E_{i}(t')\rangle$ to  be real and set $\delta =0$.}
    \begin{equation}
       \mathcal{H}_{ij}^\lab{int} = \begin{dcases}
           \frac{i \langle E_i(t) | \dot{\mathcal{H}}_\lab{D}(t) | E_j(t) \rangle}{E_i(t) - E_j(t)} \exp\left(i \int^t \!\ud t' \,(E_i(t') - E_j(t'))\right) & i \neq j\, , \\
           \qquad \qquad\,\, 0 & i = j\, .
          \end{dcases} \label{eq:interactionPotential}
    \end{equation}
    Clearly, this term is relevant only when two or more energy levels become nearly degenerate. Furthermore, since eigenvalues repel, this interaction is only relevant for a short amount of time. This can be seen quantitatively for the two-state system from \S\ref{sec:twoState}, where the interaction Hamiltonian is explicitly
     \begin{equation}
      \mathcal{H}_{+-}^\lab{int}(t) = -\frac{i \gamma}{4 \es \eta} \frac{(\tau + \sqrt{1 + \tau^2})^{2 i \eta^2/\gamma}}{1 + \tau^2} \,e^{(2 i \eta^2/\gamma) \tau \sqrt{1 + \tau^2}}\,,\quad \text{with} \quad \tau = \frac{\gamma t}{2 \eta} \,.
    \end{equation}
     As pictured in Fig.~\ref{fig:interactionPotential}, this interaction is well localized in time about the  resonance and becomes irrelevant on a timescale  $\Delta t \sim 2 \eta/\gamma$, cf.~(\ref{eq:lzTimeScale}).

    The fact that this interaction quickly turns off allows us to define a notion of stationary asymptotic states---the instantaneous eigenstates with their dynamical phases extracted---and thus a scattering matrix
    \begin{equation}
      S = U^\lab{int}(\infty, -\infty)\,,
    \end{equation}   
    where $U^\lab{int}(t, t_0)$ is the time-evolution operator associated to the interaction Hamiltonian~$\mathcal{H}^{\lab{int}}$. In fact,  since the interaction is negligible between resonances, we can define a scattering matrix associated to each resonance,
    \begin{equation}
      S_i = U^\lab{int}(t_i + \Delta t_\lab{w}, t_i - \Delta t_\lab{w})\,,
    \end{equation}
where $t_i$ is the time at which the $i$-th resonance occurs and $\Delta t_\lab{w}$ is some multiple of the characteristic interaction timescale (\ref{eq:lzTimeScale}). For virtually all resonant transitions, this interaction timescale is much shorter than the time it takes for a binary to move to the next resonance. To good approximation, we may then think of the scattering matrix $S_i$ as evolving the cloud from the asymptotic past of the $i$-th resonance to its asymptotic future. As a result, each scattering event is effectively isolated from its neighbouring resonances and the cloud's evolution throughout the inspiral is well described by combining the $S-$matrices for all previous resonances (see Fig.~\ref{fig:inspiral}),\footnote{We must note that this separation into isolated events discards potentially crucial phase information, especially if we only consider the S-matrix magnitudes in anything other than the interaction picture. This phase information would be needed to characterize possible coherence effects. Fortunately, if the resonances are widely separated, it is very unlikely that any type of coherence can be achieved and it is thus a useful approximation to simply drop this information. However, since such coherent behavior is more likely for nearby resonances, it may be necessary to consider these resonances as a single event.}
    \begin{equation}
      U^\lab{int}(t, -\infty) \approx \prod_{i=1}^{k} S_i\,, \mathrlap{\qquad t_{k+1} > t > t_{k}\,,}
    \end{equation}
    for any time $t$ between the $k$-th and $(k+1)$-th resonance.

  There is one more simplifying approximation we can make. Note that the instantaneous eigenstates asymptote to dressed frame states far from the resonance, cf.~(\ref{eq:instantBasisExpansion}) and (\ref{eq:dressedBasisExpansion}). This implies that, unless the system occupies multiple instantaneous eigenstates which oscillate among one another at some fixed frequency as in \S\ref{sec:MultiState}, the \emph{magnitudes} of the S-matrix elements in the dressed frame basis will have well-defined limits far from the resonance. In fact, their magnitudes will equal those of the instantaneous eigenstate basis. This is useful, since it then is not necessary to compute the interaction Hamiltonian (\ref{eq:interactionPotential}) at all if we are only interested in the relative populations of the states that the cloud occupies after a resonance. We can instead work entirely in the dressed frame basis, in which the Hamiltonian takes a simple form, and analyze the resonance there.

  \subsubsection*{General S-matrix} 

    Since we will only need to consider the orbit shortly before and after a resonance, we may take the orbital frequency $\Omega(t)$ to be roughly linear in time (\ref{eq:linearizedOmeg}). This relies on the assumption that the resonance bandwidth, $\Delta \Omega \sim \eta$, is much smaller than the orbital frequency itself, cf.~(\ref{eqn:Omega-Quasi-Circular}), which we argued in \S\ref{sec:TimeDependentTidalMoments} is true for all the transitions we consider. Similarly, we may also assume that the level mixings $\eta_{ab}(t)$ are time-independent.

Using (\ref{eq:linearizedOmeg}), the dressed frame Hamiltonian (\ref{eq:multiStateDressedHam}) can be written as
    \begin{equation}
      \left(\mathcal{H}_\lab{D}\right)_{ab} = \mathcal{A}_{ab} + \mathcal{B}_a \delta_{ab} t \,,\label{eq:lzHam}
    \end{equation}
    where we have defined 
    \begin{equation}
    \begin{aligned}
      \mathcal{A}_{ab} &\equiv \delta_{ab} \left(E_a \mp m_a \Omega_0\right) + \eta_{ab} \, ,\\
      \mathcal{B}_{a} &\equiv \mp m_a \gamma\,.
      \end{aligned}
    \end{equation}
    This is known as the \emph{generalized Landau-Zener problem} \cite{Brundobler:1993smat}. We will be specifically interested in the magnitudes of the S-matrix in this basis, defined by
    \begin{equation}
      |S_{ab}| = \lim_{t \to \infty} |U_{ab}(t, -t)|\,,
    \end{equation}
    where $U_{ab}$ is the time-evolution operator associated with (\ref{eq:lzHam}). 
    Unfortunately, analytic solutions for this problem are only known in special cases. However, a small amount of progress can be made by noting that the Schr\"{o}dinger equation for this system is invariant under a collection of symmetry transformations \cite{Brundobler:1993smat}, and thus so are the S-matrix elements. The invariant combinations of parameters made from $\mathcal{A}_{ab}$ and $\mathcal{B}_{a}$ are the generalized Landau-Zener parameters
    \begin{equation}
      z_{ab} \equiv \frac{|\mathcal{A}_{ab}|^2}{|\mathcal{B}_{a} - \mathcal{B}_{b}|} = \frac{\eta_{ab}^2}{\gamma |\Delta m_{ab}|}\,, \label{eq:genLZ}
    \end{equation}
    which extends the original definition (\ref{eqn:LZParam}) to the general, multi-state system. The S-matrix elements must then be functions of these parameters and their combinations. 
    
    \section{Summary}
    
In this chapter, we studied the evolutions of the boson clouds when they are parts of binary systems. We showed that the gravitational perturbation sourced by a binary companion greatly enriches the dynamics of the system, by inducing mixing between different energy eigenstates of the cloud. Interestingly, just like the hydrogen atom under the influence of an external perturbation, these level mixings must obey certain selection rules, which are different for the scalar and vector clouds. At certain critical resonance frequencies, the level mixings become non-perturbative and are significantly enhanced. In order to investigate the dynamics of the clouds at these resonance frequencies, it was crucial that we carefully incorporated the gradual increase of the orbital frequency due to the shrinking motion of the orbit. In that case, we showed that the associated dynamics are analogous to the Landau-Zener transition in quantum mechanics. 

\vskip 2pt

We found that the transitions in the scalar case typically involve only two well-separated energy eigenstates, while those for the vector cloud can simultaneously involve multiple nearly-degenerate states. As a result, in the scalar case, the Landau-Zener transition typically forces the initial cloud to completely populate the state that it couples to. On the other hand, for the vector cloud, multiple levels can be occupied in the final state. The simultaneous occupation of multiple end states would result in neutrino-like oscillations between different vector eigenstates --- an effect that persists even after the cloud exits the resonance bands. Furthermore, a key parameter that dictates the nature of the transition is the Landau-Zener parameter (\ref{eqn:LZParam}), which quantifies the adiabaticity of the transition. As demonstrated in Figs.~\ref{fig:ad2by2} and \ref{fig:3state}, an adiabatic transition would lead to a smooth transition to the final states, while a non-adiabatic transition would result in further oscillatory features in the final states. 

\vskip 2pt

The phenomenological implications of these rich dynamics will be explored in detail in the next chapter. Crucially, we shall find that the backreaction of these effects on the orbit are sensitive to the mass and intrinsic spins of the boson fields. Furthermore, we saw in Figs.~\ref{fig:2by2eigen} and \ref{fig:threeStateEigVals} that the dynamics at each of these resonances can be described effectively by an S-matrix. As the binary experiences multiple resonances throughout its inspiral, the total evolution of the cloud can therefore be succinctly described by multiplying successive scattering matrices. Motivated by its various analogy with particle collider physics, we therefore refered to these types of binary systems as ``gravitational colliders."

\chapter{Observational Signatures} \label{sec:signatures}

I have described the foundations underlying the dynamics of the gravitational atoms when they are parts of binary systems. This chapter is dedicated to exploring the rich phenomenologies that arise from these ``gravitational collider". In particular, I will describe the effects of these dynamics on the gravitational waves emitted by the cloud and the binary system. Crucially, I will also explain how these signals can carry vital information about the masses and intrinsic spins of the ultralight bosons. This chapter is also based on the works in~\cite{Baumann:2018vus, Baumann:2019ztm}.

\section{Overview and Outline}

\begin{figure}[t]
        \centering
        \includegraphics[scale=0.9, trim=0 0 0 0]{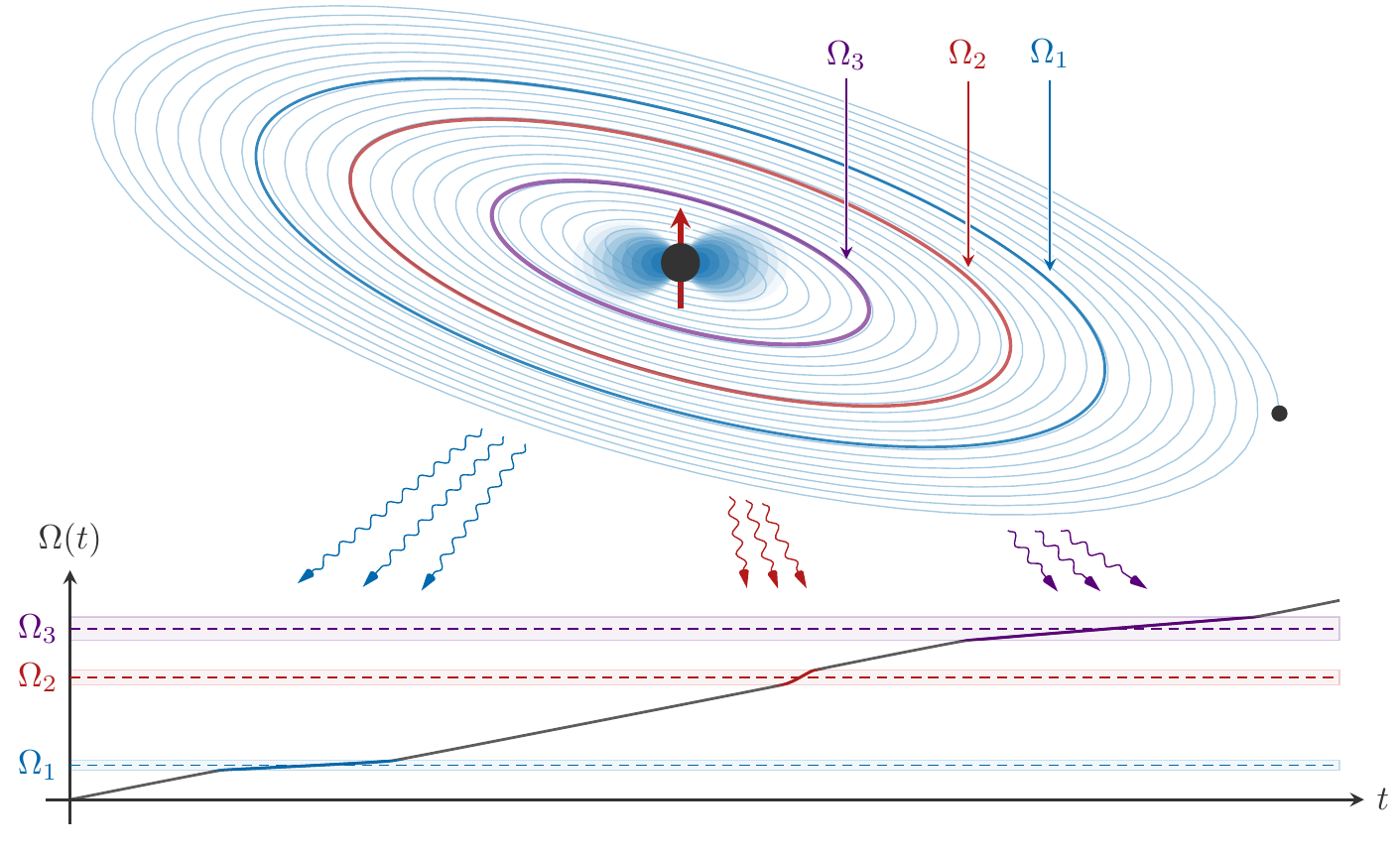}
    \caption{Illustration of the gravitational atom in a binary system. During the inspiral, the companion induces level mixings between different states of the cloud. This mixing is strongest when the orbital frequency $\Omega(t)$, which slowly increases due to gravitational-wave emission, matches the energy gap between the coupled states. During these resonant transitions, which occur at the frequencies $\Omega_i$, the change in the state of the cloud produces a backreaction on the binary's orbital dynamics. These effects either induce floating or sinking orbits across these resonance bands, thus leaving distinct imprints in the gravitational waves emitted by the binary. Moreover, minute time-dependent finite-size effects can further distinguish the nature of the boson cloud, including its  intrinsic spin.} \label{fig:AtomInBinary}
\end{figure}

One of the central goals in this chapter is to include the backreaction of the cloud's dynamics on the orbit, which was not studied in the previous chapter. As we will see, this backreaction can significantly affect the binary's orbital motion and associated gravitational-wave signals. For instance, any change in the cloud's angular momentum (which is most prominent during a resonance) must be balanced by a change in the orbital angular momentum of the binary, so that the total angular momentum is conserved. Depending on the orientation of the orbit and the nature of the transition, the binary can either absorb angular momentum from the cloud or release it. The resulting backreaction can either lead to a floating, sinking, or kicked orbit. In all cases, the induced changes in the orbital motion produce significant dephasings in the gravitational waves emitted by the binary, relative to the evolution without a cloud.\footnote{For intermediate and extreme mass ratio inspirals, with the cloud carried by the larger black hole, the backreaction on the orbit can be a dramatic effect.  If angular momentum is transfered from the cloud to the binary, long-lived floating orbits are naturally induced, creating a new source of continuous monochromatic gravitational waves.  In contrast, when the binary transfers angular momentum to the  cloud, it can experience a sudden kick instead. This greatly enriches the dynamics of the binary system, for instance by producing eccentric orbits. While these kicks are difficult to model quantitatively, they are distinctive qualitative signatures of the existence of a gravitational atom in a binary system.} See Fig.~\ref{fig:AtomInBinary} for an illustration of evolution of the orbital frequency as the binary transitions through multiple successive Landau-Zener transitions. Importantly, these dephasings are correlated with the positions of the resonances and therefore probe the spectral properties of the cloud. 

\vskip 4pt

Even though the backreaction on the orbit is a smoking gun for the existence of boson clouds, many degeneracies between the scalar and vector case still remain, since the transfer of angular momentum is not very sensitive to the internal structure of the clouds. To break these degeneracies, we need to probe the clouds through their time-dependent finite-size effects on the gravitational waveforms. This is possible because, as we found in Chapter~\ref{sec:signatures}, the many nearly degenerate states of the vector field can result in neutrino-like oscillations between different eigenstates, with characteristic frequencies set by the energy splitting between the states. Crucially, this behaviour is absent in the scalar case. The oscillations persist after the resonance and are therefore a key characteristic of clouds carrying intrinsic spin. Although these effects are difficult to model in detail, they are correlated with the position of the resonances, and therefore represent robust signatures of the boson clouds.

\vskip 4pt

The continuous and monochromatic gravitational waves emitted by the cloud would also be affected when the clouds are in binary systems. For instance, the Landau-Zener transition can result in a complete population of a decaying mode in the spectrum, resulting in a depletion (or significant attenuation) of the energy density stored in the cloud. Furthermore, this monochromatic signal can experience Doppler modulation due to the orbital motion of the binary. In short, both the signals from the cloud and the binary are unique imprints of the cloud and provide powerful new probes of physics beyond the Standard Model. Our findings motivate revisiting current waveform models and continuous-wave searches in order to target the unique signatures of ultralight particles in binary gravitational-wave searches. 

\vskip 4pt

The outline of this chapter is as follow: in Section~\ref{sec:Backreaction}, I describe the backreaction of time-dependent boson clouds on the orbital motion. I also provide an estimate of the finite-size effects of the cloud, as well as their time dependence induced by the Landau-Zener transitions. In Section~\ref{sec:cloud}, I discuss how the binary system affects the continuous gravitational-wave emission from the cloud. In Section~\ref{sec:unravel}, I explore the observational consequences of boson clouds in a binary system, and study the effects that can help us to unravel their atomic structure. Finally, in Appendix~\ref{app:AMTransfer}, I provide further details on the angular momentum transfer between the cloud and the orbit.
 
 \vspace{-1pt}
 
\section{Signals from the Binary} 
\label{sec:Backreaction}

In the previous chapter, we saw that Landau-Zener transitions can significantly redistribute energy and angular momentum between different states of the cloud.
The mass and spin of the cloud in a general superposition of states, at small $\alpha$, are given by
\beq
\begin{aligned}
M_c(t) &= M_{c, 0} \left(  |c_1|^2 +  |c_2|^2 + \cdots +  |c_N|^2 \right)  , 
\\
\textbf{S}_c(t) &  = S_{c, 0} \left( m_1 |c_1|^2 +  m_2 |c_{2}|^2 + \cdots + m_{N} |c_{N}|^2 \right) \hat{\textbf{z}} \, ,\label{eqn:Scloud}
\end{aligned}
\eeq
where  $S_{c, 0} \equiv M_{c, 0}/ \mu$ and $\hat{\textbf{z}}$ is a unit vector along the spin-axis of the black hole, cf.~Fig.~\ref{fig:BinaryPlane}, and the $c_i$ account for the population of the different states. We have ignored the $\hat{\mb{x}}$ and $\hat{\mb{y}}$ components of the spin, as they decouple for equatorial orbits. Any change in the energy and angular momentum of the cloud must be balanced by an associated change of the binding energy and orbital angular momentum of the binary's orbit. As we will see, this greatly enriches the dynamics of the system and the gravitational-wave signals emitted from the binary. 

\newpage

\subsection{Floating, Sinking, and Kicked Orbits}
\label{sec:floatdive}

To analyze the effect an LZ transition has on the binary orbit, it is most convenient to consider the transfer of angular momentum between the cloud and the binary. 
Ignoring the intrinsic spins of the black holes, conservation of 
angular momentum implies 
\beq
\frac{\d}{\d t} \left(\textbf{L} + \textbf{S}_{c}\right)  = \textbf{T}_{\rm gw} \, , \label{eqn:JConserved}
\eeq
where $\textbf{L}$ is the orbital angular momentum and $\textbf{T}_{\rm gw}$ is the torque due to gravitational-wave emission from the binary. 
Since (\ref{eqn:JConserved}) is a vectorial equation, the dynamics associated to angular momentum transfer can in general be very complicated, and depend on the relative orientations of the different angular momenta.  For simplicity, I will only consider equatorial orbits and ignore precession~\cite{Porto:2016pyg}. Hence, the magnitudes of the vectors in (\ref{eqn:JConserved}) obey the relation
\beq
 \frac{\d}{\d t}\left(L \pm S_{c} \right) = T_{\rm gw} \, , \label{eqn:JCons1}
\eeq
where, as usual, the upper (lower) sign denotes co-rotating (counter-rotating) orbits. 
Specializing further to quasi-circular orbits, we find that the orbital frequency evolves as (see Appendix~\ref{app:AMTransfer} for a derivation)  
\beq
\begin{aligned}
\frac{\d \Omega}{\d t}  & = \gamma \left( \frac{\Omega}{\Or}  \right)^{11/3} \! \\
& \pm  3 R_J \,\Or \!\left( \frac{\Omega}{\Or} \right)^{4/3}  \! \frac{\d }{\d t} \left[ m_1 |c_1|^2 + m_{2} |c_{2}|^2 + \cdots + m_{N} |c_{N}|^2  \right]  , \label{eqn:CircularBackreaction}
\end{aligned}
\eeq
where $\gamma$ was defined in (\ref{eqn:circleRate}), and $R_J$ is the ratio of the spin of the cloud to the orbital angular momentum of the binary at the resonance frequency,
\beq
R_J \equiv \left( \frac{ S_{c, 0}}{M^2} \right) \frac{(1+q)^{1/3}}{q} (M \Or)^{1/3} \, . \label{eq:ratioAngMom}
\eeq  
Far from the transition, the occupation densities in (\ref{eqn:CircularBackreaction}) are constant and the instantaneous frequency during the inspiral is well-approximated by the standard quadrupole formula, cf.~(\ref{eqn:Omega-Quasi-Circular}). During the transition, on the other hand, the angular momentum transfer between states with different $m$ exerts an additional torque on the binary. Crucially, the direction of the torque depends on various factors, including the orientation of the orbit, the signs of the azimuthal quantum numbers, and whether the occupation densities are growing or decaying in time. For simplicity, we set $m_{2} = m_{3} = \cdots = m_{N} $,\footnote{This turns out to be the case for any transition induced by the gravitational quadrupole $\ell_* = 2$.} such that near the resonance (\ref{eqn:CircularBackreaction}) simplifies to
\beq
\begin{aligned}
\frac{\d \Omega}{\d t}  \simeq \gamma  \mp 3\hskip 1pt \Delta m \hskip 2pt R_J  \, \Or\, \frac{\d |c_1 (t)|^2}{\d t} \, ,  \label{eqn:CircularBackreaction2}
\end{aligned}
\eeq
where we have defined $\Delta m \equiv m_2 - m_1$ and used conservation of the occupation densities during the transitions.

\begin{figure}[t]
\centering
\makebox[\textwidth][c]{ \includegraphics[scale=0.78, trim = 0 0 0 0]{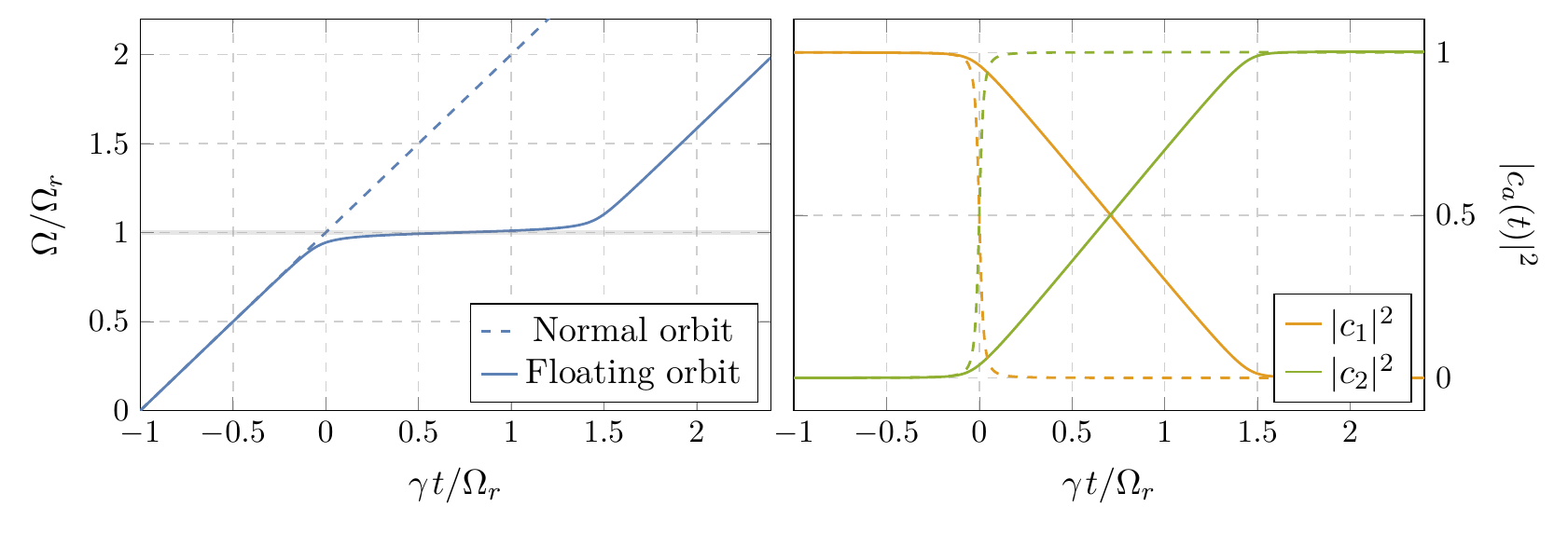}}
\caption{Evolution of the orbital frequencies (\textit{left}) and the occupation densities (\textit{right}) of an unbackreacted orbit ({\it dashed lines}) and a floating orbit ({\it solid lines}). Shown is a transition near the resonance frequency $\Omega_r = 5\mu \alpha^2/144$ with $\alpha = 0.07$ and $q=0.1$.  We also assumed that the parent black hole spun maximally, $S_{c, 0}/M^2 = 1$, before it grew the cloud.
The transition begins
when the binary enters the resonance band, denoted by the thin gray band about $\Omega = \Omega_r$.} 
\label{fig:Floating}
\end{figure}

By choosing the initial condition $|c_1(-\infty)|^2 = 1$, the time derivatives of $|c_1(t)|^2$ must decrease during the transition.  When $\Delta m < 0$, the cloud loses angular momentum to the orbit, forcing a co-rotating inspiral to stall and counter-rotating inspiral to shrink faster. 
These phenonema are reversed between co-rotating and counter-rotating orbits when $\Delta m > 0$.  To investigate the detailed dynamics of these backreaction effects, we must solve (\ref{eqn:CircularBackreaction2}) numerically. The solution in Figs.~\ref{fig:Floating} and~\ref{fig:Sinking} display the floating and sinking orbits, whose properties we describe below.

\subsubsection*{Floating orbits}

When the torque acts against the shrinking of the orbit, we find that LZ transitions naturally induce \textit{floating orbits}~\cite{Press:1972zz}.\footnote{This floating mechanism is distinct from the one studied in~\cite{Zhang:2018kib}, which assumes $\gamma=0$. It is also different from the analysis in~\cite{Cardoso:2011xi}, which relies on modifications of General Relativity.} This is shown in Fig.~\ref{fig:Floating} for a two-state transition, where the orbital frequency increases more slowly in the resonance band.
More precisely, the rate of change in the frequency of gravitational waves emitted by the binary, $f_{\rm gw}$, during the transition satisfies (cf. Appendix~\ref{app:AMTransfer})
\beq
\left( \frac{\d f_{\rm gw}}{\d t} \right)_{\rm float} =  \frac{1}{1 + 3  R_J  \, |\Delta m \Delta E| / (4 \eta)} \left(  \frac{\gamma}{\pi} \right)  , \label{eqn:floatingChirp}
\eeq
which is clearly smaller than the unperturbed rate, $\gamma / \pi$. Interestingly, in the large backreaction limit, (\ref{eqn:floatingChirp}) asymptotes to zero and never turns negative.\footnote{While the chirp rate (\ref{eqn:floatingChirp}) was derived assuming perfect adiabaticity, this observation remains true even in numerical simulations. However, if the system is tuned to be nearly non-adiabatic, there may be small oscillations in the instantaneous frequency. The parameters for which these oscillations occur are not physically realizable and so we do not pursue them further.} This means that the orbit can at most float, but never grows during a transition.
Furthermore, the frequency of the gravitational waves emitted during the floating phase, 
$f_{\rm float}$, is directly related to the resonance frequency
\beq
f_{\lab{float}} = \frac{\Or}{\pi} = \frac{1}{\pi} \left| \frac{\Delta E}{\Delta m} \right|   \, . \label{eqn:floatingMean}
\eeq
A measurement of this approximately monochromatic gravitational-wave signal thus provides a direct probe of the spectral properties of the boson cloud. In Section~\ref{sec:unravel}, we will describe the phenomenological consequences of these floating orbits in more detail. In particular, we will discuss how the duration of the floating depends on the parameters of the system.

The description above remains qualitatively unchanged for multi-level transitions, as they typically satisfy $m_2 = \dots = m_N$.
Moreover, since multi-level transitions necessarily involve nearly degenerate excited states, $E_2 \approx \cdots \approx E_N$, the expression (\ref{eqn:floatingMean}) remains an excellent approximation to the gravitational-wave frequency emitted by the corresponding floating orbits. The only quantitative difference is that $\eta$ must be replaced by an effective coupling $\eta_\lab{eff}$ that characterizes the multi-state transition, which is well-approximated by the Pythagorean sum of the diagonalized couplings $\tilde{\eta}_{1a}$, cf.~(\ref{eq:diagDressFrameHam}). We discuss its precise definition in Appendix~\ref{app:AMTransfer}.

Another key feature of these floating orbits is that the adiabaticity of the transition is \textit{enhanced}. 
In analogy to the LZ parameter (\ref{eqn:LZParam}), the degree of adiabaticity of the evolution can now be quantified by the parameter $z^\prime \equiv \eta^2/\dot{\Omega}$. 
Floating orbits, with $\dot{\Omega} < \gamma$, thus enhance the adiabaticity of the transition, $z^\prime > z$.  The predictions made by adiabatically following the instantaneous eigenstates of the system, such as the final-state occupation densities, are therefore robust.

\subsubsection*{Sinking and kicked orbits}

\begin{figure}[t]
\centering
\makebox[\textwidth][c]{\includegraphics[scale=0.78]{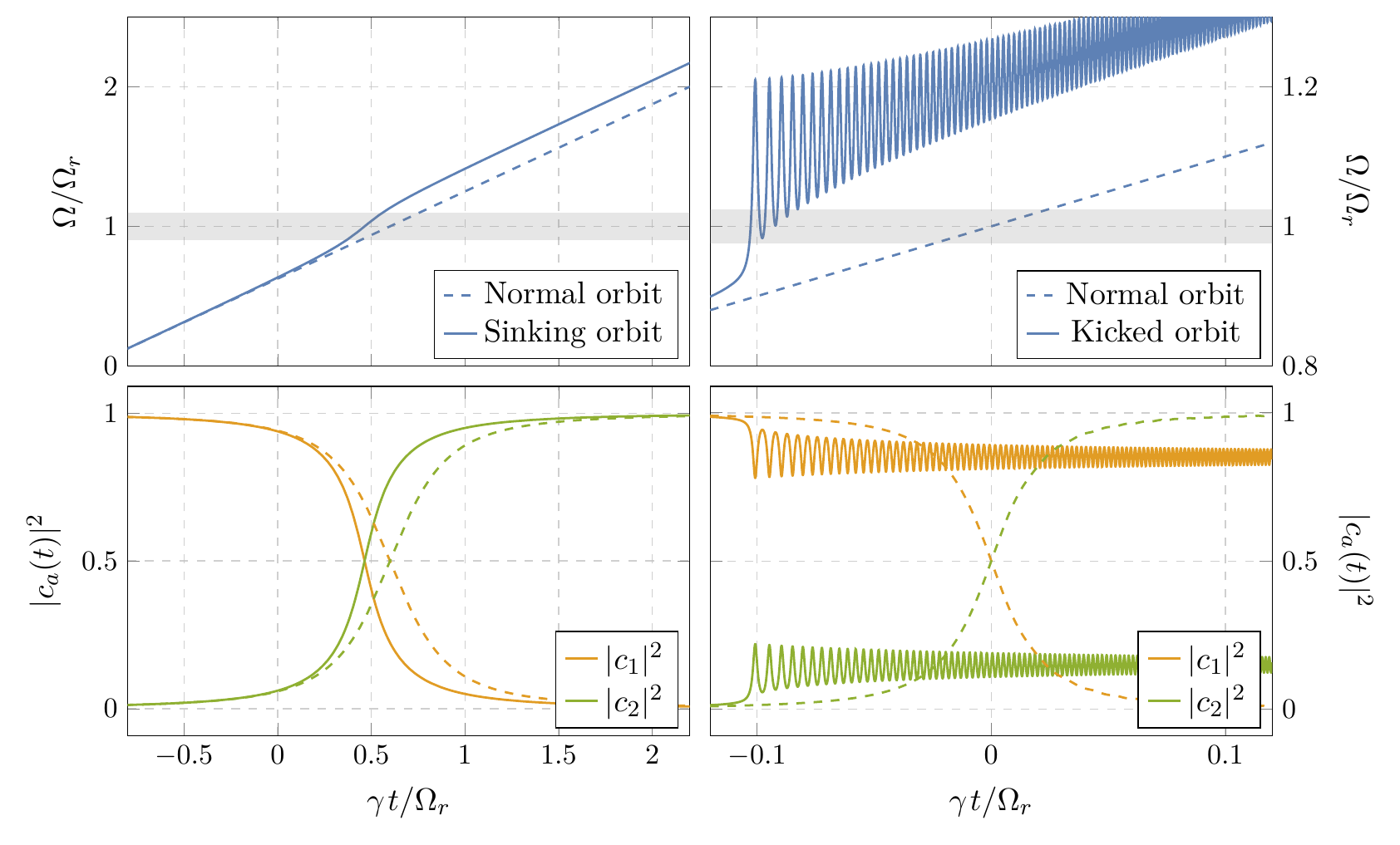}}
\caption{ Sinking (\emph{left}) and kicked (\emph{right}) orbits, with their corresponding occupation densities below. As in Figure~\ref{fig:Floating}, this transition is near the resonance $\Omega_r = 5\mu \alpha^2/144$ with $\alpha = 0.07$.  However, we flip the sign of~(\ref{eqn:CircularBackreaction2}) so that backreaction causes the orbit to either sink ($q =1$) or kick ($q = 0.1$).} 
\label{fig:Sinking}
\end{figure}

The transition can also exert a torque that increases the orbital frequency of the binary, and therefore temporarily accelerates the shrinking of the orbit.
If the effect is weak, we call it a {\it sinking orbit} (see the left panel in  Fig.~\ref{fig:Sinking}). As long as the adiabaticity of the transition is preserved, the chirp rate
of the emitted gravitational waves is\hskip 1pt\footnote{While (\ref{eqn:KickingChirp}) naively admits solutions that are divergent and negative, this simply indicates the breakdown of the adiabatic approximation in those corresponding regions of parameter space, cf.~(\ref{eqn:Non-adiabaticity}). \label{footnote:non-adiabatic}}  
\beq
 \left( \frac{\d f_{\rm gw}}{\d t} \right)_{\rm sink}  = \frac{1}{1 - 3  R_J  |\Delta m \Delta E| / (4 \eta)} \left( \frac{\gamma}{\pi} \right)   , \label{eqn:KickingChirp}
\eeq
which differs from (\ref{eqn:floatingChirp}) by an important sign. 
Compared to the unperturbed orbits, these sinking orbits spend a much shorter amount of time in the region where the LZ transition is efficient.

A sufficiently strong ``kick,'' on the other hand, can destroy any initial adiabaticity. Using $ \dot{\Omega} \lesssim \eta^2$ as a diagnostic, we estimate that adiabaticity is preserved whenever
\beq
R_J \lesssim \frac{4 \eta \left( 1 - \gamma/\eta^2 \right)}{3  \left| \Delta m \Delta E \right|  } \simeq \frac{4 \hskip 1pt  \eta }{3  \left| \Delta m \Delta E \right| } \, , \label{eqn:Non-adiabaticity}
\eeq
where the second relation assumes an initial adiabatic transition, $\gamma \ll \eta^2$. As we discussed in Section~\ref{sec:gcollider}, the resulting non-adiabatic transition yields drastically different final states, most prominently through the presence of oscillations in the occupation densities of final state. These oscillations transfer angular momentum between different states of the cloud, even after the transition. As shown in bottom panel of Fig.~\ref{fig:Sinking}, this effect also modulates the quasi-circular motion of the binary, leaving dramatic oscillatory features in the binary's gravitational-wave signal. Having said that, we caution that (\ref{eqn:CircularBackreaction2}) assumes that the binary motion is always well-described by a quasi-circular orbit. This need not be the case---strong non-adiabatic transitions in which $\dot{\Omega} \gtrsim \Omega^2$ can induce significant eccentricity (or even unbind the orbit). The full phenomenology of the kicked orbits can therefore be much richer than what was described above, and deserves a more detailed investigation. We will focus thus primarily on sinking orbits, where this backreaction is weak enough to preserve both adiabaticity and the quasi-circularity of the initial orbit.

\subsection{Finite-Size Effects}

In addition to the backreactions on the orbit during resonances, the cloud can also reveal itself in the gravitational-wave signal through subtle modifications in the waveform due to various finite-size effects. These effects were elaborated in \S\ref{sec:dynamics} and \S\ref{sec:waveforms}, where I described their impacts on the dynamics of the binary system and the associated gravitational waveform. In the following, we describe how these effects are manifested specifically for the cloud.

\subsubsection*{Spin-induced quadrupole} \label{sec:kappa}

The spinning motion of the cloud generates a hierarchy of mass multipole moments. For simplicity, we consider a cloud that occupies a \emph{single} eigenstate. Its mass quadrupole moments then experience $\alpha$-suppressed oscillations with frequency $\mu$, which are responsible for the cloud's well-studied decay via gravitational-wave emission \cite{Arvanitaki:2010sy,Yoshino:2014}.  For example, denoting the spherical harmonic decomposition of the mass multipole moments by $Q_{\ell m}$, the axisymmetric and non-axisymmetric quadrupole moments of the scalar $|2 \es 1 \es 1 \rangle$ state are 
\beq
Q_c \equiv Q_{20} = -6 M_c r_c^2 \quad \text{and} \quad Q_{2, \pm 2} \sim \, \alpha^2 Q_c \hskip 1pt e^{\pm 2 i \mu t} \, . \label{eqn:QcSingle}
\eeq
Since $\mu \gg \Omega(t)$, these oscillations average out over an orbital period and therefore do not affect the dynamics of the binary significantly. This is the generic behavior of all clouds occupying a single eigenstate---finite-size effects in these cases are dominated by their time-independent component.

In \S\ref{sec:dynamics}, we saw that these spin-induced moments would perturb the dynamics of the binary system and leave imprints in their gravitational waveforms. For the dominant axisymmetric quadrupole moment, this backreaction is often parameterized through the relation (\ref{eqn:Quad}), where $\kappa$ is the dimensionless quadrupole parameter. A rigorous computation of $\kappa_c$, which is the quadrupole parameter for the cloud, would require incorporating the backreaction on the spacetime geometry, which is beyond the scope of this work.\footnote{Exact solutions with (complex) scalar hair around spinning black holes have been studied numerically in the literature, see e.g. \cite{Herdeiro:2014goa}. In principle, these quasi-stationary spacetimes can also have $\kappa \gg 1$.} Instead, we will assume that the cloud dominates the contribution to the multipole moments of the cloud, and estimate $\kappa_c$ by comparing the cloud's mass quadrupole, $Q_c$, to $J^2/M$, where $M$ and $J$ are the {\it total} mass and angular momentum of cloud. We choose to normalize $\kappa_c$ with $(M, J)$ of the cloud, instead of $(M_c, J_c)$ of the cloud itself, because {\it i}\hskip 1pt) the quantities $(M, J)$ can be directly measured through gravitational-wave observations, and {\it ii}\hskip 2pt) they are conserved throughout the evolution (up to small losses due to gravitational-wave emission). The parameters $(M, J)$ therefore also coincide with the mass and angular momentum of the \textit{initial} black hole  prior to the formation of the cloud.

From the stress-energy tensor of the scalar field, we find for the $\ket{211}$ state
\beq
 \kappa_c(\alpha)\, \geq \, - \frac{Q_c(\alpha)}{M^3} \sim  10^3 \left(\frac{M_c(\alpha)/M}{0.1}\right)\left(\frac{0.1}{\alpha}\right)^4\, ,
\eeq 
where we have imposed the weak cosmic censorship condition, $J \leq M^2$~\cite{Penrose:1969pc}, to obtain the lower bound. 
The effect on the gravitational-wave phase then scales as
\beq
\kappa_c(\alpha) \hskip 1pt v^4 \hskip 1pt \chi^2\, \gtrsim \, 10^{-2} \left(\frac{M_c(\alpha)/M}{0.1}\right) \left(\frac{v}{\alpha/2}\right)^4  \, ,\label{kappac}
\eeq 
where we used $\chi \simeq 1$, and assumed that the initial black hole is rapidly rotating (which is required for the cloud to form in the first place). 
Notice that, in the regime of validity of the perturbative expansion, the relative velocity satisfies  $v \lesssim \alpha/2$.\footnote{For $q \lesssim 1$, the virial theorem implies $v \simeq (\alpha/2) \sqrt{r_c/R_*}$ and hence $v \lesssim \alpha/2$ in the regime of validity of the multipole expansion, $r_c < R_*$.}
 For $v > \alpha/2$, the companion experiences a smaller amount of the cloud according to Gauss' law. Even though larger relative velocities may seem favorable, the reduction in the effective mass of the cloud will dominate, leading to negligible finite-size effects once the companion enters the cloud.

\subsubsection*{Tidal deformations}

In addition to the spin-induced multipole moments, the cloud may also acquire induced multipoles in the presence of a gravitational perturbation. This tidal deformation effect is often parameterized through the relation (\ref{eqn:love}), where $\lambda$ is the dimensionless Love numebr. Since the cloud is much less compact than an isolated black hole, but carries a significant fraction of the mass and angular momentum of the system, we can,  in principle, have large Love numbers for the cloud system.  Although a detailed computation is beyond the scope of this work, on dimensional grounds we expect $\lambda_c$, which is the Love number of the cloud, to scale as
\beq
\label{tildece}
\lambda_c (\alpha)  \,\sim\, \left(\frac{M_c(\alpha)}{M}\right) \left(\frac{r_c}{2r_g}\right)^4 \,\sim\, 10^7 \left(\frac{M_c(\alpha)/M}{0.1}\right) \left(\frac{0.1}{\alpha}\right)^8  \, .
\eeq
The parameter $\lambda_c$ enters in the phase of the waveform at 5PN order, and its imprint on the signal from the binary therefore scales as
\beq
\lambda_c (\alpha) \hskip 2pt v^{10} \sim 10^{-6} \left(\frac{M_c(\alpha)/M}{0.1}\right) \left( \frac{\alpha}{0.1}\right)^2 \left( \frac{v}{\alpha/2} \right)^{10} \,,
\eeq
where $v \lesssim \alpha/2$ in the regime of validity of the multipole expansion. While this PN order seems high, the lack of standard model background in Einstein's theory offers a powerful opportunity to probe the dynamics of vacuum spacetimes in General Relativity, through precision gravitational-wave data.

\subsection{Time-Dependent Dephasings}
\label{sec:modulatingfinite}

In the previous subsection, we focused on clouds that occupy a single eigenstate. In particular, we found that these finite-size effects in these cases are time independent. However, as we saw in Section~\ref{sec:gcollider}, Landau-Zener transitions can lead to the simultaneous occupation of different energy levels for the vector field. We will therefore relate the occupation densities $|c_a(t)|^2$ to the time-dependences of the multipole moments of the cloud in this subsection.

\subsubsection*{Superpositions and oscillations}

For scalar clouds, a typical adiabatic Landau-Zener transition forces an initial state to fully transfer its population to another state with a different shape. As a result, the effective parameters $\kappa$ and~$\lambda$ evolve in time. Interesting multipolar time dependences are possible if the cloud occupies \emph{multiple} eigenstates, which is typical for vector clouds. In a general time-dependent superposition, the axisymmetric quadrupole moment is approximately
\beq
Q_c(t) = \sum_{a} |c_a (t)|^2 \, Q_{c, a} + \sum_{\substack{a \neq b}}  \delta_{m_a, m_b}\,|c_a(t) c_b(t)|  \, Q_{ab}  \cos ( \Delta E_{ab} \,t) \,  , \label{eqn:QAxisymmetric}
\eeq
where we have discarded terms that are either subleading in $\alpha$ or rapidly oscillating. The first term represents the weighted sum of the individual axisymmetric moments $Q_{c, a}$ of each state~$|a \rangle$, while the second represents additional contributions that arise from \textit{interference effects}, which occur between different states with the same azimuthal angular momenta. Crucially, the $Q_{ab}$'s are not $\alpha$-suppressed and may lead to large oscillations in $Q_c(t)$ at frequencies set by the energy differences~$\Delta E_{ab}$. Furthermore, there can be similar interference effects in the non-axisymmetric moments if the superposition involves states with different azimuthal quantum numbers, as may happen after a kick-induced non-adiabatic transition.

There are, in fact, two sources of oscillatory behavior in (\ref{eqn:QAxisymmetric}). As we discussed in \S\ref{sec:MultiState}, a mildly non-adiabatic multi-state transition can excite long-lived oscillations in the occupation densities $|c_{a}(t)|$, with frequencies set by the gravitational perturbation $\eta_{ab}$. It is clear from (\ref{eqn:QAxisymmetric}) that the multipole moment will inherit these oscillations. On the other hand, even if the $|c_a(t)|$ are constant, there are still oscillations in the multipole moment since the system occupies multiple, non-degenerate energy eigenstates, each with their own spatial profile. These are similar to neutrino oscillations,\footnote{The oscillations are neutrino-like in the following sense: after the resonance, the system occupies an instantaneous eigenstate (like a ``flavor eigenstate" for neutrinos) which, away from the interaction, becomes a superposition of energy eigenstates (like ``mass eigenstates" for neutrinos).} 
and are present in the final states of both adiabatic and non-adiabatic multi-state transitions. Notice that their frequencies depend only on the energy differences $\Delta E_{ab}$, which have been calculated~\cite{Baumann:2019eav}.

While we have only explicitly demonstrated this for the spin-induced quadrupole moment, these qualitative features also apply for the higher-order spin-induced moments and the tidally-induced moments. In summary, whenever the cloud occupies a superposition of energy eigenstates, the mass moments and tidally-induced moments oscillate with unsuppressed amplitude at frequencies that are slow compared to the orbital frequency.\footnote{In principle, these oscillating quadrupoles are also new sources of gravitational waves emitted by the cloud. While we leave the study of their emission rates to future work, we note that their frequencies are typically very small and so we do not expect that this is an efficient decay channel for the cloud.}   A detailed study of the impact that  these dynamical finite-size effects have on the orbit, and ultimately on the waveform, is beyond the scope of this work. Yet, it is clear that they serve as important relics that can help us decode the nature of an LZ transition that occurred in the past.  As we will discuss in \S\ref{sec:final}, correlating these oscillatory finite-size effects with the dephasing induced by floating and sinking orbits will serve as a powerful probe of the mass and spin of the ultralight particles.

\subsubsection*{Cloud depletion}

Since a transition can populate decaying states, the energy and angular momentum of the cloud
 can also be reabsorbed by the central black hole over the typical decay timescales of these modes.
The mass of the cloud approximately evolves as
\beq
\begin{aligned}
\frac{M_{c} (t)}{M_{c, 0} } &= \exp \left( \sum_a 2 \, \Gamma_a \!\int^t_{0}\! \d t^\prime \,|c_a(t^\prime)|^2 \right) \equiv \exp \big[ - 2 \mathcal{A}(t, 0) \big]  \, , \label{eqn:CloudMassRatio}
\end{aligned}
\eeq
where the quantity $\mathcal{A}$ is called the depletion estimator. Since the spin-induced and tidally-induced moments scale linearly with the mass of the cloud, cf. (\ref{kappac}) and (\ref{tildece}), this decay can introduce additional time dependence in the various effects discussed above, which in turn also affect the waveforms.  
As we have seen in Chapter~\ref{sec:spectraAtom}, the depletion is generally slower than the oscillation timescale discussed in (\ref{eqn:QAxisymmetric}). During the transition, they may therefore appear as modulations of the dominant finite-size effects. Importantly, this depletion persists even after the cloud has passed the resonance and can then be observed even away from resonances, thereby providing another unique signature of the boson clouds.\footnote{The total amount of depletion during a resonance was estimated in \cite{Baumann:2018vus} and \cite{Berti:2019wnn}, assuming a constant resonance frequency. As we show in this thesis, LZ transitions lead to a much longer depletion time which overwhelms the results in \cite{Baumann:2018vus, Berti:2019wnn} in almost all scenarios.}

\section{Signals from the Cloud}
\label{sec:cloud}

We described in \S\ref{sec:Hatom} that the real boson cloud is a source of continuous, nearly monochromatic gravitational-wave signal~\cite{Arvanitaki:2010sy, Yoshino:2014}. We elaborate on the detectability of this signal here, and discuss how it can develop new time dependences when it is in a binary system.

\begin{figure}[t!]
\centering
\includegraphics[scale=0.9, trim = 0 0 0 0]{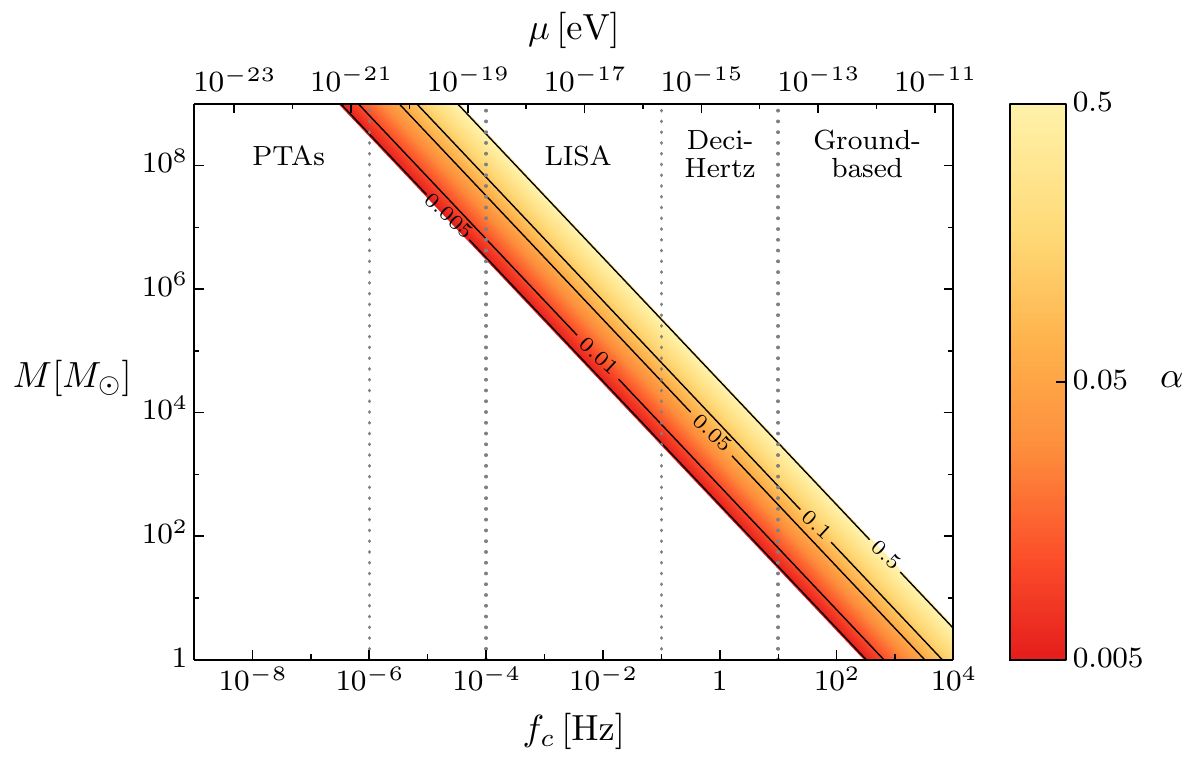}
\caption{Frequency of the monochromatic signal for  physically motivated ranges of $\alpha$ and $M$. The vertical dotted lines illustrate the typical observational bands of various gravitational-wave observatories.} 
\label{fig: faPlot}
\end{figure}

The frequency of these gravitational waves is determined by the mass of the boson field,
\begin{align}
f_{c} \simeq \frac{\mu}{\pi} = 484 \, \text{Hz} \, \left(\frac{\mu}{10^{-12} \text{ eV}} \right) . \label{eqn: GW annihilation frequency} 
\end{align}
This is shown in Fig.~\ref{fig: faPlot} for physically motivated values of $\alpha$ and $M$, and compared to the frequency bands of ground-based~\cite{TheLIGOScientific:2014jea, TheVirgo:2014hva, Somiya:2011np, Sathyaprakash:2012jk, Evans:2016mbw} and space-based~\cite{AmaroSeoane:2012km, Kawamura:2011zz, Graham:2017pmn} detectors, as well as pulsar timing arrays (PTAs)~\cite{Kramer:2013kea, McLaughlin:2013ira, Hobbs:2013aka, Smits:2008cf}. We see that ground-based experiments probe  $M \lesssim 3 \times 10^3 M_\odot$, while LISA would give us access to $M \gtrsim 3 \times 10^3 M_\odot$.

On the other hand, the root-mean-square strain amplitude depends on the intrinsic spin of the boson, cf. more discussion in \S\ref{sec:Hatom}. For concreteness, we consider the strain amplitude of the dominant scalar $\ket{211}$ mode. From (\ref{eqn: Schwarzschild strain 2}), we find that~\cite{Brito:2014wla, Yoshino:2014} 
\begin{align}
h_c \simeq 2 \times 10^{-26} \left( \frac{M}{3 M_{\odot}} \right) \left( \frac{M_c(\alpha)/M}{0.1} \right) \left( \frac{\alpha}{0.07 }\right)^6  \left( \frac{10 \,  \text{kpc}}{d} \right), \label{eqn: GW annihilation strain 1}
\end{align}
where $d$ is the (Euclidean) distance of the source. Since the signal is emitted continuously, its detectability depends on the total observation time $T_{\rm obs}$. The signal-to-noise ratio (SNR) of the cloud signal is
\begin{align}
\text{SNR} = h_c \, \langle F^2_+ \rangle^{1/2}  \frac{T_{\rm obs}^{1/2}}{\sqrt{S_n(f_c)} } \, ,
\end{align}
where $S_n(f_c)$ is the (one-sided) noise spectral density at the frequency $f_c$, and $\langle F^2_+ \rangle = \langle F^2_\times \rangle$ is the angular average 
 of the square of the detector pattern functions $F_{+, \times}$ for each gravitational-wave polarization.   
Using  
(\ref{eqn: GW annihilation strain 1}),  we get
\beq
\text{SNR} \simeq   13 \, \langle F^2_+ \rangle^{1/2} \left( \frac{T_{\rm obs}}{1 \text{ yr}} \right)^{1/2} \left( \frac{10^{-23} \text{ Hz}^{-\frac{1}{2}}}{\sqrt{S_n(f_c)}} \right) \left( \frac{M}{3 M_{\odot}} \right)  \left( \frac{M_c/M}{0.1} \right) \left(\frac{\alpha}{0.07} \right)^6 \left( \frac{10 \,  \text{kpc}}{d} \right) ,  \label{eqn: SNR  annihilation} 
\eeq
which is strongly dependent on $\alpha$. 
We see that cloud systems with $\alpha < 0.07$ and $M < 10^2\, M_\odot$ may be detectable only within our own galaxy, while those with larger values of $\alpha$, and bigger masses, may be observed at extra-galactic distances. 
Further discussion and detailed forecasts can be found in~\cite{Arvanitaki:2014wva, Arvanitaki:2016qwi, Brito:2017zvb,Riles:2017evm}.

\subsection{Resonant Attenuation} 

When the cloud is part of a binary system, the cloud may deplete over time, according to:
\beq
 \frac{h_c(t)}{h_c(t_0)} = \frac{M_c(t)}{M_c(t_0)} \sim \exp \big[ - 2\mathcal{A}(t, t_0) \big] \, , \label{eqn: cloud mass ratio}
\eeq
where $\mathcal{A}$ is the depletion estimator introduced in (\ref{eqn:CloudMassRatio}). 
This decay is most prominent if the Landau-Zener transition forces the initial cloud to fully populate a final decaying state. Depending on the state that is being occupied, the depletion can occur over sufficiently short timescales for moderate values of $\alpha$. Observing this time-dependent change in the monochromatic signal would therefore provide a tentative hint of the presence of the cloud and that it had experienced a Landau-Zener transition in its earlier binary evolution.

\subsection{Doppler Modulation}

In the presence of a companion, the gravitational-wave frequency $f_c$ will also be modulated by the orbital motion.\footnote{This is the same effect that famously led to the discovery of the Hulse-Taylor binary pulsar, where the frequency of the emitted electromagnetic radiation was found to be modulated by the orbital motion~\cite{Hulse:1974eb}.} In particular, if the orbital plane of the binary is oriented such that the cloud has a non-vanishing radial velocity along the line-of-sight, the orbital motion will induce a periodic Doppler shift of the gravitational-wave signal,
\beq
\frac{\Delta f_c}{f_c} = \frac{v_{r}}{c} = \left( \frac{M_* }{M_{\rm tot}} \right) \frac{R_* \Omega }{c} \sin \iota \, ,
\eeq
where $v_{r}$ is the radial component of the velocity of the cloud along the line-of-sight, and $\iota$ is its angle relative to the normal of the orbit. While detecting the periodic modulation of the frequency is experimentally (and computationally) challenging, this Doppler effect would be very convincing evidence that the cloud is part of a binary system. At the same time, it will open a new avenue to infering parameters of a binary system at low orbital frequencies, by monitoring the signal from the cloud with continuous searches. Similar searches are being performed for neutron stars in binaries~\cite{Aasi:2014erp}.

\section{Unraveling the Atomic Structure}
\label{sec:unravel}

So far, we have studied simplified models for the evolution of boson clouds in black hole binaries. This allowed us to identify a number of interesting dynamical effects that, in principle, can have important observational consequences. In this section, we investigate under which conditions these effects are accessible to current and future gravitational-wave observations.  Although a comprehensive study of the rich phenomenology is beyond the scope of this thesis, we will identify a few robust observational signatures that are worth exploring further.

We will begin, in \S\ref{sec:initial}, with a discussion of the most likely state of the cloud {\it before} 
it experiences any resonances. 
This initial state is prepared via superradiance and is sufficiently long-lived to be subsequently observed with ground-based and space-based observatories. We will entertain the possibility that all of the interesting physics of the resonances and the associated backreaction on the orbit occurs within the sensitivity of gravitational-wave experiments. In \S\ref{sec:single}, we start by determining the regions in parameter space for which resonant transitions can occur within the frequency bands of present and future detectors. 
For this range of parameters, we will then determine which effects can be observed due to the existence of one or several resonant transitions. 
In~\S\ref{sec:final}, we discuss the states of the boson cloud {\it after} each resonance transition and explain how they can be probed during the inspiral.

\subsection{Initial States of the Cloud}
\label{sec:initial}

As described in \S\ref{sec:Hatom}, superradiant growth occurs when the angular velocity of the black hole is larger than the angular phase velocity of a quasi-bound state. For future convenience, we repeat the superradiance condition here,
\beq
\Omega_\lab{H} > \frac{\omega}{m} \, . \label{eqn:superradiant}
\eeq
The resulting boson cloud will first populate the dominant growing mode, with $m=1$, until the superradiance condition saturates at $\Omega_\lab{H} = \omega/m \approx \mu$. 
This mode remains stable for a long time, as it only
decays via gravitational-wave emission on a much longer timescale. After that, the fastest growing $m=2$ mode will continue to superradiantly drain mass and angular momentum from the black hole, until it too becomes stable when $\Omega_\lab{H} = \omega/2$.
While this process can, in principle, continue towards larger values of $m$, each subsequent growth
occurs over timescales that are much longer than the previous one, cf.~(\ref{eqn:ScalarRate}) and (\ref{eqn:VectorRates}), so that 
only the first few $m$ modes can be produced on astrophysical timescales.

Both the mass $ M_{c, 0}$ and the angular momentum $S_{c} \equiv m \hskip 1pt S_{c, 0}$  contained in each growing mode increase during superradiance, until  $\Omega_\lab{H} = \omega/m$. The spin of the black hole at saturation is 
\beq
\frac{a}{M} = \frac{4 m (M \omega)}{m^2 + 4 (M \omega)^2} = \frac{4 \alpha}{m} + \mathcal{O}(\alpha^3) \, . \label{eqn:BHspinSat}
\eeq
Using angular momentum conservation and the relation $M_{c, 0} = \omega \hskip 1pt S_{c, 0}$~\cite{Bekenstein:1973mi}, we can estimate the final mass and angular momentum stored in each mode. The results are summarized in Table~\ref{table:Superrad}. These estimates are rather conservative, as they ignore the re-absorption of the lower $m$ modes, which could spin up the black hole and further enhance the amplitude of the cloud. Within this approximation, we find that $M_{c,0}$ and $S_{c, 0}$ are suppressed by a power of $\alpha$ for modes with $m > 1$.  In the following, we will refer to the $m =1$ modes as ``ground states,'' and call the $m >1$ modes ``excited states.'' However, we stress that this terminology is only meant to distinguish between the growing modes, and that the ground state is not the lowest frequency mode of the cloud (see Figs.~\ref{fig:ScalarSpectra} and \ref{fig:VectorSpectra}). 

\begin{table}[t!]
\begin{center}
\begin{tabular}{ | m{6.7em} || m{4em} | m{4em}| m{4em} |  m{4em} | } 
\hline
 & $m=1$ & $m=2$ & $m=3$ & $\cdots$ \\ 
\hline
Scalar: \hskip 1pt$ | n \es \ell \es m \rangle $ & $| 2 \es 1 \es 1\rangle$ & $|3 \es 2 \es 2\rangle$ & $|4 \es 3 \es 3\rangle$ & $\cdots$ \\ 
\hline
Vector: $|n \es \ell \es j\es  m\rangle $  & $|1 \es 0\es 1\es 1\rangle$ & $|2 \es 1 \es 2 \es 2\rangle$ &$|3\es 2 \es 3 \es 3\rangle$ & $\cdots$\\ 
\hline
$M_{c, 0}/M$ & $\alpha - 4 \alpha^2$ & $ \alpha^2$ & $2\alpha^2/9$ & $\cdots$ \\ 
\hline
$S_{c, 0}/M^2$ & $1- 4\alpha$ & $ \alpha$ & $2\alpha/9 $ & $\cdots$ \\ 
\hline 
\end{tabular}
\caption{ List of the dominant growing modes for scalar and vector fields. Shown also are the mass $M_{c, 0}$ and angular momentum $S_{c, 0}$ extracted by each mode in the  limit $\alpha \ll 1$. For the $m=1$ modes, we assumed that the black hole was initially maximally spinning. 
} \label{table:Superrad}
\end{center} 
\end{table}

\subsubsection*{Ground states}

It is natural to first consider the fastest growing modes, with $m=1$.
These are the states $|2\es 1\es 1\rangle$ and $|1 \es 0 \es 1 \es 1\rangle$ for the scalar and vector clouds, respectively. In the  limit $\alpha \ll 1$, the typical growth timescales of these states are 
\beq
\begin{aligned}
\Gamma^{-1}_{211} & \simeq \frac{10^6 \,\es \lab{yrs}}{\tilde a} \left( \frac{M}{60 M_\odot} \right) \left( \frac{0.019}{\alpha} \right)^9 \mathrlap{\qquad\,\,\, \text{(scalar)}\,,} \\ 
\Gamma^{-1}_{1011} & \simeq \frac{10^6 \,\es \lab{yrs}}{\tilde a} \left( \frac{M}{60 M_\odot} \right)  \left( \frac{0.0033}{\alpha} \right)^7 \mathrlap{\qquad \text{(vector)}\,,} \label{eqn:m=1GrowthRates}
\end{aligned}
\eeq
where $\tilde{a} \equiv a/M \leq 1$ is the dimensionless spin of the black hole. 
We see that for $\alpha \gtrsim 0.019$ (scalar) and $\alpha \gtrsim 0.0033$ (vector) the clouds grow quickly on the timescale that is shorter than the typical merger time ($\sim 10^{6}\,\es \lab{yrs}$). 
Since these clouds emit continuous, monochromatic gravitational waves, they gradually deplete over timescales of order~\cite{Arvanitaki:2010sy, Baumann:2018vus, Yoshino:2014, Baryakhtar:2017ngi, East:2017mrj, East:2018glu} 
\beq
\begin{aligned} 
T_{211} & \simeq 10^8 \, \text{yrs} 
\left( \frac{M}{60 M_\odot} \right) \left( \frac{0.07}{\alpha} \right)^{15} \mathrlap{\qquad \ \text{(scalar)}\, ,} \\[2pt] 
T_{1011} & \simeq 10^8 \, \text{yrs} 
\left( \frac{M}{60 M_\odot} \right) \left( \frac{0.01}{\alpha} \right)^{11} \mathrlap{\,\,\qquad \text{(vector)}\, ,} \label{eqn:m=1Lifetimes}
\end{aligned}
\eeq
where we have assumed that $M_c \simeq \alpha M$.
Requiring these modes to be stable on astrophysical timescales, $T \gtrsim 10^8$ years, puts upper bounds on the values of $\alpha$. For stellar-mass black holes, $M \sim 60 M_\odot$, we impose $\alpha \lesssim 0.07 $ (scalar) and $ \alpha \lesssim 0.01$ (vector), while for supermassive black holes, $M \sim 10^6 M_\odot$, the range can increase to $\alpha \lesssim 0.18 $ (scalar) and  $\alpha \lesssim 0.04$ (vector).

\begin{figure}
		\centering
		\includegraphics[scale=0.82, trim=5 0 0 0]{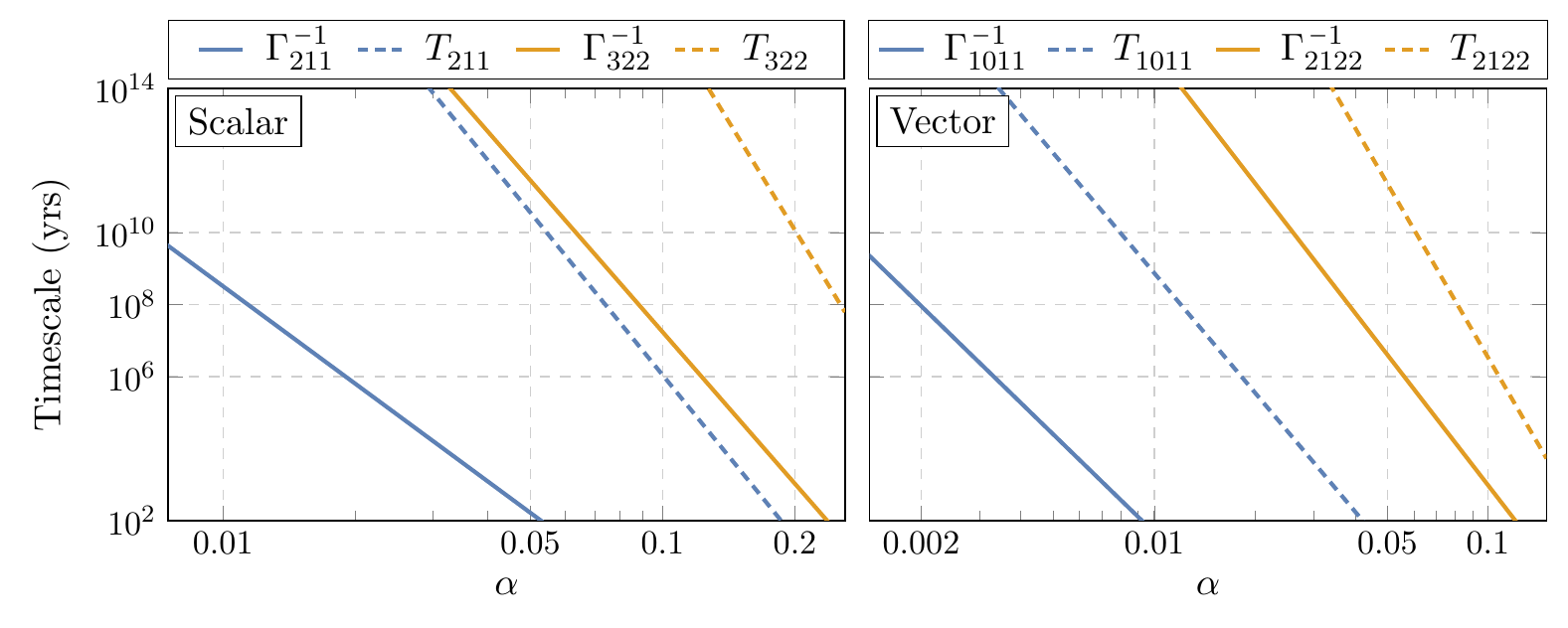}
		\caption{Relevant timescales for the ground states and first excited states of the scalar (\emph{left}) and vector (\emph{right}), as functions of $\alpha$ and for fixed $M = 60 M_\odot$. Note that both the growth time and the lifetime scale linearly with $M$. We have indicated the timescales that are shorter than typical merger times ($\sim10^{6}$ years) 
and general astrophysical processes ($\sim10^8$ years), as well as the age of the Universe ($\sim10^{10}$ years).    \label{fig:growthDecay}}
\end{figure}

\subsubsection*{Excited states}

To probe larger values of $\alpha$, we must instead consider excited states with $m\geq 2$, which are much longer lived. These excited states also take much longer to form than the ground states. For example, the typical growth timescales for 
the scalar $|3 \es 2\es 2\rangle$ and vector $|2\es 1\es 2\es 2\rangle$ modes are \beq
\begin{aligned}
 \Gamma^{-1}_{322} & \simeq   \frac{10^6 \,\text{yrs}}{\tilde a} \left( \frac{M}{60 M_\odot} \right) \left( \frac{0.11}{\alpha} \right)^{13}  \mathrlap{\qquad \ \ \,  \text{(scalar)}\, ,}   \\ 
\Gamma^{-1}_{2122} & \simeq \frac{ 10^{6} \, \text{yrs}}{\tilde a} \left( \frac{M}{60 M_\odot} \right)  \left( \frac{0.046}{\alpha} \right)^{11}   \mathrlap{\,\,\qquad \text{(vector)} \, .} \hskip 20pt \label{eqn:GrowthRates}
\end{aligned}
\eeq
At the same time, these excited states are also much more stable than the ground states, depleting via gravitational-wave emission over the timescales
\beq
	\begin{aligned}
	T_{322} &\simeq 10^{8}\,\lab{yrs} \left(\frac{M}{60 M_\odot}\right) \left(\frac{0.22}{\alpha}\right)^{20}  \mathrlap{\,\,\,\,\,\qquad \text{(scalar)}\,,} \\
		T_{2122} &\simeq 10^8 \, \lab{yrs} \left(\frac{M}{60 M_\odot}\right)\left(\frac{0.08}{\alpha}\right)^{16} \mathrlap{\,\,\qquad\,\,\, \text{(vector)}\,.} \hskip 20pt
\end{aligned} \label{eqn:m=2Lifetimes}
\eeq
Demanding the cloud to be stable on astrophysical timescales, $T \gtrsim 10^8 \, \lab{years}$, leads to  $\alpha \lesssim 0.22$ (scalar) and $\alpha \lesssim 0.08$ (vector) for $M \sim 60 M_\odot$, and  $\alpha \lesssim 0.44$ (scalar) and 
 $\alpha \lesssim 0.19$ (vector) for $M \sim 10^6 M_\odot$.\footnote{While the gravitational-wave emission rate of the scalar $|3 \es 2\es 2\rangle$ mode is known~\cite{Yoshino:2014}, the analogous rate for the vector $|2 \es 1 \es 2 \es 2 \rangle$ mode is not. Nevertheless, since the emission rate for the $\ell = 0$ modes of both the scalar and vector states share the same $\alpha$-scaling, we assume that a similar relation also applies for the scalar and vector excited states. In other words, we assume that $|2 \es 1 \es 2 \es 2 \rangle$ decays at the same rate as $|2 \es 1 \es 1 \rangle$. The only difference between the lifetimes $T_{211}$ and $T_{2122}$ in (\ref{eqn:m=1Lifetimes}) and (\ref{eqn:m=2Lifetimes}) is then that we take the initial mass of the cloud to be $M_{c,0} = \alpha M$ in the former and $M_{c, 0} = \alpha^2 M$ in the latter.} 
Beyond these limits, we must consider excited states with  larger $m$.  However, we do not expect these higher excited states to be qualitatively different, and so we will only focus on the phenomenology of the $m=1$ and $m=2$ states.

The cloud's history is summarized in Figure~\ref{fig:growthDecay}. The black hole will first grow the ground state (blue, solid). We require that this process takes place on a
short enough timescale ($\lesssim 10^{6}\,\lab{yrs}$) to be observable, which sets a lower bound on $\alpha$.   This ground state will then decay via gravitational-wave emission (blue, dashed). We find an upper bound on $\alpha$ by demanding that this state is suitably long-lived ($\gtrsim 10^{8} \, \lab{yrs}$). For larger $\alpha$, we are more likely to observe the cloud in an excited state, which grows (yellow, solid) and decays (yellow, dashed) via the same mechanisms, but on longer timescales.

\subsection{Resonant Transitions }
\label{sec:ResonanceSignals}

The states presented above can live long enough to be accessible to gravitational-wave observatories. Resonant transitions can then occur {\it in band} during the binary's inspiral phase.
As a consequence, there will be a series of distinct signatures which can reveal the nature of the gravitational~atom.  
We will first describe the observational characteristics of a single transition, and then discuss how observing multiple, sequential transitions can be used to further elucidate the properties of the boson clouds.

\subsubsection*{Single transition}
\label{sec:single}

\begin{figure}[t]
		\centering
		\includegraphics[scale=0.9, trim=0 10 0 5]{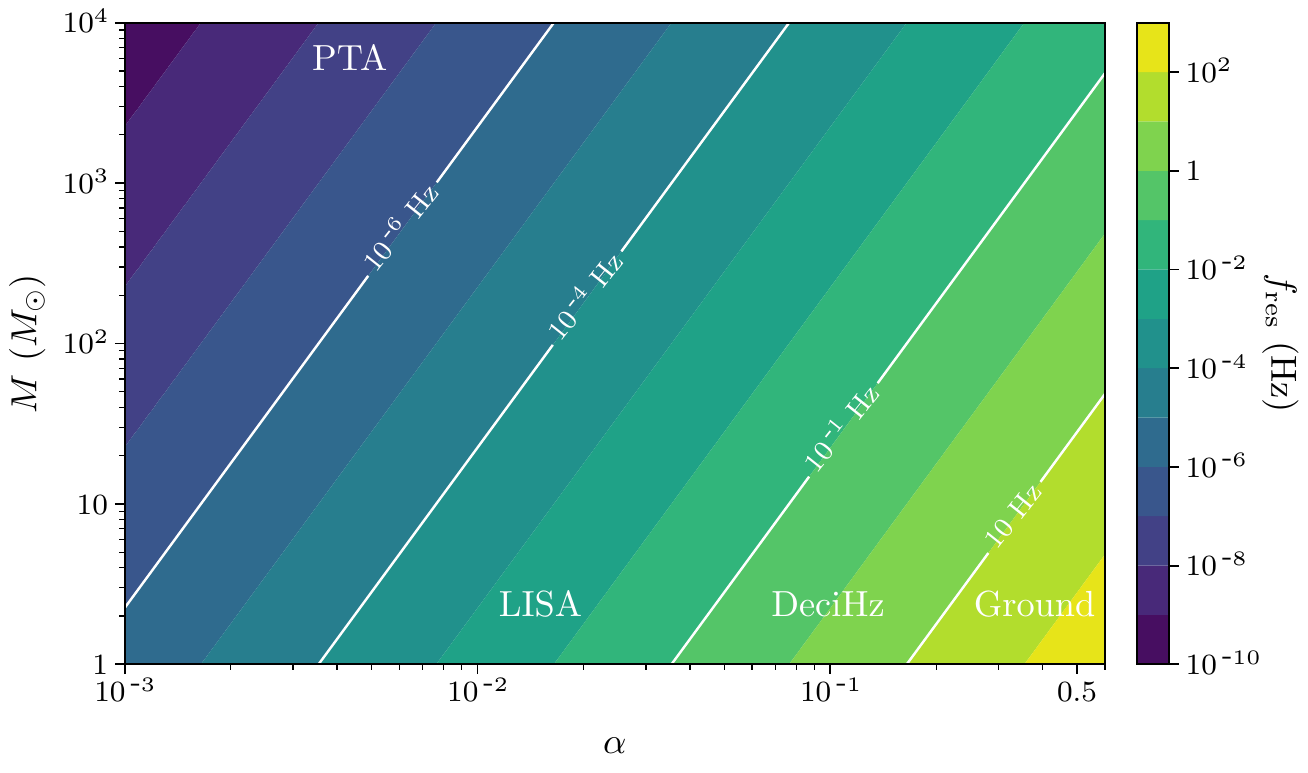}
		\caption{Resonance frequency of the Bohr transition $|2\es 1 \es 1 \rangle \to |3 \es 1 \es\es {\protect \minus 1}\rangle$, as a function of $M$ and~$\alpha$. This is representative for all Bohr transitions, since the frequency (\ref{eq:resonanceFrequencies}) is relatively insensitive to the choice of  transition, 	
under which $f_\lab{res} \to \varepsilon_\lab{B} f_\lab{res}$. Depending on the values of $M$ and $\alpha$, the resonance may fall in the band of pulsar timing arrays (PTA), the LISA observatory, proposed deciHertz experiments, or currently operating ground-based LIGO/Virgo detectors.   \label{fig:resonanceFrequencies}}
\end{figure}

Consider a transition between states with principal quantum numbers $n_a$ and $n_b$. 
The orbital frequency at the resonance, $\Omega_{r}$, implies the following frequency of the emitted gravitational waves
\beq
f_{\rm res} = \frac{\Omega_{r}}{\pi} = 1.3 \times 10^{-2} \, \text{Hz} \,\left( \frac{60 M_\odot}{M} \right) \left( \frac{\alpha}{0.07} \right)^3 \varepsilon_{\lab{B}} \, , \label{eq:resonanceFrequencies}
\eeq
where we have introduced the transition-dependent quantity
\beq
	\varepsilon_{\lab{B}} \equiv  \frac{36}{5} \frac{2}{|\Delta m|} \!\left|\frac{1}{n_
	a^2} - \smash{\frac{1}{n_{b}^{2}}} \right| ,  \label{eqn:VarepsilonBohr}
\eeq
and chosen the normalization such that
$\varepsilon_{\lab{B}} \equiv1$ for the first Bohr transition of the scalar ground state, $|2 \es 1 \es 1 \rangle \to |3 \es 1 \, \minus{1}\rangle$. The resonance frequency for this specific transition, as a function of $M$ and $\alpha$, is shown in Fig.~\ref{fig:resonanceFrequencies}. Indicated are the ranges for which the signal falls into the frequency bands of current and future gravitational-wave observations.  

   \begin{figure}[t]
  	\centering
    \includegraphics[scale=0.87]{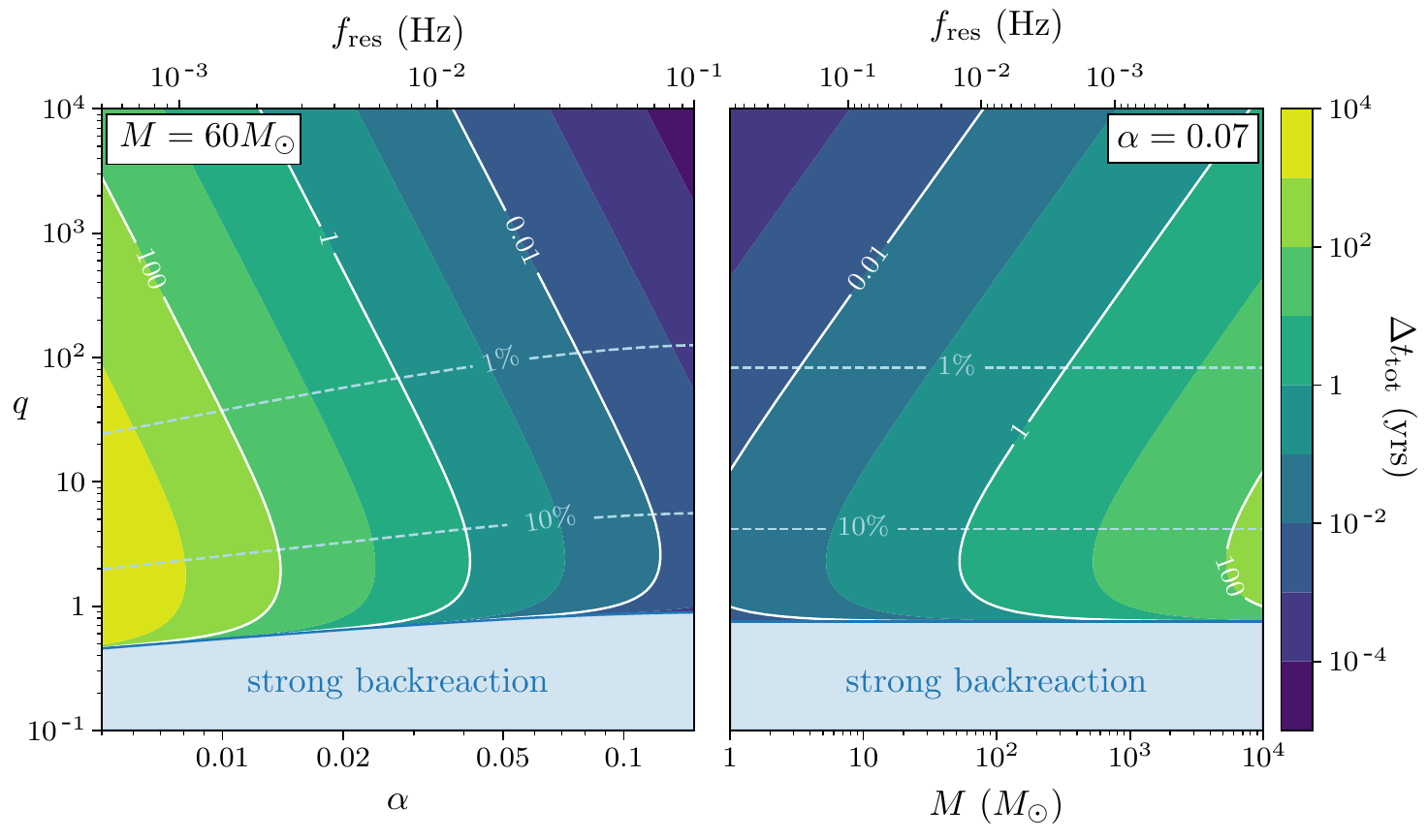}
    \caption{Total time $\Delta t_\lab{tot}$ spent in the sinking Bohr transition $|2 \es 1 \es 1 \rangle \to |3 \es 1 \es\es {\protect \minus} 1\rangle$. On the left, we plot this as function of $\alpha$ and $q$, for fixed mass $M = 60 M_\odot$.  On the right, we instead fix $\alpha = 0.07$ and plot the total time as a function of $M$ and $q$. For both, we provide the corresponding resonance frequency $f_\lab{res}$ on the top axis. 
 We plot constant contours of the ratio $|\Delta t_\lab{c}/\Delta t|$ as dashed light blue 
  lines. The cloud's backreaction on the orbit becomes stronger as $q$ decreases, shortening the time it takes for the binary to move through the transition. In blue, we indicate the region in parameter space where the adiabatic condition (\ref{eqn:Non-adiabaticity}) is violated and the backreaction is strong enough that we lose predictive control and the estimation (\ref{eq:totalTime}) is inapplicable. 
    \label{fig:211Kick}}
\end{figure}

An important additional characteristic is the time spent within the resonance band.
Given the resonance bandwidth of $\Delta \Omega \sim 2 \eta$, and using (\ref{eqn:floatingChirp}) and (\ref{eqn:KickingChirp}), we can estimate this as  
 \beq
 \Delta t_{\rm tot} \,\simeq\, \Delta t \pm \Delta t_c\, ,\label{eq:totalTime}
 \eeq 
 where $+/-$ represents floating/sinking orbits, respectively. Here, $\Delta t \simeq 2 \eta / \gamma$ is the time it takes the inspiral to move through the resonance in the absence of backreaction, whereas $\Delta t_c \simeq 3R_J  | \Delta m \Delta E |  / 2\gamma$ is the additional contribution from the presence of the cloud,  
 \begin{align}
 \Delta t  & \simeq\, 4 \, \text{yrs} \,  \left(\frac{M }{60 M_\odot} \right) \frac{1}{(1+q)^{2/3}} \left( \frac{0.07}{\alpha}\right)^8 \left(\frac{R_{ab}}{0.3} \right)  \varepsilon_\lab{B}^{-8/3} \, , \label{eqn:NoBackreactionTime} \\[4pt]
 \Delta t_{c}  &  \simeq \, 1 \, \text{yr} \, \left(\frac{M }{60 M_\odot} \right) \frac{(1+q)^{2/3}}{q^2}   \left( \frac{S_{c, 0}}{M^2} \right)  \left( \frac{0.07}{\alpha}\right)^7  \left(\frac{|\Delta m|}{2} \right)^2 \varepsilon_\lab{B}^{-7/3}\, . \label{eqn:BackreactionTime}
 \end{align}
 In (\ref{eqn:NoBackreactionTime}), we used (\ref{eqn:Eta}) to extract the value of $\eta_{ab}$ as a function of $R_{ab}$. To assess the strength of the backreaction, it is useful to take the ratio between these two timescales, 
\beq
\frac{\Delta t_{c}}{\Delta t} \simeq \frac{1}{4} \frac{(1+q)^{4/3}}{q^2} \left(\frac{\alpha}{0.07}\right)  \left( \frac{S_{c, 0}}{M^2} \right) . \label{eq:timeRatio}
\eeq 
In all cases, we find that the backreaction crucially depends on the angular momentum stored in the cloud, $S_{c,0}$, as well as the mass ratio, $q$. Notice that, for excited states, $S_{c,0}$ is suppressed by a power of $\alpha$, and the backreaction is typically weaker.  
Alternatively, we can also estimate the size of the backreaction through the difference in the number of orbital cycles spent in the resonance band with and without the cloud 
\beq
\begin{aligned}
\Delta N_c &=  f_{\rm res} \Delta t_{c}  \,\simeq\, 10^5 \, \frac{(1+q)^{2/3}}{q^2}  \left(\frac{S_{c, 0}}{M^2}\right)  \left( \frac{0.07}{\alpha} \right)^{4}  \left(\frac{|\Delta m|}{2} \right) \varepsilon^{-4/3}_\lab{B} \, . \label{eqn:DeltaN}
\end{aligned}
\eeq
Since binary searches are sensitive to $\gtrsim \mathcal{O}(1)$ difference in the number of orbital cycles between signal and template waveforms, the estimate (\ref{eqn:DeltaN}) suggests that backreaction on the orbit can dramatically alter the gravitational-wave signal, introducing a substantial dephasing with respect to the waveform model without the cloud. The effect is especially prominent for $q \ll 1$, yielding a significant dephasing for intermediate (IMRIs) and extreme mass ratio inspirals (EMRIs), where a small black hole perturbs a larger cloud.\footnote{Requiring these resonant transitions to be adiabatic, cf.~(\ref{eqn:HigherOrderLZ}), weakly constrains the mass ratio $q \gtrsim \alpha^5$.} 

\begin{figure}[t]
  	\centering
    \includegraphics[scale=0.9]{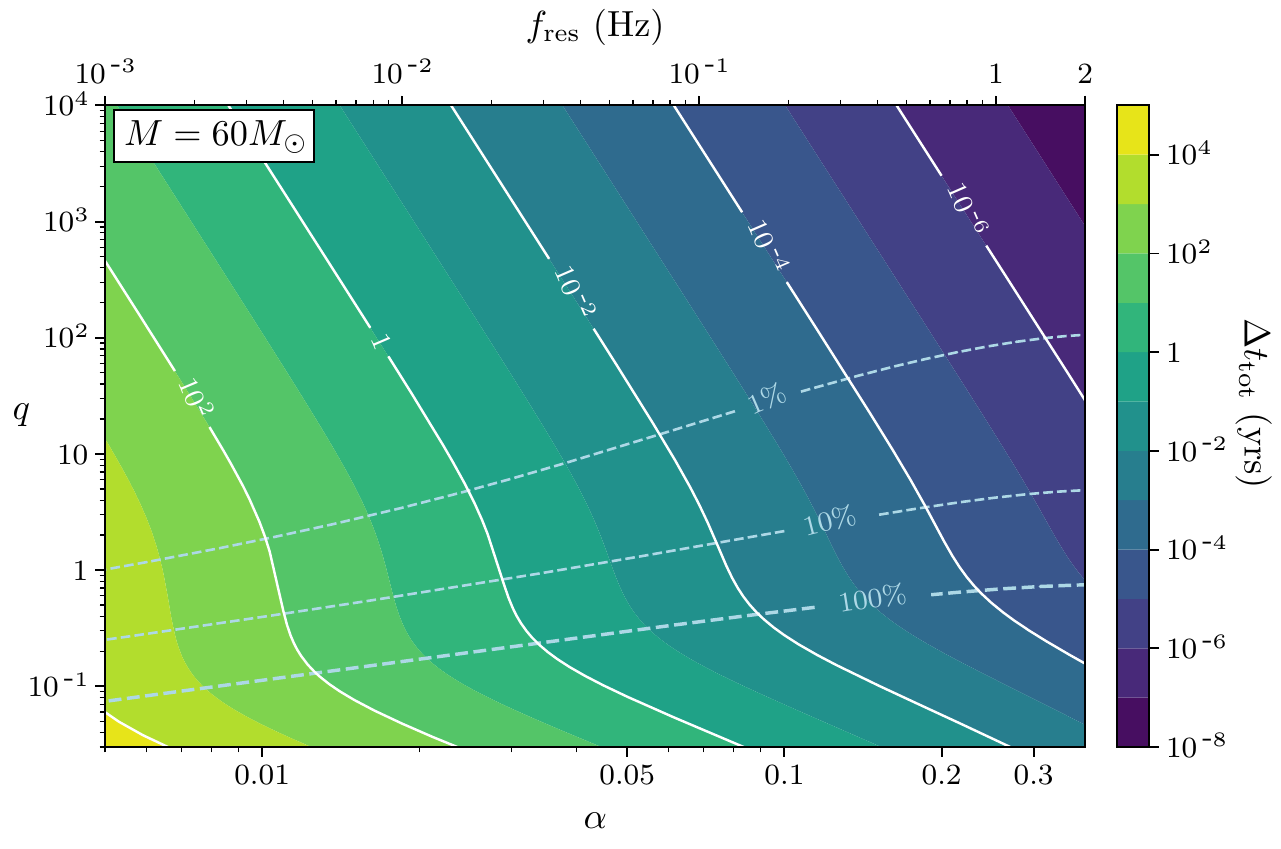}
    \caption{Total time $\Delta t_\lab{tot}$ spent in the floating Bohr transition $|3 \es 2 \es  2 \rangle \to |2 \es 0 \es 0 \rangle$, for a parent black hole of mass $M = 60 M_\odot$, as a function of $\alpha$, $f_\lab{res}$, and $q$. Increasing the fiducial mass $M$ reduces the resonance frequency and enhances the floating time, according to $f_\lab{res} \to (60 M_\odot/M) f_\lab{res}$  and ${\Delta t_\lab{tot} \to (M/60 M_\odot) \Delta t_\lab{tot}}$, for fixed $\alpha$. We plot constant contours of the ratio $\Delta t_\lab{c}/\Delta t$ as dashed light blue lines, 
    indicating where the inspiral spends an additional $1\%$, $10\%$, and $100\%$ amount of time in the transition. \label{fig:322Float}}
\end{figure}

The dephasing due to the backreaction on the orbit is a robust signature of boson clouds in binary systems. 
However, whether the orbit floats or sinks depends on its orientation (co-rotating or counter-rotating with respect to the black hole's spin), and the nature of the transition; see the discussion below (\ref{eqn:CircularBackreaction2}). For example, for the transition $|2\es 1 \es 1 \rangle \to |3 \es 1 \es\es {\protect \minus 1}\rangle$, which occurs for counter-rotating orbits, the cloud absorbs angular momentum from the orbit, which therefore shrinks faster, thus reducing
the time it takes for the binary to move through the transition. We plot the total time it takes for the inspiral to cross this resonance in Fig.~\ref{fig:211Kick}. Notice that backreaction dominates for $q \ll 1$, when the cloud contains a large fraction of the total angular momentum. Especially in this limit, there is a danger that the system moves through the transition too quickly, violating the condition (\ref{eqn:Non-adiabaticity}) and hence destroying the validity of~(\ref{eq:totalTime}). In that case, the binary's orbit can receive a significant kick from the cloud (cf.~Fig.~\ref{fig:Sinking}), potentially making the orbit highly eccentric. As discussed in Section~\ref{sec:Backreaction}, the cloud's dynamics then becomes highly non-adiabatic and depends sensitively on the backreacted dynamics of the orbit. A precise characterization of this interesting region in parameter space thus requires either different analytic techniques or direct numerical simulations.

For excited states, on the other hand, Bohr transitions can also occur for co-rotating orbits. For example, during the transition $|3 \es 2 \es 2 \rangle \to |2 \es 0 \es 0 \rangle$,\footnote{Since we only focus on Bohr transitions, we will assume that the initial frequency of this binary is higher than the resonance frequencies of the hyperfine $|3 \es 2 \es 2\rangle \to |3\es 2 \es 0\rangle$ and fine $| 3\es 2 \es 2\rangle \to |3\es 0\es 0\rangle$ transitions, such that they were missed by the binary inspiral.} the cloud loses angular momentum, thereby producing a floating orbit. Fig.~\ref{fig:322Float} illustrates the total time it takes for the binary inspiral to move through this transition.
 As a reference we choose the fiducial value $M=60M_\odot$, but the plot can be easily scaled to any value of $M$. 
 The thick dashed line (labeled 100\%) denotes parameters for which the presence of the cloud doubles the time spent in the resonance region. During this transition, the inspiral floats and the orbital frequency $\Omega(t)$ remains roughly constant at the resonance value~(\ref{eqn:floatingMean}). The backreaction time again dominates for $q \ll 1$ but, unlike for the sinking orbits, this regime is still under analytic control.
 As before, this effect is particularly relevant for IMRIs and EMRIs, in which case the float can be extremely long-lived---much longer than the typical timescale of gravitational-wave observatories ($\sim 10$ years). They may thus serve as another source of continuous monochromatic gravitational waves.

As we have demonstrated, for favorable values of $M$ and $M_*$, 
the cloud provides large corrections to the inspiral at a frequency that falls within the bands of 
future gravitational-wave observatories. A measurement of this resonance frequency, together with a measurement of $q$ and $M$, would then provide a strong constraint on the allowed values of $\alpha$, and thus the boson's mass~$\mu$. However, it is important to note that the determination of a single resonance frequency does not uniquely fix the boson mass, since additional information is needed to identify which specific transition occurred. This degeneracy is broken if we observe multiple transitions.

\subsubsection*{Multiple transitions}

\begin{figure}[t!]
\centering
	\includegraphics[scale=0.83, trim=12 0 0 0]{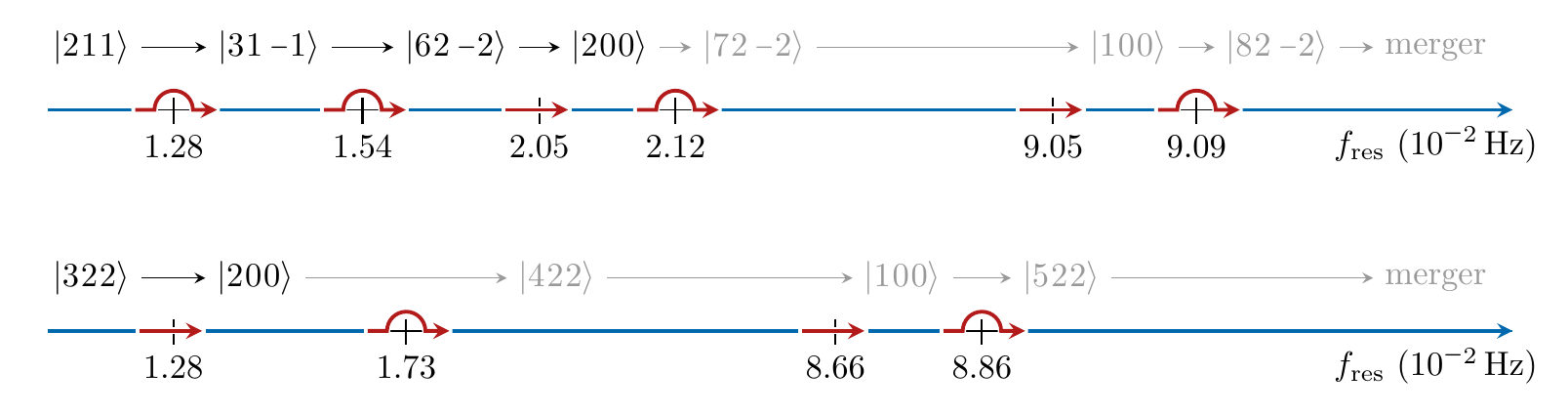}
\caption{Evolution of the $|2 \es 1 \es 1\rangle$ ({\it top}) and $|3 \es 2 \es 2 \rangle$ ({\it bottom}) states, during a counter-rotating and co-rotating inspiral, respectively (for $\alpha=0.07$, $q=1$ and $M=60M_\odot$).  Each history contains a series of floating ({\protect\tikz[baseline=-3pt] \protect\draw[cornellRed, line width=1.1, >=stealth, ->] (-0.3, 0) -- (0.2, 0);}) and sinking ({\protect\tikz[baseline=-1pt] \protect\draw[cornellRed, line width=1.1, >=stealth, ->] (-0.4, 0)--(-0.2,0) arc (180 : 0: 0.2)--(0.45, 0);}) orbits, separated by periods of ``normal'' inspiral evolution ({\protect\tikz[baseline=-3pt] \protect\draw[cornellBlue, line width=1.1, >=stealth] (-0.2, 0) -- (0.2, 0);}), with only weak perturbative mixing. As we discuss in the main text, the $|2 \es 0 \es 0 \rangle$ state has a large decay width. Unless the binary moves quickly to the next transition, which occurs only for $q \gg 1$, this forces the cloud to deplete before experiencing the next resonance.  We indicate this by the reduced opacity of the states after $|2 \es 0\es 0 \rangle$.}
\label{fig:scalarTree}
\end{figure}

As seen in Figs.~\ref{fig:211Kick} and~\ref{fig:322Float}, the binary passes through the transition relatively quickly for moderate values of $\alpha$ and $q$.  
Since subsequent resonance frequencies may be nearby, this region of parameter space suggests that we may observe \emph{multiple} resonant transitions in a given observational window.
In Fig.~\ref{fig:scalarTree}, we show two representative
 examples  for the evolution of the scalar cloud, starting from the ground state $|2 \es 1 \es 1\rangle$ and first excited state $|3 \es 2 \es 2 \rangle$, respectively. Notice that, because the gravitational atom has a finite \emph{ionization energy}, cf. (\ref{eqn:scalarspectrum}) and (\ref{eqn:vectorspectrumGeneral}), there is a maximum frequency at which resonant transitions can occur,
	\begin{equation}
		f_\lab{max} \simeq  0.2\,\, \lab{Hz} \left(\frac{60 M_\odot}{M}\right) \left(\frac{\alpha}{0.07}\right)^3. \label{eqn:fResMax}
	\end{equation}
Together with the selection rules discussed in \S\ref{sec:gravlevelmix}, this explains why the transition trees\footnote{The dominant $\ell_*=2$ perturbation mediates all of the resonances displayed in these transition trees, except for the $|3 \es 1\, \es \minus 1\rangle \to |6 \es 2 \, \es \minus 2\rangle$ and $|3100\rangle \to | 62j\, \es \minus 1 \rangle$ resonances, which are instead induced by the weaker $\ell_*=3$ perturbation. This is because, after experiencing an earlier resonance in these cases, the binary has an orbital frequency that already exceeds all resonance frequencies that can be excited from the newly prepared $| n=3, \ell=1 \rangle$ states by $\ell_*=2$. A further transition is nevertheless still possible through $\ell_*=3$, which supports perturbations with  $|\Delta m|=1$ (see Appendix~\ref{app:TidalMoments}) and hence induces transitions at higher frequencies, cf. (\ref{eqn:VarepsilonBohr}).}    pictured in Figs.~\ref{fig:scalarTree} and~\ref{fig:vectorTree} terminate at a particular \emph{end state}.\footnote{In principle, there is also a continuum of states above this ionization energy where the cloud becomes unbound from the black hole. However,  since the gravitational perturbation is very weak, it is unlikely these states can be appreciably occupied by the sort of resonant transitions we described in \S\ref{sec:ResonanceSignals}. } A more detailed numerical analysis may then be needed to fully incorporate all of the relevant physics.

As a rough guide as to whether or not multiple transitions can be observed, we estimate the time it takes the inspiral to move from one resonance frequency $f_{{\rm res}}^{(i)}$ to another~$f_{{\rm res}}^{(j)}$. Assuming quasi-circular adiabatic evolution, we find  
\beq
\Delta T_{i \to j} = 2.5 \, \text{yrs} \, \left( \frac{M}{60M_\odot} \right) \frac{(1+q)^{1/3}}{q} \left( \frac{0.07}{\alpha} \right)^8 \left( \varepsilon_{\lab{B}, (i)}^{-8/3} - \varepsilon_{\lab{B},(j)}^{-8/3} \right) , \label{eq:timeSpent}
\eeq
where $\varepsilon_{\lab{B}, (i)}$ represents the parameter (\ref{eqn:VarepsilonBohr}) associated to the $i$-th transition. For example, we have $\varepsilon_{\lab{B},(1)} = 1$ and  
$\varepsilon_{\lab{B},(2)} = 6/5$ for the transitions $|2 \es 1 \es 1 \rangle \to |3 \es 1\, \es \minus 1\rangle$ and  $|3 \es 1\, \es \minus 1\rangle \to |6 \es 2 \, \es \minus 2\rangle$, respectively. Comparing \eqref{eq:timeSpent} with (\ref{eqn:NoBackreactionTime}), we find that, as long as the binary moves through a single Bohr transition in a reasonable amount of time (say $1$ year), we should expect to observe a second Bohr transition on a similar timescale. Furthermore, since the rate at which the binary sweeps through the frequency accelerates, we should generically expect to observe several transitions.

Observing the dephasing of sequential Bohr transitions allows us to break degeneracies among the different parameters.
 For instance, let us consider successive Bohr transitions, labeled $i$ and $i+1$, between states with principal quantum numbers $n_a$, $n_b$, and $n_c$. The azimuthal angular momentum differences between the states are $\Delta m_{ab}$ and $\Delta m_{bc}$. The ratio of the resonance frequencies then is
\begin{equation}
\frac{f_{\lab{res}}^{(i+1)}}{f_{\lab{res}}^{(i)} }= \frac{\varepsilon_{\lab{B},(i+1)}}{\varepsilon_{\lab{B},(i)} }= \left| \frac{\Delta m_{ab}}{\Delta m_{bc}} \right|  \left(\frac{n_a}{n_c}\right)^2 \left| \frac{n^2_c - n_b^2}{n_b^2 - n_a^2} \right| , \label{eqn:ratiosFres}
\end{equation}
which crucially depends only on integer quantities.\footnote{Note that the presence of fine or hyperfine corrections does not affect our conclusions.} An observed sequence of these ratios then provides a fingerprint with which a particular transition history can be identified. For instance, the sequence of successive frequency ratios \{1.2, 1.33, 1.03, 4.27, 1.005\} of the counter-rotating history in Fig.~\ref{fig:scalarTree}  
 is clearly different from that of the co-rotating history, 
 \{1.35, 5, 1.02\}. Even though some of the ratios in these two histories are nearly equal, they never appear in the same order, and so we may use this sequence as a unique identifier for each history.  
This, combined with a measurement of the black hole mass $M$, can then be used to infer the boson mass $\mu$. However, there is a caveat. As illustrated in Figs.~\ref{fig:scalarTree} and \ref{fig:vectorTree}, the cloud may deplete considerably before reaching the next resonance. Fortunately, as we discuss next, this can also be used as a unique signature of the boson cloud.

\subsubsection*{Cloud depletion}

Throughout the above discussion, we have implicitly assumed that the cloud does not appreciably evolve away from the resonant transitions. 
However, in the later stages of the evolution, the cloud may occupy a rapidly decaying mode and is quickly reabsorbed back into the black hole, before arriving at the next resonance. 
After that, the effects of the cloud on the gravitational-wave signal become negligible.

 Using \eqref{eqn:CloudMassRatio}, we can estimate the number of $e$-folds of decay between two successive transitions. At leading-order in $\alpha$, we find\hskip 1pt\footnote{In order to obtain the $\alpha$-scaling in \eqref{eqn:efold} , we also used the fact that the central black hole is slowly spinning after experiencing superradiance, cf. (\ref{eqn:BHspinSat}).}
\begin{equation}
  |\Gamma| \Delta T_{i \to j} = \mathcal{C}\, \frac{(1+q)^{1/3}}{q} \alpha^{2j + 2 \ell - 2} \left(\varepsilon_{\lab{B}, (i)}^{-8/3} - \varepsilon_{\lab{B}, (j)}^{-8/3}\right)  , \label{eqn:efold}
\end{equation}
where $\Gamma$ is the decay rate of the state that is occupied after the $i$-th transition. Typically, the dimensionless coefficient $\mathcal{C} \lesssim \mathcal{O}(0.1)$ for the states of interest, and can depend sensitively on the state's quantum numbers.
Depletion is significant when the ratio in (\ref{eqn:efold}) is greater than $1$.

\begin{figure}[t!]
	\centering
		\includegraphics[scale=0.83, trim=12 0 0 0]{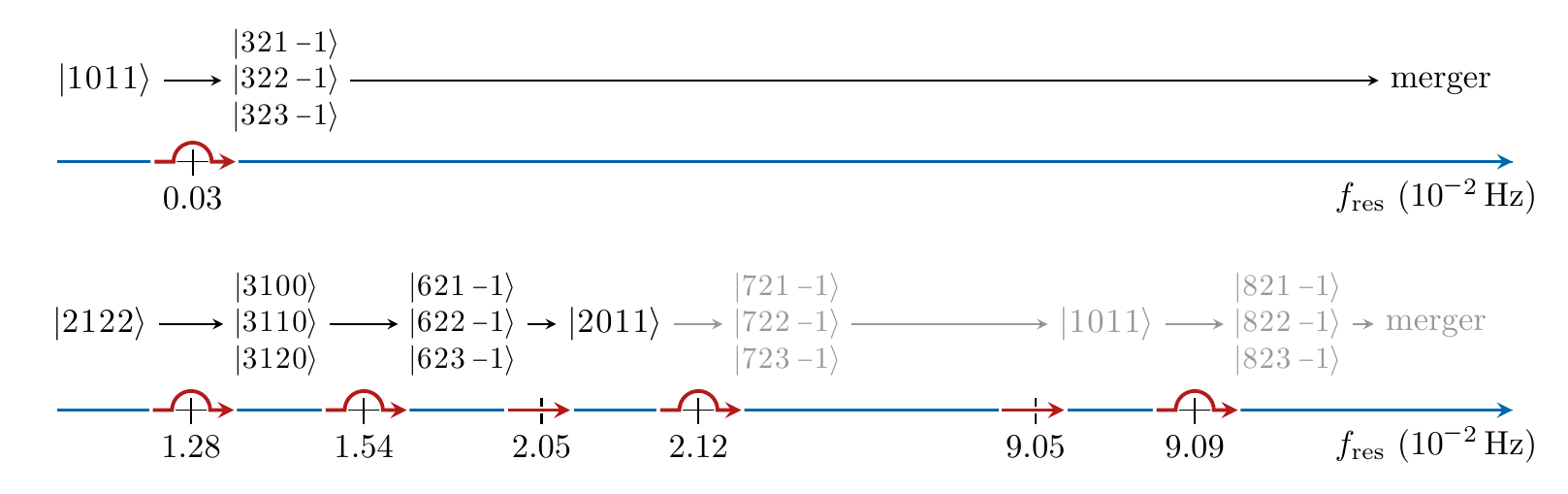}
		\caption{
		Evolution of the $|1 \es 0 \es 1 \es 1\rangle$ ({\it top}) and $|2 \es 1 \es 2 \es 2\rangle$ ({\it bottom}) states during a counter-rotating inspiral (for $\alpha = 0.01$ and $\alpha=0.07$, respectively). In both cases, $q=1$ and $M=60M_\odot$. Notice that the ground state of the vector cloud $|1 \es 0 \es 1 \es 1 \rangle$ only experiences a single transition. The excited state $|2 \es 1 \es 2 \es 2 \rangle$, on the other hand, mimics the history of the scalar ground state $|2 \es 1 \es 1 \rangle$. Yet, we see that vector transitions involve superpositions of many  states. 
The large decay width of the state  $|2 \es 0 \es 1 \es 1 \rangle$ makes it unlikely for the cloud to survive after that point, unless $q \gg 1$. As in Fig.~\ref{fig:scalarTree}, we indicate this by reducing the opacity of the states after the $|2 \es 0 \es 1 \es 1 \rangle$ state.}
		\label{fig:vectorTree}
\end{figure}

For scalar fields, the fastest decaying modes have $\ell=0$, in which case the estimator in \eqref{eqn:efold} scales inversely with $\alpha$. A cloud that populates these states will therefore almost always deplete before it reaches the next resonance. The cloud only survives when $q \gtrsim \alpha^{-3}$, namely when the cloud is on the smaller black hole, in which case the system moves quickly enough to the next transition. Similarly, for vector clouds, a fast decaying state with $\ell=0$ and  $j=1$ becomes occupied along the transition tree in Fig.~\ref{fig:vectorTree}.\footnote{Notice that, because we are assuming that the central black hole has a spin that saturates at the superradiance condition, the  state $|2011 \rangle$ turns into a decaying mode, despite having $m>0$.} Unless $q \gg 1$, it therefore suffers the same fate as the scalar cloud. However, a vector cloud still experiences more resonances over a wider range of mass ratios than a scalar cloud.

Although a rapid depletion of the cloud prevents us from exploring later resonant transitions, it is a unique feature of gravitational atoms in binary systems~\cite{Baumann:2018vus} that helps to distinguish them from other exotic compact objects like boson stars, e.g.~\cite{Sennett:2017etc}. Using~(\ref{eqn:efold}), we see that, for $q \ll 1$, a cloud that populates a fast-decaying state will typically deplete before it reaches the next resonance. Nevertheless, these IMRIs and EMRIs are precisely the binaries for which we only expect to see a single transition in a reasonable observational period (\ref{eq:timeSpent}). Furthermore, they are also the binaries for which backreaction is most significant. This suggests that, to probe these types of binary systems, we will typically need to hunt for long floats or strongly kicked orbits, rather than multiple correlated transitions.

\subsection{Post-Resonance Evolution}
\label{sec:final}

As argued above, observing multiple successive resonances provides a detailed mapping of the spectral properties of the cloud, allowing us to extract the mass $\mu$ of the ultralight boson. However, since this fingerprint depends only on the energy differences between states, it can fail to distinguish between scalar and vector clouds, which mainly differ by the number of states involved in a transition. Fig.~\ref{fig:vectorTree} shows transitions of the vector cloud, starting from the ground state $|1 \es 0 \es 1 \es 1\rangle$  and the first excited state $|2 \es 1 \es 2 \es 2\rangle$, respectively.
While the dynamics of the vector ground state is rather unique (involving only a single transition), distinguishing between scalar and vector clouds for other histories is more difficult.  In particular, the vector state $|2 \es 1 \es 2 \es 2 \rangle$ in Fig.~\ref{fig:vectorTree} has a  similar transition history as the scalar state $|2 \es 1 \es 1 \rangle$ in Fig.~\ref{fig:scalarTree}. 
It is thus non-trivial  to disentangle these two histories using only information about their resonance frequencies.  As we discuss here, studying finite-size effects can help us break this degeneracy.\footnote{In principle, this degeneracy can also be lifted through measurements of $\partial_0 \Omega$ and $\Delta t_{\rm tot}$ across a resonance, as the mode $|2 \es 1 \es 2 \es 2 \rangle$ carries less angular momentum than the mode $|2 \es 1 \es 1 \rangle$ (for the same value of $\alpha$), and this should have a weaker effect on the orbit.}

\subsubsection*{Finite-size effects}

 There is a major qualitative difference between the evolution of scalar and vector clouds that we can exploit to distinguish them: vector transitions typically involve more than two states, while scalar transitions do not. As discussed in \S\ref{sec:modulatingfinite}, the result of these multi-state vector transitions is a superposition of states, each with their own shape and characteristic frequency. 
 There are then large oscillations in the shape of the vector cloud, at frequencies which depend on the energy differences $\Delta E_{ab}$ between the occupied states. Detecting these oscillatory finite-size effects is thus a smoking gun of particles with spin.\footnote{We expect the same phenomena to apply also to higher-spin particles. Namely, for a given $n$, $\ell$ and $m$, there will be a larger number of  states, and the finite-size effects can oscillate at a larger number of frequencies.}

  We can illustrate this effect quantitatively using  the first Bohr transition of the vector excited state, $|2 \es 1 \es 2 \es 2 \rangle \to |3 \es 1 \es j \es 0 \rangle$. After the transition, the state evolves into a superposition with has negligible population in the $|3 \es 1 \es 1 \es 0 \rangle$ mode:\hskip 1pt\footnote{The gravitational perturbation only mediates degenerate subspace transitions between $|3 \es 1 \es 0 \es 0 \rangle$ and $|3 \es 1 \es 2 \es 0 \rangle$. As a consequence, the final state is an equally weighted sum of these two states, independent of both $q$ and $\alpha$.}
  \begin{equation}
    |\psi(t) \rangle \approx \frac{1}{\sqrt{2}}\hskip 2pt e^{-i E_{3 1 0 0} t} \left(|3 \es 1 \es 0 \es 0 \rangle +  e^{-i \Delta E_{2 0} t } |3 \es 1 \es 2 \es 0 \rangle\right) ,
  \end{equation}
where we have defined $\Delta E_{20} \equiv E_{3120} - E_{3100}$.
  The dominant energy difference is set by the fine-structure splitting, cf.~(\ref{eqn:vectorspectrumGeneral}) and~(\ref{eqn:vectorspectrumfh}), and so the frequency of these oscillations is roughly
    \begin{equation} 
    \frac{\Delta E_{20}}{2 \pi} \approx  6 \times 10^{-5} \, \lab{Hz} \, \left(\frac{60 M_\odot}{M}\right) \left(\frac{\alpha}{0.07}\right)^5\,. \label{eq:firstBohrSplit}
  \end{equation}
At the same time, the cloud's axisymmetric quadrupole moment inherits these oscillations, which can be parameterized as
  \begin{equation}
    \kappa_c(t) = \kappa_0 \left(1 - 2 \sqrt{2} \cos(\Delta E_{2 0} \,t)\right)\,,
  \end{equation}
  with $\kappa_c$ defined in \S\ref{sec:kappa} and we have ignored corrections that are suppressed by $\alpha$. The $\kappa_0$ value represents the quadrupole moment of the state $|3\es 1\es 0 \es 0 \rangle + i |3 \es 1 \es 2 \es 0 \rangle$. For $\alpha = 0.07$ and $M = 60 M_\odot$, the cloud's mass distribution oscillates between bulging at the black hole's spin axis and bulging along its equatorial plane at a period of roughly $4.6$ hours. Since it then takes $1.2$ years to evolve towards the next transition, cf.~(\ref{eq:timeSpent}), we could potentially observe $2.2 \times 10^3$ of these cycles. 
If, in addition, we include the relatively fast depletion of the $|3 \es 1 \es 0 \es  0 \rangle$ mode, these oscillations decay on a timescale of about a month.

  Because the frequency of these oscillations is so heavily $\alpha$-suppressed, observing them is a challenge. For example, a single oscillation of the $|3\es 2 \es j \, \es \minus 1 \rangle$ superposition produced by the vector ground state can take decades for $\alpha = 0.01$ and $M = 60 M_\odot$. This is far outside present observational timescales.   Similarly, the binary might merge or encounter another resonance before the cloud has a chance to complete a single cycle. This is typically the case for the end state of a transition tree, like the $|8 \es 2 \es j \, \es \minus 1 \rangle$ superposition for the vector excited state in Fig.~\ref{fig:vectorTree}. Though we can avoid these issues by considering supermassive black holes and/or higher excited states where larger values of $\alpha$ are allowed, we are more likely to detect
  these oscillations for the earlier Bohr transitions. For instance, while the superpositions $|6 \es 2 \es j \, \es \minus 1 \rangle$ and $|7 \es 2 \es j \, \es \minus 1 \rangle$ oscillate $\sim \!40$ times slower than the $|3 \es 1 \es j \es 0 \rangle$ with frequency (\ref{eq:firstBohrSplit}), they still execute hundreds of cycles, which may be detected with high-precision templates.

Since finite-size effects enter at higher post-Newtonian orders, they are difficult to measure during the early stages of the inspiral when the binary's relative velocity is small \cite{Porto:2016pyg}. However, boson clouds can have much larger multipole moments compared to black holes in isolation \cite{Baumann:2018vus}, which can greatly enhance our chances of detecting them through gravitational-wave precision measurements, even before the system achieves high velocities in the merger phase. As we argued in \S\ref{sec:modulatingfinite}, the axisymmetric moments of the cloud roughly scale as $Q_c \sim M_c r_c^2$, where $M_c$ and $r_c$ are the mass and Bohr radius of the cloud, respectively. For an excited initial state, the dimensionless quadrupole moment is then of the order $\kappa_c \sim (M_c/M)\tilde{a}^{-2} \alpha^{-4}$, which is much larger than the corresponding moment for the pure Kerr black hole, $\kappa = 1$. The effects of these large, fluffy clouds are amplified near the merger, and we might expect that we can infer detailed information about the shape of the end states in a transition tree  by accurately modeling the finite-size effects in this phase.

\subsubsection*{Decaying shapes}

In \S\ref{sec:modulatingfinite}, we described how finite-size effects can receive further time-dependent changes when the cloud occupies decaying states. While this depletion typically occurs over much longer timescales than the orbital period, it can be significant when the decaying states have small angular momentum. 
For examples, the decay times of the $|2\es 0\es 0\rangle$ and $|2\es 0\es 1\es 1\rangle$ modes---the main depletion channels for the histories depicted in Figs.~\ref{fig:scalarTree} and \ref{fig:vectorTree}---are
\beq
\begin{aligned}
\left| \Gamma_{200}^{-1} \right| & \simeq 1 \, \text{yr} \left( \frac{M}{60 M_\odot} \right) \left( \frac{0.014}{\alpha} \right)^6 \mathrlap{\qquad\,\,\, \text{(scalar)}\,,} \\ 
\left| \Gamma_{2011}^{-1} \right| & \simeq 1 \, \text{yr} \left( \frac{M}{60 M_\odot} \right) \left( \frac{0.045}{\alpha} \right)^8 \mathrlap{\qquad\,\,\, \text{(vector)}\,.}
\end{aligned}
\eeq
We see that, even for moderate values of $\alpha$, these depletion effects can be significant and are observable within the typical lifetimes of gravitational-wave observatories. As pointed out in~\cite{Baumann:2018vus}, this change in the contribution from finite-size effects can also strongly indicate the presence of a boson cloud.

\newpage
\section{Summary}

In this chapter, we combined our findings in the previous two chapters and explored the phenomenological implications of the gravitational collider. Firstly, we incorporated the backreaction of the Landau-Zener transition on the binary orbit, which was omitted in our analysis in Chapter~\ref{sec:Collider}. We found that the transfer of angular momentum between the cloud and the orbit during each resonance can cause large corrections to the gravitational-wave signal from the binary. Notably, there is a dephasing with respect to the frequency evolution without a cloud, arising from transient floating and sinking orbits, cf. Figs.~\ref{fig:Floating} and \ref{fig:Sinking}. For a floating orbit, the orbital frequency of the binary is approximately constant and has a mean value that is given by the resonance frequency. Since the adiabaticity of the orbital evolution is enhanced in this case, the corresponding Landau-Zener transition involves a smooth and prolonged transition between the initial and the final states of the cloud. On the other hand, if the backreaction generates a sinking orbit, the orbital frequency of the binary temporarily increases across the resonance frequency. If this backreaction is sufficiently strong, the Landau-Zener transition can turn an initially adiabatic orbit to evolve non-adiabatically. As a result, the oscillatory features that we discovered in the previous chapter could be imprinted on the binary motion, significantly enriching the orbit evolution of the binary. The orbit could also receive a strong kick from the cloud, which could potentially generate large orbit eccentricities and even unbind the orbit.

\vskip 2pt

Furthermore, we saw that time-dependent finite-size effects provide additional information about the available states in the gravitational atom which, due to the extended nature of the cloud, may be observed during the early inspiral phase. Similarly, strong mixing with decaying modes during the resonant transition can also deplete the cloud as it approaches merger. The induced depletion of the cloud can have important phenomenological consequences, both for the monochromatic signal emitted by the cloud and the finite-size effects imprinted in the waveforms of the binary signal. The characteristic time dependence of the signals thus becomes a very distinctive feature of the dynamics of boson clouds in binary systems, which would not be present in other scenarios, e.g. a boson star. Figures~\ref{fig:scalarTree} and~\ref{fig:vectorTree} illustrate a few such examples for the evolutions of the scalar and vector clouds in binary systems. Crucially, observing these signatures would allow us to infer the microscopic properties of the ultralight boson fields. In particular, the resonance frequencies determine the boson's mass and we must observe the final state of a transition to distinguish particles of different spin. A precise reconstruction of a gravitational-wave signal can therefore help us not only to detect new ultralight bosons, but also distinguish between the scalar and vector cases.

\chapter{Searching for Dark Compact Objects} \label{sec:search}

So far, we focused on the dynamics and the signatures of gravitational atoms in binary systems. However, other types of dark compact objects can also arise in many scenarios for physics beyond the Standard Model. In order to search for any of these compact objects, we would need to perform matched filtering on the observational data with their associated template waveforms.\footnote{Coherent burst search methods~\cite{Klimenko:2015ypf, Lynch:2015yin, Cornish:2014kda} have been developed to detect transient gravitational waves in a model-independent way. Nevertheless, they only capture loud and short-duration events, such as the near-merger regime of binary coalescences, and are insensitive to weak and long-duration binary inspirals~\cite{Abbott:2016ezn,Abbott:2019heg, LIGOScientific:2018mvr}.}  Nevertheless, modeling accurate waveforms for each of these general\footnote{In this chapter, I use the word `general' to refer to any binary system involving at least one non-standard astrophysical compact object. In contrast, we use the word `standard' to refer to binary systems involving only black holes and neutron stars.} binary signals is a laborious task. Fortunately, the complicated microphysics of all of these objects are encoded in various universal finite-size effects, which we discussed in some detail in Section~\ref{sec:CBC}. In this chapter, I describe work based on the collaborative work~\cite{Chia:2020psj}, where we assess the extent to which existing template banks can be used to search for gravitational-wave signals emitted by general binary coalescences.

\section{Overview and Outline}

To address the problem described above, we compute the so-called effectualness~\cite{Damour:1997ub, Buonanno:2009zt} of existing template banks to general waveforms. The effectualness describes how much signal-to-noise ratio is retained when we compute the overlap between a signal and the best-fitting template waveform in a bank. In addition to the usual mass and spin parameters, these general waveforms incorporate the effects of the spin-induced quadrupole moment and the compactness of the binary components. A detection of these general signals would therefore represent a discovery of new physics in binary systems. Our template bank is designed to resemble those used by the LIGO/Virgo collaboration \cite{Brown:2012qf,DalCanton:2017ala,Roy:2017oul,2018arXiv181205121M}, demonstrating that, if these new signals exist, they could remain undetected. When constructing our template bank, we follow closely the geometric-placement method presented in~\cite{Roulet:2019hzy}. Our work is complementary to~\cite{Venumadhav:2019tad, Venumadhav:2019lyq} which further optimized the LIGO/Virgo search pipelines. Instead, we hope to broaden the searches beyond these standard binary black hole and binary neutron star signatures.

We emphasize that our work is in contrast to several proposed tests of new physics in binary systems~\cite{Krishnendu:2017shb, Krishnendu:2018nqa, Kastha:2018bcr, Kastha:2019brk}, which seek to measure or constrain plausible parametric deviations in observed waveforms. This \textit{a priori} assumes a successful detection of the new binary system. Detection is typically achieved through matched filtering with current template banks, which necessarily means that the waveform deviations are small. Our focus is instead on the detectability of these plausible new binary signals, including those that incur large deviations from the binary black hole template waveforms.

\begin{figure}[t]
        \centering
        \includegraphics[scale=0.81, trim=18 0 0 0]{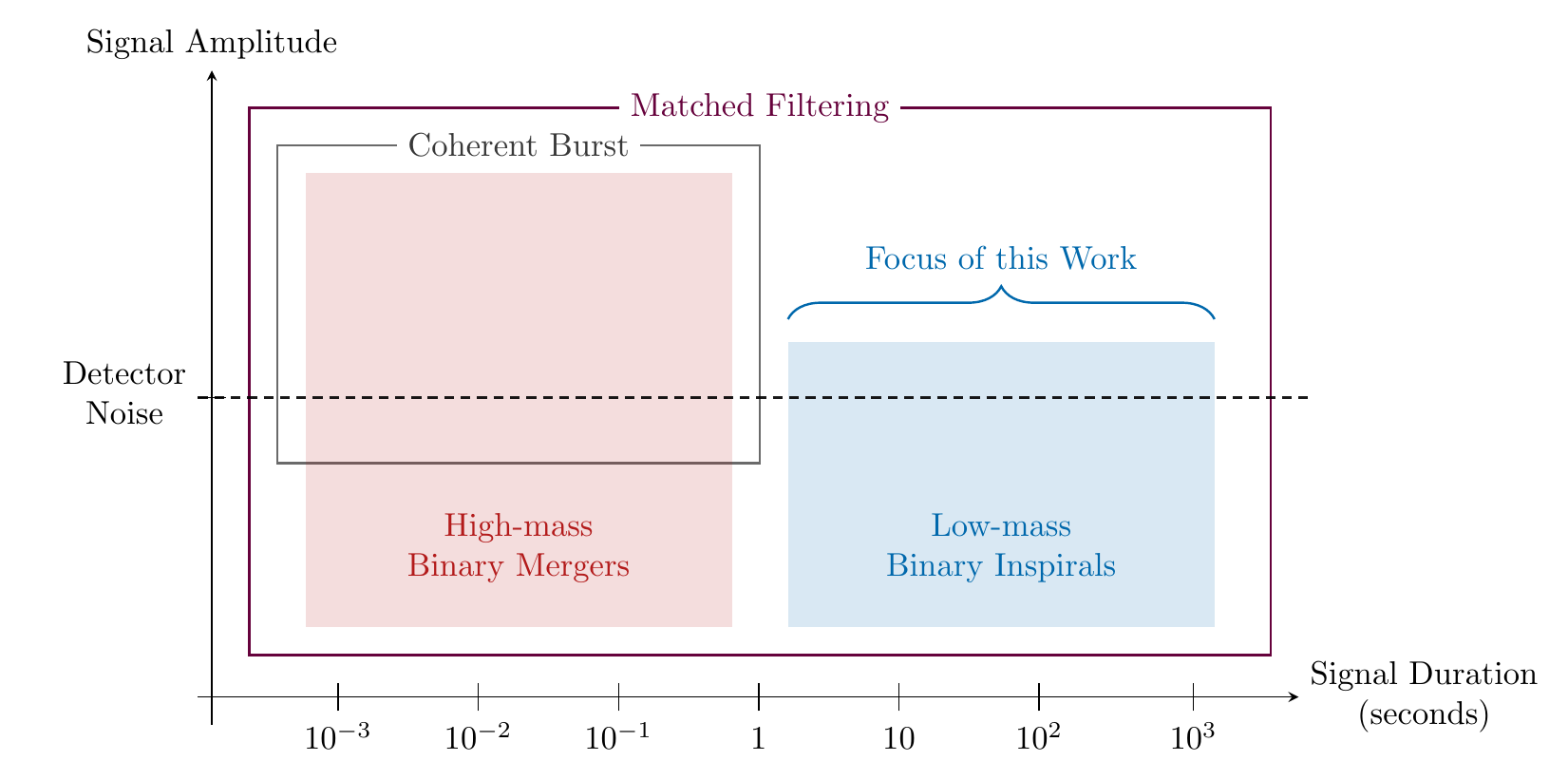}
\caption{Overview of the different methods used to search for gravitational waves emitted by a general binary system with the LIGO and Virgo detectors. The signals of low-mass binary systems are typically weaker and last longer than those of high-mass systems. The horizontal dashed line schematically illustrates the detector noise level, below which coherent burst searches become quickly insensitive. In this chapter we focus on low-mass binary inspirals which are \textit{only} detectable with matched filtering. } 
    \label{fig:search_overview}
\end{figure}

The ability to generate accurate template waveforms is a crucial prerequisite to achieving our goal. We therefore restrict ourselves to general low-mass binary systems, where the total mass of the binary is $M \lesssim 10 \, M_\odot$, for the following reasons:
\begin{itemize}

    \item In the LIGO and Virgo detectors, the inspiral regime only dominates for low-mass binary systems. The binary inspiral is an interesting regime because analytic results of the PN dynamics are readily available. This provides us with a well-defined framework to construct precise waveforms that incorporate additional physics, such as finite-size effects.

    \item The inspiraling signals of these systems last up to several minutes (corresponding to hundreds or thousands of cycles) and are typically very weak. They are therefore hard to detect with coherent burst searches. Matched filtering is the optimal and only realistic avenue to search for them  (cf. Fig.~\ref{fig:search_overview} for a comparison of these different search techniques).
    
\end{itemize}
\noindent By assuming that the PN dynamics are valid up until the merger regime, we have implicitly ignored other plausible effects that may occur even in the early inspiraling regime such as: Roche-lobe mass transfer~\cite{Paczynski1971}; third-body perturbation~\cite{Kozai, LIDOV1962719, Lim:2020cvm}; floating, sinking, or kicked orbits~\cite{Baumann:2019ztm}; dark matter environmental effects~\cite{Blas:2016ddr, Edwards:2019tzf, Kavanagh:2020cfn}; new fifth forces~\cite{Hook:2017psm, Kopp:2018jom, Wong:2019yoc}; and strong gravitational dynamics~\cite{Bezares:2017mzk, Palenzuela:2017kcg, Bezares:2018qwa}. Despite these limitations, our general waveforms still capture a wide class of new types of binary systems which have been overlooked in the literature.

The structure of this chapter is as follows: In Section~\ref{sec:binarywaveform}, I discuss the construction the waveform for a general binary inspiral. Specifically, the impact of various finite-size effects on the waveform will be incorporated. In Section~\ref{sec:testbank}, I describe the development of a template bank that is representative of those used in standard search pipelines. I then show the effectualness of our general waveform to this template bank.

\section{General Inspiral Waveforms} \label{sec:binarywaveform}

In this section, we construct the template waveforms for general binary inspirals. We first motivate several classes of dark compact objects that could exist in our Universe (\S\ref{sec:darkcompact}). We then review how an astrophysical object's multipole structure, imprints itself on the phase of the gravitational waves emitted by the binary system (\S\ref{sec:phase}). Finally, we describe how an object's size, parameterized through its compactness, affects the cutoff frequency of our template waveforms (\S\ref{sec:peakfreq}). These waveforms are generalizations of those constructed for binary systems with black holes and neutron stars. Their effectualness to existing template banks will be studied in Section~\ref{sec:testbank}.

\subsection{Dark Compact Objects} \label{sec:darkcompact}
In Section~\ref{sec:weak-coupling}, I discussed two broad classes of dark matter based on their mass ranges, which in turn determine whether they behave as ``waves" or ``particles" over astrophysical length scales. In the former case, the Compton wavelengths of the fields can be larger than the sizes of astrophysical black holes and can therefore form gravitational atoms. In the latter case, other types of dark compact objects can arise. Here I briefly discuss a few such examples.

Particle-like dark matter can accrete around black holes and form high-density minispikes, with overdensities that peak near the black hole horizon~\cite{Gondolo:1999ef, Sadeghian:2013laa}. In the simplest scenario where the accretion process is uninterruped by other astrophysical processes, these spikes would evolve from an initial dark matter halo distribution, such as the NFW profile, to a profile that is several orders of magnitude denser. Since the accretion occurs over long astrophysical timescales, in reality, astrophysical baryonic effects and self-annihilations of the dark matter particles could flatten these spikes~\cite{Merritt:2002vj, Sadeghian:2013laa, Ferrer:2017xwm}.  Despite these uncertainties, the density profiles of these spikes could still be much larger than the average dark matter density in our Universe.  Like the gravitational atom, these minispikes make black holes interesting probes of dark matter. Indeed, one of the best ways of probing the true radial profiles of these spikes is through their signatures in binary systems, which could potentially allow us to infer the microscopic properties of the dark matter particles~\cite{Hannuksela:2019vip, Kavanagh:2020cfn}.

A myriad of bosonic configurations can arise in physics beyond the Standard Model scenarios that propose the existence of additional massive boson fields in our Universe. These objects differ from the gravitational atoms in that they have regular boundary conditions at their centers, while the gravitational atoms satisfy the ``only-ingoing" condition at the black hole horizon. These objects are often called boson stars if they are made up of complex fields~\cite{Kaup1968, Ruffini1969, Breit:1983nr, Colpi1986, Liebling:2012fv} (see e.g.~\cite{Schunck:2003kk} for a review), and are stabilized by the fact that they possess conserved charges. On the other hand, if the objects are made of real fields, they can form solitons~\cite{Lee1987-2, Lee1987-3, Lee1987-4, Lynn:1988rb} or oscillons~\cite{Seidel1991, Copeland:1995fq, Braaten:2015eeu}, with the oscillons being denser, more compact, and therefore subject to more strong relativistic effects. Unlike boson stars, these real bosonic configurations are time-dependent and therefore dissipates through gravitational-wave radiation. All of these boson states can be formed either through the Jean's instability of a relic initial density in the early Universe or through  gravitational cooling processes, see e.g.~\cite{Bianchi:1990mha, Seidel:1993zk}. Depending on the precise microphysical properties of the underlying bosons, these configurations would have distinctive mass-radius relations. By measuring the mass, radius, and other finite-size effects in theirn binary waveforms, we could potentially detect the presence of these compact objects and even infer the microscopic properties of the boson fields. 

\newpage

Finally, we briefly mention that primordial black holes could exist in our Universe~\cite{HawkingPBH}. Unlike astrophysical black holes that are formed through collapses of stars, primordial  black holes are proposed to had been formed in the early Universe, such as through critical collapse of curvature perturbations due to sharp features in the primordial curvature power spectrum (see e.g.~\cite{Sasaki:2018dmp} for a review). Importantly, the masses of primordial black holes do not have to obey the Tolman-Volkoff-Oppenheimer lower bound $\gtrsim 3 M_\odot$, which only applies to black holes that are formed through stellar collapses~\cite{Tolman1939}. As such, while it can be difficult to distinguish between primordial black holes and those of astrophysical origins if they are heavy, a detection of low-mass black holes could indicate the presence of these putative new objects. Since the finite-size effects of black holes are known in detail, e.g. the spin-induced quadrupole moment $\kappa=1$ and tidal Love number $\lambda=0$, cf. Section~\ref{sec:CBC}, one can further distinguish primordial black holes from other putative low-mass compact objects through precise measurements of their finite-size effects in binary waveforms.

\subsection{Dephasing from Quadrupole Moment} \label{sec:phase}

In Section~\ref{sec:CBC}, we described how a general astrophysical object can source a series of multipole moments~\cite{Geroch:1970cd, Hansen:1974zz, Thorne:1980ru}. For objects that are spherically symmetric when they are not spinning, the dominant finite-size effect is given by the spin-induced quadrupole moment, which is often parameterized through the relation (\ref{eqn:Quad}). There, the $\kappa$ parameter is the dimensionless quadrupole parameter and $\chi$ is the dimensionless spin of the object. Importantly, the value of $\kappa$ depends sensitively on the detailed properties of the object. It is instructive to recall what the typical values of $\kappa$ are for known examples. In particular, we know that Kerr black holes  have $\kappa = 1$~\cite{Hansen:1974zz, Thorne:1980ru}, while $2\lesssim \kappa \lesssim 10$ for neutron stars, with the precise value depending on the nuclear equation of state~\cite{Laarakkers:1997hb, Pappas:2012ns}. For more speculative objects, such as superradiant boson clouds and boson stars, $\kappa$ can be as large as $\sim 10^2 -10^3$. It can even develop oscillatory time-dependences or vary significantly with $\chi$~\cite{Baumann:2018vus, Baumann:2019ztm, Ryan:1996nk, Herdeiro:2014goa}. Absent a specific compact-object model in mind, we will henceforth treat $\kappa$ as a free constant parameter, with the requirement that $\kappa \geq 1$.

\begin{figure}[t!]
        \centering
        \includegraphics[width=1.0\textwidth, scale=0.9]{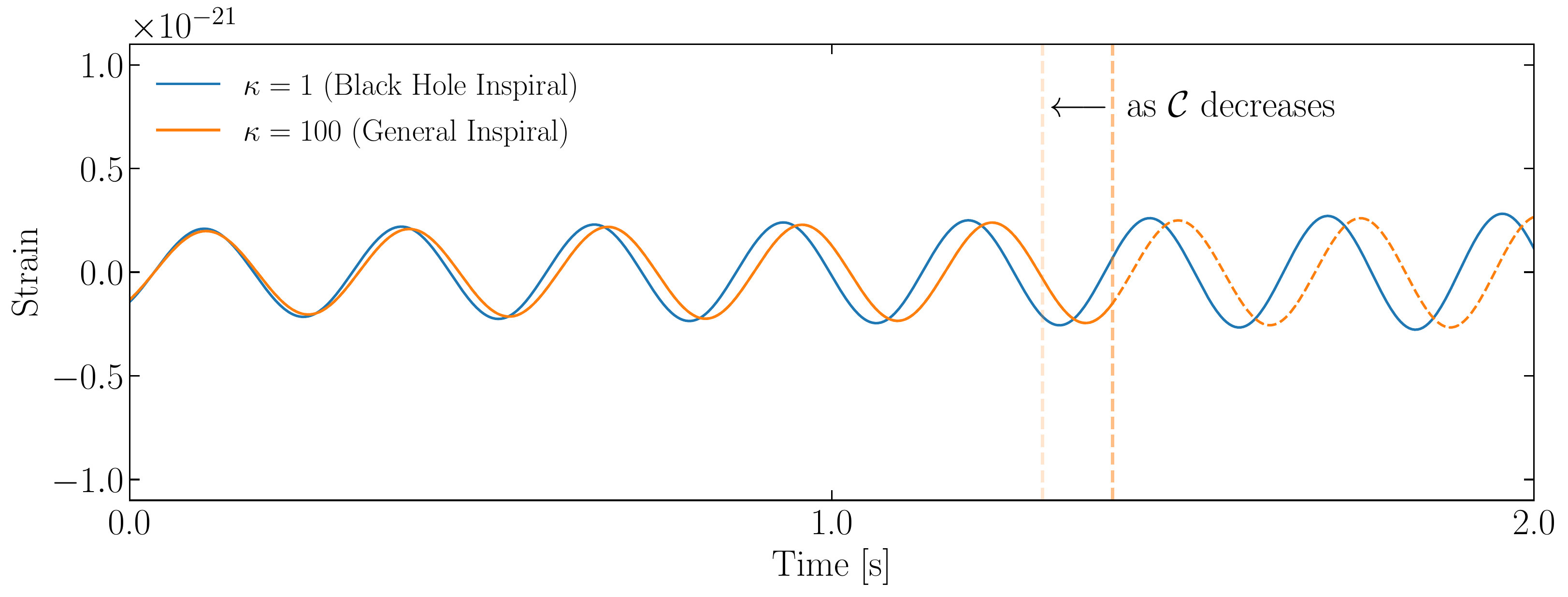}
\caption{Comparison between the waveforms of a binary black hole inspiral and a general binary inspiral. In both cases, the binary components' masses and spins are $m_1 = m_2 = 2 M_\odot$ and $\chi_1 = \chi_2 = 0.8$. No amplitude modulation is observed as we assume that the components' spins are aligned with the orbital angular momentum of the binary. One of the binary components is assumed to be a black hole $(\kappa_1 = 1)$, while the other can be a more general object $(\kappa_2 \equiv \kappa \geq 1)$. The vertical dashed line represents the cutoff of our general waveform, which happens earlier in the inspiral (lower frequency) for binary components which are less compact (i.e. as $\mathcal{C}$ decreases).}
    \label{fig:waveform}
\end{figure}

When the object is part of a binary system, the precise effect of $Q$ on the gravitational-wave signal is known in the early-inspiral, post-Newtonian regime of the coalescence~\cite{Barker75, Poisson:1997ha, nrgrs, Porto:2008jj, Hergt:2010pa, Buonanno:2012rv, Bohe:2015ana, Marsat:2014xea, Krishnendu:2017shb}. This effect was discussed in Section~\ref{sec:CBC}; for future convenience, we repeat some of the key points here. To simplify our analysis, we restrict ourselves to binary orbits which are quasi-circular, and assume that the binary components' spins are parallel to the (Newtonian) orbital angular momentum vector of the binary. In the Fourier domain, the gravitational-wave strain is~\cite{Cutler:1994ys} 
\beq
\tilde{h} (f; \bm{p}) = \mathcal{A}(\bm{p}) \, f^{-7/6} \, e^{i \psi (f; \bm{p})} \, \theta \big( f_{\rm cut} (\bm{p}) - f \big)\, , \label{eqn:waveform}
\eeq
where $f$ is the gravitational-wave frequency, $\bm{p}$ is the set of intrinsic parameters of the binary, $\mathcal{A}$ is the strain amplitude,\footnote{The amplitude is independent of $f$ at leading Newtonian order. We will ignore higher-order PN corrections to $\mathcal{A}$, as they do not substantially affect the overlap between different waveforms. As a result, the constant $\mathcal{A}$ disappears in the normalized inner product (see Section~\ref{sec:testbank} later).} $\psi$ is the phase, $\theta$ is the Heaviside theta function, and $f_{\rm cut}$ is the cutoff frequency of our general waveform. Schematically, the phase evolution reads
\beq
\begin{aligned}
\psi (f; \bm{p})  = & \,\, 2 \pi f t_c - \phi_c - \frac{\pi}{4} + \frac{3}{128 \hskip 1pt \nu \hskip 1pt v^5} \Big( \psi_{\rm NS} + \psi_{\rm S} \Big) \, , \label{eqn:TaylorF2phase}
\end{aligned}
\eeq 
where $t_c$ is the time of coalescence, $\phi_c$ is the phase of coalescence, $\nu = m_1 m_2 / M^2$ is the symmetric mass ratio, $M=m_1 + m_2$ is the total mass of the binary, and $v = (\pi M f)^{1/3}$ is the circular orbital velocity. The quantities $\psi_{\rm NS}$ and $\psi_{\rm S}$ represent the non-spinning and spinning phase contribution, respectively. Because the overlap between waveforms are especially sensitive to phase coherence~\cite{Cutler:1992tc}, we will retain terms in the phase up to 3.5PN order --- these terms are fully known in the literature; see for example~\cite{Arun:2004hn, Wade:2013hoa, Mishra:2016whh, Krishnendu:2017shb}. The quadrupole parameter $\kappa$ in (\ref{eqn:Quad}) contributes to $\psi_{\rm S}$ through the interaction between $Q$ and the tidal field sourced by the binary companion. It first appears at 2PN order~\cite{Poisson:1997ha, Wade:2013hoa}
\beq
\psi_{\rm S} \supset - 50 \sum_{i=1}^2  \, \left( \frac{m_i}{M} \right)^2 \kappa_i \hskip 1pt  \chi_i^2 \, v^4 \, , \label{eqn:Kappa2PN}
\eeq
where the subscript $i=1, 2$ represents each of the binary components. It also appears in the phase at 3PN order, though in a slightly complicated manner~\cite{Krishnendu:2017shb}
\beq
\begin{aligned}
\psi_{\rm S} \supset \frac{5}{84} \sum_{i=1}^2 \sum_{j \neq i}    \, \bigg[ & 15609  \left( \frac{m_i}{M}\right)^2  \\
& + 27032 \hskip 1pt \frac{m_i m_j}{M^2} + 9407 \hskip 1pt \left( \frac{m_j}{M} \right)^2 \bigg] \left( \frac{m_i}{M} \right)^2  \kappa_i \hskip 1pt \chi_i^2 \, v^6 \, . \label{eqn:Kappa3PN}
\end{aligned}
\eeq
We have simplified (\ref{eqn:Kappa2PN}) and (\ref{eqn:Kappa3PN}) by enforcing $\chi_i$ to be aligned with the orbital angular momentum ($0 < \chi_i \leq 1$) or anti-aligned with it ($-1 \leq \chi_i < 0$). By incorporating these dephasing effects into template waveforms, we can potentially detect the presence of new astrophysical objects through observations of a binary's inspiral. As we will show, if the dephasing is large enough, these signals could even be missed by current LIGO/Virgo template-bank searches (see Fig.~\ref{fig:waveform} for an illustration of a dephased waveform).

In principle, the gravitational waves emitted during the binary's merger regime also provide information about the quadrupolar structure of the objects. However, the detailed dynamics of this regime are often sensitive to the microphysics of the objects and can only be resolved accurately through numerical relativity simulations. To preserve analytic control over our waveform (\ref{eqn:waveform}), we will ignore the merger regime throughout this chapter. This is achieved in practice by restricting ourselves to low-mass binary systems with $M \lesssim 10 M_\odot$, whose merger frequencies typically lie above the upper bound of the observational windows of ground-based detectors (see \S\ref{sec:peakfreq} later for more detailed discussions). Crucially, these binary systems would inspiral within the detector bands over a large number of orbiting cycles, making matched filtering with our general inspiral waveform the optimal way of searching for them, cf. Fig.~\ref{fig:search_overview}.

While we have only focused on the object's quadrupole moment so far, other types of finite-size effects, such as the object's higher-order spin-induced moments and tidal deformabilities~\cite{Chia:2020yla, Binnington:2009bb, Damour:2009vw, Landry:2015zfa, Pani:2015hfa}, can also contribute to the phase (\ref{eqn:TaylorF2phase}). Nevertheless, these additional terms only start to appear at $3.5$PN~\cite{Levi:2014gsa, Marsat:2014xea, Krishnendu:2017shb} and $5$PN~\cite{nrgr, Flanagan:2007ix} orders, respectively, and are therefore subdominant compared to (\ref{eqn:Kappa2PN}) and (\ref{eqn:Kappa3PN}). In particular, these higher-PN terms are suppressed in the early inspiral of the coalescence, and only become non-negligible in the strong gravity, near merger regime.\footnote{Since the leading-order term in (\ref{eqn:TaylorF2phase}) scales as $\sim v^{-5}$, roughly speaking, orbital parameters that appear at $\lesssim 2.5$PN order affect the phase predominantly in the inspiralling stage, when the number of inspiraling cycles is large, while those with $\gtrsim 2.5$PN become more prominent near merger; see for example~\cite{Harry:2018hke}.} By focusing on the binary's early inspiral regime we can therefore ignore these higher-order effects. For concreteness, we will set these quantities to their corresponding values for black holes~\cite{Hansen:1974zz, Thorne:1980ru, Landry:2015zfa, Pani:2015hfa}.\footnote{Note however that we retain the $\kappa$-dependence in the 3.5PN phasing term~\cite{Krishnendu:2017shb} in Section~\ref{sec:testbank}.}

\subsection{Cutoff Frequency from Small Compactness} \label{sec:peakfreq}

While the parameter $\kappa$, described in \S\ref{sec:phase}, characterizes the deformation of an astrophysical object's shape, it does not carry information about its size. This is instead described by the compactness parameter
\beq
\mathcal{C} \equiv \frac{m}{r} \, , \label{eqn:compactness}
\eeq
where $r$ is the equatorial radius of the object. Black holes, which are the most compact known astrophysical objects, have $0.5 \leq \mathcal{C} \leq 1$, where the lower and upper bound corresponds to the compactness of a Schwarzschild and an extremal Kerr black hole respectively.
For neutron stars, $\mathcal{C}$ can range between $\sim 0.1 -0.3 $~\cite{Ozel:2016oaf, Abbott:2018exr}. On the other hand, objects that arise in many BSM scenarios can be much more diffuse, with $\mathcal{C} \ll 0.1$ (for example see~\cite{AmaroSeoane:2010qx, Croon:2018ybs}).

The compactness of each binary component can significantly affect the cutoff frequency of the waveform (\ref{eqn:waveform}) because together they dictate the binary separation at which the merger occurs. For binary systems with highly-compact objects, such as black holes and neutron stars, the merger frequencies must be deduced through detailed numerical relativity studies, as the full non-linear dynamics of the binary merger must be taken into account (see for example~\cite{Bernuzzi:2015rla, Bohe:2016gbl} for the fitting formulae for the merger frequencies of binary black hole and binary neutron star waveforms). Roughly speaking, these strong-gravity effects become important when the orbital separation, $R$, is smaller than the binary's innermost stable circular orbit (ISCO), $r_{\rm ISCO} \approx 6M$~\cite{Blanchet:2006zz, Schafer:2009dq}. However, if the binary components have sufficiently small $\mathcal{C}$'s, the binary would have already merged at $R < r_{\rm ISCO}$. In this case, the strong-gravity regime is not reached at merger, making the analytic PN approximation still a valid description of the dynamics.

Pledging ignorance to the merger dynamics of binary systems with small-compactness objects, we will terminate our waveform (\ref{eqn:waveform}) when the binary touches, i.e. when $R = r_1 + r_2$.\footnote{Some papers use $R=r_{\rm ISCO}$ as a merger condition for low-compactness binary systems. However, the notion of an ISCO ceases to exist for objects with $\mathcal{C} \lesssim 1/6 \approx 0.17$, as this fictitious ISCO would be located in the interior of the object. This leads to a factor of $\approx 6$ underestimation in $f_{\rm cut}$, which can substantially reduce the frequency range over which the SNR could be accummulated. } Using Kepler's third law, our cutoff frequency is therefore
\beq
f_{\rm cut} = \frac{1}{\pi} \sqrt{\frac{m_1 + m_2}{\left(m_1 / \mathcal{C}_1 + m_2 / \mathcal{C}_2\right)^3}}   \hskip 8pt , \label{eqn:cutoff}
\eeq
where we have ignored PN corrections to (\ref{eqn:cutoff}). Furthermore, we have neglected the influence of the binary components' quadrupoles and higher-order multipole moments on (\ref{eqn:cutoff}), which can be important when the binary separation is small. These additional corrections would deepen the gravitational potential of the binary, thereby increasing the cutoff frequency towards larger values~\cite{Blanchet:2006zz, Schafer:2009dq}. In other words, (\ref{eqn:cutoff}) underestimates the actual touching frequency of the binary, and may thus be viewed as a conservative cutoff of our waveform frequency. The impact of this cutoff on our waveform is schematically illustrated in Fig.~\ref{fig:waveform}. In the special case where both of the binary components have the same compactness, $\mathcal{C}_1 = \mathcal{C}_2 = \mathcal{C}$, (\ref{eqn:cutoff}) becomes
\beq
f_{\rm cut} \simeq  1440 \, \text{Hz}  \left( \frac{\mathcal{C}}{0.2} \right)^{3/2} \left(\frac{4 M_\odot}{M}\right)   \, . \label{eqn:cutoff2}
\eeq
For low-mass binary systems ($ M \lesssim 10 M_\odot$), the observational lower bound of ground-based detectors, $f \gtrsim 10 \, \text{Hz} $, implies that we can probe binary inspirals with $\mathcal{C} \gtrsim 10^{-3}$. Although the touching condition (\ref{eqn:cutoff2}) is inaccurate for binaries with $\mathcal{C} \gtrsim 1/6 \approx 0.17$ (see discussion above), their actual merging frequencies are greater than $f \gtrsim 10^3 \, \text{Hz}$, which is beyond the upper bound of the detector sensitivity bands. The precise values of $f_{\rm cut}$ in these cases are therefore immaterial, as the inspiral signal in the observational band remains unchanged.

Finally, we note that the parameters $\chi, \kappa$ and $\mathcal{C}$ of a given astrophysical object are in principle related to each other. For instance, by requiring the speed of a test mass on the equatorial surface to be smaller than its escape velocity, we can obtain a mass-shedding bound that relates $\mathcal{C}$ with the maximum value of $\chi$. Furthermore, through simple dimensional analysis $Q \propto m r^2$, we find that $\kappa \chi^2 \propto \mathcal{C}^{-2}$. These imply that the dephasing from (\ref{eqn:Kappa2PN}) and (\ref{eqn:Kappa3PN}) are actually correlated with the compactness of the object for a given equation of state. Absent a detailed astrophysical model in mind, we will treat $\chi, \kappa$, and $\mathcal{C}$ as independent parameters, although we emphasize that their implicit correlation can perhaps be exploited in future work, similar to how universal relations~\cite{Yagi:2013bca} are used to simplify analyses of binary neutron star signals.

\section{Detectability of General Inspirals} \label{sec:testbank}

In this section, we study the detectability of our newly proposed waveforms through current matched-filtering searches.
We first construct a binary black hole template bank that is representative of those used by the LIGO/Virgo collaboration (\S\ref{sec:templateBank}). We then investigate how reliable this template bank is at recovering our general inspiral waveforms, specifically by computing its \textit{effectualness} (\S\ref{sec:match}). As a prerequisite to these analyses, we introduce the inner product between two arbitrary waveforms, $h_1$ and $h_2$, defined as~\cite{Cutler:1994ys} 
\beq
  \left(h_1|h_2\right) \equiv 4 \, \mathrm{Re} \int^{\infty}_{0} \d f \, \frac{ \tilde{h}_1(f) \tilde{h}^*_2(f)}{S_n(f)}\, , \label{eqn:innerprod}
\eeq
where $\tilde{h}_1, \tilde{h}_2$ are their Fourier representations, and $S_n$ is the (one-sided) noise spectral density. For future convenience, we denote the normalized inner product by
\beq
\left[h_1|h_2\right] \equiv \frac{\left(h_1|h_2\right)}{\sqrt{\left(h_1|h_1\right)\left(h_2|h_2\right)}}\, . \label{eqn:dotprodnorm}
\eeq
Throughout this work, we use the \texttt{aLIGO\_MID\_LOW}~\cite{LIGOnoise} detector specification for $S_n$, which is representative of the first LIGO observational run, O1. When evaluating the frequency integral (\ref{eqn:innerprod}), we use the lower and upper cutoff frequencies $f_l=30 \mathrm{\, Hz}$ and $f_u=512 \mathrm{\, Hz}$. These choices reflect the fact that low mass binary inspirals accumulate a minimum of 95\% of their signal-to-noise ratio (SNR) within this frequency range. Finally, although represented as a continuous integral, (\ref{eqn:innerprod}) is in practice evaluated discretely in frequency. We therefore specify a sampling rate of $1024\mathrm{\, Hz}$ and take the maximum time spent in band to be $94 \mathrm{\, s}$.\footnote{This corresponds to the time it takes a binary with component masses $m_1=m_2=1 \, M_{\odot}$, which is the smallest mass we consider in this chapter, to inspiral between the stated frequency cutoffs.} These choices give us the grid of possible coalescence times to maximize over when computing the effectualness later.

\subsection{Binary Black Hole Template Bank}
\label{sec:templateBank}

We now construct a template bank that is representative of those used by the LIGO/Virgo Collaboration \cite{Brown:2012qf,DalCanton:2017ala,Roy:2017oul,2018arXiv181205121M}. In order to do so, we use the Taylor F2 waveform model for binary black holes~\cite{Arun:2004hn, Wade:2013hoa, Mishra:2016whh} with intrinsic parameters $\bm{p}_{ \tmp} = \{m_1, \, m_2, \,\chi_1, \,\chi_2\}$, where the spins $\chi_1, \chi_2$ are parallel to the orbital angular momentum of the binary. This waveform model is exactly the same as that in (\ref{eqn:waveform}), except we now specify the $\kappa$ parameters to be unity~\cite{Hansen:1974zz, Thorne:1980ru} and neglect the cutoff frequency introduced by the $\mathcal{C}$'s for black holes. Crucially, since we are only interested in the signals emitted during the inspiraling regime, we do not have to use the phenomenological~\cite{Khan:2015jqa} or the effective-one-body waveform~\cite{Bohe:2016gbl} models, which include numerical relativity waveforms near the binary merger.

Matched-filtering searches involve computing the inner product (\ref{eqn:dotprodnorm}) between the data and a set of template waveforms. Practically, this requires a discretized sampling of the parameter space $\bm{p}_{\tmp}$ in the form of a template bank. We adopt the geometric-placement technique described in~\cite{Roulet:2019hzy}, although many other methods exist; see for example~\cite{Harry:2009ea,Ajith:2012mn,Brown:2012qf}. 
We only sketch this method here and refer the reader to the original work for further details. The key feature of this formalism lies in the following decomposition of the waveform phase
\beq
 \psi (f; \bm{p}_{\tmp}) = \overline{\psi} (f) + \sum_{\alpha=0}^{N} c_\alpha (\bm{p}_{\tmp}) \, \psi_\alpha (f) \, , \label{eqn:Phase_geometric}
\eeq
where $\overline{\psi} $ is an average behaviour of the phase, which is chosen for convenience, $c_\alpha$ is a set of basis coefficients that only depend on $\bm{p}_{\rm bbh}$, and $\psi_\alpha$ is a set of orthonormal basis functions that satisfy $\langle \psi_\alpha | \psi_\beta \rangle  = \delta_{\alpha \beta}$, with $\langle \cdot | \cdot \rangle$ being an inner product that is slightly modified from (\ref{eqn:innerprod})~\cite{Roulet:2019hzy}. Crucially, this new inner product is designed such that the mismatch distance\footnote{The match between two waveforms $h_1$ and $h_2$ is obtained by maximizing (\ref{eqn:dotprodnorm}) over the time and phase of coalescence, i.e. $ m_{12} \equiv \max _{t_{c}, \phi_{c}} \hskip 2pt \left[h_1 \, | \, h_2 \right] $. The mismatch distance is then defined as $\sqrt{1-m_{12}}$.} between neighbouring waveforms in parameter space translates to a Euclidean distance in the space of $c_\alpha$. After the phases of a random sample of waveforms are extracted, the functions $\psi_\alpha$ can then be computed through a singular-value decomposition of a matrix that consists of these extracted phases~\cite{Roulet:2019hzy}. The template bank is finally constructed by grid sampling the set of basis coefficients $c_\alpha $. Because the metric is Euclidean in the space of $c_\alpha$ (at least for neighbouring waveforms), an optimal bank is therefore easily obtained by sampling a regular grid in this space. The dimension of the basis, $N$, and the grid spacing, $\Delta c_\alpha$, therefore dictate the resolution of our template bank.

To create the basis functions $\psi_\alpha$, we randomly sample $4\times10^4$ parameter combinations and generate waveforms for each. We use the parameter ranges $1.0 M_\odot \leq m_i \leq 3.0 M_\odot$ and $|\chi_i| \leq 0.8$ for component masses and dimensionless spins respectively. These parameter ranges in turn dictate the ranges of $c_\alpha$. This mass range is motivated by the fact that the resulting binary systems have relatively large chirp masses, and at the same time, would inspiral over long periods of time in the detector band (e.g. $\gtrsim 10\,$s). A study which includes a wider mass range, especially subsolar-mass objects, is certainly possible, though it would not alter the qualitative conclusions of this work (see \S\ref{sec:match} for a more detailed discussion). We find that taking $N=3$ in (\ref{eqn:Phase_geometric}) is sufficient to describe the behaviour of the phases of these low-mass binary systems (cf. Fig.~\ref{fig:resphase}). Finally, we take the grid spacing $\Delta c_\alpha = 0.13$\footnote{Not all combinations of $c_\alpha$ give physically realizable waveforms. We therefore use a fudge factor~\cite{Roulet:2019hzy} of $\zeta=0.01$.}, which leads to a total of $ 572,558$ templates in our bank. As we shall see in \S\ref{sec:match}, these choices lead to a very well-sampled template bank. Note that we did not seek to minimize the number of templates, but instead ensured that our bank's coverage is sufficient to assess the loss of effectualness for our general waveforms. 

\newpage

\subsection{Effectualness to Inspiral Waveforms} 
\label{sec:match}

We are now ready to test how well a binary black hole template bank can be used to detect the general inspiral waveforms that we constructed in Section~\ref{sec:binarywaveform}. For concreteness, we assume that one of the binary components is a black hole, $\kappa_1 = 1$ and $0.5 \leq \mathcal{C}_1 \leq 1$, while the other is a general compact object, whose finite-size parameters are labeled by $\kappa_2 \equiv \kappa$ and $\mathcal{C}_2 \equiv \mathcal{C}$. Since it is important that we distinguish the intrinsic parameters of the binary black hole template waveforms from those of the general waveforms, we will denote them by  $\bm{p}_{\tmp}$ and $\bm{p}_{\gen}$ respectively, with the latter being $\bm{p}_{\gen} = \{ m_1, \, m_2, \, \chi_1, \, \chi_2, \, \kappa, \, \mathcal{C} \}$.

The primary tool for assessing the template bank's effectiveness at recovering our general waveforms is its effectualness~\cite{Damour:1997ub, Buonanno:2009zt}. More precisely, this is obtained by maximizing the inner product (\ref{eqn:dotprodnorm}) between the template and general waveforms over their relative time of coalescence $t_c$, phase of coalescence $\phi_c$, and the intrinsic parameters $\bm{p}_{\tmp}$ of every template in the bank:
\beq
\varepsilon \left( \bm{p}_{\tmp}, \bm{p}_{\gen} \right) \equiv \max _{t_{c}, \phi_{c}, \{\bm{p}_{\tmp} \} } \hskip 2pt \left[h (\bm{p}_{\tmp}) \, | \, h (\bm{p}_{\gen}) \right]  \, , \label{eqn:effectualness}
\eeq
where $\{ \bm{p}_{\tmp} \}$ denotes the list of template parameter combinations. In other words, (\ref{eqn:effectualness}) quantifies the overlap between the general waveform and the best-fitting template waveform in the bank. To obtain a rough idea of how a reduced effectualness translates into a loss in signal detectability, we note that existing searches adopt an SNR detection threshold of 8, while a typical binary system detected thus far has an SNR ranging from $10$ to $15$~\cite{LIGOScientific:2018mvr}. For a signal with true SNR of $12.5$, a template bank with effectualness $ \varepsilon < 8/12.5 \approx 0.64$ would reduce the observed SNR to values below the detection threshold, thereby leading to missed events. This can instead be phrased as a reduced sensitive volume of $1-\varepsilon^3 \gtrsim 0.74$~\cite{Lindblom:2008cm}. A commonly adopted target in LIGO and Virgo searches is $\varepsilon > 0.97$, which leads to a $10\%$ loss in sensitive volume.

While the maximization of (\ref{eqn:effectualness}) over $t_c$ and $\phi_c$ can be performed efficiently (through fast Fourier transform), the iterative computation over the list $\{  \bm{p}_{\tmp} \}$ is much more computationally expensive. One of the benefits of the geometric-placement method described in \S\ref{sec:templateBank} is that this brute-force iteration can be replaced by a simple search for the \textit{best-fitting point, $ \left\{ \hskip 1pt c_\alpha (\bm{p}_{\tmp}) \hskip 1pt \right\}_{\mathrm{best}}$}, in the bank~\cite{Roulet:2019hzy}. This is achieved by projecting $c_\alpha(\bm{p}_{\gen}) = \langle \psi (\bm{p}_{\gen}) - \overline{\psi} \, | \, \psi_\alpha\rangle$, where $\psi (\bm{p}_{\gen})$ is the phase of the general waveform, while $\overline{\psi}$ and $\psi_\alpha$ are the average phase and basis functions constructed for our bank in (\ref{eqn:Phase_geometric}), respectively. The best-fitting point  $ \left\{ \hskip 1pt c_\alpha (\bm{p}_{\tmp}) \hskip 1pt \right\}_{\mathrm{best}}$ is the closest $c_\alpha(\bm{p}_{\tmp})$ to $c_\alpha(\bm{p}_{\gen})$, as measured by their Euclidean distance. The effectualness can then be evaluated straight-forwardly using these nearby parameters and maximizing over $t_c$ and $\phi_c$.

\begin{figure}[t!]
        \centering
        \includegraphics[width=\textwidth]{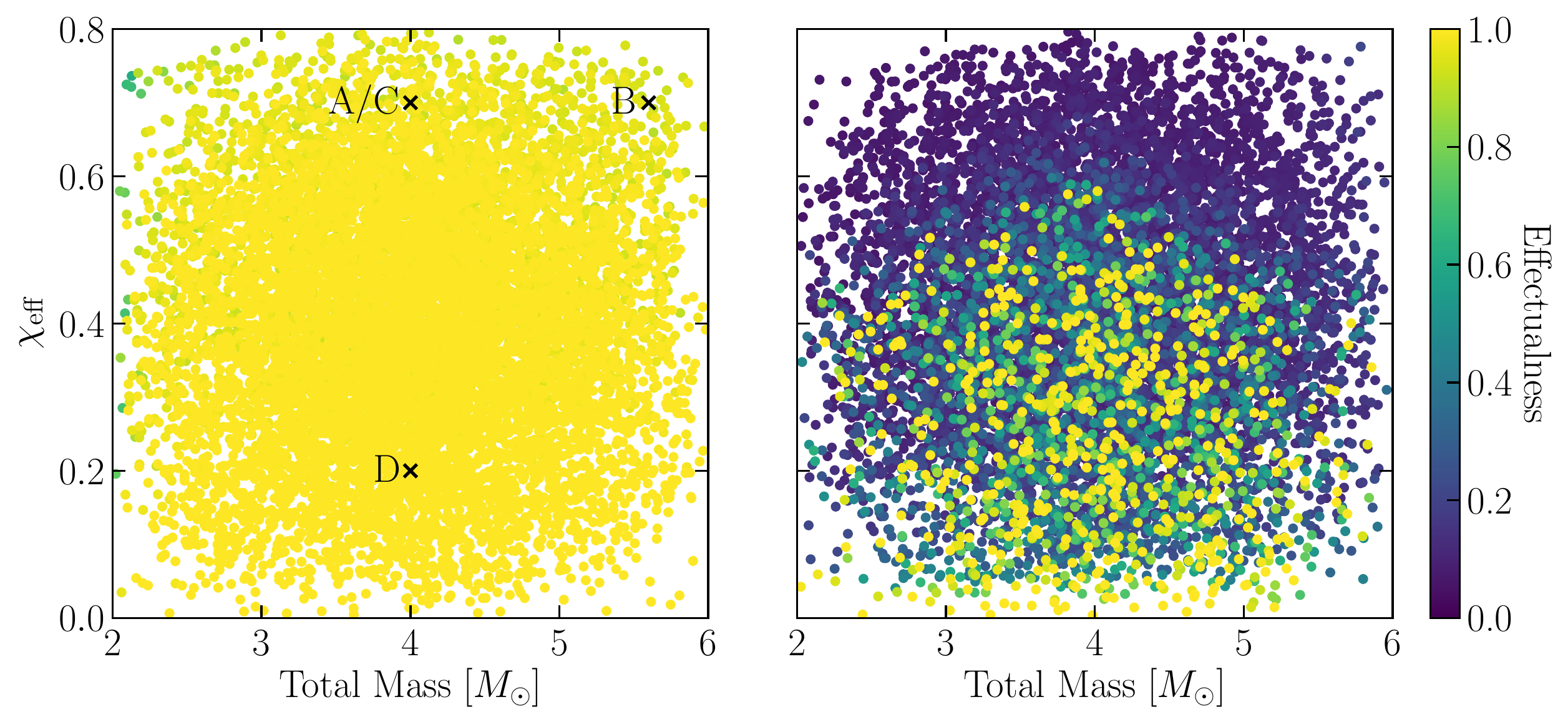}
\caption{
Effectualness of our template bank for binary black hole inspiral waveforms (\textit{left}) and general inspiral waveforms with $\kappa=500$ and $\mathcal{C}=0.1$ (\textit{right}). The effectualness is the ratio of the observed SNR to the true SNR of a signal. Comparing these panels we see a drastic loss in effectualness from the finite-size effects. For convenience, we have indicated the masses and spins of scenarios A$-$D that are considered in Table~\ref{table:scenarios} and Fig.~\ref{fig:matchvkap} in the \textit{left} panel. Note that all points are plotted from least to most effectual --- the points with the highest effectualness are therefore the most visible.}     \label{fig:tbankeffectualness} 
\end{figure}

To test the validity of this procedure, we randomly sample an independent set of $10^4$ binary black hole waveforms within the same parameter ranges used to generate $\psi_\alpha$ in \S\ref{sec:templateBank}. We then compute our template bank's effectualness to these waveforms with the prescription above. The result is shown in the \textit{left} panel of Fig.~\ref{fig:tbankeffectualness}, where the effectualness is plotted as a function of the binary total mass $M$ and the effective mass-weighted spin, $\chi_{\rm eff} = \left( m_1 \chi_1 + m_2 \chi_2 \right) / M $.  We find that 99\% of the random templates have $\varepsilon > 0.9$ and 84\% have $\varepsilon > 0.97$. We also find that the effectualness decreases slightly as $\chi_{\rm eff}$ increases, indicating that the basis functions are less able to capture the high-spin behaviour. In essence, this figure demonstrates both the validity of our method of evaluating (\ref{eqn:effectualness}) and our construction of a highly effectual bank for detecting binary black hole signals.

Taking the \textit{left} panel of Fig.~\ref{fig:tbankeffectualness} as a baseline (optimal) effectualness of our template bank, we can compare the bank's effectualness to a general inspiral waveform. For concreteness, we generate $10^4$ general inspiral waveforms within the same mass and spin ranges, fixing $\kappa=500$ and $\mathcal{C} = 0.1$. The effectualness is shown in the \textit{right} panel of Fig.~\ref{fig:tbankeffectualness}, where we see that a large spin-induced quadrupole moment can significantly decrease the effectualness of the bank. This is especially true in the large-spin limit, since the phase contributions (\ref{eqn:Kappa2PN}) and (\ref{eqn:Kappa3PN}) are proportional to $ \kappa_i \chi_i^2$.  Statistically, we find that only 6\% of the random templates have $\varepsilon > 0.9$ and 30\% have $\varepsilon > 0.2$.  This degradation in effectualness is entirely due to the spin-induced quadrupole, as the cutoff frequency of the waveform for $\mathcal{C}=0.1$ is greater than $f_u$, thereby having no effect on our analysis. 

\vskip 4pt

\begin{table}[h!]
\centering
\begin{tabular}{|c|ccccc| l|}
\hline
Scenario & $m_{1} \,[\mathrm{M_{\odot}}]$ & $m_2\,[\mathrm{M_{\odot}}]$ & $\chi_1$ & $\chi_2$ & $\mathcal{C}$ & Description\\
\hline
A & 2.0 & 2.0 & 0.7 & 0.7 & 0.1 &  Fiducial case \\
B & 2.8 & 2.8 & 0.7 & 0.7 & 0.1 &  Heavier total mass \\
C & 3.0 & 1.0 & 0.7 & 0.7 & 0.1 &  Lighter general object \\
D & 2.0 & 2.0 & 0.2 & -0.2 & 0.1 &  Reduced$+$anti-aligned spins \\
E & 2.0 & 2.0 & 0 & 0 & 0.01 &  Reduced compactness \\
\hline
\end{tabular}
\caption{List of representative binary configurations for Fig.~\ref{fig:matchvkap}. The parameters $\{ m_1, \chi_1 \}$ describe the black hole, while $\{ m_2, \chi_2, \kappa, \mathcal{C}\}$ are the parameters of the general astrophysical object.}
\label{table:scenarios}
\end{table}%

\begin{figure}[t!]
        \centering
        \includegraphics[scale=0.58]{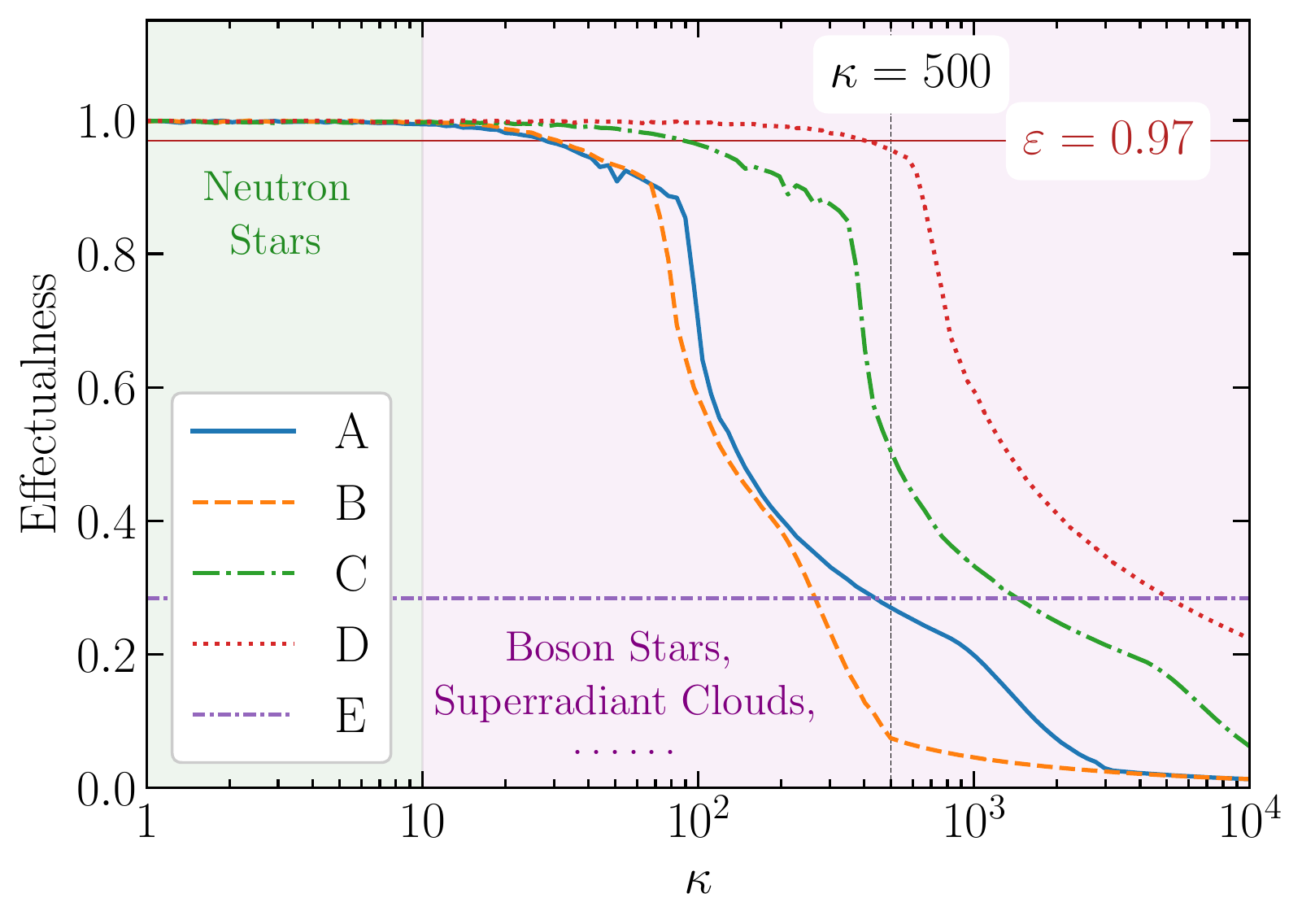}
\caption{Effectualness of the various scenarios listed in Table~\ref{table:scenarios} as a function of $\kappa$. The vertical dashed line denotes systems with $\kappa=500$ and is included for comparison with the \textit{right} panel of Fig.~\ref{fig:tbankeffectualness}. We also include the reference line $\varepsilon=0.97$, which is a commonly adopted requirement for template banks in actual searches. For convenience, we indicate the approximate ranges of $\kappa$ for neutron stars and various BSM objects in green and purple respectively. Kerr black holes have $\kappa=1$.}
    \label{fig:matchvkap}
\end{figure}

Instead of fixing $\kappa$ and $\mathcal{C}$, it is also interesting to examine the bank's effectualness as a function of these parameters. For concreteness, we consider five qualitatively-distinct scenarios, which are listed in Table~\ref{table:scenarios}. For each case, we treat $\kappa$ as a free parameter and calculate the effectualness within the range $1 \leq \kappa \leq 10^4$. Note that scenario E has no spins and is therefore unaffected by varying $\kappa$; this scenario is meant to show the effect of reducing the compactness of an object, which terminates the waveform in the detector sensitivity band. The results are presented in Fig.~\ref{fig:matchvkap}, where we find that each scenario with non-vanishing spins (A$-$D) shows a universal behaviour with the following three distinct regions as a function of $\kappa$:
\begin{enumerate}
    \item No loss in effectualness at low values of $\kappa$. For scenario A this occurs for $1 \leq \kappa \lesssim 20$.
    
    \item We see a series of different rates of declining effectualness as a function of $\kappa$, possibly indicating the varying rates of importance of the various higher-order PN contributions such as (\ref{eqn:Kappa2PN}) and (\ref{eqn:Kappa3PN}). For scenario A we see this at $20\lesssim \kappa \lesssim 3\times10^3$.
    
    \item Finally we see a flattening of the effectualness. This flattening occurs when $\kappa$ is so large that the 2PN term (\ref{eqn:Kappa2PN}) becomes the dominant contribution to the phase evolution. At this point, maximising over $t_c$ and $\phi_c$ always finds a small region of frequency space where the overlap between the two waveforms is nearly vanishing.
\end{enumerate}
We tested many additional scenarios and found this overall behaviour to be universal, showing the three distinct regions described above. Note that the value of $\kappa$ at which each scenario enters the three regions and the length spent there differs greatly as can be seen in Fig.~\ref{fig:matchvkap}. Below we give some qualitative arguments to compare the differences between the scenarios A$-$E.

Importantly, in the first region we see no noticeable reduction in the effectualness for $ \kappa \lesssim 20$, even for reasonably highly-spinning objects. As mentioned in \S\ref{sec:phase}, this range overlaps with the expected values of $\kappa$ for neutron stars~\cite{Laarakkers:1997hb, Pappas:2012ns}. It is for this reason that $\kappa$ can be safely ignored when searching for black hole - neutron star systems with binary black hole templates, although variations from $\kappa=1$ must be accounted for during parameter estimation \cite{Abbott:2018exr}. Binary neutron star systems would have additional contributions to their phase evolution since both objects now contribute to the phase with $\kappa \gtrsim 1$. LIGO and Virgo typically only consider slowly spinning neutron stars, $\chi \leq 0.4$ \cite{Ajith:2012mn}, where the effect of $1 < \kappa \lesssim 10$ is small --- binary black hole templates are therefore still suitable at the search level.

For larger values of $\kappa$, the effectualness quickly drops below the normal requirement of $\varepsilon \geq 0.97$ for template banks. Scenarios A and B show similar behaviour up to $\kappa \approx 200$ at which point they start to diverge. To understand this behaviour, we first note that the prefactors of the $v-$dependence in (\ref{eqn:Kappa2PN}) and (\ref{eqn:Kappa3PN}) are unchanged for equal-mass-ratio binary systems, regardless of their total mass. However, $v$ retains some dependence on the total mass --- $v\propto M^{1/3}$. The various PN terms therefore scale differently with $M$, causing a different rate of loss of effectualness. The difference between scenarios A and C can again be easily understood by examining the prefactors of (\ref{eqn:Kappa2PN}) and (\ref{eqn:Kappa3PN}). Since our general object is the lighter of the two components in scenario C (see Table~\ref{table:scenarios}), the mass dependencies of these terms dictate that the general object contributes less to the phase evolution. This reduced contribution can be compensated for by a larger value of $\kappa$, producing an overall shift to the right in Fig.~\ref{fig:matchvkap} from scenarios A to C. Similarly, if we were to fix the total mass and choose the heavier component to be our general object, we would see an overall shift from scenario A to the left. Finally, scenario D has significantly smaller spins, reducing the overall phase contribution from the spin-induced quadrupole.

For scenario E we see a significant reduction in effectualness, even for small values of $\kappa$ (the horizontal line merely reflects our choice of vanishing spins). This is because the frequency cutoff set by (\ref{eqn:cutoff}) is $f_{\rm cut} \approx 44 \, \mathrm{Hz}$, which lies inside the sensitivity band of ground-based detectors. For comparison, the cutoffs for scenarios A$-$D lie above $f_u=512 \, \mathrm{Hz}$ in our analysis. More generally, we find that  $\mathcal{C} \gtrsim 0.05$ gives a cutoff frequency of $f_{\rm cut} \gtrsim f_u$.\footnote{ The precise range of $\mathcal{C}$ depends on the mass of the binary components, and can be calculated more accurately through (\ref{eqn:cutoff}).} This loss in effectualness cannot be compensated for by adding additional waveforms to the bank, unlike for scenarios A$-$D. Instead it represents a truncation of the waveform and therefore a reduction in SNR. Since the introduction of $f_{\rm cut}$ merely represents our ignorance of the actual merger dynamics of the binary, the effectualness for objects with small values of $\mathcal{C}$ can potentially be improved. For instance, as discussed in \S\ref{sec:peakfreq}, PN corrections to (\ref{eqn:cutoff}) would increase $f_{\rm cut}$ towards higher values. Alternatively, model-dependent numerical relativity simulations can be performed to fully extract the merger waveforms. 

It is important to assess whether the loss of effectualness in Fig.~\ref{fig:matchvkap} is due to the limited range of component masses considered in the bank. We study this by repeating the procedure outlined in \S\ref{sec:templateBank} but instead consider a component mass range $1.0 M_\odot \leq m_i \leq 5.0 M_\odot$. We find no difference in the initial reductions of effectualness for all scenarios (this corresponds to the $\kappa \lesssim 100$ and $\varepsilon \gtrsim 0.85$ region in scenario A). As $\kappa$ increases to larger values, the effectualness still drops rapidly (as observed for $\kappa \gtrsim 100$ in scenario A) although the rates of reduction slightly decrease. This slight improvement occurs because, in this large-$\kappa$ region, the best-fitting points $ \left\{ \hskip 1pt c_\alpha (\bm{p}_{\tmp}) \hskip 1pt \right\}_{\mathrm{best}}$ are located inside and outside the bounds of $c_\alpha(\bm{p}_{\tmp})$ in the bigger and smaller template bank, respectively. In contrast, the earlier reduction is robust to the increased component mass range because $ \left\{ \hskip 1pt c_\alpha (\bm{p}_{\tmp}) \hskip 1pt \right\}_{\mathrm{best}}$ lies within the bounds of $c_\alpha(\bm{p}_{\tmp})$ for both banks. Despite this slight dependence on the size of the template bank parameter space, the effectualness maintains its monotonically-decreasing trend with increasing $\kappa$ for all scenarios. This robust behaviour suggests that waveforms with large $\kappa$ cannot be mimicked by binary black hole waveforms with vastly wrong intrinsic parameters.

\begin{figure}[t!]
        \centering
        \includegraphics[width=\textwidth, trim= 20 0 0 0]{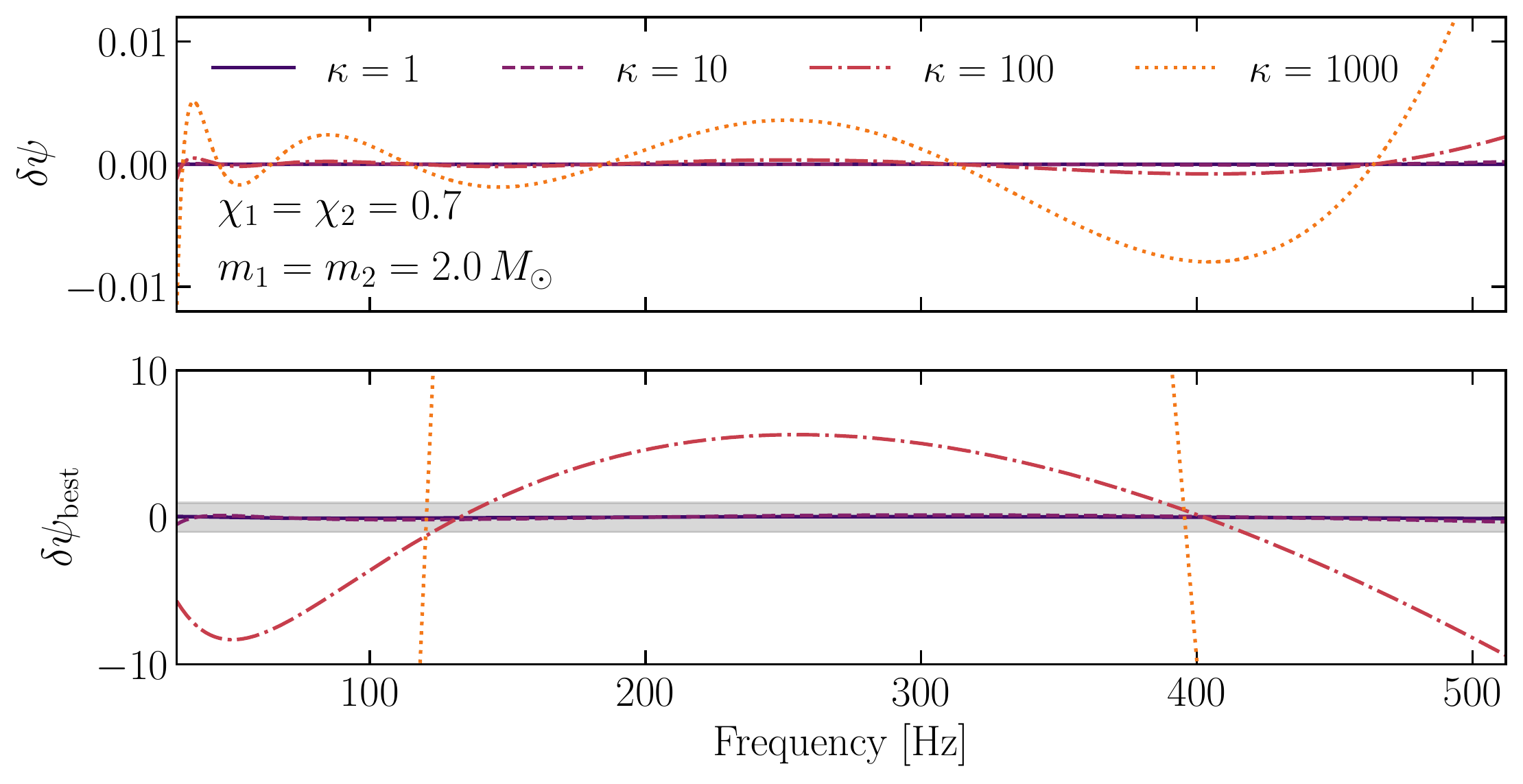}
\caption{The residual phases of scenario A at various values of $\kappa$. In the \textit{top} panel, the magnitudes of $\delta \psi$ are smaller than unity, indicating that the basis functions $\psi_\alpha$ can sufficiently describe the phases of the general waveforms. The gray band in the \textit{bottom} panel indicates the range  $ -1 \leq \delta \psi_{\mathrm{best}} \leq 1$. The phase residual $\delta \psi_{\mathrm{best}}$ for $\kappa \gtrsim 100$ clearly exceeds order unity, resulting in a reduced effectualness of the binary black hole template bank.}
    \label{fig:resphase}
\end{figure}

While the effectualness is a good measure of the differences between the signal and template waveforms, it is an integrated quantity with a less clear interpretation. We therefore examine two phase residuals:
\beq
\begin{aligned}
 \delta \psi  & = \psi(\bm{p}_{\gen}) - \left[\overline{\psi} + \sum_{\alpha=0}^{N=3} c_\alpha (\bm{p}_{\gen}) \, \psi_\alpha \right]\, , \\
 \delta \psi_{\mathrm{best}} & = \psi(\bm{p}_{\gen}) - \left[\overline{\psi} + \sum_{\alpha=0}^{N=3} \left\{ \hskip 1pt c_\alpha (\bm{p}_{\tmp}) \hskip 1pt \right\}_{\mathrm{best}} \, \psi_\alpha \right]\, ,\label{eqn:Phase_residual}
 \end{aligned}
\eeq
where $c_\alpha(\bm{p}_{\gen}) = \langle \psi (\bm{p}_{\gen}) - \overline{\psi} \, | \, \psi_\alpha\rangle$ (see \S\ref{sec:templateBank}). The residual phase $\delta \psi$ is a measure of the basis functions $\psi_\alpha$'s ability to capture the phase evolution of a general waveform. Instead, $\delta \psi_{\mathrm{best}}$ quantifies the phase deviation between the general waveform and its best-fitting waveform in our bank.

These residual phases are shown in Fig.~\ref{fig:resphase} at various values of $\kappa$ for scenario A. In the \textit{top} panel, the fact that $\left| \delta \psi \right| \ll 1$ for all values of $\kappa$ indicates that our basis functions, with $N=3$ in (\ref{eqn:Phase_geometric}), are able to describe the phase of the general waveform to a high degree of accuracy. The reduction in effectualness in Fig.~\ref{fig:matchvkap} is therefore a consequence of the large separations between $c_\alpha(\bm{p}_{\gen})$ and $\left\{ \hskip 1pt c_\alpha (\bm{p}_{\tmp}) \hskip 1pt \right\}_{\mathrm{best}}$. In the \textit{bottom} panel, we see that $\left| \delta \psi_{\mathrm{best}} \right| \ll 1$ for $\kappa=1 $ and $\kappa=10 $ --- it is for this reason that we can still detect these binary systems with standard binary black hole waveforms, in agreement with Fig.~\ref{fig:matchvkap}.  For $\kappa =100$, we start to see deviations exceeding order unity, $\left| \delta \psi_{\mathrm{best}} \right| \gtrsim \mathcal{O}(1)$, resulting in a reduced effectualness. Crucially, the \textit{bottom} panel of Fig.~\ref{fig:resphase} illustrates that our general inspiral waveforms are \textit{not degenerate} with a binary black hole template waveform with the wrong intrinsic parameters; if this were the case, we would see zero phase residuals.

In a nutshell, Figs.~\ref{fig:tbankeffectualness},~\ref{fig:matchvkap}, and \ref{fig:resphase} clearly indicate that, for values of $\kappa \lesssim 20$, binary black hole template banks are still able to detect general astrophysical objects. On the other hand, for $\kappa \gtrsim 20$, there are large parts of parameter space where the binary black hole templates cannot be used to recover these general signals. This is especially true if the general object is highly spinning and has a larger mass when compared with its binary companion. Moving forward, new template banks must be constructed with $\kappa$ as a one-parameter extension to the standard waveforms. Since $\mathcal{C}$ is simply a truncation of the waveform, it is not necessary to include it as an additional parameter.

\section{Summary}

In this work, we examined whether binary black hole template banks can be used to search for the gravitational waves emitted by a binary coalescence of general dark compact objects. We focused on the early inspiral regime of low-mass binary systems. This restriction had multiple benefits. Firstly, the inspiral is a regime where analytic results of the PN dynamics are readily available. This provided us with a well-defined framework to construct our general waveforms, where the physics contributing to waveform deformations can be clearly interpreted. Secondly, ground-based detectors are able to probe these inspirals over hundreds or thousands of cycles, thereby allowing for a precise characterization of the physics at play. Inspiral signals in LIGO/Virgo observations therefore represent a new avenue to probe BSM physics and novel astrophysical phenomena.

\vskip 4pt

We focused on binary systems with components that can have large spin-induced quadrupole moments and/or small compactness. Figure~\ref{fig:matchvkap} clearly demonstrates that as the quadrupole term becomes large, its phase contribution to the waveform becomes significant. Binary black hole template banks are thus insufficient for searching for these general dark compact objects. More precisely, we find that the effectualness of these template banks are quickly reduced for $\kappa \gtrsim 20$ of highly-spinning objects (see scenario A in Table~\ref{table:scenarios}). This range of $\kappa$ coincides with an interesting part of the parameter space where compact objects in various BSM scenarios may exist~\cite{Baumann:2018vus, Baumann:2019ztm, Ryan:1996nk, Herdeiro:2014goa}. Figure~\ref{fig:resphase} further shows that these signatures are not degenerate with binary black hole template waveforms with the wrong intrinsic parameters. It is therefore essential that extended template banks are created in order to search for these novel signatures. As a byproduct of our analysis, we recovered the result that the effectualness remains high for smaller values of~$\kappa$. Binary black hole waveforms can therefore be used to search for binary systems with neutron stars, as is currently done by the LIGO/Virgo collaboration~\cite{TheLIGOScientific:2017qsa}. 

\vskip 4pt

In addition, we considered the impact of an object's compactness on the merger frequency of the binary. Since a detailed description of the merger dynamics is model-dependent, we truncated the waveform through a frequency cutoff. For objects with small-compactness, this cutoff is set by the point at which the binary components touch. This truncation only has a significant effect on the effectualness when the cutoff frequency is within the sensitivity bands of ground-based detectors. As a fiducial guide, our estimate showed that this is the case for $\mathcal{C} \lesssim 0.05$ in low-mass binary systems. This loss in effectualness can be compensated for through more detailed modeling of the binary merger dynamics. Overall, our findings therefore showed that it is essential that extended template banks are created in order to search for these novel signatures.

\chapter{Conclusions and Outlook} \label{sec:Conclusions}

The detection of gravitational waves marked the beginning of a new era for multi-messenger astronomy. It also raises the interesting question whether precision 
gravitational wave observations can become a new tool for fundamental physics. We often associate new physics with short-distance (or high-energy) modifications of the Standard Model.  The decoupling between physics at short and long distances then provides an immediate challenge for using the long-wavelength gravitational waves produced by the dynamics of macroscopic objects to probe physics at much shorter distances. This, however, ignores the possibility that new physics can be both very light and weakly coupled, which allows for coherent effects on astrophysical length scales. This is the case for ultralight bosons in  black hole binaries, whose Compton wavelengths are larger than the typical sizes of the constituents. The extended nature of the associated boson clouds enhances the effects due to the internal structure of the compact objects, mitigating the decoupling challenge. Furthermore, various types of new compact objects can arise in theories that propose the presence of new degrees of freedom in our Universe. The dynamics of these objects when they are parts of binary systems are often different from those of binary black holes and neutron stars. As a consequence, the observation of gravitational waves from binary systems has opened a new window into physics beyond the Standard Model at the weak-coupling frontier.

\subsubsection*{Summary}

In this thesis, I focused on the gravitational-wave signatures of boson clouds that are superradiantly amplified around rotating black holes. These clouds are often called ``gravitational atoms" because they resemble the proton-electron structure of the hydrogen atom. In Chapter~\ref{sec:spectraAtom}, I first presented a detailed analysis of various properties of the scalar and vector gravitational atoms, such as their energy spectra and instability rates. Importantly, our detailed analytic and numeric computations have unveiled all qualitative features of the boson clouds. For instance, we computed the fine and hyperfine splittings of the scalar and vector gravitational atoms, thereby completing our understanding of the relative orderings of all of the eigenstates in the spectra. As I showed in Chapter~\ref{sec:Collider}, these subtle spectral features play crucial roles in the dynamics of the clouds when they are perturbed by a binary companion. For example, since the vector atom contains many more nearly-degenerate eigenstates in its spectrum, the level mixings induced by the companion often involve multiple states simultaneously coupled to one another. On the other hand, for the scalar atom, it is typically the case that only two well-separated energy eigenstates mix with each other. At certain critical orbital frequencies of the binary, these level mixings become non-perturbative and are resonantly enhanced. The dynamics of the clouds at these resonances are analogous to the Landau-Zener transitions of quantum mechanics and all the physics can be encoded in an S-matrix. By carefully considering the backreaction of these transitions on the orbit, we found in Chapter \ref{sec:signatures} that the Landau-Zener transitions can generate distinctive floating and sinking orbits, which significantly affect the gravitational-waves emitted by the binary system. These Landau-Zener transitions can also induce interesting time-dependent evolutions in the shapes of the cloud, which would be observable as modulated finite-size effects in the phases of the waveforms.

We refered the discipline of searching for ultralight bosons in binary systems as ``gravitational collider physics" because it is in many ways analogous to the discipline of ordinary particle collider physics. In particular, the dynamics of a boson cloud at the resonances mentioned above is similar to a scattering process, with all the relevant physics described by an S-matrix. Furthermore, in an ordinary collider, a particle’s mass is determined by the energy at which the particle appears as a resonant excitation, while its spin is measured via the angular dependence of the final state. Similarly, in the gravitational collider, the position of the resonances determine the boson’s mass, and we must observe the final state of a transition to distinguish particles of different spin. Fortunately, the properties of this state are accessible through the imprint of finite-size effects on the waveform. Hence, a precise reconstruction of a gravitational-wave signal can help us not only to detect new ultralight bosons, but also determine their mass and intrinsic spin. The discovery potential of gravitational-wave observations thus necessitates the development of sufficiently accurate template waveforms, which include the characteristic features of boson clouds in binary systems that we have uncovered.

While the primary focus of this thesis was on the gravitational atoms, in Chapter~\ref{sec:search}, I also discussed the imprints of general dark compact objects on the gravitational waveforms in binary systems. Without specializing to a specific model of compact object, we parameterized the putative new physics by allowing the spin-induced quadrupole moment and the compactness of the binary components to take arbitary values. In a nutshell, we found that there exists a wide region in the quadrupole-compactness parameter space where existing binary black hole templates are insufficient to detect these new types of binary systems. Since current gravitational-wave search pipelines exclusively use binary black hole waveforms, the non-detection of new compact objects so far could simply be a consequence of not widening our existing matched-filtering searches. This work therefore motivates revisiting existing search strategies in order to detect the presence of putative new dark compact objects through their gravitational-wave emissions in binary systems.

\subsubsection*{Open questions}

The findings in this thesis raise a number of interesting follow-up questions and warrant future investigations. For instance, we mentioned in Chapter~\ref{sec:spectraAtom} that a separable equation of motion for the vector magnetic modes in a Kerr geometry is still not known. In that case, we were able to obtain analytic expressions for the magnetic spectra through educated guessworks, and confirmed their accuracies through comparisons with numeric results, which we computed without relying on separability. We wish to check our conjectured expressions once a separable form of the vector magnetic equation is derived in the future.  Futhermore, the ultralight bosons in our analysis only coupled via gravity. In principle, there could also be self-interactions of the field, couplings to other ultralight bosons, or direct couplings to ordinary matter. It would be interesting to explore if and how these extra interactions would affect the stability and dynamics of the cloud~\cite{Arvanitaki:2010sy, Yoshino:2012kn}, including whether new types of gravitational-wave signatures would be generated.

Our studies of the gravitational collider physics in Chapters~\ref{sec:Collider} and \ref{sec:signatures} can also be extended in several ways. Firstly, for simplicity, we have mostly studied quasi-circular equatorial orbits.  We wish to extend our analysis to general inclined and elliptical orbits, in which case the quasi-periodic driving force provided by the companion has additional frequency components; see e.g.~\cite{Zhang:2019eid, Berti:2019wnn}, Appendix~\ref{app:TidalMoments}, and Appendix~\ref{app:adFloTheo}. Secondly, to maintain maximal theoretical control, we have focused on the early inspiral phase of the binary's evolution. After a sequence of resonant transitions, the system evolves into a final state that can be probed sensitively during the late stages of the inspiral. It is unclear to what degree this regime can still be describe analytically, or if we have to resort to numerical simulations. For instance, additional effects such as dynamical friction~\cite{Zhang:2019eid} can become important when the binary companion enters the cloud.  Most of our quantitative analysis has been restricted to the regime of weak backreaction.  In these cases, 
the effects of the cloud on the orbital dynamics can be treated perturbatively and the 
dephasing during the resonances can be predicted analytically. This regime of weak backreaction is most relevant for black holes of roughly equal masses.  There is, however, also great interest in the case of extreme mass ratio inspirals. If the cloud is around the more massive black hole, the backreaction on the dynamics of the small companion will be very large. The companion will either float for a very long time, or receive a very strong kick that is likely to induce significant eccentricity of the orbit. While the floating behavior is still under analytical control, our approximations break down for strong kicks. These very interesting cases deserve a more dedicated analysis.

Our analysis in Chapter~\ref{sec:search} showed that many new signatures of binary systems could be missed by current search pipelines.  Although we focused on finite-size effects, many other types of physical phenomena can affect the frequency evolution of a binary. In the future, we wish to incorporate these additional dynamics into more general waveforms. Furthermore, we aim to search for these novel signatures in the data collected in the O1$-$O3 observation runs. Using the same procedure and grid spacing as in \S\ref{sec:templateBank}, we estimate that an order-of-magnitude more templates would be required to search for these new signals, though we leave a more refined study to future work. While a detection in these data would certainly indicate a signature of new physics, non-observations can also be used to place meaningful bounds on the space of astrophysical objects that exist in our Universe.

\subsubsection*{Outlook}

We are already seeing a large influx of gravitational-wave data from the LIGO/Virgo Collaboration. As I summarized in Section~\ref{sec:detectors}, all of the observing runs so far have brought about new scientific opportunities and have significantly advanced our understanding of astrophysics. Having said that, these observations have not yet fulfilled their potential of shedding light on particle physics. This could be the case because we do not yet have the appropriate tools, such as the right template waveforms, to detect these putative new signals, even though they are present in the collected data. Gravitational-wave science is a precision science; it is therefore crucial that we develop new search strategies to detect new signals. At the same time, we must broaden our theoretical ideas for the types of gravitational-wave signatures that could be sourced by physics in the dark sector.

Our quests to develop new search techniques and explore new theoretical ideas are especially urgent because future gravitational-wave detectors will measure signals with better precisions and over a much larger range of frequencies. The prospects of using these observations to probe new sources in the dark sector are therefore very bright. Now is absolutely an opportune time to explore new ways of maximizing the discovery potentials of these gravitational-wave observations. While the ideas presented in this thesis are far from the only interesting avenues of probing new physics with gravitational waves, they represent concrete steps towards our goal of utilizing these measurements to search for physics beyond the Standard Model and novel astrophysical phenomena.

\appendix
\chapter{Details of the Spectral Analysis} \label{app:SpectralDetails}

This appendix contains further details about the analytic and numeric computations for the spectra of the gravitational atoms, which were discussed in Chapter~\ref{sec:spectraAtom}.
	
\section{Tensor Spherical Harmonics} \label{app:harmonics}
	
We now describe the spherical harmonic decomposition of tensor fields in black hole backgrounds, focusing in particular on the construction of vector spherical harmonics.

\subsubsection*{Tensor representations of SO(3)}

Besides the time-like Killing vector $k_t = -i \partial_t$, the spherically symmetric Schwarzschild black hole enjoys three additional Killing vectors, which in Boyer-Lindquist 
coordinates take the form
	 \begin{equation}
			\begin{aligned}
				k_x &= -i \left(y \frac{\partial}{\partial z} -  z \frac{\partial}{\partial y}\right)= i \left(\sin \phi\, \frac{\partial}{\partial \theta} + \cot \theta \cos \phi\, \frac{\partial}{\partial \phi}\right) , \\
				k_y &= -i \left(z \frac{\partial}{\partial x} - x \frac{\partial}{\partial z}\right) = i \left(-\cos \phi \,\frac{\partial}{\partial \theta} + \cot \theta \sin \phi\, \frac{\partial}{\partial \phi}\right) , \\
				k_z &= -i \left(x \frac{\partial}{\partial y} - y \frac{\partial}{\partial x}\right) = -i \frac{\partial}{\partial \phi}\, .
				\label{eq:so3algebra}
			\end{aligned}
	\end{equation}
	These Killing vectors are the
	generators of the group $\lab{SO}(3)$ and therefore satisfy the commutation relations
	\begin{equation}
		\left[k_i, k_j\right] = \pounds_{i} k_j = i \epsilon_{ijk} k_k\,,
	\end{equation}
  where we use $i,j,k = 1, 2, 3$ to represent $x, y, z$, respectively, and have employed the shorthand $\pounds_{i} = \pounds_{k_i}$ for the Lie derivative along $k_i$. This algebra has a quadratic Casimir,
	\begin{equation}
		\pounds^2 = \pounds_x^2 + \pounds_y^2 + \pounds_z^2 \, ,
	\end{equation}
	that commutes with the generators $k_i$. Irreducible representations of this group can then be labeled by their eigenvalues under $\pounds^2$ and one of the generators $\pounds_i$.
It is conventional to define the representation  $|j, m \rangle$ by
	\begin{equation}
	\begin{aligned}
		\pounds^2 |j, m \rangle &= j(j+1)|j, m \rangle\, , \\
		 \pounds_z | j, m \rangle &= m |j, m \rangle \,.
		 \end{aligned}
	\end{equation}
  For now, $|j, m\rangle$ denotes an arbitrary (tensor) representation of this algebra with total angular momentum $j$ and total azimuthal angular momentum $m$.

We may also define the vector fields
	\begin{equation}
		k_{\pm} = k_{x} \pm i k_y = e^{\pm i\phi} \left(\pm \frac{\partial}{\partial \theta} + i \cot \theta \frac{\partial}{\partial \phi}\right)  , 
	\end{equation}
	and the corresponding operators $\pounds_{\pm} = \pounds_{k_\pm}$. These operators raise/lower the azimuthal angular momentum
	\begin{equation}
	\begin{aligned}
		\pounds_{\pm} k_{\mp} &= [k_{\pm}, k_{\mp}] = \pm 2 k_{z} \,, \\
		 \pounds_z k_\pm &= [k_z, k_{\pm}] = \pm k_{\pm} \,.
		\end{aligned}
	\end{equation}	
  A physical, finite-dimensional representation requires that 
	\begin{equation}
		\pounds_\pm | j, \pm j \rangle = 0\,, \label{eq:loweringDef}
	\end{equation}
	since we could otherwise violate the requirement that $\pounds^2 - \pounds_z^2 \geq 0$. With the normalization condition $\langle j, \pm j | j, \pm j \rangle = 1$, (\ref{eq:loweringDef}) defines the states $|j, \pm j\rangle$.  The other states are then defined in the \emph{Condon-Shortley phase convention}\footnote{Which convention should be used depends on the convention for the Clebsch-Gordan coefficients. We use this convention since it is the one used by Mathematica.} by 
	\begin{equation}
	\begin{aligned}
		\pounds_+ |j,  m \rangle &= \sqrt{(j - m)(j + m + 1)}\, |j, m+1 \rangle\, , \\
		 \pounds_- |j, m \rangle &= \sqrt{(j + m)(j - m + 1)} \, |j, m-1\rangle\,.
	\end{aligned}
	\end{equation}
	Two representations of $\lab{SO}(3)$ can be combined into a single representation with definite total and azimuthal angular momentum using the Clebsch-Gordan coefficients,
	\begin{equation}
		|j, m; \ell, s \rangle = \sum_{m_\ell = - \ell}^{\ell} \sum_{m_s = -s}^{s} |(s\, m_s)\, (\ell\, m_\ell)\rangle \langle (s\, m_s)\, (\ell\, m_\ell) | j, m; \ell, s \rangle\,.
	\end{equation}
	This yields a well-defined prescription for generating tensor spherical harmonics of any rank. That is, given a tensor with an arbitrary number of spatial indices, we can define the basis of tensors $|s, m_s \rangle$ by first solving 
	\begin{equation}
	\begin{aligned}
		\pounds_+\hskip 2pt \chi^{s,s}_{ijk\dots} &= 0 \, ,\\ \pounds_z\hskip 2pt \chi^{s,s}_{ijk\dots} &= s\hskip 2pt \chi^{s, s}_{ijk\dots}\,,
		\end{aligned}
	\end{equation}
  and then using $\pounds_-$, with the normalization convention above, to define the other $|s, m_s\rangle$. Note that these conditions immediately imply that $\pounds^2 \chi^{s, s}_{ijk\dots} = s(s+1) \chi^{s,s}_{ijk\dots}$.
	We may then combine $|s, m_s\rangle$ with a \emph{scalar} representation of $|\ell, m_\ell\rangle$, i.e.~the scalar spherical harmonics $Y_{\ell\hspace{0.75pt} m_\ell}(\theta, \phi)$,
	  to generate the tensor spherical harmonics
  \begin{equation}
    T_{ijk\dots}^{j,m; \ell, s}(\theta, \phi) = \sum_{m_\ell = -\ell}^{\ell} \sum_{m_s = -s}^{s} \langle (s\, m_s)\, (\ell\, m_\ell) | j, m; \ell, s \rangle\, Y_{\ell \hspace{0.75pt} m_\ell}(\theta, \phi)\, \chi^{s,m_s}_{ijk\dots}\,. \label{eq:tensorSphericalHarmonics}
  \end{equation}
This procedure yields a tensor representation of $\lab{SO}(3)$. Because our ultimate goal is to expand vector fields in $\mathbb{R}^3$
, the $T_{ijk\dots}^{j, m; \ell,s}$ must actually form a representation of $\lab{O}(3)$, i.e. the group $\lab{SO}(3)$ together with inversions. Since the scalar spherical harmonics have definite parity $Y_{\ell m_\ell}(\pi-\theta, \phi+\pi) = (-1)^{\ell} Y_{\ell m_\ell}(\theta, \phi)$, we can choose $\chi^{s, m_s}_{ijk\dots}$ to also have definite parity, so that (\ref{eq:tensorSphericalHarmonics}) defines a tensor representation of $\lab{O}(3)$. Below, we will explicitly construct a family of vector spherical harmonics. In that case, the choice of parity determines whether (\ref{eq:tensorSphericalHarmonics}) yields \emph{vector} or \emph{pseudo-vector} spherical harmonics.

	\subsubsection*{Vector spherical harmonics} 
	\label{app:vsh} 

To form a basis of vector spherical harmonics, 
we first construct the basis of vectors $\chi_{1, m_s}^{i}$ by the defining relations
    \begin{equation}
    \begin{aligned}
      \pounds_{+} \chi_{1, 1}^i &= 0 \, , \\ \pounds_z \chi_{1, 1}^i &= \chi_{1, 1}^i \,.
      \end{aligned}
    \end{equation}
    These equations have the general solution
    \begin{equation}
      \chi_{1,1}^{i} \,\partial_i = \frac{e^{i \phi}}{\sqrt{2}}\Big[F_{r}(r) \sin \theta\, \partial_r + F_{\phi}(r)(\partial_\theta + i \cot \theta \,\partial_\phi) + F_{\theta}(r)(\cos \theta \, \partial_\theta + i \csc \theta\, \partial_\phi)\Big]\,,
    \end{equation}
    where the $F_{i}$ are, in general, undetermined functions of $r$. Imposing that $\chi_{1,1}^i$ transforms like a \emph{vector} under inversions, leads to $F_{\phi} = 0$. Taking, instead, $F_r = F_\theta = 0$ would yield a basis of \emph{pseudo-vector} spherical harmonics.
We may then define $\sqrt{2} \chi_{1, 0}^i = \pounds_- \chi_{1,1}^i$ and $\sqrt{2}\chi_{1, \sminus 1}^{i} = \pounds_{-} \chi_{1,0}^i$, to find the complete basis
    \begin{equation}
      \begin{aligned}
        \chi_{1,1}^{i}\,\partial_i &= \frac{e^{i \phi}}{\sqrt{2}}\Big[F_{r}(r) \sin \theta\, \partial_r  + F_{\theta}(r)(\cos \theta \, \partial_\theta + i \csc \theta\, \partial_\phi)\Big] \, ,\\
        \chi_{1, 0}^{i} \, \partial_i &= - F_{r}(r) \cos \theta \,\partial_r + F_\theta(r) \sin \theta \, \partial_\theta\, , \\
        \chi_{1, \sminus 1}^{i}\, \partial_i &= \frac{e^{-i \phi}}{\sqrt{2}}\Big[\minus F_{r}(r) \sin \theta\, \partial_r  + F_{\theta}(r)(\minus \cos \theta \, \partial_\theta + i \csc \theta\, \partial_\phi)\Big]\, .
      \end{aligned}
      \label{eq:vecBasis}
    \end{equation}
In flat space, we can take $F_r = -1$ and $F_\theta = -1/r$, so that these basis vectors are covariantly constant, $\nabla_\mu \chi_{1, m_s}^\nu = 0$.\footnote{If we can find basis vectors that are covariantly constant, then it is possible to unambiguously disentangle spin and orbital angular momentum, so that $\ell$ can be a good quantum number.}  These harmonics then reduce to those commonly used in the literature; see e.g.~\cite{Thorne:1980ru}. This is impossible in the Kerr geometry. It will instead be more convenient to choose $F_r$ and $F_\theta$ to simplify the boundary conditions of the vector field at the horizon and at spatial infinity, as discussed in Appendix~\ref{app:boundaryConditions}. 	For each choice of $F_{r}(r)$ and $F_{\theta}(r)$, we then have a set of \emph{vector spherical harmonics}
		\begin{equation}
			Y^i_{\ell, j m} = \sum_{m_s = -1}^{1} \langle (1\, m_s)\, (\ell\, m - m_s) | j, m \rangle \,Y_{\ell, m - m_s}(\theta,\phi)\, \chi_{1, m_s}^{i} \,, \label{eq:vectorSphericalHarmonics}
		\end{equation}
		which have definite angular momentum,
		\begin{equation}
		\begin{aligned}
			\pounds^2 Y^i_{\ell, j m} &= j(j+1) Y^i_{\ell, j m} \, ,\\ \pounds_z Y^i_{\ell, j m} &= m Y^i_{\ell,j m}\,,
			\end{aligned}
		\end{equation}
    and by construction satisfy
    \beq
    \begin{aligned}
      \pounds^2 \Big(e^{-i \omega t} G(r) Y^i_{\ell, j m} \Big) &= j(j+1) e^{-i \omega t} G(r) Y^i_{\ell, j  m}\,, \\
       \pounds_z \Big(e^{-i \omega t} G(r) Y^i_{\ell, j m} \Big) &= m e^{-i \omega t} G(r) Y^i_{\ell,j  m}\, .
    \end{aligned}	
    \eeq
The Clebsch-Gordan coefficients are only non-vanishing for $j \geq 0$, $|m| \leq j$, and $j+1 \geq \ell \geq |j-1|$.
   Any smooth spatial vector field can be decomposed into these harmonics,
      \begin{equation}
        V^i(t, r, \theta, \phi) = \sum_{j = 0}^{\infty} \,\sum_{\ell = |j -1|}^{j +1} \,\sum_{m = -j}^{j}  V_{\ell, j m}(t, r) \,Y^i_{\ell, jm}(\theta, \phi)\, .
      \end{equation} 
    By construction, the harmonics (\ref{eq:vectorSphericalHarmonics}) have definite parity, and acquire a factor of $(-1)^{\ell+1}$ under the transformation $(\theta, \phi) \to (\pi - \theta, \phi + \pi)$. 
   A general odd-parity vector field $E^i$ 
can thus be expanded as
   \begin{equation}
   	E^i = \sum_{j, \ell, m} E_{\ell, jm}(t,r) Y^i_{\ell, j m}(\theta, \phi)\,,
   \end{equation}
   where $j$ runs over all positive integers, 
   $\ell$ is even and takes values between $j\pm 1$, 
   and $m$ runs from $-j$ to $j$. Similarly, a general even-parity vector field $B^i$ can be expanded as
   \begin{equation}
   	B^i = \sum_{j, \ell, m} B_{\ell, jm}(t, r) \,Y^i_{\ell,j m}(\theta, \phi) \, ,
   \end{equation}
where $j$ runs over all non-negative integers, $\ell$ runs over all odd numbers between $j\pm 1$, and $m$ runs from $-j$ to~$j$. 
In both cases, if the field has definite azimuthal angular momentum $m_z$, the sum over $m$ is restricted to $m_z$ and hence can be dropped.

    A major benefit of this basis of vector fields (\ref{eq:vecBasis}) is that it also acts simply on the scalar spherical harmonics. We may define the operators
      \begin{align}
         \mathcal{D}_\pm Y_{\ell m_\ell} & = \frac{e^{\pm i \phi}}{\sqrt{2}} \left(\pm \cos \theta \, \partial_\theta + i \csc \theta \, \partial_\phi\right)Y_{\ell m_\ell}(\theta, \phi) \nonumber \\
      &= \sqrt{\frac{(\ell+1)^2 (\ell\mp m_\ell)(\ell\mp m_\ell-1)}{2(2 \ell + 1)(2 \ell-1)}} Y_{\ell-1, m_\ell \pm 1} \label{eq:dpm} \\
      & + \sqrt{\frac{\ell^2(\ell\pm m_\ell +1)(\ell\pm m_\ell +2)}{2(2\ell+3)(2 \ell+1)}} Y_{\ell+1, m_\ell \pm 1}\,, \nonumber \\[6pt]
      \mathcal{D}_0 Y_{\ell m_\ell}  &=\sin \theta \,\partial_\theta Y_{\ell m_\ell }(\theta, \phi) \label{eq:d0} \\[2pt]
      &=  \sqrt{\frac{\ell^2 (\ell+m_\ell + 1)(\ell-m_\ell +1)}{(2\ell+1)(2\ell+3)}} Y_{\ell+1, m_\ell} -\sqrt{\frac{(\ell+1)^2 (\ell+m_\ell)(\ell-m_\ell)}{(2 \ell+1)(2 \ell-1)}}Y_{\ell-1, m_\ell}\, , \nonumber
      \end{align}
    so that the angular legs of each of (\ref{eq:vecBasis}) can be rewritten using these operators,
    \begin{equation}
      \begin{aligned}
        \chi_{1,1}^i \,\partial_i &= \frac{1}{\sqrt{2}} e^{i \phi} \sin \theta F_r \,\partial_r + F_\theta \,\mathcal{D}_+ \,,\\
        \chi_{1, 0}^i \,\partial_i &= -F_r \cos\theta \, \partial_r + F_\theta \,\mathcal{D}_0 \, ,\\
        \chi_{1, \sminus 1}^i \,\partial_i &= \minus \frac{1}{\sqrt{2}}e^{-i \phi} F_r \sin \theta \, \partial_r + F_\theta\, \mathcal{D}_-\,.
      \end{aligned}
    \end{equation}
    While not strictly necessary, it will prove extremely convenient in our decomposition of the Proca equation to write all angular derivatives in terms of $\pounds^2$, $\pounds_z$, $\mathcal{D}_\pm$, and $\mathcal{D}_0$, as derivatives like $\partial_\theta Y_{\ell, m}$ are not easily expressible in terms of other spherical harmonics.

    Finally, it will also be useful to construct a basis of one-form harmonics $\chi_i^{1, m_s}\, \ud x^i$, which are dual to the vector fields (\ref{eq:vecBasis}), in the sense that
    \begin{equation}
      \bar{\chi}^{1, m_s}_i \chi_{1, m_s'}^{i} = \delta^{m_s}_{m_s'} \quad\text{and} \quad \bar{\chi}^{1 m_s}_{i} \chi_{1, m_s}^{j} = \delta_i^j\,,
    \end{equation} 
    where the bar denotes complex conjugation.
    Explicitly, these form fields are
    \beq
    \begin{aligned}
          \chi^{1, 1}_{i} \,\ud x^i &= \frac{1}{\sqrt{2}}e^{i\phi}\left(F_r^{-1} \, \sin \theta \, \ud r + F_\theta^{-1}\left( \cos \theta \, \ud \theta + i \sin \theta\, \ud \phi\right)\right) , \\
          \chi^{1,0}_{i} \,\ud x^i &= -F_r^{-1} \cos \theta\, \ud r + F_\theta^{-1} \cos \theta \, \ud \theta\,, \\
          \chi^{1,\sminus 1}_i \, \ud x^i &= \frac{1}{\sqrt{2}} e^{-i \phi} \left( \minus F_r^{-1} \sin \theta \, \ud r + F_\theta^{-1}\left( - \cos \theta \, \ud \theta + i \sin \theta \, \ud \phi\right)\right).
      \end{aligned} 
      \eeq
      We can then write the set of {\it one-form spherical harmonics} as
      \begin{equation}
        Y_i^{\ell, j m} = \sum_{m_s = -1}^{1} \langle (1\, m_s)\, (\ell\, m - m_s) | j, m \rangle \,Y_{\ell, m - m_s}(\theta,\phi) \,\chi^{1, m_s}_{i}\,, \label{eq:formSphericalHarmonics}
      \end{equation}
   which satisfy the orthonormality condition,
      \begin{equation}
        \int_{\lab{S}^2}\!\ud \Omega\,\,  \bar{Y}^{\ell, jm}_{i} (\theta,\phi) \,Y^i_{\ell', j'm'}(\theta, \phi) = \delta_{j'}^{j} \delta^\ell_{\ell'} \delta^{m}_{m'}\, ,
      \end{equation}
      where $\int_{\lab{S}^2} \!\ud \Omega = \int_{0}^{2\pi}\!\ud \phi\, \int_{0}^{\pi}\!\sin \theta\, \ud \theta$ denotes integration over the sphere.

\section{Details of the Analytical Treatment}
\label{app:details}  

In Section~\ref{sec:analytic}, we derived the quasi-bound state spectra for massive scalar and vector fields on the Kerr geometry, using the method of match asymptotic expansion. In this appendix, we provide some of the technical details that we left out in Section~\ref{sec:analytic}. 

\subsubsection*{Separable ansatz for vectors}

In the coordinates (\ref{equ:Kerr2}), the polarization tensor $B^{\mu \nu}$ in (\ref{eqn: Proca ansatz}) reads 
\begin{align}
& B^{\mu \nu} =  
\frac{1}{\Sigma q_r} \left( 
\begin{array}{cccc}
0 & i q_a \lambda^{-1} r & 0 & 0 \\
- i q_a \lambda^{-1} \hskip 1pt r & \Delta & 0 & - i \alpha \hskip 0.5pt \tilde{a} \lambda^{-1} r \\ 
0 & 0 & 0 & 0 \\
0 & i \alpha \hskip 0.5pt \tilde{a} \lambda^{-1} \hskip 1pt r & 0 & - \alpha^2 \tilde{a}^2 \Delta^{-1} 
\end{array} \right) \nonumber \\
& + \frac{q_a}{ \Delta  q_r q_\theta } \left( 
\begin{array}{cccc}
-1 + \alpha^2 \tilde{a}^2 \lambda^{-2} \hskip 5pt &  0 \hskip 5pt & \hskip 5pt  0 &  \hskip 5pt \alpha  \hskip 0.5pt \tilde{a}  \lambda^{-2}   \\
0  &  0  \hskip 5pt & \hskip 5pt 0 & 0 \\
0  & 0  \hskip 5pt & \hskip 5pt 0 & 0 \\
\alpha \hskip 0.5pt  \tilde{a} \lambda^{-2}  & 0 \hskip 5pt & \hskip 5pt 0 & 0
\end{array} \right)  \label{eqn:BtensorFull} \\[4pt]
& + \frac{1}{\Sigma q_\theta} \left( 
\begin{array}{cccc}
- 2\alpha^3   \tilde{a}^2 \Delta^{-1} r \hskip 1pt \sin^2 \theta   &  \hskip 5pt 0  & \hskip 5pt  i \alpha^2 \tilde{a}^2 \lambda^{-1} \cos \theta \sin \theta & \hskip 5pt  - 2 \alpha^2 \tilde{a} \Delta^{-1} r \\
0 & \hskip 5pt  0 & \hskip 5pt  0 &\hskip 5pt  0 \\
- i \alpha^2 \tilde{a}^2 \lambda^{-1} \cos \theta \sin \theta & \hskip 5pt  0 & \hskip 5pt  1 & \hskip 5pt  - i \alpha \tilde{a} \lambda^{-1} \cot \theta \\
- 2 \alpha^2 \tilde{a} \Delta^{-1} r & \hskip 5pt  0 & \hskip 5pt  i \alpha \tilde{a} \lambda^{-1} \cot \theta & \hskip 5pt  \csc^2 \theta \end{array} \right) \, , \nonumber
\end{align}
where the following quantities have been introduced
\begin{equation}
\begin{aligned}
q_r & \equiv 1 + \lambda^{-2} r^2 \, , \\
q_\theta & \equiv 1 - \alpha^2 \tilde{a}^2  \lambda^{-2} \cos^2 \theta \, ,\\
q_a & \equiv r^2 + \alpha^2 \tilde{a}^2  \, .
\end{aligned}
\end{equation}
Expanding $B^{\mu \nu}$ to leading order in $\alpha$, which is equivalent to taking the flat-space limit, leads to the result (\ref{eqn: ansatz a0 explicit}).

In \cite{Dolan:2018dqv}, the following special limit was taken\footnote{To accommodate scenarios with non-vanishing black hole spin, we take $\alpha \to 0$, instead of $\tilde{a} \to 0$ as in~\cite{Dolan:2018dqv}.}
\begin{align}
\lim_{\alpha \to 0} \left( \alpha \tilde{a}  \lambda^{-1} \right) = \chi  \, ,  \label{eqn: Dolan magnetic mode limit}
\end{align}
with $\chi =  m \pm 1$, and the separation constant $\lambda$ formally vanishes. It was then found that the ansatz~(\ref{eqn: Proca ansatz}) in this limit captures a subset of the magnetic-type modes of the vector field.  We will now investigate these special solutions in more detail. The vector field in this case becomes
\begin{align}
A^\mu_{\indlab{0}} = \frac{1}{r^2\left( 1 - \chi^2 \cos^2 \theta \right)}\begin{pmatrix}
\hphantom{1} 0\hphantom{1}  & \hphantom{1} 0\hphantom{1} & 0 & 0 \\
0 & 0 & 0 & 0 \\
 0 & 0 & 1 & -i \chi \cot \theta \\
 0 & 0 & i \chi \cot \theta & \csc^{2} \theta \\
\end{pmatrix}  \!\begin{pmatrix} \hphantom{|}\partial_t \hphantom{|}\\ \partial_r \\ \partial_\theta \\ \partial_\phi \end{pmatrix} \!Z_0 \, , \label{eqn: Dolan magnetic mode A}
\end{align}
where only the angular gradient terms are non-vanishing. The leading-order angular equation then is
\beq
\begin{aligned}
\frac{\ud^2 S_0}{\d \theta^2} \,+ & \left( \cot \theta +  2 \tan \theta +  \frac{\tan \theta}{ \chi \cos \theta - 1} - \frac{\tan \theta}{\chi \cos \theta + 1} \right) \frac{\d S_0}{\d \theta} \\
& \hskip 5pt + \left( m \chi - \frac{m^2}{\sin^2 \theta} + \frac{m \sec \theta}{\chi \cos \theta-1} + \frac{m \sec \theta}{\chi \cos \theta+ 1}\right)S_0 = 0 \, . \label{eqn: Dolan magnetic mode angular}
\end{aligned}
\eeq
Although (\ref{eqn: Dolan magnetic mode angular}) is not the general Legendre equation, the associated Legendre functions $P_{\pm (m \pm 1), m}$ are solutions to this equation, with the upper (lower) sign denoting solutions with positive (negative) $m$.
Substituting these solutions into (\ref{eqn: Dolan magnetic mode A}), and taking the appropriate scale factors $\{1, r, r \sin \theta \}$ into account, we find that 
\beq
A^i_{\indlab{0}} \propto Y^i_{j, j m} \, ,   \label{eqn: Dolan magnetic mode B}
\eeq
with $ j = |m|$. The relation in (\ref{eqn: Dolan magnetic mode B}) shows that the limit  (\ref{eqn: Dolan magnetic mode limit}) recovers a special type of magnetic mode, subject to the restriction $ j = |m|$. It remains an open problem to obtain separated equations for all magnetic modes in the Kerr background, and it would be nice to understand if the other magnetic modes can also be recovered from the ansatz (\ref{eqn: Proca ansatz}) or to show conclusively that this is not possible.

\subsubsection*{Higher-order corrections}

In Section~\ref{sec:analytic}, we showed that the equations of motion at higher orders in $\alpha$ can be solved iteratively by treating the lower-order solutions as source terms; cf.~\eqref{eqn:Jdef}. In the following, we discuss in more detail how these inhomogeneous equations  can be solved through the method of variation of parameters.

Imposing the relevant boundary conditions at the horizon and at infinity, the near and far-zone solutions of the scalar field are 
\beq
\begin{aligned}
R^{\rm near}_i(z) & = \mathcal{C}_{i}^{\rm near} n_c(z) - n_c(z) \int^z_0 \d t \, \frac{J_i^z(t) n_d(t)}{W_n(t)} + n_d(z) \int^z_0 \d t \, \frac{J_i^z(t) n_c(t)}{W_n(t)} \, , \\
R^{\rm far}_i(x) & = \mathcal{C}_{i}^{\rm far} f_c(x)  + f_c(x) \int^\infty_x \d t \, \frac{J_i^x(t) f_d(t)}{W_f(t)} - f_d(x) \int^\infty_x \d t \, \frac{J_i^x(t) f_c(t)}{W_f(t)} \, ,  \label{eqn:VOP2}
\end{aligned}
\eeq
where we defined the homogeneous solutions of the leading-order equations of motion as
\beq
\begin{aligned}
n_c(z) & \equiv  \left( \frac{z}{z+1} \right)^{i P_+ } {}_2 F_1(-\ell, \ell+1, 1-2 i P_+, 1+z)  \, , \\
n_d(z) & \equiv z^{i P_+ }( z + 1)^{ i P_+} 
  {}_2F_1\left( - \ell + 2 i P_+, \ell + 1 + 2 i P_+, 1 + 2 i P_+, 1 + z \right) \, , \\[4pt]
f_c(x) & \equiv e^{-x/2} x^{\ell}\, U( \ell+1 -\nu_0, 2+2\ell, x) \, , \\[4pt]
f_d(x) & \equiv e^{-x/2} x^{\ell}\, {}_1 F_1( \ell+1 -\nu_0, 2+2\ell, x) \, ,
\end{aligned}
\eeq
and introduced the Wronskian $W_{n} \equiv W\left[ n_c, n_d \right] = n_c n^\prime_d -n_d n^\prime_c $ (and similarly for $W_{f}$), with the prime denoting a derivative with respect to the argument of the function. In general, the integrals in~(\ref{eqn:VOP2}) cannot be solved analytically. However, to perform the matched asymptotic expansions, we are only interested in the asymptotic expansions of these integrals.  To avoid clutter, we will work with the $z$ and $x$-coordinates for the near and far-zone solutions, instead of converting them to the matching coordinate $\xi$. Recall that the limit $\alpha \to 0$  of the asymptotic expansions in terms of $\xi$ is equivalent to taking the limits $z\to \infty$ and $x \to 0$.

We first consider the integrals for the far-zone solutions. It is convenient to rewrite these integrals in the following general form 
\begin{align}
\int^\infty_x \d t \, \frac{J_i^x(t) f (t)}{W_f(t)} \equiv \int^\infty_x \d t \, e^{-t} H_i^x(t) \, ,
\end{align}
where the exponential factor regulates all divergences in the limit $t \to \infty$. To obtain the asymptotic series in the limit $x \to 0$, we Taylor expand $H_i^x(t)$ for  $t\to 0$. The result can be organized in the following way~\cite{bender1999advanced}
\beq
\begin{aligned}
\int^\infty_x \d t \, e^{-t} H_i^x(t)  = & \int^\infty_x \d t  \, e^{-t} \sum_{k=N_{\rm min}}^{-1} \sum_{m=0}^\infty  a_{k,m} \hskip 1pt t^k \log^m(t) \\
& +  \int^\infty_0 \d t  \, e^{-t} \Bigg[ H_i^x(t) - \sum_{k=N_{\rm min}}^{-1} \sum_{m=0}^\infty  a_{k,m} \hskip 1pt t^k \log^m(t)\Bigg]   \\
& -  \int^x_0 \d t  \, e^{-t}  \sum_{k=0}^{\infty} \sum_{m=0}^\infty  a_{k,m} \hskip 1pt t^k \log^m(t)\, , \label{eqn:xintegral}
\end{aligned}
\eeq
where $a_{k, m}$ are constants, and $N_{\rm min}$ is the smallest power that appears in the series. The first line in (\ref{eqn:xintegral}) consists of terms that converge as    $t \to \infty$, the second line is a constant, while the third line consists of terms that diverge as $t \to \infty$, but converge as $t \to 0$. This is why we have rewritten the integral limits $\int^\infty_x \to \int^\infty_0 - \int^x_0$ in the last line. Written in this form, (\ref{eqn:xintegral}) can be integrated term by term to give the asymptotic series as $ x \to 0$.

Similarly, in the near zone, we encounter integrals of the form 
\begin{align}
\int_0^z \d t \, \frac{J^z_i (t) n(t)}{W_n(t)} \equiv \int_0^z \d t \, e^{-1/t} \hskip 1pt H_i^z(t) \, ,
\end{align}
where the exponential factor regulates potential divergences for $t \to 0$. To obtain the asymptotic expansion as $z \to \infty$, we Taylor expand $H^z_i(t)$ for $t \to \infty$. The integral  then becomes 
\beq
\begin{aligned}
\int_0^z \d t \, e^{-1/t} \hskip 1pt H_i^z(t)  = & \int_0^z \d t \, e^{-1/t} \hskip 1pt \sum^{N_{\rm max}}_{k=-1} \sum_{m=0}^\infty  a_{k,m} \hskip 1pt t^k \log^m(t)  \\
& + \int_0^\infty \d t \, e^{-1/t} \hskip 1pt \Bigg[ H_z(t) -  \sum^{N_{\rm max}}_{k=-1} \sum_{m=0}^\infty  a_{k,m} \hskip 1pt t^k \log^m(t)  \Bigg]  \\
& -\int_z^\infty \d t \, e^{-1/t} \hskip 1pt
\sum^{-2}_{k=-\infty} \sum_{m=0}^\infty  a_{k,m} \hskip 1pt t^k \log^m(t) \, , \label{eqn:zintegral}
\end{aligned}
\eeq
where $N_{\rm max}$ is the maximum power of the integrand (which varies depending on the form of~$J^z_i $). The reasoning behind the organization of (\ref{eqn:zintegral}) is similar to that for (\ref{eqn:xintegral}): the first line converges as $t \to 0$, the second line is a constant, and the third line is convergent as $t \to \infty$. To evaluate the integrals in \eqref{eqn:zintegral}, it is convenient to perform the coordinate transformation $s \equiv 1/t$, such that they become directly analogous to the integrals in (\ref{eqn:xintegral}).

\subsubsection*{Ordinary perturbation theory} 

In Chapter~\ref{sec:spectraAtom}, we  argued that the $\ell > 0$ modes can be treated by extrapolating the far-zone solutions towards the outer horizon, $x \to 0$, and solve the spectra using ordinary perturbation theory.
Here, we will provide further details of that claim, We will also present results for the higher-order electric angular problem that were omitted in~\S\ref{sec:ProcaLO}.

To compute the higher-order eigenvalues, we take the inner product of the general equations of motion (\ref{eqn:Jdef}) with the zeroth-order eigenstates 
\beq
0 = \int w \, X^*_0 \, \square^{\indlab{0}}  X_i = \int w \, X^*_0 \, J^X_i \, , \label{eqn:ExpEval}
\eeq
where $w$ is the Sturm-Louiville weight factor and the integral is performed with respect to the relevant coordinate. The left-hand side vanishes after using the hermiticity of $\square^{\indlab{0}}$ and the leading-order equations of motion. The eigenvalues, which are contained in $J_i^X$, are then computed at every order by performing this integral. At each order in $\alpha$, we may back-substitute the eigenvalues and equations of motion of the previous order to simplify $J_i^X$.

The angular equations for the scalar field are trivially solved, since $J_i^\theta = 0$ up to the order of interest, cf.~(\ref{eqn:ScalarHigherAngular}). For the vector field, on the other hand, the higher-order angular equations contain additional $\theta$-dependent terms on the right-hand side of (\ref{equ:ProcaS}), which induce new cross couplings in the angular eigenstates, cf.~(\ref{eqn: Electric angular eigenstate order2}). The coefficients in (\ref{eqn: Electric angular eigenstate order2}) are
\beq
\begin{aligned}
b_{j-2} & =  - \left[ \frac{(j^2 - m^2)[(j-1)^2 - m^2]}{(2j-3)(2j-1)^2(2j+1)} \right]^{1/2} \frac{(j+1 - \lambda_0)}{\lambda_0^2(2j-1)} \, ,  \\[4pt]
b_{j+2} & = - \left[\frac{[(j+1)^2 - m^2] [(j+2)^2 - m^2] }{(2j+1)(2j+3)^2(2j+5)} \right]^{1/2}  \!\!\frac{( j + \lambda_0)}{\lambda_0^2(2j+3)} \, , 
\end{aligned}
\eeq
and  $c_{j \pm 2} = 2 m \lambda_0^{-5} \, b_{j \pm 2} $. These expressions are valid for both $\lambda_0 = \lambda_0^\pm$. Up to order $\alpha^3$, the angular eigenvalues for the $j=\ell \pm 1$ modes are
\beq  
\begin{aligned}
\lambda = \lambda_0 - \frac{\alpha \tilde{a} m}{\lambda_0} &- \left(\frac{\lambda_0}{2 n^2( 2 \lambda_0 - 1)}-\frac{\tilde{a}^2 (\lambda_0 + 1) (\lambda_0^2 - m^2)}{\lambda_0^3(2 \lambda_0 + 1)} \right)\alpha^2  \\
&+ \left(\frac{1}{n^2 (2 \lambda_0 - 1)} + \frac{\tilde{a}^2 ( 2 + \lambda_0)(\lambda_0^2 - m^2)}{\lambda_0^5 (2 \lambda_0 + 1)}\right) \tilde{a}\es m \es \alpha^3\, . \label{eqn:lambdaPlusMinus}
\end{aligned}
\eeq
These results are substituted into the radial equations to solve for the energy eigenvalues at higher orders.

For the radial equations, we restrict ourselves to the far-zone radial equations 
and impose regular boundary conditions at $x=0$. The latter assumption allows us to remove derivative terms in $J_i^x$ through integrations by parts. For a scalar field, the leading-order far-zone solution can be written as
\beq
R^{\text{far}}_0(x) =  \sqrt{\frac{(n-\ell-1)!}{2 n (n+\ell)!}} \, e^{-x/2} x^\ell L^{(2\ell+1)}_{n-\ell-1}(x) \, , \label{eqn: scalar radial far LO 2}
\eeq
where $L_k^{(\rho)}$ is the associated Laguerre polynomial. 
The overall coefficient has been fixed by requiring the integral of the 
square of the mode function to be unity. 
Each term in the inner product (\ref{eqn:ExpEval}) then has the following generic form
\beq
\langle x^{-s} \rangle \equiv \frac{(n-\ell-1)!}{2 n (n+\ell)!} \int_0^\infty \d x \, e^{-x} x^{2\ell+2-s}\, \left( L^{(2\ell+1)}_{n-\ell-1} \right)^2 \, , \label{eqn:farexpval}
\eeq
where $J_i^x \supset x^{-s}$, with positive integer $s$. We may then compute the energy eigenvalues through the following identities
\beq
\begin{alignedat}{2}
\left\langle \frac{1}{x} \right\rangle & = \frac{1}{2n} \, , &&  \hskip 4pt \left\langle \frac{1}{x^3} \right\rangle = \frac{1}{2\ell (2\ell+1) (2\ell+2)}
 \, , \\
\left\langle \frac{1}{x^2} \right\rangle & = \frac{1}{2n (2\ell+1)}    \, , && \hskip 4pt  \left\langle \frac{1}{x^4} \right\rangle   = \frac{3 n^2-\ell ^2-\ell }{n  (2 \ell -1)(2 \ell) (2 \ell +1)(2\ell +2) (2 \ell +3)} \, , 
\end{alignedat}
 \label{eqn:Identities1}
\eeq
which are valid for all $\ell$. The fact that $\langle x^{-3} \rangle$ and $\langle x^{-4} \rangle$ diverge for $\ell =0$ reflects the sensitivity of these modes to the near region of the black hole. In general, the presence of these terms indicates a breakdown of ordinary perturbation theory, and calculating the eigenvalues requires the more rigorous matched asymptotic expansion. However, in the case of a scalar field, these terms have coefficients that are proportional to $m$, which vanish for $\ell=0$, so that the naively divergent terms do not contribute to the spectrum. Ordinary perturbation theory, with the assumption of a regular boundary condition at the horizon, is therefore sufficient for the scalar spectrum.

Although the far-zone radial equations of the scalar and vector fields are different, the inner product (\ref{eqn:farexpval}) is also valid for the vector case. This is because the additional power of $x$ that appears in the radial solution for a vector field cancels, as the Sturm-Liouville weight factor is now $1$, instead of $x^2$ for the scalar case. The identities (\ref{eqn:Identities1}) are therefore also valid for the vector field. 
Unlike in the scalar case, however, the coefficients of the $\langle x^{-3} \rangle$ and $\langle x^{-4} \rangle$  terms do not vanish for the vector $\ell=0$ modes, so ordinary perturbation theory is expected to yield divergent results. Indeed, these divergences can be written as divergent boundary terms in the integral (\ref{eqn:farexpval}). Remarkably, however, if these boundary terms are discarded, we obtain results that agree with those obtained through the more rigorous matched asymptotic expansion. Since an implicit assumption in deriving the rules (\ref{eqn:Identities1}) is precisely that any boundary terms are negligible, substituting them into the divergent operators can still provide finite results. Indeed, we find that the coefficients of the $\langle x^{-3} \rangle$ and $\langle x^{-4} \rangle$ terms are precisely such that the divergences for $\ell=0$ cancel in the sum. As a consequence of this cancellation, ordinary perturbation theory gives the correct results even for the fine and hyperfine splittings of vector spectrum. 
Interestingly, two wrongs \emph{can} make a right.

  \section{Details of the Numerical Treatment} 
  \label{app:numericalDetails}
  
  A large part of the analysis presented in Section~\ref{sec:numeric} was focused on achieving accurate numeric results for the quasi-bound state spectrum without a separable ansatz. In \S\ref{sec:nosep}, we presented a schematic outline for how one translates the unseparated Proca equation and Lorenz constraint into a nonlinear eigenvalue problem that can be readily solved on a computer. In the interest of pedagogy, we kept technical details there to a minimum, and will instead provide them in this appendix.

  We first detail the decomposition of the Proca equation into its temporal, radial, and angular components using the vector spherical harmonics defined in Appendix~\ref{app:vsh}, and describe how to choose these harmonics such that the quasi-bound state boundary conditions are satisfied. We then detail the construction of the finite-dimensional matrix, whose nonlinear eigenvalues determine the bound state spectrum. Finally, in Appendix~\ref{app:cheb}, we provide a brief introduction to Chebyshev interpolation with an emphasis on its convergence properties and numeric implementation.  

\subsubsection*{Decomposition of the Proca equation} 
\label{app:decompProca}

Our numerical analysis relies on a $1+1+2$ decomposition of the Proca equation into its temporal, radial and angular components.  While such a decomposition is straightforward for a scalar field, the vector index complicates things. It will therefore be useful to first rewrite the Proca equation as a set of coupled scalar equations, reminiscent of the decomposed Klein-Gordon equation (\ref{eq:startingPoint}), with operators that act simply on the scalar spherical harmonics.

  To this end, we introduce a basis of vector fields  $e\indices{_a^\mu} \partial_\mu$ and dual form fields $f\indices{^a_\mu} \, \ud x^\mu$, where $a = 0, 1, 2, 3$. These bases are dual to one another in the sense that
    \begin{equation}
    \begin{aligned}
      \bar{e}\indices{_a^\mu} f\indices{^b_\mu} &= \delta^a_b \,,\\
       \bar{e}\indices{_a^\mu} f\indices{^a_\nu}  &= \delta_\nu^\mu \,, \label{eqn:Tetrad}
      \end{aligned}
    \end{equation}
    where the bar denotes complex conjugation. Taking the vector fields to have definite angular momentum, we may write
    \beq
      \begin{aligned}
          e\indices{_0^\mu} \partial_\mu & = F_t(r) \, \partial_t\,, \quad \ && e\indices{_1^\mu} \partial_\mu = \chi_{1, 1}^i \, \partial_i \, , \\ 
          \ e\indices{_2^\mu} \,\partial_\mu & = \chi_{1, 0}^i \, \partial_i\,, 
          && e\indices{_3^\mu} \,\partial_\mu  = \chi_{1, \sminus 1}^{i} \,\partial_i\,, \label{eq:vecFieldDefs}
           \end{aligned}
           \eeq
           and
           \beq
         \begin{aligned}
       f\indices{^0_\mu} \ud x^\mu &= F_t^{-1} \,\ud t\, , \quad  &&f\indices{^1_\mu}\ud x^\mu = \chi_i^{1,1}\,\ud x^i\,, \\ 
       f\indices{^2_\mu} \ud x^\mu  & = \chi_i^{1, 0} \, \ud x^i \, , \quad && f\indices{^3_\mu} \ud x^\mu = \chi_{i}^{1 ,\sminus 1} \, \ud x^i\,. \label{eq:formFieldDefs}
\end{aligned}
\eeq
where we have used the forms and vectors defined in the previous section.  By design, these forms are well-behaved under the Kerr isometries,
    \begin{equation}
      \pounds_t f\indices{^a_\mu} = 0\,, \quad \pounds_z f\indices{^0_\mu} = 0\,, \quad \text{and} \quad \pounds_z f\indices{^i_\mu} = (2 - i) f\indices{^i_\mu}\,, \quad \text{where} \quad i = 1, 2, 3\, ,
    \end{equation}
    and are eigenstates of the angular momentum operator, $\pounds^2 f\indices{^0_\mu} = 0$ and $\pounds^2 f\indices{^i_\mu} = 2 f\indices{^i_\mu}$. Similar relations hold for the basis of vectors, $e\indices{_a^\mu}$. Finally, it is useful to note that
    \beq
    \begin{aligned}
      \partial_\mu &= \bar{f}\indices{^a_\mu}e\indices{_a^\nu}\partial_\nu \\[2pt]
      &=  i \bar{f}\indices{^{\,0}_\mu} F_t \pounds_t + \delta_\mu^r \partial_r +  \bar{f}\indices{^{\,1}_\mu} F_\theta \mathcal{D}_+ + \bar{f}\indices{^{\,2}_\mu} F_\theta \mathcal{D}_0 + \bar{f}\indices{^{\,3}_\mu} F_\theta \mathcal{D}_-\,.
    \end{aligned}
    \eeq
    Our goal is to rewrite the Proca equation,\footnote{At a technical level, the Proca equation with a lowered index is much simpler than with the upper index and so we focus exclusively on this form.} $\nabla^2 A_\mu =A_\mu$,
    and the Lorenz condition, $\nabla^\mu A_\mu = 0$, into a form that is (roughly) a system of coupled scalar differential equations.  
    We assume that $A_\mu$ is in a state of definite frequency and azimuthal angular momentum,
    \begin{equation}
    \begin{aligned}
      \pounds_t A_\mu &= - \omega A_\mu \, , \\ \pounds_z A_\mu &= +m A_\mu\,,
      \end{aligned}
    \end{equation}
    and write   $A_\mu = A_a f\indices{^a_\mu}$.

    We begin by noting that, for any scalar $\Phi$, the purely radial derivative operator $\partial_r(\Delta \partial_r)$ can be rewritten using the Laplacian $\nabla^2$ and isometry generators $\pounds^2$, $\pounds_t$ and $\pounds_z$:
    \beq
    \begin{aligned}
        & \partial_r\left(\Delta \partial_r \Phi\right) = \\
        & \left[\Sigma \nabla^2 + \pounds^2 - \left(\Sigma + \frac{2 \alpha r(r^2 + \alpha^2 \tilde{a}^2)}{\Delta}\right)\pounds_t^2 - \frac{\alpha^2 \tilde{a}^2}{\Delta} \pounds_z^2 - \frac{4 \alpha^2 \tilde{a} r}{\Delta} \pounds_t \pounds_z\right] \Phi\,.
        \end{aligned}
    \eeq
    For the Klein-Gordon equation, this almost immediately yields the decomposition into temporal, radial, and angular degrees of freedom we are after. The Proca equation requires a bit more work, but eventually can be written as
      \begin{align}
        0\, =\, \ & \bar{e}\indices{_b^\mu}\Delta^{-1} \Sigma \left(\nabla^2 - 1\right)\left(A_a f\indices{^a_\mu}\right) \nonumber \\
          =  \ & \bigg[\frac{1}{\Delta} \partial_r(\Delta \partial_r) - \frac{1}{\Delta} \left(\pounds^2 + \alpha^2 \tilde{a}^2 \left(1 - \omega^2\right) \cos^2 \theta \right)  - (1 - \omega^2) + \frac{P_+^2}{(r - r_+)^2}  \nonumber \\
        &+ \frac{P_-^2}{(r -r_-)^2} - \frac{A_+}{(r - r_+)(r_+ - r_-)} + \frac{A_-}{(r -r_-)(r_+ - r_-)}\bigg] A_b \label{eq:procaEqDecomp} \\[4pt]
        &+ \mathcal{S}\indices{_b^a} A_a + \mathcal{Q}\indices{_b^a} \pounds_z A_a + \mathcal{R}\indices{_b^a} \partial_r A_a + \mathcal{P}\indices{_b^a} \mathcal{D}_+ A_a + \mathcal{Z}\indices{_b^a} \mathcal{D}_0 A_a + \mathcal{M}\indices{_b^a} \mathcal{D}_- A_a \nonumber \,, 
      \end{align} 
      where we have introduced the following `mixing matrices':
        \begin{align}
          \mathcal{S}\indices{_b^a} &= \frac{1}{\Delta} \bar{e}\indices{_b^\mu}\Big[\Sigma \nabla^2 f\indices{^a_\mu} - \frac{\alpha^2 \tilde{a}^2}{\Delta} \pounds_z^2 f\indices{^a_\mu} + \frac{4 \alpha^2 \tilde{a} \omega r}{\Delta} \pounds_z f\indices{^a_\mu} - 2 i \omega F_t \Sigma \,\bar{f}\indices{^{\,0}_\rho} \nabla^\rho f\indices{^a_\mu} \Big] \, , \nonumber \\[0.6ex]
          \mathcal{Q}\indices{_b^a} &=  - \frac{2 \alpha^2 \tilde{a}^2}{\Delta^2} \bar{e}\indices{_b^\mu} \pounds_z f\indices{^a_\mu}\, , \qquad  \quad \ \mathcal{R}\indices{_b^a} = 2 \Sigma \Delta^{-1}\,  \bar{e}\indices{_b^\mu} \nabla^\rho f\indices{^a_\mu} \delta^r_\rho\, , \\[0.6ex]
          \mathcal{P}\indices{_b^a} &= 2\Sigma \Delta^{-1}\,  \bar{e}\indices{_b^\mu} \nabla^\rho f\indices{^a_\mu} \bar{f}\indices{^{\,1}_\rho} F_\theta\, ,  \quad \mathcal{Z}\indices{_b^a} = 2\Sigma\Delta^{-1}\,  \bar{e}\indices{_b^\mu} \nabla^\rho f\indices{^a_\mu} \bar{f}\indices{^{\,2}_\rho} F_\theta\, , \nonumber \\[0.75ex]
          \mathcal{M}\indices{_b^a} &= 2 \Sigma \Delta^{-1}\,  \bar{e}\indices{_b^\mu} \nabla^\rho f\indices{^a_\mu} \bar{f}\indices{^{\,3}_\rho} F_\theta \nonumber\,.
        \end{align} \label{eq:mixingMatrices}
      Similarly, the Lorenz condition can be written as
      \begin{equation}
        0 \,=\, \mathcal{T}^0 A_0 + \mathcal{S}^i A_i + \mathcal{R}^i \partial_r A_i + \mathcal{P}^i \mathcal{D}_+ A_i + \mathcal{Z}^i \mathcal{D}_0 A_i + \mathcal{M}^i \mathcal{D}_- A_i\,, \label{eq:lorenzEqDecomp}
      \end{equation}
      where we have defined the `mixing vectors':
      \begin{equation}
        \begin{aligned}
          \mathcal{S}^i &= \nabla^\mu f\indices{^i_\mu} - i \omega F_t f\indices{^i_\mu} g^{\mu \lambda} \bar{f}\indices{^{\,0}_\lambda}\,, \hskip 4pt \mathcal{R}^i = f\indices{^i_\mu} g^{\mu r}\,, \hskip 4pt \mathcal{P}^i = f\indices{^i_\mu} g^{\mu \lambda} \bar{f}\indices{^{\,1}_\lambda} F_\theta\,, \\[0.8ex]
           \mathcal{Z}^i &= f\indices{^i_\mu} g^{\mu \lambda} \bar{f}\indices{^{\,1}_\lambda} F_\theta\,, \hskip 4pt  \mathcal{M}^i = f\indices{^i_\mu} g^{\mu \lambda} \bar{f}\indices{^{\,1}_\lambda} F_\theta \,, \hskip 4pt \mathcal{T}^0 = i F_t^{-1}\big(m g^{t \phi} - \omega g^{tt}\big).
        \end{aligned} \label{eq:mixingVectors}
      \end{equation}
      These mixing matrices and vectors encode how the temporal and spatial components couple to one another and how the Kerr geometry distinguishes scalar and vector fields.

  \subsubsection*{Boundary conditions} 
  \label{app:boundaryConditions}

    So far, the radial functions $F_r$, $F_\theta$, $F_t$ have not been specified. 
    We are free to choose these functions in such a way that all of the components $A_a$ 
scale in the same way as $r \to r_+$ and $r \to \infty$. This will make it easy to impose the ingoing boundary conditions at the outer horizon and the decaying boundary conditions at spatial infinity.

    Let us first concentrate on the behavior at the event horizon, which will be easiest to analyze using the standard form of the Proca equation $\nabla^2 A_\mu = A_\mu$ instead of the decomposition (\ref{eq:procaEqDecomp}). We demand that $A_\mu$ is an eigenstate of both frequency $\pounds_t A_\mu = - \omega A_\mu$ and angular momentum, $\pounds_z A_\mu = m A_\mu$, and so the coefficients $A_\mu$ can all be written as $A_\mu = A_\mu(r, \theta)\,e^{-i \omega t + i m \phi}$. We can then solve the Lorenz condition for $A_0(r, \theta)$ and eliminate it from the Proca equation to attain equations of motion that only involve the $A_i$. While the system of equations is quite complicated, for what follows, we only need to consider the asymptotic behavior of these equations as $r \to r_+$. We will thus use $(\,\dots)$ to denote functions that are constant in $r$ (though they generally depend on $\theta$, $r_\pm$, $m$ and $\omega$) and  whose precise form are irrelevant to the discussion.

    With the benefit of hindsight, we take
    \begin{equation}
      A_r = \left(\frac{r - r_-}{r- r_+}\right) \tilde{A}_r\,, \quad A_\theta = (r - r_-) \tilde{A}_\theta\,, \quad A_\phi = (r - r_-) \tilde{A}_\phi \,.
    \end{equation}
    With these factors peeled off, the near-horizon behavior for the $r$, $\theta$, $\phi$ components of the Proca equation are 
      \begingroup \allowdisplaybreaks
      \begin{align}
        0 \,=\,\ & \partial_r^2 \tilde{A}_r + \frac{1}{r - r_+} \partial_r \tilde{A}_r + \frac{P_+^2}{(r -r_+)^2} \tilde{A}_r + \frac{(\,\dots)}{r - r_+} \partial_\theta \tilde{A}_\theta + \frac{(\,\dots)}{r - r_+} \partial_\theta^2 \tilde{A}_r   \nonumber \\
        & + (\,\dots) \partial_r \tilde{A}_\theta +  \frac{(\,\dots)}{r - r_+} \partial_\theta \tilde{A}_r+ \frac{(\,\dots)}{r -r_+} \tilde{A}_\phi  + \frac{(\,\dots)}{r-r_+} \tilde{A}_\theta + \cdots \, ,\\
        0 \,=\, \ & \partial_r^2 \tilde{A}_\theta +\frac{1}{r - r_+} \partial_r \tilde{A}_\theta +  \frac{P_+^2}{(r - r_+)^2} \tilde{A}_\theta  + \frac{(\,\dots)}{r - r_+} \partial_\theta \tilde{A}_r +\frac{(\,\dots)}{r - r_+} \partial_\theta \tilde{A}_\theta   \nonumber \\
        &+ \frac{(\,\dots)}{r - r_+} \partial_\theta^2 \tilde{A}_\theta +  \frac{(\,\dots)}{r - r_+} \tilde{A}_\phi+ (\,\dots) \partial_r \tilde{A}_r +(\,\dots) \tilde{A}_r + \cdots\, ,\\
        0 \,=\ \, & \partial_r^2 \tilde{A}_\phi + \frac{1}{r - r_+} \partial_r \tilde{A}_\phi + \frac{P_+^2}{(r - r_+)^2} \tilde{A}_\phi + \frac{(\,\dots)}{r -r_+} \tilde{A}_\theta + (\,\dots) \partial_r \tilde{A}_\theta + (\,\dots) \partial_\theta \tilde{A}_r  \nonumber \\
        & + \left(\,\dots\right)\left(\partial_r^2 \tilde{A}_r + \frac{1}{r - r_+} \partial_r \tilde{A}_r + \frac{P_+^2}{(r -r_+)^2}\tilde{A}_r \right) + (\,\dots) \partial_r \partial_\theta \tilde{A}_\theta   \nonumber \\
        &  + \frac{(\,\dots)}{r - r_+} \partial_\theta \tilde{A}_\theta + \frac{(\,\dots)}{r - r_+} \partial_\theta \tilde{A}_\phi + (\,\dots) \partial_\theta^2 \tilde{A}_\theta + \frac{(\,\dots)}{(r - r_+)} \partial_\theta^2 \tilde{A}_\phi + \cdots\, .
      \end{align}
      \endgroup
      We see that most singular terms imply that $\tilde{A}_r$, $\tilde{A}_\theta$, $\tilde{A}_\phi$ all go as $(r - r_+)^{\pm i P_+}$ as $r \to r_+$. In a similar way, we can also examine the Lorenz condition as $r \to r_+$, 
      \beq
      \begin{aligned}
        A_0 & = (\,\dots)(r -r_+) \tilde{A}_r \\
        & + (r-r_+) \partial_r \tilde{A}_r + (\,\dots)(r-r_+) \partial_\theta \tilde{A}_\theta + (\,\dots)(r-r_+) \tilde{A}_\theta + (\,\dots) \tilde{A}_\phi\,,
      \end{aligned}
      \eeq
      and we see that $A_0$ also approaches $(r - r_+)^{\pm i P_+}$ as $r \to r_+$. With these scalings in mind, we take the radial functions to be
      \begin{equation}
        F_t = 1\,, \quad F_r = \frac{r - r_+}{r - r_-}\,, \quad F_\theta = \frac{1}{r - r_-}\,, \label{eq:decompRadialFuncs}
      \end{equation} 
      so that $A_a \sim (r- r_+)^{\pm i P_+}\left(1 + \mathcal{O}(r -r_+)\right)$ as $r \to r_+$ for each $a$.

      To analyze how the coefficients $A_a$ behave as $r \to \infty$, we turn to the decomposed Proca equation (\ref{eq:procaEqDecomp}), where the mixing matrices are computed with the radial functions (\ref{eq:decompRadialFuncs}). 
      Ignoring the mixing matrices, for a moment, (\ref{eq:procaEqDecomp}) predicts that the $A_a$ scale in the same way as the scalar solution~(\ref{eqn: scalar radial infinity}):
      \begin{equation}
        A_a \sim r^{-1 - \nu + 2 \alpha^2/\nu} e^{-\sqrt{1 - \omega^2}(r - r_+)}\left(1 + \mathcal{O}(r^{-1})\right) .
      \end{equation}
      This conclusion will not be affected by including the mixing matrices, as long as they 
       decay faster at spatial infinity than $r^{-1}$ and, by explicit computation, we confirm that $\mathcal{S}\indices{_0^0} = 0$, $\{\mathcal{S}\indices{_0^i}, \mathcal{S}\indices{_i^0}, \mathcal{R}\indices{_0^0}, \mathcal{R}\indices{_i^j}\}$ scale as $r^{-2}$, $\{\mathcal{S}\indices{_i^j}, \mathcal{R}\indices{_0^i}, \mathcal{R}\indices{_i^0}, \mathcal{P}\indices{_0^0}, \mathcal{P}\indices{_i^j}, \mathcal{Z}\indices{_0^0}, \mathcal{Z}\indices{_i^j}, \mathcal{M}\indices{_0^0}, \mathcal{M}\indices{_i^j}\}$ scale as $r^{-3}$, while the rest decay as $r^{-4}$.

      We have chosen the radial functions (\ref{eq:decompRadialFuncs}), so that the components $A_a$ all have the same asymptotic behavior as $r \to r_+$ and $r \to \infty$. We then peel off this asymptotic behavior by writing
      \begin{equation}
        A_a(r) = \left(\frac{r - r_+}{r - r_-}\right)^{i P_+} \!\!\!\left(r - r_-\right)^{-1+\nu-2 \alpha^2/\nu} e^{-\alpha(r - r_+)/\nu} B_a(r)\,, \label{eq:asympBehaviorVec}
      \end{equation}
      and work with the functions $B_a$, which, for the modes of interest, approach constants as $r \to r_+$ and $r \to \infty$. In the numerics, this allows us to impose the boundary conditions for the quasi-bound states by simply expanding $B_a$ in a set of functions that approach constant values at the boundaries.

    \subsubsection*{Constructing the matrix equation}
      To convert (\ref{eq:procaEqDecomp}) into a finite-dimensional matrix equation, we first write \eqref{eq:asympBehaviorVec} as
      \beq
       \begin{aligned}
        A_a(r) & = \left(\frac{r - r_+}{r - r_-}\right)^{i P_+} \! \\
        & \hskip 20pt \times (r - r_-)^{-1 + \nu - 2 \alpha^2/\nu}e^{-\alpha(r - r_+)/\nu} B_a(\zeta(r)) \equiv F(r) B_a(\zeta(r))\,,
      \end{aligned}
      \eeq
    which is similar to the scalar case (\ref{eq:pseudoSpecPeelOff}), with $\zeta(r)$ a map from $[r_+, \infty)$ to $[-1, 1]$. The spatial components $i = 1, 2, 3$ of the Proca equation (\ref{eq:procaEqDecomp}) can then be written as
      \begin{equation}
        \mathcal{D}\indices{_i^k}[B_k] + \mathcal{D}\indices{_i^0}[B_0] = 0\,, \label{eq:procaSchematic}
      \end{equation}
      where
      \begin{align}
         \mathcal{D}\indices{_i^k} & [B_k] \,  =\,\  \Bigg(\partial_\zeta^2 + \frac{1}{\zeta'(r)}\left(\frac{1}{r - r_+} + \frac{1}{r - r_-} + \frac{2 F'(r)}{F(r)} + \frac{\zeta''(r)}{\zeta'(r)}\right) \partial_\zeta \Bigg) B_i \nonumber \\
        & - \frac{1}{\zeta'(r)^2 \Delta} \left(\pounds^2 + \alpha^2 \tilde{a}^2 \left(1 - \omega^2\right) \cos^2 \theta \right)B_i + \frac{1}{\zeta'(r)^2} \Bigg(\frac{F'/F}{r-r_+}  \nonumber
\\
        & + \frac{F'/F}{r - r_-}  + \frac{F''}{F} - (1 - \omega^2) + \frac{P_+^2}{(r - r_+)^2} + \frac{P_-^2}{(r -r_-)^2}  \\[4pt]
        &   - \frac{A_+}{(r - r_+)(r_+ - r_-)} + \frac{A_-}{(r -r_-)(r_+ - r_-)}\Bigg) B_i  + \tilde{\mathcal{S}}\indices{_i^k} B_k     \nonumber \\[4pt]
       & + \tilde{\mathcal{Q}}\indices{_i^k}\, \pounds_z B_k + \tilde{\mathcal{R}}\indices{_i^k}\, \partial_\zeta B_k + \tilde{\mathcal{P}}\indices{_i^k}\, \mathcal{D}_+ B_k + \tilde{\mathcal{Z}}\indices{_i^k}\, \mathcal{D}_0 B_k + \tilde{\mathcal{M}}\indices{_i^k} \,\mathcal{D}_- B_k \, ,   \nonumber \\[14pt]
       & \mathcal{D}\indices{_i^0}[B_0] \,=\,\  \tilde{\mathcal{S}}\indices{_i^0} B_0 + \tilde{\mathcal{Q}}\indices{_i^0} \pounds_z B_0 + \tilde{\mathcal{R}}\indices{_i^0} \partial_\zeta B_i \nonumber \\
       & \hskip 40pt + \tilde{\mathcal{P}}\indices{_i^0} \,\mathcal{D}_+ B_0 + \tilde{\mathcal{Z}}\indices{_i^0} \,\mathcal{D}_0 B_0 + \tilde{\mathcal{M}}\indices{_i^0} \,\mathcal{D}_- B_0\, .
      \end{align}
 In the above, we have defined the transformed mixing matrices 
      \begin{equation}
        \begin{aligned}
        \tilde{\mathcal{S}}\indices{_b^a} &= \frac{1}{\zeta'^{\hskip 1pt 2}} \mathcal{S}\indices{_b^a} + \frac{F'}{ \zeta'^{\hskip 1pt 2} F} \mathcal{R}\indices{_b^a}\,, \quad \tilde{\mathcal{R}}\indices{_b^a} = \mathcal{R}\indices{_b^a}/\zeta'\,, \quad \tilde{\mathcal{Q}}\indices{_b^a} = \mathcal{Q}\indices{_b^a}/\zeta'^{\hskip 1pt 2} \,,\\[1ex]
         \tilde{\mathcal{P}}\indices{_b^a} &= \mathcal{P}\indices{_b^a}/\zeta'^{\hskip 1pt 2} \,,\quad \tilde{\mathcal{Z}}\indices{_b^a} = \mathcal{Z}\indices{_b^a}/\zeta'^{\hskip 1pt 2} \,,\quad \tilde{\mathcal{M}}\indices{_b^a} = \mathcal{M}\indices{_b^a}/\zeta'^{\hskip 1pt 2}\,,
      \end{aligned}
      \end{equation}
      where $\zeta' = \partial_r \zeta$.  Similarly, the Lorenz condition can be written as
      \begin{equation}
        \mathcal{D}\indices{_0^0}[B_0] + \mathcal{D}\indices{_0^i}[B_i] = 0\,,\label{eq:lorenzConstraintSchematic}
      \end{equation}
      where
      \begin{align}
    \mathcal{D}\indices{_0^0}[B_0] &= \mathcal{T}^0 B_0\,, \\
        \mathcal{D}\indices{_0^i}[B_i] &=\tilde{\mathcal{S}}^i  B_i + \tilde{\mathcal{R}}^i \partial_\zeta B_i + \mathcal{P}^i \mathcal{D}_+ B_i + \mathcal{Z}^i \mathcal{D}_0 B_i + \mathcal{M}^i \mathcal{D}_- B_i\,,
      \end{align}
      with $ \tilde{\mathcal{S}}^i = \mathcal{S}^i + F' \mathcal{R}^i/F$ and $\tilde{\mathcal{R}}^i = \zeta' \, \mathcal{R}^i$.

      We then expand the  temporal component of the field into scalar harmonics and the spatial components into one-form harmonics, 
      \begin{align}
       B_0 &= \sum_{j'} B_{j' m} Y_{j' m}\, , \\
        B_i &= \sum_{\ell', j'} B_{\ell',j' m} \,Y_i^{\ell',j' m} \, .
      \end{align}
      As discussed in Appendix~\ref{app:vsh}, the summation ranges of $j'$ and $\ell'$ depend on the parity of the mode we are solving for. 
      We then project the Proca equation (\ref{eq:procaSchematic}) onto the vector harmonics and the Lorenz constraint (\ref{eq:lorenzConstraintSchematic}) onto the scalar harmonics, to obtain
      \begin{align}
        \sum_{\ell',j'} \mathcal{D}^{\lab{ss}}_{\ell j\,|\,\ell' j'}[B_{\ell', j' m}(\zeta)] + \sum_{j'} \mathcal{D}^{\lab{st}}_{\ell j\, |\, j'}[B_{j'm}(\zeta)] &= 0\,, \label{eq:procaProject} \\
          \sum_{\ell',j'} \mathcal{D}^{\lab{ts}}_{j\, |\, \ell' j'}[B_{\ell', j' m}(\zeta)] + \sum_{j'} \mathcal{D}^{\lab{tt}}_{j\, | \, j'}[B_{j' m}(\zeta)] &= 0\,. \label{eq:lorenzProject}
      \end{align}
      Explicitly, we find
      \begin{equation}
        \begin{aligned}
          \mathcal{D}^\lab{ss}_{\ell j\, |\, \ell' j'}[B_{\ell',j' m}(\zeta)] &= \int_{\lab{S}^2}\!\ud \Omega\,\, \bar{Y}^{i}_{\ell,j m} \, \mathcal{D}\indices{_i^k}\big[B_{\ell',j'm}(\zeta) Y^{\ell',j' m}_k\big]\, , \\
          \mathcal{D}^{\lab{st}}_{\ell j\,|\,j'}[B_{j'm}(\zeta)] & = \int_{\lab{S}^2}\!\ud \Omega\, \, \bar{Y}^{i}_{\ell,jm} \, \mathcal{D}\indices{_i^0}\big[B_{j'm}(\zeta) Y_{j' \,m}\big]\, , \\
          \mathcal{D}^{\lab{t s}}_{j\,|\,\ell'j'}[B_{\ell',j'm}(\zeta)] &= \int_{\lab{S}^2}\!\ud \Omega \,\, \bar{Y}_{jm}\,\mathcal{D}\indices{_0^k}\big[B_{\ell', j' m}(\zeta) Y^{\ell',j' m}_{k}\big]\, , \\
          \mathcal{D}^{\lab{tt}}_{j\,|\,j'}[B_{j'm}(\zeta)] &= \int_{\lab{S}^2}\!\ud \Omega\,\,\bar{Y}_{jm}\,\mathcal{D}\indices{_0^0}\big[B_{j'm}(\zeta) Y_{j'm}\big]\,.
        \end{aligned}\label{eq:sphericalMatrices}
      \end{equation}
      Because we defined the $f\indices{^a_\mu}$ to have definite angular momentum, the harmonics (\ref{eq:vectorSphericalHarmonics}) and \ref{eq:formSphericalHarmonics}) take simple forms in this tetrad basis. For instance, the components of the one-form harmonics are
      \begin{equation}
        Y^{\ell, j m}_{i} = \langle (1 \, m_i) \, ( \ell\,\, m - m_i) | j \, m \rangle \,Y_{\ell, m- m_i}(\theta, \phi)\,,
      \end{equation}
      where $m_i = 2 - i$. In practice, the overlaps (\ref{eq:sphericalMatrices}) can be computed efficiently, as each term in the rewritten Proca equation (\ref{eq:procaEqDecomp}) operates very simply on these harmonics.

    A vector field
     is a solution of 
       the Proca equation and the Lorenz constraint if and only if (\ref{eq:procaProject}) and (\ref{eq:lorenzProject}) are satisfied for all $j$ and $\ell$. Following the scalar case in \S\ref{sec:nosep}, we may approximate the radial functions as
      \begin{equation}
        B_{\ell',j'm}(\zeta) = \sum_{k = 0}^{N} B_{\ell',j'm}(\zeta_k) \,p_k(\zeta) \qquad \text{and} \qquad B_{j' m} = \sum_{k = 0}^{N} B_{j' m}(\zeta_k) \,p_k(\zeta)\,,
      \end{equation}
      where $p_k(\zeta)$ are the cardinal polynomials associated to the points $\{\zeta_k\}$. By substituting these approximations into (\ref{eq:procaProject}) and (\ref{eq:lorenzProject}), sampling each equation at the $\{\zeta_n\}$, and truncating the angular expansion, we convert this system of equations into a finite-dimensional matrix equation. This matrix, whose structure is depicted in Fig.\,\ref{fig:vectorMatrixStructure}, can then be passed to a nonlinear eigenvalue solver to determine the bound state spectrum.

  \subsubsection*{Chebyshev interpolation} \label{app:cheb} 

    Our numerical techniques rely on approximating the scalar and vector field configurations using finite and discrete sets of data. This discretization is a necessary step towards approximating the Klein-Gordon and Proca equations---i.e.~partial \emph{differential} equations---as finite-dimensional matrix equations. 
    How we discretize matters immensely, and a careless choice can cause the numerics to fail outright. This section
     explains both what Chebyshev interpolation is and why we choose it, and provides various technical details needed for the numerics outlined in \S\ref{sec:Separable}. For an excellent introduction to the subject, see \cite{Trefethen:2013ata}.

    Any smooth function $f(\zeta)$ on the interval $\zeta \in [-1, 1]$ has a unique representation in terms of Chebyshev polynomials,
    \begin{equation}
      f(\zeta) = \sum_{k = 0}^{\infty} a_k T_k(\zeta)\,,\label{eq:chebExpApp}
    \end{equation}
    where the  Chebyshev polynomials of  degree $k$ are  $T_k(\cos \theta) = \cos k \theta$. 
    The Chebyshev coefficients are then given by
    \begin{equation}
      a_k = \frac{2}{\pi} \int_{-1}^{1} \!\ud \zeta\, \frac{T_k(\zeta)\, f(\zeta)}{\sqrt{1 - \zeta^2}}\, ,\label{eq:chebCoefficients}
    \end{equation}
    for $k \geq 1$, while the right-hand side is multiplied by a factor of $1/2$ when $k = 0$. If we define an expanded function on the unit circle, 
    $F(z) = F(1/z) = f(\zeta)$, with $\zeta =  (z+z^{-1})/2$, then the Chebyshev expansion (\ref{eq:chebExpApp}) is nothing more than the unique Laurent series for $F(z)$ and (\ref{eq:chebCoefficients}) is simply a translation of the Cauchy integral formula.

    With this expansion in hand, we may define a degree $N$ polynomial approximation to $f(\zeta)$ by truncating the sum,
    \begin{equation}
      \tilde{f}_{N}(\zeta) = \sum_{k = 0}^{N} a_k T_k(\zeta)\,. \label{eq:truncApprox}
    \end{equation}
    While this approximation is guaranteed to become exact as $N \to \infty$, its accuracy at finite $N$ depends on how quickly the Chebyshev coefficients decay. This, in turn, depends on the analytic structure of $f(\zeta)$ and thus $F(z)$. Since $f(\zeta)$ is analytic on the interval $\zeta \in [-1, 1]$, the expanded function $F(z)$ is necessarily analytic within an annulus about the unit circle, the largest of which we denote $\rho^{-1} \leq |z| \leq \rho$. The Laurent series for $F(z)$ diverges outside of this annulus, and so its size determines the asymptotic behavior of the Chebyshev coefficients, i.e.~$a_k < \mathcal{C} \rho^{-k}$~as~$k \to \infty$ for some constant $\mathcal{C}$. Since $|T_k(\zeta)| \leq 1$ on the interval, the truncation error scales as
    \begin{equation}
      |f(\zeta)  - \tilde{f}_{N}(\zeta)| \leq \sum_{k = N+1}^{\infty} \left|a_k T_k(\zeta)\right| \leq \frac{\mathcal{C} \rho^{-N}}{\rho - 1}\,, \quad \text{as}\quad N \to \infty\,. \label{eq:convergenceProps}
    \end{equation}
    The Chebyshev approximation (\ref{eq:truncApprox}) thus converges to $f(\zeta)$ exponentially quickly, at a rate set by the singularity closest to the unit circle $|z| = 1$. If $F(z)$ is entire, the approximation converges even faster.

\begin{figure}
      \begin{center}
        \includegraphics[scale=0.9]{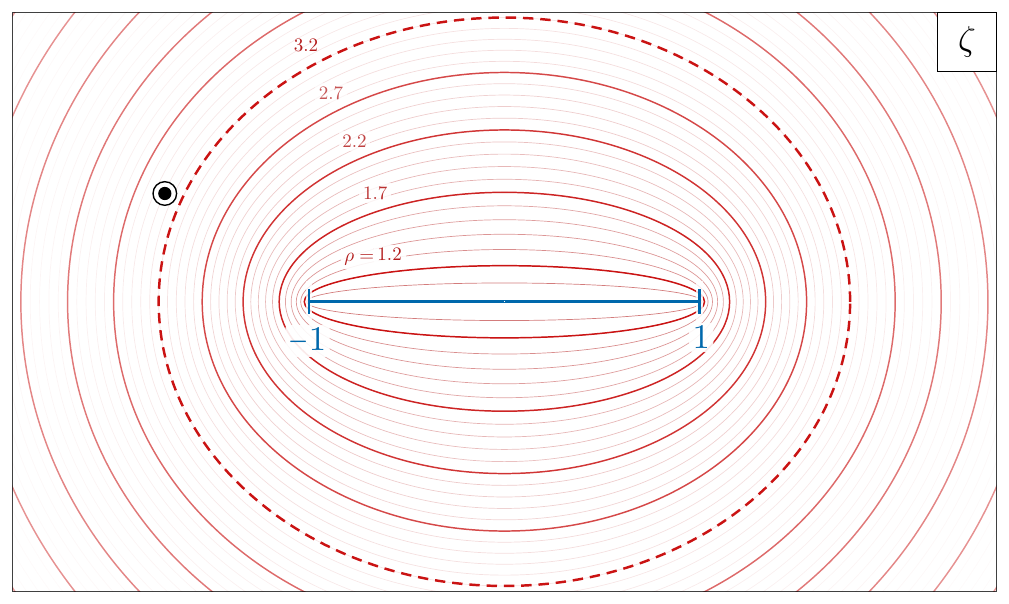}
        \caption{The interval $\zeta \in [-1, 1]$ and Bernstein ellipses of various sizes. The distance between the closest singularity of the function $f(\zeta)$ (pictured as \mySingularity) and the finite interval determines how quickly the Chebyshev expansion and interpolation converge. As illustrated here, truncation errors roughly decay as $(3.2)^{-N}$. \label{fig:bernstein}}
      \end{center} 
    \end{figure}

    Of course, we are mainly interested in the analytic structure of $f(\zeta)$, and not its expanded counterpart $F(z)$. However, we can relate the two by noting that $\zeta = (z+z^{-1})/2$ maps a circle of radius $\rho$ into an ellipse with foci at $\zeta = \pm 1$ defined by
    \begin{equation}
      \zeta(\theta) = \frac{1}{2}\left(\rho + \frac{1}{\rho}\right) \cos \theta + \frac{i}{2} \left(\rho - \frac{1}{\rho}\right) \sin \theta\,. \label{eq:bernsteinEllipse}
    \end{equation}
    This is called the Bernstein ellipse of radius $\rho$, depicted in Fig.~\ref{fig:bernstein}.  From the above logic, the size of the largest such ellipse inside which $f(\zeta)$ is analytic then determines how quickly the Chebyshev expansion (\ref{eq:chebExpApp}) converges, and thus the number of terms needed to approximate $f(\zeta)$ to a desired accuracy.

   In practice, we will use a different degree $N$ polynomial approximation, called the \emph{interpolant}
    \begin{equation}
      f_{N}(\zeta) = \sum_{k = 0}^{N} c_k T_k(\zeta)\,,
    \end{equation}
    that will inherit the convergence properties of the projection (\ref{eq:truncApprox}), yet is more convenient and numerically robust. The $N+1$ coefficients $c_k$ are defined by requiring that $f_{N}(\zeta_n) = f(\zeta_n)$ at a set of $N+1$ interpolation points $\{\zeta_n\}$. We may rewrite this interpolant in terms of the degree~$N$~\emph{Lagrange polynomials} 
    \begin{equation}
      f_{N}(\zeta) = \sum_{k = 0}^{N} f(\zeta_k) \,p_k(\zeta)\,, \label{eq:approxLagrange}
    \end{equation}
    which are defined with respect to the interpolation points as 
    \begin{equation}
      p_{k}(\zeta_n) = \delta_{nk}\,. \label{eqn:CardialPDef}
    \end{equation}
    Explicitly, the Lagrange polynomials are given by
    \begin{equation}
      p_n(\zeta) = \frac{\prod_{k \neq n} (\zeta - \zeta_k)}{\prod_{k \neq n} (\zeta_n - \zeta_k)} = \frac{p(\zeta)\, w_n}{\,(\zeta - \zeta_n)}\,, \label{eq:lagrangePolynomials}
    \end{equation}
    where we have defined both the degree $N+1$ \emph{node polynomial}
    \begin{equation}
      p(\zeta) = \prod_{k = 0}^{N} (\zeta - \zeta_n)\,
    \end{equation}
    and the \emph{weights} $w_n^{-1} = p'(\zeta_n)$.

    Whether or not this interpolant accurately approximates $f(\zeta)$ away from the points $\{\zeta_n\}$ depends crucially on how the interpolation points are distributed on the interval. One can show that the difference between the exact function and its approximation is given by \cite{Trefethen:2013ata,Boyd:2001cfp}
    \begin{equation}
      f(\zeta) - f_N(\zeta) = \frac{f^{(N+1)}(\zeta)\, p(\zeta)}{(N+1)!}\,.
    \end{equation}
    If the interpolation points are such that the maximum value of $|p(\zeta)|$ on the interval does not grow with its degree, then,  for large enough $N$, this difference is guaranteed to be small everywhere on the interval. Given that the Chebyshev polynomial $T_{N+1}(\zeta)$ is both extremely simple in its trigonometric form and is bounded on the interval by $\pm 1$ for all $N$, we will take $p(\zeta) = T_{N+1}(\zeta)/2^{N}$ and sample it at its zeros
    \begin{equation}
      \zeta_{k} = \cos \left(\frac{\pi(2 k + 1)}{2(N+1)}\right),\qquad k = 0, \dots, N\,, \label{eq:chebNodes}
    \end{equation}  
    with corresponding weights
    \begin{equation}
      w_k = \sin\left(\frac{2 \pi k(N+2) + \pi}{2(N+1)}\right), \qquad k = 0, \dots, N\,.
    \end{equation}
    The interpolant inherits the convergence properties (\ref{eq:convergenceProps}) of the truncation if we interpolate with the \emph{Chebyshev nodes} (\ref{eq:chebNodes}),\footnote{In fact, any set of interpolation points that are distributed according to the equilibrium distribution, $\rho(\zeta) = (\pi \sqrt{1 - \zeta^2})^{-1}$, will share the convergence properties of the truncation (\ref{eq:truncApprox}). See \cite{Trefethen:2013ata} for more details.} and so (\ref{eq:approxLagrange}) converges to $f(\zeta)$ at rate set by the largest Bernstein ellipse.

    In practice, evaluating the interpolant (\ref{eq:approxLagrange}) away from the interpolation points using (\ref{eq:lagrangePolynomials}) is a numerical disaster \cite{Berrut:2004bli,Trefethen:2011six,Trefethen:2013ata}. Fortunately, the \emph{second barycentric form},
    \begin{equation}
      p_n(\zeta) = \frac{\lambda_n}{\zeta - \zeta_n}\,\Bigg/\sum_{k = 0}^{N} \frac{\lambda_k}{\zeta - \zeta_k}\,,
    \end{equation}
    is numerically robust~\cite{Higham:2004num,Berrut:2004bli} and computationally efficient. This formula can be used to compute the derivative matrices $p_k'(\zeta_n)$ and $p_k''(\zeta_n)$ that appear throughout the text in a stable way. Explicitly, we take
    \begin{align}
      p^{\hskip 1pt \prime}_k(\zeta_n) &= \begin{dcases}
        \frac{w_k/w_n}{\zeta_n - \zeta_k} & n \neq k \\
        -\sum_{k\neq n} p^{\hskip 1pt \prime}_k(\zeta_n) & n = k
      \end{dcases} \, , \label{eq:cardDeriv} \\[4pt]
      p^{\hskip 1pt \prime \prime}_k(\zeta_n) &= \begin{dcases}
        2 p^{\hskip 1pt \prime}_{k}(\zeta_{n})p^{\hskip 1pt \prime}_{n}(\zeta_{n}) - \frac{2 p^{\hskip 1pt \prime}_{k}(\zeta_{n})}{\zeta_{n} - \zeta_k} & n \neq k \\
        -\sum_{k \neq n} p^{\hskip 1pt \prime \prime}_k(\zeta_n) & n = k 
      \end{dcases}\,. \label{eq:cardDDeriv}
    \end{align}
    Numerically, it is extremely important \cite{Baltensperger:2000imp} to set the diagonal elements of these differentiation matrices such that the sum of all rows vanish identically. This forces the differentiation matrices to annihilate constant functions on the interval and can have an outsized effect on accuracy.

\chapter{Gravitational Perturbations} \label{app:GP}

In this appendix, we provide further details of the gravitational perturbations introduced in  \S\ref{sec:gravlevelmix}. We first present an alternative perspective of the derivation of (\ref{eqn:h00Expansion}) by considering a \textit{three-body analogy} (Appendix~\ref{sec:threebodyanalog}). We then show how a fictitous dipole in the gravitational potential, generated by a change of coordinates, cancels out (Appendix~\ref{sec:fictitious}). In addition, we generalize the tidal moments to inclined orbits (Appendix~\ref{app:TidalMoments}) and present the next-to-leading order corrections to the gravitational potential (Appendix~\ref{app:HigherOrder}).

\section{Three-Body Analogy} \label{sec:threebodyanalog}

The eigenfunctions of the cloud are determined by the Hamiltonian of a test particle of mass~$\mu$ in the Schr\"odinger equation, cf.~(\ref{eqn:NonRelScalar}) and (\ref{eqn:NonRelVector}). This is analogous to the Hamiltonian of the electron in a hydrogen atom: while the electron wavefunction, as a solution of the Schr\"odinger equation, has a wave-like distribution around its nucleus, it is treated as a point particle at the level of the Hamiltonian.  In the rest of this appendix, we will adopt this particle picture of the Hamiltonian. 

\begin{figure}[thb]
\begin{center}
\includegraphics[scale=0.5]{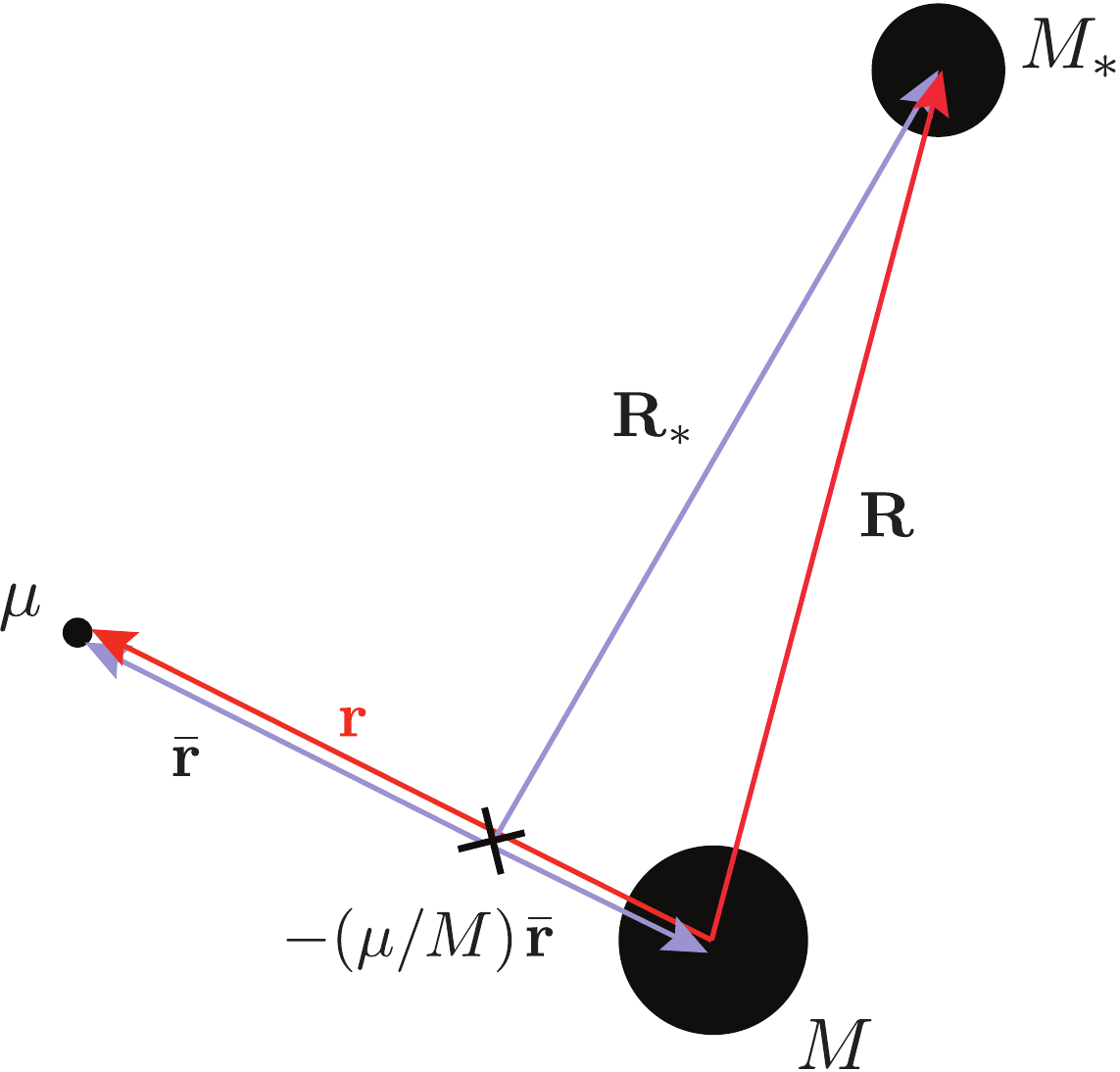}
\caption{Coordinates of the effective three-body system.} 
\label{fig:coordinates}
\end{center}
\end{figure}

In the presence of a companion of mass $M_*$, the cloud will accelerate due to the external gravitational field. The Hamiltonians in the Schr\"odinger equations~(\ref{eqn:NonRelScalar}) and (\ref{eqn:NonRelVector}) must hence be modified to include both the kinetic term of $M$ as well as the contributions from $M_*$. It is convenient to introduce the center-of-mass coordinate, $\boldsymbol{\rho} \equiv (M \r_1+\mu\r_2)/(M+\mu)$,
so that the total Hamiltonian $H_{\rm tot}$ of the three-body system can be written as 
\beq
H_{\rm tot} 
= \bigg[ \frac{\p_\rho^2}{2 (M + \mu)} + \frac{\p_r^2}{2\hat{\mu}}  + V_c(|\r|) \bigg] + \bigg[   \frac{\p_*^2}{2M_*} + V_*(\bar \r, \R_*) \bigg]\, ,
\eeq 
where $\r \equiv \r_2 - \r_1$ is the relative spatial separation between $\mu$ and $M$, and we introduced the reduced mass $\hat{\mu} \equiv M\mu/(M+\mu)$, the momenta  $\p_\rho \equiv (M+\mu) \dot{\boldsymbol{\rho}}$ and $\p_r \equiv \hat{\mu} \dot{\r}$, and the coordinates ${\bf R_*}\equiv \{R_*,\Theta_*,\Phi_*\}$ and $\bar \r \equiv \{\bar r, \bar \theta, \bar \phi\}$ relative to the center-of-mass (cf. Fig.~\ref{fig:coordinates}). Notice that we have used a different notation for the center-of-mass coordinates compared to that in the main text here. Working in the Newtonian limit, the external potential $V_*$ is given by 
\begin{align}
V_*(\bar \r, \R_*) & = - \frac{M_* M}{|\R_* + \mu \bar \r/ M  |} - \frac{M_* \mu}{|\R_* -  \bar \r  |}   \nonumber \\
& = - M_*  \sum_{\ell_* m_*} \frac{4\pi}{2\ell_* + 1} \left( \frac{M (-\mu/M)^{\ell_* }+ \mu}{R_*} \right) \left( \frac{\bar r}{R_*} \right)^{\ell_*} \, \nonumber \\
& \hskip 100pt \times Y^*_{\ell_* m_*}(\Theta_*, \Phi_*) \,Y_{\ell_* m_*}(\bar \theta, \bar \phi)   \, .  \label{eqn: Newtonian perturbation new 1}
\end{align}
Substituting $\bar \r = (1+\mu/M)^{-1}\, \r$, and expanding in $\mu/M\ll 1$, we find 
\begin{align}
 V_*(r, R_*) &= - \frac{M_* (M+\mu) }{R_*} \nonumber \\
 & - \frac{M_* \mu}{R_*}\,  \sum_{\ell_* \ge 2} \sum_{|m_*| \le \ell_*} \frac{4\pi}{2\ell_*+1} \left(\frac{r}{R_*}\right)^{\ell_*}\, Y^*_{\ell_* m_*}(\Theta_*,\Phi_*) \,Y_{\ell_* m_*}(\theta,\phi)\, , \nonumber\\[4pt]
&\equiv  V_{*,0}(R_*) + \sum_{\ell_* \ge 2} V_{*, \ell_*}(r,R_*)\, .   \label{equ:dV2} 
\end{align}
The leading monopole term $V_{*, 0}$ determines the acceleration of the center-of-mass of the cloud system.  It doesn't depend on the relative separation $\r$, so it doesn't lead to a shift in the energy levels or the mode functions of the system.  The remaining terms are a sum over harmonics, starting with the quadrupole $\ell_* =2$. Importantly, the dipole contribution $\ell_* =1$ vanishes in the center-of-mass frame.

\section{Fictitious Dipole} \label{sec:fictitious}

In the previous subappendix, we have shown that the contribution from the dipole vanishes in the centre-of-mass frame. By virtue of the equivalence principle, this property must hold for all coordinate systems. More generally, this is a manifestation of the fact that a constant gravitational gradient --- in this case the dipole term induced by the external gravitational field produced by the companion~$M_*$ --- is physically unobservable. We now show explicitly that this is indeed the case for the coordinate system centered at $M$.

Consider expressing $H_{\rm tot} $ in terms of $(\r_1, \r)$ instead of $(\boldsymbol{\rho}, \r)$,
\beq
\begin{aligned}
H_{\rm tot} & =  \left[ \frac{\p_*^2}{2M_*} + V_*(\r, \R) \right] \\
& +  \left[  \left( 1 + \frac{\mu}{M} \right) \left( \frac{\textbf{p}_1^2}{2 M} + \frac{\textbf{p}_r^2}{2 \hat{\mu}}  \right)  + \frac{\mu}{ M \hat{\mu}} \textbf{p}_1 \cdot \textbf{p}_r  +   V_c(|\r|)   \right] \, ,
\end{aligned}
\eeq
where the gravitational potential in this choice of coordinates reads (cf.~Fig.\ref{fig:coordinates})
\beq
V_*(\r, \R)  = - \frac{M_* M}{|\R|} - \frac{M_* \mu}{|\R -  \r  |}   \, .
\eeq
Expanding $V_*$ for $r \ll R$ produces a dipole term
\beq
- \frac{M_* \mu}{|\R- \textbf{r}|} \supset - \left( \frac{M_* \mu}{R}  \right) \left( \frac{\hat{\textbf{n}} \cdot \textbf{r}}{R} \right)  , \hskip 10pt  \label{eqn: gravitational dipole}
\eeq
where $\hat{\textbf{n}}$ is the unit vector along $\textbf{R}$, and we have used $|\textbf{R} - \textbf{r} | = R - \hat{\textbf{n}} \cdot \textbf{r} + \mathcal{O}(r^2)$. 
We will now show that this dipole is cancelled by the kinetic mixing term $ \mu \, \textbf{p}_1 \cdot \textbf{p}_r / M \hat{\mu} $. This is manifested most transparently in the $M_* / R \gg V_*$ limit, in which the gravitational attraction between $M$ and $\mu$ is negligible, compared to the force exerted by $M_*$.  In this limit, $M$ and $\mu$ free fall separately under the gravitational influence of $M_*$:
\beq
\dot{\textbf{r}}^2_1 = \frac{2 M_*}{R} \, , \qquad \dot{\textbf{r}}^2_2 = \frac{2 M_*}{|\textbf{R} - \textbf{r}|}  \, , \label{eqn: energy balance 1 2}
\eeq
and the angle between $\dot{\textbf{r}}_1$ and $\dot{\textbf{r}}_2$, denoted by $\gamma$, is given by
\beq
\cos \gamma = \frac{R^{ 2} + |\textbf{R} - \textbf{r} |^2 - r^2}{2 R |\textbf{R} - \textbf{r}|} \, .
\eeq
The dipole arising from the kinetic mixing becomes,
\beq
\begin{aligned}
 \frac{\mu}{ M \hat{\mu}} \textbf{p}_1 \cdot \textbf{p}_r  & = \mu (\dot{\textbf{r}}_1 \cdot \dot{\textbf{r}}_2 - \dot{\textbf{r}}^2_1  ) \\ & = \mu \left(   \frac{M_* (R^{ 2} + |\textbf{R} - \textbf{r} |^2 - r^2)}{(R |\textbf{R} - \textbf{r}|)^{3/2}  } - \frac{2 M_*}{R} \right)  \\[4pt]
& =  \mu \left[   \frac{M_*}{R} \left(2 + \frac{ \hat{\textbf{n}} \cdot \textbf{r}}{R} \right)  - \frac{2M_*}{R} + \mathcal{O}\left( \frac{r}{R} \right)^2 \right] \\[4pt]
& = + \left( \frac{M_* \mu }{R } \right)  \left( \frac{ \hat{\textbf{n}} \cdot \textbf{r}}{R}  \right) +  \mathcal{O}\left( \frac{r}{R} \right)^2 \, .
\end{aligned}
\eeq
As advertised, this precisely cancels the dipole contribution in (\ref{eqn: gravitational dipole}).

\section{Tidal Moments of Inclined Orbits} 
\label{app:TidalMoments}

We wish to describe the tidal moments $\mathcal{E}_{\ell_* m_*}$ in (\ref{eqn:h00Expansion}) for general orbits.  
In the usual angular spherical coordinates $\{\Theta_*, \Phi_*\}$ (see Fig.~\ref{fig:BinaryPlane}), they are
\beq
\mathcal{E}_{\ell_* m_*}(\Theta_*, \Phi_* ) = \frac{4\pi}{2\ell_* + 1} Y^*_{\ell_* m_*} (\Theta_* , \Phi_* ) \, , \label{eqn:QuadrupoleTidalMomentOld0}
\eeq
where have suppress all explicit dependences on time $t$.  
Although (\ref{eqn:QuadrupoleTidalMomentOld0}) provides a closed-form expression for the tidal moments for all $\ell_*$, these angular coordinates do not naturally adapt to the orbital motion of the binary. Instead, it is more convenient to express the tidal moments in terms of the inclination angle and the true anomaly $\{ \iota_*, \varphi_*\}$ described in \S\ref{sec:gravlevelmix}: 
\beq
\mathcal{E}_{\ell_* m_* }(\iota_*, \varphi_*) = \sum_{ m_\varphi=-\ell_*}^{\ell_*} \varepsilon^{(m_\varphi)}_{\ell_* m_*}(\iota_* ) \, e^{-i m_\varphi \varphi_* } \, , \label{eqn:mgammaDef}
\eeq
where the reduced tidal moments, $\varepsilon_{\ell_* m_*}^{(m_\varphi)}$, can be obtained by using the normalized binary separation vector\footnote{This expression for $\hat{\textbf{n}}$ is obtained by applying the coordinate transformation (\ref{eqn:Angles}) to its equivalent expression in the usual spherical coordinates, $\hat{\textbf{n}} = (\sin \Theta_* \cos \Phi_*, \sin \Theta_* \sin \Phi_*, \cos \Phi_*)$ .} $\hat{\textbf{n}} = (\cos \iota_* \cos \varphi_*, \sin \varphi_*, \sin \iota_* \cos \varphi_*)$ in the symmetric-trace-free tensor representation of the gravitational perturbation (see e.g.~\cite{Taylor:2008xy, Thorne:1980ru}). The sizes of these reduced moments affect the magnitudes of the the overlap (\ref{eqn:etaDef}), while the oscillatory term $e^{-i m_\varphi \varphi_* }$ are shown explicitly in (\ref{eqn:etaDef}). Interestingly, the true anomaly $\varphi_*$ only appears in the exponents of the oscillatory terms. On the other hand, the inclination angle $\iota_*$ determines the strength of the perturbation and implicitly affects the summation over $m_\varphi$ in (\ref{eqn:mgammaDef}). As we shall see, this means that $\iota_*$ also determines the polarization of the perturbation.

It is instructive to illustrate the effect of $\iota_*$ on the tidal moments through an explicit example. Let us therefore consider the quadrupole $\ell_*=2$, where the tidal moments are
\beq
\begin{aligned}
\mathcal{E}_{2 \mp  2} & = \frac{1}{2}\sqrt{\frac{6 \pi }{5}}    \left( \cos \iota_*  \cos \varphi_* \pm  i \sin \varphi_* \right)^2 \, , \\
\mathcal{E}_{2 \mp 1} & = \pm \sqrt{\frac{6 \pi }{5}}  \sin \iota_* \cos \varphi_* \left( \cos \iota_*  \cos
   \varphi_* \pm  i  \sin \varphi_* \right), \\
   \mathcal{E}_{2 0} & =- \frac{1}{4}\sqrt{\frac{\pi}{5}}  
   \left( 1 - 3 \cos 2 \varphi_*  + 6 \cos^2 \varphi_*  \cos 2 \iota_* \right) \, . \label{eqn:Equadrupole}
\end{aligned}
\eeq
For equatorial orbits, $\iota_*=0$, this become 
\beq
\begin{aligned}
\mathcal{E}_{2 m_*}(\iota_*=0) & = \frac{1}{2} \sqrt{\frac{6\pi}{5}} \left\{ e^{ + 2 i \varphi_*}, \, 0 , \, -\sqrt{2/3} , \, 0 , \, e^{-2 i \varphi_*} \right\}  , \label{eqn:ElectricEquator}
\end{aligned}
\eeq
where the list runs from $m_* \in \{ -2, \cdots \hskip -2pt, +2 \}$. Since the tidal moments vanish for $m_* = \pm 1$, states with $\Delta m_{ab} = \pm 1$ decouple from each other---see the selection rules (S1) and (V1) in~\S\ref{sec:gravlevelmix}. On the other hand, all  of the non-vanishing components in (\ref{eqn:ElectricEquator}) consists of a \textit{single} oscillatory term $\propto e^{- i m_* \varphi_*}$, with $m_\varphi = m_*$. These properties generalize to all values of $\ell_*$. In particular, for even (odd) values of $\ell_*$, the odd (even) $m_*$-th component vanishes, and each of the non-vanishing component is a single oscillatory term with $m_\varphi = m_*$. For instance, for $\ell_*=3$, we find 
\beq
\begin{aligned}
\mathcal{E}_{3 m_*}(\iota_*=0) & = \frac{1}{2} \sqrt{\frac{5\pi}{7}} \\
& \times \left\{  e^{ + 3 i \varphi_*}, \, 0 , \, - \sqrt{3/5} e^{+ i \varphi_* }  , \, 0 , \,   \sqrt{3/5} e^{- i \varphi_* }  , \, 0 , \, - e^{- 3 i \varphi_* } \right\} \, , \label{eqn:ElectricEquatorOctupole}
\end{aligned}
\eeq
where states with $\Delta m_{ab} = \pm 1$ are now connected through the weaker octupolar perturbation. This property can be seen most transparently by setting $\iota_*=0$ in (\ref{eqn:Angles}), where $\Phi_* = \varphi_*$ and the spherical harmonics in (\ref{eqn:QuadrupoleTidalMomentOld0}) obeys $Y^*_{\ell_* m_*} \propto e^{- i m_* \Phi_*} \propto e^{- i m_* \varphi_*}$. Since the tidal moments for equatorial orbits induce perturbations with definite frequency $m_* \varphi_*$, these perturbations are said to be \textit{circularly-polarized}.

The properties described above no longer hold for general inclined orbits with $\iota_* \neq 0$. For example, for $\iota_*=\pi/3$, (\ref{eqn:Equadrupole}) becomes
\beq
\begin{aligned}
\mathcal{E}_{2 \mp 2}(\iota_*=\pi/3) & = \frac{1}{32}\sqrt{\frac{6\pi}{5}} \left( 9 e^{\pm 2 i \varphi_*} + e^{\mp 2 i \varphi_*} - 6 \right)  , \\
\mathcal{E}_{2 \mp 1}(\iota_*=\pi/3) & = \pm \frac{1}{16} \sqrt{\frac{18\pi}{5}}\left( 3 e^{\pm 2 i \varphi_*} - e^{\mp 2 i \varphi_*} + 2  \right)   , \\
\mathcal{E}_{2 0}(\iota_*=\pi/3) & = \frac{1}{48}\sqrt{\frac{9\pi}{5}} \left( 9 e^{+ 2 i \varphi_*} + 9 e^{- 2 i \varphi_*} + 2 \right)  . \label{eqn:ElectricInclined}
\end{aligned} 
\eeq
Not only are all of the tidal components now non-vanishing, they are given by a superposition of $\propto e^{\pm 2 i \varphi_*}$ and constant terms. 
This feature can also be generalized to all $\ell_*$ in inclined orbits, where the multipole moments consist of superpositions of oscillatory terms $\propto e^{- i m_\varphi \varphi_*}$, with $ m_\varphi$ summed over all permissible odd (even) values for odd (even) values of $\ell_*$. In these cases, the off-diagonal terms in the Schr\"odinger frame Hamiltonian, such as (\ref{eqn:Two-State-Schr-H}), consist of a summation over different frequencies.  
Furthermore, unlike for equatorial orbits, these tidal moments now necessarily contain oscillatory terms with opposite values of $\pm | m_\varphi |$. This is because the motion of an inclined orbit, \textit{projected} onto the equatorial plane, can be represented by a superposition of co-rotating and counter-rotating orbits. Since the amplitudes of these oscillatory terms are different (except for $m_*=0$, which is azimuthal about the spin-axis of the cloud), the projected equatorial orbit still has definite orientation. Inclined orbits therefore generate \textit{elliptically-polarized} gravitational perturbations.

Finally, in the special limit $\iota_*=\pi/2$, (\ref{eqn:Equadrupole}) turns into
\beq
\begin{aligned}
& \mathcal{E}_{2 m_*}(\iota_*=\pi/2)  = \frac{1}{2} \sqrt{\frac{6\pi}{5}}\\
& \hskip 10pt \times  \left\{ - \sin ^2 \varphi_*, \,  i \sin 2\varphi_*, \,  (3 \cos 2 \varphi_* +1) / \sqrt{6}, \,   i \sin 2 \varphi_* ,\, - \sin ^2\varphi_* \right\}  . \label{eqn:ElectricOrthogonal}
\end{aligned}
\eeq
The amplitudes of both $\pm |m_\varphi|$ oscillatory terms now have the same magnitude. In this case, the orbital motion projected onto the equatorial plane is just given by an oscillating perturbation  along a straight line. By having the same magnitudes, the projected co-rotating and counter-rotating orbits cancel the net motion along the direction orthogonal to this straight line. The gravitational perturbation generated in this case is hence \textit{linearly-polarized}. Our explicit example for the $\ell_*=2$ tidal moment, from (\ref{eqn:ElectricEquator}) to (\ref{eqn:ElectricOrthogonal}), represent the continuous behavior of $\mathcal{E}_{\ell_* m_*}$ of all $\ell_*$ under a rotation about $\iota_*$.

\section{Higher-Order Couplings} \label{app:HigherOrder}

In \S\ref{sec:gravlevelmix}, we ignored the $\alpha$-corrections to the gravitational potentials (\ref{eqn:ScalarPert}) and (\ref{eqn:VectorPert}). However, these higher-order couplings can still mediate interesting transitions, especially if they mediate resonances that are forbidden by the selection rules of their leading-order counterparts. 
We will now show the expressions of these additional couplings, leaving a detailed investigation of their effects for future work.

It is useful to first estimate the order in $\alpha$ at which the $\alpha$-expansion should be truncated. As a useful guide, we require the Landau-Zener parameter (\ref{eqn:LZParam}) to be greater than order unity, such that the nature of the transitions are adiabatic. 
Denoting the higher-order overlaps by $\eta_\beta \equiv \alpha^\beta \eta$, where $\eta$ is the leading-order overlap, cf.~(\ref{eqn:etaDef}), we find that 
\beq
\frac{\eta_\beta^2}{\gamma} \simeq 2.8 \times 10^4 \frac{q}{(1+q)^{5/3}}\left(\frac{R_{ab}}{0.3} \right)^2 \left( \frac{|\Delta m_{ab}|}{2}\right)^{5/3} \left| \frac{n_a^2 n^{2}_b}{n^{ 2}_a - n_b^2  } \right|^{5/3} \left( \frac{0.07}{\alpha}\right)^5 \alpha^{2 \beta}   \, , \label{eqn:HigherOrderLZ}
\eeq
where we assumed the transitions are Bohr-like, and have used (\ref{eqn:gamma}) and (\ref{eqn:Eta}). By demanding adiabaticity in (\ref{eqn:HigherOrderLZ}), i.e.~$\eta_\beta^2/\gamma \gtrsim 1$, we find $\beta \lesssim 2$. This conclusion remains unchanged for fine and hyperfine transitions. In what follows, we will therefore expand the gravitational interactions up to order $\alpha^2$.

For the scalar cloud, the gravitational perturbation up to order $\alpha^2$ is
\beq
\begin{aligned}
V_*  = - \frac{1}{4} \mu \bar{h}^{00}  - i \bar{h}^{0i} \partial_i + \mathcal{O}(\alpha^3) \, , \label{eqn:ScalarPert4}
\end{aligned}
\eeq
where the first term is the same as that in (\ref{eqn:ScalarPert}), and the second term arises from the (trace-reversed) gravitomagnetic perturbation $\bar{h}^{0i}$. The latter metric perturbation scales as $\bar{h}^{0i} \sim v \hskip 1pt \bar{h}^{00}$, where $v$ is the typical velocity of the binary. Since $\partial_i \psi \sim \mu \alpha \psi$ and $v \lesssim \alpha$ (see Footnote~\ref{footnote:alpha-scaling}), this new term is $\alpha^2$-suppressed compared to its leading-order counterpart. On the other hand, for the vector cloud, we find that 
\beq
\begin{aligned}
V_*^{il} = &  -\frac{1}{4} \delta^{il} \mu  \bar{h}^{00}   - \frac{\bar{h}^{00}}{ 2\mu}\partial^i \partial^l   - \frac{\partial_r \bar{h}^{00}}{4\mu} {\epsilon^{i}}_{rp} {\epsilon^{l}}_{pq} \partial_q \\
& + \frac{i}{2} \partial^l \bar{h}^{0i} + \frac{i}{2} \bar{h}^{0l} \partial^i  - i  \delta^{il} \bar{h}^{0p} \partial_p   \, , \label{eqn:VSpatial4}
\end{aligned}
\eeq
where $\epsilon_{ijk}$ is the Levi-Civita symbol, and we replace all of the temporal components $\psi^0$ with the Lorenz condition  $\partial_i \psi_i + i \mu\psi_0 =\mathcal{O}(\alpha^3) $. To obtain both (\ref{eqn:ScalarPert4}) and (\ref{eqn:VSpatial4}), we utilized the fact that the gravitomagnetic perturbation is divergence-less, i.e. $\partial_i \bar{h}^{i0}=0$. We also simplified these results by ignoring higher-order corrections that contain time-derivatives, $\partial_0 \psi \sim \mu \alpha^2 \psi$, as the types of resonances that they mediate are already accessible by the leading-order potentials. This is to be contrasted with terms with spatial gradients, which change the angular structures of the field and therefore  induce resonances that obey different selection rules.

\chapter{Landau-Zener Transition}\label{app:LZ}

As discussed in detail in Section~\ref{sec:gcollider}, a key feature of the evolution of boson clouds in the inspiral of a binary are resonant Landau-Zener transitions.
In this appendix, we collect details of these transitions that are relevant to, but not necessarily suitable for, the main text. Specifically, in Appendix~\ref{app:twoState}, we first review the exact solution of the two-state Landau-Zener transition, cf.~\cite{Landau,Zener}, which provides intuition for understanding the more complicated multi-state transitions. In Appendix~\ref{app:adFloTheo}, we then describe adiabatic Floquet theory. This is necessary for generalizing the dressed frame presented in Section~\ref{sec:gcollider} to connect the dynamics of a general binary-cloud system to those of a series of LZ transitions.

\section{The Two-State Model} 
\label{app:twoState}

  The Schr\"{o}dinger equation for the two-state LZ transition presented in \S\ref{sec:twoState} can be reduced to the Weber equation of a quantum harmonic oscillator, and is thus exactly solvable. In the following, we will review this solution.

  The dressed frame coefficients $d_1(t)$ and $d_2(t)$ evolve according to the Hamiltonian (\ref{eqn:Two-State-Schr-H}): 
  \begin{equation}
     \begin{aligned}
       i \dot{d}_1 &= +\tilde{\gamma} t d_1/2 + \eta d_2\, , \\
      i \dot{d}_2 &= -\tilde{\gamma} t d_2/2 + \eta d_1\,,
     \end{aligned} \label{eq:weberCoupled}
  \end{equation}
where we defined $\tilde{\gamma} \equiv |\Delta m| \gamma$. These equations can be combined into 
the Weber equations for either coefficient,
  \begin{equation}
    \begin{aligned}
     \ddot{d}_1 + \tfrac{1}{4}\left( (\tilde{\gamma} t)^2 + 2 i \tilde{\gamma} + 4\eta^2 \right) d_1 &= 0\,, \\
\ddot{d}_2 + \tfrac{1}{4} \left((\tilde{\gamma} t)^2 - 2 i \tilde{\gamma} + 4 \eta^2 \right) d_2 &= 0\, ,
    \end{aligned}
  \end{equation}
whose solutions are the parabolic cylinder functions \cite{NIST:DLMF},
   \begin{equation}
   \begin{aligned}
     d_{1}(t) &= C_1\, \lab{D}_{-i z}\left((-1)^{1/4}\sqrt{\tilde{\gamma}} t\right) + C_2\, \lab{D}_{i z - 1}\left((-1)^{3/4}\sqrt{\tilde{\gamma}} t\right) , \\ 
      d_{2}(t) &= \frac{(-1)^{1/4} }{\sqrt{z}} C_2 \,\lab{D}_{i z}\left((-1)^{3/4} \sqrt{\tilde{\gamma}} t\right) \\
      & + (-1)^{1/4} \sqrt{z}\, C_1\, \lab{D}_{-i z-1}\left((-1)^{1/4} \sqrt{\tilde{\gamma}} t\right) ,
   \end{aligned} \label{eq:generalSolution}
  \end{equation}
   where we have defined $z \equiv \eta^2/\tilde{\gamma}$ and related the two solutions using (\ref{eq:weberCoupled}). We are ultimately interested in finding the total population contained in the second state $|d_2(t)|^2$ in the asymptotic future, assuming that the system fully occupied the first state in the asymptotic past, ${|d_{1}(-\infty)|^2 = 1}$. Imposing this initial condition, the undetermined coefficients in  (\ref{eq:generalSolution}) are (up to an arbitrary choice of phase)
   \begin{equation}
      C_1 = \exp\Big(\frac{3 \pi }{4 } z\Big) \quad \text{and} \quad C_2 = \frac{i \sqrt{2 \pi}}{\Gamma(i z )}\exp\Big(\frac{\pi }{4} z\Big)\,.
   \end{equation}
   We thus find that
   \begin{equation}
   \begin{aligned}
    |d_{1}(\infty)|^2 &= e^{-2 \pi z} \, ,\\
     |d_{2}(\infty)|^2 &= 1 - e^{-2\pi z} \,. \label{eq:twoStateResult1}
     \end{aligned}
  \end{equation}
  We see that for large values of the Landau-Zener parameter, $z= \eta^2/\tilde{\gamma} \gg 1$, 
  the ``probability'' that the system remains in the state $|1 \rangle$ is exponentially small
  and the state $|2\rangle$ becomes almost fully occupied.

  Likewise, if we impose $|d_{2}(-\infty)|^2 = 1$, then the coefficients in (\ref{eq:generalSolution}) are (up to an arbitrary phase)
  \begin{equation}
    C_1 = 0 \quad \text{and} \quad C_2 = z \exp\Big( \hskip -2pt -\frac{\pi}{4} z \Big)\, ,
  \end{equation}
and the asymptotic occupations  are the converse of (\ref{eq:twoStateResult1}):
  \begin{equation}
  \begin{aligned}
    |d_{1}(\infty)|^2 &= 1 - e^{-2 \pi z}\, , \\
    |d_{2}(\infty)|^2 &= e^{-2 \pi z}\, .
    \end{aligned}
    \label{eq:twoStateResult2}
  \end{equation}
  As discussed in \S\ref{sec:Smatrix}, the information contained in (\ref{eq:twoStateResult1}) and (\ref{eq:twoStateResult2}) can be encoded in an S-matrix, $|\psi(\infty)\rangle = S |\psi(-\infty)\rangle$. We will often only consider the modulus of the S-matrix elements
  \begin{equation}
    |S| = \begin{pmatrix} e^{-\pi z} & \displaystyle \sqrt{1 - e^{-2 \pi z}} \\ \displaystyle \sqrt{1 - e^{-2 \pi z}}  & e^{-\pi z}  \end{pmatrix},
  \end{equation}
which have well-defined asymptotic limits. 

  \section{Adiabatic Floquet Theory} 
  \label{app:adFloTheo}
  
    Throughout Section~\ref{sec:gcollider}, it was convenient to work in a dressed frame that rotated along with the orbital motion of the gravitational perturbation, making it clear that the system could evolve adiabatically. However, states are generically connected by perturbations that oscillate at multiple frequencies, especially when we consider more general inspiral configurations with eccentric and inclined orbits, cf.~Appendix~\ref{app:TidalMoments}. In that case, there are multiple dressed frames and one would have to awkwardly switch between them depending on which resonance one is interested in. Fortunately, the Floquet Hamiltonian provides a natural generalization of the dressed frame to arbitrary inspiral configurations, and we now review its construction.

    Floquet theory relies on the existence of a compact variable or phase to trivialize some part of the dynamics. Throughout, we will concentrate on the Hamiltonian studied in the main text, 
    \begin{equation}
      \mathcal{H}_{ab}(t, \varphi_*(t)) = E_a \delta_{ab} + \sum_{m_\varphi \in \mathbb{Z}} \eta_{ab}^{(m_\varphi)}(t) e^{-i m_\varphi \varphi_*(t)}\,, \label{eq:appMainHam}
    \end{equation}
    where we now emphasize its dependence on the angular variable $\varphi_*(t)$. This Hamiltonian describes generic orbits, with inclination and eccentricity. Intuitively, it would be useful to expand in the Fourier modes $e^{i k \varphi_*(t)}$, as this will trivialize the oscillatory part of the dynamics. This is how Floquet theory is typically presented---if a Hamiltonian is time-periodic, $\mathcal{H}(t) = \mathcal{H}(t + 2\pi)$, we may expand the wavefunction as $\psi(x, t) = \sum_{k \in \mathbb{Z}} \psi_k(x, t) e^{i k t}$ to derive a ``Floquet Hamiltonian'' for the modes $\psi_k(x, t)$ that does not depend on time. However, it is not clear what the Fourier modes are for the phase $\varphi_*(t)$, which can be an arbitrary non-monotonic function of time. It will thus be useful to present an alternative construction of the Floquet counterpart of (\ref{eq:appMainHam}), which is more thoroughly reviewed in \cite{Guerin:2003}.

    It will be useful to extend (\ref{eq:appMainHam}) to a family of Hamiltonians $\mathcal{H}_{ab}(t, \varphi_*(t) + \theta)$, parameterized by a new phase $\theta \in [0, 2 \pi)$, each with its own time evolution operator
    \begin{equation}
      i \es \partial_t U(t, t_0; \theta) = \mathcal{H}(t, \varphi_*(t) + \theta) U(t, t_0; \theta)\,.
    \end{equation}
    This is useful because we can then extend our original Hilbert space $\mathscr{H}$ to $\mathscr{K} = \mathscr{H} \otimes \mathscr{L}$, where $\mathscr{L} = L_2(S_1, \ud \theta/2 \pi)$ is the space of square-integrable $2 \pi$-periodic functions. This space is generated by $e^{i k \theta}$, with $k \in \mathbb{Z}$. That is, we now think of $\theta$ as an additional variable in our problem. The dressed-frame coefficients then become functions of this variable, $c_a (t) \to c_a(t, \theta)$, and inner products between two states are defined by
    \begin{equation}
      \langle \psi_1 | \psi_2 \rangle = \sum_a \int_{0}^{2 \pi}\!\frac{\ud \theta}{2 \pi} c_{1, a}^*(t, \theta) c_{2, a}(t, \theta)\,,
    \end{equation}
    where $|\psi_i\rangle \equiv \sum_a c_{i, a}(t, \theta) |a \rangle$.
    Time evolution is again generated by $U(t, t_0; \theta)$, such that 
    \begin{equation}
      c_a(t, \theta) = U_{ab}(t, t_0; \theta) c_{b}(t_0, \theta)\,,
    \end{equation}
    where the time-evolution operator has been extended to the new Hilbert space $\mathscr{K}$ by treating its dependence on $\theta$ as multiplication.

    The reason this is useful is that we may now define a phase translation operator
    \begin{equation}
      \mathcal{T}_{\phi} = \exp(\phi\, \partial_\theta)\,,
    \end{equation}    
    which acts on a state $\psi(t, \theta)$ like $\mathcal{T}_\phi \psi(t, \theta) = \psi(t, \theta + \phi)$. Crucially, this operator allows us to remove the dependence of $\varphi_*(t)$ of the Hamiltonian (\ref{eq:appMainHam}):
    \begin{equation}
      \mathcal{H}(t, \varphi_*(t) + \theta) = \mathcal{T}_{\varphi_{*}(t)} \mathcal{H}(t, \theta) \mathcal{T}_{-\varphi_{*}(t)} \, . \label{eq:dressedFrameExt}
    \end{equation}
    This allows us to rewrite the full time evolution operator in terms of the \emph{Floquet time evolution operator}
    \begin{equation}
      U(t, t_0; \theta) = \mathcal{T}_{\varphi_*(t)} U_{\mathcal{K}}(t, t_0) \mathcal{T}_{-\varphi_*(t_0)}\,,
    \end{equation}
    which evolves $i \,\partial_t U_{\mathcal{K}}(t, t_0) = \mathcal{K}(t, \theta) U_{\mathcal{K}}(t, t_0)$ according to the \emph{Floquet Hamiltonian}
   \begin{equation}
      \mathcal{K}_{ab}(t, \theta) =  \delta_{ab} \big(E_a -i \dot{\varphi}_*(t) \partial_\theta\big) + \sum_{m_\varphi \in \mathbb{Z}} \eta_{ab}^{(m_\varphi)}(t)\, e^{-i m_\varphi \theta}\,. \label{eq:floquetHam}
    \end{equation}
    The transformation (\ref{eq:dressedFrameExt}) and the Hamiltonian (\ref{eq:floquetHam}) generalize the dressed frame transformation (\ref{eq:multiDressedUnitary}) and the Hamiltonian (\ref{eq:multiStateDressedHam}), respectively. All of the ``fast'' motion due to $\varphi_*(t)$ has now been removed from (\ref{eq:floquetHam}), which evolves slowly in time.

    Given a solution of the Schr\"{o}dinger equation 
    \begin{equation}
      i \,\partial_t c_a(t, \theta) = \mathcal{K}_{ab}(t, \theta) c_{b}(t, \theta)\,,
    \end{equation}
    we can then generate a solution for the original Hamiltonian by simple substitution, $c_{a}(t) = c_{a}(t, \varphi_*(t))$. If we expand $c_{a}(t, \theta)$ into the Floquet eigenbasis $e^{i k \theta}$, 
    \begin{equation}
      c_{a}(t, \theta) = \sum_{k \in \mathbb{Z}} c_{a}^{(k)}(t) e^{i k \theta}\,,
    \end{equation}
    then it is clear that this substitution is simply the Fourier expansion in $\varphi_*(t)$ we originally wanted.

   It is particularly useful to analyze the dynamics in this Floquet basis, in which the Floquet Hamiltonian is
   \beq
   \begin{aligned}
      \mathcal{K}_{ab}^{f_1, f_2}(t) & =  \int_{0}^{2 \pi}\!\frac{\ud \theta}{2\pi} \, e^{i f_1 \theta} \mathcal{K}_{ab}(t, \theta) e^{-i f_2 \theta} \\
      & = \delta_{ab} \delta_{f_1, f_2} \left(E_a - f_2 \dot{\varphi}_*(t)\right) + \eta_{ab}^{(f_1 - f_2)}(t)\,.
   \end{aligned}
   \eeq
   Clearly, this is of the same form as the dressed Hamiltonian we considered in Section~\ref{sec:gcollider}, and the same decoupling arguments apply here. Indeed, we may consider the instantaneous (quasi)-energy eigenstates and describe the cloud's evolution throughout the entire inspiral as a series of isolated scattering events. Crucially, if the system is in a state with definite Floquet number, the translation in $\theta$ does not affect populations, $|c_a(t, \varphi_*(t))|^2 = |c_{a}(t, \theta)|^2$. We can avoid working with the Schr\"{o}dinger-frame coefficients entirely, if we are only interested in the population contained within a state.

\chapter{Backreaction on the Orbit} \label{app:AMTransfer}

Our analysis of the backreaction in \S~\ref{sec:floatdive} relied crucially on the balance of angular momentum between the orbit and the cloud. In this appendix, we provide a more detailed analysis of this balance. We first derive the cloud-orbit coupling through conservation of angular momentum, and then analyze the orbit's behavior during an adiabatic transition. We find that angular momentum conservation during the transition can cause the orbit to either float or kick.

\section{Orbital Dynamics}

  The motion of the binary companion satisfies (see e.g.~\cite{Blanchet:2013haa, Galley:2016zee})
   \begin{align}
 &     \ddot{R}_* - R_* \Omega^2  = -\frac{(1+q)M}{R_*^2}  \nonumber \\
 & \hskip 30pt + \frac{64 q (1+q) M^3}{15} \frac{\dot{R}_*}{R_*^4} + \frac{16 q M^2 }{5} \frac{\dot{R}^3_*}{R_*^3} + \frac{16 q M^2 }{5 R_*} \dot{R}_* \Omega^2\,, \hskip 10pt \label{eq:eomNewtonLaw}\\
&      R_*^2 \dot{\Omega} + 2 R_* \dot{R}_* \Omega = -\frac{24 q (1+q) M^3}{5 R_*^2} \Omega - \frac{8 q M^2}{5 R_*} \dot{R}_*^2 \Omega - \frac{8 q M^2}{5} R_* \Omega^3\,, \label{eq:eomAngMomCons}
    \end{align}
    where we have included the leading Newtonian and radiation reaction forces. Both the radial separation, $R_* = R_*(t)$, and instantaneous frequency, $\Omega = \Omega(t)$, are functions of time and, since the orbital plane does not precess, they fully describe the orbital motion. We should interpret (\ref{eq:eomAngMomCons}) as a statement of angular momentum conservation: the orbital angular momentum of the binary on the left-hand side decreases due to gravitational-wave emission on the right-hand side.

    Fortunately, most of the terms in (\ref{eq:eomNewtonLaw}) and (\ref{eq:eomAngMomCons}) can be dropped for the large, quasi-circular orbits we consider. Let us consider such an orbit of characteristic size $R_{r}$ and frequency $\Omega_{r}$. We take the frequency to be near a Bohr resonance, $\Omega_{r} \sim \mu \alpha^2$. Because the orbit is quasi-circular, we have $\Omega_{r}^2 R_{r}^3 = (1+q)M$, and so the dimensionless quantity 
    \begin{equation}
      \varpi \equiv \frac{2 q^{1/5}  R_{r} \Omega_{r}}{(1+q)^{2/5}} \propto \frac{q^{1/5} \alpha}{(1+q)^{1/15}}\,, \label{eq:alphaSuppApp}
    \end{equation}
    is suppressed $\alpha \ll 1$.  
    Because the orbit changes on the timescale of order $\Omega_r/\gamma$, where $\gamma$ was defined in~(\ref{eqn:circleRate}), it will also be useful to define the dimensionless time
    \begin{equation}
      \tau \equiv \frac{\gamma t}{\Omega_r}\,.
    \end{equation}
 Finally, we define the dimensionless radius and frequency 
    \begin{equation}
      R_*(t) = R_{r} \es r_*(\tau) \quad \text{and} \quad \Omega_*(t) = \Omega_{r} \es \omega_*(\tau)\, ,
    \end{equation}
    so that the equations of motion become 
    \begin{align}
      0 &= 1 - \omega_*^2 r_*^3 + \frac{\varpi^{10}}{50} \left( 18 r^2 r_*'' - \frac{(4 + 3 \omega_*^2 r_*^3)r_*'}{r_*^2}\right)   - \frac{27 \varpi^{20}}{1250} \frac{r_*'^{3}}{r_*} \, , \\
      0 &= \omega_*' + \frac{2 \omega_* r_*'}{r^*} + \frac{\omega_*(3 + \omega_*^2 r_*^3)}{12 r_*^4} + \frac{3 \varpi^{10}}{100}\frac{\omega_* r_*'^{2}}{r_*^3}\,,
    \end{align}
    where primes denote derivatives with respect to the time $\tau$.
    Clearly, the terms dressed by (high) powers of $\varpi$  are suppressed\footnote{This suppression is lost for extremely large mass ratios. However, for $\alpha \lesssim 0.2$, such mass ratios~$q \gtrsim 10^{5}$ are far beyond those that we consider in this thesis, and so we can safely ignore these terms.} for $\alpha \ll 1$, and we may ignore them. In that case, (\ref{eq:eomNewtonLaw}) reduces to
    \begin{equation}
      \Omega^2 R_*^3 = (1+q)M\,, \label{eq:eomNewtonLaw2}
    \end{equation}
    while (\ref{eq:eomAngMomCons}) becomes\hskip 1pt\footnote{Strictly, this is only true in the adiabatic limit, where we also ignore angular momentum lost due to gravitational-wave emission of the cloud. Fortunately, these effects are negligible on the timescales we are interested in.}
    \begin{align}
      R_*^2 \dot{\Omega} + 2 R_* \dot{R}_* \Omega &= -\frac{24 q (1+q) M^3}{5 R_*^2} \Omega  - \frac{8 q M^2}{5} R_* \Omega^3\,. \label{eq:eomAngMomCons2}
    \end{align}
Including the angular momentum of the cloud (\ref{eqn:Scloud}), the last equation receives an extra contribution 
     \begin{align}
      R_*^2 \dot{\Omega} + 2 R_* \dot{R}_* \Omega \pm \frac{(1+q)\dot{S}_\lab{c}(t)}{q M} &= -\frac{24 q (1+q) M^3}{5 R_*^2} \Omega  - \frac{8 q M^2}{5} R_* \Omega^3\,, \label{eq:eomAngMomCons2X}
    \end{align}
    where the upper (lower) sign denotes co-rotating (counter-rotating) orbits. Using (\ref{eq:eomNewtonLaw2}), the orbital angular momentum $L = q M R_r^2 \Omega_r/(1+q)$, and the cloud's angular momentum (\ref{eqn:Scloud}),  this equation can then be rewritten as
   \beq
    \begin{aligned}
     \frac{\ud \Omega}{\ud t} & = \gamma \left(\frac{\Omega}{\Omega_r}\right)^{11/3}\!\! \\
     & \pm  3 R_J \es \Omega_r \left(\frac{\Omega}{\Omega_r}\right)^{4/3} \frac{\ud}{\ud t}\left[m_1 |c_1|^2 + m_2 |c_2|^2 + \dots + m_N |c_N|^2\right]\,, \label{eq:angMomCons2}
    \end{aligned}
    \eeq
    where we have introduced $R_J$, the ratio of the cloud and orbital angular momenta (\ref{eq:ratioAngMom}).

Assuming $m_2 = \dots = m_N$ and $|c_1|^2 +  \dots + |c_N|^2 = 1$, we may write (\ref{eq:angMomCons2}) as
    \begin{equation}
  \frac{\ud \Omega}{\ud t} = \gamma \left(\frac{\Omega}{\Omega_r}\right)^{11/3} \!\! \pm 3 R_J \Delta m \Omega_r \left(\frac{\Omega}{\Omega_r}\right)^{4/3}\! \left(-\frac{\ud |c_1|^2}{\ud t}\right), \label{eq:dotOmegaInterp}
    \end{equation}
    where $\Delta m =m_2 - m_1$.  To get an intuition for the physics behind this equation, let us assume that the orbit is co-rotating, so that $+$ sign is appropriate above.  The population $|c_{1}|^2$ depletes during the transition, so its time derivative will be negative. If $\Delta m < 0 $, the cloud \emph{loses} angular momentum, and we see that the frequency will increase more slowly---the angular momentum lost by the cloud resupplies the orbital angular momentum, which depletes due to gravitational-wave emission. Conversely, if $\Delta m > 0$, the cloud \emph{gains} angular momentum during the transition, causing the orbit to speed up and shrink faster.

\section{Backreaction Time}

We would like to understand how long the transition takes when we include the cloud's backreaction. Assuming $|c_1|^2$ to be a function only of $\Omega$, we may write (\ref{eq:dotOmegaInterp}) as
    \begin{equation}
\frac{\ud \Omega}{\ud t}= \gamma \left(\frac{\Omega}{\Omega_r}\right)^{11/3} \left[1 \pm 3 R_J \Delta m \Omega_r \left(\frac{\Omega}{\Omega_r}\right)^{4/3} \frac{\ud |c_1|^2}{\ud \Omega}\right]^{-1}\,. \label{eq:omegaDotEquation}
    \end{equation}
    It will be convenient to define the effective coupling $\eta_\lab{eff}$ by
    \begin{equation}
      \left.\frac{\ud |c_1|^2}{\ud \Omega}\right|_{\Omega_r} \!\!\equiv -\frac{|\Delta m|}{4 \eta_\lab{eff}}\,, \label{eq:etaEffDef}
    \end{equation}
     such that $\eta_\lab{eff} = |\eta|$ for the two-state system (\ref{eq:dressedFrameHam}). An exact form for $\eta_\lab{eff}$ for multi-state systems is difficult to find. However, numerical experiments show that $\eta_\lab{eff}^2 \approx  \tilde{\eta}_{1 2}^2 + \tilde{\eta}_{1 3}^2 + \cdots + \tilde{\eta}_{1 N}^2$ is a good approximation, where the parameters $\tilde{\eta}_{1 a}$ are the diagonalized couplings between the initial state and the degenerate subspace, cf.~(\ref{eq:diagDressFrameHam}).

   Near $\Omega = \Omega_r$, the equation of motion~(\ref{eq:omegaDotEquation}) simplifies to
    \begin{equation}
  \frac{\ud \Omega}{\ud t} = \frac{\gamma}{1 \mp 3 R_J \Delta m |\Delta m| \, \Omega_r/(4 \eta_{\lab{eff}})} +  \mathcal{O}(\Omega - \Omega_r)\,,
    \end{equation}
which integrates to
    \begin{equation}
      \Omega(t) \approx \Omega_r + \frac{\gamma t}{1 \mp 3 R_J |\Delta m| \Delta m \,\Omega_r/(4 \eta_\lab{eff})}\, + \mathcal{O}(t^2)\,. \label{equ:C16}
    \end{equation}
    As we explained before, if the orbit is co-rotating ($-$ sign) and $\Delta m < 0$, the cloud will \emph{lose} angular momentum to the  orbit, and the instantaneous frequency will increase more slowly. If $\Delta m >0$ however, the cloud will \emph{receive} angular momentum from the orbit, and the instantaneous frequency will grow more quickly.
The state of the cloud changes rapidly when the instantaneous frequency is in the resonance band $|\Omega(t) - \Omega_r| \lesssim \eta_\lab{eff}$. The time  it takes to cross this band decomposes into two pieces, $\Delta t_\lab{tot} = \Delta t \pm \Delta t_c$, where we have defined both the time it takes the binary to cross the resonance band without including backreaction, $\Delta t$, and the time added or subtracted by the cloud, $\Delta t_c$. In terms of the parameters in (\ref{equ:C16}), we get
   \begin{equation}
		\Delta t = \frac{2 \eta_\lab{eff}}{\gamma}  \quad {\rm and} \quad 
		\Delta t_c = \frac{3 R_J |\Delta m^2 \Omega_r|}{2 \gamma}\,.
	\end{equation}
	We discuss the observable consequences of this backreaction in Sections~\ref{sec:Backreaction} and~\ref{sec:unravel}.

\backmatter
 		 \selectlanguage{english}
         
\summary{Summary / Samenvatting}{In 1915, Einstein formulated General Relativity as the law of gravity. Some of the most fascinating predictions of the theory include the existence of black holes and gravitational waves in our Universe. Remarkably, both black holes and gravitational waves are directly observed by the LIGO observatories in 2015, exactly 100 years since the inception of the theory. Over the past several years, the current network of gravitational-wave observatories have detected the coalescences of multiple binary black hole and binary neutron star systems. These detections have already transformed our understanding of astrophysics in many ways. Future detectors will achieve better sensitivities and observe over a larger range of frequencies, therefore unveiling much more about our Universe.

Another landmark achievement in physics in the last century is the development of the Standard Model of particle physics. This theory is extremely successful at predicting the outcomes of virtually all particle physics experiments, such as those conducted at the Large Hadron Collider in CERN. Having said that, it is now an established fact that the Standard Model does not offer a complete description of our Universe. For instance, cosmological and astrophysical observations have shown that most matter in the Universe is a mysterious form of dark matter, which can only be described by physics beyond the Standard Model. In this thesis, I exploit gravitational wave observations to probe this dark side of the universe. Specifically, I investigate the effect dark matter has on the gravitational waves emitted by merging black holes.

One of the most promising ways of probing dark matter with black holes is through a phenomenon called black hole superradiance. This is a process in which any ultralight dark matter bosons in our Universe would be spontaneously amplified around rotating black holes, thereby forming large boson clouds around these black holes. Because these boson clouds are large and carry significant amounts of energy, they can perturb their environments and leave important astrophysical signatures. In Chapter~\ref{sec:spectraAtom}, I presented the detailed analytic and numeric computations for various properties of these boson clouds. Since superradiance can occur for boson fields of any intrinsic spin, I described the computations for clouds that are made up of scalar and vector fields. These highly-precise results reveal all of the qualitative features of the clouds, which serve as crucial inputs for detailed studies of their observational implications.

The boson clouds are easiest to detect when they are parts of binary systems. Specifically, their presence in binary systems can significantly perturb the binaries' trajectories, thereby affecting the associated gravitational-wave emissions. The detailed dynamics of the clouds and the backreaction on the orbits were investigated in Chapters~\ref{sec:Collider} and~\ref{sec:signatures}. Interestingly, it was found that the mathematics that describe these binaries' dynamical evolution are similar to those used to describe scattering processes in particle collider physics. I also explored the types of imprints the clouds could leave on the binaries' gravitational waveforms. Furthermore, I described how these imprints are sensitive to the masses and intrinsic spins of the underlying bosons --- direct observations of these waveform signatures would therefore not only imply the presence of the bosons in Nature, but also allow us to measure the microscopic properties of these new particles. This way of probing physics beyond the Standard Model makes binary black hole systems novel dark matter detectors.

The boson clouds described above are by no means the only interesting probes of dark matter in binary systems. Other types of astrophysical objects also arise in many physics beyond the Standard Model scenarios. Like the boson clouds, the dynamics of these new objects in binary systems could also be highly non-trivial ---  their associated gravitational waveforms are therefore also rich sources of information about the putative new physics at play. In Chapter~\ref{sec:search}, I presented an analysis that quantifies the extent to which we could detect these new binary signals with existing search methods, such as template bank searches with binary black hole waveforms. I concluded by describing how new search strategies and the construction of new template waveforms would be needed in order to not miss these signals in the observational data. This motivates further advancements in current search strategies in order to fulfill our quest towards probing physics beyond the Standard Model using gravitational wave observations.

\newpage

In 1915 formuleerde Einstein de algemene relativiteitstheorie als de wet van de zwaartekracht. Enkele van de meest fascinerende voorspellingen van de theorie zijn het bestaan van zwarte gaten en zwaartekrachtsgolven in ons heelal. Opmerkelijk is dat zowel de zwarte gaten als de zwaartekrachtsgolven in 2015, precies 100 jaar na het onstaan van de theorie, rechtstreeks door de LIGO-observatoria werden waargenomen. Het huidige netwerk van observatoria voor zwaartekrachtsgolven heeft de afgelopen jaren de botsingen van meerdere binaire zwarte gaten en neutronensterren geobserveerd. Deze waarnemingen hebben ons begrip van astrofysica al op vele manieren veranderd. Toekomstige detectoren zullen gevoeliger zijn, over een groter frequentiebereik observeren en zodoende veel meer over ons universum onthullen.

Een andere mijlpaal in de natuurkunde van de vorige eeuw is de ontwikkeling van het standaardmodel voor deeltjesfysica. Deze theorie is zeer succesvol in het voorspellen van de uitkomsten van vrijwel alle deeltjesfysica-experimenten, zoals die zijn uitgevoerd met de Large Hadron Collider in CERN. Afgezien daarvan is het nu een vaststaand feit dat het standaardmodel geen volledige beschrijving van ons universum biedt. Zo hebben kosmologische en astrofysische waarnemingen aangetoond dat de meeste materie in het heelal een mysterieuze vorm van donkere materie is, die alleen kan worden beschreven door de natuurkunde buiten het standaardmodel. In dit proefschrift maak ik gebruik van zwaartekrachtsgolfwaarnemingen om deze donkere kant van het universum te onderzoeken. Mijn onderzoek richt zich met name op het effect dat donkere materie heeft op de zwaartekrachtsgolven die worden uitgezonden door samensmeltende zwarte gaten.

Een van de meest veelbelovende manieren om donkere materie met zwarte gaten te onderzoeken, is via een fenomeen dat zwarte gat superradiantie wordt genoemd. Dit is een proces waarbij alle ultralichte donkere materie bosonen in ons heelal spontaan worden versterkt rond roterende zwarte gaten, waardoor grote bosonwolken rond deze zwarte gaten worden gevormd. Omdat deze bosonwolken groot zijn en aanzienlijke hoeveelheden energie bevatten, kunnen ze hun omgeving verstoren en belangrijke astrofysische kenmerken achterlaten. In hoofdstuk~\ref{sec:spectraAtom} presenteer ik gedetailleerde analytische en numerieke berekeningen voor verschillende eigenschappen van deze bosonwolken. Aangezien superradiantie kan optreden voor bosonvelden van elke intrinsieke spin, beschrijf ik de berekeningen voor wolken die bestaan uit scalaire en vectorvelden. Deze zeer precieze resultaten onthullen alle kwalitatieve kenmerken van de wolken, wat cruciaal is voor gedetailleerd onderzoek naar hun observatie-implicaties.

De bosonwolken zijn het best te detecteren wanneer ze deel uitmaken van binaire systemen. In het bijzonder kan hun aanwezigheid in binaire systemen de trajecten van de objecten aanzienlijk verstoren, waardoor de bijbehorende emissie van zwaartekrachtsgolven wordt be{\"i}nvloed. De gedetailleerde dynamiek van de wolken en de terugreactie op de banen worden onderzocht in de hoofdstukken~\ref{sec:Collider} en~\ref{sec:signatures}. Interessant is dat de wiskunde die de dynamische evolutie van deze binaire systemen beschrijft, vergelijkbaar is met de wiskunde die wordt gebruikt om verstrooiingsprocessen in de fysica van deeltjesbotsingen te beschrijven. Ook heb ik onderzocht welke soorten kenmerken de wolken kunnen achterlaten op de zwaartekrachtsgolfvormen van de binaire systemen. Verder beschrijf ik hoe deze kenmerken gevoelig zijn voor de massa’s en intrinsieke spins van de onderliggende bosonen - directe waarnemingen van deze golfvormsignaturen zouden daarom niet alleen de aanwezigheid van de bosonen in de natuur impliceren, maar ons ook in staat stellen de microscopische eigenschappen van deze nieuwe deeltjes te meten. Deze manier om natuurkunde buiten het standaardmodel te onderzoeken, maakt binaire systemen met zwarte gaten nieuwe detectoren voor donkere materie.

De hierboven beschreven bosonwolken zijn zeker niet de enige interessante detectoren van donkere materie in binaire systemen. Andere soorten astrofysische objecten komen ook voor in veel natuurkunde buiten het standaardmodel scenario’s. Net als de bosonwolken kan de dynamiek van deze nieuwe objecten in binaire systemen ook zeer niet-triviaal zijn - hun bijbehorende zwaartekrachtsgolfvormen zijn daarom rijke bronnen van informatie over de vermeende nieuwe natuurkunde die in het spel is. In hoofdstuk~\ref{sec:search} presenteer ik een analyse die kwantificeert in hoeverre we deze nieuwe binaire sig- nalen kunnen detecteren met bestaande zoekmethoden, zoals via sjabloonbibliotheken met golfvormen van binaire zwarte gaten. Ten slotte beschrijf ik hoe nieuwe zoekstrategie{\"e}n en de constructie van nieuwe sjabloongolfvormen nodig zijn om deze signalen niet te missen in de observationele data. Dit motiveert verdere vooruitgang in de huidige zoekstrategie{\"e}n om onze zoektocht naar natuurkunde buiten het standaardmodel met behulp van zwaartekrachtsgolfwaarnemingen te vervolgen.

\textit{Vertaald door Lotte ter Haar.}}
\acknowledgements{Acknowledgements}{I am incredibly fortunate to have Daniel Baumann as my academic advisor. Daniel is an exceptional mentor who is always generous with his time and supportive of my career development.
Among his many influences on me, his teachings on the art of doing science have been most significant. From ensuring that I communicate my science effectively to emphasizing the importance of doing the right research at the right time, Daniel's guidance has been critical to my upbringing as a physicist.
All in all, he has led by example and shown me the attributes of an excellent physicist. 

\vskip 2pt

I am also very grateful to Rafael Porto, who in many ways took on an informal role as my second advisor. He had been influential in encouraging me to venture into gravitational-wave physics, and I have since learned a lot from him about the field. Rafael's excitement about physics is always palpable during our discussions. He would never hesitate to provide me the most candid of advices when I needed them, which is something that I have always appreciated.  

\vskip 2pt

I must thank my wonderful collaborator, John Stout, for his significant contributions to this thesis. Among his many strengths, John's patience, depth of thought, and Mathematica wizardry never cease to amaze me. We discussed physics from day to (mid)night, through which I have learned a lot from him. Though I occasionally regret his irreversible influence on me into compulsive daily-intake of coffee, I am glad to have known him as a great friend.

\vskip 2pt

Halfway through my PhD, Samaya Nissanke joined the university as a new faculty member. She has since become the light that brightens up the department through her compassion and openness to scientific research. Samaya's arrival has also attracted a large influx of experts in gravitational-wave physics to the department, many whom I have interacted regularly. Above all, I am grateful for her continuous support and encouragement.

\newpage

I'd also like to thank Asimina Arvanitaki and Savas Dimopoulos for sharing their insights in particle physics and inspiring me with their scientific creativity. Their insistences on me to provide numbers and do quick checks on the observability of an idea have certainly had an impact on the way I do science. 

\vskip 2pt

To my collaborators: Christoffel Doorman, Thomas Edwards, Tanja Hinderer, and Lotte ter Haar -- thank you all for the enjoyable times we have had doing science together. I have learned a lot from all of you.

\vskip 2pt

For stimulating physics discussions, I'd also like to thank Mustafa Amin, Masha Baryakhtar, Gianfranco Bertone, Diego Blas, Beatrice Bonga, Katy Clough, David Cyncynates, Liang Dai, Sam Dolan, Will East, Marcos Garcia, Tudor Giurgic$\check{a}$-Tiron, Daniel Green, Junwu Huang, Badri Krishnan, Robert Lasenby, Eugene Lim, Krista Lynne, David Nichols, Roberto Oliveri, Frans Pretorius, Bob Wagoner, Dan Wilkins, Helvi Witek, Huan Yang, Matias Zaldarriaga, Jun Zhang, and Aaron Zimmerman. 

\vskip 2pt

Many other people have made my time in Amsterdam a great pleasure. To the cosmology group members: Matteo Biagetti, Carlos Duaso Pueyo,  Garrett Goon, Austin Joyce, Hayden Lee, Guilherme Pimentel, and  Benjamin Wallisch; thank you all for the great friendships. Group lunches and speakers' dinners will always be memorable times of my PhD. I'd also like to thank my fellow graduate students, to name a few: Gr\'egoire Mathys, Antonio Rotundo, Ka Wa Tsang, and Dong-Gang Wang, for making my graduate experience far more enjoyable.

\vskip 2pt

My graduate experience had been greatly enriched by my extended visits to the Henri Poincar\'e Institute, Perimeter Institute, and the Stanford Institute for Theoretical Physics. I thank these institutes for their kind hospitality and for providing intellectually-stimulating environments. Special thanks to King's College London for always offering me a desk during my visits to London.  

\vskip 2pt

As an engineering freshman eight years ago, I could not had foreseen myself switching to physics, let alone pursuing a doctorate in this field. For this I am grateful to Andrew Jardine and Matthew Wingate for supporting my transitions to the Cavendish Laboratory and DAMTP at Cambridge. %

\vskip 2pt

My family has played a pivotal role in my personal growth and development. A career in academic research is uncommon in our culture, yet they have shown unwavering support in my choices. To my mother, Audrey Chin, thank you for your unconditional love and for being my pillar of support. It is hard to overstate how important you have been to my upbringing. To my siblings, Shin Miin Chia and Shin Li Chia, thank you for all the memorable times we have spent together. 

\newpage

To my fianc\'ee, Zhao Feng Ooi, thank you for your love and emotional support throughout these years. Your infectious cheerfulness and optimism are my eternal source of joy. You have made me a better person. I look forward to our adventure ahead and spending the rest of my life with you.

\vskip 2pt

More than anyone else, I am indebted to my late father, Ming Chew Chia, without whom none of the opportunities I have had would be possible. His perseverance against all odds has always been my source of inspiration. I cannot thank him enough for the love and care he had provided to our family.

}

\newpage
\phantomsection
\bibliographystyle{utphys}
\bibliography{thesis}
\end{document}